\makeatletter
\def\arxivroot{}
\IfFileExists{arxiv/imsart.cls}{\def\arxivroot{arxiv/}}{}
\edef\input@path{{\arxivroot}}
\makeatother

\documentclass[aos,preprint]{imsart}
\usepackage{tabularx,array,algorithm,algorithmicx,algpseudocode,makecell}
\usepackage[utf8]{inputenc}
\usepackage{textgreek}

\usepackage{fix-cm}
\usepackage{mathrsfs}
\RequirePackage{amsthm,amsmath,amsfonts,amssymb,amscd}
\RequirePackage{mathtools}
\RequirePackage{mathrsfs}
\RequirePackage{bm}
\RequirePackage{xcolor}
\RequirePackage{graphicx}
\RequirePackage{adjustbox}
\RequirePackage{booktabs}
\RequirePackage{multirow}
\RequirePackage{enumitem}
\RequirePackage[caption=false]{subfig}
\RequirePackage{tikz-cd}
\RequirePackage[numbers,sort&compress]{natbib}
\RequirePackage[colorlinks=true,citecolor=blue,urlcolor=blue,linkcolor=blue]{hyperref}
\usepackage{fix-cm}
\usepackage{booktabs,longtable}
\usepackage{placeins}
\usepackage{comment}
\startlocaldefs

\newif\ifdraft
\draftfalse
\ifdraft
  \newcommand{\todo}[1]{{\color{red}#1}}
  
\else
  \newcommand{\todo}[1]{}
  
\fi

\theoremstyle{plain}
\newtheorem{theorem}{Theorem}[section]

\newtheorem{lemma}[theorem]{Lemma}
\newtheorem{corollary}[theorem]{Corollary}
\newtheorem{proposition}[theorem]{Proposition}

\theoremstyle{definition}
\newtheorem{definition}[theorem]{Definition}
\newtheorem{condition}[theorem]{Condition}

\newtheorem*{defn*}{Definition}
\newtheorem*{example*}{Example}

\theoremstyle{definition}
\newtheorem{remark}[theorem]{Remark}

\DeclareMathOperator*{\argmin}{arg\,min}

\newcommand{\q}{\mathsf{q}}

\newcommand{\wh}{\widehat}

\def\1v{\mathbf{1}}
\def\0v{\mathbf{0}}

\newcommand{\fre}{Fr\'{e}chet}

\endlocaldefs

\graphicspath{{\arxivroot}}
\let\arxivbibliography\bibliography
\renewcommand{\bibliography}[1]{\arxivbibliography{\arxivroot#1}}
\let\arxivbibliographystyle\bibliographystyle
\renewcommand{\bibliographystyle}[1]{\arxivbibliographystyle{\arxivroot#1}}
\makeatletter
\def\nocontentsline{%
  \let\@@addcontentsline\addcontentsline
  \def\addcontentsline##1##2##3{\let\addcontentsline\@@addcontentsline}%
}
\makeatother

\makeatletter
\def\user@hy@title{Complex-Valued One-Factor Model and its Application to ENDOR Spectroscopy}
\makeatother
\begin{document}
\addtocontents{toc}{\protect\setcounter{tocdepth}{-1}}

\begin{frontmatter}

\title{Complex-Valued One-Factor Model and its Application to ENDOR Spectroscopy}
\runtitle{Complex-Valued One-Factor Model}

\makeatletter
\def\author@num{3}
\makeatother

\begin{aug}
\author[A]{\fnms{Jaesung}~\snm{Park}}
\author[B]{\fnms{Sungkyu}~\snm{Jung}}
\author[C]{
\fnms{Yvo}~\snm{Pokern}}
\address[A]{Institute for Data Innovation in Science, Seoul National University, Seoul, Republic of Korea}

\address[B]{Department of Statistics,
Seoul National University, Seoul, Republic of Korea}

\address[C]{Department of Statistical Science,
University College London, London, United Kingdom}
\end{aug}

\begin{abstract}
Estimating a common signal from repeated complex-valued measurements requires separating amplitude and phase variation from measurement error. We propose a \texttt{plug-in} procedure for a complex-valued one-factor model with unknown anisotropic noise. The procedure alternates loading and noise covariance estimation without specifying a parametric distribution for the factor scores. We establish strong consistency, stochastic expansions, and central limit theorems under suitable conditions. Explicit asymptotic covariance formulas identify when using directional differences in noise precision improves efficiency over covariance-based estimation. These results also provide a practical inference procedure for electron--nuclear double resonance spectroscopy: pointwise spectrum confidence intervals account for noise covariance estimation and phase selection without bootstrap resampling. In simulations, the proposed intervals generally achieve near-nominal coverage at substantially lower computational cost than the profile-likelihood bootstrap procedures considered, which show undercoverage at low signal-to-noise ratios. Analyses of two published datasets illustrate improvements in spectrum estimation and show how precision gains depend on the noise and factor-score structure.
\end{abstract}



\end{frontmatter}

\maketitle	

\section{Introduction}\label{sec:Intro}

Repeated complex-valued measurements often contain a common signal whose level, amplitude, and phase vary across observations. Recovering this signal requires separating these variations from measurement error. In electron–nuclear double resonance (ENDOR) spectroscopy, phase drift between repeated measurements can cause signal cancellation under simple averaging. Uncertainty estimates are also needed to interpret weak or broad spectral features. \citet{pokern2021statistical} introduced a complex drift model that accounts for observation-specific levels and rotation-scaling factors, supporting spectrum estimation and uncertainty quantification.

We study a complex-valued one-factor model with a common complex loading vector, complex observation-specific levels, complex factor scores, and measurement error. In this model, the complex factor scores represent amplitude and phase variation, while unequal variances and correlation between the real and imaginary error components allow the measurement error standard deviation to vary across directions. We estimate both the loading vector and the measurement error covariance under this model without specifying a parametric distribution for the factor scores.

The central task is to estimate the signal with minimal error and to quantify the uncertainty of this error. We show that a natural loading estimator based on the leading eigenvector of the empirical Hermitian second moment and projected residual moments (Section~\ref{subsec:cov-estimator}), which we refer to as the \texttt{cov}-estimator, is consistent but does not account for measurement error anisotropy and hence often yields sub-optimal estimation error. With a fixed observation dimension, joint profile-likelihood estimation of the loading and noise covariance matrix faces a Neyman--Scott-type incidental parameter problem~\cite{neyman1948consistent} and is inconsistent for the loading class and measurement error covariance jointly (Theorem~\ref{thm: PL strong inconsistency}). We therefore combine method-of-moments estimation with profile-likelihood loading updates. The \texttt{cov} estimator provides consistent initial estimates, which we follow up with an iterative \texttt{plug-in} procedure that alternates between updating the loading using the estimated measurement error covariance, giving greater weight to low-variance directions, and recomputing the covariance from residual moments (Section~\ref{subsec:plugin-estimator}).

Under mild conditions, we establish strong consistency and central limit theorems for the \texttt{cov} and \texttt{plug-in} estimators, with explicit formulas for their asymptotic covariance matrices. These formulas show how relative efficiency depends on the model geometry (Theorem~\ref{thm:trace-comparison}). When the same loading is observed at different phases, its components have different orientations relative to the dominant direction of the measurement error. The \texttt{plug-in} method accounts for these differences in precision by first estimating the noise covariance matrix and using it to update the loading estimates. The stochastic expansions stabilize after the first \texttt{plug-in} update: further iterations leave the loading expansion unchanged through first order and the noise covariance expansion unchanged through second order, for any fixed number of iterations.

For ENDOR analysis, we recover a real-valued spectrum using the maximum method, which selects the phase maximizing the energy of the real part of the loading (Section~\ref{sec: maximum method}). We establish strong consistency and asymptotic normality of the estimated spectrum, with explicit covariance formulas accounting for measurement error covariance estimation and phase selection (Section~\ref{subsec:endor-maximum-asymptotics}). These formulas yield pointwise standard errors and confidence intervals without bootstrap resampling. To account for auxiliary parameters that enter the cost function and are estimated from the same data, we also develop a generalized \fre~mean framework with random cost functions.

Simulations evaluate loading and spectrum estimation accuracy, confidence interval coverage probability, and computational cost across a range of model settings. The results support the asymptotic approximations and show that \texttt{plug-in} generally achieves near-nominal coverage at substantially lower computational cost than the bootstrap procedures considered. Bootstrap intervals based on joint profile likelihood show significant under-coverage at low signal-to-noise ratios, even with bootstrap bias correction as employed in \cite{pokern2021statistical}.

We also analyze two published ENDOR datasets~\cite{pokern2021statistical,PNASdata}. In one dataset, the \texttt{plug-in} procedure produces a visibly less noisy spectrum than both simple averaging and the \texttt{cov} estimator, together with substantially shorter confidence intervals than those provided by the \texttt{cov} estimator. In the other dataset, \texttt{cov} and \texttt{plug-in} estimators give similar estimates and confidence interval lengths. These differences agree with our efficiency analysis and illustrate how our theoretical results support confidence interval construction and help explain differences between methods.

The paper is organized as follows. Section~\ref{sec:model} introduces the model and discusses parameter identifiability. Section~\ref{sec:estimators} presents the proposed estimators and relates them to existing methods. Section~\ref{sec:Theory} studies their asymptotic properties, including the stabilization of stochastic expansions under \texttt{plug-in} iterations, and compares their relative efficiency. Section~\ref{sec:endor-application} develops spectrum estimation and confidence intervals for ENDOR data and evaluates the methods through simulations and real-data analyses. Results for generalized \fre~means with random cost functions appear in Appendix~\ref{sec: GFM}.

\section{Model}\label{sec:model}
This section introduces the complex-valued one-factor model, starting with its uncentered formulation and identifying constraints. We then remove the observation-specific level by orthogonal centering and describe the remaining phase indeterminacy through a complex projective parameter space. 

\subsection{A complex-valued one-factor model}
\label{subsec:complex-one-factor-model}

Let \(n\) denote the number of observations. We initially record
each complex-valued observation at \(p+1\) variables. Throughout our asymptotic analysis, \(p\ge 3\) is held fixed as \(n\to\infty\). For \(i=1,\ldots,n\) and \(j=0,\ldots,p\), consider
\begin{equation}
\label{model:uncentered-complex-factor}
Y_{i,j}
=
\psi_i
+
\phi_i\kappa_j^{(0)}
+
\varepsilon_{i,j}.
\end{equation}
Here, \(\psi_i,\phi_i\in\mathbb C\) are the unobserved
observation-specific level and factor score, respectively,
\(\kappa^{(0)}=(\kappa_0^{(0)},\ldots,\kappa_p^{(0)})^\top
\in\mathbb C^{p+1}\) is the common complex-valued loading vector,
and \(\varepsilon_{i,j}\) is a complex-valued error. Writing \(Y_i=(Y_{i,0},\ldots,Y_{i,p})^\top \in \mathbb{C}^{p+1}\), \(\varepsilon_i=(\varepsilon_{i,0},\ldots,\varepsilon_{i,p})^\top \in \mathbb{C}^{p+1}\), and \(\mathbf{1}_{p+1}\in \mathbb{C}^{p+1}\) for the vector of ones, the model is equivalently \(Y_i=\psi_i\mathbf{1}_{p+1}+\phi_i\kappa^{(0)}+\varepsilon_i\).

For a complex number \(c=a+ib\in\mathbb C\) and a complex vector
\(z=(z_1,\ldots,z_p)^\top\in\mathbb C^{p}\), denote
\(
\operatorname{vec}(c)
:=
( \operatorname{Re}(c), \operatorname{Im}(c) )^\top
\in\mathbb R^2\),
\(
\operatorname{vec}(z)
:=
\begin{pmatrix}
\operatorname{vec}(z_1)^{\top}&
\cdots & 
\operatorname{vec}(z_p)^{\top}
\end{pmatrix}^{\top}
\in\mathbb R^{2p}.
\)
We assume that \(\varepsilon_{i,j}\) are independent and
identically distributed across \((i,j)\) as
\(\mathcal{CN}(0,\Sigma^{(0)})\), by which we mean
\(\operatorname{vec}(\varepsilon_{i,j})\sim N_2(0,\Sigma^{(0)})\).
Here, \(\Sigma^{(0)}\in\mathrm{SPD}(2)\), where
\(\mathrm{SPD}(2)\) denotes the set of \(2\times2\) real symmetric positive-definite matrices. We do not assume isotropic noise; that is, \(\Sigma^{(0)}\) need not be proportional to \(\mathrm{Id}_2\). We also assume that \((\phi_i)_{i\geq1}\) are independent and identically
distributed and independent of all errors, with
\(0<\mathbb E[|\phi_1|^2]<\infty\). No specific distributional assumptions for \(\phi_i\) and \(\psi_i\) are imposed. 

Our primary interest is estimating \(\kappa^{(0)}\), and we also estimate \(\Sigma^{(0)}\) from the observed data \(Y_1,\ldots,Y_n\). To specify these estimation targets, we first address the unidentifiability in model~\eqref{model:uncentered-complex-factor}. Indeed, for any \(c\in\mathbb C\), we have
\[
\psi_i\mathbf{1}_{p+1}
+
\phi_i\kappa^{(0)}
=
(\psi_i-c\phi_i)\mathbf{1}_{p+1}
+
\phi_i
\bigl(\kappa^{(0)}+c\mathbf{1}_{p+1}\bigr).
\]
We therefore require
\((\mathbf{1}_{p+1})^\top\kappa^{(0)}
=0\), which separates \(\kappa^{(0)}\) from \(\psi_{i}\)s. There is also a scale indeterminacy, since
\(\phi_i\kappa^{(0)}=(c^{-1}\phi_i)(c\kappa^{(0)})\)
for any \(c\in\mathbb C\setminus\{0\}\).
We fix the magnitude by imposing \(\|\kappa^{(0)}\|=1\),
leaving a unit-modulus phase ambiguity. Here, $\|\cdot\|$ denotes the standard norm of a complex vector. We formalize the
corresponding equivalence classes after the following Helmert transformation removes the $\psi_{i}$s and reduces the loading to \(p\) complex coordinates.

\subsection{Centering via the Helmert transformation}\label{subsec:eliminating-observation-level}
We center each observation by subtracting its within-observation average, which estimates the observation-specific level \(\psi_i\).
Define \(\overline Y_{i,\cdot}:=(p+1)^{-1}\sum_{j=0}^pY_{i,j}\) and
\(\overline\varepsilon_{i,\cdot}:=(p+1)^{-1}
\sum_{j=0}^p\varepsilon_{i,j}\).
By the centering constraint \(\mathbf{1}_{p+1}^{\top}\kappa^{(0)}=0\), we have \(\overline Y_{i,\cdot}=\psi_i+\overline\varepsilon_{i,\cdot}\) and hence
\begin{equation}
\label{eq:ordinary-row-centering}
Y_i
-
\overline Y_{i,\cdot}\mathbf{1}_{p+1}
=
\phi_i\kappa^{(0)}
+
\varepsilon_i
-
\overline\varepsilon_{i,\cdot}\mathbf{1}_{p+1}.
\end{equation}
The error coordinates on the right-hand side of \eqref{eq:ordinary-row-centering} are correlated. We therefore express the centered observations in an orthonormal basis of the
\(p\)-dimensional centered subspace, yielding \(p\) coordinates with independent errors as follows.

Let \(H\in\mathbb R^{p\times(p+1)}\) be a Helmert submatrix \cite[Section~2.5]{DrydenMardia2016}, whose rows form an
orthonormal basis of \((\operatorname{span}_{\mathbb R}\{\mathbf{1}_{p+1}\})^\perp\) in \(\mathbb R^{p+1}\). Thus, \(H\) satisfies
\begin{equation}
\label{eq:Helmert-identities}
H\mathbf{1}_{p+1}=0,
\quad
HH^\top=\mathrm{Id}_p,
\quad
H^\top H
=
\mathrm{Id}_{p+1}
-
\frac{1}{p+1}
\mathbf{1}_{p+1}\mathbf{1}_{p+1}^\top.
\end{equation}
Setting \(\widetilde Y_i:=HY_i\),
\(\widetilde\kappa^{(0)}:=H\kappa^{(0)}\), and
\(\widetilde\varepsilon_i:=H\varepsilon_i\) yields \(\widetilde Y_i=\phi_i\widetilde\kappa^{(0)}
+\widetilde\varepsilon_i\). Moreover, applying \(H\) directly is equivalent to
first performing the ordinary centering in
\eqref{eq:ordinary-row-centering}, since
\(H(Y_i-\overline Y_{i,\cdot}\mathbf{1}_{p+1})=HY_i\). The centering constraint \(\mathbf{1}_{p+1}^{\top}\kappa^{(0)}=0\) and~\eqref{eq:Helmert-identities} imply
\(\|\widetilde\kappa^{(0)}\|=\|\kappa^{(0)}\|=1\) and
\(H^\top\widetilde\kappa^{(0)}=\kappa^{(0)}\).
Thus, an estimate of the transformed loading can be mapped back
by \(H^\top\) to estimate the original loading. Moreover, the transformation also preserves the independent Gaussian error structure: the coordinates
\(\widetilde\varepsilon_{i,1},\ldots,\widetilde\varepsilon_{i,p}\)
remain independent and identically distributed as \(\mathcal{CN}(0,\Sigma^{(0)})\).
An explicit choice of the Helmert matrix and a verification of these
properties are provided in~\ref{supp:sec: app model} of the Supplementary Material.

With a slight abuse of notation, we drop the tildes and work throughout
the remainder of the paper with the centered complex-valued one-factor model
\begin{equation}
\label{model:centered-complex-factor}
Y_i
=
\phi_i\kappa^{(0)}
+
\varepsilon_i,
\quad
i=1,\ldots,n,
\end{equation}
where \(Y_i=(Y_{i,1},\ldots,Y_{i,p})^\top\in\mathbb C^p\),
\(\|\kappa^{(0)}\|=1\), and \(\varepsilon_{i,j}\stackrel{\mathrm{i.i.d.}}{\sim} \mathcal{CN}(0,\Sigma^{(0)})\). Accordingly, the observed \(n\times(p+1)\) data matrix is first transformed by the Helmert matrix, and its \(n\times p\) representation is treated as the data in model~\eqref{model:centered-complex-factor}.

\subsection{Identifiability and complex projective space}\label{subsec:complex-projective-space}

Under model~\eqref{model:centered-complex-factor}, the unit-norm constraint removes the scale indeterminacy but leaves a phase
indeterminacy. Indeed, for every \(\lambda\in\mathbb R\), we have
\(\phi_i\kappa^{(0)} =(e^{-i\lambda}\phi_i)(e^{i\lambda}\kappa^{(0)})\). Since the distribution of \(\phi_i\) is left unspecified, this re-parameterization generates the same family of distributions for the observations. Consequently, no particular phase of \(\kappa^{(0)}\) is identifiable from the model. Hence, the natural parameter space is the complex projective space, defined as follows.

For \(z\in\mathbb C^p\), let \(z^*\) denote its conjugate transpose and \(\|z\|=(z^*z)^{1/2}\). On \(\mathcal S^{2p-1}:=\{\kappa\in\mathbb C^p:\|\kappa\|=1\}\), define \(\kappa\sim\kappa'\) if and only if \(\kappa'=e^{i\lambda}\kappa\) for some \(\lambda\in\mathbb R\). The quotient is \(\mathbb{CP}^{p-1}:=\mathcal S^{2p-1}/\!\sim\), with \([\kappa]:=\{e^{i\lambda}\kappa:\lambda\in\mathbb R\}\). Thus, our primary target estimand in model~\eqref{model:centered-complex-factor} is
\([\kappa^{(0)}]\in\mathbb{CP}^{p-1}\), rather than a particular representative \(\kappa^{(0)} \in \mathcal{S}^{2p-1}\). The space \(\mathbb{CP}^{p-1}\) is a compact complex manifold of complex dimension \(p-1\), and hence a
compact smooth manifold of real dimension \(2p-2\)
\citep[Example~1.9]{lee2024introduction}. 

For comparison, in a real-valued one-factor model, unit normalization of the loading vector leaves only the sign ambiguity \(\kappa\sim-\kappa\), and the loading direction lies in \(\mathbb{RP}^{p-1}\). In \eqref{model:centered-complex-factor}, this two-point ambiguity is replaced by the continuous phase action \(\kappa\mapsto e^{i\lambda}\kappa\), leading to \(\mathbb{CP}^{p-1}\).

We equip \(\mathbb{CP}^{p-1}\) with the phase-aligned Euclidean distance
\begin{equation}
\label{eq:projective-distance}
d\bigl([\kappa],[\widetilde\kappa]\bigr)
:=
\min_{\lambda\in\mathbb R}
\bigl\|\kappa-e^{i\lambda}\widetilde\kappa\bigr\|
=
\sqrt{2-2|\kappa^*\widetilde\kappa|}.
\end{equation}
This distance is independent of the chosen representatives and satisfies
\(d=2\sin(d_{\mathrm{FS}}/2)\), where \(d_{\mathrm{FS}}([\kappa],[\widetilde\kappa]) =\arccos|\kappa^*\widetilde\kappa|\) is the Fubini--Study geodesic distance \citep[Definition~2.31]{bonifant2020group}. We say that \(\kappa\) and \(\widetilde\kappa\) are in \emph{optimal position} if \(d([\kappa],[\widetilde\kappa])=\|\kappa-\widetilde\kappa\|\), or, equivalently, if \(\widetilde\kappa^*\kappa\) is real and nonnegative. When \(\widetilde\kappa^*\kappa\neq0\), the unique representative of \([\widetilde\kappa]\) in optimal position with \(\kappa\) is
\(\widetilde\kappa_{\mathrm{opt}}  =(\widetilde\kappa^*\kappa/|\widetilde\kappa^*\kappa|) \widetilde\kappa\). When \(\widetilde\kappa^*\kappa=0\), every representative of \([\widetilde\kappa]\) is in optimal position with \(\kappa\). See~\ref{supp:sec: technical details for estimators} of the Supplementary Material for further details.

\section{Estimation}
\label{sec:estimators}
This section introduces the two estimation procedures studied in this paper. We first define a moment-based estimator of the loading class $[\kappa^{(0)}]$ and the noise covariance $\Sigma^{(0)}$. We then introduce an iterated plug-in estimator that uses the covariance-based estimator as its initial step and estimates the loading and noise covariance matrix alternately.

\subsection{Covariance-based estimator}
\label{subsec:cov-estimator}

We construct method-of-moments estimators of \([\kappa^{(0)}]\) and \(\Sigma^{(0)}\) from the empirical Hermitian and transpose second-moment matrices
\begin{equation}\label{eq:sample-covariance-matrices}
\widehat\Sigma^{\texttt{Her}}
:=
\frac{1}{n}\sum_{i=1}^{n}Y_iY_i^*
\in\mathbb C^{p\times p},
\quad
\widehat\Sigma^{\texttt{Tra}}
:=
\frac{1}{n}\sum_{i=1}^{n}Y_iY_i^\top
\in\mathbb C^{p\times p}.
\end{equation}
Let \(\Sigma^{\texttt{Her}}:=\mathbb E[Y_1Y_1^*]\) and
\(\Sigma^{\texttt{Tra}}:=\mathbb E[Y_1Y_1^\top]\) denote
the corresponding population second-moment matrices. Since the noise coordinates \(\varepsilon_{1,j}\) are i.i.d. across \(j\) and independent of \(\phi_1\), we have
\begin{equation}
\label{eq:population-complex-moments}
\Sigma^{\texttt{Her}}
=
\mathbb E[|\phi_1|^2]\,
\kappa^{(0)}(\kappa^{(0)})^*
+
\tau^2\mathrm{Id}_p,
\quad
\Sigma^{\texttt{Tra}}
=
\mathbb E[\phi_1^2]\,
\kappa^{(0)}(\kappa^{(0)})^\top
+
\gamma\mathrm{Id}_p,
\end{equation}
where
\begin{equation}
\label{eq:tau-gamma}
\tau^2=(\Sigma^{(0)})_{11}+(\Sigma^{(0)})_{22},
\quad
\gamma=(\Sigma^{(0)})_{11}-(\Sigma^{(0)})_{22}+2i(\Sigma^{(0)})_{12}.
\end{equation}

The matrix $\Sigma^{\texttt{Her}}$ in \eqref{eq:population-complex-moments} has a rank-one spiked structure:
its leading eigenspace is
\(\operatorname{span}_{\mathbb C}\{\kappa^{(0)}\}\) with eigengap \(\mathbb E[|\phi_1|^2]>0\). Accordingly, let
\(\widehat\kappa^{\texttt{cov}}\) be any (unit-norm) leading eigenvector of
\(\widehat\Sigma^{\texttt{Her}}\), and define
\(\widehat{[\kappa]}^{\texttt{cov}}
:=
[\widehat\kappa^{\texttt{cov}}]
\in\mathbb{CP}^{p-1}\).

We estimate \(\Sigma^{(0)}\) from the components of the two empirical moments orthogonal to the loading and its complex conjugate. Throughout this paper, we assume that \(\kappa^{(0)}\) and \(\overline{\kappa^{(0)}}\) are linearly independent over \(\mathbb{C}\), or equivalently \(\operatorname{Re}(\kappa^{(0)})\) and \(\operatorname{Im}(\kappa^{(0)})\) are linearly independent over \(\mathbb{R}\). This condition is invariant under the choice of representative of \([\kappa^{(0)}]\). If this condition fails, a phase reparameterization makes \(\kappa^{(0)}\) real. Analogues of the estimators proposed in this Section can be constructed for this case; see Remark~\ref{rmk:real loading kappa}. We do not pursue this case here.

In the ENDOR application, failure of this condition means that
the spectrum and wave components are proportional and thus encode only a single frequency profile, making the desired spectrum indistinguishable from the undesirable wave-like artifact. The assumption is therefore natural for the application considered here. 

For any
\(\kappa\in\mathcal S^{2p-1}\) such that \(\operatorname{Re}(\kappa)\) and \(\operatorname{Im}(\kappa)\) are linearly independent over $\mathbb{R}$, define
\begin{equation}\label{eq:orthogonal-residual-projector}
\Pi_\perp(\kappa)
:=
\mathrm{Id}_p
-
\operatorname{proj}_{\operatorname{span}_{\mathbb C}
\{\kappa,\overline{\kappa}\}},
\end{equation}
where \(\operatorname{proj}\) denotes the orthogonal projection with
respect to the standard inner product on \(\mathbb C^p\). Thus,
\(
\Pi_\perp(\kappa)\kappa
=
\Pi_\perp(\kappa)\overline{\kappa}
=
0\),
\(
\operatorname{rank}_{\mathbb C}
\{\Pi_\perp(\kappa)\}
=
p-2
\). Since the projected subspace is invariant under complex conjugation,
\(\Pi_\perp(\kappa)\) has real entries. Viewed as a real-linear
operator on \( \mathbb R^{2p}\cong \mathbb C^p \), it has rank \(2p-4\).

For \(p\geq 3\), we define 
\begin{equation*}
\widehat\tau^2(\kappa)
:=
\frac{1}{p-2}
\operatorname{Tr}\left\{
\Pi_\perp(\kappa)\widehat\Sigma^{\texttt{Her}}
\right\},
\quad
\widehat\gamma(\kappa)
:=
\frac{1}{p-2}
\operatorname{Tr}\left\{
\Pi_\perp(\kappa)\widehat\Sigma^{\texttt{Tra}}
\right\}.
\end{equation*}
Using the correspondence in \eqref{eq:tau-gamma}, define the moment based estimator of $\Sigma^{(0)}$ as
\begin{equation}\label{eq:moment-covariance-map}
\mathcal M(\kappa)
:=
\frac{1}{2}
\begin{pmatrix}
\widehat\tau^2(\kappa)
+\operatorname{Re}\{\widehat\gamma(\kappa)\}
&
\operatorname{Im}\{\widehat\gamma(\kappa)\}
\\[0.3em]
\operatorname{Im}\{\widehat\gamma(\kappa)\}
&
\widehat\tau^2(\kappa)
-\operatorname{Re}\{\widehat\gamma(\kappa)\}
\end{pmatrix}.
\end{equation}
The complex span of
\(\kappa\) and \(\bar{\kappa}\)
is invariant under \(\kappa\mapsto e^{i\lambda}\kappa\). Hence
\(\Pi_\perp\), \(\widehat\tau^2\), \(\widehat\gamma\), and
\(\mathcal M\) are phase invariant, so we can write \(\mathcal M([\kappa])\) without ambiguity. We define
\(\widehat\Sigma^{\texttt{cov}}
:=
\mathcal M(\widehat{[\kappa]}^{\texttt{cov}})\) and refer to
\((\widehat{[\kappa]}^{\texttt{cov}},
\widehat\Sigma^{\texttt{cov}})\) as the \texttt{cov}-estimator. 
Theorem~\ref{thm:cov-estimator-strong-consistency} establishes the strong consistency and well-definedness of the \texttt{cov}-estimator.

\begin{remark}[Choice of residual projection]
\label{rmk:alternative-projection}
The alternative projector \(P_\perp(\kappa):=\mathrm{Id}_p-\kappa\kappa^*\) has complex rank \(p-1\), compared with \(p-2\) for \(\Pi_\perp(\kappa)\), and thus retains one additional complex residual direction. We nevertheless use \(\Pi_\perp(\kappa)\) for two reasons. First, \(\Pi_\perp(\kappa)\) preserves the Gram-matrix structure of \(\mathcal M(\kappa)\), ensuring positive semidefiniteness for every sample. Under model~\eqref{model:centered-complex-factor}, if \(n\geq2p\), it is almost surely positive definite simultaneously for all \(\kappa\) with linearly independent real and imaginary parts. The corresponding \(P_\perp\)-based moment estimator can be indefinite, especially when $\Sigma^{(0)}$ is highly anisotropic.

Second, for a \(\sqrt n\)-consistent loading estimator, loading estimation error enters the covariance estimator expansion only at second order when using \(\Pi_\perp\), whereas \(P_\perp\) generally introduces a first-order contribution. This property supports stabilizing the iterated \texttt{plug-in} expansions. Detailed theoretical and numerical comparisons are given in~\ref{supp:sec:projected-covariance-definiteness} and~\ref{sec:triangle-projector-comparison} of the Supplementary Material.
\end{remark}

\subsection{Iterated plug-in estimator}
\label{subsec:plugin-estimator}

Although \(\widehat{[\kappa]}^{\texttt{cov}}\) is strongly consistent, it uses only \(\widehat{\Sigma}^{\texttt{Her}}\) and therefore does not exploit the Gaussian error structure for the anisotropic $\Sigma^{(0)}$. To incorporate this information, we construct a profile-likelihood loading update using \(\widehat{\Sigma}^{\texttt{cov}}\).

We treat the unobserved factor scores
\(\phi_1,\ldots,\phi_n\) as nuisance parameters. Under model~\eqref{model:centered-complex-factor}, the conditional log-likelihood for \((\phi_{1}, \ldots, \phi_{n},\kappa,\Sigma)\) is, up to an additive
constant,
\begin{equation}
\label{eq:loglik-full}
l_{Y_1,\ldots,Y_n}(\phi_{1}, \ldots, \phi_{n},\kappa,\Sigma)
=
-\frac{np}{2}\log\det(\Sigma)
-
\frac{1}{2}\sum_{i=1}^{n}
\|Y_i-\phi_i\kappa\|_\Sigma^2.
\end{equation}
Here,
\(\|z\|_\Sigma^2
:=\sum_{j=1}^p
\operatorname{vec}(z_j)^\top\Sigma^{-1}\operatorname{vec}(z_j)\)
for \(z\in\mathbb C^p\) and \(\Sigma\in\mathrm{SPD}(2)\). Since the \(\phi_i\)s are unobserved, a natural approach is to profile them out. For any \(y\in\mathbb C^p\), \(\kappa\in\mathcal S^{2p-1}\), and \(\Sigma\in\mathrm{SPD}(2)\), define
\(\widehat\phi(y,\kappa,\Sigma)
:=
\operatorname*{argmin}_{\phi\in\mathbb C}
\|y-\phi\kappa\|_\Sigma^2\).
The corresponding single-observation profile loss is
\(
\rho(y,[\kappa],\Sigma)
:= \min_{\phi\in \mathbb{C}}\|y-\phi\kappa\|_\Sigma^2=
\|
y-
\widehat\phi(y,\kappa,\Sigma)\kappa
\|_\Sigma^2.
\)
Since \(\phi\) ranges over \(\mathbb C\), any phase change
in \(\kappa\) can be absorbed into \(\phi\), leaving the minimum unchanged. Thus, \(\rho\) is well-defined on \(\mathbb{C}^{p} \times \mathbb{CP}^{p-1}\times \mathrm{SPD}(2)\). An explicit formula for the unique minimizer
\(\widehat\phi(y,\kappa,\Sigma)\) is given in~\ref{supp:sec: technical details for estimators} of the Supplementary Material.

The negative profile log-likelihood, scaled by \(2/n\), is
\begin{equation}
\label{eq:profile-loglik}
-\frac{2}{n}
l_{Y_1,\ldots,Y_n}^{\mathrm{PL}}([\kappa],\Sigma)
=
p\log\det(\Sigma)
+
\frac{1}{n}\sum_{i=1}^n
\rho(Y_i,[\kappa],\Sigma).
\end{equation}
With \(p\) fixed, jointly minimizing \eqref{eq:profile-loglik} over \(\mathbb{CP}^{p-1}\times\mathrm{SPD}(2)\) leads to a Neyman--Scott-type incidental parameter problem: the number of profiled scores grows with \(n\), whereas each score is estimated from a single \(p\)-dimensional observation. This can prevent consistent estimation of the common parameter pair \(([\kappa],\Sigma)\). In the classic example of \cite{neyman1948consistent}, the maximum likelihood estimator of the common variance converges to a value strictly below the true variance since the number of observations per individual mean remains fixed.
Under the assumptions of Theorem~\ref{thm: PL strong inconsistency} below, the joint profile estimator is inconsistent for the true parameter pair.

We therefore replace the joint minimization of \eqref{eq:profile-loglik} by successive moment-based covariance and profile-loading updates. The \texttt{cov}-estimator initializes the procedure. For
the first step, define
\begin{equation}
\label{eq:iterated-plugin-initialization}
\begin{aligned}
\widehat{[\kappa]}^{[1]}
\in
\operatorname*{argmin}_{[\kappa]\in\mathbb{CP}^{p-1}}
\frac{1}{n}\sum_{i=1}^n
\rho(
Y_i,[\kappa],\widehat\Sigma^{\texttt{cov}}
), \quad \widehat\Sigma^{[1]}
:=
\mathcal M(
\widehat{[\kappa]}^{[1]}).
\end{aligned}
\end{equation}
For each integer \(s\geq2\), define recursively
\begin{equation}
\label{eq:iterated-plugin-update}
\begin{aligned}
\widehat{[\kappa]}^{[s]}
\in
\operatorname*{argmin}_{[\kappa]\in\mathbb{CP}^{p-1}}
\frac{1}{n}\sum_{i=1}^n
\rho(
Y_i,[\kappa],\widehat\Sigma^{[s-1]}
), \quad 
\widehat\Sigma^{[s]}:=
\mathcal M(
\widehat{[\kappa]}^{[s]}
).
\end{aligned}
\end{equation}
All updates are computed from the same sample \(Y_1,\ldots,Y_n\). For \(s\geq1\), we call the pair
\( (\widehat{[\kappa]}^{[s]},
\widehat\Sigma^{[s]}
)\) the $s$-step \texttt{plug-in} estimator. Throughout the asymptotic analysis, \(s\) is held fixed as \(n\to\infty\).

With \(\Sigma\) fixed at \(\widehat\Sigma^{[s-1]}\)
(or \(\widehat\Sigma^{\texttt{cov}}\) for \(s=1\)), we compute the loading estimate \(\widehat{[\kappa]}^{[s]}\) by alternately optimizing \eqref{eq:loglik-full} over
\(\phi_{i}\)s and \([\kappa]\). The update formulas and implementation details are given in~\ref{supp:subsec:fixed-covariance-loading-algorithm} of the Supplementary Material.

Our simulations show negligible gains from further loading--covariance updates, so we use \(s=1\) unless stated otherwise. Section~\ref{sec:Theory} provides a theoretical justification for this choice.

\begin{remark}
\label{rmk:real loading kappa}
If \(\kappa^{(0)}\) is real up to a global phase, i.e., $\operatorname{Re}(\kappa^{(0)})$ and $\operatorname{Im}(\kappa^{(0)})$ are linearly dependent, analogous estimators can be constructed by restricting the loading to real unit vectors. The initial loading estimate is a leading eigenvector of \(\operatorname{Re}(\widehat\Sigma^{\texttt{Her}})\).
An iterated plug-in procedure can then alternate residual covariance estimation based on \(\mathrm{Id}_p-\kappa\kappa^\top\), with normalization \(p-1\), and weighted least-squares loading updates using the current
covariance estimate. We leave the analysis of these estimators outside the scope of the present paper.
\end{remark}

\noindent\textit{Real-valued representation and relation to previous literature.} For \(\kappa\in\mathbb C^p\), define the real loading matrix \(M(\kappa):=\begin{pmatrix}\operatorname{vec}(\kappa) & \operatorname{vec}(i\kappa)\end{pmatrix}\in\mathbb R^{2p\times2}\). Then model~\eqref{model:centered-complex-factor} has the real-valued representation
\begin{equation}\label{eq:real-factor-representation}
\operatorname{vec}(Y_i)=M(\kappa^{(0)})\operatorname{vec}(\phi_i)+\operatorname{vec}(\varepsilon_i), \quad \operatorname{vec}(\varepsilon_i)\sim N_{2p}\!\left(0,\mathrm{Id}_p\otimes\Sigma^{(0)}\right).
\end{equation}
This is a real-valued two-factor model with specific constraints on both the loadings and the error covariance. When \(\Sigma^{(0)}\) is diagonal, the full error covariance is diagonal, as in classical factor analysis \citep{anderson1956statistical}.

If \(\operatorname{vec}(\phi_i)\) is Gaussian, joint estimation of the loading class and error covariance can be formulated through Gaussian marginal likelihood under the stated loading and covariance constraints \citep[cf.][]{joreskog1970general}. If, in addition, \(\Sigma^{(0)}=\sigma^2\mathrm{Id}_2\), the real-valued model is a constrained special case of probabilistic principal component analysis with at most two real factors \citep{tipping1999probabilistic}.

With known \(\Sigma^{(0)}\), profiling out the scores gives a weighted least-squares fit over \(\mathbb{CP}^{p-1}\) without specifying their distribution \citep{wiechers2023drift}, providing a structured counterpart of the real-valued weighted fit of \citet{wentzell1997maximum}. Under isotropic noise, this profile fit coincides with the \texttt{cov} estimate. In the zero-mean proper complex Gaussian one-factor model with isotropic noise considered by \citet{stoica2009maximum}, the same line is the maximum likelihood estimate of the signal subspace.

With unknown, possibly anisotropic noise covariance and an unspecified score distribution, fitting the model second moments to their sample counterparts under the stated constraints provides another route to joint estimation \citep[Section~3]{shapiro2007statistical}. A related iterative approach in \citet{narasimhan2008model} alternates real subspace fitting with estimating a structured noise covariance from constraint residuals. Those real-valued formulations address loading indeterminacy by identifying restrictions or estimating the subspace, but do not specialize to complex-line loadings in \(\mathbb{CP}^{p-1}\). Starting from \texttt{cov}, our \texttt{plug-in} alternates between weighted least-squares fitting over \(\mathbb{CP}^{p-1}\) and updating the common \(2\times2\) error covariance from the second moments of residuals projected by \(\Pi_\perp(\kappa)\).

\section{Theory}\label{sec:Theory}
We establish strong consistency, stochastic expansions, and central limit theorems for the \texttt{cov} and iterated \texttt{plug-in} estimators. Across all fixed \(s\geq1\), the \texttt{plug-in} loading iterates share a first-order expansion, and the covariance iterates share an expansion through second order. We then compare total asymptotic loading variances to determine when \texttt{plug-in} has an efficiency gain compared to \texttt{cov}.

\subsection{Asymptotic analysis of the \texttt{cov}-estimator}
\label{subsec:asymptotic-analysis-cov-estimator}

We first establish the strong consistency of the \texttt{cov} estimator from Section~\ref{subsec:cov-estimator}. The proof uses the almost sure convergence \(\widehat\Sigma^{\texttt{Her}}\to\Sigma^{\texttt{Her}}\), the positive eigengap of \(\Sigma^{\texttt{Her}}\), the Davis--Kahan-type theorem for complex Hermitian matrices in Section~\ref{supp:sec:daviskahan-proof} of the Supplementary Material, and the continuity of the moment map \(\mathcal{M}\).

\begin{theorem}
\label{thm:cov-estimator-strong-consistency}
Almost surely, the pair
\((\widehat{[\kappa]}^{\texttt{cov}},\widehat\Sigma^{\texttt{cov}})\)
is uniquely well-defined for all sufficiently large \(n\), and
\(
d(
\widehat{[\kappa]}^{\texttt{cov}},
[\kappa^{(0)}]
)\to 0\),
\(\|
\widehat\Sigma^{\texttt{cov}}-\Sigma^{(0)}
\|_F\to 0
\) as \(n\to\infty\).
\end{theorem}

To study the loading estimator in Euclidean coordinates, we choose the optimally aligned representative
\(
\widehat\kappa^{\texttt{cov,opt}}\in \widehat{[\kappa]}^{\texttt{cov}}
\). Theorem~\ref{thm:cov-estimator-strong-consistency} ensures that this representative is almost surely unique for all sufficiently large \(n\), since \(\bigl|(\widehat\kappa^{\texttt{cov}})^*\kappa^{(0)}\bigr| \to1\) almost surely.

For the expansion below, define the empirical moment fluctuations \(E_H:=\widehat\Sigma^{\texttt{Her}}-\Sigma^{\texttt{Her}}\), \(E_T:=\widehat\Sigma^{\texttt{Tra}}-\Sigma^{\texttt{Tra}}\), and \(P_{\perp,0}:=\mathrm{Id}-\kappa^{(0)}{ \kappa^{(0)}}^{*} \). We use the grouped real representation \(\widetilde{\operatorname{vec}}(z):=(\operatorname{Re}(z)^\top,\operatorname{Im}(z)^\top)^\top\in\mathbb R^{2p}\), a fixed permutation of \(\operatorname{vec}(z)\).

Perturbing the leading eigen-equation of \(\widehat\Sigma^{\texttt{Her}}\) around its population counterpart gives the stochastic expansion below. Asymptotic normality immediately follows from the multivariate central limit theorem.

\begin{theorem}
\label{thm:cov-loading-expansion-clt}
Suppose \(\mathbb E[|\phi_1|^4]<\infty\).
Then \(\widehat\kappa^{\texttt{cov,opt}}\) admits the expansion
\begin{equation}
\label{eq:cov-loading-stochastic-expansion}
\widehat\kappa^{\texttt{cov,opt}}
=
\kappa^{(0)}
+
L_{\kappa,n}^{\texttt{cov}}
+
\mathcal O_p(n^{-1}),
\end{equation}
where
\begin{equation}
\label{eq:cov-loading-first-term}
L_{\kappa,n}^{\texttt{cov}}
:=
\frac{1}{\mathbb E[|\phi_1|^2]}
P_{\perp,0}E_H\kappa^{(0)}
=
\mathcal O_p(n^{-1/2}).
\end{equation}
Consequently,
\begin{equation}
\label{eq:cov-loading-clt}
\sqrt n
\left\{
\widetilde{\operatorname{vec}}
(\widehat\kappa^{\texttt{cov,opt}})
-
\widetilde{\operatorname{vec}}(\kappa^{(0)})
\right\}
\xrightarrow{d}
N_{2p}(0,V^{\texttt{cov}}),
\end{equation}
with
\begin{equation}
\label{eq:cov-loading-asymptotic-covariance}
V^{\texttt{cov}}
:=\operatorname{Var}\!\left[
\widetilde{\operatorname{vec}}
\left(P_{\perp,0}Y_1Y_1^*\kappa^{(0)}\right)
\right]
/\{\mathbb E[|\phi_1|^2]\}^2
.
\end{equation}
\end{theorem}
An explicit formula for \(V^{\texttt{cov}}\) is provided in~\ref{supp:subsec:explicit-cov-loading-asymptotic-covariance} of the Supplementary Material. Since \((\kappa^{(0)})^*L_{\kappa,n}^{\texttt{cov}}=0\), the first-order fluctuation lies in the horizontal space representing \(T_{[\kappa^{(0)}]}\mathbb{CP}^{p-1}\). Under the model assumptions, \(V^{\texttt{cov}}\) has rank \(2p-2\) and is therefore singular in the ambient space \(\mathbb R^{2p}\).

Applying Theorem~\ref{thm:cov-loading-expansion-clt} to \eqref{eq:moment-covariance-map} gives a second-order expansion of \(\widehat\Sigma^{\texttt{cov}}\). We write \(\Pi_{\perp,0}:=\Pi_\perp(\kappa^{(0)})\), \(\mathcal U(z):=[\operatorname{Re}(z),\operatorname{Im}(z)]\in\mathbb R^{p\times2}\) for \(z\in \mathbb{C}^{p}\), \(U_0:=\mathcal U(\kappa^{(0)})\), and
\(
(\widehat\tau^{2,\texttt{cov}},\widehat\gamma^{\texttt{cov}})
:=
\bigl(
\widehat\tau^2(\widehat{[\kappa]}^{\texttt{cov}}),
\widehat\gamma(\widehat{[\kappa]}^{\texttt{cov}})
\bigr)
\).

\begin{theorem}\label{thm:cov-noise-covariance-expansion}
Suppose \(\mathbb E[|\phi_1|^4]<\infty\). Then \(\widehat\Sigma^{\texttt{cov}}\) admits the second-order
stochastic expansions
\begin{align}
\widehat\tau^{2,\texttt{cov}}-\tau^2
&=
L_{\tau,n}^{\texttt{cov}}
+
Q_{\tau,n}^{\texttt{cov}}
+
\mathcal O_p(n^{-3/2}),
\label{eq:cov-tau-second-order-expansion}
\\
\widehat\gamma^{\texttt{cov}}-\gamma
&=
L_{\gamma,n}^{\texttt{cov}}
+
Q_{\gamma,n}^{\texttt{cov}}
+
\mathcal O_p(n^{-3/2}).
\label{eq:cov-gamma-second-order-expansion}
\end{align}
The first-order terms, both \(\mathcal O_p(n^{-1/2})\), are
\begin{equation}
\label{eq:cov-noise-first-order-terms}
L_{\tau,n}^{\texttt{cov}}
:=\frac{\operatorname{Tr}(\Pi_{\perp,0}E_H)}{p-2},
\quad
L_{\gamma,n}^{\texttt{cov}}
:=\frac{\operatorname{Tr}(\Pi_{\perp,0}E_T)}{p-2}.
\end{equation}
The second-order terms \(Q_{\tau,n}^{\texttt{cov}}, Q_{\gamma,n}^{\texttt{cov}}\) are both \(\mathcal O_p(n^{-1})\).
\end{theorem}

The centered first-order terms determine the asymptotic distributions, while the expectations of the second-order terms give the leading \(n^{-1}\) bias under suitable additional integrability conditions. Explicit expressions for the second-order terms are given in~\ref{supp:subsec:second-order-noise-covariance-expansion} of the Supplementary Material.

\subsection{Asymptotic analysis of the plug-in estimators}
\label{subsec:asymptotic-analysis-plugin-estimators}
We now study the asymptotic behavior of the iterated \texttt{plug-in} estimators introduced in Section~\ref{subsec:plugin-estimator}. At iteration \(s\), \(\widehat{[\kappa]}^{[s]}\) is an
\(M\)-estimator based on a cost function parameterized by an auxiliary covariance estimated from the same sample: \(\widehat\Sigma^{\texttt{cov}}\) for \(s=1\) and
\(\widehat\Sigma^{[s-1]}\) for \(s\geq2\). This estimate enters every summand, making the cost function itself random and dependent on the observations. To accommodate this dependence, Appendix~\ref{sec: GFM} develops a generalized \fre~mean framework with random cost functions and establishes consistency and stochastic expansions. Starting from the strongly consistent \texttt{cov} estimator, we apply these results recursively, together with continuity of~\eqref{eq:moment-covariance-map}.

\begin{theorem}
\label{thm:iterated-plugin-strong-consistency}
For any fixed integer \(s\geq1\), almost surely, the pair
\((\widehat{[\kappa]}^{[s]},\widehat\Sigma^{[s]})\)
is well-defined for all sufficiently large \(n\), and \(
d(
\widehat{[\kappa]}^{[s]},
[\kappa^{(0)}]
)\to 0\),
\(\|
\widehat\Sigma^{[s]}-\Sigma^{(0)}
\|_F\to 0
\) as \(n\to\infty\).
\end{theorem}

As before, we use the optimally aligned representative \(\widehat\kappa^{[s],\texttt{opt}}\in\widehat{[\kappa]}^{[s]}\), with respect to \(\kappa^{(0)}\). For each fixed \(s\geq1\), strong consistency ensures that the optimally aligned representative within the selected class \(\widehat{[\kappa]}^{[s]}\) is almost surely unique for all sufficiently large \(n\).

To state the loading expansion and central limit theorem, we first introduce the derivatives of the cost function and the influence function. We use the smooth extension \(\widetilde\rho\) of \(\rho\), defined by
\(\widetilde\rho(y,\kappa,\Sigma):=\rho\bigl(y,[\kappa/\|\kappa\|],\Sigma\bigr)\)
for \(\kappa\in\mathbb C^p\setminus\{0\}\). The derivative operators \(\nabla_\kappa\), \(\nabla_{\kappa\kappa}^2\), and \(\nabla_{\kappa\Sigma}^2\) are applied to \(\widetilde\rho\) using \(\widetilde{\operatorname{vec}}(\kappa)\) for loading coordinates and \(\operatorname{vech}(\Sigma)\) for covariance coordinates, with \(\operatorname{vech}\begin{pmatrix}a&b\\b&c\end{pmatrix}=(a,b,c)^\top\).
Under the assumption \(\mathbb{E}[|\phi_{1}|^{4}] < \infty\), Theorem~\ref{thm:cov-noise-covariance-expansion} gives
\begin{equation}
\label{eq:covariance-influence-representation}
\operatorname{vech}(\widehat\Sigma^{\texttt{cov}}-\Sigma^{(0)})
=
\frac1n\sum_{i=1}^n\xi_\Sigma(Y_i)
+\mathcal O_p(n^{-1}),
\end{equation}
where, for \(y\in\mathbb C^p\),
\begin{equation}
\label{eq:covariance-influence-function}
\xi_\Sigma(y)
:=
\frac12
\bigl(
\ell_\tau(y)+\operatorname{Re}\ell_\gamma(y),
\operatorname{Im}\ell_\gamma(y),
\ell_\tau(y)-\operatorname{Re}\ell_\gamma(y)
\bigr)^\top,
\end{equation}
with
\[
\ell_\tau(y)
:=
\frac{1}{p-2}
\operatorname{Tr}\left[
\Pi_{\perp,0}
\left\{
yy^*-\Sigma^{\texttt{Her}}
\right\}
\right],
\quad
\ell_\gamma(y)
:=
\frac{1}{p-2}
\operatorname{Tr}\left[
\Pi_{\perp,0}
\left\{
yy^\top-\Sigma^{\texttt{Tra}}
\right\}
\right].
\]
The next two theorems are proved jointly by induction over the iterations. At \(s=1\), we apply Theorem~\ref{thm: clt_gfm} in Appendix~\ref{sec: GFM}, using \eqref{eq:covariance-influence-representation} as the auxiliary expansion, to obtain the loading expansion. Applying \eqref{eq:moment-covariance-map} to the resulting loading estimate gives the second-order covariance expansion, with the same first-order influence function \(\xi_\Sigma\). This covariance expansion supplies the auxiliary input for the next application of the theorem, and the argument repeats for any fixed number of iterations.

\begin{theorem}
\label{thm:iterated-plugin-loading-expansion-clt}
Assume \(0<\mathbb E[|\phi_1|^4]<\infty\), and fix an integer
\(s\geq1\). Then
\begin{equation}
\label{eq:iterated-plugin-common-expansion}
\widehat\kappa^{[s],\texttt{opt}}
=
\kappa^{(0)}
+
L_{\kappa,n}^{\texttt{plug-in}}
+
\mathcal O_p(n^{-1}),
\end{equation}
where \(L_{\kappa,n}^{\texttt{plug-in}}\in\mathbb C^p\) is \(\mathcal O_p(n^{-1/2})\) and is defined through the effective score \(\psi_\kappa^{\texttt{plug-in}}:\mathbb C^p\to\mathbb R^{2p}\) by
\begin{equation}
\label{eq:plugin-common-loading-first-order-term}
\begin{aligned}
\widetilde{\operatorname{vec}}
\left(L_{\kappa,n}^{\texttt{plug-in}}\right)
&:=
-J^+\frac{1}{n}\sum_{i=1}^n
\psi_\kappa^{\texttt{plug-in}}(Y_i),
\\
\psi_\kappa^{\texttt{plug-in}}(y)
&:=
\nabla_\kappa
\rho(y,[\kappa^{(0)}],\Sigma^{(0)})
+
C_0\xi_\Sigma(y).
\end{aligned}
\end{equation}
Here,
\begin{equation}
\label{eq:plugin-J-C-definition}
\begin{aligned}
J:=
\mathbb E\left[
\nabla_{\kappa\kappa}^2
\rho(Y_1,[\kappa^{(0)}],\Sigma^{(0)})
\right],\quad 
C_0:=
\mathbb E\left[
\nabla_{\kappa\Sigma}^2
\rho(Y_1,[\kappa^{(0)}],\Sigma^{(0)})
\right],
\end{aligned}
\end{equation}
and \(J^+\) denotes the Moore--Penrose inverse of \(J\). The effective score is centered and square integrable under the finite fourth-moment assumption. Hence,
\begin{equation}
\label{eq:iterated-plugin-common-clt}
\sqrt n
\left\{
\widetilde{\operatorname{vec}}
\left(
\widehat\kappa^{[s],\texttt{opt}}
\right)
-
\widetilde{\operatorname{vec}}
\left(
\kappa^{(0)}
\right)
\right\}
\xrightarrow{d}
N_{2p}
\left(
0,V^{\texttt{plug-in}}
\right),
\end{equation}
where
\begin{equation}
\label{eq:plugin-asymptotic-covariance}
V^{\texttt{plug-in}}
:=
J^+
\operatorname{Var}
\left[
\psi_\kappa^{\texttt{plug-in}}(Y_1)
\right]
J^+.
\end{equation}
\end{theorem}
The effective score \(\psi_\kappa^{\texttt{plug-in}}\) combines the profile-loss score under the true covariance with the contribution from preliminary covariance estimation. Here, \(C_0\) measures the local sensitivity of the loading score to covariance perturbations, while \(-J^+\) maps the combined score to the first-order loading perturbation. The matrix \(J\) vanishes in the radial and phase directions and is positive definite on the horizontal space, where \(J^+\) acts as its inverse.

As in the \texttt{cov} case, \(L_{\kappa,n}^{\texttt{plug-in}}\) lies in the horizontal space representing \(T_{[\kappa^{(0)}]}\mathbb{CP}^{p-1}\). Under the model assumptions, \(V^{\texttt{plug-in}}\) has rank \(2p-2\) and is singular in \(\mathbb R^{2p}\). An explicit formula for \(V^{\texttt{plug-in}}\) is given in~\ref{supp:sec:asymptotic-efficiency-comparison} of the Supplementary Material.

We next consider the noise covariance estimator. To state this result, write \((\widehat\tau^{2,[s]},\widehat\gamma^{[s]}):=\bigl(\widehat\tau^2(\widehat{[\kappa]}^{[s]}),\widehat\gamma(\widehat{[\kappa]}^{[s]})\bigr)\).
\begin{theorem}\label{thm:iterated-plugin-noise-covariance-expansion}
Assume \(0<\mathbb E[|\phi_1|^4]<\infty\), and fix an integer \(s\geq1\). Then
\begin{align}
\widehat\tau^{2,[s]}-\tau^2
&=
L_{\tau,n}^{\texttt{cov}}
+
Q_{\tau,n}^{\texttt{plug-in}}
+
\mathcal O_p(n^{-3/2}),
\label{eq:iterated-plugin-tau-second-order-expansion}
\\
\widehat\gamma^{[s]}-\gamma
&=
L_{\gamma,n}^{\texttt{cov}}
+
Q_{\gamma,n}^{\texttt{plug-in}}
+
\mathcal O_p(n^{-3/2}).
\label{eq:iterated-plugin-gamma-second-order-expansion}
\end{align}
The first-order terms are defined in \eqref{eq:cov-noise-first-order-terms}, and the common second-order terms are both \(\mathcal O_p(n^{-1})\).
\end{theorem}

Theorems~\ref{thm:iterated-plugin-loading-expansion-clt} and \ref{thm:iterated-plugin-noise-covariance-expansion} show that all fixed \texttt{plug-in} loading iterates share the same first-order term, while their covariance estimates share the same expansion through second order. The first covariance update replaces \((Q_{\tau,n}^{\texttt{cov}},Q_{\gamma,n}^{\texttt{cov}})\) by \((Q_{\tau,n}^{\texttt{plug-in}},Q_{\gamma,n}^{\texttt{plug-in}})\), which remain unchanged at subsequent fixed iterations. Thus, for any fixed integers \(s,t\geq1\), we have
\[
\begin{aligned}
\|\widehat\kappa^{[s],\texttt{opt}}-\widehat\kappa^{[t],\texttt{opt}}\|
&=\mathcal O_p(n^{-1}),\\
\|\widehat\Sigma^{[s]}-\widehat\Sigma^{[t]}\|_F
&=\mathcal O_p(n^{-3/2}),\quad \|\widehat\Sigma^{[s]}-\widehat\Sigma^{\texttt{cov}}\|_F
=\mathcal O_p(n^{-1}).
\end{aligned}
\]

Since the first-order terms are centered, the second-order expectations determine the leading \(n^{-1}\) bias when additional moment conditions make the remainders \(o(n^{-1})\) in expectation. The initial \texttt{cov} and \texttt{plug-in} covariance estimators may therefore have different leading biases, whereas all fixed \texttt{plug-in} covariance iterates \(s\geq1\) share the same leading bias.

Explicit expressions for \((Q_{\tau,n}^{\texttt{plug-in}},Q_{\gamma,n}^{\texttt{plug-in}})\) are given in Section~\ref{supp:sec:plugin-covariance-moment-map-expansion} of the Supplementary Material. In the simulations reported in Supplementary~\ref{subsec:endor-simulation-additional-results}, the one-step \texttt{plug-in} estimator generally has smaller absolute empirical bias in estimating \(\operatorname{tr}(\Sigma^{(0)})\) than \texttt{cov}. Further analysis is needed to determine whether this improvement extends to the absolute bias of individual covariance entries or the Frobenius norm of the covariance bias matrix.

\subsection{Asymptotic efficiency of the loading estimators}
\label{subsec:asymptotic-efficiency-comparison}

We compare the total asymptotic variability of the optimally aligned \texttt{cov} and \texttt{plug-in} loading estimators through \(\operatorname{Tr}(V^{\texttt{cov}})\) and \(\operatorname{Tr}(V^{\texttt{plug-in}})\). To clarify how the model affects relative efficiency, we express the trace comparison in terms of five quantities: the noise anisotropy \(d_\Sigma\), the loading imbalance \(d_\kappa\), the second-moment anisotropy of the factor scores \(\rho_\phi\), the factor–noise alignment coefficient \(\lambda_\phi\), and the signal-to-noise ratio \(\eta\).

We first define the noise anisotropy by
\begin{equation}
\label{eq:def-d-sigma}
d_\Sigma:=\frac{\lambda_{\max}(\Sigma^{(0)})-\lambda_{\min}(\Sigma^{(0)})}{\lambda_{\max}(\Sigma^{(0)})+\lambda_{\min}(\Sigma^{(0)})}=\frac{|\gamma|}{\tau^2}\in[0,1).
\end{equation}
This quantity measures the noise energy imbalance between the major and minor principal axes of \(\Sigma^{(0)}\). When \(d_\Sigma=0\), the noise is isotropic. As \(d_\Sigma\) approaches one, nearly all the noise energy lies along the major axis.

Throughout the comparison, we use a common phase convention for the loading and factor scores. When \(d_\Sigma>0\), this convention lets us describe the orientation of the signal relative to the major axis of the noise covariance. Treating the loading coordinates \(\kappa_1^{(0)},\ldots,\kappa_p^{(0)}\) as points in the complex plane, we apply a common phase rotation to align a major axis of these coordinates with the real axis. We rotate each factor score by the same angle in the opposite direction to preserve the signal \(\phi_i\kappa^{(0)}\). Under this convention, the loading imbalance is
\begin{equation}
\label{eq:def-d-kappa}
d_\kappa:=\left\|\operatorname{Re}(\kappa^{(0)})\right\|^2-\left\|\operatorname{Im}(\kappa^{(0)})\right\|^2=(\kappa^{(0)})^\top\kappa^{(0)}\in[0,1).
\end{equation}
When \(d_\kappa=0\), the real and imaginary components have equal energy at every phase. As \(d_\kappa\) approaches one, nearly all the loading energy lies along the major axis. When \(d_\kappa>0\), the real axis is the unique major axis of the aligned loading coordinates. Each nonzero factor score \(\phi_i\) rotates this axis through \(\arg(\phi_i)\), so the major axis of \(\phi_i\kappa_1^{(0)},\ldots,\phi_i\kappa_p^{(0)}\) has direction \(\arg(\phi_i)\) modulo \(\pi\).

We measure the second-moment anisotropy of the factor scores by
\begin{equation}
\label{eq:def-rho-phi}
\rho_\phi:=\frac{\lambda_{\max}(\Omega_\phi)-\lambda_{\min}(\Omega_\phi)}{\lambda_{\max}(\Omega_\phi)+\lambda_{\min}(\Omega_\phi)}=\frac{|\mathbb E[\phi_1^2]|}{\mathbb E[|\phi_1|^2]}\in[0,1],
\end{equation}
where \(\Omega_\phi:=\mathbb E\!\left[\operatorname{vec}(\phi_1)\operatorname{vec}(\phi_1)^\top\right]\) is the uncentered second-moment matrix of the rephased scores. The coefficient \(\rho_\phi\) measures the energy imbalance between its major and minor principal axes. When \(\rho_\phi=0\), \(\Omega_\phi\) is isotropic. When \(\rho_\phi=1\), the factor scores lie almost surely on a fixed line through the origin.

When $d_\kappa>0$, the phase convention above uniquely determines $\Omega_\phi$. When $d_\kappa=0$, we fix an arbitrary representative of the loading class and use the corresponding factor scores. If $d_\Sigma>0$ and $\rho_\phi>0$, let $\theta_\phi\in[0,\pi/2]$ be the smaller angle between the major principal axes of $\Omega_\phi$ and $\Sigma^{(0)}$, and define the factor--noise alignment coefficient by
\begin{equation}\label{eq:def-lambda-phi}
\lambda_\phi:=\cos(2\theta_\phi)\in[-1,1].
\end{equation}
Thus, $\lambda_\phi=1$ corresponds to coincident major axes, and $\lambda_\phi=-1$ corresponds to perpendicular major axes. We set $\lambda_\phi=0$ by convention if either matrix is isotropic. When $d_\kappa=0$, $\lambda_\phi$ may depend on the chosen representative, but the loading trace formulas in Theorem~\ref{thm:trace-comparison} are independent of this choice.

Finally, set \(\eta:=\mathbb E[|\phi_1|^2]/\operatorname{Tr}(\Sigma^{(0)})\in(0,\infty)\), the ratio of signal strength to noise variance.

\begin{theorem}
\label{thm:trace-comparison}
Suppose \(\mathbb E[|\phi_1|^4]<\infty\). For every \(\eta>0\), 
\begin{align}
\operatorname{Tr}(V^{\texttt{cov}})
&=(p-1)\eta^{-1}+R^{\texttt{cov}}\eta^{-2},
\label{eq:cov-loading-trace-decomposition}
\\
\operatorname{Tr}(V^{\texttt{plug-in}})
&=\underbrace{\left\{
\frac{1-d_\kappa^2d_\Sigma^2}{1-d_\kappa^2d_\Sigma^2\rho_\phi^2}
+(p-2)\frac{1-d_\Sigma^2}{1-d_\Sigma^2\rho_\phi^2}
\right\}}_{A^{\texttt{plug-in}}}\eta^{-1}
+R^{\texttt{plug-in}}\eta^{-2}.
\label{eq:plugin-loading-trace-decomposition}
\end{align}
Here \(R^{\texttt{cov}}=(p-1)+d_\Sigma^2(1-d_\kappa^2)\). Only \(R^{\texttt{plug-in}}\) can depend on \(\lambda_\phi\). For fixed \(p,d_\kappa,d_\Sigma,\rho_\phi\), \(R^{\texttt{plug-in}}\) is independent of \(\eta\) and uniformly bounded over \(\lambda_\phi\in[-1,1]\). Moreover, the following hold.
\begin{enumerate}
\item[(i)] If \(d_\Sigma=0\) (isotropic noise), then \(V^{\texttt{cov}}=V^{\texttt{plug-in}}\), so their traces coincide.

\item[(ii)] Suppose \(d_\Sigma>0\). For fixed \(p,d_\kappa,d_\Sigma,\rho_\phi\), as \(\eta\to\infty\),
\begin{equation}
\label{eq:high-snr-trace-ratio}
\frac{\operatorname{Tr}(V^{\texttt{plug-in}})}{\operatorname{Tr}(V^{\texttt{cov}})}=\frac{A^{\texttt{plug-in}}}{p-1}+O(\eta^{-1})
\end{equation}
uniformly over \(\lambda_\phi\in[-1,1]\). The leading ratio satisfies
\begin{equation}
\label{eq:high-snr-leading-ratio-bounds}
\frac{1-d_\Sigma^2}{1-d_\Sigma^2\rho_\phi^2}\leq\frac{A^{\texttt{plug-in}}}{p-1}\leq\frac{1-d_\kappa^2d_\Sigma^2}{1-d_\kappa^2d_\Sigma^2\rho_\phi^2}\leq1.
\end{equation}
If \(\rho_\phi<1\), the leading ratio lies strictly between its two bounds, as does the full trace ratio for all sufficiently large \(\eta\), uniformly over \(\lambda_\phi\). In particular, the leading ratio is strictly less than one. If \(\rho_\phi=1\), both bounds equal one and \(\operatorname{Tr}(V^{\texttt{plug-in}})-\operatorname{Tr}(V^{\texttt{cov}})=(R^{\texttt{plug-in}}-R^{\texttt{cov}})\eta^{-2}\).

\item[(iii)] If \(d_\Sigma>0\) and \(\rho_\phi=0\), then \(\operatorname{Tr}(V^{\texttt{plug-in}})<\operatorname{Tr}(V^{\texttt{cov}})\) for every \(\eta>0\).
\end{enumerate}
\end{theorem}

Parts~(i) and~(iii) give two comparisons that hold at every signal-to-noise ratio. Under isotropic noise, the two estimators have the same asymptotic covariance matrix. Under anisotropic noise, \texttt{plug-in} has a strictly smaller trace whenever the factor scores have isotropic second moments. Part~(ii) extends this gain to every fixed \(\rho_\phi<1\) at sufficiently high SNR. The leading ratio \(A^{\texttt{plug-in}}/(p-1)\) is the same for all values of \(\lambda_\phi\), and the approximation in \eqref{eq:high-snr-trace-ratio} holds uniformly. At finite SNR, \(\lambda_\phi\) can affect the comparison through the \(\eta^{-2}\) term of the \texttt{plug-in} trace.

For a fixed nonzero loading coordinate \(\kappa_j^{(0)}\), variation in the phase of \(\phi_i\) changes the orientation of the signal \(\phi_i\kappa_j^{(0)}\) relative to the noise axes. The \texttt{cov} estimator does not account for differences in directional noise variance, whereas \texttt{plug-in} uses the estimated noise covariance to give greater weight to low-variance directions.

The larger \(d_\Sigma\) is, the greater the ratio of the noise variances along the major and minor axes. For fixed \(\rho_\phi<1\), this greater difference in precision makes weighting more useful and lowers the high-SNR leading trace ratio.

For a fixed nonzero \(\kappa_j^{(0)}\), smaller \(\rho_\phi\) means that the signal energy \(\phi_i\kappa_j^{(0)}\) is more balanced across directions. Precision weighting can then combine complementary information from low-noise directions, reducing total loading variability at high SNR. As \(\rho_\phi\) approaches one, the signal energy concentrates near a single line, and this benefit diminishes. At \(\rho_\phi=1\), the signal orientation relative to the noise ellipse is fixed modulo \(\pi\). The leading high-SNR trace ratio then equals one, although higher-order terms may differ.

At a fixed score magnitude, if \(d_\kappa=0\), rotating the loading leaves the total signal energy projected onto each noise axis unchanged. As \(d_\kappa\) increases to one, more loading energy lies along a single axis, so a change in score phase can shift a larger share of the signal energy between the high- and low-variance noise directions. This helps explain the additional gain from weighting as \(d_\kappa\) increases. Its effect on the leading ratio becomes smaller as \(p\) grows, since the term involving \(d_\kappa\) has weight \(1/(p-1)\).

The factor--noise alignment \(\lambda_\phi\) affects which combinations of the real and imaginary loading components are measured precisely, while their relative phases determine how observations provide complementary information. Changing the overall alignment redistributes precision between loading components. Since the loading trace sums errors in both components, this redistribution leaves its leading high-SNR term unchanged.

\section{Application to ENDOR spectroscopy}
\label{sec:endor-application}

We apply the proposed methods to spectrum estimation in ENDOR spectroscopy. After introducing the experimental setting and the drift model, we study the profile likelihood method and spectrum recovery, derive asymptotic results, and compare estimators using simulated and real data.

\subsection{ENDOR spectroscopy and the complex drift model}
Structural biology aims to understand the geometry of bio-molecules with applications, e.g. in drug design \cite{Anderson2003, Un2026}. To this end, there is an array of experimental techniques in mutually supportive use such as X-ray crystallography, cryogenic electron microscopy and spectroscopic methods. The latter include nuclear magnetic resonance \cite{Riek2016}, electron paramagnetic resonance (EPR) \cite{Goldfarb2018, Remmel2024} and their combination: electron-nuclear double resonance (ENDOR) \cite{Feher1956, Gemperle1991}. Magnetic resonance techniques have in common that they place the chemical sample of molecules under study in a strong magnetic field, then expose it to one or several pulses of electromagnetic energy, and record any electromagnetic energy emitted by the sample in response, which is known as the `echo' signal. In ENDOR spectroscopy, a sequence of microwave (MW) pulses is used to excite any unpaired electrons, which typically only occur at a single site of a bio-molecule as almost all of the very large number of electrons form pairs whence ENDOR spectroscopy is highly selective. In contrast to EPR, ENDOR additionally uses one or several pulses of radio frequency (RF) energy to selectively excite one type of nucleus, e.g. hydrogen, oxygen, or fluorine nuclei, and the type of nucleus excited depends on the frequency $\nu$ chosen for the RF pulse. For simplicity, we will assume a fixed MW frequency throughout but see \cite{Gemperle1991} for more details.

The MW echo signal is then recorded with reference to the MW pulses sent into the chemical sample and decomposed into one component that is in phase with the reference signal (stored as the real part of a complex number $X$) and another component that has a phase difference of 90 degrees relative to the reference signal (stored as the imaginary part of $X$). These two components correspond to the $\cos$ and $\sin$ terms in a real Fourier decomposition of a signal of a single, known frequency. 

This experiment is repeated for a grid of RF frequencies $\{\nu_0,\ldots, \nu_p\}$, where we consider $p\in\mathbb{N}$ fixed. Covering this range once is known as a \emph{scan} and yields an observation $X=(X_0,\ldots, X_p)^T \in \mathbb{C}^{p+1}$. In order to achieve a signal-to-noise ratio sufficient for tuning the experimental apparatus, several scans $X_{i}^{(s)}$ enumerated by $s \in\{1,\ldots, S\}$ are additively combined to yield a single row $i$ of the data matrix $Y$, $Y_{i,j}=\sum_{s=1}^S X^{(s)}_{i,j}$, known as a \emph{batch}. Finally, the experiment is repeated for a total of $n$ batches, so that $Y\in\mathbb{C}^{n\times {(p+1)}}$ results. In this subsection, we return to the original frequency-grid
coordinates and write \(Y_{i,j}\) for the untransformed observations,
with \(j=0,\ldots,p\).

Prior to \cite{pokern2021statistical}, the standard approach to processing these data \cite{Epel2003, Rizzato2014} was the following procedure. Firstly, all batches are averaged to obtain \(Z_j=n^{-1}\sum_{i=1}^n Y_{i,j}\) yielding $Z=(Z_0,\ldots,Z_p)^\top\in\mathbb{C}^{p+1}$. This is followed by a \emph{phase correction} where \(\lambda\in[0,2\pi)\) is manually chosen so that the non-normalized spectrum
\(\tilde I_\text{grid}=\operatorname{Re}(e^{i\lambda}Z)\) is visually assessed to contain the strongest possible signal. We use the subscript $_\text{grid}$ to emphasize that $\tilde I_\text{grid}\in\mathbb{R}^{p+1}$ uses non-Helmertized coordinates. Finally, normalization is applied to yield the estimated spectrum

\begin{align} \label{eq:spectrumMaxNormalization}
    \widehat{I}_{\text{grid},j} = \frac{\tilde{I}_{\text{grid},j} - \min_{j'\in\{0,\ldots,p\}} \tilde{I}_{\text{grid},j'}}{\max_{j'\in\{0,\ldots,p\}} \tilde{I}_{\text{grid},j'}- \min_{j'\in\{0,\ldots,p\}} \tilde{I}_{\text{grid},j'}}.
\end{align}
This approach achieves three objectives: firstly, by subtraction in the numerator and denominator in \eqref{eq:spectrumMaxNormalization}, it eliminates an additive offset (also known as baseline \cite{Epel2003}) $\psi\in\mathbb{C}$ in the microwave echo signal, which arises as the echo signal is a sum of the signals originating from the many molecules in the chemical sample: some that have interacted with the RF pulse and some that have not. Secondly, by dividing the numerator by the denominator in \eqref{eq:spectrumMaxNormalization},  it also removes an unknown multiplicative factor $\phi \in\mathbb{C}$ which arises since the MW coupling of the chemical sample to the MW receiver is very sensitive: small changes in the experimental apparatus lead to a changed fraction and phase of the MW echo signal emitted by the chemical sample reaching the receiver. Thirdly, it selects a particular direction $e^{i\lambda}$ in the complex plane along which the spectrum lies. The orthogonal direction in the complex plane also contains a signal $W_\text{grid}$ referred to as `the Wave' in \cite{pokern2021statistical} which, contrary to the spectrum, differs depending on the spectrometer used and is likely an artifact produced by the spectrometer's internal arrangement.

An alternative data processing approach that avoids manual phase correction is to set
\begin{align}
    \label{eq:absoluteValueModel}
    \tilde{Z}_{j} = \frac{1}{n} \sum_{i=1}^n |Y_{i,j}|
\end{align}
and use $\tilde{Z}_j$ in equation \eqref{eq:spectrumMaxNormalization} in place of $\tilde{I}_{\text{grid},j}$. We refer to this approach as the \texttt{absolute value} method and it has been used in spectroscopy, e.g. \cite{meyer2022}, even though neither the materials and methods section nor the supporting information of this publication makes this clear \cite{meyer2026priv}.

In \cite{pokern2021statistical}, the standard approach using Equation (40) and a manually chosen phase correction was re-cast as the equivalent \texttt{averaging} model:
\begin{align}\label{eq:averagingModel}
    Z &= \psi\mathbf 1_{p+1} + \phi (I_\text{grid} + iW_\text{grid}).
\end{align}
Due to the use of minimum and maximum in \eqref{eq:spectrumMaxNormalization}, this formula is obviously sensitive to noise, so a different standardization was introduced in \cite{pokern2021statistical} imposing $\sum_{j=0}^p (I_{\text{grid},j}+iW_{\text{grid},j})\stackrel{!}{=}0$ and $\sum_{j=0}^p |I_{\text{grid},j} + i W_{\text{grid},j}|^2\stackrel{!}{=}1$. The model \eqref{eq:averagingModel} exhibits $2(p+1)$  real degrees of freedom on the left hand side and $2+2+2(p+1)-2-1$ real degrees of freedom on the right-hand side, with the disparity lifted by the criterion used to choose $\mathrm{arg}(\phi)$. Hence, this is a deterministic model that does not account for measurement error naturally arising in the experimental observation process.

Writing $\kappa_\text{grid}:= I_\text{grid}+iW_\text{grid}$ turns \eqref{eq:averagingModel} into $Z=\psi \mathbf 1_{p+1} + \phi \kappa_\text{grid}$, which shows that estimation of $\kappa_\text{grid}$ exhibits the identifiability issues discussed in Section \ref{sec:model} leading us to the parameter space 
$
\left\{
[\kappa]\in\mathbb{CP}^{p}:
\mathbf1_{p+1}^{\top}\kappa=0
\right\}
$ in a first step. In a second step, we consider how to systematically choose $\mathrm{arg}(\phi)$, which corresponds to choosing $\lambda$ in  $I_\text{grid}=\operatorname{Re}(e^{i\lambda} \kappa_\text{grid})$ given a representative $\kappa_\text{grid}$ of $[\kappa_\text{grid}]$. One possible approach is the 
 \emph{maximum method} which we will detail in Section \ref{sec: maximum method}: it chooses $\lambda$ such as to maximize $\|I_{\text{grid}}\|$. Naturally, there are other options which we will discuss in Section \ref{sec:Conclusion}. 
 


For any such method to stand a chance of success, however, it must be possible to distinguish the spectrum $I_\text{grid}$ from the wave $W_\text{grid}$ and in experiments this is invariably the case whence we may assume that \(\operatorname{Re}(\kappa^{(0)}_\text{grid})\) and \(\operatorname{Im}(\kappa^{(0)}_\text{grid})\) are linearly independent in \(\mathbb R^{{p+1}}\) at least for the true value $[\kappa^{(0)}_\text{grid}]$ underlying any estimated $[\kappa_\text{grid}]$.

Separate application of \eqref{eq:averagingModel} to each batch reveals that $\psi$ and $\phi$ are batch-dependent which is explained by the thermal drift experienced by the experimental apparatus over the course of the experiment that often runs over several hours, affecting the sensitive coupling of MW energy from the chemical sample to the receiver. Hence, \cite{pokern2021statistical} introduced a \emph{homoscedastic drift model} that allowed for separate $\psi$ and $\phi$ for each batch, leading to parameters $\psi,\phi \in \mathbb{C}^n$ and non-zero noise $\varepsilon_{i,j} \stackrel{i.i.d.}{\sim}\mathcal{CN}(0,\Sigma)$ with $\Sigma \in \mathrm{SPD}(2)$:
\begin{align}
    \label{eq:homoscedasticDrift}
    Y_{i,j} &= \psi_i + \phi_i \kappa_{\text{grid},j} + \varepsilon_{i,j}\quad \varepsilon_{i,j} \stackrel{i.i.d.}{\sim} \mathcal{CN}(0,\Sigma),
\end{align}
which is indeed identical to the uncentered complex one-factor model \eqref{model:uncentered-complex-factor}. Since all reasonable experiments couple a finite, non-zero amount of energy from the chemical sample to the MW receiver, we shall assume \(0<\mathbb E[|\phi_{1}|^2]<\infty\) throughout.

The homoscedastic drift model \eqref{eq:homoscedasticDrift} was estimated using profile likelihood estimation of $\psi,\phi \in\mathbb{C}^n, \, [\kappa_\text{grid}] \in \mathbb{CP}^{p}, \Sigma \in \mathrm{SPD}(2)$ and has been found to fit the data unusually well \cite{pokern2021statistical}. It has since been applied in several practical ENDOR studies \cite{Wiechers2023, Hiller2022, Remmel2024} and has demonstrated improvements in signal-to-noise ratio in the estimated spectrum gained over the \texttt{averaging} model, as noted in \cite{pokern2021statistical}. This is because the homoscedastic drift model counteracts the signal cancellation experienced by the \texttt{averaging} model when the microwave echo signal exhibits substantial phase drift. Indeed, the spectrum obtained by the \texttt{averaging} model consists of nothing but noise in the extreme case when the $\phi_i$ in the corresponding homoscedastic drift model fit satisfy $\sum_{i=1}^{n} \phi_{i}=0$. Finally, \cite{pokern2021statistical} also found the PL estimator to be biased (as is partially proved in \cite{wiechers2023drift}), which necessitated bootstrap bias correction for both $\phi$ and $\Sigma$.

We adopt the profile-likelihood approach of \cite{pokern2021statistical} for the centered model \eqref{model:centered-complex-factor} in Helmert coordinates. In the remainder of this section, we show that the resulting joint PL estimator is inconsistent for the true parameter pair. Furthermore, we establish consistency and asymptotic normality of the estimated spectrum obtained via the maximum method. Next, we compare the finite-sample performance of the PL estimator, the bootstrap-corrected PL estimator, the covariance estimator, and the plug-in method using simulated data with a range of realistic parameter settings. Finally, we apply these estimators to two real datasets from the spectroscopic literature.

\subsection{Inconsistency of the joint profile likelihood estimator}
\label{sec: inconsistency PL}

We show that the joint profile likelihood estimator \((\wh{[\kappa]}^{\texttt{PL}},\wh{\Sigma}^{\texttt{PL}})\), defined as a minimizer of \eqref{eq:profile-loglik} over \(\mathbb{CP}^{p-1}\times\operatorname{SPD}(2)\) is inconsistent for the true parameter pair. In this subsection and the next subsections, we use the Helmert coordinates introduced in Section~\ref{subsec:eliminating-observation-level}. For the centered spectrum and wave \(I_{\text{grid}},W_{\text{grid}}\in\mathbb R^{p+1}\) of the preceding subsection, we now write \(\kappa=H(I_{\text{grid}}+iW_{\text{grid}})\in\mathbb C^p\). The Helmert map identifies the centered loading classes on the original grid with \(\mathbb{CP}^{p-1}\), preserves the projective distance \(d\) in
\eqref{eq:projective-distance}, and linear independence of the real and imaginary parts.

\begin{theorem}
\label{thm: PL strong inconsistency}
Almost surely, the empirical joint profile-likelihood objective has minimizers for all sufficiently large \(n\), and their set remains outside a fixed neighborhood of the true pair \(([{\kappa^{(0)}}],\Sigma^{(0)})\).
\end{theorem}

Theorem~\ref{thm: PL strong inconsistency} follows since the true pair \(([\kappa^{(0)}],\Sigma^{(0)})\) does not minimize the population profile objective; see~\ref{supp:sec:joint-profile-likelihood-inconsistency} of the Supplementary Material. It rules out a joint \(\sqrt n\)-CLT centered at the true pair but does not establish inconsistency of the loading component alone. To the best of our knowledge, no closed-form characterization of the population minimizers is available. Hence, classical Wald-type inference for the true parameter pair requires further analysis of the discrepancy between the population minimizers and the true parameters.

To address the covariance bias, \cite{pokern2021statistical} employed a parametric bootstrap bias correction for \(\widehat\Sigma^{\texttt{PL}}\). A multiplicative bias correction for the magnitude of $\phi$ was used additionally.

In Subsection \ref{subsec:endor-simulation-analysis} below, we apply this bias-correction scheme to the PL fit of the centered model \eqref{model:centered-complex-factor}. Our simulations show under-coverage of the resulting confidence intervals, especially at low signal-to-noise ratios. We also compare this procedure numerically with the \texttt{cov} and iterated \texttt{plug-in} estimators, whose consistency and stochastic expansions provide practically usable approximations at realistic sample sizes even without bootstrap bias correction.

\subsection{Spectrum--wave recovery by the maximum method}
\label{sec: maximum method}

The estimators in Section~\ref{sec:estimators} identify the loading only as an equivalence class \([\kappa]\in\mathbb{CP}^{p-1}\). Recovering the spectrum and wave \(I_{\text{grid}},W_{\text{grid}}\in\mathbb{R}^{p+1}\) on the original frequency grid therefore requires selecting a phase and applying \(H^\top\) to the real and imaginary parts of the resulting representative. Since \(H\) is real and \(HH^\top=\mathrm{Id}_p\) (see \eqref{eq:Helmert-identities}), this transformation preserves norms and inner products, and
\[
\left\|\operatorname{Re}\left(e^{i\lambda}H^\top\kappa\right)\right\|^2
=
\left\|\operatorname{Re}\left(e^{i\lambda}\kappa\right)\right\|^2.
\]
Thus, the phase maximization below can be performed directly in Helmert coordinates.

To separate the spectrum and wave, we introduce the \emph{maximum method}, which is motivated by the expectation that the spectrum dominates the wave component, as in many ENDOR experiments \citep{pokern2021statistical, Wiechers2023}. It selects the phase maximizing the norm of the real component. Precisely, for any \(\kappa\in\mathcal S^{2p-1}\), define
\begin{equation}
\label{eq:maximum-phase-definition}
\lambda^{\texttt{max}}(\kappa)
\in
\operatorname*{argmax}_{\lambda\in[0,2\pi)}
\left\|
\operatorname{Re}\left(e^{i\lambda}\kappa\right)
\right\|^2.
\end{equation}
We denote the real and imaginary parts of \(e^{i\lambda^{\texttt{max}}(\kappa)}\kappa\) by \(I^{\texttt{max}}(\kappa)\) and \(W^{\texttt{max}}(\kappa)\), respectively. Both are vectors in \(\mathbb R^p\) expressed in
Helmert coordinates. The following lemma characterizes the maximizing phase and its equivariance under the choice of representative.

\begin{lemma}
\label{lem: max method projection}
For \(\kappa\in\mathcal S^{2p-1}\) with
\(\kappa^\top\kappa\neq0\), the maximizing phase is uniquely
determined modulo \(\pi\) by
\[
\lambda^{\texttt{max}}(\kappa)
\equiv-\operatorname{Arg}(\kappa^\top\kappa) /2
\pmod{\pi}.
\]
The resulting spectrum and wave satisfy
\[
\|I^{\texttt{max}}(\kappa)\|^2
>
\|W^{\texttt{max}}(\kappa)\|^2,
\quad
\langle I^{\texttt{max}}(\kappa),W^{\texttt{max}}(\kappa)\rangle=0.
\]
Moreover,
\(\lambda^{\texttt{max}}(e^{i\alpha}\kappa)
\equiv\lambda^{\texttt{max}}(\kappa)-\alpha\pmod{\pi}\)
for every \(\alpha\in\mathbb R\), so the pair is determined by
\([\kappa]\) up to a common sign.
If \(\kappa^\top\kappa=0\), every phase is a maximizer.
\end{lemma}

For \(\kappa^\top\kappa\neq0\), we resolve the remaining sign
ambiguity by requiring
\(
\left[H^\top I^{\texttt{max}}(\kappa)\right]_{j_c}>0
\), provided that this component is nonzero, where \(j_c\in\{0,\ldots,p\}\) is a fixed index corresponding to a central frequency on the original grid. This convention reflects the physical expectation of a
positive central resonance in the ENDOR spectrum. Let \(\kappa^{\texttt{max}}\in[\kappa]\) denote the resulting unique representative, with
\(I^{\texttt{max}}=\operatorname{Re}(\kappa^{\texttt{max}})\)
and
\(W^{\texttt{max}}=\operatorname{Im}(\kappa^{\texttt{max}})\).

Throughout the following analysis, we choose the phase of
\(\kappa^{(0)}=I^{(0)}+iW^{(0)}\) so that it satisfies the
maximum-alignment condition in Lemma~\ref{lem: max method projection} and the sign
convention, with
\(d_\kappa=\|I^{(0)}\|^2-\|W^{(0)}\|^2>0\) and
\([H^\top I^{(0)}]_{j_c}>0\).
The factor scores are re-phased accordingly to preserve
\(\phi_i\kappa^{(0)}\).

To transfer the optimally aligned loading asymptotics to the recovered spectrum and wave, choose near \(\kappa^{(0)}\) a continuous real-valued branch of
\(\lambda^{\texttt{max}}\) modulo \(\pi\), with
\(\lambda^{\texttt{max}}(\kappa^{(0)})=0\), and define
\begin{equation}
\label{eq:maximum-alignment-map}
\mathcal A^{\texttt{max}}(\kappa)
:=
s^{\texttt{max}}(\kappa)
e^{i\lambda^{\texttt{max}}(\kappa)}\kappa,
\quad
s^{\texttt{max}}(\kappa)
:=
\operatorname{sgn}
\left(
\left[
H^\top
\operatorname{Re}
\left\{
e^{i\lambda^{\texttt{max}}(\kappa)}\kappa
\right\}
\right]_{j_c}
\right).
\end{equation}
In a sufficiently small neighborhood of \(\kappa^{(0)}\),
\(s^{\texttt{max}}=1\) and \(\mathcal A^{\texttt{max}}\) is smooth.
By Lemma~\ref{lem: max method projection}, it maps each representative
in this neighborhood to the unique representative satisfying both
conventions.

We denote the maximum-aligned loading estimators by
\(\widehat\kappa^{\texttt{cov,max}}\) and
\(\widehat\kappa^{[s],\texttt{max}}\), \(s\geq1\).
Whenever the optimally aligned representatives lie in this neighborhood,
their real-valued spectrum and wave components satisfy
\[
\begin{aligned}
\widehat\kappa^{\texttt{cov,max}}
&=
\widehat I^{\texttt{cov,max}}
+i\widehat W^{\texttt{cov,max}}
=
\mathcal A^{\texttt{max}}
\left(\widehat\kappa^{\texttt{cov,opt}}\right),\\
\widehat\kappa^{[s],\texttt{max}}
&=
\widehat I^{[s],\texttt{max}}
+i\widehat W^{[s],\texttt{max}}
=
\mathcal A^{\texttt{max}}
\left(\widehat\kappa^{[s],\texttt{opt}}\right).
\end{aligned}
\]

\subsection{Asymptotic analysis of the maximum aligned spectrum}
\label{subsec:endor-maximum-asymptotics}

Using the continuity and local smoothness of \(\mathcal A^{\texttt{max}}\), we transfer the consistency and asymptotic normality of the optimally aligned loading estimators to the maximum-aligned spectrum estimators.

\begin{theorem}
\label{thm:maximum-aligned-strong-consistency}
Assume $\kappa^{(0)}$ satisfies the maximum-alignment and sign conventions above. For any fixed integer \(s\geq1\), almost surely, the corresponding maximum-aligned representatives are uniquely defined for all sufficiently large \(n\), and
\[
\widehat I^{\texttt{cov,max}}, ~\widehat I^{[s],\texttt{max}}
\rightarrow I^{(0)}, \quad
\widehat W^{\texttt{cov,max}}, ~\widehat W^{[s],\texttt{max}}
\rightarrow W^{(0)}
\quad\text{as } n\to\infty.
\]
\end{theorem}

\begin{theorem}
\label{thm:maximum-aligned-loading-expansion-clt}
Assume the conditions of Theorem~\ref{thm:maximum-aligned-strong-consistency}
and \(\mathbb E[|\phi_1|^4]<\infty\). Fix an integer \(s\geq1\). The Jacobian of \(\kappa\mapsto\operatorname{Re}\{\mathcal A^{\texttt{max}}(\kappa)\}\) at \(\kappa^{(0)}\), in grouped real coordinates, is
\begin{equation}
\label{eq:maximum-alignment-jacobian}
A_I^{\texttt{max}}
:=
\begin{pmatrix}
\mathrm{Id}_p+\dfrac{W^{(0)}(W^{(0)})^\top}{d_\kappa}
&
\dfrac{W^{(0)}(I^{(0)})^\top}{d_\kappa}
\end{pmatrix} \in \mathbb{R}^{p \times 2p}.
\end{equation}
Then
\begin{equation}
\label{eq:maximum-aligned-common-expansion}
\begin{aligned}
\widehat I^{\texttt{cov,max}}-I^{(0)}
&=
A_I^{\texttt{max}}
\widetilde{\operatorname{vec}}
\left(L_{\kappa,n}^{\texttt{cov}}\right)
+\mathcal O_p(n^{-1}),
\\
\widehat I^{[s],\texttt{max}}-I^{(0)}
&=
A_I^{\texttt{max}}
\widetilde{\operatorname{vec}}
\left(L_{\kappa,n}^{\texttt{plug-in}}\right)
+\mathcal O_p(n^{-1}).
\end{aligned}
\end{equation}
Consequently,
\begin{align}
\sqrt n
\left(
\widehat I^{\texttt{cov,max}}-I^{(0)}
\right)
&\xrightarrow{d}
N_p\left(0,V_I^{\texttt{cov,max}}\right),
\label{eq:maximum-aligned-cov-clt}
\\
\sqrt n
\left(
\widehat I^{[s],\texttt{max}}-I^{(0)}
\right)
&\xrightarrow{d}
N_p\left(0,V_I^{\texttt{plug-in,max}}\right)
\label{eq:maximum-aligned-common-clt}
\end{align}
where
\begin{equation}
\label{eq:maximum-aligned-asymptotic-covariances}
V_I^{m,\texttt{max}}
:=
A_I^{\texttt{max}}V^m(A_I^{\texttt{max}})^\top,
\quad
m\in\{\texttt{cov},\texttt{plug-in}\}.
\end{equation}
\end{theorem}
Recall that on the original frequency grid, the spectrum estimators are \(H^\top\widehat I^{\texttt{cov,max}}\) and \(H^\top\widehat I^{[s],\texttt{max}}\), with target \(H^\top I^{(0)}\). Their consistency and asymptotic normality follow by applying the linear map \(H^\top\) to the results above. The corresponding asymptotic covariance matrices are
\(H^\top V_I^{m,\texttt{max}}H\),
\(m\in\{\texttt{cov},\texttt{plug-in}\}\). Moreover, since \(HH^\top=\mathrm{Id}_p\),
\(
\operatorname{Tr}
\left(H^\top V_I^{m,\texttt{max}}H\right)
=
\operatorname{Tr}\left(V_I^{m,\texttt{max}}\right)
\). Thus, the trace comparisons below also apply on the original frequency grid. 

The same transformation gives the wave estimators on the original frequency grid. Joint asymptotic results for the spectrum and wave, together with explicit formulas for \(V_I^{m,\texttt{max}}\), are provided in Section~\ref{supp:sec:maximum-method} of the Supplementary Material.

\subsection{Asymptotic efficiency of the maximum-aligned spectrum estimators}
\label{subsec:endor-spectrum-efficiency}
Now we compare the total asymptotic variability of \(\widehat I^{\texttt{cov,max}}\) and \(\widehat I^{[s],\texttt{max}}\) through \(\operatorname{Tr}(V_I^{\texttt{cov,max}})\) and \(\operatorname{Tr}(V_I^{\texttt{plug-in,max}})\). The comparison applies to every \(s\geq1\), since these iterations share the same first-order asymptotic distribution.

\begin{theorem}
\label{thm:maximum-spectrum-trace-comparison}
Under the conditions of Theorem~\ref{thm:maximum-aligned-loading-expansion-clt}, assume \(0<d_\kappa<1\). For every \(\eta>0\), the traces decompose exactly as
\begin{align}
\operatorname{Tr}(V_I^{\texttt{cov,max}})
&=A_I^{\texttt{cov}}\eta^{-1}+R_I^{\texttt{cov}}\eta^{-2},
\label{eq:cov-maximum-spectrum-trace-decomposition}
\\
\operatorname{Tr}(V_I^{\texttt{plug-in,max}})
&=A_I^{\texttt{plug-in}}\eta^{-1}+R_I^{\texttt{plug-in}}\eta^{-2},
\label{eq:plugin-maximum-spectrum-trace-decomposition}
\end{align}
where
\begin{align*}
A_I^{\texttt{cov}}
&:=\frac{p-2}{2}(1+d_\Sigma\rho_\phi\lambda_\phi)
+\frac{(1-d_\kappa)(1-d_\kappa d_\Sigma\rho_\phi\lambda_\phi)}{4}
+\frac{(1+d_\kappa)(1+d_\kappa d_\Sigma\rho_\phi\lambda_\phi)}{4d_\kappa^2},
\\
A_I^{\texttt{plug-in}}
&:=\frac{1-d_\Sigma^2}{1-d_\Sigma^2\rho_\phi^2}\frac{p-2}{2}(1+d_\Sigma\rho_\phi\lambda_\phi)+\frac{1-d_\kappa^2d_\Sigma^2}{1-d_\kappa^2d_\Sigma^2\rho_\phi^2}
\left\{A_I^{\texttt{cov}}-\frac{p-2}{2}(1+d_\Sigma\rho_\phi\lambda_\phi)\right\}.
\end{align*}
The finite remainder coefficients \(R_I^{\texttt{cov}},R_I^{\texttt{plug-in}}\) are independent of \(\eta\) and uniformly bounded over \(\lambda_\phi\in[-1,1]\) for fixed \(p,d_\kappa,d_\Sigma,\rho_\phi\). Moreover, the following hold.
\begin{enumerate}
\item[(i)] If \(d_\Sigma=0\) (isotropic noise), then \(V_I^{\texttt{cov,max}}=V_I^{\texttt{plug-in,max}}\), so their traces coincide.

\item[(ii)] Suppose \(d_\Sigma>0\). For fixed \(p,d_\kappa,d_\Sigma,\rho_\phi\), as \(\eta\to\infty\),
\begin{equation}
\label{eq:high-snr-maximum-spectrum-trace-ratio}
\frac{\operatorname{Tr}(V_I^{\texttt{plug-in,max}})}{\operatorname{Tr}(V_I^{\texttt{cov,max}})}
=\frac{A_I^{\texttt{plug-in}}}{A_I^{\texttt{cov}}}+O(\eta^{-1})
\end{equation}
uniformly over \(\lambda_\phi\in[-1,1]\). The leading ratio satisfies
\begin{equation}
\label{eq:high-snr-maximum-spectrum-leading-ratio-bounds}
\frac{1-d_\Sigma^2}{1-d_\Sigma^2\rho_\phi^2}
\leq\frac{A_I^{\texttt{plug-in}}}{A_I^{\texttt{cov}}}
\leq\frac{1-d_\kappa^2d_\Sigma^2}{1-d_\kappa^2d_\Sigma^2\rho_\phi^2}
\leq1.
\end{equation}
If \(\rho_\phi<1\), all three inequalities are strict, and the full trace ratio also lies strictly between the lower and upper bounds for all sufficiently large \(\eta\), uniformly over \(\lambda_\phi\). If \(\rho_\phi=1\), the leading ratio and both bounds equal one, and the \(\eta^{-2}\) terms determine the comparison.

\item[(iii)] If \(d_\Sigma>0\) and \(\rho_\phi=0\), then \(\operatorname{Tr}(V_I^{\texttt{plug-in,max}})<\operatorname{Tr}(V_I^{\texttt{cov,max}})\) for every \(\eta>0\).
\end{enumerate}
\end{theorem}

As in the loading efficiency comparison, Parts~(i) and~(iii) show that isotropic noise yields identical asymptotic covariance matrices, whereas anisotropic noise and factor scores with isotropic second moments yield a strict trace reduction at every signal-to-noise ratio. Part~(ii) extends this gain to every fixed \(\rho_\phi<1\) at sufficiently high signal-to-noise ratios, uniformly over \(\lambda_\phi\).

Unlike the leading loading MSE ratio in \eqref{eq:high-snr-trace-ratio}, the leading spectrum MSE ratio in \eqref{eq:high-snr-maximum-spectrum-trace-ratio} can depend on \(\lambda_\phi\). Spectrum estimation retains the real part of the maximum-aligned loading, so which loading components are measured precisely matters. When \(\lambda_\phi\) is close to one, the major principal axes of \(\Omega_\phi\) and \(\Sigma^{(0)}\) are nearly aligned. Thus, more energy of \(\phi_i I^{(0)}\), the spectrum contribution to the full signal \(\phi_i\kappa^{(0)}\), lies along the high-variance noise direction. For \(d_\Sigma>0\) and \(\rho_\phi>0\), with the remaining parameters fixed, the leading high-SNR terms of both \(\operatorname{Tr}(V_I^{\texttt{cov,max}})\) and \(\operatorname{Tr}(V_I^{\texttt{plug-in,max}})\) increase with \(\lambda_\phi\), but not necessarily at the same relative rate.

Despite this dependence, the leading ratio is a weighted average of the two bounds in \eqref{eq:high-snr-maximum-spectrum-leading-ratio-bounds}, neither of which depends on \(\lambda_\phi\), with weight
\(
{(p-2)(1+d_\Sigma\rho_\phi\lambda_\phi)}/{2A_I^{\texttt{cov}}}
\)
on the lower bound. For fixed remaining parameters, this weight tends to one uniformly over \(\lambda_\phi\) as \(p\) increases.
The two bounds also coalesce as \(d_\kappa\to1\).
Thus, either a large \(p\) or a \(d_\kappa\) close to one makes the leading ratio close to the lower bound, with little dependence on \(\lambda_\phi\).
As \(\rho_\phi\to1\), both bounds instead approach one, and the leading-order efficiency gain vanishes.

\subsection{Simulation study}
\label{subsec:endor-simulation-analysis}

Using a simulation setting motivated by the real ENDOR measurements in \cite{pokern2021statistical}, we assess whether the strong consistency and central limit theorem of loading and spectrum estimates (Theorems~\ref{thm:iterated-plugin-strong-consistency}, \ref{thm:maximum-aligned-strong-consistency}, and \ref{thm:maximum-aligned-loading-expansion-clt}) translate into more accurate loading and spectrum estimates and reliable confidence intervals. We compare the finite-sample performance of \texttt{cov}, iterated \texttt{plug-in}, and joint profile likelihood across different model configurations in terms of loading and spectrum MSEs, the coverage probabilities and average lengths of pointwise 95\% spectrum confidence intervals. For joint profile likelihood, we consider parametric bootstrap procedures based on the original fitted model (\texttt{PL}) and on bias-corrected noise covariance and factor-score scale estimates (\texttt{PL-BC}) as employed in \cite{pokern2021statistical}. We use the maximum method to extract spectra. For a maximum-aligned loading estimate \(\widehat\kappa^{\texttt{max}}\) in Helmert coordinates, the estimated spectrum on the original frequency grid is \(\widehat I_{\text{grid}}^{\texttt{max}} =H^\top\operatorname{Re}(\widehat\kappa^{\texttt{max}})\).

\medskip
\noindent\textit{Simulation setting.}
We use \(p+1=71\) equally spaced frequency points \(\nu_j=-15+30j/70\), \(j=0,\ldots,70\), and define the true spectrum \(I_{\mathrm{grid}}^{(0)}\) and wave \(W_{\mathrm{grid}}^{(0)}\) by
\[
I_{\mathrm{grid},j}^{(0)}
=
\sqrt{\frac{1+d_\kappa}{2}}\,
\frac{g_j-\bar g}
{\left\{\sum_{\ell=0}^{70}(g_\ell-\bar g)^2\right\}^{1/2}},
\quad
W_{\mathrm{grid},j}^{(0)}
=
\sqrt{\frac{1-d_\kappa}{2}}\,
\frac{\cos(0.5\nu_j+\pi/2)}
{\left\{\sum_{\ell=0}^{70}\cos^2(0.5\nu_\ell+\pi/2)\right\}^{1/2}},
\]
where
\[
g_j
=
\sum_{\delta\in\{-1,1\}}
\left\{
1.2\exp\!\left[-\frac{(\nu_j-0.8\delta)^2}{2(0.5)^2}\right]
+
0.5\exp\!\left[-\frac{(\nu_j-10\delta)^2}{2(0.6)^2}\right]
\right\}
\]
and \(\bar g=71^{-1}\sum_{j=0}^{70}g_j\). Both the peak structure defined by \(g_j\), with two main peaks and two weaker outer peaks, and the cosine-like form of \(W_{\mathrm{grid}}^{(0)}\) are motivated by the ENDOR measurements in \cite{pokern2021statistical}. Both grid vectors are centered and mutually orthogonal. By construction, the Helmert-transformed loading
\[
\kappa^{(0)}
=H\kappa_{\text{grid}}^{(0)}
=H\bigl(I_{\text{grid}}^{(0)}+iW_{\text{grid}}^{(0)}\bigr)
=I^{(0)}+iW^{(0)}
\in\mathbb C^{70}
\]
has unit norm and satisfies strict maximum alignment for \(d_\kappa>0\).

We normalize the total noise variance to one and control the SNR \(\eta\) through the factor-score energy. The factor scores \(\phi_i\), \(i=1,\ldots,n\), are generated independently as
\[
\phi_i=R_i e^{i\theta_i},
\quad
R_i=\frac{\sqrt{\eta}\,U_i}{\sqrt{1+0.1^2/3}},
\quad
U_i\sim\operatorname{Unif}(0.9,1.1),
\quad
\theta_i\sim\operatorname{Unif}(\mu-a/2,\mu+a/2),
\]
with mutually independent amplitudes and phases. For \(0<a\le2\pi\), measured in radians, this gives \(\mathbb E[|\phi_1|^2]=\eta\) and \(\rho_\phi=|\sin(a)|/a\). At fixed \(\eta\), the angular width \(a\) controls the factor-score second-moment anisotropy without changing the total energy. These choices provide a simplified model of the amplitude and phase variation observed in the factor scores fitted to the PNAS data \citep{pokern2021statistical}, with uniformity and amplitude--phase independence imposed for a controlled simulation design.

We generate \(Y_i\in\mathbb C^{70}\) from model~\eqref{model:centered-complex-factor} with
\(\Sigma^{(0)}=\frac{1}{\mathrm{CN}+1}\operatorname{diag}(\mathrm{CN},1)\), where \(\mathrm{CN}\ge1\) is the noise condition number and \(d_\Sigma=(\mathrm{CN}-1)/(\mathrm{CN}+1)\). Since \(\mathbb E[\phi_1^2]=\eta e^{2i\mu}\sin(a)/a\) and the noise major axis lies along the real axis, \(\lambda_\phi=\operatorname{sgn}(\sin a)\cos(2\mu)\) when \(d_\Sigma\rho_\phi>0\); otherwise, we set \(\lambda_\phi=0\).

The simulation parameters are informed by our reanalysis of the protein ENDOR data released with \citet{pokern2021statistical}. We use 19 measurement conditions, excluding the BDPA reference and overlapping sensitivity subsets. Using the \texttt{plug-in} estimator with \(s=1\) and estimated second moments for the factor scores, the estimated ranges are \(\mathrm{CN}=1.64\)--\(32.07\), \(d_\kappa=0.695\)--\(0.966\), \(\rho_\phi=0.402\)--\(0.998\), \(\lambda_\phi=-0.975\)--\(0.989\), and \(\eta=11.38\)--\(505.83\). To relate phase variation to our uniform-angle model, we define the equivalent width \(a_{\mathrm{eq}}\in(0,\pi)\) by \(\sin(a_{\mathrm{eq}})/a_{\mathrm{eq}}=\rho_\phi\). These widths range from \(6.23^\circ\) to \(121.54^\circ\); they match the estimated second-moment anisotropy and do not assume that the measured phases follow a uniform distribution.

\begin{figure}[tbp]
\centering
\includegraphics[width=0.9\linewidth]{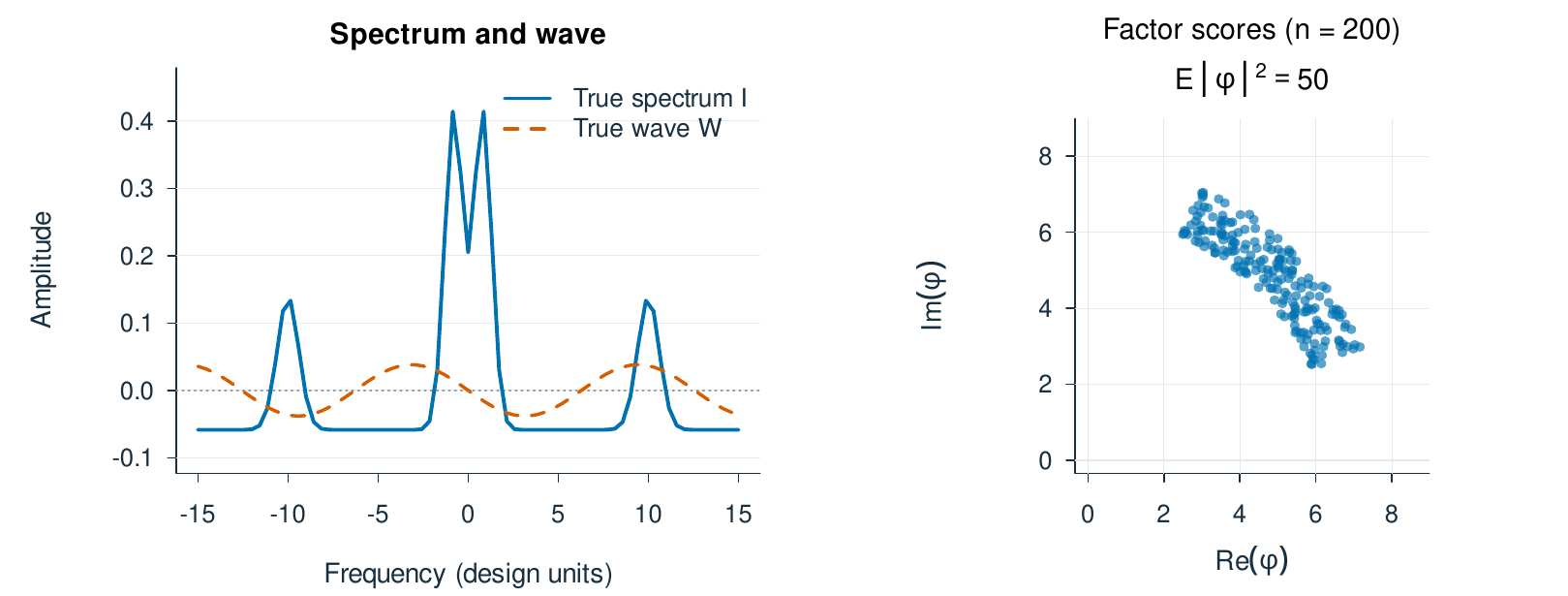}
\caption{Baseline ENDOR-like simulation setting.
Left: the true spectrum \(I_{\mathrm{grid}}^{(0)}\) and wave \(W_{\mathrm{grid}}^{(0)}\),
with squared norms \(0.95\) and \(0.05\), respectively,
and \(\langle I_{\mathrm{grid}}^{(0)},W_{\mathrm{grid}}^{(0)}\rangle=0\).
Right: a realization of \(n=200\) independent factor scores
with \(\mu=a=\pi/4\) and
\(\mathbb E[|\phi_1|^2]=50\).}
\label{fig:endor-simulation-design}
\end{figure}

We use \( (\mathrm{CN},d_\kappa,a,\lambda_\phi,\eta)=(6,0.9,\pi/4,0,50) \) as the baseline setting, with \(\mu=\pi/4\) and \(\rho_\phi=2\sqrt{2}/\pi\approx0.9003\). The choices \(\mathrm{CN}=6\), \(d_\kappa=0.9\), and \(\eta=50\) are close to the empirical medians \(6.30\), \(0.902\), and \(43.17\), respectively. The width \(a=45^\circ\) is also close to the median equivalent width of \(38.12^\circ\). We choose \(\lambda_\phi=0\) as a reference orientation with zero coupling between the noise and score second-moment anisotropies. This baseline represents typical parameter scales in the protein data. Figure~\ref{fig:endor-simulation-design} illustrates spectrum, wave, and factor scores in this setting.

We vary one parameter at a time around the baseline, with sweeps spanning \(\mathrm{CN}\in[1,50]\), \(d_\kappa\in[0.01,0.995]\), \(a\in[5^\circ,360^\circ]\), \(\lambda_\phi\in[-1,1]\), and \(\eta\in[5,500]\). The extensions beyond the estimated data ranges assess specific limiting regimes. The noise sweep includes the isotropic reference \(\mathrm{CN}=1\) and stronger anisotropy at \(\mathrm{CN}=50\). Small values of \(d_\kappa\), down to \(0.01\), make the spectrum and wave energies nearly equal and test the maximum method when its phase is weakly identified; \(d_\kappa=0.995\) approaches the opposite limit of a negligible wave component. The angular sweep includes nearly fixed phases at \(a=5^\circ\), broader phase variation beyond the equivalent data range, and full-circle phases at \(a=360^\circ\). Both \(a=180^\circ\) and \(a=360^\circ\) give isotropic score second moments, with \(\rho_\phi=0\); over the full sweep, \(\rho_\phi=|\sin(a)|/a\) is not monotone in \(a\). The endpoints \(\lambda_\phi=\pm1\) examine aligned and orthogonal principal axes of the noise and score second-moment matrices. Finally, \(\eta=5,10\) extend below the estimated SNR range to assess estimation and interval calibration under weaker signals, whereas \(\eta=500\) approximates the largest estimated SNR.

The design comprises 184 configurations across sample sizes \(n\in\{100,200,400,800\}\). For each configuration, we generate \(R=200\) independent datasets and apply all methods to each dataset. We compute \texttt{plug-in} estimates for \(s=1,\ldots,4\), with \(s=1\) used for the primary comparison.

\medskip
\noindent\textit{Mean squared errors of loading and spectrum.} For each method and each simulation, we evaluate the squared loading estimation error \(d(\widehat{[\kappa]},[\kappa^{(0)}])^2\) and the squared spectrum estimation error
\(\|\widehat I_{\mathrm{grid}}^{\texttt{max}}-I_{\mathrm{grid}}^{(0)}\|^2\). The reported MSEs are the averages of these losses over the \(R=200\) Monte Carlo simulations. 

We first assess whether further \texttt{plug-in} updates or joint profile likelihood improve loading and spectrum MSEs over \texttt{plug-in} with \(s=1\) (Table~\ref{tab:endor-mse-iteration-comparison}). Further \texttt{plug-in} updates yield negligible MSE gains and sometimes increase empirical MSE. This is consistent with Theorems~\ref{thm:iterated-plugin-loading-expansion-clt} and~\ref{thm:maximum-aligned-loading-expansion-clt}, which imply, for every fixed \(s\ge2\),
\[
\left\|\widehat\kappa^{[s],\texttt{opt}}-\widehat\kappa^{[1],\texttt{opt}}\right\|
=
\mathcal O_p(n^{-1}),
\quad
\left\|\widehat I^{[s],\texttt{max}}-\widehat I^{[1],\texttt{max}}\right\|
=
\mathcal O_p(n^{-1}).
\]
Together with the comparable MSEs of \texttt{PL}, these results support using \(s=1\) for the comparison with \texttt{cov}.

\begin{table}[tbh]
\centering
\caption{Relative Monte Carlo MSE changes (\%) from \texttt{plug-in} with \(s=1\). Entries give \([\min\Delta_m,\max\Delta_m]\) across 184 settings, where \(\Delta_m=100(\mathrm{MSE}_m/\mathrm{MSE}_{s=1}-1)\).}
\label{tab:endor-mse-iteration-comparison}
\begin{tabular}{lcc}
\hline
Method & Loading (\%) & Spectrum (\%) \\
\hline
\texttt{plug-in}, \(s=2\) & $[0.00000,\; +1.58770]$ & $[-0.09648,\; +1.04512]$ \\
\texttt{plug-in}, \(s=3\) & $[0.00000,\; +1.63337]$ & $[-0.09735,\; +1.07573]$ \\
\texttt{plug-in}, \(s=4\) & $[0.00000,\; +1.63472]$ & $[-0.09736,\; +1.07664]$ \\
\texttt{PL} & $[-0.00119,\; +2.28939]$ & $[-0.24940,\; +1.51353]$ \\
\hline
\end{tabular}
\end{table}

\begin{figure}[p]
\centering
\includegraphics[
  width=\linewidth,
  height=.9\textheight
]{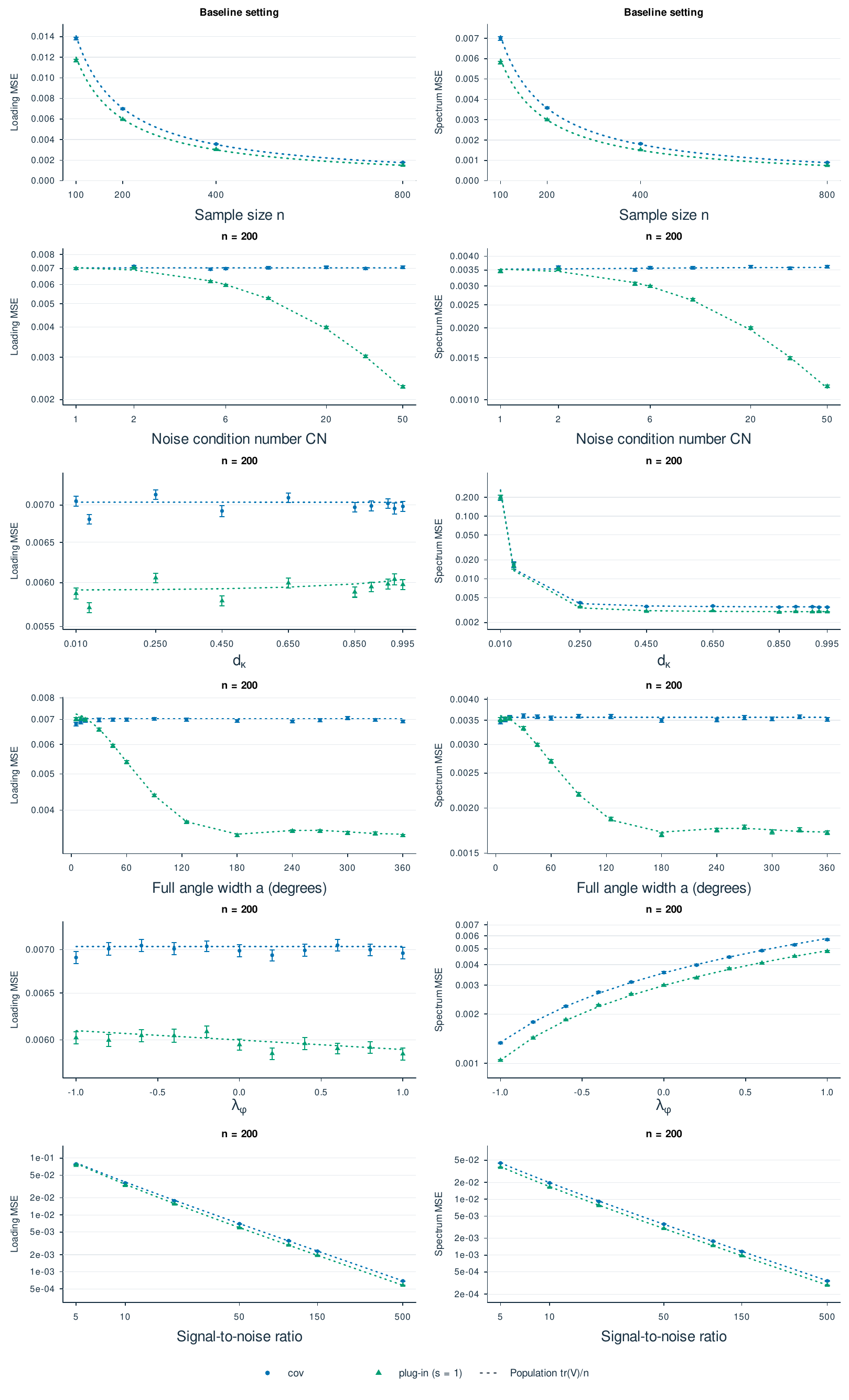}
\caption{Monte Carlo loading MSEs (left) and spectrum MSEs (right) for \texttt{cov} and \texttt{plug-in} with \(s=1\), based on \(R=200\) datasets per setting.
Rows vary \(n\), \(\mathrm{CN}\), \(d_\kappa\), \(a\),
\(\lambda_\phi\), and \(\eta\), respectively, with the
remaining model parameters at their baseline values.
Sample size is \(n=200\) except in the first row.
Points and error bars show empirical MSEs and
\(\pm1\) Monte Carlo standard error. Dotted curves show the first-order MSE approximations obtained from the full asymptotic covariance formulas evaluated at the generating parameters.}
\label{fig:endor-simulation-mse}
\end{figure}

Figure~\ref{fig:endor-simulation-mse} compares the loading and spectrum MSEs of \texttt{cov} and \texttt{plug-in} with \(s=1\) across six parameter sweeps. The corresponding first-order approximations are \(n^{-1}\operatorname{Tr}(V^m)\) and \(n^{-1}\operatorname{Tr}(V_I^{m,\texttt{max}})\), respectively, for \(m\in\{\texttt{cov},\texttt{plug-in}\}\).

In the first row, under the baseline setting, both MSEs decrease approximately as \(n^{-1}\) and closely follow the asymptotic predictions even at \(n=100\). This supports the accuracy of the first-order approximations at moderate sample sizes. The \texttt{plug-in} estimator has lower MSE for both targets throughout the sample-size comparison.

In the second row, the methods have nearly identical MSEs under isotropic noise (\(\mathrm{CN}=1\)), consistent with Theorems~\ref{thm:trace-comparison}.(i)
and~\ref{thm:maximum-spectrum-trace-comparison}.(i).
The growing advantage of \texttt{plug-in} as noise anisotropy increases agrees with the high-SNR comparisons in Theorems~\ref{thm:trace-comparison}.(ii)
and~\ref{thm:maximum-spectrum-trace-comparison}.(ii).

The third row shows little variation in loading MSE with \(d_\kappa\). In the leading trace coefficients in \eqref{eq:cov-loading-trace-decomposition} and~\eqref{eq:plugin-loading-trace-decomposition}, \(d_\kappa\) affects only one bounded term for \texttt{plug-in}; the remaining \(p-2\) terms and the \texttt{cov} coefficient are independent of \(d_\kappa\). Spectrum MSE, however, rises sharply near \(d_\kappa=0\), where nearly equal spectrum and wave energies make maximum alignment sensitive to estimation error. This sensitivity is reflected in the \(d_\kappa^{-2}\) terms in \eqref{eq:cov-maximum-spectrum-trace-decomposition}
and~\eqref{eq:plugin-maximum-spectrum-trace-decomposition}.

In the fourth row, increasing \(a\) up to \(180^\circ\) decreases \(\rho_\phi\) and widens the MSE advantage of \texttt{plug-in}, consistent with Theorems~\ref{thm:trace-comparison}.(ii) and~\ref{thm:maximum-spectrum-trace-comparison}.(ii). Beyond \(180^\circ\), \(\rho_\phi\) first increases and then decreases to zero at \(360^\circ\). At the baseline alignment \(\lambda_\phi=0\), the leading-order advantage therefore weakens and then strengthens again. At \(a=5^\circ\) and \(10^\circ\), the methods have nearly identical MSEs, with slightly larger loading MSE for \texttt{plug-in}. Here \(\rho_\phi\) is close to one, so the leading advantage is small and the \(\eta^{-2}\) terms can affect the comparison.

In the fifth row, loading MSE changes little with \(\lambda_\phi\), which enters the loading traces only through the \(\eta^{-2}\) terms. Spectrum MSE increases as larger \(\lambda_\phi\) places more energy of \(\phi_i I^{(0)}\) along the high-variance noise axis. This contrast agrees with the trace decompositions in Theorems~\ref{thm:trace-comparison} and~\ref{thm:maximum-spectrum-trace-comparison}.

In the last row, both axes use logarithmic scales. At high SNR, the loading and spectrum MSEs lie approximately along straight lines with slope \(-1\), indicating that both decrease approximately as \(\eta^{-1}\). This agrees with the leading terms in the trace decompositions of Theorems~\ref{thm:trace-comparison} and~\ref{thm:maximum-spectrum-trace-comparison}.

\medskip
\noindent\textit{Spectrum confidence intervals.}
We next compare the reliability of asymptotic spectrum confidence intervals from \texttt{cov} and \texttt{plug-in} with bootstrap intervals from \texttt{PL} and \texttt{PL-BC}, particularly at low SNR. We consider \(\eta\in\{50,20,10,5\}\) and \(n\in\{100,200,400,800\}\), with the other model parameters at their baseline values.

For \texttt{cov} and \texttt{plug-in}, we evaluate the closed-form asymptotic covariance matrices \(V_I^{m,\texttt{max}}\) at estimated parameters, giving
\(\widehat V_I^{m,\texttt{max}}\),
\(m\in\{\texttt{cov},\texttt{plug-in}\}\). The estimated covariance matrices on the original 71-point grid are
\(
H^\top\widehat V_I^{m,\texttt{max}}H/n\). Pointwise standard errors are the square roots of their diagonal entries. See~\ref{supp:subsec:spectrum-covariance-estimation} of the Supplementary Material for the details.

For \texttt{PL} and \texttt{PL-BC}, we adopt the parametric bootstrap approach of \cite{pokern2021statistical} for the centered model \eqref{model:centered-complex-factor}. The PL fit is defined by minimizing \eqref{eq:profile-loglik}. For \texttt{PL}, bootstrap datasets are generated from the fitted signal \(\widehat\phi_i^{\texttt{PL}}\widehat\kappa^{\texttt{PL}}\) with independent noise coordinates distributed as \(\mathcal{CN}(0,\widehat\Sigma^{\texttt{PL}})\). Here \(\widehat\phi_i^{\texttt{PL}}=\widehat\phi(Y_i,\widehat\kappa^{\texttt{PL}},\widehat\Sigma^{\texttt{PL}})\) is the profiled factor-score estimate. The bootstrap generating models below are expressed in the \(p\)-dimensional Helmert coordinates.

The fitted scores absorb part of the measurement error, which can inflate the estimated factor-score energy. To address this inflation and downward bias in the noise covariance, \texttt{PL-BC} applies additive bias correction to \(\widehat\Sigma^{\texttt{PL}}\) and multiplicative bias correction to the fitted scores (Section~4.3 of the Supplementary Information to \cite{pokern2021statistical}). The corrected parameters generate the final bootstrap datasets, while the original loading and spectrum point estimates are retained.

For both procedures, standard errors are the coordinatewise sample standard deviations of the maximum-aligned spectra on the original 71-point grid, obtained by applying \(H^\top\) to the spectrum coordinates from each of the final \(B\) bootstrap refits. We use \(B=1000\) replications per stage, requiring \(B\) refits for \texttt{PL} and \(3B\) for \texttt{PL-BC}, including its two bias-correction stages.

At each frequency \(\nu_j\), \(j=0,\ldots,p\), all methods use
a pointwise \(95\%\) confidence interval of the form
\begin{equation}
\label{eq:endor-simulation-pointwise-ci}
\widehat I_{\mathrm{grid},j}^{\texttt{max}} \pm z_{0.975}\,\widehat{\operatorname{se}}\bigl(
    \widehat I_{\mathrm{grid},j}^{\texttt{max}}\bigr), 
\quad z_{0.975}\simeq1.96.
\end{equation}

\begin{figure}[tbh]
\centering
\includegraphics[
  width=\linewidth,height=.7\textheight
]{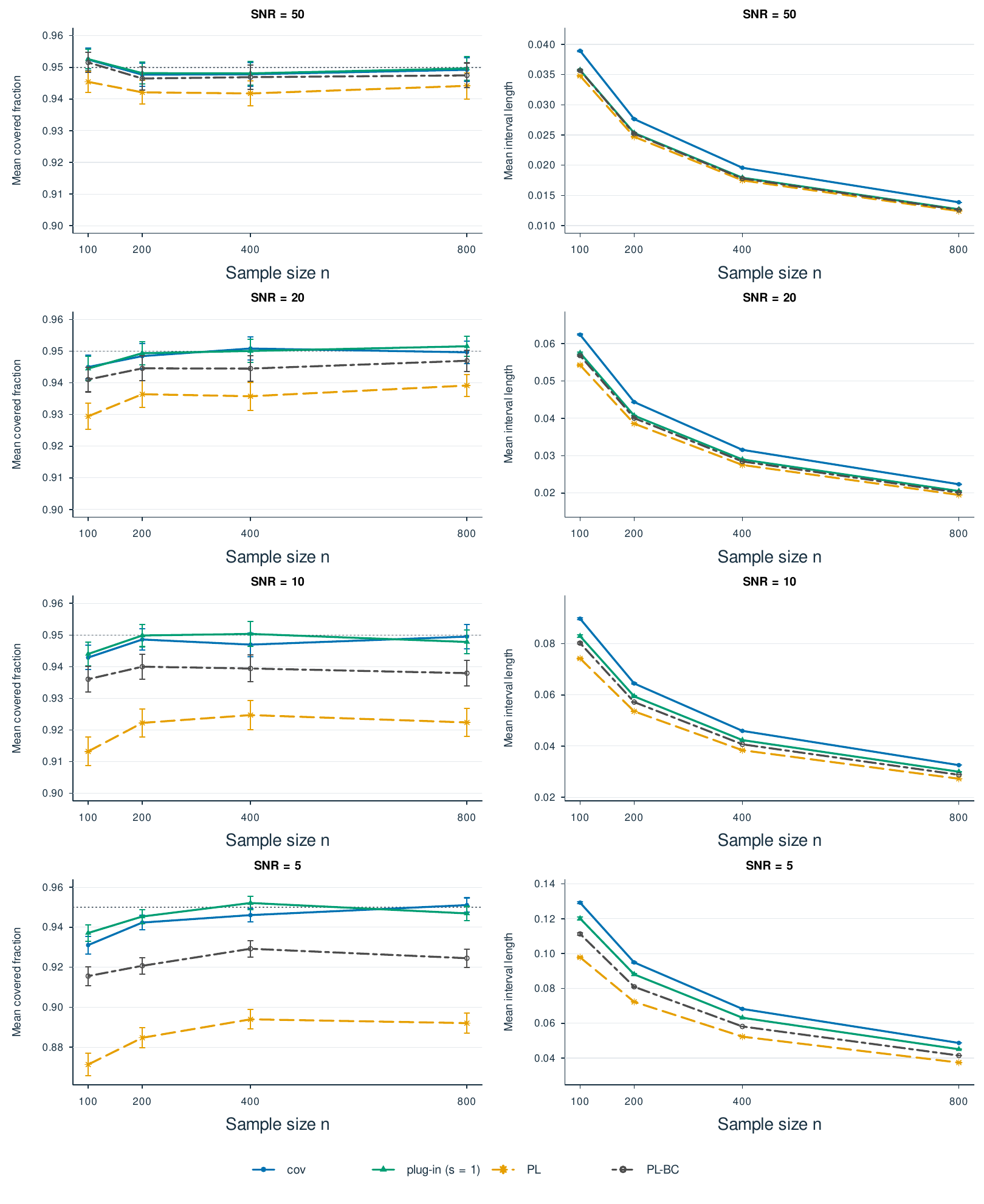}
\caption{Mean coverage (left) and mean length (right) of pointwise 95\% spectrum confidence intervals for \texttt{cov}, \texttt{plug-in} (\(s=1\)), \texttt{PL}, and \texttt{PL-BC}, averaged over 71 frequencies and \(R=200\) independent datasets per setting. Error bars show \(\pm 1.96\)  Monte Carlo standard errors (MCSEs), computed across the independent datasets after averaging over frequencies within each dataset. Rows correspond to \(\eta=50,20,10,5\), from top to bottom, with other model parameters at their baseline values. Horizontal dotted lines mark the nominal coverage of \(0.95\).}
\label{fig:endor-simulation-snr-ci}
\end{figure}

Figure~\ref{fig:endor-simulation-snr-ci} shows that \texttt{cov} and \texttt{plug-in} generally attain mean coverage close to \(0.95\), supporting the normal approximations at moderate sample sizes. Some under-coverage remains at the lowest SNR and smallest sample size, particularly for \texttt{cov}, but diminishes as \(n\) increases. The \texttt{plug-in} intervals are shorter than those of \texttt{cov}, with comparable or better coverage.

At \(\eta=50\), \texttt{PL-BC} achieves nearly nominal coverage, whereas \texttt{PL} shows mild under-coverage. However, at lower SNR, both methods exhibit greater under-coverage across sample sizes, with shorter intervals than \texttt{cov} and \texttt{plug-in}. Bias correction helps but remains insufficient: at \(\eta=5\) and \(n=800\), mean coverage is approximately \(0.89\) for \texttt{PL} and \(0.92\) for \texttt{PL-BC}.

A normal approximation provides a quantitative link between shorter intervals and under-coverage. If the estimation error is approximately centered normal and the estimated pointwise standard errors are a factor \(c\) of the true standard errors, the resulting coverage is approximately \(2\Phi(cz_{0.975})-1\), where \(\Phi\) is the standard normal cumulative distribution function. For \(c=0.9\), the effective cutoff is approximately \(1.764\), giving coverage of \(0.922\) instead of \(0.95\). This approximation closely matches the simulation comparisons examined here. This agreement is consistent with underestimated standard errors, although mean length ratios provide only a rough proxy for pointwise scaling.

The observed under-coverage and short confidence intervals may reflect an inflated SNR in the bootstrap generating model. Profiling $\phi_{i}$s can absorb observational noise into the fitted factor scores, inflating the estimated signal strength \(n^{-1}\sum_{i=1}^n|\widehat\phi_i^{\texttt{PL}}|^2\). Together with observed downward bias in the noise covariance, this can yield underestimated spectrum standard errors. Absorbed noise has a larger relative effect at low SNR, even if the relative covariance bias changes only slightly. This inflated SNR may also limit \texttt{PL-BC}: bootstrap biases of the covariance and total factor-score energy estimates under the fitted model may poorly approximate those under the true generating model. The score correction in \texttt{PL-BC} rescales all fitted scores by a common positive factor. It therefore preserves their phases and the normalized second moment \(\sum_i(\widehat\phi_i^{\texttt{PL}})^2/\sum_i|\widehat\phi_i^{\texttt{PL}}|^2\). If noise in the fitted phases distorts this moment, the distortion persists in the bootstrap generating model and may contribute to the remaining under-coverage. Our simulations do not isolate the size of this effect.

Previous studies have shown that the choice of parameter estimates used to generate bootstrap samples can affect inferential accuracy in models with many nuisance parameters \cite{zhao2025adaptively,benussi2026accurate}. In a related study of fixed-effects PCA, \citet{josse2016confidence} generated bootstrap samples from the fitted low-rank model using a residual-based estimate of the common noise variance, a nuisance parameter for inference on the row scores. The resulting confidence regions for the row scores also showed undercoverage at low SNR in simulations. These findings motivate examining whether the estimated factor scores used in our bootstrap procedure contribute to the observed undercoverage, although this connection remains to be established.

\medskip
\noindent\textit{Computation time and summary.}
Table~\ref{tab:endor-computation-time} compares fitting and
confidence interval construction times at the baseline SNR
\(\eta=50\).
The \texttt{cov} estimator uses eigendecomposition and closed-form moment calculations, while \texttt{plug-in} uses the block coordinate updates described in~\ref{supp:subsec:fixed-covariance-loading-algorithm} of the Supplementary Material. For the \texttt{PL} fit, we adapt the alternating optimization scheme in Section~4.1 of the SI Appendix to \cite{pokern2021statistical} to the objective \eqref{eq:profile-loglik}. The \texttt{cov} estimator is the fastest to fit. The \texttt{plug-in} estimator with $s=1$ has lower mean fitting times than PL at $n=100,200,400$, whereas PL is faster at $n=800$. Further \texttt{plug-in} iterations increase fitting time without appreciable MSE gains.

The largest differences arise in interval construction. For \texttt{cov} and \texttt{plug-in}, this requires evaluating closed-form covariance formulas, whereas the bootstrap procedures require repeated model fitting. \texttt{cov} and \texttt{plug-in} require only a few milliseconds, compared with tens to hundreds of seconds for the bootstrap procedures. Similar patterns are observed at the other SNR levels considered;
detailed results are provided in the Supplementary Material~\ref{subsec:endor-simulation-additional-results}. Together with the MSE and coverage results, these timings favor \texttt{plug-in} with \(s=1\) as the best overall balance of estimation accuracy, interval coverage, and computational cost in the settings considered.

\begin{table}[tbp]
\centering
\caption{Mean elapsed times \(\pm\) one Monte Carlo standard error over \(R=200\) datasets at the baseline setting (\(\eta=50\)). Panel (a) includes \texttt{cov} initialization and all \(s\) updates for \texttt{plug-in}, but excludes bias correction and interval construction. Panel (b) excludes the original fit. Bootstrap timings use \(B=1000\) per stage and include both correction stages for \texttt{PL-BC}.}
\label{tab:endor-computation-time}
\begingroup
\footnotesize
\scriptsize
\setlength{\tabcolsep}{1.5pt}
\renewcommand{\arraystretch}{1.0}

\begin{tabular}{@{}crrrrrr@{}}
\multicolumn{7}{c}{\textit{(a) Point estimation (milliseconds)}} \\[3pt]
\toprule
\multirow{2}{*}{$n$}
& \multirow{2}{*}{\texttt{cov}}
& \multicolumn{4}{c}{\texttt{plug-in}}
& \multirow{2}{*}{\texttt{PL}} \\
\cmidrule(lr){3-6}
& & $s=1$ & $s=2$ & $s=3$ & $s=4$ & \\
\midrule
100 & $3.65\pm0.06$ & $8.83\pm0.11$ & $13.93\pm0.16$ & $18.96\pm0.21$ & $28.30\pm0.50$ & $99.19\pm0.81$ \\
200 & $5.17\pm0.34$ & $35.30\pm2.33$ & $83.09\pm5.58$ & $98.91\pm5.46$ & $107.23\pm5.51$ & $100.16\pm0.61$ \\
400 & $4.63\pm0.06$ & $74.27\pm5.26$ & $94.68\pm5.21$ & $110.20\pm5.14$ & $126.29\pm5.20$ & $110.28\pm1.05$ \\
800 & $28.67\pm1.74$ & $226.28\pm6.86$ & $265.73\pm6.88$ & $306.28\pm7.23$ & $346.29\pm7.36$ & $183.44\pm2.87$ \\
\bottomrule
\end{tabular}

\par\smallskip

\begin{tabular}{@{}crrrr@{}}
\multicolumn{5}{c}{\textit{(b) Confidence interval construction (seconds)}} \\[3pt]
\toprule
$n$
& \texttt{cov}
& \texttt{plug-in} ($s=1$)
& \texttt{PL}
& \texttt{PL-BC} \\
\midrule
100 & $0.00278\pm0.00004$ & $0.00273\pm0.00004$ & $83.67\pm1.13$ & $246.00\pm3.03$ \\
200 & $0.00280\pm0.00004$ & $0.00265\pm0.00004$ & $101.50\pm2.01$ & $297.49\pm5.80$ \\
400 & $0.00279\pm0.00004$ & $0.00275\pm0.00004$ & $106.92\pm1.28$ & $315.37\pm3.39$ \\
800 & $0.00698\pm0.00239$ & $0.00453\pm0.00007$ & $161.09\pm2.32$ & $472.90\pm6.23$ \\
\bottomrule
\end{tabular}
\endgroup
\end{table}

\subsection{Real-data analysis of ENDOR spectra}
\label{subsec:pnas-endor}

We analyze two $263$-GHz $^1$H Davies ENDOR datasets from the D2 isotopologue of the tyrosyl radical $Y_{122}^{\bullet}$ in the $\beta_2$ subunit of \textit{Escherichia coli} ribonucleotide reductase \cite{pokern2021statistical}. In this isotopologue, the two hydrogen atoms at the $\beta$-methylene sites are replaced by deuterium. We use the field positions denoted $g_y$ and $g_x$ in the original study (SI Appendix, Figure~S5), from version~1 of the public data archive \cite{PNASdata}. The datasets contain $n=186$ and $n=300$ batches, respectively, each measured at $p+1=360$ frequencies. For each dataset, we fit \eqref{model:centered-complex-factor} in \(p=359\) Helmert coordinates and report spectra and pointwise confidence intervals on the original grid of \(p+1=360\) frequencies. We compare the maximum-spectrum estimates and pointwise $95\%$ confidence intervals from \texttt{cov}, \texttt{plug-in} with $s=1$, \texttt{PL}, and \texttt{PL-BC}.

In the original study, goodness-of-fit tests based on residuals from the fitted model~\eqref{model:uncentered-complex-factor}
did not reject normality of either the real or imaginary components for these two datasets
\cite{pokern2021statistical}; see its SI Appendix, Section~7 and Table~S1. These results support the Gaussian error assumption underlying model~\eqref{model:uncentered-complex-factor}. 

\begin{figure}[tbp]
\centering
\includegraphics[width=\linewidth]
{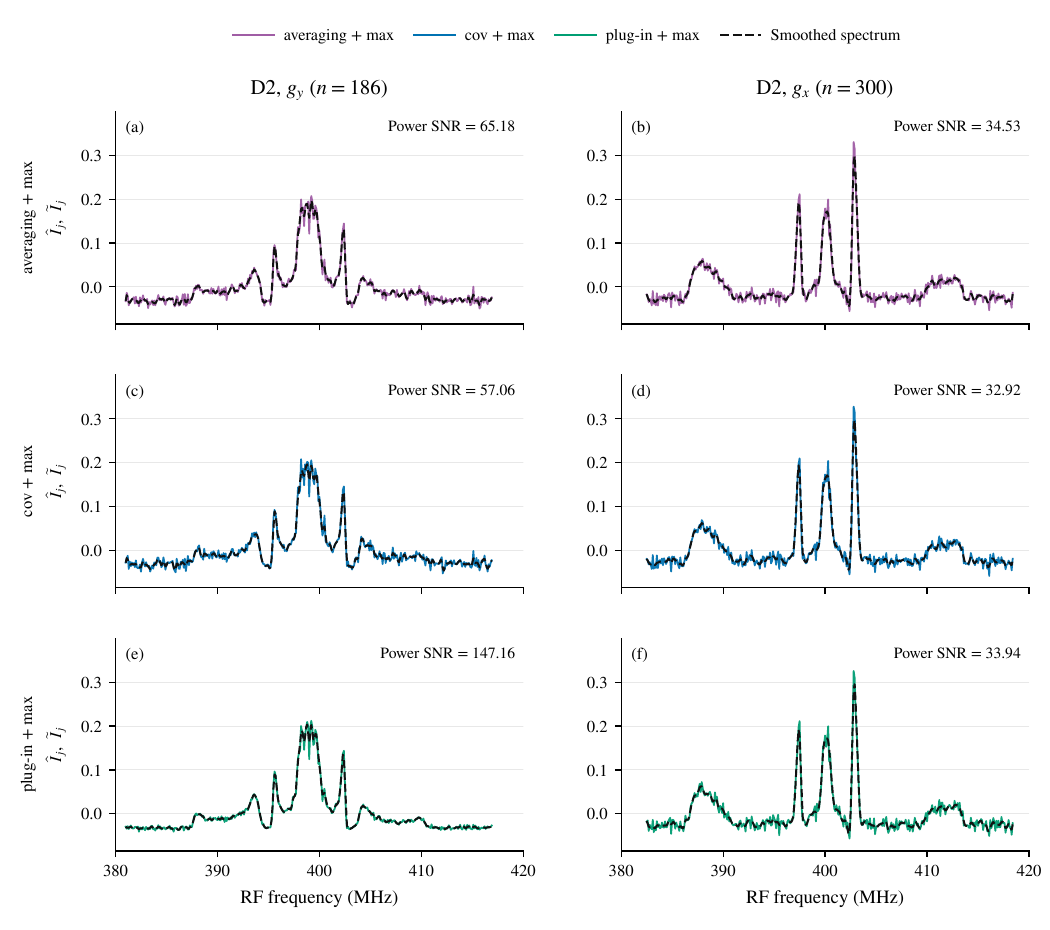}
\caption{Estimated and smoothed maximum spectra for the published D2 ENDOR data
\cite{pokern2021statistical,PNASdata}.
Columns show $g_y$ (left) and $g_x$ (right); rows show \texttt{averaging}, \texttt{cov}, and \texttt{plug-in}, each followed by the maximum rotation. Colored solid curves show the estimated spectra, and black dashed curves show their smoothed spectra. The common smoothing bandwidths are $0.1251$ MHz for $g_y$ and $0.1529$ MHz for $g_x$. Each panel reports the power SNR defined in the text.}
\label{fig:pnas-spectra-snr}
\end{figure}

Figure~\ref{fig:pnas-spectra-snr} compares the
maximum-spectrum estimates from \texttt{averaging}, \texttt{cov}, and \texttt{plug-in} with their smoothed versions. For \texttt{averaging}, we average the original complex observations across batches at each frequency, center the resulting vector over frequencies, and normalize it to unit Euclidean norm. We then apply the same maximum method and sign convention as for the other estimators. The \texttt{PL} spectrum is nearly identical to the \texttt{plug-in} spectrum, and \texttt{PL-BC} retains the \texttt{PL} point estimate by construction; we therefore omit these two methods from the figure.

We use Gaussian-weighted local quadratic smoothing with a common
bandwidth across methods within each dataset. Details of bandwidth
selection are given in~\ref{supp:subsec:endor-bandwidth-selection} of the Supplementary Material.
For each method \(m\), let \(\widehat I_{m,j}:=(\widehat I_{\text{grid}}^{m,\texttt{max}})_j\) denote the estimated spectrum at the measured frequency \(\nu_j\). Let \(\widetilde I_{m,j}\) denote the corresponding value after smoothing and centering over frequencies. To assess spectral quality, we use the centered smooth as a proxy for signal and the residual as a proxy for noise. We define the descriptive power SNR as
\[
\text{Power}~\mathrm{SNR}_{m}
=
\frac{\sum_{j=0}^{p}\widetilde I_{m,j}^{\,2}}
     {\sum_{j=0}^{p}(\widehat I_{m,j}-\widetilde I_{m,j})^2}.
\]

A larger value indicates stronger smooth spectral features relative to the remaining fluctuations. For $g_y$, the \texttt{plug-in} power SNR is $147.16$,
compared with $57.06$ for \texttt{cov} and $65.18$ for \texttt{averaging}. The higher power SNR reflects smaller irregular fluctuations around a broadly similar smooth spectrum. For $g_x$, the corresponding values are $33.94$, $32.92$, and $34.53$, indicating comparable spectral
quality under this measure. This descriptive power SNR differs from the model SNR $\eta$ and depends on the smoothing bandwidth and spectral shape. Since the remaining fluctuations may contain narrow spectral features as well as noise, it does not measure error relative to the unknown true spectrum.

\begin{figure}[tbp]
\centering
\includegraphics[width=\linewidth]
{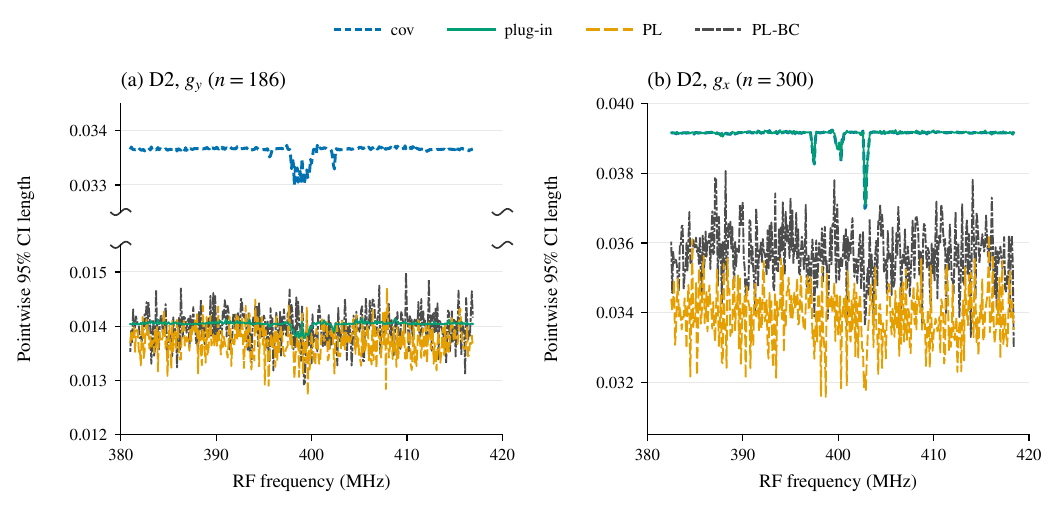}
\caption{Pointwise $95\%$ confidence-interval lengths
for the published D2 ENDOR data
\cite{pokern2021statistical,PNASdata}:
(a) $g_y$ and (b) $g_x$.
Curves show $\ell_{m,j}=U_{m,j}-L_{m,j}$ for \texttt{cov}, \texttt{plug-in}, \texttt{PL}, and \texttt{PL-BC}. The vertical axis in panel (a) has a break between 0.0155 and 0.0325; all observed CI lengths remain visible.}
\label{fig:pnas-example}
\end{figure}

For each method $m\in\{\texttt{cov},\texttt{plug-in},
\texttt{PL},\texttt{PL-BC}\}$, we construct pointwise $95\%$ confidence intervals on
the original frequency grid as in \eqref{eq:endor-simulation-pointwise-ci}. For \texttt{cov} and \texttt{plug-in}, we obtain standard errors from the estimated asymptotic covariance matrices. For \texttt{PL} and \texttt{PL-BC}, we use parametric-bootstrap standard errors, with the latter based on bias-corrected nuisance parameter estimates. All confidence intervals are based on the unsmoothed estimates. Denoting the lower and upper endpoints at frequency $j$ by $L_{m,j}$ and $U_{m,j}$, respectively, we compare the root-mean-square (RMS) lengths of the pointwise confidence intervals,
\[
R_m
=
\left\{
\frac{1}{p+1}
\sum_{j=0}^{p}(U_{m,j}-L_{m,j})^2
\right\}^{1/2}.
\]
For \texttt{cov} and \texttt{plug-in}, the ratio of these RMS lengths equals the square root of the ratio of their estimated spectrum covariance traces. Table~\ref{tab:pnas-example} reports the parameter estimates and RMS CI lengths, while Figure~\ref{fig:pnas-example} shows the CI lengths across frequency.

\begin{table}[tbp]
\centering
\caption{Estimated parameters and confidence-interval lengths for the published D2 ENDOR data \cite{pokern2021statistical,PNASdata}. Panel (a) reports \texttt{plug-in} parameter estimates. Panel (b) reports root-mean-square (RMS) pointwise $95\%$ confidence-interval lengths on the original frequency grids.}
\label{tab:pnas-example}
\small
\begin{tabular}{@{}lrrrrrr@{}}
\toprule
\multicolumn{7}{l}{\textit{(a) Data characteristics and \texttt{plug-in} parameter estimates}}\\
\addlinespace[3pt]
Dataset & $n$ & $\widehat d_\kappa$ & $\widehat d_\Sigma$ & $\widehat\rho_\phi$ & $\widehat\lambda_\phi$ & $\widehat\eta$ (SNR)\\
\midrule
D2, $g_y$ & 186 & 0.9446 & 0.9395 & 0.6173 & $-0.3567$ & 29.94 \\
D2, $g_x$ & 300 & 0.8873 & 0.8033 & 0.9774 & $-0.0300$ & 17.33 \\
\bottomrule
\end{tabular}

\par\vspace{8pt}

\begin{tabular}{@{}lrrrr@{}}
\toprule
\multicolumn{5}{l}{\textit{(b) RMS CI lengths $R_m$}}\\
\addlinespace[3pt]
Dataset & \texttt{cov} & \texttt{plug-in} & \texttt{PL} & \texttt{PL-BC}\\
\midrule
D2, $g_y$ & 0.033633 & 0.014041 & 0.013731 & 0.013981 \\
D2, $g_x$ & 0.039128 & 0.039132 & 0.033928 & 0.035689 \\
\bottomrule
\end{tabular}

\par\vspace{5pt}

\begin{minipage}{0.98\linewidth}
\footnotesize
\texttt{PL} uses $1{,}000$ parametric-bootstrap replicates, whereas \texttt{PL-BC} uses $1{,}000$ replicates at each of its two bias-correction stages and the final uncertainty-estimation stage.
\end{minipage}
\end{table}

The $g_y$ dataset has the higher estimated model SNR and the lower estimated $\rho_\phi$ of the two examples (Table~\ref{tab:pnas-example}). It also has strongly anisotropic noise, with $\widehat d_\Sigma=0.9395$.
The RMS CI lengths follow the pattern
\[
R_{\texttt{cov}}
\gg R_{\texttt{plug-in}}
\approx R_{\texttt{PL}}
\approx R_{\texttt{PL-BC}}.
\]
The ratio $R_{\texttt{plug-in}}/R_{\texttt{cov}}$
is closely approximated by the high-SNR trace formula in Theorem~\ref{thm:maximum-spectrum-trace-comparison}. Evaluated at the \texttt{plug-in} parameter estimates, the square root of the leading-order spectrum trace ratio is approximately $0.421$, compared with the computed RMS CI-length ratio of $0.417$.

The $g_x$ dataset has a lower estimated model SNR and $\widehat\rho_\phi=0.9774$, close to one. Here, \texttt{cov} and \texttt{plug-in} yield nearly identical RMS CI lengths, while \texttt{PL-BC} and \texttt{PL} give shorter intervals:
\[
R_{\texttt{cov}}
\approx R_{\texttt{plug-in}}
> R_{\texttt{PL-BC}}
> R_{\texttt{PL}}.
\]
The small difference between \texttt{cov} and
\texttt{plug-in} is consistent with Theorem~\ref{thm:maximum-spectrum-trace-comparison}: the leading-order efficiency gain diminishes as $\rho_\phi$ approaches one, while finite-SNR terms can affect the comparison.

The normal approximation in Section~\ref{subsec:endor-simulation-analysis} helps interpret the observed interval lengths. Taking \texttt{plug-in} as an approximately calibrated reference, an RMS length ratio \(c\) corresponds to an approximate coverage of \(2\Phi(cz_{0.975})-1\), provided that estimation bias is negligible, the estimators have comparable sampling variability, and their pointwise interval widths are approximately proportional across frequencies. For \(g_x\), the length ratios of \texttt{PL} and \texttt{PL-BC} relative to \texttt{plug-in} are \(0.867\) and \(0.912\), giving illustrative coverages of \(91.1\%\) and \(92.6\%\), respectively. The corresponding values for \(g_y\) are \(94.5\%\) and \(94.9\%\). Even at the same model SNR, the relative effect of noise on the fitted score second moments can vary with the model geometry. Thus, the shorter intervals for \(g_x\) suggest a possible coverage loss despite the near-nominal coverage observed in simulations at a comparable SNR. Actual coverage remains unknown for both datasets.

The consistency and asymptotic normality results (Theorems~\ref{thm:maximum-aligned-strong-consistency} and~\ref{thm:maximum-aligned-loading-expansion-clt}), the asymptotic efficiency gains over \texttt{cov} at high SNR, and the near-nominal coverage support using \texttt{plug-in} for spectrum estimation and inference.

\section{Conclusion and Future work}\label{sec:Conclusion}
We proposed a \texttt{plug-in} procedure for the complex-valued one-factor model that alternates estimation of the loading and the anisotropic noise covariance. Under suitable moment and identifiability conditions, we established strong consistency, derived stochastic expansions, and obtained central limit theorems with explicit asymptotic covariance formulas. The comparison with the \texttt{cov} estimator shows that exploiting directional differences in noise can improve estimation efficiency. By giving greater weight to low-variance directions, the \texttt{plug-in} procedure combines information from observations at different phases according to their precision. The gain depends on noise anisotropy, factor-score second moments, and other model features.

These theoretical results provide a practical procedure for spectrum inference. The normal approximation from the loading CLT is accurate even at relatively small sample sizes, so we can compute pointwise standard errors and confidence intervals directly from the asymptotic covariance of the spectrum estimator without bootstrap resampling. Indeed, in our simulations with $n=100$ and $p=70$, confidence intervals from the \texttt{plug-in} procedure generally achieved near-nominal coverage at substantially lower computational cost than the bootstrap procedures for \texttt{PL} and \texttt{PL-BC}. The coverage advantage was particularly clear at low signal-to-noise ratios, where both bootstrap profile-likelihood procedures showed clear under-coverage.

One direction for future work is to apply model~\eqref{model:uncentered-complex-factor} directly to repeated measurements of landmark positions in $\mathbb{R}^2$ in order to estimate a common planar shape. The model accommodates translations, rotations, scale changes, and measurement error, with the loading class representing the common true shape. Two key differences to other approaches to shape estimation are that no parametric assumptions are made regarding the distribution of rotations and scale changes and that the additive measurement error for landmarks need not exhibit rotational symmetry. This application also motivates further work on intrinsic confidence regions for the shape that do not depend on the choice of representative.

For some ENDOR datasets, the physical spectrum \(I\) and wave \(W\) have comparable norms and need not be orthogonal. Thus, the phase rotation used by the maximum method to make the two components orthogonal can mix part of the wave into the extracted spectrum. To address this problem, \citet{pokern2021statistical} introduced the minimum method, which uses the shape of the instrumental wave component to guide phase selection. This motivates extending our asymptotic analysis and confidence interval construction to the minimum method while retaining the same loading estimators.



Finally, the model can be extended to allow the noise covariance to vary across observations and depend on the observation-specific level \(\psi_i\) or factor score \(\phi_i\) which has also been observed in some spectrometers \cite{wiechers2023drift} owing to strong phase noise contributions of their microwave sources.

\begin{appendix}

\section{Generalized \fre~mean with random cost function}\label{sec: GFM}
This appendix develops general asymptotic results for generalized \fre~means whose empirical cost functions depend on an estimated auxiliary parameter, possibly obtained from the same observations. We establish strong consistency, asymptotic normality, and a refined stochastic expansion, which provide the theoretical tools for analyzing the iterated \texttt{plug-in} estimators.

\subsection{Generalized \fre~mean with random cost function}\label{sec: GFM definition}

We introduce the formal definition of the generalized \fre~mean with a random cost function as follows. Let $T$ be a topological space in which the data take values; we refer to $T$ as the data space. Throughout, we assume that the underlying probability space
$(\Omega,\mathcal A,\mathcal P)$ is complete, passing to its
completion if necessary.
Let $X:(\Omega,\mathcal A,\mathcal P)\to T$
be a $T$-valued random element. Assume that $X$ follows a distribution $P$ on $T$, and that $X_{1}, X_{2}, \ldots, X_{n}$ are i.i.d.\ samples from $P$. 

The parameter of interest takes values in a metric space $(M,d_{M})$, which we call the primary parameter space. To accommodate complex statistical models where the estimation relies on an unknown auxiliary parameter, we introduce a second metric space $(Q,d_{Q})$ to serve as the space of these auxiliary parameters.

Cost function \(\rho: T\times M \times Q \rightarrow \mathbb{R}\)
measures the cost of representing a data point $t\in T$ by a primary parameter $m\in M$ under the auxiliary configuration $q\in Q$. Throughout, we assume that $\rho(X,m,q)$ is integrable for all $m\in M$ and $q\in Q$, i.e., $\rho(X,m,q)$ is measurable and $\mathbb{E}|\rho(X,m,q)| < \infty$.

On the auxiliary parameter space $Q$, we assume the existence of a true parameter $q_{0}\in Q$. Our estimand is the primary parameter in $M$ that minimizes the expected cost parametrized by this true $q_{0}$. Since $q_{0}$ is unknown in practice, we adopt a \texttt{plug-in} approach by substituting it with its strongly consistent estimator $\hat{q}_{n}$. Precisely, we assume that $\lim_{n\rightarrow \infty} d_{Q}(\hat{q}_{n}, q_{0}) = 0$ almost surely. The estimator $\hat{q}_{n}$ may be independent of $(X_{1},\dots, X_{n})$, but it is typically constructed from the same data and hence dependent on $(X_{1},\dots, X_{n})$. When \(\hat q_n=q_0\) for all \(n\), the random cost function reduces to the usual fixed cost function, so the framework and results below include the
conventional theory of generalized \fre~means \cite{huckemann2011intrinsic, schotz2022strong} as a special case.

Guided by this plug-in principle, given the consistent estimator $\hat{q}_n$, we define the empirical plug-in cost function and its population counterpart evaluated at the true parameter $q_0$ as:
\[
    F_n(m) := \frac{1}{n} \sum_{i=1}^n \rho(X_i, m, \hat{q}_n) \quad \text{and} \quad F(m) := \mathbb{E}[\rho(X, m, q_0)].
\]
The empirical and population generalized \fre~mean sets, which represent our practical estimator and our target estimand, respectively, are given by 
\[
\hat{E}_{n} := \argmin_{m\in M} F_{n}(m) \quad \text{and} \quad E_{0} := \argmin_{m\in M} F(m).
\]

\medskip
\noindent\textit{Application to the \texttt{plug-in} estimator.}
For the loading updates in Section~\ref{subsec:plugin-estimator}, take
\(T=\mathbb C^p\), \(M=\mathbb{CP}^{p-1}\),
\(Q=\mathrm{SPD}(2)\), and \(X_i=Y_i\), with
\(q_0=\Sigma^{(0)}\).
The cost function is the profile loss
\(\rho(y,[\kappa],\Sigma)
=\min_{\phi\in\mathbb C}\|y-\phi\kappa\|_\Sigma^2\).
At each fixed iteration \(s\geq1\), set
\(\hat q_n=\widehat\Sigma^{\texttt{cov}}\) for \(s=1\) and
\(\hat q_n=\widehat\Sigma^{[s-1]}\) for \(s\geq2\).
The loading update then satisfies
\(\widehat{[\kappa]}^{[s]}\in\hat E_n\), so it is a generalized \fre~mean with a random cost function. Here, the cost is random through the covariance input \(\hat q_n\), which is estimated from the same observations \(Y_1,\ldots,Y_n\).

\medskip
\noindent\textit{Relation to previous literature.}
Classical generalized \fre~means minimize empirical averages of a fixed cost function over a fixed parameter space \cite{huckemann2011intrinsic,schotz2022strong}. In contrast, our framework allows the cost function to depend on an estimated auxiliary parameter, possibly obtained from the same observations. This differs from generalized \fre~means with random minimizing domains \cite{park2026generalized}, where the cost function is fixed but the domain of minimization may depend on the data. Thus, the two frameworks address different sources of randomness: here, the minimizing domain remains fixed, while the cost function is data-dependent.

Related data dependence arises in backward nested families of descriptors \cite{huckemann2018backward}, where previously estimated descriptors determine subsequent admissible sets and projected-data cost functions. Although the admissible sets may be fixed in special cases, this framework relies on a recursive nested descriptor structure. In contrast, our framework allows an arbitrary consistent auxiliary estimator to enter the cost function, without requiring nested descriptors or projection maps. Our framework is also related to classical two-step estimation \cite{murphy1985twostep,newey1994large}, which uses preliminary parameter estimates in subsequent estimation procedures and is typically formulated in Euclidean parameter spaces. For smooth manifold-valued parameters, the local asymptotic analysis in Section~\ref{sec: GFM CLT} combines this two-step principle with constrained M-estimation in Euclidean coordinates \cite{geyer1994constrained}.

\subsection{Strong Consistency of Generalized \fre~mean}\label{sec: GFM consistency}
In this subsection, we aim to clarify the general conditions when $\hat{E}_{n}$ is consistent with $E_{0}$ in the following sense:
\begin{equation}\label{eq: BP consistency}
\mathcal{P} \left(
 \left\{
  \omega\in \Omega :
  \lim_{n\rightarrow \infty}
  \sup_{m\in \hat{E}_{n}(\omega)} d_{M}(m, E_{0}) =0
 \right\}
\right)=1.
\end{equation}
Following \cite{Bhattacharya2003}, when \eqref{eq: BP consistency} is satisfied, $\hat{E}_{n}$ is termed a strongly consistent estimator of $E_{0}$ in the Bhattacharya-Patrangenaru sense (or BP consistent with $E_{0}$). For brevity, we will often refer to $\hat{E}_n$ as simply BP consistent when $E_{0}$ is apparent in the context.

Under BP consistency, almost surely, every sequence \(e_n\in\hat E_n\)
satisfies \(d_M(e_n,E_0)\to 0\). If \(E_0\) is a singleton, this implies convergence to its unique element as in the following Lemma (see also \cite{huckemann2011intrinsic, schotz2022strong, park2026generalized}).

\begin{lemma}\label{lem: singletone}
If $\hat{E}_{n}$ is BP consistent with $E_{0}$ and $E_{0}$ consists of a single element $\{e_{0}\}$, then almost surely, any sequence $e_{n}\in \hat{E}_{n}$ will converge to $e_{0}$.\end{lemma}

\begin{theorem}
\label{thm: BP-SC}
Under Conditions~1--7 in~\ref{supp:sec:gfm} of the Supplementary Material,
\(\widehat E_n\) is BP consistent with \(E_0\), in the sense of
\eqref{eq: BP consistency}.
\end{theorem}

The proof and details of the regularity conditions and their roles are provided in~\ref{supp:sec:gfm} of the Supplementary Material.

\subsection{Central Limit Theorem for Generalized Fr\'{e}chet Mean with Random cost function}
\label{sec: GFM CLT}

Let \(M\) and \(Q\) be differentiable manifolds of dimensions
\(D_M\) and \(D_Q\). Assume that \(E_0=\{m_0\}\).
The empirical argmin set \(\widehat E_n\) need not be a singleton;
we work with any measurable selection \(\widehat m_n:\Omega\to M\) satisfying \(\widehat m_n\in\widehat E_n\).

We use a Euclidean representation to state the asymptotic distribution of the estimator in coordinates of interest. Such a representation may require some geometric constraints, for example from normalization or phase
alignment. Let \(\mathcal{F}:U_{m_0}\to\mathbb R^K\) be a smooth embedding of a neighborhood of \(m_0\). We describe its image locally by \(h(\mu)=0\), where \(h\) maps an open neighborhood of \(\mathcal{F}(m_0)\) in \(\mathbb R^K\) into \(\mathbb R^P\), with \(P=K-D_M\). Assume that \(\mathcal{F}(U_{m_0})\) is relatively open in the embedded submanifold \(S=\{\mu:h(\mu)=0\}\). For the auxiliary parameter, let \(\mathcal{G}:V_{q_0}\to\mathbb R^{D_Q}\) be a local chart at \(q_0\). Write \(\mu_0=\mathcal{F}(m_0)\), \(\widehat\mu_n=\mathcal{F}(\widehat m_n)\), \(\nu_0=\mathcal{G}(q_0)\), and \(\widehat\nu_n=\mathcal{G}(\widehat q_n)\). The estimators belong to the coordinate neighborhoods with
probability tending to one. Outside this event, we assign arbitrary
fixed values to their coordinate representations.

On \(T\times \mathcal{F}(U_{m_0})\times \mathcal{G}(V_{q_0})\), define
\begin{equation}
\label{eq: mapped cost function}
    \ell(t,\mu,\nu)
    :=
    \rho\bigl(t,\mathcal{F}^{-1}(\mu),\mathcal{G}^{-1}(\nu)\bigr).
\end{equation}
Assume that \(\ell\) admits a jointly measurable local extension
\(\tilde{\ell}\) to an open neighborhood of \((\mu_0,\nu_0)\) in
\(\mathbb R^K\times\mathbb R^{D_Q}\), agreeing with \(\ell\)
on its original domain. All derivatives and derivative-envelope
conditions below refer to \(\tilde{\ell}\).

Write the auxiliary linear representation as
\begin{equation}
\label{eq: aux_linear}
    \sqrt n(\widehat\nu_n-\nu_0)
    =
    \frac1{\sqrt n}\sum_{i=1}^n\xi(X_i)+o_p(1),
\end{equation}
where \(\mathbb E[\xi(X)]=0\) and
\(\mathbb E[\|\xi(X)\|^2]<\infty\).
Let \(H\in\mathbb R^{K\times P}\) and
\(\lambda_0\in\mathbb R^P\) satisfy
\begin{equation}
\label{eq: jacobian_H}
    H:=Dh(\mu_0)^\top,
    \quad
    \mathbb E[\nabla_\mu\tilde{\ell}(X,\mu_0,\nu_0)]
    +H\lambda_0=0.
\end{equation}
Define
\begin{equation}
\label{eq: hessian_J}
    J
    :=
    \mathbb E[\nabla_{\mu\mu}^2\tilde{\ell}(X,\mu_0,\nu_0)]
    +
    \sum_{j=1}^P\lambda_{0,j}\nabla^2h_j(\mu_0),
\end{equation}
and let \(B_{11}\) be the upper-left \(K\times K\) block of
\(B^{-1}\), where
\(B:=\begin{pmatrix}J&H\\H^\top&0\end{pmatrix}\).
The effective score is
\begin{equation}
\label{eq: effective_score}
\begin{aligned}
    Z(X)
    &:=
    \nabla_\mu\tilde{\ell}(X,\mu_0,\nu_0)
    -
    \mathbb E[\nabla_\mu\tilde{\ell}(X,\mu_0,\nu_0)]
    +
    \mathbb E[\nabla_{\mu\nu}^2\tilde{\ell}(X,\mu_0,\nu_0)]
    \xi(X).
\end{aligned}
\end{equation}

\begin{theorem}
\label{thm: clt_gfm}
Assume Conditions~8--12 of~\ref{supp:sec:gfm} of the Supplementary Material and
\(\mathbb E[\|Z(X)\|^2]<\infty\). Then
\begin{equation}
\label{eq: first_order_expansion}
    \widehat\mu_n-\mu_0
    =
    -B_{11}\frac1n\sum_{i=1}^n Z(X_i)
    +o_p(n^{-1/2}),
\end{equation}
and, with \(V:=\operatorname{Var}\{Z(X)\}\),
\begin{equation}
\label{eq: clt_result}
    \sqrt n(\widehat\mu_n-\mu_0)
    \xrightarrow{d}
    N(0,B_{11}VB_{11}).
\end{equation}
If Condition~13 of~\ref{supp:sec:gfm} additionally holds, then
\begin{equation}
\label{eq: quadratic_stochastic_expansion}
    \widehat\mu_n-\mu_0
    =
    -B_{11}\frac1n\sum_{i=1}^n Z(X_i)
    +O_p(n^{-1}).
\end{equation}
\end{theorem}

Condition~13 is needed only for the \(O_p(n^{-1})\) remainder, which permits recursive application to a fixed number of \texttt{plug-in} updates. The proof and a detailed discussion of the regularity conditions and their roles are provided in
\ref{supp:sec:gfm} of the Supplementary Material.

\medskip
\noindent\textit{Application to optimally positioned loading estimators.}
For a fixed iteration \(s\), take
\(\widehat m_n=\widehat{[\kappa]}^{[s]}\).
Near \([\kappa^{(0)}]\), let
\(\mathcal{F}([\kappa])=\widetilde{\operatorname{vec}}(\kappa^{\texttt{opt}})\),
where \(\kappa^{\texttt{opt}}\) is the unit representative of
\([\kappa]\) optimally positioned with \(\kappa^{(0)}\).
Thus,
\(\widehat\mu_n=
\widetilde{\operatorname{vec}}(\widehat\kappa^{[s],\texttt{opt}})\),
while \(\mathcal{G}(\Sigma)=\operatorname{vech}(\Sigma)\) gives the auxiliary
coordinates. Here \(K=2p\), \(P=2\), and the constraint manifold
\(S=h^{-1}(0)\) is defined by
\[
    h\bigl(\widetilde{\operatorname{vec}}(\kappa)\bigr)
    =
    \begin{pmatrix}
        \frac12(\|\kappa\|^2-1)\\
        \operatorname{Im}\{(\kappa^{(0)})^*\kappa\}
    \end{pmatrix}.
\]
The first constraint fixes the norm, and the second makes the inner product with \(\kappa^{(0)}\) real. Together with the local restriction
\(\operatorname{Re}\{(\kappa^{(0)})^*\kappa\}>0\), the second condition selects the optimally positioned representative. For \(\kappa\in\mathbb C^p\setminus\{0\}\) and \(\Sigma\in\mathrm{SPD}(2)\), define the ambient extension of profile loss
\[
    \tilde{\rho}(y,\kappa,\Sigma)
    :=
    \min_{\phi\in\mathbb C}\|y-\phi\kappa\|_\Sigma^2,
\]
which agrees with \(\rho(y,[\kappa],\Sigma)\) on unit representatives.
Its coordinate form is
\[
    \tilde{\ell}\bigl(
        y,\widetilde{\operatorname{vec}}(\kappa),
        \operatorname{vech}(\Sigma)
    \bigr)
    =
    \tilde{\rho}(y,\kappa,\Sigma).
\]
The mapped cost \(\ell\) is the restriction of \(\tilde{\ell}\) to
\(\mathcal{F}(U_{m_0})\times \mathcal{G}(V_{q_0})\) in its parameter arguments.
The extension permits differentiation in the ambient Euclidean space,
including directions outside \(S\).

The measurability of \(\widehat E_n\) and sufficient conditions for
the existence of a measurable selection
\(\widehat m_n\in\widehat E_n\) are discussed in~\ref{supp:sec:gfm-measurable-minimizers} of the Supplementary Material.
\end{appendix}

\begin{acks}[Acknowledgments]
The authors thank Stephan Huckemann, Benjamin Eltzner, and Henrik Wiechers for their foundational contributions to the Helmert-based formulation of the drift model, including identifying the population minimizer at the true noise covariance.
\end{acks}

\begin{funding}
Jaesung Park and Sungkyu Jung were supported by the Samsung Science and Technology Foundation (SSTF-BA2002-03) and the National Research Foundation of Korea (RS-2023-00301976, RS-2024-00333399).

Jaesung Park and Yvo Pokern were funded by the Deutsche Forschungsgemeinschaft (DFG, German Research Foundation) under Germany's Excellence Strategy – EXC-2047/2 – 390685813.

Yvo Pokern was also supported by the Deutsche Forschungsgemeinschaft (DFG, German Research Foundation) CRC 1456, Project Id 432680300. 
\end{funding}

\bibliographystyle{imsart-nameyear}
\bibliography{library}

@Article{Bhattacharya2003,
  author    = {Bhattacharya, Rabi and Patrangenaru, Vic},
  journal   = {The Annals of Statistics},
  title     = {{Large Sample Theory of Intrinsic and Extrinsic Sample Means on Manifolds. I}},
  year      = {2003},
  issn      = {00905364},
  month     = "feb",
  number    = {1},
  pages     = {1--29},
  volume    = {31},
  publisher = {Institute of Mathematical Statistics},
  url       = {http://www.jstor.org/stable/3448366},
}

@book{durrett2019probability,
  author    = {Durrett, Rick},
  title     = {Probability: Theory and Examples},
  edition   = {5th},
  series    = {Cambridge Series in Statistical and Probabilistic Mathematics},
  volume    = {49},
  year      = {2019},
  publisher = {Cambridge University Press},
  doi       = {10.1017/9781108591034}
}

@article{korf2001random,
  title={Random lsc functions: an ergodic theorem},
  author={Korf, Lisa A and Wets, Roger J-B},
  journal={Mathematics of Operations Research},
  volume={26},
  number={2},
  pages={421--445},
  year={2001},
  publisher={INFORMS}
}

@article{schotz2022strong,
  author  = {Sch{\"o}tz, Christof},
  title   = {Strong laws of large numbers for generalizations of {Fr{\'e}chet} mean sets},
  journal = {Statistics},
  volume  = {56},
  number  = {1},
  pages   = {34--52},
  year    = {2022},
  doi     = {10.1080/02331888.2022.2032063},
  note    = {Correction: Statistics 59(4), 1084 (2025), \url{https://doi.org/10.1080/02331888.2025.2523605}}
}

@article{huckemann2018backward,
  author  = {Huckemann, Stephan F. and Eltzner, Benjamin},
  title   = {Backward nested descriptors asymptotics with inference
             on stem cell differentiation},
  journal = {The Annals of Statistics},
  year    = {2018},
  volume  = {46},
  number  = {5},
  pages   = {1994--2019},
  doi     = {10.1214/17-AOS1609}
}

@article{evans2024limit,
  title={Limit theorems for {F}r{\'e}chet mean sets},
  author={Evans, Steven N and Jaffe, Adam Q},
  journal={Bernoulli},
  volume={30},
  number={1},
  pages={419--447},
  year={2024},
  publisher={Bernoulli Society for Mathematical Statistics and Probability}
}

@article{Riek2016,
title = {The Exact {NOE} as an alternative in ensemble structure determination},
journal = {Biophys. J.},
volume = {110},
number = {1},
pages = {113--126},
year = {2016},
issn = {0006-3495},
doi = {https://doi.org/10.1016/j.bpj.2015.11.031},
url = {https://www.sciencedirect.com/science/article/pii/S000634951501214X},
author = {Beat Vögeli and Simon Olsson and Peter Güntert and Roland Riek}
}

@article{Remmel2024,
  author  = {Remmel, Laura and Meyer, Andreas and Ackermann, Katrin
             and Hagelueken, Gregor and Bennati, Marina and Bode, Bela E.},
  title   = {Pulsed {EPR} methods in the {Angstrom} to nanometre scale
             shed light on the conformational flexibility of a fluoride
             riboswitch},
  journal = {Angewandte Chemie International Edition},
  year    = {2024},
  volume  = {63},
  number  = {49},
  pages   = {e202411241},
  doi     = {10.1002/anie.202411241}
}

@article{Gemperle1991,
   author = {Gemperle, Claudius and Schweiger, Arthur},
   title = {Pulsed electron-nuclear double resonance methodology},
   journal = {Chem. Rev.},
   volume = {91},
   number = {7},
   pages = {1481--1505},
   ISSN = {0009-2665},
   DOI = {10.1021/cr00007a011},
   year = {1991},
   type = {Journal Article}
}

@article{Feher1956,
  author  = {Feher, G.},
  title   = {Observation of nuclear magnetic resonances via the
             electron spin resonance line},
  journal = {Physical Review},
  year    = {1956},
  volume  = {103},
  number  = {3},
  pages   = {834--835},
  doi     = {10.1103/PhysRev.103.834}
}

@book{Goldfarb2018,
title = {EPR Spectroscopy: Fundamentals and Methods},
editor={D. Goldfarb and S. Stoll},
publisher = {Wiley},
year = {2018},
}

@article{Un2026,
  author  = {Un, Sun and Garda, Zolt{\'a}n and Mihovilcevic, Damila
             and Tibiletti, Tania and T{\'o}th, {\'E}va},
  title   = {{Mn(II)} {$^{19}$F} electron--nuclear double resonance
             spectroscopy as a structural tool: Resolution of the
             structural diversity of a {Mn(II)} dual {$^{19}$F/$^{1}$H}
             {MRI} probe},
  journal = {Journal of the American Chemical Society},
  year    = {2026},
  volume  = {148},
  number  = {23},
  pages   = {23686--23697},
  doi     = {10.1021/jacs.6c00899}
}

@article{Anderson2003,
  author ={Amy C. Anderson},
  title = {The process of structure-based drug design},
  journal = {Chemistry \& Biology},
  volume = {10},
  issue = {9},
  pages = {787--797},
  year = {2003}
}

@article{Epel2003,
  author  = {Epel, B. and Arieli, D. and Baute, D. and Goldfarb, D.},
  title   = {Improving {W}-band pulsed {ENDOR} sensitivity---random
             acquisition and pulsed special {TRIPLE}},
  journal = {Journal of Magnetic Resonance},
  year    = {2003},
  volume  = {164},
  number  = {1},
  pages   = {78--83},
  doi     = {10.1016/S1090-7807(03)00191-5},
url = {https://www.sciencedirect.com/science/article/pii/S1090780703001915}
}

@article{Rizzato2014,
  author = {Roberto Rizzato and Marina Bennati},
  title = {Enhanced sensitivity of electron-nuclear double resonance (ENDOR) by cross polarisation and relaxation},
  journal ={Phys. Chem. Chem. Phys.},
  volume = {16},
  pages = {7681--7685},
  year ={2014},
}

@article{huckemann2011intrinsic,
  author  = {Huckemann, Stephan F.},
  title   = {Intrinsic inference on the mean geodesic of planar shapes
             and tree discrimination by leaf growth},
  journal = {The Annals of Statistics},
  year    = {2011},
  volume  = {39},
  number  = {2},
  pages   = {1098--1124},
  doi     = {10.1214/10-AOS862}
}

@inproceedings{ziezold1977expected,
	title={On expected figures and a strong law of large numbers for random elements in quasi-metric spaces},
	author={Ziezold, Herbert},
	booktitle={Transactions of the Seventh Prague Conference on Information Theory, Statistical Decision Functions, Random Processes and of the 1974 European Meeting of Statisticians},
	pages={591--602},
	year={1977},
	organization={Springer}
}

@article{yu2015useful,
  title={A useful variant of the Davis--Kahan theorem for statisticians},
  author={Yu, Yi and Wang, Tengyao and Samworth, Richard J},
  journal={Biometrika},
  volume={102},
  number={2},
  pages={315--323},
  year={2015},
  publisher={Oxford University Press}
}

@article{wiechers2023drift,
  title={Drift models on complex projective space for electron-nuclear double resonance},
  author={Wiechers, Henrik and Zobel, Markus and Bennati, Marina and Tkach, Igor and Eltzner, Benjamin and Huckemann, Stephan and Pokern, Yvo},
  journal={arXiv preprint arXiv:2307.12414},
  year={2023}
}

@article{pokern2021statistical,
  title={Statistical analysis of ENDOR spectra},
  author={Pokern, Yvo and Eltzner, Benjamin and Huckemann, Stephan F and Beeken, Clemens and Stubbe, JoAnne and Tkach, Igor and Bennati, Marina and Hiller, Markus},
  journal={Proceedings of the National Academy of Sciences},
  volume={118},
  number={27},
  pages={e2023615118},
  year={2021},
  publisher={National Academy of Sciences}
}

@article{park2026generalized,
  author  = {Park, Jaesung and Jung, Sungkyu},
  title   = {Generalized {Fr{\'e}chet} means with random minimizing
             domains and its strong consistency},
  journal = {Biometrika},
  year    = {2026},
  volume  = {113},
  number  = {1},
  pages   = {asag002},
  doi     = {10.1093/biomet/asag002}
}

@article{neyman1948consistent,
  author  = {Neyman, Jerzy and Scott, Elizabeth L.},
  title   = {Consistent estimates based on partially consistent observations},
  journal = {Econometrica},
  year    = {1948},
  volume  = {16},
  number  = {1},
  pages   = {1--32},
  doi     = {10.2307/1914288}
}

@book{stewart1990matrix,
  title={Matrix Perturbation Theory},
  author={Stewart, G.W. and Sun, J.},
  isbn={9780126702309},
  lccn={lc90033378},
  series={Computer Science and Scientific Computing},
  url={https://books.google.co.kr/books?id=l78PAQAAMAAJ},
  year={1990},
  publisher={Elsevier Science}
}

@article{hoffman1953variation,
  title={The variation of the spectrum of a normal matrix},
  author={Hoffman, Alan J and Wielandt, Helmut W },
  journal={Duke Math. J},
  volume={20},
  number={1},
  pages={37--39},
  year={1953},
  publisher={World Scientific},
  doi={10.1215/S0012-7094-53-02004-3}
}

@incollection{newey1994large,
  author    = {Newey, Whitney K. and McFadden, Daniel},
  title     = {Large sample estimation and hypothesis testing},
  booktitle = {Handbook of Econometrics},
  editor    = {Engle, Robert F. and McFadden, Daniel L.},
  volume    = {4},
  chapter   = {Chapter~36},
  pages     = {2111--2245},
  publisher = {Elsevier},
  year      = {1994},
  doi       = {10.1016/S1573-4412(05)80005-4}
}

@book{lee2024introduction,
  author    = {Lee, John M.},
  title     = {Introduction to Complex Manifolds},
  series    = {Graduate Studies in Mathematics},
  volume    = {244},
  publisher = {American Mathematical Society},
  address   = {Providence, RI},
  year      = {2024},
  doi       = {10.1090/gsm/244}
}

@article{bonifant2020group,
  author  = {Bonifant, Araceli and Milnor, John},
  title   = {Group actions, divisors, and plane curves},
  journal = {Bulletin of the American Mathematical Society},
  volume  = {57},
  number  = {2},
  pages   = {171--267},
  year    = {2020},
  doi     = {10.1090/bull/1681}
}

@article{Wiechers2023,
title = {Bayesian optimization to estimate hyperfine couplings from {$^{19}$F} {ENDOR} spectra},
journal = {J. Magn. Reson.},
volume = {353},
pages = {107491},
year = {2023},
issn = {1090-7807},
doi = {https://doi.org/10.1016/j.jmr.2023.107491},
url = {https://www.sciencedirect.com/science/article/pii/S109078072300126X},
author = {H. Wiechers and A. Kehl and M. Hiller and B. Eltzner and S. Huckemann and A. Meyer and I. Tkach and M. Bennati and Y. Pokern },
}

@article{Hiller2022,
  author  = {Hiller, M. and Tkach, I. and Wiechers, H. and Eltzner, B.
             and Huckemann, S. and Pokern, Y. and Bennati, M.},
  title   = {Distribution of {H$^{\beta}$} hyperfine couplings in a
             tyrosyl radical revealed by {263 GHz} {ENDOR} spectroscopy},
  journal = {Applied Magnetic Resonance},
  year    = {2022},
  volume  = {53},
  pages   = {1015--1030},
  doi     = {10.1007/s00723-021-01411-5}
}

@misc{PNASdata,
  author = {Pokern, Yvo},
  title = {{ENDOR 263}},
  year = {2021},
  howpublished = {UCL Research Data Repository},
  note = {Version 1. Source code and data},
  doi = {10.5522/04/14217068.v1},
  url = {https://doi.org/10.5522/04/14217068.v1}
}

@article{zhao2025adaptively,
  author  = {Zhao, Qian and Cand{\`e}s, Emmanuel J.},
  title   = {An Adaptively Resized Parametric Bootstrap for Inference in High-Dimensional Generalized Linear Models},
  journal = {Statistica Sinica},
  year    = {2025},
  volume  = {35},
  number  = {1},
  pages   = {91--110},
  doi     = {10.5705/ss.202022.0296}
}

@article{benussi2026accurate,
  author  = {Benussi, Davide and Salvan, Alessandra and Sartori, Nicola},
  title   = {Accurate inference on a parameter of interest via parametric bootstrap in two-way crossed fixed effects models with sparse discrete data},
  journal = {Statistical Methods \& Applications},
  year    = {2026},
  volume  = {35},
  pages   = {103--127},
  doi     = {10.1007/s10260-026-00833-4}
}

@inproceedings{anderson1956statistical,
  author    = {Anderson, T. W. and Rubin, Herman},
  title     = {Statistical Inference in Factor Analysis},
  booktitle = {Proceedings of the Third Berkeley Symposium
               on Mathematical Statistics and Probability},
  editor    = {Neyman, Jerzy},
  volume    = {5},
  pages     = {111--150},
  year      = {1956},
  publisher = {University of California Press},
  address   = {Berkeley, CA}
}

@article{joreskog1970general,
  author  = {J{\"o}reskog, Karl G.},
  title   = {A General Method for Analysis of Covariance Structures},
  journal = {Biometrika},
  volume  = {57},
  number  = {2},
  pages   = {239--251},
  year    = {1970},
  doi     = {10.1093/biomet/57.2.239}
}

@article{tipping1999probabilistic,
  author  = {Tipping, Michael E. and Bishop, Christopher M.},
  title   = {Probabilistic Principal Component Analysis},
  journal = {Journal of the Royal Statistical Society:
             Series B (Statistical Methodology)},
  volume  = {61},
  number  = {3},
  pages   = {611--622},
  year    = {1999},
  doi     = {10.1111/1467-9868.00196}
}

@article{stoica2009maximum,
  author  = {Stoica, Petre and Jansson, Magnus},
  title   = {On Maximum Likelihood Estimation in Factor
             Analysis---An Algebraic Derivation},
  journal = {Signal Processing},
  volume  = {89},
  number  = {6},
  pages   = {1260--1262},
  year    = {2009},
  doi     = {10.1016/j.sigpro.2009.01.002}
}

@article{murphy1985twostep,
  author  = {Murphy, Kevin M. and Topel, Robert H.},
  title   = {Estimation and Inference in Two-Step
             Econometric Models},
  journal = {Journal of Business \& Economic Statistics},
  year    = {1985},
  volume  = {3},
  number  = {4},
  pages   = {370--379},
  doi     = {10.1080/07350015.1985.10509471}
}

@article{geyer1994constrained,
  author  = {Geyer, Charles J.},
  title   = {On the Asymptotics of Constrained
             {M}-Estimation},
  journal = {The Annals of Statistics},
  year    = {1994},
  volume  = {22},
  number  = {4},
  pages   = {1993--2010},
  doi     = {10.1214/aos/1176325768}
}

@article{josse2016confidence,
  author  = {Josse, Julie and Wager, Stefan and Husson, Fran{\c{c}}ois},
  title   = {Confidence Areas for Fixed-Effects {PCA}},
  journal = {Journal of Computational and Graphical Statistics},
  year    = {2016},
  volume  = {25},
  number  = {1},
  pages   = {28--48},
  doi     = {10.1080/10618600.2014.950871}
}

@article{meyer2022,
    author = {Meyer, Andreas and Kehl, Annemarie and Cui, Chang and Reichardt, Fehmke A. K. and Hecker, Fabian and Funk, Lisa-Marie and Ghosh, Manas K. and Pan, Kuan-Ting and Urlaub, Henning and Tittmann, Kai and Stubbe, JoAnne and Bennati, Marina},
    title = {$^{19}$F Electron-Nuclear Double Resonance Reveals
Interaction between Redox-Active Tyrosines across the α/β
Interface of E. coli Ribonucleotide
Reductase},
    journal = {Journal of the American Chemical Society},
    volume = {144},
    number = {25},
    pages = {11270--11282},
    year = {2022},
    month = {06},
    issn = {0002-7863},
    doi = {10.1021/jacs.2c02906},
    url = {https://doi.org/10.1021/jacs.2c02906},
    eprint = {https://pubs.acs.org/jacsat/article-pdf/144/25/11270/3921288/ja2c02906.pdf},
}

@book{srivastava1998course,
  author    = {Srivastava, S. M.},
  title     = {A Course on Borel Sets},
  series    = {Graduate Texts in Mathematics},
  volume    = {180},
  publisher = {Springer},
  address   = {New York},
  year      = {1998},
  isbn      = {978-0-387-98412-4},
  doi       = {10.1007/b98956}
}

@book{molchanov2017theory,
  author    = {Molchanov, Ilya},
  title     = {Theory of Random Sets},
  edition   = {2},
  series    = {Probability Theory and Stochastic Modelling},
  volume    = {87},
  publisher = {Springer},
  address   = {London},
  year      = {2017},
  isbn      = {978-1-4471-7349-6},
  doi       = {10.1007/978-1-4471-7349-6}
}

@article{wentzell1997maximum,
  author  = {Wentzell, Peter D. and Andrews, Darren T. and Hamilton, David C. and Faber, Klaas and Kowalski, Bruce R.},
  title   = {Maximum likelihood principal component analysis},
  journal = {Journal of Chemometrics},
  year    = {1997},
  volume  = {11},
  number  = {4},
  pages   = {339--366},
  doi     = {10.1002/(SICI)1099-128X(199707)11:4<339::AID-CEM476>3.0.CO;2-L}
}

@article{narasimhan2008model,
  author  = {Narasimhan, Shankar and Shah, Sirish L.},
  title   = {Model identification and error covariance matrix estimation from noisy data using {PCA}},
  journal = {Control Engineering Practice},
  year    = {2008},
  volume  = {16},
  number  = {1},
  pages   = {146--155},
  doi     = {10.1016/j.conengprac.2007.04.006}
}

@incollection{shapiro2007statistical,
  author    = {Shapiro, Alexander},
  title     = {Statistical Inference of Moment Structures},
  booktitle = {Handbook of Latent Variable and Related Models},
  editor    = {Lee, Sik-Yum},
  publisher = {North-Holland},
  year      = {2007},
  pages     = {229--260},
  doi       = {10.1016/S1871-0301(06)01011-0}
}

@book{DrydenMardia2016,
  author    = {Dryden, Ian L. and Mardia, Kanti V.},
  title     = {Statistical Shape Analysis, with Applications in {R}},
  edition   = {2},
  year      = {2016},
  publisher = {John Wiley \& Sons},
  address   = {Chichester},
  doi       = {10.1002/9781119072492}
}

@misc{meyer2026priv,
  author       = {Andreas Meyer},
  howpublished = {Private communication},
  year         = {2026},
  title        = {},
  note         = {Email to Yvo Pokern, November 12\textsuperscript{th}, 2026}
}

\clearpage
\setcounter{section}{0}
\setcounter{subsection}{0}
\setcounter{subsubsection}{0}
\setcounter{theorem}{0}
\setcounter{figure}{0}
\setcounter{table}{0}
\setcounter{equation}{0}
\renewcommand{\thesection}{S\arabic{section}}
\renewcommand{\thefigure}{S\arabic{figure}}
\renewcommand{\thetable}{S\arabic{table}}
\numberwithin{equation}{section}
\renewcommand{\theHsection}{supp.\arabic{section}}
\renewcommand{\theHfigure}{supp.\arabic{figure}}
\renewcommand{\theHtable}{supp.\arabic{table}}
\renewcommand{\theHequation}{supp.\arabic{section}.\arabic{equation}}
\let\condition\relax
\makeatletter
\let\c@condition\relax
\let\thecondition\relax
\let\theHcondition\relax
\makeatother
\theoremstyle{plain}
\newtheorem{condition}{Condition}
\setcounter{condition}{0}
\def\theHcondition{supp.\arabic{condition}}

\addtocontents{toc}{\protect\setcounter{tocdepth}{2}}
\makeatletter
\renewcommand{\@pnumwidth}{2.5em}
\renewcommand{\@tocrmarg}{4em}
\renewcommand*{\l@section}{\@dottedtocline{1}{0pt}{2.3em}}
\renewcommand*{\l@subsection}{\@dottedtocline{2}{2.3em}{3.2em}}
\makeatother

\begin{frontmatter}

\title{Supplementary Material for ``Complex-Valued One-Factor Model and its Application to ENDOR Spectroscopy''}
\runtitle{Supplementary Material}

\begin{abstract}
This supplementary material provides proofs, technical derivations, and numerical details for the main paper on complex-valued one-factor models and their application to ENDOR spectroscopy. We establish properties of the Helmert transformation, derive profile-likelihood updates, and present a complex Hermitian version of the Davis--Kahan theorem. We provide detailed proofs of strong consistency, stochastic expansions, and asymptotic normality for the covariance-based and iterated plug-in estimators, together with general asymptotic results for generalized \fre~means with random cost functions. We also derive explicit asymptotic covariance formulas and efficiency comparisons, prove the inconsistency of the joint profile-likelihood estimator, and establish the theoretical results for maximum-aligned spectrum inference. Finally, we present additional simulation results, numerical
diagnostics, and implementation details for ENDOR spectrum analysis.
\end{abstract}

\end{frontmatter}

\tableofcontents

\section{Centering via the Helmert transformation}
\label{supp:sec: app model}

We justify the Helmert reduction used in
Section~\ref{subsec:eliminating-observation-level}.
Throughout this section, we use the original,
uncentered variables and the assumptions of
model~\eqref{model:uncentered-complex-factor}.
In particular,
\[
    \kappa^{(0)}\in\mathbb C^{p+1},
    \quad
    \mathbf 1_{p+1}^{\top}\kappa^{(0)}=0,
    \quad
    \|\kappa^{(0)}\|=1,
\]
and the errors \(\varepsilon_{i,j}\)s are independent and identically distributed as \(\mathcal{CN}(0,\Sigma^{(0)})\).

The following lemma gives an explicit Helmert matrix and
verifies that the transformation preserves both the loading
norm and the independent Gaussian error structure.

\begin{lemma}[Helmert reduction]
\label{supp:lem: Helmert reduction}
Let \(e_1,\dots,e_{p+1}\) be the standard basis vectors of
\(\mathbb R^{p+1}\), and, for \(j=1,\dots,p\), define
\[
    h_j
    :=
    \frac{1}{\sqrt{j(j+1)}}
    \left(
        \sum_{m=1}^{j}e_m-je_{j+1}
    \right).
\]
Let \(H\in\mathbb R^{p\times(p+1)}\) be the matrix whose
\(j\)-th row is \(h_j^\top\). Then
\[
    H\mathbf 1_{p+1}=0,
    \quad
    HH^\top=\mathrm{Id}_p,
    \quad
    H^\top H
    =
    \mathrm{Id}_{p+1}
    -
    \frac{1}{p+1}
    \mathbf 1_{p+1}\mathbf 1_{p+1}^\top.
\]

For \(Y_i=(Y_{i,0},\dots,Y_{i,p})^\top\) and
\(\varepsilon_i=(\varepsilon_{i,0},\dots,\varepsilon_{i,p})^\top\),
set
\[
    \widetilde Y_i=HY_i,\quad
    \widetilde\kappa^{(0)}=H\kappa^{(0)},\quad
    \widetilde\varepsilon_i=H\varepsilon_i.
\]
Then
\[
    \widetilde Y_i
    =
    \phi_i\widetilde\kappa^{(0)}+\widetilde\varepsilon_i,
    \quad
    \widetilde\varepsilon_{i,j}
    \stackrel{\mathrm{i.i.d.}}{\sim}
    \mathcal{CN}(0,\Sigma^{(0)}),
\]
for \(i=1,\dots,n\) and \(j=1,\dots,p\).

Furthermore, \(H\) restricts to a complex-linear isometric
isomorphism from
\[
    \{z\in\mathbb C^{p+1}:
      \mathbf 1_{p+1}^{\top}z=0\}
\]
onto \(\mathbb C^p\), with inverse \(H^\top\).
Consequently,
\[
    H^\top\widetilde\kappa^{(0)}=\kappa^{(0)},
    \quad
    \|\widetilde\kappa^{(0)}\|
    =
    \|\kappa^{(0)}\|
    =
    1.
\]
\end{lemma}

\begin{proof}[Proof of Lemma~\ref{supp:lem: Helmert reduction}]
The definition of the Helmert vectors gives
\[
    \mathbf 1_{p+1}^{\top}h_j=0,
    \quad
    h_j^\top h_k=\delta_{jk},
    \quad j,k=1,\ldots,p.
\]
Thus the rows of \(H\) form an orthonormal basis of
\(\{x\in\mathbb R^{p+1}:
\mathbf 1_{p+1}^{\top}x=0\}\).
It follows that \(HH^\top=\mathrm{Id}_p\) and that
\(H^\top H\) is the orthogonal projector onto this subspace,
which proves the stated identities.

Since \(H\mathbf 1_{p+1}=0\), we have
\[
\begin{aligned}
    \widetilde Y_i
    &=
    HY_i \\
    &=
    H\bigl(
        \psi_i\mathbf 1_{p+1}
        +\phi_i\kappa^{(0)}
        +\varepsilon_i
    \bigr) \\
    &=
    \phi_iH\kappa^{(0)}+H\varepsilon_i \\
    &=
    \phi_i\widetilde\kappa^{(0)}+\widetilde\varepsilon_i.
\end{aligned}
\]
Also,
\[
    H\bigl(
        Y_i-\bar Y_{i,\cdot}\mathbf 1_{p+1}
    \bigr)
    =
    HY_i,
\]
so applying \(H\) directly to \(Y_i\) is equivalent to first
subtracting its within-observation mean.

Using the vectorization convention of Section~\ref{subsec:complex-one-factor-model},
\[
    \operatorname{vec}(\varepsilon_i)
    \sim
    \mathcal N_{2(p+1)}
    \left(
        0,
        \mathrm{Id}_{p+1}\otimes\Sigma^{(0)}
    \right).
\]
Since \(H\) is real,
\[
    \operatorname{vec}(\widetilde\varepsilon_i)
    =
    (H\otimes\mathrm{Id}_2)
    \operatorname{vec}(\varepsilon_i).
\]
The transformed vector is therefore Gaussian, with covariance
\[
\begin{aligned}
    &(H\otimes\mathrm{Id}_2)
    (\mathrm{Id}_{p+1}\otimes\Sigma^{(0)})
    (H^\top\otimes\mathrm{Id}_2)\\
    &\quad
    =
    (HH^\top)\otimes\Sigma^{(0)}
    =
    \mathrm{Id}_p\otimes\Sigma^{(0)}.
\end{aligned}
\]
Its \(p\) two-dimensional blocks are consequently independent,
each with distribution \(\mathcal N_2(0,\Sigma^{(0)})\).
Hence
\(\widetilde\varepsilon_{i,j}
\sim\mathcal{CN}(0,\Sigma^{(0)})\).
Independence across observations, and independence from
the factor scores, are preserved by the deterministic
transformation.

Finally, for any \(z\in\mathbb C^{p+1}\) satisfying
\(\mathbf 1_{p+1}^{\top}z=0\), 
\[
    H^\top Hz
    =
    \left(
        \mathrm{Id}_{p+1}
        -
        \frac{1}{p+1}
        \mathbf 1_{p+1}\mathbf 1_{p+1}^\top
    \right)z
    =
    z.
\]
Consequently,
\[
    \|Hz\|^2
    =
    z^*H^\top Hz
    =
    \|z\|^2.
\]
Conversely, for every \(u\in\mathbb C^p\),
\[
    H(H^\top u)=u,
    \quad
    \mathbf 1_{p+1}^\top H^\top u
    =
    (H\mathbf 1_{p+1})^\top u
    =
    0.
\]
Thus the restriction of \(H\) is an isometric isomorphism
with inverse \(H^\top\). 
\end{proof}
With a slight abuse of notation, we henceforth drop the tildes and
write the transformed model as~\eqref{model:centered-complex-factor}.

\section{Technical details for profile likelihood}
\label{supp:sec: technical details for estimators}

This section addresses two aspects of the profile-likelihood
procedure. We first derive the conditional factor-score estimator
and show that the resulting profile loss is well defined on
\(\mathbb{CP}^{p-1}\). We then describe how to compute a
profile-loading update when the covariance matrix is fixed.

\subsection{Profiling over the factor score and phase invariance}
\label{supp:subsec:factor-score-profiling}

For fixed \(y\in\mathbb C^p\),
\(\Sigma\in\mathrm{SPD}(2)\), and
\(\kappa\in\mathcal S^{2p-1}\), we investigate 
\[
\min_{\phi\in\mathbb C}\|y-\phi\kappa\|_\Sigma^2.
\]
We establish that the minimizer is unique and that the minimized
value is unchanged when \(\kappa\) is replaced by
\(e^{i\lambda}\kappa\).

For a complex number \(c=a+ib\in\mathbb C\), define
\[
    M(c)
    :=
    \begin{pmatrix}
        a & -b\\
        b & a
    \end{pmatrix},
    \quad
    \mathfrak c
    \begin{pmatrix}
        x\\
        y
    \end{pmatrix}
    :=
    x+iy.
\]
For \(z,w\in\mathbb C^q\) and \(A\in\mathrm{SPD}(2)\), define
\[
z\diamond_A w
:=
\sum_{j=1}^{q}
M(z_j)^\top A^{-1}M(w_j),
\quad
z\bullet_A w
:=
\sum_{j=1}^{q}
M(z_j)^\top A^{-1}\operatorname{vec}(w_j).
\]

The following elementary identities connect complex multiplication with its
real two-dimensional matrix representation and will be used below. They follow
directly from the definitions of \(\operatorname{vec}\) and \(M\), so their proof is
omitted.

\begin{lemma}
\label{supp:lem: complex number vectorization}
For \(z,w\in\mathbb C\), we have
\begin{enumerate}
    \item \(M(z)^\top=M(\overline z)\);
    \item \(\operatorname{vec}(zw)=M(z)\operatorname{vec}(w)=M(w)\operatorname{vec}(z)\);
    \item \(M(zw)=M(z)M(w)=M(w)M(z)\).
\end{enumerate}
\end{lemma}

Now we provide three auxiliary lemmas. Lemma~\ref{supp:lem: diamond_invertible} first establishes the positive
definiteness needed to derive the unique conditional MLE of \(\phi\)
in Lemma~\ref{supp:lem: phi-hat-formula}. Lemma~\ref{supp:lem: phi-hat-equivariance}
then shows that the conditional MLE is phase equivariant, whereas the
corresponding fitted signal and profiled cost are phase invariant.

\begin{lemma}\label{supp:lem: diamond_invertible}
For every \(\kappa\in \mathbb{C}^{p} \setminus \{0\}\) and
\(\Sigma\in\mathrm{SPD}(2)\), the matrix
\(\kappa\diamond_\Sigma\kappa\) is symmetric positive definite and
hence invertible.
\end{lemma}

\begin{proof}[Proof of Lemma~\ref{supp:lem: diamond_invertible}]
By definition,
\[
    \kappa\diamond_\Sigma\kappa
    =
    \sum_{j=1}^{p}
    M(\kappa_j)^\top\Sigma^{-1}M(\kappa_j).
\]
Each summand is symmetric and positive semidefinite, so the sum is also symmetric and positive semidefinite. Since  \(\kappa\in \mathbb{C}^{p} \setminus \{0\}\), there exists an index
\(j_0\in\{1,\dots,p\}\) such that \(\kappa_{j_0}\neq0\). For any
nonzero \(v\in\mathbb R^2\), the matrix \(M(\kappa_{j_0})\) is
invertible since
\[
    \det M(\kappa_{j_0})=|\kappa_{j_0}|^2>0.
\]
Hence \(M(\kappa_{j_0})v\neq0\). Since \(\Sigma^{-1}\) is positive
definite,
\[
\begin{aligned}
    v^\top(\kappa\diamond_\Sigma\kappa)v
    &=
    \sum_{j=1}^{p}
    \bigl(M(\kappa_j)v\bigr)^\top
    \Sigma^{-1}
    \bigl(M(\kappa_j)v\bigr) \\
    &\ge
    \bigl(M(\kappa_{j_0})v\bigr)^\top
    \Sigma^{-1}
    \bigl(M(\kappa_{j_0})v\bigr)
    >0.
\end{aligned}
\]
Therefore, \(\kappa\diamond_\Sigma\kappa\) is positive definite and
hence invertible.
\end{proof}

\begin{lemma}\label{supp:lem: phi-hat-formula}
For every \(y\in\mathbb C^p\),
\(\Sigma\in\mathrm{SPD}(2)\), and
 \(\kappa\in \mathbb{C}^{p} \setminus \{0\}\), the function
\[
    \phi\longmapsto\|y-\phi\kappa\|_\Sigma^2
\]
has a unique minimizer over \(\mathbb C\). Denoting this minimizer by
\(\widehat\phi(y,\kappa,\Sigma)\), we have
\begin{equation}
\label{supp:eq: phi-hat}
    \widehat\phi(y,\kappa,\Sigma)
    =
    \mathfrak c\!\left(
        (\kappa\diamond_\Sigma\kappa)^{-1}
        (\kappa\bullet_\Sigma y)
    \right).
\end{equation}
\end{lemma}

\begin{proof}[Proof of Lemma~\ref{supp:lem: phi-hat-formula}]
By Lemma~\ref{supp:lem: complex number vectorization},
\[
\begin{aligned}
    \|y-\phi\kappa\|_\Sigma^2
    ={}&
    \|y\|_\Sigma^2
    -
    2\operatorname{vec}(\phi)^\top
    (\kappa\bullet_\Sigma y)
    \\
    &+
    \operatorname{vec}(\phi)^\top
    (\kappa\diamond_\Sigma\kappa)
    \operatorname{vec}(\phi).
\end{aligned}
\]

As established in Lemma~\ref{supp:lem: diamond_invertible}, the $2 \times 2$ matrix $\kappa \diamond_\Sigma \kappa$ is positive definite. Thus, $\|y - \phi \kappa\|_{\Sigma}^{2}$ is a strictly convex quadratic function of the real vector $\operatorname{vec}(\phi) \in \mathbb{R}^2$. Hence it has a unique
global minimizer.

Differentiating with respect to
\(u:=\operatorname{vec}(\phi)\) gives
\[
    \nabla_u\|y-\phi\kappa\|_\Sigma^2
    =
    -2(\kappa\bullet_\Sigma y)
    +
    2(\kappa\diamond_\Sigma\kappa)u.
\]
Setting the gradient equal to zero yields
\[
    \operatorname{vec}
    \bigl(\widehat\phi(y,\kappa,\Sigma)\bigr)
    =
    (\kappa\diamond_\Sigma\kappa)^{-1}
    (\kappa\bullet_\Sigma y).
\]
Applying \(\mathfrak c\) to both sides gives~\eqref{supp:eq: phi-hat}.
\end{proof}

\begin{lemma}\label{supp:lem: phi-hat-equivariance}
For every \(y\in\mathbb C^p\),
\(\Sigma\in\mathrm{SPD}(2)\),
\(\kappa\in\mathcal S^{2p-1}\), and \(\lambda\in\mathbb R\), the
unique minimizer in~\eqref{supp:eq: phi-hat} satisfies
\begin{equation}
\label{supp:eq:phi-equivariance}
    \widehat\phi(y,e^{i\lambda}\kappa,\Sigma)
    =
    e^{-i\lambda}\widehat\phi(y,\kappa,\Sigma).
\end{equation}
Consequently,
\begin{equation}
\label{supp:eq:fitted-signal-invariance}
    \widehat\phi(y,e^{i\lambda}\kappa,\Sigma)
    (e^{i\lambda}\kappa)
    =
    \widehat\phi(y,\kappa,\Sigma)\kappa.
\end{equation}
In particular, the profile loss defined in
Section~\ref{subsec:plugin-estimator},
\begin{equation}
\label{supp:eq:projective-profile-cost}
    \rho(y,[\kappa], \Sigma)
    =
    \left\|
        y-\widehat\phi(y,\kappa,\Sigma)\kappa
    \right\|_\Sigma^2
\end{equation}
is independent of the representative \(\kappa\in[\kappa]\) and is
therefore well-defined on \(\mathbb{CP}^{p-1}\).
\end{lemma}

\begin{proof}[Proof of Lemma~\ref{supp:lem: phi-hat-equivariance}]
For every \(\phi\in\mathbb C\),
\[
    \|y-\phi(e^{i\lambda}\kappa)\|_\Sigma^2
    =
    \|y-(e^{i\lambda}\phi)\kappa\|_\Sigma^2.
\]
Since \(\phi\mapsto e^{i\lambda}\phi\) is a bijection of
\(\mathbb C\), uniqueness of the minimizer in
Lemma~\ref{supp:lem: phi-hat-formula} gives
\[
    \widehat\phi(y,e^{i\lambda}\kappa,\Sigma)
    =
    e^{-i\lambda}\widehat\phi(y,\kappa,\Sigma).
\]
Multiplying by \(e^{i\lambda}\kappa\) proves
\eqref{supp:eq:fitted-signal-invariance}.
The invariance of the profiled cost follows immediately.
\end{proof}

We also verify the characterization of optimal position used in
Section~\ref{subsec:complex-projective-space}.

\begin{lemma}[Projective distance and optimal position]
\label{supp:lem:optimal-position}
Let \(\kappa,\widetilde\kappa\in\mathcal S^{2p-1}\).
The distance in~\eqref{eq:projective-distance} satisfies
\[
    d([\kappa],[\widetilde\kappa])
    =
    \sqrt{2-2|\kappa^*\widetilde\kappa|}
\]
and is independent of the chosen representatives.
Moreover, \(\kappa\) and \(\widetilde\kappa\) are in optimal
position if and only if
\(\widetilde\kappa^*\kappa\) is real and nonnegative.

For fixed \(\kappa\), if \(\widetilde\kappa^*\kappa\neq0\),
the unique representative of \([\widetilde\kappa]\) in
optimal position with \(\kappa\) is
\[
    \widetilde\kappa_{\mathrm{opt}}
    =
    \frac{\widetilde\kappa^*\kappa}
         {|\widetilde\kappa^*\kappa|}
    \widetilde\kappa.
\]
If \(\widetilde\kappa^*\kappa=0\), every representative of
\([\widetilde\kappa]\) is in optimal position with \(\kappa\).
\end{lemma}

\begin{proof}
Since both vectors have unit norm, for every \(\lambda\in\mathbb R\),
\[
    \|\kappa-e^{i\lambda}\widetilde\kappa\|^2
    =
    2-2\operatorname{Re}
    \left(e^{-i\lambda}\widetilde\kappa^*\kappa\right)
    \geq
    2-2|\widetilde\kappa^*\kappa|.
\]
If \(\widetilde\kappa^*\kappa\neq0\), equality holds if and only if
\[
    e^{i\lambda}
    =
    \frac{\widetilde\kappa^*\kappa}
         {|\widetilde\kappa^*\kappa|}.
\]
If \(\widetilde\kappa^*\kappa=0\), equality holds for every
\(\lambda\). This proves the distance formula and the assertions
about the optimally positioned representatives.

Taking \(\lambda=0\), the original representatives are in
optimal position if and only if
\[
    \operatorname{Re}(\widetilde\kappa^*\kappa)
    =
    |\widetilde\kappa^*\kappa|,
\]
which is equivalent to \(\widetilde\kappa^*\kappa\) being real
and nonnegative.
Finally, replacing either vector by another representative
does not change \(|\widetilde\kappa^*\kappa|\), so the distance
is independent of the chosen representatives.
\end{proof}

\subsection{Computing the profile-loading estimator for a fixed covariance}
\label{supp:subsec:fixed-covariance-loading-algorithm}

Let \(\Sigma\in\mathrm{SPD}(2)\) be fixed. We describe a numerical
procedure for minimizing
\[
\frac1n\sum_{i=1}^n\rho(Y_i,[\kappa],\Sigma)
\]
over \([\kappa]\in\mathbb{CP}^{p-1}\). As in the conditional
updates of \cite{pokern2021statistical}, the procedure alternates between
the factor scores and the loading while keeping \(\Sigma\) fixed.

Write
\(\phi=(\phi_1,\ldots,\phi_n)^\top\in\mathbb C^n\) and
\(Y_{\cdot,j}=(Y_{1,j},\ldots,Y_{n,j})^\top\in\mathbb C^n\).
The following result gives the loading update when \(\Sigma\)
and the factor scores are fixed.

\begin{proposition}[Conditional loading update]
\label{supp:prop:conditional-loading-update}
Let \(\Sigma\in\mathrm{SPD}(2)\) and
\(\phi\in\mathbb C^n\setminus\{0\}\) be fixed.
The function
\[
\kappa\longmapsto
\sum_{i=1}^n\|Y_i-\phi_i\kappa\|_\Sigma^2,
\quad \kappa\in\mathbb C^p,
\]
has a unique minimizer \(\kappa^{\mathrm{LS}}\), given coordinatewise by
\begin{equation}
\label{supp:eq:conditional-loading-update}
\kappa_j^{\mathrm{LS}}
=
\mathfrak c\!\left[
(\phi\diamond_\Sigma\phi)^{-1}
(\phi\bullet_\Sigma Y_{\cdot,j})
\right],
\quad j=1,\ldots,p.
\end{equation}
If \(\kappa^{\mathrm{LS}}\neq0\), setting
\[
\kappa^{\mathrm{new}}
=
\frac{\kappa^{\mathrm{LS}}}{\|\kappa^{\mathrm{LS}}\|},
\quad
\phi_i^{\mathrm{new}}
=
\|\kappa^{\mathrm{LS}}\|\phi_i
\quad(i=1,\ldots,n)
\]
gives \(\|\kappa^{\mathrm{new}}\|=1\) and preserves every fitted signal:
\(\phi_i^{\mathrm{new}}\kappa^{\mathrm{new}}
=\phi_i\kappa^{\mathrm{LS}}\).
\end{proposition}

\begin{proof}
The objective separates over the loading coordinates:
\[
    \sum_{i=1}^n\|Y_i-\phi_i\kappa\|_\Sigma^2
    =
    \sum_{j=1}^p\sum_{i=1}^n
    \operatorname{vec}(Y_{i,j}-\phi_i\kappa_j)^\top
    \Sigma^{-1}
    \operatorname{vec}(Y_{i,j}-\phi_i\kappa_j).
\]
For each \(j\), apply the quadratic minimization argument of
Lemma~\ref{supp:lem: phi-hat-formula} in dimension \(n\),
with response \(Y_{\cdot,j}\), design vector \(\phi\),
and scalar coefficient \(\kappa_j\).
Since \(\phi\neq0\), the coefficient matrix
\(\phi\diamond_\Sigma\phi\) is positive definite, and the unique
minimizer satisfies
\[
    \operatorname{vec}(\kappa_j^{\mathrm{LS}})
    =
    (\phi\diamond_\Sigma\phi)^{-1}
    (\phi\bullet_\Sigma Y_{\cdot,j}).
\]
This proves \eqref{supp:eq:conditional-loading-update}
and uniqueness.
The stated normalization and rescaling preserve
\(\phi_i\kappa^{\mathrm{LS}}\) for every \(i\), and therefore
leave the objective value unchanged.
\end{proof}

For the first plug-in loading update, we set \(\Sigma=\widehat\Sigma^{\texttt{cov}}\). For each subsequent step
\(s\geq2\), we set \(\Sigma=\widehat\Sigma^{[s-1]}\).
Each fixed-covariance fit starts from the SVD initialization in
Algorithm~\ref{supp:alg:fixed-covariance-loading}, including at later
plug-in steps. Within each fit, the alternating updates do not increase
the profile objective. However, convergence to a global or local minimizer is
not guaranteed as the number of alternating updates increases.

To specify the stopping rule, define the implemented loss by
\begin{equation}
\label{supp:eq:fixed-covariance-implemented-loss}
\ell(\kappa,\phi;\Sigma)
=
\frac1n\sum_{i=1}^n\|Y_i-\phi_i\kappa\|_\Sigma^2
+
p\log\det\Sigma.
\end{equation}
The determinant term is constant during a fixed-covariance fit, but is retained in the relative loss test. Define the score normal-equation residuals by
\begin{equation}
\label{supp:eq:score-normal-equation-residual}
\begin{aligned}
g_{\phi,i}(\kappa,\phi;\Sigma)
&=
(\kappa\diamond_\Sigma\kappa)\operatorname{vec}(\phi_i)
-
\kappa\bullet_\Sigma Y_i,
\quad i=1,\ldots,n,\\
g_\phi(\kappa,\phi;\Sigma)
&=
\bigl(g_{\phi,1}^{\top},\ldots,g_{\phi,n}^{\top}\bigr)^{\top}
\in\mathbb R^{2n}.
\end{aligned}
\end{equation}
Thus \(g_{\phi,i}=(n/2)\nabla_{\operatorname{vec}(\phi_i)}\ell\), and \(\|g_\phi\|_2^2=\sum_{i=1}^n\|g_{\phi,i}\|_2^2\), without averaging over observations. The residual is evaluated after the loading update and score rescaling, before the scores are updated again.

In the simulations, we use \(\varepsilon_\ell=10^{-8}\), \(\varepsilon_\phi=10^{-6}\), and \(m_{\max}=1000\). Both stopping conditions must hold. The final score--loading--score pass is performed after the loop and does not change its convergence flag. A zero or nonfinite loading update is recorded as a numerical failure. The joint \texttt{PL} solver uses the same tolerances and iteration limit.

\begin{algorithm}[!h]
\caption{Profile-loading estimation for fixed \(\Sigma\)}
\label{supp:alg:fixed-covariance-loading}
\begin{algorithmic}[1]
\Require Nonzero data matrix \(\mathsf Y=(Y_1,\ldots,Y_n)\in\mathbb C^{p\times n}\), \(\Sigma\in\mathrm{SPD}(2)\), tolerances \(\varepsilon_\ell,\varepsilon_\phi>0\), and positive integer \(m_{\max}\).

\State Compute a leading singular triplet \((d_1,u_1,v_1)\) of \(\mathsf Y\).
\State \(\kappa\gets u_1\), \(\phi\gets d_1\overline{v_1}\).
\State \(\ell_{\mathrm{old}}\gets\ell(\kappa,\phi;\Sigma)\), \(\mathrm{converged}\gets\mathrm{false}\).

\For{\(m=1,\ldots,m_{\max}\)}
  \State \(\phi_i\gets\widehat\phi(Y_i,\kappa,\Sigma)\) for \(i=1,\ldots,n\).
  \State Compute \(\kappa_j^{\mathrm{LS}}\) from \eqref{supp:eq:conditional-loading-update} for \(j=1,\ldots,p\).
  \State \(r\gets\|\kappa^{\mathrm{LS}}\|\), \(\kappa\gets\kappa^{\mathrm{LS}}/r\), \(\phi\gets r\phi\).
  \State \(\ell_{\mathrm{new}}\gets\ell(\kappa,\phi;\Sigma)\), \(g\gets g_\phi(\kappa,\phi;\Sigma)\).
  \State \(\delta_\ell\gets|\ell_{\mathrm{new}}-\ell_{\mathrm{old}}|/(1+|\ell_{\mathrm{old}}|)\).
  \State \(\ell_{\mathrm{old}}\gets\ell_{\mathrm{new}}\).

  \If{\(\delta_\ell<\varepsilon_\ell\) and \(\|g\|_2<\varepsilon_\phi\)}
    \State \(\mathrm{converged}\gets\mathrm{true}\).
    \State \textbf{break}.
  \EndIf
\EndFor

\State \(\phi_i\gets\widehat\phi(Y_i,\kappa,\Sigma)\) for \(i=1,\ldots,n\).
\State Compute \(\kappa_j^{\mathrm{LS}}\) from \eqref{supp:eq:conditional-loading-update} for \(j=1,\ldots,p\).
\State \(\kappa\gets\kappa^{\mathrm{LS}}/\|\kappa^{\mathrm{LS}}\|\).
\State \(\phi_i\gets\widehat\phi(Y_i,\kappa,\Sigma)\) for \(i=1,\ldots,n\).
\State \Return \([\kappa]\), \(\phi\), and \(\mathrm{converged}\).
\end{algorithmic}
\end{algorithm}

\newpage
\section{Complex Hermitian Davis--Kahan theorem}
\label{sec:daviskahan}
\label{supp:sec:daviskahan-proof}

This section presents a complex Hermitian version of the Davis--Kahan perturbation theorem used in the asymptotic analysis of the \texttt{cov} estimator. The result extends \citet[Theorem~2]{yu2015useful} from real symmetric matrices to complex Hermitian matrices, and the proof follows their argument.

We begin with a preliminary lemma regarding the Frobenius norm of a complex matrix.

\begin{lemma}\label{supp:lem: frobenius norm lemma}
For any $A\in\mathbb{C}^{m\times n}$, $U\in\mathbb{C}^{m\times p}$ and $W\in\mathbb{C}^{n\times q}$,  the following hold:
\begin{enumerate}
    \item[(i)] If $U^*U=\mathrm{Id}_p$ and $W^*W=\mathrm{Id}_q$, then $\|U^*AW\|_F \le \|A\|_F$.
    \item[(ii)] If $UU^*=\mathrm{Id}_{m}$ and $WW^*=\mathrm{Id}_{n}$, then $\|U^*AW\|_F = \|A\|_F$.
\end{enumerate}
\end{lemma}

\begin{proof}
For the first statement, we can extend $U$ and $W$ to form unitary matrices $U_2=(U\,|\,U_1)$ and $W_2=(W\,|\,W_1)$. Then we get
\begin{align*}
\|A\|_F^2
&=
\operatorname{Tr}(A^*A)
\\
&=
\operatorname{Tr}
\left(
W_2^*A^*U_2U_2^*AW_2
\right)
\\
&=
\left\|
U_2^*AW_2
\right\|_F^2
\\
&\ge
\left\|
U^*AW
\right\|_F^2.
\end{align*}
The second statement follows immediately from the assumptions $UU^*=\mathrm{Id}_{m}$ and $WW^*=\mathrm{Id}_{n}$:
\begin{align*}
\left\|
U^*AW
\right\|_F^2
&=
\operatorname{Tr}
\left(
W^*A^*UU^*AW
\right)
\\
&=
\operatorname{Tr}
\left(
W^*A^*AW
\right)
\\
&=
\operatorname{Tr}
\left(
WW^*A^*A
\right)
\\
&=
\operatorname{Tr}(A^*A)
\\
&=
\|A\|_F^2.
\end{align*}
\end{proof}

The following theorem gives bounds for the principal angles and the distance between suitably aligned orthonormal bases.

\begin{theorem}[Davis–Kahan theorem for complex Hermitian matrices]
\label{thm: Davis-Kahan theorem complex version}
Let $H,\widehat H\in\mathbb{C}^{p\times p}$ be Hermitian matrices with eigenvalues $\lambda_1\ge\cdots\ge\lambda_p$ and $\widehat\lambda_1\ge\cdots\ge\widehat\lambda_p$, respectively. Fix $1\le r\le s\le p$ and assume that
\[
\min(\lambda_{r-1}-\lambda_r,\ \lambda_s-\lambda_{s+1})>0,
\]
with the conventions $\lambda_0:=+\infty$ and $\lambda_{p+1}:=-\infty$. Let $d:=s-r+1$, and $V=(v_r,\dots,v_s)\in\mathbb{C}^{p\times d}$ and $\widehat V=(\widehat v_r,\dots,\widehat v_s)\in\mathbb{C}^{p\times d}$ be matrices with orthonormal columns satisfying $H v_j=\lambda_j v_j$ and $\widehat H\widehat v_j=\widehat\lambda_j\widehat v_j$ for $j=r,\dots,s$.

Then,
\begin{equation}\label{eq: grassmanian angle distance bound}
\|\sin\Theta(V,\widehat V)\|_{F}
\;\le\;
\frac{2\min\big(d^{1/2}\|\widehat H-H\|_{\text{op}},\ \|\widehat H-H\|_F\big)}
{\min(\lambda_{r-1}-\lambda_r,\ \lambda_s-\lambda_{s+1})}
\end{equation}
where
\(
\Theta(V,\widehat V)
:=
\operatorname{diag}(\theta_1,\ldots,\theta_d),
\)
with \(\theta_1,\ldots,\theta_d\in[0,\pi/2]\) denoting the principal angles between the column spaces of \(V\) and \(\widehat V\), and
\(
\sin\Theta(V,\widehat V)
:=
\operatorname{diag}(\sin\theta_1,\ldots,\sin\theta_d).
\) Moreover, there exists a unitary $d\times d$ matrix $\widehat O$ such that
\begin{equation}\label{eq: Frobenius norm bound}
\|\widehat V\widehat O - V\|_F
\;\le\;
\frac{2^{3/2}\min\big(d^{1/2}\|\widehat H-H\|_{\text{op}},\ \|\widehat H-H\|_F\big)}
{\min(\lambda_{r-1}-\lambda_r,\ \lambda_s-\lambda_{s+1})}.
\end{equation}
\end{theorem}

\begin{proof}[Proof of Theorem~\ref{thm: Davis-Kahan theorem complex version}]
If \(r=1\) and \(s=p\), both conclusions are immediate because
the two column spaces equal \(\mathbb C^p\); hence we assume
\((r,s)\neq(1,p)\). Define
\[
\delta
:=
\min\bigl(
    \lambda_{r-1}-\lambda_r,\,
    \lambda_s-\lambda_{s+1}
\bigr).
\]
Then \(0<\delta<\infty\).

Let $\Lambda=\mathrm{diag}(\lambda_r,\dots,\lambda_s)$ and $\widehat\Lambda=\mathrm{diag}(\widehat\lambda_r,\dots,\widehat\lambda_s)$. Since $\widehat V$ contains the eigenvectors of $\widehat H$, we have
\[
0 = \widehat H\widehat V - \widehat V\widehat\Lambda
= (\widehat H-H)\widehat V + \widehat V(\Lambda-\widehat\Lambda) + H\widehat V - \widehat V\Lambda.
\]
Rearranging the terms and applying the triangle inequality, we obtain
\begin{align}
\|\widehat V\Lambda-H\widehat V\|_F
&\le \|(\widehat H-H)\widehat V\|_F + \|\widehat V(\Lambda-\widehat\Lambda)\|_F \nonumber\\
&\le d^{1/2}\|\widehat H-H\|_{\operatorname{op}} + \|\Lambda-\widehat\Lambda\|_F \nonumber\\
&\le 2d^{1/2}\|\widehat H-H\|_{\operatorname{op}}.
\label{supp:eq: operator norm inequality}
\end{align}
The second inequality uses
\[
    \|(\widehat H-H)\widehat V\|_F
    \leq
    \|\widehat H-H\|_{\operatorname{op}}
    \|\widehat V\|_F
    =
    d^{1/2}\|\widehat H-H\|_{\operatorname{op}},
\]
together with
\[
    \|\widehat V(\Lambda-\widehat\Lambda)\|_F
    =
    \|\Lambda-\widehat\Lambda\|_F,
\]
since \(\widehat V^*\widehat V=\mathrm{Id}_d\).
The final inequality follows from Weyl's inequality
for the ordered eigenvalues of Hermitian matrices:
\[
    \max_{1\leq j\leq p}
    |\widehat\lambda_j-\lambda_j|
    \leq
    \|\widehat H-H\|_{\operatorname{op}};
\]
see \citet[Chapter~IV]{stewart1990matrix}.

Alternatively, we can bound the terms using the Frobenius norm:
\begin{align}
\|\widehat V\Lambda-H\widehat V\|_F
&\le \|(\widehat H-H)\widehat V\|_F + \|\widehat V(\Lambda-\widehat\Lambda)\|_F \nonumber\\
&\le \|\widehat H-H\|_F + \|\Lambda-\widehat\Lambda\|_F \nonumber\\
&\le 2\|\widehat H-H\|_F.
\label{supp:eq: frobenius norm inequality}
\end{align}
The second inequality follows from
Lemma~\ref{supp:lem: frobenius norm lemma}.
For the final inequality, the Hermitian form of the
Hoffman--Wielandt inequality gives
\[
    \|\Lambda-\widehat\Lambda\|_F^2
    =
    \sum_{k=r}^s|\lambda_k-\widehat\lambda_k|^2
    \leq
    \sum_{k=1}^p|\lambda_k-\widehat\lambda_k|^2
    \leq
    \|H-\widehat H\|_F^2;
\]
see \citep{hoffman1953variation}.

Combining~\eqref{supp:eq: operator norm inequality} and~\eqref{supp:eq: frobenius norm inequality} yields the upper bound:
\begin{equation}\label{supp:eq: Vhat bound}
\|\widehat V\Lambda-H\widehat V\|_F
\le
2\min\big(d^{1/2}\|\widehat H-H\|_{\operatorname{op}},\ \|\widehat H-H\|_F\big).
\end{equation}

Next, complete the columns \(v_r,\ldots,v_s\) of \(V\)
to an orthonormal eigenbasis \(v_1,\ldots,v_p\) of \(H\),
with \(Hv_j=\lambda_jv_j\) for \(j=1,\ldots,p\). Using the orthonormal eigenbasis \(\{v_j\}_{j=1}^p\) of \(H\), Parseval's identity gives
\begin{align*}
\|\widehat V\Lambda-H\widehat V\|_F^2
&= \sum_{j=1}^p \|v_j^*(\widehat V\Lambda-H\widehat V)\|_F^2\\
&= \sum_{j=1}^p \Big\|\big((\lambda_r-\lambda_j)v_j^*\widehat v_r,\dots,(\lambda_s-\lambda_j)v_j^*\widehat v_s\big)\Big\|_F^2\\
&= \sum_{j=1}^p \sum_{k=r}^s (\lambda_k-\lambda_j)^2 |v_j^*\widehat v_k|^2.
\end{align*}
Partitioning the sum over $j$ and dropping the non-negative terms where $j \in \{r,\dots,s\}$ yields a lower bound:
\begin{align*}
\|\widehat V\Lambda-H\widehat V\|_F^2
&= \sum_{j \in \{r,\dots,s\}} \sum_{k=r}^s (\lambda_k-\lambda_j)^2 |v_j^*\widehat v_k|^2 + \sum_{j \notin \{r,\dots,s\}} \sum_{k=r}^s (\lambda_k-\lambda_j)^2 |v_j^*\widehat v_k|^2 \\
&\ge \sum_{j \notin \{r,\dots,s\}} \sum_{k=r}^s (\lambda_k-\lambda_j)^2 |v_j^*\widehat v_k|^2.
\end{align*}
Since the eigenvalues are ordered, the distance $|\lambda_k - \lambda_j|$ for $k \in \{r,\dots,s\}$ and $j \notin \{r,\dots,s\}$ is bounded below by the eigengap $\delta = \min(\lambda_{r-1}-\lambda_r,\ \lambda_s-\lambda_{s+1})$. Substituting $(\lambda_k - \lambda_j)^2 \ge \delta^2$ gives:
\begin{align*}
\sum_{j \notin \{r,\dots,s\}} \sum_{k=r}^s (\lambda_k-\lambda_j)^2 |v_j^*\widehat v_k|^2 \ge \delta^2 \sum_{j \notin \{r,\dots,s\}} \sum_{k=r}^s  |v_j^*\widehat v_k|^2
\end{align*}

To relate this to principal angles, let $\widehat V^*V = U_C C W_C^*$ be a singular value decomposition, where $C = \mathrm{diag}(\cos\theta_1,\dots,\cos\theta_d)$ contains the cosines of the principal angles. Let $V_\perp$ denote the matrix of the remaining eigenvectors $v_j$ ($j \notin \{r,\dots,s\}$). Using \(\mathrm{Id}_p=VV^*+V_\perp V_\perp^*\), we get:
\begin{align*}d &= \|\widehat{V}\|_F^2 \\
&= \operatorname{Tr}(\widehat{V}^* \widehat{V}) \\
&= \operatorname{Tr}\big(\widehat{V}^* (VV^* + V_\perp V_{\perp}^{*}) \widehat{V}\big) \\
&= \operatorname{Tr}(\widehat{V}^* V V^* \widehat{V}) + \operatorname{Tr}(\widehat{V}^* V_\perp V_\perp^* \widehat{V}) \\ &= \|V^* \widehat{V}\|_F^2 + \|V_\perp^* \widehat{V}\|_F^2.\end{align*}

The summation term in our derivation is precisely $\|V_\perp^* \widehat{V}\|_F^2$:
\begin{equation*}
\sum_{j \notin \{r,\dots,s\}} \sum_{k=r}^s |v_j^*\widehat v_k|^2 = \|V_\perp^* \widehat{V}\|_F^2.
\end{equation*}

By the unitary invariance of the Frobenius norm, we evaluate $\|V^* \widehat{V}\|_F^2$:
\begin{equation*}
\|V^* \widehat{V}\|_F^2 = \|\widehat{V}^* V\|_F^2 = \|U_C C W_C^*\|_F^2 = \|C\|_F^2 = \sum_{k=1}^d \cos^2\theta_k.
\end{equation*}

Substituting this back yields:
\begin{align*}\|V_\perp^* \widehat{V}\|_F^2
&= d - \|V^* \widehat{V}\|_{F}^2  \\
&=d - \sum_{k=1}^d \cos^2\theta_k \\
&= \sum_{k=1}^d \sin^2\theta_k = \|\sin\Theta(V, \widehat V)\|_F^2.\end{align*}
 Returning to our lower bound, we obtain:
\begin{align*}
\|\widehat V\Lambda-H\widehat V\|_{F}^2
&\ge \min(\lambda_{r-1}-\lambda_r,\lambda_s-\lambda_{s+1})^2 \sum_{j\notin \{r,\dots,s\}} \Big\|\big(v_{j}^{* }\widehat {v}_r,\dots,v_j^{*}\widehat {v}_{s}\big)\Big\|_{F}^2 \\
&= \min(\lambda_{r-1}-\lambda_r,\lambda_s-\lambda_{s+1})^2 \|\sin\Theta(\widehat V,V)\|_F^2.
\end{align*}
Combining this with~\eqref{supp:eq: Vhat bound} establishes~\eqref{eq: grassmanian angle distance bound}.

To show~\eqref{eq: Frobenius norm bound}, set
\(\widehat O=U_CW_C^*\), which is unitary. Then
\begin{align}
\label{supp:eq: principal angle inequality}
    \|\widehat V\widehat O-V\|_F^2
    &=
    2d
    -
    2\operatorname{Re}
    \operatorname{Tr}(\widehat O^*\widehat V^*V)
    \nonumber\\
    &=
    2\sum_{k=1}^d(1-\cos\theta_k)
    \nonumber\\
    &\leq
    2\sum_{k=1}^d(1-\cos^2\theta_k)
    \nonumber\\
    &=
    2\|\sin\Theta(\widehat V,V)\|_F^2,
\end{align}
where the inequality uses \(0\leq\cos\theta_k\leq1\).
Taking square roots and applying
\eqref{eq: grassmanian angle distance bound}
proves~\eqref{eq: Frobenius norm bound}.
\end{proof}

As an immediate corollary, we obtain the result for the one-dimensional case, $s=r=j$.

\begin{corollary}\label{cor: Davis-Kahan theorem complex version for one eigenvector}
Let $H,\widehat H\in\mathbb{C}^{p\times p}$ be Hermitian matrices with eigenvalues $\lambda_1\ge\cdots\ge\lambda_p$ and $\widehat\lambda_1\ge\cdots\ge\widehat\lambda_p$, respectively. Fix $1\le j\le p$ and assume that
\[
\min(\lambda_{j-1}-\lambda_j,\ \lambda_j-\lambda_{j+1})>0,
\]
with the conventions $\lambda_0:=+\infty$ and $\lambda_{p+1}:=-\infty$. If the unit-norm eigenvectors $v_{j},\widehat v_{j}\in\mathbb{C}^p$ satisfy $H v_{j}=\lambda_j v_{j}$ and $\widehat H\widehat{v}_{j}=\widehat\lambda_j\widehat{v}_{j}$, then
\[
\sin(v_{j},\widehat{v}_{j})=\sqrt{1-|v_j^*\widehat v_j|^2}
\;\le\;
\frac{2\|\widehat H-H\|_{\text{op}}}
{\min(\lambda_{j-1}-\lambda_j, \lambda_j-\lambda_{j+1})}.
\]
Moreover, there exists $\hat\theta \in[0,2\pi)$ such that
\[
\|v_{j} - e^{i\hat\theta}\widehat v_{j}\|_{2}
\;\le\;
\frac{2^{3/2}\|\widehat H-H\|_{\text{op}}}
{\min(\lambda_{j-1}-\lambda_j,\ \lambda_j-\lambda_{j+1})}.
\]
\end{corollary}

\section{Proofs and auxiliary results for the
\texttt{cov}-estimator}
\label{supp:sec:cov-estimator-proofs}

This section collects the proofs and auxiliary results for the
\texttt{cov}-estimator. Specifically, we prove the strong consistency
result in Theorem~\ref{thm:cov-estimator-strong-consistency}, the loading
expansion and asymptotic normality in
Theorem~\ref{thm:cov-loading-expansion-clt}, and the second-order noise
covariance expansion in
Theorem~\ref{thm:cov-noise-covariance-expansion}. We begin with strong
consistency under the second-moment condition and then impose the stronger
moment conditions required for the stochastic expansions.

\subsection{Strong consistency of the
\texttt{cov}-estimator}
\label{supp:subsec:cov-strong-consistency}

This subsection proves the strong consistency of the
\texttt{cov}-estimator stated in
Theorem~\ref{thm:cov-estimator-strong-consistency}. We first establish
the almost sure convergence of the empirical complex moment matrices and
the estimated loading direction, and then propagate these convergences
through the residual projection and the moment covariance map.

\begin{proof}[Proof of
Theorem~\ref{thm:cov-estimator-strong-consistency}]
Under model~\eqref{model:centered-complex-factor}, the observations
are independent and identically distributed, and
\[
    \mathbb E[\|Y_1\|^2]
    =
    \mathbb E[|\phi_1|^2]+p\tau^2
    <
    \infty.
\]
Consequently, for all \(j,k=1,\ldots,p\),
\[
    \mathbb E[|Y_{1,j}Y_{1,k}|]
    \leq
    \bigl\{
        \mathbb E[|Y_{1,j}|^2]
        \mathbb E[|Y_{1,k}|^2]
    \bigr\}^{1/2}
    <
    \infty.
\]
The same bound applies to
\(\mathbb E[|Y_{1,j}\overline{Y_{1,k}}|]\).
Applying the strong law of large numbers to the real and imaginary
parts of each entry of the sample moment matrices in
\eqref{eq:sample-covariance-matrices} therefore gives their almost
sure entrywise convergence to
\(\Sigma^{\texttt{Her}}\) and
\(\Sigma^{\texttt{Tra}}\), respectively. Since the matrix dimensions
are finite, this entrywise convergence immediately yields convergence
in the operator norm, i.e.,
$$
    \left\|
        \widehat\Sigma^{\texttt{Her}}
        -
        \Sigma^{\texttt{Her}}
    \right\|_{\operatorname{op}}
    \longrightarrow 0,
    \quad
    \left\|
        \widehat\Sigma^{\texttt{Tra}}
        -
        \Sigma^{\texttt{Tra}}
    \right\|_{\operatorname{op}}
    \longrightarrow 0
$$
as \(n\to\infty\) almost surely.

Let \(\lambda_1\geq\lambda_2\) and
\(\widehat\lambda_1\geq\widehat\lambda_2\) denote the two largest
eigenvalues of \(\Sigma^{\texttt{Her}}\) and
\(\widehat\Sigma^{\texttt{Her}}\), respectively. Since
$$
    \lambda_1-\lambda_2
    =
    \mathbb E[|\phi_1|^2]
    >0,
$$
Weyl's eigenvalue perturbation inequality gives
$$
\begin{aligned}
    \widehat\lambda_1-\widehat\lambda_2
    &\geq
    \lambda_1-\lambda_2
    -
    2
    \left\|
        \widehat\Sigma^{\texttt{Her}}
        -
        \Sigma^{\texttt{Her}}
    \right\|_{\operatorname{op}}>0
\end{aligned}
$$
almost surely for all sufficiently large \(n\). Thus, the largest eigenvalue of \(\widehat\Sigma^{\texttt{Her}}\) is eventually simple, and hence
\(\widehat{[\kappa]}^{\texttt{cov}}\) is eventually uniquely defined.

Applying Corollary~\ref{cor: Davis-Kahan theorem complex version for one eigenvector}
with \(j=1\), together with the definition of the projective distance
in~\eqref{eq:projective-distance}, gives
$$
\begin{aligned}
    d\left(
        \widehat{[\kappa]}^{\texttt{cov}},
        [\kappa^{(0)}]
    \right)
    &\leq
    \frac{
        2^{3/2}
        \left\|
            \widehat\Sigma^{\texttt{Her}}
            -
            \Sigma^{\texttt{Her}}
        \right\|_{\operatorname{op}}
    }{
        \mathbb E[|\phi_1|^2]
    }
    \longrightarrow 0
\end{aligned}
$$
almost surely.

For all sufficiently large \(n\), let
\(\widehat\kappa^{\texttt{cov,opt}}\in
\widehat{[\kappa]}^{\texttt{cov}}\)
be the representative in optimal position with \(\kappa^{(0)}\). This representative is eventually unique since
$$
    \left|
        \left(
            \widehat\kappa^{\texttt{cov,opt}}
        \right)^*
        \kappa^{(0)}
    \right|
    \longrightarrow 1
$$
almost surely. By the definition of optimal position,
$$
\begin{aligned}
    \left\|
        \widehat\kappa^{\texttt{cov,opt}}
        -
        \kappa^{(0)}
    \right\|
    &=
    d\left(
        \widehat{[\kappa]}^{\texttt{cov}},
        [\kappa^{(0)}]
    \right)
    \longrightarrow 0
\end{aligned}
$$
almost surely.

Define
$$
    \widehat U^{\texttt{cov,opt}}
    :=
    \mathcal U
    \left(
        \widehat\kappa^{\texttt{cov,opt}}
    \right).
$$
Recall that
$$
    U_0=\mathcal U(\kappa^{(0)}),
    \quad
    \Pi_{\perp,0}
    =
    \Pi_\perp(\kappa^{(0)}).
$$
The preceding convergence and the continuity of \(\mathcal U\) imply
$$
    \widehat U^{\texttt{cov,opt}}
    \longrightarrow
    U_0
$$
almost surely. By the assumed linear independence of
\(\operatorname{Re}(\kappa^{(0)})\) and
\(\operatorname{Im}(\kappa^{(0)})\), the matrix \(U_0\)
has full column rank.
Hence \(\widehat U^{\texttt{cov,opt}}\) also has full column rank
almost surely for all sufficiently large \(n\). Moreover,
$$
    \left\{
        (\widehat U^{\texttt{cov,opt}})^\top
        \widehat U^{\texttt{cov,opt}}
    \right\}^{-1}
    \longrightarrow
    (U_0^\top U_0)^{-1}
$$
almost surely. Since
\[
    \operatorname{span}_{\mathbb C}
    \{\kappa,\overline{\kappa}\}
    =
    \operatorname{span}_{\mathbb C}
    \{\operatorname{Re}(\kappa),\operatorname{Im}(\kappa)\},
\]
and \(\mathcal U(\kappa)\) has real entries, the orthogonal
projection in \eqref{eq:orthogonal-residual-projector}
can be expressed using the Gram matrix of \(\mathcal U(\kappa)\).
Consequently,
$$
\begin{aligned}
    \widehat\Pi_\perp^{\texttt{cov}}
    &:=
    \Pi_\perp
    \left(
        \widehat\kappa^{\texttt{cov,opt}}
    \right)
    \\
    &=
    \mathrm{Id}_p
    -
    \widehat U^{\texttt{cov,opt}}
    \left\{
        (\widehat U^{\texttt{cov,opt}})^\top
        \widehat U^{\texttt{cov,opt}}
    \right\}^{-1}
    (\widehat U^{\texttt{cov,opt}})^\top
    \longrightarrow
    \Pi_{\perp,0}
\end{aligned}
$$
almost surely. As noted in
Section~\ref{subsec:cov-estimator},
\(\widehat\Pi_\perp^{\texttt{cov}}\) does not depend on the choice of
representative of
\(\widehat{[\kappa]}^{\texttt{cov}}\).

Recall the notation introduced in
Section~\ref{subsec:asymptotic-analysis-cov-estimator}:
\[
    \widehat\tau^{2,\texttt{cov}}
    =
    \widehat\tau^2
    \left(
        \widehat\kappa^{\texttt{cov,opt}}
    \right),
    \quad
    \widehat\gamma^{\texttt{cov}}
    =
    \widehat\gamma
    \left(
        \widehat\kappa^{\texttt{cov,opt}}
    \right).
\]
By the definition of \(\Pi_{\perp,0}\),
$$
    \Pi_{\perp,0}\kappa^{(0)}
    =
    0,
    \quad
    \operatorname{Tr}(\Pi_{\perp,0})
    =
    p-2.
$$
Therefore, the population moment identities in
\eqref{eq:population-complex-moments} give
$$
    \frac{1}{p-2}
    \operatorname{Tr}
    \left(
        \Pi_{\perp,0}
        \Sigma^{\texttt{Her}}
    \right)
    =
    \tau^2,
    \quad
    \frac{1}{p-2}
    \operatorname{Tr}
    \left(
        \Pi_{\perp,0}
        \Sigma^{\texttt{Tra}}
    \right)
    =
    \gamma.
$$
Since the map
\((A,P)\mapsto\operatorname{Tr}(PA)\) is continuous in both arguments,
the continuous mapping theorem yields
$$
\begin{aligned}
    \widehat\tau^{2,\texttt{cov}}
    &=
    \frac{1}{p-2}
    \operatorname{Tr}
    \left(
        \widehat\Pi_\perp^{\texttt{cov}}
        \widehat\Sigma^{\texttt{Her}}
    \right)\xrightarrow{\mathrm{a.s.}}
    \frac{1}{p-2}
    \operatorname{Tr}
    \left(
        \Pi_{\perp,0}
        \Sigma^{\texttt{Her}}
    \right)
    =
    \tau^2.
\end{aligned}
$$
Similarly, for the transpose moment component,
$$
\begin{aligned}
    \widehat\gamma^{\texttt{cov}}
    &=
    \frac{1}{p-2}
    \operatorname{Tr}
    \left(
        \widehat\Pi_\perp^{\texttt{cov}}
        \widehat\Sigma^{\texttt{Tra}}
    \right)\xrightarrow{\mathrm{a.s.}}
    \frac{1}{p-2}
    \operatorname{Tr}
    \left(
        \Pi_{\perp,0}
        \Sigma^{\texttt{Tra}}
    \right)
    =
    \gamma.
\end{aligned}
$$
Finally, the inverse map
$$
    (\tau^2,\gamma)
    \longmapsto
    \frac{1}{2}
    \begin{pmatrix}
        \tau^2+\operatorname{Re}(\gamma)
        &
        \operatorname{Im}(\gamma)
        \\[0.3em]
        \operatorname{Im}(\gamma)
        &
        \tau^2-\operatorname{Re}(\gamma)
    \end{pmatrix},
$$
corresponding to the relationship in
\eqref{eq:tau-gamma}, is continuous. Therefore, by another application of the continuous mapping theorem,
\[
    \left\|
        \widehat\Sigma^{\texttt{cov}}
        -
        \Sigma^{(0)}
    \right\|_F
    \xrightarrow{\mathrm{a.s.}}0.
\]
By construction, \(\widehat\Sigma^{\texttt{cov}}\) is real symmetric.
Moreover,
\[
    \left\|
        \widehat\Sigma^{\texttt{cov}}-\Sigma^{(0)}
    \right\|_{\operatorname{op}}
    \leq
    \left\|
        \widehat\Sigma^{\texttt{cov}}-\Sigma^{(0)}
    \right\|_F
    \xrightarrow{\mathrm{a.s.}}0.
\] Since
\(\lambda_{\min}(\Sigma^{(0)})>0\), Weyl's inequality gives
$$
    \lambda_{\min}
    \left(
        \widehat\Sigma^{\texttt{cov}}
    \right)
    \geq
    \lambda_{\min}(\Sigma^{(0)})
    -
    \left\|
        \widehat\Sigma^{\texttt{cov}}
        -
        \Sigma^{(0)}
    \right\|_{\operatorname{op}}
    >0
$$
almost surely for all sufficiently large \(n\). Hence
\(\widehat\Sigma^{\texttt{cov}}\in\mathrm{SPD}(2)\) almost surely for
all sufficiently large \(n\). Therefore, the complete
\texttt{cov}-estimator is well-defined almost surely for all
sufficiently large \(n\), which completes the proof.
\end{proof}

\subsection{Empirical moment bounds, loading expansion, and
asymptotic normality}
\label{supp:subsec:cov-moment-bounds-loading-expansion}

The stochastic behavior of the \texttt{cov}-estimator is driven by the
empirical Hermitian and transpose moment fluctuations \(E_H\) and \(E_T\). We first establish moment bounds for these fluctuations. 

\begin{lemma}
\label{supp:lem:cov-moment-bounds}
Assume model~\eqref{model:centered-complex-factor} with true parameters
\(\kappa^{(0)}\in\mathcal S^{2p-1}\) and
\(\Sigma^{(0)}\in\mathrm{SPD}(2)\). Recall from
Section~\ref{subsec:asymptotic-analysis-cov-estimator} that
\[
    E_H
    =
    \widehat\Sigma^{\texttt{Her}}
    -
    \Sigma^{\texttt{Her}},
    \quad
    E_T
    =
    \widehat\Sigma^{\texttt{Tra}}
    -
    \Sigma^{\texttt{Tra}}.
\]
Suppose that, for some integer \(k\geq2\), \( 0<\mathbb E[|\phi_1|^{2k}]<\infty \). Then, for every \(r=1,\ldots,k\),
\[
    \mathbb E\!\left[\|E_H\|_F^r\right]
    =
    \mathcal O(n^{-r/2}),
    \quad
    \mathbb E\!\left[\|E_T\|_F^r\right]
    =
    \mathcal O(n^{-r/2}).
\]
Consequently,
\[
    \|E_H\|_F^r
    =
    \mathcal O_p(n^{-r/2}),
    \quad
    \|E_T\|_F^r
    =
    \mathcal O_p(n^{-r/2}),
    \quad
    r=1,\ldots,k.
\]
\end{lemma}

\begin{proof}
First, for every \(r=1,\ldots,k\),
\(\mathbb E[\|Y_1\|^{2r}]<\infty\). Indeed, since
\[
    Y_1
    =
    \phi_1\kappa^{(0)}
    +
    \varepsilon_1
\]
and \(\|\kappa^{(0)}\|=1\),
\[
    \|Y_1\|^{2r}
    \leq
    2^{2r-1}
    \left(
        |\phi_1|^{2r}
        +
        \|\varepsilon_1\|^{2r}
    \right).
\]
The first term has finite expectation since \(r\leq k\), whereas the
Gaussian noise has moments of every order.

Write
\[
    E_H
    =
    \frac{1}{n}
    \sum_{i=1}^n Z_{H,i},
    \quad
    E_T
    =
    \frac{1}{n}
    \sum_{i=1}^n Z_{T,i},
\]
where
\(
    Z_{H,i}
    :=
    Y_iY_i^*
    -
    \mathbb E[Y_1Y_1^*]\) and \(Z_{T,i}
    :=
    Y_iY_i^\top
    -
    \mathbb E[Y_1Y_1^\top].
\)
The pairs \((Z_{H,i},Z_{T,i})\) are i.i.d.\ across \(i\), and
\[
    \mathbb E[Z_{H,i}]
    =
    \mathbb E[Z_{T,i}]
    =
    0.
\] Moreover,
\[
    \|Y_iY_i^*\|_F
    =
    \|Y_iY_i^\top\|_F
    =
    \|Y_i\|^2.
\]
Hence, for every \(r=1,\ldots,k\),
\[
    \mathbb E[\|Z_{H,1}\|_F^r]<\infty,
    \quad
    \mathbb E[\|Z_{T,1}\|_F^r]<\infty.
\]

We first consider \(r\geq2\). Applying Rosenthal's inequality
coordinatewise to the real and imaginary parts of the matrix entries, and
using the equivalence of norms in the finite-dimensional space
\(\mathbb C^{p\times p}\), gives
\[
\begin{aligned}
    \mathbb E
    \left\|
        \sum_{i=1}^n Z_{H,i}
    \right\|_F^r
    &\leq
    C_r
    \left\{
        n\mathbb E[\|Z_{H,1}\|_F^r]
        +
        \left(
            n\mathbb E[\|Z_{H,1}\|_F^2]
        \right)^{r/2}
    \right\}=
    \mathcal O(n^{r/2}),
\end{aligned}
\]
where the constant \(C_r\) may depend on \(r\) and \(p\), but not on
\(n\). Therefore,
\[
    \mathbb E[\|E_H\|_F^r]
    =
    \frac{1}{n^r}
    \mathbb E
    \left\|
        \sum_{i=1}^n Z_{H,i}
    \right\|_F^r
    =
    \mathcal O(n^{-r/2}).
\]
The same argument applied to \(Z_{T,i}\) yields
\[
    \mathbb E[\|E_T\|_F^r]
    =
    \mathcal O(n^{-r/2}),
    \quad
    r=2,\ldots,k.
\]

For \(r=1\), the result follows from the second-moment bounds
and Jensen's inequality:
\[
    \mathbb E[\|E_H\|_F]
    \leq
    \left(
        \mathbb E[\|E_H\|_F^2]
    \right)^{1/2}
    =
    \mathcal O(n^{-1/2}),
\]
and similarly,
\[
    \mathbb E[\|E_T\|_F]
    \leq
    \left(
        \mathbb E[\|E_T\|_F^2]
    \right)^{1/2}
    =
    \mathcal O(n^{-1/2}).
\]

Finally, Markov's inequality gives, for every \(r=1,\ldots,k\),
\[
    \|E_H\|_F^r
    =
    \mathcal O_p(n^{-r/2}),
    \quad
    \|E_T\|_F^r
    =
    \mathcal O_p(n^{-r/2}),
\]
which completes the proof.
\end{proof}

We next derive a quantitative perturbation expansion for the loading
estimator. The result strengthens the stochastic expansion stated in
Theorem~\ref{thm:cov-loading-expansion-clt} by giving an explicit deterministic
quadratic bound for the remainder in terms of \(\|E_H\|_F\).

\begin{lemma}[Quantitative loading expansion]
\label{supp:lem:cov-quantitative-loading-expansion}
Assume model~\eqref{model:centered-complex-factor} with true parameters
\(\kappa^{(0)}\in\mathcal S^{2p-1}\) and
\(\Sigma^{(0)}\in\mathrm{SPD}(2)\), satisfying
\(0<\mathbb E[|\phi_1|^4]<\infty\). Let \(\widehat\kappa^{\texttt{cov,opt}}\) be a representative of \(\widehat{[\kappa]}^{\texttt{cov}}\) in optimal position with
\(\kappa^{(0)}\), and recall that
\[
    P_{\perp,0}
    =
    \mathrm{Id}_p
    -
    \kappa^{(0)}(\kappa^{(0)})^*.
\] On the event
\(
    \left(
        \widehat\kappa^{\texttt{cov,opt}}
    \right)^*
    \kappa^{(0)}
    \neq0
\), this representative is unique.

Then
\begin{equation}
\label{supp:eq:cov-quantitative-loading-expansion}
\widehat\kappa^{\texttt{cov,opt}}
=
\kappa^{(0)}
+
\frac{1}{\mathbb E[|\phi_1|^2]}
P_{\perp,0}E_H\kappa^{(0)}
+
R_{\kappa,n}^{\texttt{cov}},
\end{equation}
where
\begin{equation}
\label{supp:eq:cov-loading-remainder-bound}
\left\|
    R_{\kappa,n}^{\texttt{cov}}
\right\|
\leq
\frac{
    4(\sqrt{2}+1)
}{
    \{\mathbb E[|\phi_1|^2]\}^2
}
\|E_H\|_F^2.
\end{equation}
In particular,
\(R_{\kappa,n}^{\texttt{cov}} = \mathcal O_p(n^{-1}).
\)

Equivalently, its grouped real-coordinate representation is
\begin{equation}
\label{supp:eq:cov-quantitative-loading-real-expansion}
\begin{aligned}
\widetilde{\operatorname{vec}}
\left(
    \widehat\kappa^{\texttt{cov,opt}}
\right)
&=
\widetilde{\operatorname{vec}}
\left(
    \kappa^{(0)}
\right)
+
\frac{1}{\mathbb E[|\phi_1|^2]}
\widetilde{\operatorname{vec}}
\left(
    P_{\perp,0}E_H\kappa^{(0)}
\right)
+
\widetilde{\operatorname{vec}}
\left(
    R_{\kappa,n}^{\texttt{cov}}
\right).
\end{aligned}
\end{equation}
\end{lemma}

\begin{proof}
By the definition of the covariance-based loading estimator,
\[
    \widehat\Sigma^{\texttt{Her}}
    \widehat\kappa^{\texttt{cov,opt}}
    =
    \widehat\lambda_1
    \widehat\kappa^{\texttt{cov,opt}},
    \quad
    \Sigma^{\texttt{Her}}\kappa^{(0)}
    =
    \lambda_1\kappa^{(0)},
\]
where
\( \lambda_1
    =
    \mathbb E[|\phi_1|^2]
    +
    \tau^2 \)
is the leading eigenvalue of \(\Sigma^{\texttt{Her}}\), and
\(\widehat\lambda_1\) is the leading eigenvalue of
\(\widehat\Sigma^{\texttt{Her}}\). Let
\[
    \delta_\kappa
    :=
    \widehat\kappa^{\texttt{cov,opt}}
    -
    \kappa^{(0)},
    \quad
    \delta_\lambda
    :=
    \widehat\lambda_1-\lambda_1.
\]
Then
\[
    \left(
        \Sigma^{\texttt{Her}}+E_H
    \right)
    \left(
        \kappa^{(0)}+\delta_\kappa
    \right)
    =
    \left(
        \lambda_1+\delta_\lambda
    \right)
    \left(
        \kappa^{(0)}+\delta_\kappa
    \right).
\]
Expanding both sides and using
\(\Sigma^{\texttt{Her}}\kappa^{(0)}
=\lambda_1\kappa^{(0)}\) gives
\[
    \Sigma^{\texttt{Her}}\delta_\kappa
    +
    E_H\kappa^{(0)}
    +
    E_H\delta_\kappa
    =
    \lambda_1\delta_\kappa
    +
    \delta_\lambda\kappa^{(0)}
    +
    \delta_\lambda\delta_\kappa.
\]

By the optimal positioning of
\(\widehat\kappa^{\texttt{cov,opt}}\) and \(\kappa^{(0)}\),
\(
    \left(
        \widehat\kappa^{\texttt{cov,opt}}
    \right)^*
    \kappa^{(0)}
    \in\mathbb R.
\)
Thus,
\(
    \delta_\kappa^*\kappa^{(0)}
    =
    \left(
        \widehat\kappa^{\texttt{cov,opt}}
    \right)^*
    \kappa^{(0)}
    -
    1
    \in\mathbb R.
\)
Since
\[
    \left\|
        \widehat\kappa^{\texttt{cov,opt}}
    \right\|^2
    =
    \|\kappa^{(0)}\|^2
    =
    1,
\]
expanding
\(\|\kappa^{(0)}+\delta_\kappa\|^2=1\) yields
\begin{equation}
\label{supp:eq:cov-loading-radial-component}
    (\kappa^{(0)})^*\delta_\kappa
    =
    \delta_\kappa^*\kappa^{(0)}
    =
    -\frac{1}{2}\|\delta_\kappa\|^2.
\end{equation}
Thus, the component of \(\delta_\kappa\) along
\(\kappa^{(0)}\) is of quadratic order.

Premultiplying the perturbed eigenvalue equation by
\(P_{\perp,0}\) and using
\(P_{\perp,0}\kappa^{(0)}=0\), we obtain
\[
\begin{aligned}
    P_{\perp,0}
    \Sigma^{\texttt{Her}}\delta_\kappa
    +
    P_{\perp,0}E_H\kappa^{(0)}
    +
    P_{\perp,0}E_H\delta_\kappa
    &=
    \lambda_1P_{\perp,0}\delta_\kappa
    +
    \delta_\lambda
    P_{\perp,0}\delta_\kappa.
\end{aligned}
\]
Under model~\eqref{model:centered-complex-factor}, \(
    \Sigma^{\texttt{Her}}
    =
    \mathbb E[|\phi_1|^2]
    \kappa^{(0)}(\kappa^{(0)})^*
    +
    \tau^2\mathrm{Id}_p,
\)
which implies
\(
    P_{\perp,0}\Sigma^{\texttt{Her}}
    =
    \tau^2P_{\perp,0}.
\)
Substituting this identity and
\(\lambda_1=\mathbb E[|\phi_1|^2]+\tau^2\) into the preceding equation
and rearranging gives
\begin{equation}
\label{supp:eq:cov-projected-eigenvector-equation}
\mathbb E[|\phi_1|^2]
P_{\perp,0}\delta_\kappa
=
P_{\perp,0}E_H\kappa^{(0)}
+
P_{\perp,0}E_H\delta_\kappa
-
\delta_\lambda
P_{\perp,0}\delta_\kappa.
\end{equation}

By~\eqref{supp:eq:cov-loading-radial-component},
\[
\begin{aligned}
    P_{\perp,0}\delta_\kappa
    &=
    \delta_\kappa
    -
    \kappa^{(0)}
    (\kappa^{(0)})^*
    \delta_\kappa=
    \delta_\kappa
    +
    \frac{1}{2}
    \|\delta_\kappa\|^2
    \kappa^{(0)}.
\end{aligned}
\]
Substituting this expression into
\eqref{supp:eq:cov-projected-eigenvector-equation} and solving for
\(\delta_\kappa\) yields
\[
    \delta_\kappa
    =
    \frac{1}{\mathbb E[|\phi_1|^2]}
    P_{\perp,0}E_H\kappa^{(0)}
    +
    R_{\kappa,n}^{\texttt{cov}},
\]
where
\[
\begin{aligned}
    R_{\kappa,n}^{\texttt{cov}}
    &:=
    \frac{1}{\mathbb E[|\phi_1|^2]}
    P_{\perp,0}
    \left(
        E_H-\delta_\lambda\mathrm{Id}_p
    \right)
    \delta_\kappa
    -
    \frac{1}{2}
    \|\delta_\kappa\|^2
    \kappa^{(0)}.
\end{aligned}
\]
This proves the identity in
\eqref{supp:eq:cov-quantitative-loading-expansion}. It remains to
establish the quantitative remainder bound.

Apply
Corollary~\ref{cor: Davis-Kahan theorem complex version for one eigenvector}
with
\[
    H=\Sigma^{\texttt{Her}},
    \quad
    \widehat H=\widehat\Sigma^{\texttt{Her}},
    \quad
    j=1,
    \quad
    v_1=\kappa^{(0)},
    \quad
    \widehat v_1
    =
    \widehat\kappa^{\texttt{cov,opt}}.
\]
Since the leading eigenvalue of
\(\Sigma^{\texttt{Her}}\) is
\(\mathbb E[|\phi_1|^2]+\tau^2\) and all the remaining eigenvalues are
\(\tau^2\), the eigengap is
\(\mathbb E[|\phi_1|^2]\). Moreover, since
\(\widehat\kappa^{\texttt{cov,opt}}\) is in optimal position with
\(\kappa^{(0)}\), we obtain
\begin{equation}
\label{supp:eq:cov-loading-Davis-Kahan-bound}
    \|\delta_\kappa\|
    =
    \left\|
        \widehat\kappa^{\texttt{cov,opt}}
        -
        \kappa^{(0)}
    \right\|
    \leq
    \frac{
        2^{3/2}\|E_H\|_{\operatorname{op}}
    }{
        \mathbb E[|\phi_1|^2]
    }.
\end{equation}
Weyl's eigenvalue perturbation inequality gives
\begin{equation}
\label{supp:eq:cov-loading-Weyl-bound}
    |\delta_\lambda|
    \leq
    \|E_H\|_{\operatorname{op}}.
\end{equation}

Using the triangle inequality and the submultiplicativity of the
operator norm, we obtain
\[
\begin{aligned}
    \left\|
        R_{\kappa,n}^{\texttt{cov}}
    \right\|
    &\leq
    \frac{1}{\mathbb E[|\phi_1|^2]}
    \|P_{\perp,0}\|_{\operatorname{op}}
    \left(
        \|E_H\|_{\operatorname{op}}
        +
        |\delta_\lambda|
    \right)
    \|\delta_\kappa\|
    +
    \frac{1}{2}
    \|\delta_\kappa\|^2
    \|\kappa^{(0)}\|.
\end{aligned}
\]
Since
\(
    \|P_{\perp,0}\|_{\operatorname{op}}
    =
    \|\kappa^{(0)}\|
    =
    1
\), substituting
\eqref{supp:eq:cov-loading-Davis-Kahan-bound} and
\eqref{supp:eq:cov-loading-Weyl-bound} gives
\[
    \left\|
        R_{\kappa,n}^{\texttt{cov}}
    \right\|
    \leq
    \frac{
        4(\sqrt{2}+1)
    }{
        \{\mathbb E[|\phi_1|^2]\}^2
    }
    \|E_H\|_{\operatorname{op}}^2.
\]
Since
\[
    \|E_H\|_{\operatorname{op}}
    \leq
    \|E_H\|_F,
\]
the bound in~\eqref{supp:eq:cov-loading-remainder-bound} follows.

By Lemma~\ref{supp:lem:cov-moment-bounds} with \(k=2\),
\(
    \|E_H\|_F
    =
    \mathcal O_p(n^{-1/2}) \), and hence
\(
    R_{\kappa,n}^{\texttt{cov}}
    =
    \mathcal O_p(n^{-1}).
\)
The event on which the optimally positioned representative is unique
occurs almost surely for all sufficiently large \(n\) by
Theorem~\ref{thm:cov-estimator-strong-consistency}. Finally,
\eqref{supp:eq:cov-quantitative-loading-real-expansion}
follows by applying the real-linear isometry
\(\widetilde{\operatorname{vec}}\) to
\eqref{supp:eq:cov-quantitative-loading-expansion}.
\end{proof}

The preceding lemmas provide the quantitative expansion and remainder
control needed to show Theorem~\ref{thm:cov-loading-expansion-clt}. It remains to identify the leading term as an average of i.i.d.\ random vectors and apply the
multivariate central limit theorem.

\begin{proof}[Proof of Theorem~\ref{thm:cov-loading-expansion-clt}]
The stochastic expansion in
\eqref{eq:cov-loading-stochastic-expansion}, together with its stated
remainder order, follows directly from
Lemma~\ref{supp:lem:cov-quantitative-loading-expansion}. In particular,
\[
    \widehat\kappa^{\texttt{cov,opt}}
    -
    \kappa^{(0)}
    =
    \frac{1}{\mathbb E[|\phi_1|^2]}
    P_{\perp,0}E_H\kappa^{(0)}
    +
    R_{\kappa,n}^{\texttt{cov}},
\]
where
\[
    \sqrt n\,
    \widetilde{\operatorname{vec}}
    \left(
        R_{\kappa,n}^{\texttt{cov}}
    \right)
    =
    \mathcal O_p(n^{-1/2})
    =
    o_p(1).
\]

Moreover,
\[
    E_H
    =
    \frac{1}{n}
    \sum_{i=1}^n
    \left\{
        Y_iY_i^*
        -
        \Sigma^{\texttt{Her}}
    \right\}.
\]
Expanding
\(Y_i=\phi_i\kappa^{(0)}+\varepsilon_i\) and using \( \Sigma^{\texttt{Her}} = \mathbb{E}[|\phi_{1}|^{2}] \kappa^{(0)} (\kappa^{(0)})^* + \tau^{2}\mathrm{Id}_p\), \(P_{\perp,0}\kappa^{(0)}=0\), we obtain
\[
    P_{\perp,0}E_H\kappa^{(0)}
    =
    \frac{1}{n}
    \sum_{i=1}^n u_i,
\]
where
\begin{equation}
\label{supp:eq:cov-loading-influence-vector}
    u_i
    :=
    P_{\perp,0}
    \left(
        \overline{\phi_i}\varepsilon_i
        +
        \varepsilon_i\varepsilon_i^*
        \kappa^{(0)}
    \right).
\end{equation}
Consequently,
\begin{equation}
\label{supp:eq:cov-loading-influence-representation}
\begin{aligned}
    \sqrt n
    \left\{
        \widetilde{\operatorname{vec}}
        \left(
            \widehat\kappa^{\texttt{cov,opt}}
        \right)
        -
        \widetilde{\operatorname{vec}}
        \left(
            \kappa^{(0)}
        \right)
    \right\}
    &=
    \frac{
        1
    }{
        \sqrt n\,
        \mathbb E[|\phi_1|^2]
    }
    \sum_{i=1}^n
    \widetilde{\operatorname{vec}}(u_i)
    +
    o_p(1).
\end{aligned}
\end{equation}

The random vectors \(u_i\) are i.i.d. We first verify that they are
centered. By the independence of \(\phi_1\) and \(\varepsilon_1\),
together with \(\mathbb E[\varepsilon_1]=0\),
\(
    \mathbb E[
        \overline{\phi_1}\varepsilon_1
    ]
    =
    0
\). Moreover,
\(
    \mathbb E[
        \varepsilon_1\varepsilon_1^*
    ]
    =
    \tau^2\mathrm{Id}_p.
\)
Therefore,
\[
\begin{aligned}
    \mathbb E[u_1]
    &=
    P_{\perp,0}
    \left\{
        \mathbb E[
            \overline{\phi_1}\varepsilon_1
        ]
        +
        \mathbb E[
            \varepsilon_1\varepsilon_1^*
        ]
        \kappa^{(0)}
    \right\}=
    P_{\perp,0}
    \left(
        \tau^2\kappa^{(0)}
    \right)
    =
    0.
\end{aligned}
\]
Hence \(\mathbb E
    \left[
        \widetilde{\operatorname{vec}}(u_1)
    \right]
    =
    0.
\)

The random vector \(u_1\) also has a finite second moment. Indeed, using
\(\|P_{\perp,0}\|_{\operatorname{op}}=1\) and
\(\|\kappa^{(0)}\|=1\),
\[
    \|u_1\|
    \leq
    |\phi_1|\|\varepsilon_1\|
    +
    \|\varepsilon_1\|^2.
\]
Therefore,
\[
\begin{aligned}
    \mathbb E[\|u_1\|^2]
    &\leq
    2\mathbb E[
        |\phi_1|^2
        \|\varepsilon_1\|^2
    ]
    +
    2\mathbb E[
        \|\varepsilon_1\|^4
    ]
    <
    \infty,
\end{aligned}
\]
where we used the independence of \(\phi_1\) and \(\varepsilon_1\),
the finiteness of \(\mathbb E[|\phi_1|^2]\), and the finite fourth
moment of the Gaussian noise.

Hence, by the multivariate central limit theorem,
\[
    \frac{1}{\sqrt n}
    \sum_{i=1}^n
    \widetilde{\operatorname{vec}}(u_i)
    \xrightarrow{d}
    \mathcal N_{2p}
    \left(
        0,
        \operatorname{Cov}
        \left\{
            \widetilde{\operatorname{vec}}(u_1)
        \right\}
    \right).
\]
Combining the multivariate central limit theorem with
\eqref{supp:eq:cov-loading-influence-representation} and applying
Slutsky's theorem gives
\[
    \sqrt n
    \left\{
        \widetilde{\operatorname{vec}}
        \left(
            \widehat\kappa^{\texttt{cov,opt}}
        \right)
        -
        \widetilde{\operatorname{vec}}
        \left(
            \kappa^{(0)}
        \right)
    \right\}
    \xrightarrow{d}
    \mathcal N_{2p}
    \left(
        0,
        V^{\texttt{cov}}
    \right).
\] Indeed,
\[
    u_1
    =
    P_{\perp,0}
    \left(
        Y_1Y_1^*
        -
        \Sigma^{\texttt{Her}}
    \right)
    \kappa^{(0)},
\]
so the limiting covariance is
\[
    \frac{
        \operatorname{Var}
        \left\{
            \widetilde{\operatorname{vec}}(u_1)
        \right\}
    }{
        \{\mathbb E[|\phi_1|^2]\}^2
    }
    =
    V^{\texttt{cov}},
\]
with \(V^{\texttt{cov}}\) as defined in
\eqref{eq:cov-loading-asymptotic-covariance}.
This proves~\eqref{eq:cov-loading-clt} and completes the proof.
\end{proof}

\subsection{Explicit form of
\texorpdfstring{$V^{\texttt{cov}}$}{Vcov}}
\label{supp:subsec:explicit-cov-loading-asymptotic-covariance}

For the subsequent efficiency analysis, we require an explicit expression
for \(V^{\texttt{cov}}\). We first give the fourth-order moment
identities for the possibly non-circular complex Gaussian noise.

\begin{lemma}[Complex Gaussian fourth-order moments]
\label{supp:lem:complex-normal-fourth-moments}
Let
\(\varepsilon=(\varepsilon_1,\ldots,\varepsilon_p)^\top\)
be a \(p\)-dimensional complex random vector whose components are independent
and identically distributed as
\(\mathcal{CN}(0,\Sigma)\) for $\Sigma\in \mathrm{SPD}(2)$. Define
\[
    \tau^2
    :=
    \mathbb E[|\varepsilon_j|^2], \quad
    \gamma
    :=
    \mathbb E[\varepsilon_j^2].
\]

For any deterministic matrix \(A\in\mathbb C^{p\times p}\), the following
fourth-order moment identities hold:
\begin{align}
    \mathbb E[
        \varepsilon\varepsilon^*
        A
        \varepsilon\varepsilon^*
    ]
    &=
    \tau^4 A
    +
    \tau^4\operatorname{Tr}(A)\mathrm{Id}_p
    +
    |\gamma|^2 A^\top,
\label{supp:eq:complex-fourth-moment-H}
\\
    \mathbb E[
        \varepsilon\varepsilon^*
        A
        \overline{\varepsilon}\varepsilon^\top
    ]
    &=
    \tau^4 A
    +
    \tau^4 A^\top
    +
    |\gamma|^2\operatorname{Tr}(A)\mathrm{Id}_p.
\label{supp:eq:complex-fourth-moment-T}
\end{align}
\end{lemma}

\begin{proof}
Recall that
\[
    \mathbb E[
        \varepsilon\varepsilon^*
    ]
    =
    \tau^2\mathrm{Id}_p,
    \quad
    \mathbb E[
        \varepsilon\varepsilon^\top
    ]
    =
    \gamma\mathrm{Id}_p.
\]
The real and imaginary parts of \(\varepsilon\) are jointly
centered Gaussian. By multilinearity, Isserlis' formula therefore
applies to products of the coordinates \(\varepsilon_j\) and
their complex conjugates.

Expanding the \((i,j)\)-th entry of the first matrix product gives
\[
\begin{aligned}
    \mathbb E[
        \varepsilon\varepsilon^*
        A
        \varepsilon\varepsilon^*
    ]_{ij}
    &=
    \mathbb E
    \left[
        \varepsilon_i
        \left(
            \sum_{k,l}
            \overline{\varepsilon}_k
            A_{kl}
            \varepsilon_l
        \right)
        \overline{\varepsilon}_j
    \right] \\
    &=
    \sum_{k,l}
    A_{kl}
    \mathbb E[
        \varepsilon_i
        \overline{\varepsilon}_k
        \varepsilon_l
        \overline{\varepsilon}_j
    ].
\end{aligned}
\]
Since the components \(\varepsilon_\nu\) are independent and centered, the
expectation is nonzero only when the indices are matched in pairs. We
evaluate this by considering separately \(i\neq j\) and \(i=j\).

When \(i\neq j\), the variables
\(\varepsilon_i\) and \(\overline{\varepsilon}_j\) must be paired with
the remaining variables. This occurs when \(k=i\) and \(l=j\), yielding
\[
    \mathbb E[
        |\varepsilon_i|^2
        |\varepsilon_j|^2
    ]
    =
    \tau^4
\]
multiplied by \(A_{ij}\), or when \(l=i\) and \(k=j\), yielding
\[
    \mathbb E[
        \varepsilon_i^2
        \overline{\varepsilon}_j^{\,2}
    ]
    =
    |\gamma|^2
\]
multiplied by \(A_{ji}\). Thus, for \(i\neq j\), the sum is
\[
    \tau^4A_{ij}
    +
    |\gamma|^2A_{ji}.
\]

When \(i=j\), the expectation becomes
\[
    \mathbb E[
        |\varepsilon_i|^2
        \overline{\varepsilon}_k
        \varepsilon_l
    ],
\]
which is nonzero only if \(k=l\). Therefore, we evaluate
\[
    \sum_k
    A_{kk}
    \mathbb E[
        |\varepsilon_i|^2
        |\varepsilon_k|^2
    ].
\]
For \(k\neq i\),
\[
    \mathbb E[
        |\varepsilon_i|^2
        |\varepsilon_k|^2
    ]
    =
    \tau^4,
\]
whereas Gaussian fourth-moment calculations give
\[
    \mathbb E[
        |\varepsilon_i|^4
    ]
    =
    2\tau^4+|\gamma|^2.
\]
Consequently,
\[
\begin{aligned}
    \sum_k
    A_{kk}
    \mathbb E[
        |\varepsilon_i|^2
        |\varepsilon_k|^2
    ]
    &=
    \sum_{k\neq i}A_{kk}\tau^4
    +
    A_{ii}
    \left(
        2\tau^4+|\gamma|^2
    \right) =
    \tau^4\operatorname{Tr}(A)
    +
    \tau^4A_{ii}
    +
    |\gamma|^2A_{ii}.
\end{aligned}
\]
Combining the two cases gives
\[
    \mathbb E[
        \varepsilon\varepsilon^*
        A
        \varepsilon\varepsilon^*
    ]_{ij}
    =
    \tau^4A_{ij}
    +
    |\gamma|^2A_{ji}
    +
    \tau^4\operatorname{Tr}(A)\delta_{ij}.
\]
Rewriting this expression in matrix form proves
\eqref{supp:eq:complex-fourth-moment-H}.

For the second identity, expanding its \((i,j)\)-th entry gives
\[
\begin{aligned}
    \mathbb E[
        \varepsilon\varepsilon^*
        A
        \overline{\varepsilon}\varepsilon^\top
    ]_{ij}
    &=
    \mathbb E
    \left[
        \varepsilon_i
        \left(
            \sum_{k,l}
            \overline{\varepsilon}_k
            A_{kl}
            \overline{\varepsilon}_l
        \right)
        \varepsilon_j
    \right] \\
    &=
    \sum_{k,l}
    A_{kl}
    \mathbb E[
        \varepsilon_i
        \varepsilon_j
        \overline{\varepsilon}_k
        \overline{\varepsilon}_l
    ].
\end{aligned}
\]
Applying the same independence and pairing argument, suppose first that
\(i\neq j\). The variables \(\varepsilon_i\) and \(\varepsilon_j\)
must be paired with their complex conjugates. This occurs when
\(k=i\) and \(l=j\), yielding \(\tau^4A_{ij}\), or when
\(k=j\) and \(l=i\), yielding \(\tau^4A_{ji}\). Thus, for
\(i\neq j\), the sum is
\[
    \tau^4A_{ij}
    +
    \tau^4A_{ji}.
\]

When \(i=j\), the expectation becomes
\[
    \mathbb E[
        \varepsilon_i^2
        \overline{\varepsilon}_k
        \overline{\varepsilon}_l
    ],
\]
which is nonzero only if \(k=l\). We therefore evaluate
\[
    \sum_k
    A_{kk}
    \mathbb E[
        \varepsilon_i^2
        \overline{\varepsilon}_k^{\,2}
    ].
\]
For \(k\neq i\),
\[
    \mathbb E[
        \varepsilon_i^2
        \overline{\varepsilon}_k^{\,2}
    ]
    =
    |\gamma|^2,
\]
whereas, for \(k=i\),
\[
    \mathbb E[
        \varepsilon_i^2
        \overline{\varepsilon}_i^{\,2}
    ]
    =
    \mathbb E[
        |\varepsilon_i|^4
    ]
    =
    2\tau^4+|\gamma|^2.
\]
It follows that
\[
\begin{aligned}
    \sum_k
    A_{kk}
    \mathbb E[
        \varepsilon_i^2
        \overline{\varepsilon}_k^{\,2}
    ]
    &=
    \sum_{k\neq i}
    A_{kk}|\gamma|^2
    +
    A_{ii}
    \left(
        2\tau^4+|\gamma|^2
    \right) =
    |\gamma|^2\operatorname{Tr}(A)
    +
    2\tau^4A_{ii}.
\end{aligned}
\]
Combining the two cases gives
\[
    \mathbb E[
        \varepsilon\varepsilon^*
        A
        \overline{\varepsilon}\varepsilon^\top
    ]_{ij}
    =
    \tau^4A_{ij}
    +
    \tau^4A_{ji}
    +
    |\gamma|^2\operatorname{Tr}(A)\delta_{ij}.
\]
Rewriting this expression in matrix form proves
\eqref{supp:eq:complex-fourth-moment-T}.
\end{proof}

We apply Lemma~\ref{supp:lem:complex-normal-fourth-moments}
with \(\Sigma=\Sigma^{(0)}\), so that \(\tau^2\) and \(\gamma\)
coincide with the noise moments in \eqref{eq:tau-gamma}.
Together with the influence vector in
\eqref{supp:eq:cov-loading-influence-vector}, these identities
yield the following explicit expression for \(V^{\texttt{cov}}\).

\begin{proposition}[Explicit form of \(V^{\texttt{cov}}\)]
\label{supp:lem:explicit-cov-loading-asymptotic-covariance}
Under the conditions of
Theorem~\ref{thm:cov-loading-expansion-clt},
\begin{equation}
\label{supp:eq:explicit-cov-loading-asymptotic-covariance}
V^{\texttt{cov}}
=
\frac{
    1
}{
    2\{\mathbb E[|\phi_1|^2]\}^2
}
\begin{pmatrix}
    \operatorname{Re}(\Gamma_H+\Gamma_T)
    &
    \operatorname{Im}(-\Gamma_H+\Gamma_T)
    \\[0.3em]
    \operatorname{Im}(\Gamma_H+\Gamma_T)
    &
    \operatorname{Re}(\Gamma_H-\Gamma_T)
\end{pmatrix},
\end{equation}
where
\begin{align}
\Gamma_H
:=
\mathbb E[u_1u_1^*]
&=
\left\{
    \mathbb E[|\phi_1|^2]\tau^2
    +
    \tau^4
\right\}
P_{\perp,0}
+
|\gamma|^2
P_{\perp,0}
\overline{\kappa^{(0)}}
(\kappa^{(0)})^\top
P_{\perp,0},
\label{supp:eq:cov-loading-Gamma-H}
\\
\Gamma_T
:=
\mathbb E[u_1u_1^\top]
&=
\left\{
    \gamma
    \mathbb E[
        \overline{\phi_1}^{\,2}
    ]
    +
    |\gamma|^2
    (\kappa^{(0)})^\top\kappa^{(0)}
\right\}
P_{\perp,0}
\overline{P_{\perp,0}}.
\label{supp:eq:cov-loading-Gamma-T}
\end{align}
Here, \(u_1\) is given in
\eqref{supp:eq:cov-loading-influence-vector}.
\end{proposition}

Since the real covariance of \(P_{\perp,0}\overline{\phi_1}\varepsilon_1\)
is positive definite on the horizontal space \(\{\widetilde{\operatorname{vec}}(z): (\kappa^{(0)})^*z=0\}\), and this term is uncorrelated with the quadratic noise term, \(V^{\texttt{cov}}\) is positive definite on this space and vanishes on its orthogonal complement, so
\(\operatorname{rank}(V^{\texttt{cov}})=2p-2\).

\begin{proof}[Proof of
Proposition~\ref{supp:lem:explicit-cov-loading-asymptotic-covariance}] Recall that \(\mathbb E[u_1]=0\). By the definition of $\widetilde{\operatorname{vec}}$, we get
\begin{equation}
\label{supp:eq:complex-to-real-covariance}
    \operatorname{Cov}
    \left\{
        \widetilde{\operatorname{vec}}(u_1)
    \right\}
    =
    \frac{1}{2}
    \begin{pmatrix}
        \operatorname{Re}(\Gamma_H+\Gamma_T)
        &
        ~\operatorname{Im}(-\Gamma_H+\Gamma_T)
        \\
        \operatorname{Im}(\Gamma_H+\Gamma_T)
        &
        \operatorname{Re}(\Gamma_H-\Gamma_T)
    \end{pmatrix}.
\end{equation}

We first calculate \(\Gamma_H\). By the definition of \(u_1\),
\[
\begin{aligned}
    \Gamma_H
    &=
    P_{\perp,0}
    \mathbb E
    \Big[
        \left(
            \overline{\phi_1}\varepsilon_1
            +
            \varepsilon_1\varepsilon_1^*
            \kappa^{(0)}
        \right)
        \left(
            \phi_1\varepsilon_1^*
            +
            (\kappa^{(0)})^*
            \varepsilon_1\varepsilon_1^*
        \right)
    \Big]
    P_{\perp,0}.
\end{aligned}
\]
Since the odd-order moments of the complex Gaussian noise are zero and
\(\phi_1\) and \(\varepsilon_1\) are independent, all cross terms vanish.
Therefore,
\[
\begin{aligned}
    \Gamma_H
    &=
    P_{\perp,0}
    \Big(
        \mathbb E[|\phi_1|^2]
        \mathbb E[
            \varepsilon_1\varepsilon_1^*
        ]
        +
        \mathbb E[
            \varepsilon_1\varepsilon_1^*
            \kappa^{(0)}(\kappa^{(0)})^*
            \varepsilon_1\varepsilon_1^*
        ]
    \Big)
    P_{\perp,0}.
\end{aligned}
\]
Applying
Lemma~\ref{supp:lem:complex-normal-fourth-moments} with
\(
    A
    =
    \kappa^{(0)}(\kappa^{(0)})^*
\)
and using
\(
    \operatorname{Tr}
    \left\{
        \kappa^{(0)}(\kappa^{(0)})^*
    \right\}
    =
    1
\),
we obtain
\[
\begin{aligned}
    &\mathbb E[
        \varepsilon_1\varepsilon_1^*
        \kappa^{(0)}(\kappa^{(0)})^*
        \varepsilon_1\varepsilon_1^*
    ] =
    \tau^4
    \kappa^{(0)}(\kappa^{(0)})^*
    +
    \tau^4\mathrm{Id}_p
    +
    |\gamma|^2
    \overline{\kappa^{(0)}}
    (\kappa^{(0)})^\top.
\end{aligned}
\]
Substituting this result and using
\(P_{\perp,0}\kappa^{(0)}=0\), we arrive at
\[
\begin{aligned}
    \Gamma_H
    &=
    \left\{
        \mathbb E[|\phi_1|^2]\tau^2
        +
        \tau^4
    \right\}
    P_{\perp,0}
    +
    |\gamma|^2
    P_{\perp,0}
    \overline{\kappa^{(0)}}
    (\kappa^{(0)})^\top
    P_{\perp,0}.
\end{aligned}
\]

Next, we calculate \(\Gamma_T\). By definition,
\[
\begin{aligned}
    \Gamma_T
    &=
    P_{\perp,0}
    \mathbb E
    \Big[
        \left(
            \overline{\phi_1}\varepsilon_1
            +
            \varepsilon_1\varepsilon_1^*
            \kappa^{(0)}
        \right)
        \left(
            \overline{\phi_1}\varepsilon_1^\top
            +
            (\kappa^{(0)})^\top
            \overline{\varepsilon}_1
            \varepsilon_1^\top
        \right)
    \Big]
    P_{\perp,0}^\top.
\end{aligned}
\]
Using
\(
    P_{\perp,0}^\top
    =
    \overline{P_{\perp,0}}
\)
and eliminating the odd-order cross terms as before gives
\[
\begin{aligned}
    \Gamma_T
    &=
    P_{\perp,0}
    \Big(
        \mathbb E[
            \overline{\phi_1}^{\,2}
        ]
        \mathbb E[
            \varepsilon_1\varepsilon_1^\top
        ]
        +
        \mathbb E[
            \varepsilon_1\varepsilon_1^*
            \kappa^{(0)}(\kappa^{(0)})^\top
            \overline{\varepsilon}_1
            \varepsilon_1^\top
        ]
    \Big)
    \overline{P_{\perp,0}}.
\end{aligned}
\]
Applying
Lemma~\ref{supp:lem:complex-normal-fourth-moments} with
\(
    A
    =
    \kappa^{(0)}(\kappa^{(0)})^\top
\)
gives
\[
\begin{aligned}
    &\mathbb E[
        \varepsilon_1\varepsilon_1^*
        \kappa^{(0)}(\kappa^{(0)})^\top
        \overline{\varepsilon}_1
        \varepsilon_1^\top
    ] =
    2\tau^4
    \kappa^{(0)}(\kappa^{(0)})^\top
    +
    |\gamma|^2
    \left\{
        (\kappa^{(0)})^\top\kappa^{(0)}
    \right\}
    \mathrm{Id}_p.
\end{aligned}
\]
Substituting this expression, using
\(
    \mathbb E[
        \varepsilon_1\varepsilon_1^\top
    ]
    =
    \gamma\mathrm{Id}_p,
\)
and applying \(P_{\perp,0}\kappa^{(0)}=0\), we obtain
\[
\begin{aligned}
    \Gamma_T
    &=
    \left\{
        \gamma
        \mathbb E[
            \overline{\phi_1}^{\,2}
        ]
        +
        |\gamma|^2
        (\kappa^{(0)})^\top\kappa^{(0)}
    \right\}
    P_{\perp,0}
    \overline{P_{\perp,0}}.
\end{aligned}
\]

By \eqref{eq:cov-loading-asymptotic-covariance} and the
influence representation established above,
\[
    V^{\texttt{cov}}
    =
    \frac{
        \operatorname{Cov}
        \left\{
            \widetilde{\operatorname{vec}}(u_1)
        \right\}
    }{
        \{\mathbb E[|\phi_1|^2]\}^2
    }.
\]
Substituting
\eqref{supp:eq:cov-loading-Gamma-H} and
\eqref{supp:eq:cov-loading-Gamma-T} into
\eqref{supp:eq:complex-to-real-covariance}
therefore proves
\eqref{supp:eq:explicit-cov-loading-asymptotic-covariance}.
\end{proof}

\subsection{Expansion of the real residual projection}
\label{supp:subsec:real-residual-projection-expansion}

The estimator of the noise covariance uses the real residual projection onto the orthogonal complement of the real and imaginary components of the loading. For \(\kappa\in\mathcal S^{2p-1}\) such that \(\mathcal U(\kappa)\) has full column rank, the projection \(\Pi_\perp(\kappa)\) defined in
\eqref{eq:orthogonal-residual-projector} is represented by a real symmetric \(p\times p\) matrix. Acting on \(\mathbb R^p\), this matrix is the orthogonal projector onto
\[
    \operatorname{span}_{\mathbb R}
    \{\operatorname{Re}(\kappa),\operatorname{Im}(\kappa)\}^{\perp}.
\]
Its complex-linear extension to \(\mathbb C^p\) is the orthogonal
projector onto
\[
    \operatorname{span}_{\mathbb C}
    \{\kappa,\overline \kappa\}^{\perp}.
\]
The real span is unchanged under multiplication of \(\kappa\) by a
unit-modulus complex number. Hence \(\Pi_\perp(\kappa)\) is independent of the representative of \([\kappa]\).

We first establish a deterministic perturbation expansion of
this projection. We then derive a general projection expansion
induced by a loading expansion and apply it to the
\texttt{cov}-estimator.

\begin{lemma}
\label{supp:lem: deterministic real residual projection first order expansion}
Let
\(\kappa^{(0)},\kappa\in\mathcal S^{2p-1}\), and write
\(
U:=\mathcal U(\kappa).
\)
Assume that \(U_0\) has full column rank. Then there exist constants \(\delta>0\) and \(C>0\), depending only on \(\kappa^{(0)}\), such that,
whenever
\[
\|\kappa-\kappa^{(0)}\|<\delta,
\]
the matrix \(U\) also has full column rank and
\begin{equation}
\label{supp:eq: deterministic projection first order expansion}
\begin{aligned}
    \Pi_\perp(\kappa)-\Pi_{\perp,0}
    &=
    -\mathcal U(\Pi_{\perp,0}\kappa)
    (U_0^\top U_0)^{-1}U_0^\top
    -
    U_0(U_0^\top U_0)^{-1}
    \mathcal U(\Pi_{\perp,0}\kappa)^\top
    +
    R_\Pi(\kappa,\kappa^{(0)}),
\end{aligned}
\end{equation}
where the remainder satisfies
\begin{equation}
\label{supp:eq: deterministic projection quadratic remainder bound}
    \|R_\Pi(\kappa,\kappa^{(0)})\|_{\operatorname{op}}
    \le
    C\|U-U_0\|_{F}^2=C\|\kappa-\kappa^{(0)}\|^{2}.
\end{equation}
\end{lemma}

\begin{proof}
Let
\[
    \Delta U:=U-U_0,
    \quad
    U(t):=U_0+t\Delta U,\quad 0\le t\le1.
\]
Define
\[
    \mathcal H(V)
    :=
    \mathrm{Id}_p-V(V^\top V)^{-1}V^\top
\]
on the set of full-column-rank \(p\times2\) real matrices \(V\). Then
\[
    F(t):=\mathcal H(U(t))
    =
    \mathrm{Id}_p-U(t)\{U(t)^\top U(t)\}^{-1}U(t)^\top .
\]
Thus \(F(0)=\Pi_{\perp,0}\) and \(F(1)=\Pi_\perp(\kappa)\).

Since \(U_0\) has full column rank, there
exists \(\delta>0\), depending only on \(U_0\), such that if
\(\|U-U_0\|_{F}<\delta\), \(U\) has full column rank. For every \(t\in[0,1]\),
\[
    \|U(t)-U_0\|_F
    =
    t\|\Delta U\|_F
    <
    \delta.
\]
Hence the entire path lies in the full-column-rank neighborhood
of \(U_0\). After decreasing \(\delta\) if necessary, there is a constant
\(C>0\), depending only on \(U_0\), such that
\begin{equation}
\label{supp:eq: deterministic uniform U inverse bounds}
    \sup_{0\le t\le1}\|U(t)\|_{F}\le C,
    \quad
    \sup_{0\le t\le1}
    \left\|
        \{U(t)^\top U(t)\}^{-1}
    \right\|_{\operatorname{op}}
    \le C .
\end{equation}

We first compute \(F'(0)\). Since \(U'(t)=\Delta U\), differentiating the inverse
matrix gives
\begin{equation}
\label{supp:eq: deterministic inverse first derivative}
\begin{aligned}
    \frac{d}{dt}
    &\{U(t)^\top U(t)\}^{-1}
    \\
    &=
    -\{U(t)^\top U(t)\}^{-1}
    \{\Delta U^\top U(t)+U(t)^\top\Delta U\}
    \{U(t)^\top U(t)\}^{-1}.
\end{aligned}
\end{equation}
Therefore, differentiating
\begin{align}
\label{supp:eq: F(t)}
F(t)
    =
    \mathrm{Id}_p-U(t)\{U(t)^\top U(t)\}^{-1}U(t)^\top
\end{align}
yields
\begin{align}
\label{supp:eq: F'(t)}
    F'(t)
    &=
    -\Delta U(U(t)^\top U(t))^{-1}U(t)^\top
    -U(t)(U(t)^\top U(t))^{-1}\Delta U^\top \notag\\
    &\quad
    +U(t)(U(t)^\top U(t))^{-1}
    \{\Delta U^\top U(t)+U(t)^\top\Delta U\}
    (U(t)^\top U(t))^{-1}U(t)^\top .
\end{align}
Substituting \(t=0\) into \eqref{supp:eq: F'(t)}, we get
\begin{align}
\label{supp:eq: deterministic projection derivative raw}
    F'(0)
    &=
    -\Delta U(U_0^\top U_0)^{-1}U_0^\top
    -U_0(U_0^\top U_0)^{-1}\Delta U^\top \notag\\
    &\quad
    +U_0(U_0^\top U_0)^{-1}
    \{\Delta U^\top U_0+U_0^\top\Delta U\}
    (U_0^\top U_0)^{-1}U_0^\top .
\end{align}
Using \( \Pi_{\perp,0}
    =
    \mathrm{Id}_p-U_0(U_0^\top U_0)^{-1}U_0^\top, \) we get
\begin{equation}
\label{supp:eq: deterministic projection derivative simplified}
    F'(0)
    =
    -\Pi_{\perp,0}\Delta U(U_0^\top U_0)^{-1}U_0^\top
    -
    U_0(U_0^\top U_0)^{-1}\Delta U^\top\Pi_{\perp,0}.
\end{equation}
Since \(\Delta U=U-U_0=\mathcal U(\kappa-\kappa^{(0)})\), and since
\(\Pi_{\perp,0}\) is real and \(\Pi_{\perp,0} \kappa^{(0)} = 0\), we get
\[
    \Pi_{\perp,0}\Delta U
    =
    \Pi_{\perp,0}\mathcal U(\kappa - \kappa^{(0)})
    =
    \mathcal U(\Pi_{\perp,0}\kappa - \Pi_{\perp,0}\kappa^{(0)})
    =
    \mathcal U(\Pi_{\perp,0}\kappa)
    ,
\]
and similarly,
\[
    \Delta U^\top\Pi_{\perp,0}
    =
    \mathcal U(\Pi_{\perp,0}\kappa)^\top .
\]
Hence
\begin{equation}
\label{supp:eq: deterministic projection derivative final}
    F'(0)
    =
    -\mathcal U(\Pi_{\perp,0}\kappa)
    (U_0^\top U_0)^{-1}U_0^\top
    -
    U_0(U_0^\top U_0)^{-1}
    \mathcal U(\Pi_{\perp,0}\kappa)^\top .
\end{equation}

It remains to bound the Taylor remainder. Recall that the map \(F(t)\) is the composition
of the smooth projection map \(\mathcal H\) and the affine path
\(t\mapsto U_0+t\Delta U\). Since this affine path has a second derivative of zero, every
second-order term in \(F''(t)\) must contain exactly two factors of \(\Delta U\). We now
make this explicit.

Differentiating \eqref{supp:eq: deterministic inverse first derivative} once more gives
\begin{equation}
\label{supp:eq: deterministic inverse second derivative}
\begin{aligned}
&\frac{d^2}{dt^2}
\{U(t)^\top U(t)\}^{-1} \\
&=
2\{U(t)^\top U(t)\}^{-1}
\{\Delta U^\top U(t)+U(t)^\top\Delta U\}
\{U(t)^\top U(t)\}^{-1}
\\
&\quad\quad\{\Delta U^\top U(t)+U(t)^\top\Delta U\}
\{U(t)^\top U(t)\}^{-1} \\
&
-
2\{U(t)^\top U(t)\}^{-1}
\Delta U^\top\Delta U
\{U(t)^\top U(t)\}^{-1}.
\end{aligned}
\end{equation}
By \eqref{supp:eq: deterministic uniform U inverse bounds},
\[
    \sup_{0\le t\le1}
    \left\|
        \Delta U^\top U(t)+U(t)^\top\Delta U
    \right\|_{F}
    \le
    2C\|\Delta U\|_{F}.
\]
Therefore \eqref{supp:eq: deterministic inverse first derivative} and
\eqref{supp:eq: deterministic inverse second derivative} imply
\begin{equation}
\label{supp:eq: deterministic inverse derivative bounds}
\begin{aligned}
    \sup_{0\le t\le1}
    \left\|
        \frac{d}{dt}
        \{U(t)^\top U(t)\}^{-1}
    \right\|_{F} \le C'\|\Delta U\|_{F}, ~
    \sup_{0\le t\le1}
    \left\|
        \frac{d^2}{dt^2}
        \{U(t)^\top U(t)\}^{-1}
    \right\|_{F}
    \le
    C'\|\Delta U\|_{F}^2 .
\end{aligned}
\end{equation}
for some \(C'>0\), which only depends on \(U_{0}\).

Now differentiate \eqref{supp:eq: F(t)} twice. Since \(U''(t)=0\), we obtain
\begin{equation}
\label{supp:eq: deterministic F second derivative formula}
\begin{aligned}
    F''(t)
    &=
    -2\Delta U
    \left[
        \frac{d}{dt}
        \{U(t)^\top U(t)\}^{-1}
    \right]
    U(t)^\top
    -2\Delta U
    \{U(t)^\top U(t)\}^{-1}
    \Delta U^\top \\
    &\quad
    -U(t)
    \left[
        \frac{d^2}{dt^2}
        \{U(t)^\top U(t)\}^{-1}
    \right]
    U(t)^\top
    -2U(t)
    \left[
        \frac{d}{dt}
        \{U(t)^\top U(t)\}^{-1}
    \right]
    \Delta U^\top .
\end{aligned}
\end{equation}
Each term in \eqref{supp:eq: deterministic F second derivative formula} is of
quadratic order in \(\Delta U\): the second term contains two explicit factors
of \(\Delta U\), the first and fourth terms contain one explicit factor of
\(\Delta U\) and one first derivative of the inverse matrix, and the third term
contains the second derivative of the inverse matrix. Combining
\eqref{supp:eq: deterministic uniform U inverse bounds} and
\eqref{supp:eq: deterministic inverse derivative bounds}, we get
\begin{equation}
\label{supp:eq: deterministic projection second derivative bound}
    \sup_{0\le t\le1}\|F''(t)\|_{F}
    \le
    C''\|\Delta U\|_{F}^2
\end{equation}
for some \(C''>0\), which only depends on \(U_{0}\).

By Taylor's theorem with integral remainder,
\[
    F(1)-F(0)
    =
    F'(0)
    +
    \int_0^1(1-t)F''(t)\,dt .
\]
Define
\[
    R_\Pi(\kappa,\kappa^{(0)})
    :=
    \int_0^1(1-t)F''(t)\,dt .
\]
Then \eqref{supp:eq: deterministic projection second derivative bound} implies
\[
    \|R_\Pi(\kappa,\kappa^{(0)})\|_{F}
    \le
    C''\|\Delta U\|_{F}^2
    =
    C''\|U-U_0\|_{F}^2.
\]
Since
\[
    \|R_\Pi(\kappa,\kappa^{(0)})\|_{\operatorname{op}}
    \le
    \|R_\Pi(\kappa,\kappa^{(0)})\|_{F},
\]
this shows the desired quadratic remainder bound. Finally,
\(F(1)-F(0)=\Pi_\perp(\kappa)-\Pi_{\perp,0}\), and substituting
\eqref{supp:eq: deterministic projection derivative final} gives
\eqref{supp:eq: deterministic projection first order expansion}.
\end{proof}

\begin{lemma}[Residual projection induced by a loading expansion]
\label{supp:lem:real-residual-projection-loading-expansion}
Let \(\kappa^{(0)}\in\mathcal S^{2p-1}\), and suppose that
\(U_0=\mathcal U(\kappa^{(0)})\) has full column rank.
Let \(\widehat\kappa_n\in\mathcal S^{2p-1}\) satisfy
\[
    \widehat\kappa_n
    =
    \kappa^{(0)}
    +
    L_{\kappa,n}
    +
    R_{\kappa,n},
    \quad
    L_{\kappa,n}=\mathcal O_p(n^{-1/2}),
    \quad
    R_{\kappa,n}=\mathcal O_p(n^{-1}).
\]
Define
\[
\begin{aligned}
    A_{\Pi,n}:=
    \mathcal U(\Pi_{\perp,0}L_{\kappa,n})
    (U_0^\top U_0)^{-1}U_0^\top
    +
    U_0(U_0^\top U_0)^{-1}
    \mathcal U(\Pi_{\perp,0}L_{\kappa,n})^\top.
\end{aligned}
\]
Then \(\mathcal U(\widehat\kappa_n)\) has full column rank
with probability tending to one. Set
\[
    \widehat\Pi_{\perp,n}
    :=
    \begin{cases}
        \Pi_\perp(\widehat\kappa_n),
        & \operatorname{rank}\{\mathcal U(\widehat\kappa_n)\}=2,
        \\
        \Pi_{\perp,0},
        & \text{otherwise}.
    \end{cases}
\]
There exist remainders \(R_{\Pi,n}\) and \(r_{\Pi,n}\) such that
\begin{align}
    \widehat\Pi_{\perp,n}-\Pi_{\perp,0}
    &=
    -A_{\Pi,n}+R_{\Pi,n},
    &
    \|R_{\Pi,n}\|_{\operatorname{op}}
    &=
    \mathcal O_p(n^{-1}),
    \label{supp:eq:general-residual-projection-expansion}
    \\
    \widehat\Pi_{\perp,n}\kappa^{(0)}
    &=
    -\Pi_{\perp,0}L_{\kappa,n}+r_{\Pi,n},
    &
    \|r_{\Pi,n}\|
    &=
    \mathcal O_p(n^{-1}),
    \label{supp:eq:general-projected-true-loading-expansion}
\end{align}
where \(r_{\Pi,n}=R_{\Pi,n}\kappa^{(0)}\).
Moreover, \(A_{\Pi,n}=\mathcal O_p(n^{-1/2})\).

More precisely, there exist constants \(\delta,C>0\), depending
only on \(\kappa^{(0)}\), such that, on
\(\{\|\widehat\kappa_n-\kappa^{(0)}\|<\delta\}\),
\begin{equation}
\label{supp:eq:general-projection-remainder-bound}
    \|R_{\Pi,n}\|_{\operatorname{op}}
    \leq
    C\left(
        \|R_{\kappa,n}\|
        +
        \|\widehat\kappa_n-\kappa^{(0)}\|^2
    \right).
\end{equation}
\end{lemma}

\begin{proof}
The assumed loading expansion gives
\[
    \|\widehat\kappa_n-\kappa^{(0)}\|
    =
    \mathcal O_p(n^{-1/2}).
\]
Choose \(\delta>0\) as in
Lemma~\ref{supp:lem: deterministic real residual projection first order expansion},
and write
\[
    \mathcal C_n
    :=
    \{\|\widehat\kappa_n-\kappa^{(0)}\|<\delta\}.
\]
Then \(\mathcal{P}(\mathcal C_n)\to1\), and
\(\mathcal U(\widehat\kappa_n)\) has full column rank on
\(\mathcal C_n\).

Define
\[
    R_{\Pi,n}
    :=
    \widehat\Pi_{\perp,n}-\Pi_{\perp,0}+A_{\Pi,n}.
\]
On \(\mathcal C_n\), the deterministic projection expansion and
\(\Pi_{\perp,0}\kappa^{(0)}=0\) give
\[
\begin{aligned}
    R_{\Pi,n}
    =
    R_\Pi(\widehat\kappa_n,\kappa^{(0)})
    -
    \mathcal U(\Pi_{\perp,0}R_{\kappa,n})
    (U_0^\top U_0)^{-1}U_0^\top
    -
    U_0(U_0^\top U_0)^{-1}
    \mathcal U(\Pi_{\perp,0}R_{\kappa,n})^\top.
\end{aligned}
\]
Since
\(\|\mathcal U(z)\|_{\operatorname{op}}
\leq\|\mathcal U(z)\|_F=\|z\|\),
the deterministic remainder bound proves
\eqref{supp:eq:general-projection-remainder-bound}.
The assumed stochastic orders and
\(\mathcal{P}(\mathcal C_n)\to1\) therefore imply
\(\|R_{\Pi,n}\|_{\operatorname{op}}=\mathcal O_p(n^{-1})\).
The definition of \(A_{\Pi,n}\) also gives
\(A_{\Pi,n}=\mathcal O_p(n^{-1/2})\).

Finally,
\[
    (U_0^\top U_0)^{-1}U_0^\top\kappa^{(0)}
    =
    \begin{pmatrix}1\\ i\end{pmatrix},
    \quad
    \mathcal U(\Pi_{\perp,0}L_{\kappa,n})^\top
    \kappa^{(0)}
    =
    0.
\]
The second identity follows because \(\Pi_{\perp,0}\) is real
and annihilates both the real and imaginary parts of
\(\kappa^{(0)}\). Consequently,
\[
    A_{\Pi,n}\kappa^{(0)}
    =
    \mathcal U(\Pi_{\perp,0}L_{\kappa,n})
    \begin{pmatrix}1\\ i\end{pmatrix}
    =
    \Pi_{\perp,0}L_{\kappa,n}.
\]
Multiplying
\eqref{supp:eq:general-residual-projection-expansion}
by \(\kappa^{(0)}\) proves
\eqref{supp:eq:general-projected-true-loading-expansion},
with \(r_{\Pi,n}=R_{\Pi,n}\kappa^{(0)}\).
Since \(\|\kappa^{(0)}\|=1\), its remainder has the stated order.
\end{proof}

We now apply Lemma~\ref{supp:lem:real-residual-projection-loading-expansion} to the \texttt{cov}-estimator. Although the optimally aligned representative is used to obtain a local vector expansion, the resulting real residual projection does not depend on this choice of representative.

Whenever an estimator-based quantity below is undefined, assign it
an arbitrary fixed finite value. By Theorem~\ref{thm:cov-estimator-strong-consistency}, such exceptional events have probabilities tending to zero. These assignments do not affect the stochastic-order statements below.

\begin{corollary}[Real residual projection expansion for the
\texttt{cov}-estimator]
\label{supp:cor:cov-real-residual-projection-expansion}
Assume the conditions of
Theorem~\ref{thm:cov-loading-expansion-clt}, and suppose that
\(
    U_0
    =
    \mathcal U(\kappa^{(0)})
\)
has full column rank. Denote
\(
    \widehat U^{\texttt{cov,opt}}
    =
    \mathcal U
    \left(
        \widehat\kappa^{\texttt{cov,opt}}
    \right)
\)
and
\[
    \widehat\Pi_\perp^{\texttt{cov}}
    =
    \Pi_\perp
    \left(
        \widehat\kappa^{\texttt{cov,opt}}
    \right)
    =
    \mathrm{Id}_p
    -
    \widehat U^{\texttt{cov,opt}}
    \left\{
        (\widehat U^{\texttt{cov,opt}})^\top
        \widehat U^{\texttt{cov,opt}}
    \right\}^{-1}
    (\widehat U^{\texttt{cov,opt}})^\top.
\]
Then there exist constants \(\delta>0\) and \(C>0\), depending only on
the fixed population quantities, such that whenever
\(
    \|E_H\|_F<\delta
\), the matrix \(\widehat U^{\texttt{cov,opt}}\) has full column rank and
\begin{equation}
\label{supp:eq:cov-real-residual-projection-expansion}
    \widehat\Pi_\perp^{\texttt{cov}}
    -
    \Pi_{\perp,0}
    =
    -
    A_{\Pi,n}^{\texttt{cov}}
    +
    R_{\Pi,n}^{\texttt{cov}},
\end{equation}
where \(A_{\Pi,n}^{\texttt{cov}}\) is the matrix
\(A_{\Pi,n}\) in
Lemma~\ref{supp:lem:real-residual-projection-loading-expansion}
with \(L_{\kappa,n}=L_{\kappa,n}^{\texttt{cov}}\), and
\begin{equation}
\label{supp:eq:cov-real-residual-projection-remainder}
    \left\|
        R_{\Pi,n}^{\texttt{cov}}
    \right\|_{\operatorname{op}}
    \leq
    C\|E_H\|_F^2.
\end{equation}
Consequently,
\[
    A_{\Pi,n}^{\texttt{cov}}
    =
    \mathcal O_p(n^{-1/2}),
    \quad
    R_{\Pi,n}^{\texttt{cov}}
    =
    \mathcal O_p(n^{-1}),
\]
and hence
\[
    \widehat\Pi_\perp^{\texttt{cov}}
    -
    \Pi_{\perp,0}
    =
    -
    A_{\Pi,n}^{\texttt{cov}}
    +
    \mathcal O_p(n^{-1}).
\]
\end{corollary}

\begin{proof}
Apply
Lemma~\ref{supp:lem:real-residual-projection-loading-expansion}
with
\[
    \widehat\kappa_n
    =
    \widehat\kappa^{\texttt{cov,opt}},
    \quad
    L_{\kappa,n}
    =
    \frac{1}{\mathbb E[|\phi_1|^2]}
    P_{\perp,0}E_H\kappa^{(0)},
    \quad
    R_{\kappa,n}
    =
    R_{\kappa,n}^{\texttt{cov}}.
\]
By
Lemma~\ref{supp:lem:cov-quantitative-loading-expansion},
for sufficiently small \(\|E_H\|_F\),
\[
    \|\widehat\kappa^{\texttt{cov,opt}}-\kappa^{(0)}\|
    \leq C\|E_H\|_F,
    \quad
    \|R_{\kappa,n}^{\texttt{cov}}\|
    \leq C\|E_H\|_F^2.
\]
Since
\(\Pi_{\perp,0}P_{\perp,0}=\Pi_{\perp,0}\),
the matrix \(A_{\Pi,n}\) in that lemma coincides with
\(A_{\Pi,n}^{\texttt{cov}}\).
Its quantitative remainder bound therefore yields
\[
\begin{aligned}
    \|R_{\Pi,n}^{\texttt{cov}}\|_{\operatorname{op}}
    &\leq
    C\left(
        \|R_{\kappa,n}^{\texttt{cov}}\|
        +
        \|\widehat\kappa^{\texttt{cov,opt}}
              -\kappa^{(0)}\|^2
    \right)
    \\
    &\leq
    C\|E_H\|_F^2.
\end{aligned}
\]
This proves the asserted local expansion and remainder bound.
The stochastic orders follow from
Lemma~\ref{supp:lem:cov-moment-bounds} with \(k=2\).
\end{proof}

\subsection{Second-order expansion of the noise covariance estimator}
\label{supp:subsec:second-order-noise-covariance-expansion}

This subsection proves
Theorem~\ref{thm:cov-noise-covariance-expansion}. The proof uses the
quadratic remainder bounds obtained for the loading and corresponding real residual projection, together with the root-\(n\) orders of \(E_H\) and \(E_T\).

For \(B\in\mathbb C^{p\times p}\), let
\[
    \|B\|_*
    :=
    \operatorname{Tr}\bigl\{(B^*B)^{1/2}\bigr\}
\]
denote its nuclear norm, equivalently the sum of its singular values.
We use the inequalities
\[
    |\operatorname{Tr}(AB)|
    \leq
    \|A\|_{\operatorname{op}}\|B\|_*,
    \quad
    \|B\|_*
    \leq
    \sqrt p\,\|B\|_F,
    \quad
    A,B\in\mathbb C^{p\times p}.
\]

The second-order terms in Theorem~\ref{thm:cov-noise-covariance-expansion} are defined by
\begin{equation}
\label{eq:cov-noise-second-order-terms}
\begin{aligned}
Q_{\tau,n}^{\texttt{cov}}
&:=\frac{1}{p-2}
\left\{
-\operatorname{Tr}(A_{\Pi,n}^{\texttt{cov}}E_H)
+\mathbb E[|\phi_1|^2]
\|\Pi_{\perp,0}L_{\kappa,n}^{\texttt{cov}}\|^2
\right\},
\\
Q_{\gamma,n}^{\texttt{cov}}
&:=\frac{1}{p-2}
\left\{
-\operatorname{Tr}(A_{\Pi,n}^{\texttt{cov}}E_T)
+\mathbb E[\phi_1^2]
(\Pi_{\perp,0}L_{\kappa,n}^{\texttt{cov}})^\top
(\Pi_{\perp,0}L_{\kappa,n}^{\texttt{cov}})
\right\},
\end{aligned}
\end{equation}
where
\begin{equation*}
\begin{aligned}
A_{\Pi,n}^{\texttt{cov}}
:=
\mathcal U
\left(
\Pi_{\perp,0}L_{\kappa,n}^{\texttt{cov}}
\right)
(U_0^\top U_0)^{-1}U_0^\top
+
U_0(U_0^\top U_0)^{-1}
\mathcal U
\left(
\Pi_{\perp,0}L_{\kappa,n}^{\texttt{cov}}
\right)^\top.
\end{aligned}
\end{equation*}

\begin{proof}[Proof of
Theorem~\ref{thm:cov-noise-covariance-expansion}]
Throughout the proof, \(C>0\) denotes a generic finite constant depending
only on the fixed population quantities, whose value may change from line to
line. Let \(\delta>0\) be sufficiently small that
Corollary~\ref{supp:cor:cov-real-residual-projection-expansion}
applies. Let
\[
    \mathcal A_n
    :=
    \left\{
        \|E_H\|_{F}<\delta
    \right\}.
\]
We may additionally assume that \(\delta\le1\). By
Lemma~\ref{supp:lem:cov-moment-bounds} with \(k=2\),
\[
    \|E_H\|_{F}
    =
    \mathcal O_p(n^{-1/2}),
    \quad
    \|E_T\|_{F}
    =
    \mathcal O_p(n^{-1/2}).
\]
In particular,
\[
    \mathcal{P}(\mathcal A_n)\longrightarrow1.
\]
The deterministic identities and bounds below are understood
on \(\mathcal A_n\).

We first consider \( \widehat\tau^{2,\texttt{cov}}\). Substituting
\[
    \widehat\Sigma^{\texttt{Her}}
    =
    \Sigma^{\texttt{Her}}+E_H,
    \quad
    \Sigma^{\texttt{Her}}
    =
    \mathbb E[|\phi_1|^2]
    \kappa^{(0)}(\kappa^{(0)})^*
    +
    \tau^2\mathrm{Id}_p
\]
into its definition and using
\(\operatorname{Tr}(\widehat\Pi_\perp^{\texttt{cov}})=p-2\), we obtain
\begin{align*}
     \widehat\tau^{2,\texttt{cov}}-\tau^2
    &=
    \frac{1}{p-2}
    \operatorname{Tr}
    \left(
        \widehat\Pi_\perp^{\texttt{cov}}E_H
    \right)+
    \frac{\mathbb E[|\phi_1|^2]}{p-2}
    (\kappa^{(0)})^*
    \widehat\Pi_\perp^{\texttt{cov}}
    \kappa^{(0)}.
\end{align*}
Accordingly, define
\begin{equation}
\label{supp:eq: tau remainder exact definition}
\begin{aligned}
\widehat R_{\tau,n}^{\texttt{cov}}
&:=
\frac{1}{p-2}
\operatorname{Tr}
\left[
    \left(
        \widehat\Pi_\perp^{\texttt{cov}}
        -
        \Pi_{\perp,0}
        +
        A_{\Pi,n}^{\texttt{cov}}
    \right)E_H
\right]\\
&+
\frac{\mathbb E[|\phi_1|^2]}{p-2}
\left[
    (\kappa^{(0)})^*
    \widehat\Pi_\perp^{\texttt{cov}}
    \kappa^{(0)}
    -
    \frac{
        \|\Pi_{\perp,0}E_H\kappa^{(0)}\|^2
    }{
        \{\mathbb E[|\phi_1|^2]\}^2
    }
\right].
\end{aligned}
\end{equation}
By the definitions in
\eqref{eq:cov-noise-first-order-terms} and
\eqref{eq:cov-noise-second-order-terms}, this gives identically
\begin{equation}
\label{supp:eq: exact cov tau second order decomposition}
     \widehat\tau^{2,\texttt{cov}}-\tau^2
    =
    L_{\tau,n}^{\texttt{cov}}
    +
    Q_{\tau,n}^{\texttt{cov}}
    +
    \widehat R_{\tau,n}^{\texttt{cov}}.
\end{equation}

On \(\mathcal A_n\),
Corollary~\ref{supp:cor:cov-real-residual-projection-expansion}
gives
\[
    \widehat\Pi_\perp^{\texttt{cov}}
    -
    \Pi_{\perp,0}
    =
    -A_{\Pi,n}^{\texttt{cov}}
    +
    R_{\Pi,n}^{\texttt{cov}},
    \quad
    \|R_{\Pi,n}^{\texttt{cov}}\|_{\operatorname{op}}
    \le
    C\|E_H\|_{F}^2.
\]
Applying the matrix trace inequality, we get a bound for the first term as
follows:
\begin{align}
\label{supp:eq: R tau hat first term bound}
&\left|
\operatorname{Tr}
\left[
    \left(
        \widehat\Pi_\perp^{\texttt{cov}}
        -
        \Pi_{\perp,0}
        +
        A_{\Pi,n}^{\texttt{cov}}
    \right)E_H
\right]
\right| \notag\\
&\quad=
    \left|
        \operatorname{Tr}
        \left(
            R_{\Pi,n}^{\texttt{cov}}E_H
        \right)
    \right| \notag\\
&\quad\le
    \|R_{\Pi,n}^{\texttt{cov}}\|_{\operatorname{op}}
    \|E_H\|_* \notag\\
&\quad\le
    \sqrt{p}\,
    \|R_{\Pi,n}^{\texttt{cov}}\|_{\operatorname{op}}
    \|E_H\|_{F} \notag\\
&\quad\le
    C\|E_H\|_{F}^3
\end{align}
on \(\mathcal A_n\).

We next expand the second term. By
\eqref{supp:eq:general-projected-true-loading-expansion}
and \(\Pi_{\perp,0}P_{\perp,0}=\Pi_{\perp,0}\),
\begin{equation}
\label{supp:eq: stochastic phihat kappa cov}
    \widehat\Pi_\perp^{\texttt{cov}}\kappa^{(0)}
    =
    -
    \frac{1}{\mathbb E[|\phi_1|^2]}
    \Pi_{\perp,0}E_H\kappa^{(0)}
    +
    r_{\Pi,n},
\end{equation}
where
\[
    r_{\Pi,n}
    :=
    R_{\Pi,n}^{\texttt{cov}}\kappa^{(0)}.
\]
On \(\mathcal A_n\),
\eqref{supp:eq:cov-real-residual-projection-remainder}
therefore gives
\begin{equation}
\label{supp:eq: r stochastic bound}
    \|r_{\Pi,n}\|
    \leq
    \|R_{\Pi,n}^{\texttt{cov}}\|_{\operatorname{op}}
    \leq
    C\|E_H\|_F^2.
\end{equation}

Since \(\widehat\Pi_\perp^{\texttt{cov}}\) is a real symmetric
orthogonal projection,
\[
    (\kappa^{(0)})^*
    \widehat\Pi_\perp^{\texttt{cov}}
    \kappa^{(0)}
    =
    \left\|
        \widehat\Pi_\perp^{\texttt{cov}}\kappa^{(0)}
    \right\|^2.
\]
Using \eqref{supp:eq: stochastic phihat kappa cov}, we therefore obtain
\begin{align*}
    &(\kappa^{(0)})^*
    \widehat\Pi_\perp^{\texttt{cov}}
    \kappa^{(0)}\\
    &\quad=
    \left\|
        \widehat\Pi_\perp^{\texttt{cov}}
        \kappa^{(0)}
    \right\|^2\\
    &\quad=
    \left\|
        -
        \frac{1}{\mathbb E[|\phi_1|^2]}
        \Pi_{\perp,0}E_H\kappa^{(0)}
        +
        r_{\Pi,n}
    \right\|^2\\
    &\quad=
    \frac{
        \|\Pi_{\perp,0}E_H\kappa^{(0)}\|^2
    }{
        \{\mathbb E[|\phi_1|^2]\}^2
    }
    -
    \frac{2}{\mathbb E[|\phi_1|^2]}
    \operatorname{Re}
    \left\{
        \bigl(
            \Pi_{\perp,0}E_H\kappa^{(0)}
        \bigr)^*
        r_{\Pi,n}
    \right\}
    +
    \|r_{\Pi,n}\|^2.
\end{align*}
Consequently, using \eqref{supp:eq: r stochastic bound}, we get
\begin{align}
\label{supp:eq: R tau hat second term bound}
&\left|
    (\kappa^{(0)})^*
    \widehat\Pi_\perp^{\texttt{cov}}
    \kappa^{(0)}
    -
    \frac{
        \|\Pi_{\perp,0}E_H\kappa^{(0)}\|^2
    }{
        \{\mathbb E[|\phi_1|^2]\}^2
    }
\right| \notag\\
&\quad\le
\frac{2}{\mathbb E[|\phi_1|^2]}
\left\|
    \Pi_{\perp,0}E_H\kappa^{(0)}
\right\|
\|r_{\Pi,n}\|
+
\|r_{\Pi,n}\|^2 \notag\\
&\quad\le
C\|E_H\|_{F}
\|E_H\|_{F}^2
+
C\|E_H\|_{F}^4 \notag\\
&\quad\le
C\|E_H\|_{F}^3
\end{align}
on \(\mathcal A_n\), where the last inequality uses
\(\|E_H\|_{F}<\delta\le1\). Therefore, combining
\eqref{supp:eq: R tau hat first term bound} and
\eqref{supp:eq: R tau hat second term bound}, we get
\begin{equation}
\label{supp:eq: tau remainder local bound}
    \left|
        \widehat R_{\tau,n}^{\texttt{cov}}
    \right|
    \le
    C\|E_H\|_{F}^3
\end{equation}
on \(\mathcal A_n\).

We next consider \(\widehat\gamma^{\texttt{cov}}\). Substituting
\[
    \widehat\Sigma^{\texttt{Tra}}
    =
    \Sigma^{\texttt{Tra}}+E_T,
    \quad
    \Sigma^{\texttt{Tra}}
    =
    \mathbb E[\phi_1^2]
    \kappa^{(0)}(\kappa^{(0)})^\top
    +
    \gamma\mathrm{Id}_p
\]
into its definition gives
\begin{align*}
    \widehat\gamma^{\texttt{cov}}-\gamma
    &=
    \frac{1}{p-2}
    \operatorname{Tr}
    \left(
        \widehat\Pi_\perp^{\texttt{cov}}E_T
    \right)+
    \frac{\mathbb E[\phi_1^2]}{p-2}
    (\kappa^{(0)})^\top
    \widehat\Pi_\perp^{\texttt{cov}}
    \kappa^{(0)}.
\end{align*}
Accordingly, define
\begin{align}
\widehat R_{\gamma,n}^{\texttt{cov}}
&:=
\frac{1}{p-2}
\operatorname{Tr}
\left[
    \left(
        \widehat\Pi_\perp^{\texttt{cov}}
        -
        \Pi_{\perp,0}
        +
        A_{\Pi,n}^{\texttt{cov}}
    \right)E_T
\right]\notag\\
&\quad+
\frac{\mathbb E[\phi_1^2]}{p-2}
\left[
    (\kappa^{(0)})^\top
    \widehat\Pi_\perp^{\texttt{cov}}
    \kappa^{(0)}
    -
    \frac{
        (\Pi_{\perp,0}E_H\kappa^{(0)})^\top
        (\Pi_{\perp,0}E_H\kappa^{(0)})
    }{
        \{\mathbb E[|\phi_1|^2]\}^2
    }
\right].
\label{supp:eq: gamma remainder exact definition}
\end{align}
By the definitions in
\eqref{eq:cov-noise-first-order-terms} and
\eqref{eq:cov-noise-second-order-terms}, this gives identically
\begin{equation}
\label{supp:eq: exact cov gamma second order decomposition}
    \widehat\gamma^{\texttt{cov}}-\gamma
    =
    L_{\gamma,n}^{\texttt{cov}}
    +
    Q_{\gamma,n}^{\texttt{cov}}
    +
    \widehat R_{\gamma,n}^{\texttt{cov}}.
\end{equation}

On \(\mathcal A_n\),
Corollary~\ref{supp:cor:cov-real-residual-projection-expansion}
gives
\[
    \widehat\Pi_\perp^{\texttt{cov}}
    -
    \Pi_{\perp,0}
    +
    A_{\Pi,n}^{\texttt{cov}}
    =
    R_{\Pi,n}^{\texttt{cov}},
    \quad
    \|R_{\Pi,n}^{\texttt{cov}}\|_{\operatorname{op}}
    \le
    C\|E_H\|_{F}^2.
\]
Therefore, using the trace duality inequality and the fact that
\(\|E_T\|_*\le\sqrt p\,\|E_T\|_{F}\), we obtain the following
bound for the first term on \(\mathcal A_n\):
\begin{align}
\label{supp:eq: R gamma hat first term bound}
    &\left|
        \operatorname{Tr}
        \left[
            \left(
                \widehat\Pi_\perp^{\texttt{cov}}
                -
                \Pi_{\perp,0}
                +
                A_{\Pi,n}^{\texttt{cov}}
            \right)E_T
        \right]
    \right| \notag\\
    &\quad=
    \left|
        \operatorname{Tr}
        \left(
            R_{\Pi,n}^{\texttt{cov}}E_T
        \right)
    \right| \notag\\
    &\quad\le
    \|R_{\Pi,n}^{\texttt{cov}}\|_{\operatorname{op}}
    \|E_T\|_* \notag\\
    &\quad\le
    \sqrt p\,
    \|R_{\Pi,n}^{\texttt{cov}}\|_{\operatorname{op}}
    \|E_T\|_{F} \notag\\
    &\quad\le
    C\|E_H\|_{F}^2\|E_T\|_{F}.
\end{align}

We next consider the second term. By
\eqref{supp:eq: stochastic phihat kappa cov},
\[
    \widehat\Pi_\perp^{\texttt{cov}}\kappa^{(0)}
    =
    -
    \frac{1}{\mathbb E[|\phi_1|^2]}
    \Pi_{\perp,0}E_H\kappa^{(0)}
    +
    r_{\Pi,n},
\]
where, on \(\mathcal A_n\), by
\eqref{supp:eq: r stochastic bound},
\[
    \|r_{\Pi,n}\|
    \le
    C\|E_H\|_{F}^2.
\]
Since \(\widehat\Pi_\perp^{\texttt{cov}}\) is a real symmetric
idempotent matrix,
\begin{align*}
    (\kappa^{(0)})^\top
    \widehat\Pi_\perp^{\texttt{cov}}
    \kappa^{(0)}
    =
    \left(
        \widehat\Pi_\perp^{\texttt{cov}}\kappa^{(0)}
    \right)^\top
    \left(
        \widehat\Pi_\perp^{\texttt{cov}}\kappa^{(0)}
    \right).
\end{align*}
Substituting the preceding expansion gives
\begin{align*}
    &(\kappa^{(0)})^\top
    \widehat\Pi_\perp^{\texttt{cov}}
    \kappa^{(0)}\\
    &=
    \left[
        -
        \frac{1}{\mathbb E[|\phi_1|^2]}
        \Pi_{\perp,0}E_H\kappa^{(0)}
        +
        r_{\Pi,n}
    \right]^\top
    \left[
        -
        \frac{1}{\mathbb E[|\phi_1|^2]}
        \Pi_{\perp,0}E_H\kappa^{(0)}
        +
        r_{\Pi,n}
    \right]\\
    &=
    \frac{
        \left(
            \Pi_{\perp,0}E_H\kappa^{(0)}
        \right)^\top
        \left(
            \Pi_{\perp,0}E_H\kappa^{(0)}
        \right)
    }{
        \{\mathbb E[|\phi_1|^2]\}^2
    }
    -
    \frac{2}{\mathbb E[|\phi_1|^2]}
    \left(
        \Pi_{\perp,0}E_H\kappa^{(0)}
    \right)^\top
    r_{\Pi,n}
    +
    r_{\Pi,n}^\top r_{\Pi,n}.
\end{align*}
Consequently,
\begin{align}
\label{supp:eq: R gamma hat second term bound}
&\left|
    (\kappa^{(0)})^\top
    \widehat\Pi_\perp^{\texttt{cov}}
    \kappa^{(0)}
    -
    \frac{
        \left(
            \Pi_{\perp,0}E_H\kappa^{(0)}
        \right)^\top
        \left(
            \Pi_{\perp,0}E_H\kappa^{(0)}
        \right)
    }{
        \{\mathbb E[|\phi_1|^2]\}^2
    }
\right| \notag\\
&\quad\le
\frac{2}{\mathbb E[|\phi_1|^2]}
\left\|
    \Pi_{\perp,0}E_H\kappa^{(0)}
\right\|
\|r_{\Pi,n}\|
+
\|r_{\Pi,n}\|^2 \notag\\
&\quad\le
C\|E_H\|_{F}
\|E_H\|_{F}^2
+
C\|E_H\|_{F}^4 \notag\\
&\quad=
C\|E_H\|_{F}^3
+
C\|E_H\|_{F}^4 \notag\\
&\quad\le
C\|E_H\|_{F}^3
\end{align}
on \(\mathcal A_n\), where we used
\eqref{supp:eq: r stochastic bound},
\(\|\Pi_{\perp,0}\|_{\operatorname{op}}=\|\kappa^{(0)}\|=1\),
\(\|E_H\|_{\operatorname{op}}\le\|E_H\|_{F}\), and
\(\|E_H\|_{F}<\delta\le1\).

Finally, substituting
\eqref{supp:eq: R gamma hat first term bound} and
\eqref{supp:eq: R gamma hat second term bound} into
\eqref{supp:eq: gamma remainder exact definition} gives
\begin{equation}
\label{supp:eq: gamma remainder local bound}
    \left|
        \widehat R_{\gamma,n}^{\texttt{cov}}
    \right|
    \le
    C\left(
        \|E_H\|_{F}^2\|E_T\|_{F}
        +
        \|E_H\|_{F}^3
    \right)
\end{equation}
on \(\mathcal A_n\).

Since
\[
    \|E_H\|_{F}
    =
    \mathcal O_p(n^{-1/2}),
    \quad
    \|E_T\|_{F}
    =
    \mathcal O_p(n^{-1/2}),
\]
and \(\mathcal{P}(\mathcal A_n)\to1\),
\eqref{supp:eq: tau remainder local bound} and
\eqref{supp:eq: gamma remainder local bound} imply
\[
    \widehat R_{\tau,n}^{\texttt{cov}}
    =
    \mathcal O_p(n^{-3/2}),
    \quad
    \widehat R_{\gamma,n}^{\texttt{cov}}
    =
    \mathcal O_p(n^{-3/2}).
\]
Combining these results with
\eqref{supp:eq: exact cov tau second order decomposition} and
\eqref{supp:eq: exact cov gamma second order decomposition} proves
\eqref{eq:cov-tau-second-order-expansion} and
\eqref{eq:cov-gamma-second-order-expansion}.

It remains to verify the stochastic orders of the displayed terms.
Since \(\Pi_{\perp,0}\) is fixed,
\[
    L_{\tau,n}^{\texttt{cov}},
    L_{\gamma,n}^{\texttt{cov}}
    =
    \mathcal O_p(n^{-1/2}).
\]
Moreover, by the definition of \(A_{\Pi,n}^{\texttt{cov}}\),
\[
    \|A_{\Pi,n}^{\texttt{cov}}\|_{\operatorname{op}}
    \le
    C\|E_H\|_{F}
    =
    \mathcal O_p(n^{-1/2}).
\]
Therefore,
\[
    \operatorname{Tr}
    \left(
        A_{\Pi,n}^{\texttt{cov}}E_H
    \right),
    \quad
    \operatorname{Tr}
    \left(
        A_{\Pi,n}^{\texttt{cov}}E_T
    \right)
    =
    \mathcal O_p(n^{-1}),
\]
and
\[
    \|\Pi_{\perp,0}E_H\kappa^{(0)}\|^2
    =
    \mathcal O_p(n^{-1}),
\]
as well as
\[
    \left(
        \Pi_{\perp,0}E_H\kappa^{(0)}
    \right)^\top
    \left(
        \Pi_{\perp,0}E_H\kappa^{(0)}
    \right)
    =
    \mathcal O_p(n^{-1}).
\]
It follows from
\eqref{eq:cov-noise-second-order-terms} that
\[
    Q_{\tau,n}^{\texttt{cov}},
    Q_{\gamma,n}^{\texttt{cov}}
    =
    \mathcal O_p(n^{-1}),
\]
which completes the proof.
\end{proof}

\subsection{Definiteness and consistency of projected covariance estimators}
\label{supp:sec:projected-covariance-definiteness}

This subsection establishes the finite-sample positive definiteness of the moment map \(\mathcal M\), compares it with two alternative covariance estimators, and examines the effect of loading estimation error.
Throughout, we assume model~\eqref{model:centered-complex-factor} with \(p\geq3\), \(\Sigma^{(0)}\in\mathrm{SPD}(2)\), and \(0<\mathbb E[|\phi_1|^2]<\infty\). Recall that the empirical moments \(\widehat\Sigma^{\texttt{Her}}\) and \(\widehat\Sigma^{\texttt{Tra}}\) are defined in
\eqref{eq:sample-covariance-matrices}, and \(\mathcal M\) is defined in \eqref{eq:moment-covariance-map}.

For \(\kappa\in\mathcal S^{2p-1}\), define the alternative residual projector
\[
P_\perp(\kappa):=\mathrm{Id}_p-\kappa\kappa^*.
\]
This Hermitian orthogonal projector annihilates \(\kappa\) and has rank \(p-1\), whereas
\(\Pi_\perp(\kappa)\) has rank \(p-2\) when the real and imaginary parts of \(\kappa\) are linearly independent. Thus, \(P_\perp\) retains one additional complex residual direction. We nevertheless use \(\Pi_\perp\) to preserve finite-sample positive definiteness under the conditions below and limit the effect of loading estimation error to second order.

\subsubsection{Positive definiteness under the real residual projection}
The residual Gram representation guarantees that \(\mathcal M(\kappa)\) is positive semidefinite for every sample. Under model~\eqref{model:centered-complex-factor}, we further establish almost sure positive definiteness when \(n\geq2p\), simultaneously for all loadings whose real and imaginary parts are linearly independent.
\begin{theorem}\label{thm:moment-covariance-finite-sample-definiteness}
If \(n\geq2p\), then, with probability one, \(\mathcal M(\kappa)\in\mathrm{SPD}(2)\) simultaneously for every
\(\kappa\in\mathcal S^{2p-1}\) whose real and imaginary parts are linearly independent over \(\mathbb R\).
\end{theorem}

\begin{proof}
For any such \(\kappa\), write \(u_i(\kappa)=\Pi_\perp(\kappa)\operatorname{Re}(Y_i)\) and
\(v_i(\kappa)=\Pi_\perp(\kappa)\operatorname{Im}(Y_i)\).
Since \(\Pi_\perp(\kappa)\) is real, symmetric, and idempotent,
\[
\widehat\tau^2(\kappa)
=\frac{1}{n(p-2)}\sum_{i=1}^n
\bigl\{\|u_i(\kappa)\|^2+\|v_i(\kappa)\|^2\bigr\},
\]
and
\[
\widehat\gamma(\kappa)
=\frac{1}{n(p-2)}\sum_{i=1}^n
\bigl\{\|u_i(\kappa)\|^2-\|v_i(\kappa)\|^2
+2i\,u_i(\kappa)^\top v_i(\kappa)\bigr\}.
\]
Consequently,
\begin{equation}
\label{supp:eq:Pi-covariance-residual-gram}
\mathcal M(\kappa)
=\frac{1}{n(p-2)}\sum_{i=1}^n
\begin{pmatrix}
\|u_i(\kappa)\|^2 & u_i(\kappa)^\top v_i(\kappa)\\
u_i(\kappa)^\top v_i(\kappa) & \|v_i(\kappa)\|^2
\end{pmatrix}
\succeq0
\end{equation}
for every sample.

To establish strict positive definiteness uniformly over \(\kappa\), let
\[
W_n:=\frac1n\sum_{i=1}^n
\begin{pmatrix}\operatorname{Re}(Y_i)\\\operatorname{Im}(Y_i)\end{pmatrix}
\begin{pmatrix}\operatorname{Re}(Y_i)\\\operatorname{Im}(Y_i)\end{pmatrix}^{\top}.
\]
For \(c=(c_1,c_2)^\top\in\mathbb R^2\), set
\(A_c=(c_1\mathrm{Id}_p\ \ c_2\mathrm{Id}_p)\).
Using \(A_cA_c^\top=\|c\|^2\mathrm{Id}_p\) and
\(\operatorname{Tr}\{\Pi_\perp(\kappa)\}=p-2\), we obtain
\[
\begin{aligned}
c^\top\mathcal M(\kappa)c
&=
\frac{1}{p-2}\operatorname{Tr}
\{\Pi_\perp(\kappa)A_cW_nA_c^\top\}\\
&\geq
\frac{\operatorname{Tr}\{\Pi_\perp(\kappa)\}}{p-2}
\lambda_{\min}(A_cW_nA_c^\top)\\
&=
\lambda_{\min}(A_cW_nA_c^\top)\\
&\geq
\lambda_{\min}(W_n)\|c\|^2,
\end{aligned}
\]
where the first inequality uses
\(\operatorname{Tr}(AB)\geq\operatorname{Tr}(A)\lambda_{\min}(B)\)
for \(A\succeq0\) and symmetric \(B\), and the last follows from
\(A_cA_c^\top=\|c\|^2\mathrm{Id}_p\). Thus, for every sample,
\begin{equation}
\label{supp:eq:Pi-covariance-uniform-bound}
\mathcal M(\kappa)\succeq\lambda_{\min}(W_n)\mathrm{Id}_2
\end{equation}
simultaneously for all such \(\kappa\).

Conditional on \(\phi_{i}\)s, the stacked real observations have a joint Gaussian density, with positive-definite error covariance \(\Sigma^{(0)}\otimes\mathrm{Id}_p\) for each observation. Their unconditional joint distribution is therefore absolutely continuous. When \(n\geq2p\), the stacked real observations span \(\mathbb R^{2p}\) almost surely: the determinant formed from the first \(2p\) observations is a nonzero polynomial, whose zero set has Lebesgue measure zero. Hence \(W_n\) is positive definite almost surely,
and \eqref{supp:eq:Pi-covariance-uniform-bound} proves the assertion.
\end{proof}

The simultaneous conclusion includes loadings estimated from the same sample, provided that their real and imaginary parts are linearly independent. No independence between the loading estimator and the observations is required. The condition \(n\geq2p\) is sufficient; it is not asserted to be necessary.

\subsubsection{Consistency of the one-sided moment estimator}
We establish the strong consistency and eventual positive definiteness of the alternative covariance moment estimator based on \(P_\perp(\kappa)\). At the true loading, this projector satisfies the population moment identities
\[
\frac{
\operatorname{Tr}\{
P_\perp(\kappa^{(0)})\Sigma^{\texttt{Her}}
\}
}{p-1}
=\tau^2,
\quad
\frac{
\operatorname{Tr}\{
P_\perp(\kappa^{(0)})\Sigma^{\texttt{Tra}}
\}
}{p-1}
=\gamma.
\]
These identities motivate the following construction.
For \(\kappa\in\mathcal S^{2p-1}\), define
\begin{equation}
\label{supp:eq:P-one-sided-moments}
\begin{aligned}
\widehat\tau_P^2(\kappa)
:=\frac{1}{p-1}\operatorname{Tr}
\{P_\perp(\kappa)\widehat\Sigma^{\texttt{Her}}\},\quad
\widehat\gamma_P(\kappa)
:=\frac{1}{p-1}\operatorname{Tr}
\{P_\perp(\kappa)\widehat\Sigma^{\texttt{Tra}}\},
\end{aligned}
\end{equation}
and reconstruct the covariance by
\begin{equation}
\label{supp:eq:P-one-sided-covariance}
\widehat\Sigma_P(\kappa)
:=\frac12
\begin{pmatrix}
\widehat\tau_P^2(\kappa)+\operatorname{Re}\{\widehat\gamma_P(\kappa)\}
& \operatorname{Im}\{\widehat\gamma_P(\kappa)\}\\
\operatorname{Im}\{\widehat\gamma_P(\kappa)\}
& \widehat\tau_P^2(\kappa)-\operatorname{Re}\{\widehat\gamma_P(\kappa)\}
\end{pmatrix}.
\end{equation}

\begin{proposition}\label{supp:prop:P-moment-definiteness}
Let \(\widehat\kappa_n\in\mathcal S^{2p-1}\) satisfy
\([\widehat\kappa_n]\to[\kappa^{(0)}]\) almost surely.
Then
\[
\widehat\Sigma_P(\widehat\kappa_n)
\longrightarrow\Sigma^{(0)}
\quad\text{almost surely},
\]
and this estimator is almost surely positive definite for all sufficiently large \(n\).
\end{proposition}

\begin{proof}
Projective consistency implies \(P_\perp(\widehat\kappa_n)\to P_\perp(\kappa^{(0)})\) almost surely. Together with the strong law for the empirical moments, this yields
\[
\begin{aligned}
\widehat\tau_P^2(\widehat\kappa_n)
\longrightarrow
\frac{\operatorname{Tr}
\{P_\perp(\kappa^{(0)})\Sigma^{\texttt{Her}}\}}{p-1}
=\tau^2, \quad
\widehat\gamma_P(\widehat\kappa_n)\longrightarrow
\frac{\operatorname{Tr}
\{P_\perp(\kappa^{(0)})\Sigma^{\texttt{Tra}}\}}{p-1}
=\gamma
\end{aligned}
\]
almost surely, where the equalities follow from \(P_\perp(\kappa^{(0)})\kappa^{(0)}=0\) and \(\operatorname{Tr}\{P_\perp(\kappa^{(0)})\}=p-1\).
The covariance reconstruction in \eqref{supp:eq:P-one-sided-covariance} therefore gives \(\widehat\Sigma_P(\widehat\kappa_n)\to\Sigma^{(0)}\) almost surely. By continuity of the smallest eigenvalue and \(\lambda_{\min}(\Sigma^{(0)})>0\), the estimator is almost surely positive definite for all sufficiently large \(n\). 
\end{proof}
This result does not guarantee finite-sample positive semidefiniteness of \(\widehat\Sigma_P(\kappa)\). Unlike \(\mathcal M\), the one-sided moment construction need not preserve the residual Gram structure. Indeed, the average sum of squared residuals, without complex conjugation, is
\[
\frac1n\sum_{i=1}^n
\{P_\perp(\kappa)Y_i\}^{\top}
\{P_\perp(\kappa)Y_i\}
=
\operatorname{Tr}\left\{
P_\perp(\kappa)^\top P_\perp(\kappa)
\widehat\Sigma^{\texttt{Tra}}
\right\}.
\]
Since \(P_\perp(\kappa)^\top P_\perp(\kappa)\neq P_\perp(\kappa)\)
in general, this need not equal \(\operatorname{Tr}\{P_\perp(\kappa)\widehat\Sigma^{\texttt{Tra}}\}\), which is used to estimate \(\gamma\) in \eqref{supp:eq:P-one-sided-moments}. Thus, the one-sided moment estimator need not coincide with the residual Gram estimator introduced below and can be indefinite in finite samples.

\subsubsection{Definiteness and inconsistency of the residual Gram estimator}

Alternatively, we can construct a covariance estimator using \(P_\perp(\kappa)\) that preserves the residual Gram structure. We show that this construction guarantees positive semidefiniteness in every sample but generally yields an inconsistent estimator of
\(\Sigma^{(0)}\).
Specifically, set
\(u_i^P(\kappa)=\operatorname{Re}\{P_\perp(\kappa)Y_i\}\) and
\(v_i^P(\kappa)=\operatorname{Im}\{P_\perp(\kappa)Y_i\}\), and define
\begin{equation}
\label{supp:eq:P-residual-gram-estimator}
\widehat\Sigma_{P,\mathrm{Gram}}(\kappa)
:=\frac{1}{n(p-1)}\sum_{i=1}^n
\begin{pmatrix}
\|u_i^P(\kappa)\|^2 & u_i^P(\kappa)^\top v_i^P(\kappa)\\
u_i^P(\kappa)^\top v_i^P(\kappa) & \|v_i^P(\kappa)\|^2
\end{pmatrix}.
\end{equation}

\begin{theorem}[Definiteness and limit of the \(P_\perp\) Gram estimator]
\label{supp:thm:P-Gram-limit}
For every sample and every \(\kappa\in\mathcal S^{2p-1}\),
\(\widehat\Sigma_{P,\mathrm{Gram}}(\kappa)\) is positive semidefinite.
If \(n\geq2p\), it is almost surely positive definite simultaneously
for all such \(\kappa\).

If \([\widehat\kappa_n]\to[\kappa^{(0)}]\) almost surely, then
\begin{equation}
\label{supp:eq:P-Gram-population-limit}
\begin{aligned}
\widehat\Sigma_{P,\mathrm{Gram}}(\widehat\kappa_n)
\xrightarrow{\mathrm{a.s.}}\;
\frac{\tau^2}{2}\mathrm{Id}_2+\frac{p-2+| (\kappa^{(0)})^\top\kappa^{(0)} |^2}{2(p-1)}
\begin{pmatrix}
\operatorname{Re}(\gamma)&\operatorname{Im}(\gamma)\\
\operatorname{Im}(\gamma)&-\operatorname{Re}(\gamma)
\end{pmatrix}.
\end{aligned}
\end{equation}
In particular, if the real and imaginary parts of \(\kappa^{(0)}\) are linearly independent, this estimator is consistent for \(\Sigma^{(0)}\) if and only if \(\gamma=0\).
\end{theorem}

\begin{proof}
Positive semidefiniteness follows from the Gram representation. For strict positive definiteness, consider the probability-one event on which the stacked real observations span \(\mathbb R^{2p}\). If the Gram matrix were singular for some \(\kappa\), there would
exist a nonzero \(c=(c_1,c_2)^\top\in\mathbb R^2\) such that
\[
c_1\operatorname{Re}\{P_\perp(\kappa)Y_i\}
+c_2\operatorname{Im}\{P_\perp(\kappa)Y_i\}=0
\quad\text{for every }i.
\]
The corresponding real-linear map would then vanish on all of \(\mathbb C^p\). Applying this identity to both \(y\) and \(iy\) would force \(P_\perp(\kappa)y=0\) for every \(y\), contradicting \(\operatorname{rank}\{P_\perp(\kappa)\}=p-1\). This argument applies simultaneously to all \(\kappa\).

The complex variance corresponding to \eqref{supp:eq:P-residual-gram-estimator} equals \(\widehat\tau_P^2(\kappa)\), since \(P_\perp(\kappa)^*P_\perp(\kappa)=P_\perp(\kappa)\). Its corresponding pseudo-variance is
\[
\widehat\gamma_{P,\mathrm{Gram}}(\kappa)
:=\frac{1}{p-1}\operatorname{Tr}
\{P_\perp(\kappa)^\top P_\perp(\kappa)
\widehat\Sigma^{\texttt{Tra}}\}.
\]
At the true loading, \(P_\perp(\kappa^{(0)})\kappa^{(0)}=0\)
eliminates the signal contribution, while
\[
\operatorname{Tr}\{P_\perp(\kappa)^\top P_\perp(\kappa)\}
=p-2+|\kappa^\top\kappa|^2.
\]
The strong law and continuity of \(P_\perp\) on projective space therefore give
\[
\widehat\tau_P^2(\widehat\kappa_n)\to\tau^2,
\quad
\widehat\gamma_{P,\mathrm{Gram}}(\widehat\kappa_n)
\to\gamma\,
\frac{p-2+| (\kappa^{(0)})^\top\kappa^{(0)} |^2}{p-1}
\quad\text{a.s}.
\]
Reconstructing the covariance proves
\eqref{supp:eq:P-Gram-population-limit}.
For a unit vector, equality in \(|\kappa^\top\kappa|\leq1\) holds if and only if its real and imaginary parts are linearly dependent. Under linear independence, the factor multiplying \(\gamma\) is strictly smaller than one, so the limiting covariance equals \(\Sigma^{(0)}\) if and only if \(\gamma=0\).
\end{proof}

\begin{remark}[Exact agreement of the \(\Pi_\perp\) Gram and moment estimators]
\label{supp:rem:Pi-Gram-equivalence}
Replacing \(P_\perp\) by \(\Pi_\perp\) and \(p-1\) by \(p-2\) in \eqref{supp:eq:P-residual-gram-estimator} gives exactly \(\mathcal M(\kappa)\) for every sample and every \(\kappa\) with linearly independent real and imaginary parts. Indeed, \(\Pi_\perp\) is real, so the projected real and imaginary parts coincide with \(u_i(\kappa)\) and \(v_i(\kappa)\) in \eqref{supp:eq:Pi-covariance-residual-gram}. Thus, the \(\Pi_\perp\) construction preserves the Gram structure and is strongly consistent whenever the loading estimator is strongly consistent and the real and imaginary parts of \(\kappa^{(0)}\)
are linearly independent.
\end{remark}

For the two unmodified \(P_\perp\) constructions considered here, the one-sided moments yield consistency without guaranteed finite-sample positive semidefiniteness, whereas the residual Gram construction preserves positive semidefiniteness but is generally
inconsistent.

\subsubsection{First-order expansion of the one-sided moment estimator}

We examine how loading estimation affects the first-order expansion of the covariance estimator in \eqref{supp:eq:P-one-sided-covariance}. Recall the notation
\[
    P_{\perp,0}=P_\perp(\kappa^{(0)}),
    \quad
    E_H
    =
    \widehat\Sigma^{\texttt{Her}}
    -
    \Sigma^{\texttt{Her}},
    \quad
    E_T
    =
    \widehat\Sigma^{\texttt{Tra}}
    -
    \Sigma^{\texttt{Tra}}.
\]

\begin{theorem}[First-order effect of loading estimation]
\label{supp:thm:P-moment-first-order-expansion}
Suppose \(\mathbb E[|\phi_1|^4]<\infty\).
Let \(\widehat\kappa_n\in\mathcal S^{2p-1}\) be a loading estimator whose representative  \(\widehat\kappa_n^{\texttt{opt}}\), optimally positioned with \(\kappa^{(0)}\), admits the expansion
\[
\widehat\kappa_n^{\texttt{opt}}
=
\kappa^{(0)}+L_{\kappa,n}+\mathcal O_p(n^{-1}),
\quad
L_{\kappa,n}=\mathcal O_p(n^{-1/2}).
\]
Then
\begin{align}
\widehat\tau_P^2(\widehat\kappa_n)-\tau^2
&=
\frac{1}{p-1}\operatorname{Tr}(P_{\perp,0}E_H)
+\mathcal O_p(n^{-1}),
\label{supp:eq:P-tau-first-order-expansion}
\\
\widehat\gamma_P(\widehat\kappa_n)-\gamma
&=
\frac{1}{p-1}\operatorname{Tr}(P_{\perp,0}E_T)
-\frac{\mathbb E[\phi_1^2]}{p-1}
(\kappa^{(0)})^\top L_{\kappa,n}
+\mathcal O_p(n^{-1}).
\label{supp:eq:P-gamma-first-order-expansion}
\end{align}
\end{theorem}

\begin{proof}
By phase invariance, \(P_\perp(\widehat\kappa_n)
=P_\perp(\widehat\kappa_n^{\texttt{opt}})\). Optimal positioning ensures that \((\kappa^{(0)})^*\widehat\kappa_n^{\texttt{opt}}\)
is real and nonnegative. Since both vectors have unit norm,
\[
\|\widehat\kappa_n^{\texttt{opt}}-\kappa^{(0)}\|^2
=
2-2(\kappa^{(0)})^*\widehat\kappa_n^{\texttt{opt}}.
\]
The assumed loading expansion therefore gives
\[
\begin{aligned}
(\kappa^{(0)})^*
(\widehat\kappa_n^{\texttt{opt}}-\kappa^{(0)})
&=
-\tfrac12
\|\widehat\kappa_n^{\texttt{opt}}-\kappa^{(0)}\|^2
=
\mathcal O_p(n^{-1}),
\\
(\kappa^{(0)})^*L_{\kappa,n}
&=
\mathcal O_p(n^{-1}).
\end{aligned}
\]
Taking complex conjugates also yields \(L_{\kappa,n}^*\kappa^{(0)}=\mathcal O_p(n^{-1})\).

For brevity, write \(\Delta P_n:=P_\perp(\widehat\kappa_n)-P_{\perp,0}\). Substituting the loading expansion into
\(P_\perp(\widehat\kappa_n)
=\mathrm{Id}_p-
\widehat\kappa_n^{\texttt{opt}}
(\widehat\kappa_n^{\texttt{opt}})^*\)
gives
\begin{equation}
\label{supp:eq:P-projector-perturbation}
\Delta P_n
=
-L_{\kappa,n}(\kappa^{(0)})^*
-\kappa^{(0)}L_{\kappa,n}^*
+\mathcal O_p(n^{-1}).
\end{equation}
Consequently, \(\Delta P_n=\mathcal O_p(n^{-1/2})\). Moreover, \(\operatorname{Tr}(\Delta P_n)=0\).

By Lemma~\ref{supp:lem:cov-moment-bounds} with \(k=2\),
\[
\|E_H\|_{\mathrm F},\ \|E_T\|_{\mathrm F}
=
\mathcal O_p(n^{-1/2}).
\]
Consequently, for \(E=E_H,E_T\),
\[
\bigl|\operatorname{Tr}(\Delta P_n E)\bigr|
\leq
\|\Delta P_n\|_{\mathrm F}\|E\|_{\mathrm F}
=
\mathcal O_p(n^{-1}).
\]

Using the population identities
\(\operatorname{Tr}(P_{\perp,0}\Sigma^{\texttt{Her}})
=(p-1)\tau^2\) and
\(\operatorname{Tr}(P_{\perp,0}\Sigma^{\texttt{Tra}})
=(p-1)\gamma\), we obtain the exact decompositions
\begin{align*}
(p-1)\{\widehat\tau_P^2(\widehat\kappa_n)-\tau^2\}
&=
\operatorname{Tr}(P_{\perp,0}E_H)
+\operatorname{Tr}(\Delta P_n\Sigma^{\texttt{Her}})
+\operatorname{Tr}(\Delta P_n E_H),
\\
(p-1)\{\widehat\gamma_P(\widehat\kappa_n)-\gamma\}
&=
\operatorname{Tr}(P_{\perp,0}E_T)
+\operatorname{Tr}(\Delta P_n\Sigma^{\texttt{Tra}})
+\operatorname{Tr}(\Delta P_n E_T).
\end{align*}
It remains to evaluate the terms involving the population moments.

For the Hermitian moment, \(\Sigma^{\texttt{Her}}
=\mathbb E[|\phi_1|^2]\kappa^{(0)}(\kappa^{(0)})^*
+\tau^2\mathrm{Id}_p\), so
\[
\begin{aligned}
\operatorname{Tr}(\Delta P_n\Sigma^{\texttt{Her}})
&=
\mathbb E[|\phi_1|^2]
(\kappa^{(0)})^*\Delta P_n\kappa^{(0)}
+\tau^2\operatorname{Tr}(\Delta P_n)
\\
&=
\mathbb E[|\phi_1|^2]
(\kappa^{(0)})^*\Delta P_n\kappa^{(0)}.
\end{aligned}
\]
By \eqref{supp:eq:P-projector-perturbation} and
\((\kappa^{(0)})^*\kappa^{(0)}=1\),
\[
\begin{aligned}
(\kappa^{(0)})^*\Delta P_n\kappa^{(0)}
&=
-(\kappa^{(0)})^*L_{\kappa,n}
-L_{\kappa,n}^*\kappa^{(0)}
+\mathcal O_p(n^{-1})
=
\mathcal O_p(n^{-1}).
\end{aligned}
\]
Hence \(\operatorname{Tr}(\Delta P_n\Sigma^{\texttt{Her}})
=\mathcal O_p(n^{-1})\).

For the transpose moment, \(\Sigma^{\texttt{Tra}}
=\mathbb E[\phi_1^2]\kappa^{(0)}(\kappa^{(0)})^\top
+\gamma\mathrm{Id}_p\), and the same trace-zero identity yields
\[
    \operatorname{Tr}
    \left(
        \Delta P_n\Sigma^{\texttt{Tra}}
    \right)
    =
    -\mathbb E[\phi_1^2]
    (\kappa^{(0)})^\top L_{\kappa,n}
    +
    \mathcal O_p(n^{-1}).
\]
This time, \eqref{supp:eq:P-projector-perturbation} gives
\[
\begin{aligned}
(\kappa^{(0)})^\top\Delta P_n\kappa^{(0)}
&=
-(\kappa^{(0)})^\top L_{\kappa,n}
-\{(\kappa^{(0)})^\top\kappa^{(0)}\}
L_{\kappa,n}^*\kappa^{(0)}
+\mathcal O_p(n^{-1})
\\
&=
-(\kappa^{(0)})^\top L_{\kappa,n}
+\mathcal O_p(n^{-1})
\end{aligned}
\]
since \(L_{\kappa,n}^*\kappa^{(0)}=\mathcal O_p(n^{-1})\).
In contrast, optimal positioning does not generally make
\((\kappa^{(0)})^\top L_{\kappa,n}\) of order \(n^{-1}\).
Therefore,
\[
\operatorname{Tr}(\Delta P_n\Sigma^{\texttt{Tra}})
=
-\mathbb E[\phi_1^2](\kappa^{(0)})^\top L_{\kappa,n}
+\mathcal O_p(n^{-1}).
\]
Substituting these two population contributions and the bounds on
\(\operatorname{Tr}(\Delta P_n E_H)\) and
\(\operatorname{Tr}(\Delta P_n E_T)\) into the exact decompositions
above, and dividing by \(p-1\), proves both expansions.
\end{proof}

\begin{remark}[Effect of loading estimation]
\label{supp:rem:P-first-order-loading-effect}
The loading term \(L_{\kappa,n}\) generally enters the first-order
covariance expansion through \(\widehat\gamma_P\), whereas it does
not contribute to the first-order expansion of \(\widehat\tau_P^2\).
This additional contribution vanishes when
\(\mathbb E[\phi_1^2]=0\).
For the \texttt{cov} loading estimator
\(\widehat\kappa^{\texttt{cov,opt}}\), the theorem applies with
\(L_{\kappa,n}=L_{\kappa,n}^{\texttt{cov}}\).

In contrast, for the estimators based on \(\Pi_\perp\),
Theorems~\ref{thm:cov-noise-covariance-expansion}
and~\ref{thm:iterated-plugin-noise-covariance-expansion}
show that neither the first-order expansion of \(\widehat\tau^2\)
nor that of \(\widehat\gamma\) contains the first-order loading
error. Loading estimation enters these expansions only at
second order, without requiring \(\mathbb E[\phi_1^2]=0\).
\end{remark}

The numerical comparisons in Section~\ref{sec:triangle-projector-comparison} use each covariance estimate in the first plug-in loading update, with its inverse as the weight matrix; non-positive-definite \(P_\perp\)-based estimates are corrected before inversion. The \(P_\perp\)-based procedure gave no consistent improvement in loading MSE and sometimes increased it. Loss of positive definiteness was also more common under strong
noise anisotropy and small sample size. These findings support retaining \(\Pi_\perp\) for covariance estimation and the subsequent plug-in loading update.

\section{Generalized \fre~ means with random cost functions}\label{supp:sec:gfm}
This section gives the conditions and proofs for the results in Appendix~\ref{sec: GFM} on generalized \fre~means whose cost functions
depend on an estimated auxiliary parameter. We first prove Ziezold and Bhattacharya--Patrangenaru consistency of the empirical mean sets. We then construct measurable estimators that belong to these sets eventually almost surely. Finally, we derive a stochastic expansion
that accounts for estimation of the auxiliary parameter and yields the central limit theorem. Under stronger conditions, the expansion
has an \(O_p(n^{-1})\) remainder.

\subsection{Strong consistency}
\label{supp:sec:gfm-consistency}

We prove the BP consistency result stated in
Section~\ref{sec: GFM consistency} for generalized \fre~means with an estimated auxiliary parameter. The argument uses measurable local cost bounds to first establish the following weaker limit-point consistency property.

\begin{definition}[Ziezold consistency]
\label{supp:def:gfm-ziezold-consistency}
The empirical minimizer sets \(\widehat E_n\) are said to be
\emph{Ziezold consistent} with \(E_0\) if
\begin{equation}
\label{supp:eq:gfm-ziezold-consistency}
\mathcal{P}
\left[
    \left\{
        \omega\in\Omega:
        \bigcap_{n=1}^{\infty}
        \overline{
            \bigcup_{k=n}^{\infty}
            \widehat E_k(\omega)
        }
        \subset E_0
    \right\}
\right]
=
1.
\end{equation}
\end{definition}

Ziezold consistency ensures that every accumulation point of a
sequence of empirical minimizers belongs to the population minimizer
set \(E_0\) \citep{ziezold1977expected}. It is weaker than BP
consistency because it does not exclude sequences of empirical
minimizers that diverge and therefore have no accumulation point.

The proof of BP consistency consequently consists of two stages. First, the local regularity and integrability conditions yield Ziezold consistency. Second, the Heine--Borel property and the almost-sure eventual boundedness of \(\widehat E_n\) exclude diverging sequences and upgrade Ziezold consistency to BP consistency
\citep{Bhattacharya2003,huckemann2011intrinsic,schotz2022strong, evans2024limit,park2026generalized}.

\subsubsection{Consistency conditions and their roles}
\label{supp:sec:gfm-consistency-conditions}

We state the seven conditions used to prove
Theorem~\ref{thm: BP-SC} and explain their roles.
These conditions concern the auxiliary estimator, the parameter spaces, the local behavior of the cost, and the empirical minimizer sets. The principal difference from the fixed-cost setting is that the empirical cost depends on the random auxiliary estimator \(\widehat q_n\). Consequently, local control is required jointly over the primary and auxiliary parameter spaces \(M\times Q\), rather than over \(M\) alone.

We first state the standing consistency assumption on the auxiliary estimator.

\begin{condition}[Strong consistency of the auxiliary estimator]
\label{supp:cond:gfm-auxiliary-consistency}
There exists \(q_0\in Q\) such that
\begin{equation}
\label{supp:eq:gfm-auxiliary-consistency}
    d_Q(\widehat q_n,q_0)
    \longrightarrow 0
    \quad\text{almost surely}.
\end{equation}
\end{condition}

Condition~\ref{supp:cond:gfm-auxiliary-consistency} ensures that,
almost surely, \(\widehat q_n\) eventually belongs to every
neighborhood of \(q_0\). It therefore permits the empirical cost
evaluated at \(\widehat q_n\) to be compared with the population cost
evaluated at \(q_0\). No independence between \(\widehat q_n\) and
\(X_1,\ldots,X_n\) is required.

We next impose conditions on the primary and auxiliary parameter
spaces.

\begin{condition}[Primary parameter space]
\label{supp:cond:gfm-primary-polish}
The metric space \((M,d_M)\) is separable and complete. Equivalently,
\(M\) is a Polish space.
\end{condition}

The separability in
Condition~\ref{supp:cond:gfm-primary-polish} allows us to work with a
countable dense subset of \(M\). This is needed to construct a single
probability-one event on which the required laws of large numbers hold
simultaneously over a countable family of local neighborhoods.
Together with completeness, it also provides the Polish structure
used in the measurability arguments of
Subsubsection~\ref{supp:sec:gfm-local-bound-measurability}.

\begin{condition}[Heine--Borel property]
\label{supp:cond:gfm-heine-borel}
Every closed and bounded subset of \((M,d_M)\) is compact. In other
words, \(M\) satisfies the Heine--Borel property.
\end{condition}

Condition~\ref{supp:cond:gfm-heine-borel} is not required to establish
Ziezold consistency. Its role is to upgrade Ziezold consistency to BP
consistency. In combination with the eventual boundedness condition
below, it ensures that any sequence of empirical minimizers has a
convergent subsequence and therefore cannot escape to infinity.

\begin{condition}[Local Polish structure of the auxiliary space]
\label{supp:cond:gfm-auxiliary-local-polish}
There exists \(r_0>0\) such that
\(
    \overline{B_Q(q_0,r_0)}
\)
is a separable and complete subspace of \((Q,d_Q)\).
\end{condition}

Only a neighborhood of \(q_0\), rather than the entire auxiliary
parameter space \(Q\), is required to be Polish. Indeed,
Condition~\ref{supp:cond:gfm-auxiliary-consistency} implies that
\(\widehat q_n\) eventually lies in this neighborhood almost surely.
The separability of this neighborhood is used together with that of
\(M\) to obtain countable local constructions, while its completeness
is used in the measurability argument for the local cost bounds.

We now state the regularity condition on the cost function.

\begin{condition}[Measurability and continuity of the cost]
\label{supp:cond:gfm-cost-regularity}
The cost function
\[
    \rho:T\times M\times Q\longrightarrow\mathbb R
\]
is measurable with respect to
\[
    \mathcal B(T)\otimes\mathcal B(M)\otimes\mathcal B(Q).
\]
Moreover, for every \(t\in T\), the map
\[
    (m,q)\longmapsto\rho(t,m,q)
\]
is continuous on \(M\times Q\).
\end{condition}

The joint measurability in
Condition~\ref{supp:cond:gfm-cost-regularity} guarantees that the cost
evaluated at the observations is a random variable. Its continuity in
\((m,q)\) controls simultaneous perturbations of the primary and
auxiliary parameters. It is also essential for the measurability of
the local extrema and for showing that shrinking local lower and upper
bounds converge pointwise to \(\rho(t,m,q_0)\).

To formulate the required local integrability condition, define the
following lower and upper envelopes.

\begin{definition}[Local cost bounds]
\label{supp:def:gfm-local-cost-bounds}
For \(m\in M\) and \(r>0\), define
\begin{align}
    \pi_{m,r}(t)
    &:=
    \inf_{\substack{
        m'\in B_M(m,r)\\
        q\in B_Q(q_0,r)
    }}
    \rho(t,m',q),
    \label{supp:eq:gfm-local-cost-infimum}
    \\
    \Pi_{m,r}(t)
    &:=
    \sup_{\substack{
        m'\in B_M(m,r)\\
        q\in B_Q(q_0,r)
    }}
    \rho(t,m',q).
    \label{supp:eq:gfm-local-cost-supremum}
\end{align}
\end{definition}

These envelopes control the empirical cost uniformly when both the
primary argument and the estimated auxiliary parameter vary in
neighborhoods of \(m\) and \(q_0\). Their measurability is not automatic
because the infimum and supremum are taken over uncountable sets. This
issue is addressed in
Subsubsection~\ref{supp:sec:gfm-local-bound-measurability}.

\begin{condition}[Local integrability]
\label{supp:cond:gfm-local-integrability}
For every \(m\in M\), there exists \(r_m>0\) such that, for every
\(0<r\le r_m\),
\[
    \pi_{m,r}(X)
    \quad\text{and}\quad
    \Pi_{m,r}(X)
\]
are integrable.
\end{condition}

Pointwise integrability of \(\rho(X,m,q)\) alone does not provide
sufficient control when \(m\) and \(q\) vary with \(n\).
Condition~\ref{supp:cond:gfm-local-integrability} supplies integrable
local lower and upper envelopes. These allow the strong law of large
numbers to be applied to the local bounds and provide the domination
needed when their neighborhoods shrink to \((m,q_0)\).

Finally, the BP consistency result requires the following additional
boundedness property of the empirical minimizer sets.

\begin{condition}[Almost-sure eventual nonemptiness and boundedness]
\label{supp:cond:gfm-eventual-boundedness}
With probability one, there exists \(N\ge1\) such that
\[
    \widehat E_n\neq\emptyset
    \quad\text{for every }n\ge N,
\]
and
\[
    \bigcup_{n\ge N}\widehat E_n
\]
is a bounded subset of \((M,d_M)\).
\end{condition}

Condition~\ref{supp:cond:gfm-eventual-boundedness} prevents empirical
minimizers from having diverging subsequences. Together with
Condition~\ref{supp:cond:gfm-heine-borel}, it makes the closure of the
eventual union of the empirical minimizer sets compact, and sequentially compact.

In summary,
Conditions~\ref{supp:cond:gfm-auxiliary-consistency},
\ref{supp:cond:gfm-primary-polish},
\ref{supp:cond:gfm-auxiliary-local-polish},
\ref{supp:cond:gfm-cost-regularity}, and
\ref{supp:cond:gfm-local-integrability}
yield the continuous-convergence argument needed for Ziezold
consistency. Conditions~\ref{supp:cond:gfm-heine-borel} and
\ref{supp:cond:gfm-eventual-boundedness} then exclude diverging
sequences of empirical minimizers and upgrade Ziezold consistency to
BP consistency.

\subsubsection{Measurability of the local cost bounds}
\label{supp:sec:gfm-local-bound-measurability}

The local bounds \(\pi_{m,r}(X)\) and \(\Pi_{m,r}(X)\) involve infima and suprema over uncountable parameter sets, so their measurability requires justification. We establish this property using the theory of random lower-semicontinuous functions in \cite{korf2001random}.

\begin{definition}[Random lsc function, \cite{korf2001random}]\label{supp:def:random-lsc}
Let $(S, d_{S})$ be a Polish space (i.e., a complete, separable metric space) with $\mathcal{B}$ the Borel sigma field on $S$, and $(\Omega, \mathcal{A}, \mathcal{P})$ a probability space where $\mathcal{A}$ is $\mathcal{P}$-complete. A random lsc (lower semicontinuous) function is an extended real-valued function $f: \Omega \times S \to \overline{\mathbb{R}}$ such that:
\begin{enumerate}
  \item [(i)] the function $(\omega, s) \mapsto f(\omega, s)$ is $\mathcal{A} \otimes \mathcal{B}$-measurable; and
  \item [(ii)] for every $\omega \in \Omega$, the function $s \mapsto f(\omega, s)$ is lower semicontinuous.
\end{enumerate}
\end{definition}

Throughout this subsection, \((\Omega,\mathcal A,\mathcal{P})\) denotes the underlying complete probability space.

\begin{theorem}[Scalarization of random lsc functions
{\cite[Theorem~3.4]{korf2001random}}]
\label{supp:thm:korf-wets}
Let
\[
    f:\Omega\longrightarrow\operatorname{lsc\mbox{-}fcns}(S),
\]
and write \(f(\omega,s)\) for the value of the lsc function
\(f(\omega)\) at \(s\in S\). For \(D\subset S\), define
\[
    \pi_D(\omega):=\inf_{s\in D}f(\omega,s)
    \in\overline{\mathbb R}.
\]
Then \(f\) is a random lsc function if and only if \(\pi_D\) is
\(\mathcal A\)-measurable for every \(D\) in any one of the following
collections:
\begin{enumerate}
    \item the collection of all open subsets of \(S\);
    \item the collection of all open balls
    \[
        B_S(s,r):=\{s'\in S:d_S(s',s)<r\};
    \]
    \item the collection of all open rational balls
    \(B_S(s,r)\), where \(s\) ranges over a fixed countable dense
    subset \(\mathcal R\subset S\) and \(r\in\mathbb Q_+\).
\end{enumerate}
\end{theorem}

Using this theoretical foundation, we can show that our local infimum and local supremum cost functions $\pi_{m,r}$ and $\Pi_{m,r}$ are measurable.

\begin{lemma}[Measurability of local bounds]
\label{supp:lem:gfm-local-bound-measurability}
Under Conditions~\ref{supp:cond:gfm-primary-polish}, \ref{supp:cond:gfm-auxiliary-local-polish}, and
\ref{supp:cond:gfm-cost-regularity}, for every \(m\in M\) and \(0<r<r_0\), the
extended-real-valued random variables
\[
    \pi_{m,r}(X)
    :=
    \inf_{\substack{m'\in B_M(m,r)\\q\in B_Q(q_0,r)}}
    \rho(X,m',q)
~\text{ and }~
    \Pi_{m,r}(X)
    :=
    \sup_{\substack{m'\in B_M(m,r)\\q\in B_Q(q_0,r)}}
    \rho(X,m',q)
\]
are \(\mathcal A\)-measurable. If, in addition,
Condition~\ref{supp:cond:gfm-local-integrability} holds for the given \(m\) and
\(r\), then both random variables are finite almost surely.
\end{lemma}

\begin{proof}[Proof of Lemma~\ref{supp:lem:gfm-local-bound-measurability}]
Let \((\Omega,\mathcal A,\mathcal{P})\) be the underlying complete
probability space, and let \(X:\Omega\to T\) be the observed random
element. By Condition~\ref{supp:cond:gfm-auxiliary-local-polish}, $V_Q := \overline{B_{Q}(q_0, r_0)}$ is Polish. By Condition~\ref{supp:cond:gfm-primary-polish}, the primary parameter space $M$ is a Polish space. 

To apply Theorem~\ref{supp:thm:korf-wets}, we restrict the parameter domain to the product space
\(
    S := M \times V_Q .
\)
Here, \(V_Q\) is equipped with the relative topology inherited from \(Q\), and
\(S\) is equipped with the product topology induced by the topologies of
\(M\) and \(V_Q\). Since both \(M\) and \(V_Q\) are Polish spaces, their product
\(S\) is also Polish. We denote by \(\mathcal{B} = \mathcal{B}(S)\) the Borel
\(\sigma\)-algebra generated by this product topology.

For \(s=(m',q)\in S\), define
\[
    f(\omega,s):=\rho(X(\omega),m',q).
\]
By Condition~\ref{supp:cond:gfm-cost-regularity}, \(f\) is
\(\mathcal A\otimes\mathcal B(S)\)-measurable, and
\(s\mapsto f(\omega,s)\) is continuous, hence lower
semicontinuous, for every \(\omega\). Thus \(f\) is a random lsc
function on \(S\).

To utilize Theorem~\ref{supp:thm:korf-wets}, let
\[
    D:=B_M(m,r)\times B_Q(q_0,r).
\]
Because \(r<r_0\), \(D\) is an open subset of
\(S=M\times V_Q\). Theorem~\ref{supp:thm:korf-wets} therefore implies
that
\[
    \omega\longmapsto
    \inf_{(m',q)\in D}\rho(X(\omega),m',q)
\]
is \(\mathcal A\)-measurable.

For the supremum, define
\[
    g(\omega,(m',q))
    :=
    -\rho(X(\omega),m',q).
\]
Since \(g(\omega,\cdot)\) is continuous, \(g\) is also a random lsc
function. Hence
\[
    \omega\longmapsto
    \inf_{(m',q)\in D}g(\omega,(m',q))
    =
    -\sup_{(m',q)\in D}\rho(X(\omega),m',q)
\]
is \(\mathcal A\)-measurable. Its negative is therefore
\(\mathcal A\)-measurable, which shows the assertion for
\(\Pi_{m,r}(X)\).
\end{proof}

\subsubsection{Proofs of Ziezold and BP consistency}
\label{supp:sec:gfm-consistency-proofs}

We first prove Ziezold consistency by establishing
\(F_n(\omega,m_n)\to F(m)\) whenever \(m_n\to m\), simultaneously on a single probability-one event. We then combine this result with eventual nonemptiness and boundedness and the Heine--Borel property to obtain BP consistency.
\begin{lemma}[Ziezold consistency]
\label{supp:lem:gfm-ziezold-consistency}
Assume
Conditions~\ref{supp:cond:gfm-auxiliary-consistency},
\ref{supp:cond:gfm-primary-polish},
\ref{supp:cond:gfm-auxiliary-local-polish},
\ref{supp:cond:gfm-cost-regularity}, and
\ref{supp:cond:gfm-local-integrability}.
Then \(\widehat E_n\) is Ziezold consistent with \(E_0\) in the sense
of Definition~\ref{supp:def:gfm-ziezold-consistency}.
\end{lemma}

\begin{proof}[Proof of Lemma~\ref{supp:lem:gfm-ziezold-consistency}] The proof begins by demonstrating that for almost all $\omega \in \Omega$, the following statement is true:\begin{equation}\label{supp:eq:gfm-continuous-convergence}\text{For every } m\in M \text{ and any sequence }m_{n}\in M \text{ that converges to } m, F(m) =\lim_{n\rightarrow \infty} F_{n}(\omega)(m_{n}).\end{equation}

Since $M$ is separable (Condition~\ref{supp:cond:gfm-primary-polish}), it contains a countable dense subset $D \subset M$. Condition~\ref{supp:cond:gfm-local-integrability} guarantees that for each $d \in D$, there exists $r_d > 0$ such that $\pi_{d,r}(X)$ and $\Pi_{d,r}(X)$ are integrable for any $0 < r \le r_d$. 
For each \(d\in D\), define
\[
    \mathcal R_d
    :=
    \left\{
        r\in\mathbb Q\cap(0,r_0):
        \pi_{d,r}(X)
        \text{ and }
        \Pi_{d,r}(X)
        \text{ are integrable}
    \right\}.
\]
By Condition~\ref{supp:cond:gfm-local-integrability}, each
\(\mathcal R_d\) is nonempty. For a fixed $d \in D$ and $r \in \mathcal R_d$, consider the set
$$\mathcal{A}_{d,r} = \left\{ \omega \in \Omega : \frac{1}{n} \sum_{i=1}^n \pi_{d,r}(X_i) \to \mathbb{E}\{\pi_{d,r}(X)\} \text{ and } \frac{1}{n} \sum_{i=1}^n \Pi_{d,r}(X_i) \to \mathbb{E}\{\Pi_{d,r}(X)\} \right\}.$$

By the strong consistency of the auxiliary estimator $\hat{q}_n$, there exists a set $\Omega_Q \subset \Omega$ with $\mathcal{P}(\Omega_Q) = 1$ such that $\lim_{n \to \infty} d_Q(\hat{q}_n(\omega), q_0) = 0$ for all $\omega \in \Omega_Q$. Furthermore, by the strong law of large numbers \citep[see][Theorem 2.4.1]{durrett2019probability}, the probability of each $\mathcal{A}_{d,r}$ is one. We then define $\Omega_{0}$ as the intersection of $\Omega_Q$ and all such $\mathcal{A}_{d,r}$'s:
$$\Omega_{0} = \Omega_Q \cap \left( \bigcap_{d \in D} \bigcap_{r \in \mathcal{R}_{d}} \mathcal{A}_{d,r} \right).$$
This set $\Omega_{0}$ also has probability one, as it is a countable intersection of probability-one sets. The objective is to show that \eqref{supp:eq:gfm-continuous-convergence} is valid for any $\omega \in \Omega_{0}$.

Let
\[
    \mathscr B
    :=
    \left\{
        B_M(d,r):
        d\in D,\ r \in \mathcal{R}_{d}
    \right\}.
\]
Since \(D\) and \(\mathbb Q\) are countable, the collection
\(\mathscr B\) is countable. We now show that \(\mathscr B\) is a
basis for \(M\) with arbitrarily small radii: for every \(m\in M\),
every open neighborhood \(U\) of \(m\), and every \(\delta>0\), there
exists \(B_M(d,r)\in\mathscr B\) such that
\[
    m\in B_M(d,r)\subset U,
    \quad
    r<\delta.
\]

To see this, fix \(m\in M\), an open neighborhood \(U\) of \(m\),
and \(\delta>0\). Choose \(\varepsilon>0\) such that
\[
    B_M(m,\varepsilon)\subset U.
\]
By Condition~\ref{supp:cond:gfm-local-integrability}, there exists \(r_m>0\)
such that \(\pi_{m,r}(X)\) and \(\Pi_{m,r}(X)\) are integrable for
every \(0<r\le r_m\). By replacing \(r_m\) with a smaller positive
number if necessary, we may assume that
\[
    r_m<\min\{r_0,\varepsilon,\delta\}.
\]

Choose \(\eta>0\) such that
\[
    2\eta<r_m.
\]
Since \(D\) is dense in \(M\), there exists \(d\in D\) satisfying
\[
    d_M(d,m)<\eta.
\]
Choose a rational number \(r\) such that
\[
    d_M(d,m)<r<\eta.
\]
Then \(r<\eta < r_{m}< \delta\) and
\[
    m\in B_M(d,r).
\]
Moreover, for every \(x\in B_M(d,r)\),
\[
    d_M(x,m)
    \le
    d_M(x,d)+d_M(d,m)
    <
    r+\eta
    <
    2\eta.
\]
Therefore,
\[
    B_M(d,r)
    \subset
    B_M(m,2\eta)
    \subset
    B_M(m,\varepsilon)
    \subset U.
\]

Since \(r<\eta<r_m\), we also have
\[
    B_M(d,r)\times B_Q(q_0,r)
    \subset
    B_M(m,r_m)\times B_Q(q_0,r_m).
\]
Consequently,
\[
    \pi_{m,r_m}(X)
    \le
    \pi_{d,r}(X)
    \le
    \Pi_{d,r}(X)
    \le
    \Pi_{m,r_m}(X).
\]
In particular,
\[
    |\pi_{d,r}(X)|
    \vee
    |\Pi_{d,r}(X)|
    \le
    |\pi_{m,r_m}(X)|
    +
    |\Pi_{m,r_m}(X)|.
\]
The right-hand side is integrable, so both
\(\pi_{d,r}(X)\) and \(\Pi_{d,r}(X)\) are integrable. Hence
\(r\in\mathcal R_d\), and therefore
\(B_M(d,r)\in\mathscr B\). This shows the claimed basis property.

Now we verify that~\eqref{supp:eq:gfm-continuous-convergence} holds for every
\(\omega\in \Omega_{0}\). Fix \(m\in M\) and a sequence
\(m_n\to m\).

By the strengthened basis property established above, there exist
sequences \(m^{(a)}\in D\) and
\(r^{(a)}\in\mathcal R_{m^{(a)}}\) such that, for every
\(a\in\mathbb N\),
\[
    m\in
    B_M\bigl(m^{(a+1)},r^{(a+1)}\bigr)
    \subset
    B_M\bigl(m^{(a)},r^{(a)}\bigr)
    \subset
    B_M\left(m,\frac1a\right),
\]
and
\[
    r^{(a+1)}<\frac12r^{(a)}.
\]
In particular, \(r^{(a)}\downarrow0\), and
\[
    \bigcap_{a=1}^{\infty}
    B_M\bigl(m^{(a)},r^{(a)}\bigr)
    =
    \{m\}.
\]

Given that $m_n \to m$ and $\hat{q}_n(\omega) \to q_0$ (since $\omega \in \Omega_{0}$), for each index $a$ we can find an $N^{(a)}$ where $m_n \in B_{M}(m^{(a)}, r^{(a)})$ and $\widehat{q}_n(\omega) \in B_{Q}(q_{0}, r^{(a)})$ for all $n \ge N^{(a)}$. Consequently, by the definition of the local infimum and supremum, we have $\pi_{m^{(a)}, r^{(a)}}(X_i) \le \rho(X_i, m_n, \hat{q}_n) \le \Pi_{m^{(a)}, r^{(a)}}(X_i)$ for all $n \ge N^{(a)}$.

This establishes the following bounds:\begin{align*}\mathbb{E} \{{\pi_{m^{(a)}, r^{(a)}}(X)} \}&= \lim_{n \to \infty} \frac{1}{n} \sum_{i=1}^n \pi_{m^{(a)}, r^{(a)}}(X_i) \le \liminf_{n \to \infty} F_n(\omega)(m_n), \\
\limsup_{n \to \infty} F_n(\omega)(m_n) &\le \lim_{n \to \infty} \frac{1}{n} \sum_{i=1}^n \Pi_{m^{(a)}, r^{(a)}}(X_i) = \mathbb{E}\{ {\Pi_{m^{(a)}, r^{(a)}}(X)} \}.\end{align*}

For each \(a\), let
\[
    G_a
    :=
    B_M(m^{(a)},r^{(a)})
    \times
    B_Q(q_0,r^{(a)}).
\]
By construction,
\[
    G_{a+1}\subset G_a,
    \quad
    \bigcap_{a=1}^\infty G_a
    =
    \{(m,q_0)\}.
\]
Since \((m,q_0)\in G_a\) for every \(a\), the continuity of
\(\rho(t,\cdot,\cdot)\) implies, for each \(t\in T\),
\[
    \pi_{m^{(a)},r^{(a)}}(t)
    \uparrow
    \rho(t,m,q_0),
    \quad
    \Pi_{m^{(a)},r^{(a)}}(t)
    \downarrow
    \rho(t,m,q_0).
\]

Moreover, \(G_a\subset G_1\) gives
\[
    \pi_{m^{(1)},r^{(1)}}(X)
    \le
    \pi_{m^{(a)},r^{(a)}}(X)
    \le
    \rho(X,m,q_0)
    \le
    \Pi_{m^{(a)},r^{(a)}}(X)
    \le
    \Pi_{m^{(1)},r^{(1)}}(X).
\]
Because \(r^{(1)}\in\mathcal R_{m^{(1)}}\), the integrable random
variable
\[
    \left|
        \pi_{m^{(1)},r^{(1)}}(X)
    \right|
    +
    \left|
        \Pi_{m^{(1)},r^{(1)}}(X)
    \right|
\]
dominates all three middle terms. Hence, by the dominated convergence
theorem,
\[
\begin{aligned}
    \mathbb E[
        \pi_{m^{(a)},r^{(a)}}(X)
    ]
    &\longrightarrow
    \mathbb E[\rho(X,m,q_0)],\\
    \mathbb E[
        \Pi_{m^{(a)},r^{(a)}}(X)
    ]
    &\longrightarrow
    \mathbb E[\rho(X,m,q_0)].
\end{aligned}
\]

By combining these bounds, the result is:$$\lim_{n \to \infty} F_n(\omega)(m_n) = \mathbb{E}\{\rho(X, m, q_{0})\} = F(m)$$
for an arbitrary $m\in M$ and converging sequence $m_n \to m$. This confirms that \eqref{supp:eq:gfm-continuous-convergence} is satisfied for all $\omega\in \Omega_{0}$.

The next step is to demonstrate that for all $\omega\in \Omega_{0}$, the inequality
\begin{equation}\label{supp:eq:gfm-empirical-minimum-limsup}
\limsup_{n\rightarrow \infty} \hat{l}_{n}(\omega)\le l_{0}
\end{equation}
holds, where $\hat{l}_n(\omega) = \inf_{m \in M} F_n(\omega)(m)$ and $ l_0 = \inf_{m \in M} F(m) $. For any chosen $m \in M$, the relationship $\limsup_{n\rightarrow \infty}F_n(\omega)(m) = F(m)$ is almost surely true because \eqref{supp:eq:gfm-continuous-convergence} holds for all $\omega \in \Omega_{0}$. Given that $\hat{l}_n(\omega) \le F_n(\omega)(m)$, we find that
\begin{equation}\label{supp:eq:gfm-empirical-minimum-upper-bound}
\limsup_{n \to \infty} \hat{l}_n(\omega) \le \limsup_{n \to \infty} F_n(\omega)(m) = F(m).
\end{equation}

Since the choice of $m \in M$ was arbitrary, taking the infimum over all $m\in M$ in \eqref{supp:eq:gfm-empirical-minimum-upper-bound} allows us to conclude that $\limsup_{n \rightarrow \infty} \hat{l}_n(\omega) \le l_0$. 

Finally, we show that $$\bigcap_{n=1}^{\infty} \overline{\bigcup_{k=n}^{\infty} \hat{E}_k(\omega)} \subset E_0$$ holds for all $\omega \in \Omega_{0}$. If the intersection $$\bigcap_{n=1}^{\infty} \overline{\bigcup_{k=n}^{\infty} \hat{E}_k(\omega)}$$ is empty, the result holds trivially. Therefore, assume the intersection is non-empty. Let $b$ represent an arbitrary point within this set. This implies the existence of a subsequence $m_{n_k}(\omega) \in \hat{E}_{n_k}(\omega)$ converging to $b$. 
Since $b \in M$ and $l_0 = \inf_{m \in M} F(m)$, it directly follows that $l_0 \le F(b)$. By the definition of $\hat{l}_n$, we know $\hat{l}_{n_k}(\omega) = F_{n_k}(\omega)(m_{n_k})$. By taking limits and applying \eqref{supp:eq:gfm-continuous-convergence}, we obtain
\begin{equation}\label{supp:eq:gfm-limit-point-minimum-bound}
F(b) = \lim_{k \to \infty} \hat{l}_{n_k}(\omega)  \le \limsup_{n \to \infty} \hat{l}_n(\omega).\end{equation}

When this is combined with \eqref{supp:eq:gfm-empirical-minimum-limsup} and the fact that $l_{0} \le F(b)$, we can deduce that $F(b) = l_0$, meaning $b \in E_0$. Because the point $b$ was arbitrary, $$\bigcap_{n=1}^{\infty} \overline{\bigcup_{k=n}^{\infty} \hat{E}_k(\omega)} \subset E_0.$$
\end{proof}

\begin{proof}[Proof of Theorem~\ref{thm: BP-SC}]
According to Lemma~\ref{supp:lem:gfm-ziezold-consistency}, the estimator $\hat{E}_{n}$ exhibits Ziezold consistency with $E_{0}$, provided 
Conditions~\ref{supp:cond:gfm-auxiliary-consistency}, \ref{supp:cond:gfm-primary-polish}, \ref{supp:cond:gfm-auxiliary-local-polish}, \ref{supp:cond:gfm-cost-regularity}, and \ref{supp:cond:gfm-local-integrability} are met. Therefore, the proof requires only demonstrating that if the set $\hat{E}_{n}(\omega)$ is eventually bounded and the condition $\cap_{n=1}^{\infty} \overline{\cup_{k=n}^{\infty} \hat{E}_{k}(\omega)} \subset E_{0}$ is satisfied, then it follows that:$$\lim_{n\rightarrow \infty}\sup_{m\in \hat{E}_{n}(\omega)}d_{M}(m, E_{0})=0.$$

Fix a sample path on which Ziezold consistency and eventual
nonemptiness and boundedness hold.
Choose \(e_n\in\widehat E_n\) for all sufficiently large \(n\).
By the Heine--Borel property, this sequence has a convergent
subsequence. Ziezold consistency places its limit in \(E_0\),
so \(E_0\neq\emptyset\).

We proceed by contradiction: assume there is an $\epsilon > 0$ and a subsequence $e_{n_k} \in \hat{E}_{n_k}(\omega)$ for which $d_{M}(e_{n_k}, E_0) \ge \epsilon$ for every $k$. Because $\hat{E}_n(\omega)$ is eventually bounded and the space $M$ possesses the Heine-Borel property (Condition~\ref{supp:cond:gfm-heine-borel}), the sequence $\{e_{n_k}\}$ must contain a subsequence $e_{n_{k_m}}$ that converges to some point $e \in M$. Furthermore, the condition $\bigcap_{n=1}^{\infty} \overline{\bigcup_{k=n}^{\infty} \hat{E}_k(\omega)} \subset E_0$ means that this limit point $e$ must belong to $E_0$. Consequently,
$$d_{M}(e_{n_{k_m}}, E_0) \le d_{M}(e_{n_{k_m}}, e) \to 0$$
as $m$ approaches infinity. This directly contradicts our initial assumption that $d_{M}(e_{n_k}, E_0) \ge \epsilon$ for all $k\ge 1$. We must therefore conclude that$$\lim_{n \to \infty} \sup_{m \in \hat{E}_n(\omega)} d_{M}(m, E_0) = 0$$ by contradiction, which is the desired result.
\end{proof}

\subsection{
\texorpdfstring{
Measurability of $\widehat{E}_n$ and measurable selections
$\widehat{m}_n \in \widehat{E}_n$
}{
Measurability of the empirical argmin set and its measurable selections
}}
\label{supp:sec:gfm-measurable-minimizers}

We study the measurability of the empirical argmin sets and the existence of measurable selections for the asymptotic analysis. The argument applies the Kuratowski--Ryll-Nardzewski theorem to a localized argmin correspondence that agrees with \(\widehat E_n\)
eventually almost surely.

Let \(\mathrm{CL}(M)\) denote the collection of all nonempty closed subsets
of the metric space \((M,d_M)\); that is,
\[
    \mathrm{CL}(M)
    :=
    \left\{
        C\subseteq M:
        C\neq\emptyset
        \text{ and } C \text{ is closed in } M
    \right\}.
\]
Under Condition~\ref{supp:cond:gfm-cost-regularity}, for every fixed
\(\omega\in\Omega\), the empirical cost function
\[
    F_n(\omega,m)
    :=
    \frac{1}{n}
    \sum_{i=1}^n
    \rho
    \left(
        X_i(\omega),
        m,
        \widehat q_n(\omega)
    \right)
\]
is continuous in \(m\in M\). Consequently, whenever the minimum is
attained, the empirical argmin set
\[
    \widehat E_n(\omega)
    :=
    \operatorname*{argmin}_{m\in M}
    F_n(\omega,m)
\]
is nonempty and closed, and hence belongs to \(\mathrm{CL}(M)\).

A natural and important question is then whether the set-valued map
\[
    \omega
    \longmapsto
    \widehat E_n(\omega)
    \in
    \mathrm{CL}(M)
\]
is measurable with respect to a suitable \(\sigma\)-algebra on \(\mathrm{CL}(M)\). As studied in the theory of random closed sets \citep{molchanov2017theory} and in the context of \fre~mean sets \citep{evans2024limit}, a standard measurable structure on the space of closed subsets is the Effros $\sigma$-algebra \citep{srivastava1998course}, defined by
\[
    \mathcal E(M)
    :=
    \sigma
    \left(
        \left\{
            \left\{
                C\in\mathrm{CL}(M):
                C\cap U\neq\emptyset
            \right\}
            :
            U\subseteq M \text{ is open}
        \right\}
    \right).
\]
Thus, \(\mathcal E(M)\) is the \(\sigma\)-algebra generated by the sets
\(\{C\in\mathrm{CL}(M):C\cap U\neq\emptyset\}\),
as \(U\) ranges over the open subsets of \(M\). Accordingly, \(\widehat E_n:\Omega\to\mathrm{CL}(M)\) is
\(\mathcal A/\mathcal E(M)\)-measurable if
\[
    \left\{
        \omega\in\Omega:
        \widehat E_n(\omega)\cap U\neq\emptyset
    \right\}
    \in\mathcal A
\]
for every open set \(U\subseteq M\). This notion of measurability is precisely the one required by the
Kuratowski--Ryll-Nardzewski measurable selection theorem, which states that an Effros-measurable correspondence with nonempty closed values in a Polish space admits a measurable point-valued selection. We state the result in the form used below; see \citet[Section~5.2]{srivastava1998course}.

\begin{theorem}[Kuratowski--Ryll-Nardzewski measurable selection theorem]
\label{supp:thm:gfm-KRN-measurable-selection}
Let \((\Omega,\mathcal A)\) be a measurable space and let \(M\) be a
Polish space. Suppose that
\[
    \Gamma:
    (\Omega,\mathcal A)
    \longrightarrow
    \bigl(\mathrm{CL}(M),\mathcal E(M)\bigr)
\]
is measurable. Then there exists an
\(\mathcal A/\mathcal B(M)\)-measurable selection map
\(s:\Omega\to M\) such that
\[
    s(\omega)\in\Gamma(\omega)
    \quad
    \text{for every }\omega\in\Omega.
\]
\end{theorem}

Under Condition~\ref{supp:cond:gfm-primary-polish}, the space \(M\) is Polish. Therefore, once \(\widehat E_n\) is shown to be nonempty-valued and \(\mathcal A/\mathcal E(M)\)-measurable, Theorem~\ref{supp:thm:gfm-KRN-measurable-selection} applied with \(\Gamma=\widehat E_n\) yields an \(\mathcal A/\mathcal B(M)\)-measurable selection \(\widehat m_n:\Omega\to M\) satisfying
\[
    \widehat m_n(\omega)\in\widehat E_n(\omega)
    \quad
    \text{for every }\omega\in\Omega.
\]

We now construct a localized empirical argmin correspondence that is measurable and admits measurable selections for every \(n\). The strong consistency of the auxiliary estimator ensures that this correspondence agrees with \(\widehat E_n\) eventually, almost surely.

Let
\(
    \mathcal Q_0
    :=
    \overline{B_Q(q_0,r_0)},
\)
where \(r_0>0\) is given by
Condition~\ref{supp:cond:gfm-auxiliary-local-polish}. Since \(\widehat q_n\) need not belong to
\(\mathcal Q_0\) for every \(n\), define the localized auxiliary estimator
\[
    \widetilde q_n(\omega)
    :=
    \begin{cases}
        \widehat q_n(\omega),
        & \widehat q_n(\omega)\in\mathcal Q_0,\\
        q_0,
        & \widehat q_n(\omega)\notin\mathcal Q_0.
    \end{cases}
\]
For each \(n\ge1\), let
\[
    S_n
    :=
    T^n\times\mathcal Q_0
\]
be equipped with the product \(\sigma\)-algebra
\[
    \mathcal S_n:=\mathcal B(T)^{\otimes n}
    \otimes
    \mathcal B(\mathcal Q_0).
\]

Define
\[
    \widetilde Z_n(\omega)
    :=
    \left(
        X_1(\omega),\ldots,X_n(\omega),\widetilde q_n(\omega)
    \right)
    \in S_n.
\]
For \(s=(t_1,\ldots,t_n,q)\in S_n\), set
\[
    g_n(s,m)
    :=
    \frac1n\sum_{i=1}^n\rho(t_i,m,q),
    \quad
    A_n(s)
    :=
    \operatorname*{argmin}_{m\in M}g_n(s,m).
\]
Then the localized empirical objective and argmin correspondence satisfy
\[
    \widetilde F_n(\omega,m)
    :=
    \frac1n\sum_{i=1}^n
    \rho
    \left(
        X_i(\omega),m,\widetilde q_n(\omega)
    \right)
    =
    g_n\bigl(\widetilde Z_n(\omega),m\bigr)
\]
and
\[
    \widetilde E_n(\omega)
    :=
    \operatorname*{argmin}_{m\in M}
    \widetilde F_n(\omega,m)
    =
    A_n\bigl(\widetilde Z_n(\omega)\bigr).
\]
Since \(\widetilde Z_n\) is
\((\mathcal A/\mathcal S_n)\)-measurable, it remains to show that
\(A_n\) is
\(\mathcal S_n/\mathcal E(M)\)-measurable.

\begin{theorem}
\label{supp:thm:gfm-empirical-argmin-measurability}
Assume Conditions~\ref{supp:cond:gfm-auxiliary-consistency},
\ref{supp:cond:gfm-primary-polish},
\ref{supp:cond:gfm-heine-borel},
\ref{supp:cond:gfm-auxiliary-local-polish}, and
\ref{supp:cond:gfm-cost-regularity}. Assume that, for every \(n\ge1\) and
\(s\in S_n\), the function \(g_n(s,\cdot)\) attains its minimum over
\(M\). Then, the following hold.

\begin{enumerate}
    \item[(i)] For every \(n\ge1\), 
\[
    A_n:
    \bigl(S_n,\mathcal S_n\bigr)
    \longrightarrow
    \bigl(\mathrm{CL}(M),\mathcal E(M)\bigr) \quad \text{and} \quad
    \widetilde E_n:
    (\Omega,\mathcal A)
    \longrightarrow
    \bigl(\mathrm{CL}(M),\mathcal E(M)\bigr)
\]
are measurable.

    \item[(ii)] There exists \(\Omega_0\in\mathcal A\) with
\(\mathcal{P}(\Omega_0)=1\) such that, for every \(\omega\in\Omega_0\),
\[
    \widetilde E_n(\omega)
    =
    \widehat E_n(\omega)
\]
for all sufficiently large \(n\).

    \item[(iii)] There exist
\(\mathcal A/\mathcal B(M)\)-measurable maps
\(\widehat m_n:\Omega\to M\) such that, almost surely,
\[
    \widehat m_n(\omega)
    \in
    \widehat E_n(\omega)
\]
for all sufficiently large \(n\). In other words,
\(\widehat m_n\) is almost surely eventually a measurable selection of
\(\widehat E_n\).

    \item[(iv)] Moreover, suppose \(\widehat q_n(\omega)\in\mathcal Q_0\) for every
\(\omega\in\Omega\) and \(n\ge1\). Then
\(\widehat E_n\) is
\(\mathcal A/\mathcal E(M)\)-measurable and admits an
\(\mathcal A/\mathcal B(M)\)-measurable selection for all $n\ge 1$. This includes the case without an estimated auxiliary parameter, in which
\(\widehat q_n=q_0\) for all \(n\ge 1\).
\end{enumerate}
\end{theorem}

\begin{proof} We show each assertion separately.\\

\noindent\textit{Proof of (i).} We first show the measurability of \(A_n\). The proof proceeds in four
steps.

As a first step, we establish the measurability of \(g_n\) and its
infimum over \(M\). By Condition~\ref{supp:cond:gfm-cost-regularity},
\(
    g_n:
    S_n\times M
    \longrightarrow
    \mathbb R
\)
is
\(
    \left(
        \mathcal S_n\otimes\mathcal B(M)
    \right)
    /
    \mathcal B(\mathbb R)
\)
-measurable. Indeed, for each \(i=1,\ldots,n\), the coordinate map
\[
    (t_1,\ldots,t_n,q,m)
    \longmapsto
    (t_i,m,q)
\]
is measurable with respect to the corresponding product
\(\sigma\)-algebras. Hence,
\[
    (s,m)
    \longmapsto
    \rho(t_i,m,q)
\]
is measurable by Condition~\ref{supp:cond:gfm-cost-regularity}, and so is their
finite average \(g_n\). Moreover, for every fixed \(s\in S_n\), the
function \(g_n(s,\cdot)\) is continuous on \(M\).

Since \(M\) is separable, we can choose a countable dense subset
\(
    D
    =
    \{m_1,m_2,\ldots\}
    \subseteq M.
\)
For each \(s\in S_n\), define
\[
    v_n(s)
    :=
    \inf_{m\in M}
    g_n(s,m).
\]
Since \(g_n(s,\cdot)\) is continuous and \(D\) is dense in \(M\), we get
\[
    v_n(s)
    =
    \inf_{l\ge1}
    g_n(s,m_l).
\]
For every \(l\ge1\), the function
\(s\mapsto g_n(s,m_l)\) is
\(\mathcal S_n/\mathcal B(\mathbb R)\)-measurable. Therefore,
\(v_n\) is \(\mathcal S_n/\mathcal B(\mathbb R)\)-measurable as a
countable infimum of measurable functions.

As a second step, we construct a compact exhaustion of \(M\). This will
allow us to reduce minimization over an arbitrary open subset of \(M\) to
countably many minimization problems over compact sets. Since \(M\) is a
separable metric space, it is second countable and hence Lindelöf, i.e.,
every open cover has a countable subcover. By
Condition~\ref{supp:cond:gfm-heine-borel}, \(M\) is locally compact, and thus every
\(m\in M\) has an open neighborhood \(V_m\) whose closure is compact.
Using the Lindelöf property of \(M\), we can choose a countable
subcollection
\[
    V_1,V_2,\ldots
\]
that covers \(M\). Define
\[
    K_j
    :=
    \bigcup_{l=1}^j
    \overline V_l,
    \quad
    j\ge1.
\]
Then each \(K_j\) is compact, and satisfies
\[
    K_j\subseteq K_{j+1},
    \quad
    M
    =
    \bigcup_{j=1}^\infty K_j.
\]

As a third step, we use this compact exhaustion to express any open
subset \(U\subseteq M\) as a countable union of compact subsets. With the
convention \(d_M(m,\emptyset)=\infty\), define
\[
    K_{j,k}(U)
    :=
    K_j
    \cap
    \left\{
        m\in M:
        d_M(m,M\setminus U)
        \ge
        \frac1k
    \right\},
    \quad
    j,k\ge1.
\]
Each \(K_{j,k}(U)\) is a closed subset of the compact set \(K_j\), and
is therefore compact. Moreover,
\begin{equation}
\label{supp:eq:gfm-open-set-compact-exhaustion}
    U
    =
    \bigcup_{j=1}^\infty
    \bigcup_{k=1}^\infty
    K_{j,k}(U).
\end{equation}
Indeed, if \(m\in U\), then \(m\in K_j\) for some \(j\), and the
openness of \(U\) implies
\[
    d_M(m,M\setminus U)>0,
\]
so that \(m\in K_{j,k}(U)\) for some \(k\). Conversely, if
\(m\in K_{j,k}(U)\) for some \(j,k\), then
\[
    d_M(m,M\setminus U)\ge\frac1k>0,
\]
which implies \(m\notin M\setminus U\), and hence \(m\in U\).

For each nonempty \(K_{j,k}(U)\), choose its countable dense subset
\(D_{j,k}(U)\), and define
\[
    w_{j,k}(s)
    :=
    \min_{m\in K_{j,k}(U)}
    g_n(s,m).
\]
If \(K_{j,k}(U)=\emptyset\), set \(w_{j,k}(s):=+\infty\). Since
\(K_{j,k}(U)\) is compact and \(g_n(s,\cdot)\) is continuous, the
minimum is attained and
\[
    w_{j,k}(s)
    =
    \inf_{m\in D_{j,k}(U)}
    g_n(s,m).
\]
Thus, each \(w_{j,k}\) is
\(\mathcal S_n/\mathcal B(\overline{\mathbb R})\)-measurable.
Since \(v_n(s)\) is finite, the equality
\(w_{j,k}(s)=v_n(s)\) can hold only when
\(K_{j,k}(U)\neq\emptyset\).

As a final step, we characterize the event that the argmin set
\(A_n(s)\) intersects \(U\) by comparing the global minimum \(v_n(s)\)
with the minima \(w_{j,k}(s)\) over the compact subsets of \(U\). We
claim that
\begin{equation}
\label{supp:eq:gfm-argmin-open-hit-characterization}
\left\{
    s\in S_n:
    A_n(s)\cap U\neq\emptyset
\right\}
=
\bigcup_{j=1}^\infty
\bigcup_{k=1}^\infty
\left\{
    s\in S_n:
    w_{j,k}(s)=v_n(s)
\right\}.
\end{equation}
To see the first inclusion, suppose that
\(A_n(s)\cap U\neq\emptyset\), and choose
\(m\in A_n(s)\cap U\). By
\eqref{supp:eq:gfm-open-set-compact-exhaustion}, there exist \(j,k\ge1\) such
that \(m\in K_{j,k}(U)\). Since \(m\in A_n(s)\), it minimizes
\(g_n(s,\cdot)\) over \(M\). Therefore, by the definition of \(v_n(s)\),
\[
    g_n(s,m)
    =
    \inf_{m'\in M}g_n(s,m')
    =
    v_n(s).
\]
Hence,
\[
    w_{j,k}(s)
    \le
    g_n(s,m)
    =
    v_n(s).
\]
Since \(v_n(s)\) is the infimum over all of \(M\), the reverse inequality
holds, and therefore
\[
    w_{j,k}(s)=v_n(s).
\]

Conversely, suppose that
\(w_{j,k}(s)=v_n(s)\) for some \(j,k\). Since the minimum defining
\(w_{j,k}(s)\) is attained, there exists
\(m\in K_{j,k}(U)\subseteq U\) such that
\[
    g_n(s,m)
    =
    w_{j,k}(s)
    =
    v_n(s).
\]
Thus, \(m\in A_n(s)\cap U\), proving
\eqref{supp:eq:gfm-argmin-open-hit-characterization}.

Since \(v_n\) and all \(w_{j,k}\) are measurable, the right-hand side
of \eqref{supp:eq:gfm-argmin-open-hit-characterization} belongs to
\(\mathcal S_n\). Therefore,
\[
    \left\{
        s\in S_n:
        A_n(s)\cap U\neq\emptyset
    \right\}
    \in
    \mathcal S_n
\]
for every open \(U\subseteq M\). By the definition of the Effros
\(\sigma\)-algebra,
\[
    A_n:
    (S_n,\mathcal S_n)
    \longrightarrow
    \bigl(\mathrm{CL}(M),\mathcal E(M)\bigr)
\]
is measurable.

To show measurability of $\widetilde E_n$, we consider the localized empirical correspondence. The set \(\mathcal Q_0\) is measurable in \(Q\), and hence 
\[
    \widetilde q_n:
    (\Omega,\mathcal A)
    \longrightarrow
    \left(
        \mathcal Q_0,
        \mathcal B(\mathcal Q_0)
    \right)
\]
is measurable. Since each \(X_i:\Omega\to T\) is measurable, the product
map
\[
    \widetilde Z_n:
    (\Omega,\mathcal A)
    \longrightarrow
    (S_n,\mathcal S_n),
    \quad
    \widetilde Z_n(\omega)
    =
    \left(
        X_1(\omega),
        \ldots,
        X_n(\omega),
        \widetilde q_n(\omega)
    \right)
\]
is measurable. By construction,
\[
    \widetilde E_n
    =
    A_n\circ\widetilde Z_n,
\]
and therefore
\[
    \widetilde E_n:
    (\Omega,\mathcal A)
    \longrightarrow
    \bigl(\mathrm{CL}(M),\mathcal E(M)\bigr)
\]
is measurable.\\

\noindent\textit{Proof of (ii).} Let
\(
    \Omega_0
    :=
    \left\{
        \omega\in\Omega:
        \widehat q_n(\omega)
        \longrightarrow
        q_0
    \right\}.
\)
By assumption,
\[
    \Omega_0\in\mathcal A,
    \quad
    \mathcal{P}(\Omega_0)=1.
\]
For every \(\omega\in\Omega_0\), there exists \(N(\omega)\) such that
\[
    d_Q
    \left(
        \widehat q_n(\omega),
        q_0
    \right)
    <
    r_0
    \quad
    \text{for every }n\ge N(\omega).
\]
Consequently,
\(
    \widetilde q_n(\omega)
    =
    \widehat q_n(\omega)
\)
and hence
\(
    \widetilde E_n(\omega)
    =
    \widehat E_n(\omega)
\)
for every \(n\ge N(\omega)\).\\

\noindent\textit{Proof of (iii).} Condition~\ref{supp:cond:gfm-primary-polish} implies that \(M\) is Polish. Since
each \(\widetilde E_n\) is nonempty, closed-valued, and
\(\mathcal A/\mathcal E(M)\)-measurable,
Theorem~\ref{supp:thm:gfm-KRN-measurable-selection} yields an
\(\mathcal A/\mathcal B(M)\)-measurable selection
\[
    \widehat m_n:
    \Omega
    \longrightarrow
    M
\]
such that
\[
    \widehat m_n(\omega)
    \in
    \widetilde E_n(\omega)
    \quad
    \text{for every }\omega\in\Omega.
\]
On \(\Omega_0\),
\[
    \widehat m_n(\omega)
    \in
    \widehat E_n(\omega)
\]
for all sufficiently large \(n\), completing the proof.\\

\noindent\textit{Proof of (iv).} (iv) is immediate since $\widehat{E}_{n} = \widetilde{E}_{n}$ whenever $\widehat{q}_n \in \mathcal{Q}_{0}$.

\end{proof}

\subsection{Asymptotic normality}
\label{supp:sec:gfm-asymptotic-normality}

This subsection states the regularity conditions and proves the stochastic expansions and central limit theorem in Theorem~\ref{thm: clt_gfm}. We use the setting and notation of Appendix~\ref{sec: GFM CLT}, including the local embedding \( \mathcal{F}\), the auxiliary coordinate map \(\mathcal{G}\), and the constraint map \(h\).

We distinguish the mapped cost \(\ell\) from its jointly measurable local extension \(\tilde{\ell}\) to the ambient Euclidean space. On the original mapped parameter domain within the extension neighborhood, they satisfy
\begin{equation}
\label{supp:eq:gfm-mapped-cost-function}
    \tilde{\ell}(t,\mu,\nu)
    =
    \ell(t,\mu,\nu)
    =
    \rho\bigl(t,\mathcal{F}^{-1}(\mu), \mathcal{G}^{-1}(\nu)\bigr).
\end{equation}
All derivatives and derivative-envelope conditions below refer to \(\tilde{\ell}\).

\subsubsection{Role of the central limit theorem conditions}
\label{supp:sec:gfm-clt-conditions}

To establish the asymptotic distribution of $\hat{\mu}_n$, we impose the following regularity conditions.

\begin{condition}
\label{supp:cond:gfm-geometric-constraint}
    There exists an open neighborhood $\mathcal U \subset \mathbb R^K$ of $\mu_0$ such that the constraint function
    $h=(h_1,\dots,h_P)^\top:\mathcal U\to\mathbb R^P$ is twice continuously differentiable. Moreover, we assume that the transpose of the Jacobian matrix
    \begin{equation}
    \label{supp:eq:gfm-jacobian-H}
        H := Dh(\mu_0)^\top
        =
        \big[ \nabla h_1(\mu_0), \, \nabla h_2(\mu_0), \, \dots, \, \nabla h_P(\mu_0) \big]
        \in \mathbb{R}^{K \times P}
    \end{equation}
    has rank $P$.
\end{condition}

Condition~\ref{supp:cond:gfm-geometric-constraint} describes the local
geometry of the embedded parameter space. The full-rank condition
ensures that the constraint set is locally a smooth submanifold of
codimension \(P\), with tangent space
\[
    T_{\mu_0}S
    =
    \left\{
        v\in\mathbb R^K:
        H^\top v=0
    \right\}.
\]
It also guarantees the linear independence of the constraint gradients
required to formulate the population and empirical KKT equations.

\begin{condition}
\label{supp:cond:gfm-primary-consistency}
    $\hat{\mu}_n$ converges in probability to $\mu_0$ ($\hat{\mu}_n \xrightarrow{p} \mu_0$).
\end{condition}

Condition~\ref{supp:cond:gfm-primary-consistency} localizes the
asymptotic analysis around \(\mu_0\). In particular, it ensures that
\(\widehat\mu_n\) and the intermediate points arising in the Taylor
expansions belong to the local coordinate neighborhood with probability
tending to one.

\begin{condition}
\label{supp:cond:gfm-auxiliary-linearity}
    $\hat{\nu}_n$ follows the asymptotic linear representation:
    \begin{equation}
    \label{supp:eq:gfm-auxiliary-linear-expansion}
        \sqrt{n}(\hat{\nu}_n - \nu_0) = \frac{1}{\sqrt{n}} \sum_{i=1}^n \xi(X_i) + o_p(1),
    \end{equation}
    where $\mathbb{E}[\xi(X)] = 0$ and $\xi(X)$ has a finite second moment.
\end{condition}

Condition~\ref{supp:cond:gfm-auxiliary-linearity} quantifies the
first-order stochastic variation of the estimated auxiliary parameter.
Because \(\widehat q_n\) may be constructed from the same observations
as the primary estimator, no independence between
\(\widehat\nu_n\) and the empirical cost is assumed. Instead, the
influence function \(\xi(X)\) propagates through the mixed derivative
of the mapped cost and contributes directly to the first-order
expansion of \(\widehat\mu_n\).

\begin{condition}
\label{supp:cond:gfm-cost-differentiability}
    There exists a compact convex neighborhood \(\mathcal K\) of \((\mu_0,\nu_0)\), contained in the domain of this extension, and an integrable random variable \(D(X)\), such that \((\mu,\nu)\mapsto \tilde{\ell}(X,\mu,\nu)\) is almost surely twice continuously differentiable on an open neighborhood of \(\mathcal K\), and
    \begin{equation}
    \label{supp:eq:gfm-derivative-dominance}
    {\max}\{ \sup_{(\mu,\nu) \in \mathcal{K}} \big\| \nabla_{\mu} \tilde{\ell}(X, \mu, \nu) \big\|,
        \sup_{(\mu,\nu) \in \mathcal{K}} \big\| \nabla_{\mu\mu}^2 \tilde{\ell}(X, \mu, \nu) \big\|,  \sup_{(\mu,\nu) \in \mathcal{K}} \big\| \nabla_{\mu\nu}^2 \tilde{\ell}(X, \mu, \nu) \big\| \} \le D(X) \quad \text{a.s.} .
    \end{equation}
\end{condition}

Condition~\ref{supp:cond:gfm-cost-differentiability}
permits differentiation under expectation and provides local uniform laws of large numbers for the empirical gradient, Hessian, and cross-Hessian. These results are established in Subsubsection~\ref{supp:sec:gfm-local-ulln}.

\begin{condition}
\label{supp:cond:gfm-kkt-invertibility}
Let \(\lambda_0 \in \mathbb{R}^P\) be the population Lagrange multiplier
satisfying
\[
    \mathbb{E}[\nabla_\mu \tilde{\ell}(X,\mu_0,\nu_0)] + H\lambda_0 = 0.
\]
Define the expected Hessian of the constrained Lagrangian by
\begin{equation}
\label{supp:eq:gfm-hessian-J}
    J
    :=
    \mathbb{E}[\nabla_{\mu\mu}^2 \tilde{\ell}(X,\mu_0,\nu_0)]
    +
    \sum_{j=1}^{P} \lambda_{0,j}\nabla^{2}h_{j}(\mu_{0})
    \in \mathbb{R}^{K\times K}.
\end{equation}
Assume that the \((K+P)\times(K+P)\) KKT block
matrix
\(
    B
    :=
    \begin{pmatrix}
        J & H \\
        H^\top & 0
   \end{pmatrix}
\)
is nonsingular.
\end{condition}

Condition~\ref{supp:cond:gfm-kkt-invertibility} is the local
identification condition for the constrained problem. It guarantees
that the linearized stationarity and feasibility equations have a
unique solution for the first-order perturbations of the primary
parameter and the Lagrange multiplier. Consequently, the empirical KKT
system can be inverted to obtain the stochastic expansion of
\(\widehat\mu_n\).

\begin{condition}
\label{supp:cond:gfm-quadratic-refinement}
    In addition to the preceding conditions, assume the following strengthened local regularity conditions.

    First, in addition to Condition~\ref{supp:cond:gfm-geometric-constraint}, assume that each \(h_j\) is three times continuously differentiable on a neighborhood of \(\mu_0\), with locally bounded third derivative.

    Second, in addition to Condition~\ref{supp:cond:gfm-auxiliary-linearity}, assume that
    \begin{equation}
    \label{supp:eq:gfm-auxiliary-quadratic-remainder}
        \hat{\nu}_n-\nu_0
        =
        \frac{1}{n}\sum_{i=1}^n \xi(X_i)
        +
        O_{p}(n^{-1}).
    \end{equation}

    Third, in addition to Condition~\ref{supp:cond:gfm-cost-differentiability}, assume that the map
    \[
        (\mu,\nu)\longmapsto \tilde{\ell}(X,\mu,\nu)
    \]
    is almost surely three times continuously differentiable on an open neighborhood of \(\mathcal K\). Assume further that the dominating random variable \(D(X)\) in Condition~\ref{supp:cond:gfm-cost-differentiability} has a finite second moment and there exists an integrable random variable \(D_3(X)\) such that
    \begin{equation}
    \label{supp:eq:gfm-third-derivative-dominance}
    \max\left\{
        \sup_{(\mu,\nu)\in\mathcal K}
        \big\|\nabla_{\mu\mu\mu}^3\tilde{\ell}(X,\mu,\nu)\big\|,
        \sup_{(\mu,\nu)\in\mathcal K}
        \big\|\nabla_{\mu\mu\nu}^3\tilde{\ell}(X,\mu,\nu)\big\|,
        \sup_{(\mu,\nu)\in\mathcal K}
        \big\|\nabla_{\mu\nu\nu}^3\tilde{\ell}(X,\mu,\nu)\big\|
    \right\}
    \le D_3(X)
    \quad\text{a.s.}
    \end{equation}
\end{condition}

Condition~\ref{supp:cond:gfm-quadratic-refinement} is not required for
the first-order stochastic expansion or the central limit theorem.
Its purpose is to strengthen the stochastic remainder from
\(o_p(n^{-1/2})\) to \(O_p(n^{-1})\). This requires third-order
smoothness of the constraint and cost functions, a second-moment bound
for the derivative envelope, and an \(O_p(n^{-1})\) refinement of the
auxiliary estimator's asymptotic linear representation.

\subsubsection{Local differentiation and uniform laws}
\label{supp:sec:gfm-local-ulln}

The following lemma records the differentiation-under-expectation and local uniform-law results used to control the empirical gradient, Hessian, and cross-Hessian at the random intermediate points appearing in the proof of Theorem~\ref{thm: clt_gfm}.

\begin{lemma}
\label{supp:lem:gfm-local-ulln-differentiation}
Assume Condition~\ref{supp:cond:gfm-cost-differentiability}. Then the following
statements hold.

\begin{enumerate}
\item[(i)] Differentiation with respect to \(\mu\) may be interchanged with
expectation at \((\mu_0,\nu_0)\):
\[
    \left.
    \nabla_\mu \mathbb{E}[\tilde{\ell}(X,\mu,\nu_0)]
    \right|_{\mu=\mu_0}
    =
    \mathbb{E}[\nabla_\mu\tilde{\ell}(X,\mu_0,\nu_0)].
\]

\item[(ii)] The maps
\[
    (\mu,\nu)\mapsto
    \mathbb{E}[\nabla_\mu\tilde{\ell}(X,\mu,\nu)], \quad
    (\mu,\nu)\mapsto
    \mathbb{E}[\nabla_{\mu\mu}^2\tilde{\ell}(X,\mu,\nu)], \quad
    (\mu,\nu)\mapsto
    \mathbb{E}[\nabla_{\mu\nu}^2\tilde{\ell}(X,\mu,\nu)]
\]
are continuous on \(\mathcal K\).

\item[(iii)] If
\(
    (\mu_n^*,\nu_n^*)
    \xrightarrow{p}
    (\mu_0,\nu_0)
\)
and
\(
    \mathcal P\bigl(
        (\mu_n^*,\nu_n^*)\in\mathcal K
    \bigr)
    \longrightarrow1,
\)
then
\[
    \frac1n\sum_{i=1}^n
    \nabla_\mu\tilde{\ell}(X_i,\mu_n^*,\nu_n^*)
    =
    \mathbb{E}[
        \nabla_\mu\tilde{\ell}(X,\mu_0,\nu_0)
    ]
    +
    o_p(1),
\]
\[
    \frac1n\sum_{i=1}^n
    \nabla_{\mu\mu}^2\tilde{\ell}(X_i,\mu_n^*,\nu_n^*)
    =
    \mathbb{E}[
        \nabla_{\mu\mu}^2\tilde{\ell}(X,\mu_0,\nu_0)
    ]
    +
    o_p(1),
\]
and
\[
    \frac1n\sum_{i=1}^n
    \nabla_{\mu\nu}^2\tilde{\ell}(X_i,\mu_n^*,\nu_n^*)
    =
    \mathbb{E}[
        \nabla_{\mu\nu}^2\tilde{\ell}(X,\mu_0,\nu_0)
    ]
    +
    o_p(1).
\]
The same conclusions hold jointly for any fixed finite collection of
such random points.
\end{enumerate}
\end{lemma}

\begin{proof}
Since all vectors and matrices are finite-dimensional, coordinatewise convergence implies convergence in any fixed norm.

\noindent\textit{Proof of (i).} Fix a coordinate direction \(e_j\). For sufficiently
small \(t\), the line segment between \((\mu_0, \nu_{0})\) and \((\mu_0+t e_j, \nu_{0})\) lies in
\(\mathcal K\). By the fundamental theorem of calculus,
\[
    \frac{\tilde{\ell}(X,\mu_0+t e_j,\nu_0)-\tilde{\ell}(X,\mu_0,\nu_0)}{t}
    =
    \int_0^1
    \partial_{\mu_j}\tilde{\ell}(X,\mu_0+s t e_j,\nu_0)\,ds .
\]
The absolute value of the right-hand side is bounded by \(D(X)\), which is
integrable. Hence, by the dominated convergence theorem,
\[
    \left.
    \partial_{\mu_j}\mathbb{E}[\tilde{\ell}(X,\mu,\nu_0)]
    \right|_{\mu=\mu_0}
    =
    \mathbb{E}[\partial_{\mu_j}\tilde{\ell}(X,\mu_0,\nu_0)].
\]
Applying this argument to all coordinates gives the first assertion.

\noindent\textit{Proof of (ii).}
The continuity of the expected derivative maps follows similarly. If
\((\mu_m,\nu_m)\to(\mu,\nu)\) in \(\mathcal K\), then, almost surely,
\[
    \nabla_\mu\tilde{\ell}(X,\mu_m,\nu_m)
    \to
    \nabla_\mu\tilde{\ell}(X,\mu,\nu).
\]
The norms are bounded by \(D(X)\), so the dominated convergence theorem gives
\[
    \mathbb{E}[\nabla_\mu\tilde{\ell}(X,\mu_m,\nu_m)]
    \to
    \mathbb{E}[\nabla_\mu\tilde{\ell}(X,\mu,\nu)].
\]
The same argument applies to
\(\nabla_{\mu\mu}^2\tilde{\ell}\) and \(\nabla_{\mu\nu}^2\tilde{\ell}\). Hence the three
expectation maps in the statement are continuous on \(\mathcal K\).

\noindent\textit{Proof of (iii).}
Let \(a(X,\mu,\nu)\) denote an arbitrary coordinate of one of the
three derivative processes
\[
    \nabla_\mu\tilde{\ell}(X,\mu,\nu),\quad
    \nabla_{\mu\mu}^2\tilde{\ell}(X,\mu,\nu),\quad
    \nabla_{\mu\nu}^2\tilde{\ell}(X,\mu,\nu).
\]
By Condition~\ref{supp:cond:gfm-cost-differentiability},
\(a(X,\mu,\nu)\) is continuous in \((\mu,\nu)\) almost surely and is
uniformly dominated on \(\mathcal K\) by an integrable random
variable. Since \(\mathcal K\) is compact, the standard uniform law
of large numbers for dominated continuous parametric classes
\citep[Lemma~2.4]{newey1994large} gives
\[
    \sup_{(\mu,\nu)\in\mathcal K}
    \left|
        \frac1n\sum_{i=1}^n a(X_i,\mu,\nu)
        -
        \mathbb E[a(X,\mu,\nu)]
    \right|
    =
    o_p(1).
\]

On the event \(
    \bigl\{
        (\mu_n^*,\nu_n^*)\in\mathcal K
    \bigr\},
\)
we therefore have
\[
\begin{aligned}
&\left|
    \frac1n\sum_{i=1}^n
    a(X_i,\mu_n^*,\nu_n^*)
    -
    \left.
    \mathbb E[a(X,\mu,\nu)]
\right|_{(\mu,\nu)=(\mu_n^*,\nu_n^*)}
\right|\le
\sup_{(\mu,\nu)\in\mathcal K}
\left|
    \frac1n\sum_{i=1}^n a(X_i,\mu,\nu)
    -
    \mathbb E[a(X,\mu,\nu)]
\right|
=
o_p(1).
\end{aligned}
\]
Since the probability of this event tends to one, we have
\[
\begin{aligned}
&\left|
    \frac1n\sum_{i=1}^n
    a(X_i,\mu_n^*,\nu_n^*)
    -
    \left.
    \mathbb E[a(X,\mu,\nu)]
\right|_{(\mu,\nu)=(\mu_n^*,\nu_n^*)}
\right|
=
o_p(1).
\end{aligned}
\]

By part~(ii) and
\(
    (\mu_n^*,\nu_n^*)
    \xrightarrow{p}
    (\mu_0,\nu_0),
\)
we also have
\[
    \left.
    \mathbb E[a(X,\mu,\nu)]
\right|_{(\mu,\nu)=(\mu_n^*,\nu_n^*)}
    \xrightarrow{p}
    \mathbb E[
        a(X,\mu_0,\nu_0)
    ].
\]
Consequently,
\[
    \frac1n\sum_{i=1}^n
    a(X_i,\mu_n^*,\nu_n^*)
    =
    \mathbb E[
        a(X,\mu_0,\nu_0)
    ]
    +
    o_p(1).
\]
Applying this argument to the finitely many coordinates of the
gradient, Hessian, and cross-Hessian shows the three asserted
convergences. Applying it finitely many times also shows the result
for any fixed finite collection of random points.
\end{proof}

\subsubsection{Proof of Theorem~\ref{thm: clt_gfm}}
\label{supp:sec:gfm-clt-proof}

We now prove the first-order stochastic expansion and central limit
theorem by linearizing the empirical KKT equations around the
population solution. The final step revisits the same expansions under
Condition~\ref{supp:cond:gfm-quadratic-refinement} to establish the
quadratic remainder refinement.

\begin{proof}[Proof of Theorem~\ref{thm: clt_gfm}]
Since \(\mu_0\) minimizes the population cost under
\(h(\mu)=0\), Lemma~\ref{supp:lem:gfm-local-ulln-differentiation}(i) and the full column rank of \(H\) give the population first-order condition
\begin{equation}
\label{supp:eq:gfm-population-foc}
    \mathbb E[
        \nabla_\mu\tilde{\ell}(X,\mu_0,\nu_0)
    ]
    +
    H\lambda_0
    =
    0.
\end{equation}

With probability tending to one, the mapped estimators belong to the local coordinate neighborhoods and
\(Dh(\widehat\mu_n)^\top\) has full column rank.
On this event, the empirical Lagrange multiplier theorem gives
\(\widehat\lambda_n\in\mathbb R^P\) satisfying
\begin{equation}
\label{supp:eq:gfm-empirical-foc}
    \frac1n\sum_{i=1}^n
    \nabla_\mu\tilde{\ell}
    (X_i,\widehat\mu_n,\widehat\nu_n)
    +
    Dh(\widehat\mu_n)^\top\widehat\lambda_n
    =
    0.
\end{equation}
We linearize this equation and the feasibility condition
\(h(\widehat\mu_n)=0\) around
\((\mu_0,\nu_0,\lambda_0)\).

\textbf{Step 1: Taylor expansion of $\frac{1}{n} \sum_{i=1}^n \nabla_\mu \tilde{\ell}(X_i, \hat{\mu}_n, \hat{\nu}_n)$. }

First, we expand the empirical gradient vector $\frac{1}{n} \sum_{i=1}^n \nabla_\mu \tilde{\ell}(X_i, \hat{\mu}_n, \hat{\nu}_n)$ around $(\mu_0, \nu_0)$. Since this gradient is a vector-valued function, we apply the mean value theorem to each component. For each $l \in \{1, \dots, K\}$, let $[\nabla_\mu \tilde{\ell}]_l$ denote the $l$-th component of the gradient. By the mean value theorem, there exists an intermediate point $(\tilde{\mu}_{n,l}^*, \tilde{\nu}_{n,l}^*)$ on the line segment between $(\mu_0, \nu_0)$ and $(\hat{\mu}_n, \hat{\nu}_n)$ such that the expansion holds for the $l$-th component: 
\begin{align}
\label{supp:eq:gfm-score-mvt-component} 
\frac{1}{n} \sum_{i=1}^n \big[\nabla_\mu \tilde{\ell}(X_i, \hat{\mu}_n, \hat{\nu}_n)\big]_l &= \frac{1}{n} \sum_{i=1}^n \big[\nabla_\mu \tilde{\ell}(X_i, \mu_0, \nu_0)\big]_l \nonumber \\ 
&\quad + \left[ \frac{1}{n} \sum_{i=1}^n \nabla_{\mu\mu}^2 \tilde{\ell}(X_i, \tilde{\mu}_{n,l}^*, \tilde{\nu}_{n,l}^*) \right]_{l, \cdot} (\hat{\mu}_n - \mu_0) \nonumber \\ 
&\quad + \left[ \frac{1}{n} \sum_{i=1}^n \nabla_{\mu\nu}^2 \tilde{\ell}(X_i, \tilde{\mu}_{n,l}^*, \tilde{\nu}_{n,l}^*) \right]_{l, \cdot} (\hat{\nu}_n - \nu_0), 
\end{align} 
where $[\cdot]_{l, \cdot}$ denotes the $l$-th row vector of a matrix. Stacking these $K$ scalar equations row by row yields: 
\begin{equation}
\label{supp:eq:gfm-score-mvt} 
\frac{1}{n} \sum_{i=1}^n \nabla_\mu \tilde{\ell}(X_i, \hat{\mu}_n, \hat{\nu}_n) = \frac{1}{n} \sum_{i=1}^n \nabla_\mu \tilde{\ell}(X_i, \mu_0, \nu_0) + \tilde{J}_n (\hat{\mu}_n - \mu_0) + \tilde{C}_n (\hat{\nu}_n - \nu_0), \end{equation} 
where $\tilde{J}_n \in \mathbb{R}^{K \times K}$, whose $l$-th row is the $l$-th row of $\frac{1}{n} \sum_{i=1}^n \nabla_{\mu\mu}^2 \tilde{\ell}(X_i, \tilde{\mu}_{n,l}^*, \tilde{\nu}_{n,l}^*)$, and $\tilde{C}_n \in \mathbb{R}^{K \times D_Q}$, whose $l$-th row is the $l$-th row of $\frac{1}{n} \sum_{i=1}^n \nabla_{\mu\nu}^2 \tilde{\ell}(X_i, \tilde{\mu}_{n,l}^*, \tilde{\nu}_{n,l}^*)$.

Since \(\mathcal K\) is convex neighborhood of \(\mu_{0}, \nu_{0}\)
and
\[
    \mathcal P\bigl(
        (\hat\mu_n,\hat\nu_n)\in\mathcal K
    \bigr)
    \longrightarrow1,
\]
all intermediate points
\((\tilde\mu_{n,l}^*,\tilde\nu_{n,l}^*)\),
\(l=1,\dots,K\), also belong to \(\mathcal K\) with probability
tending to one. Moreover, because each intermediate point lies on
the line segment between
\((\mu_0,\nu_0)\) and
\((\hat\mu_n,\hat\nu_n)\), the consistency of
\((\hat\mu_n,\hat\nu_n)\) implies
\[
    (\tilde\mu_{n,l}^*,\tilde\nu_{n,l}^*)
    \xrightarrow{p}
    (\mu_0,\nu_0),
    \quad l=1,\dots,K.
\]
Therefore, Lemma~\ref{supp:lem:gfm-local-ulln-differentiation}, applied to
this finite collection of intermediate points, gives
\[
    \tilde J_n
    =
    \mathbb E[
        \nabla_{\mu\mu}^2
        \tilde{\ell}(X,\mu_0,\nu_0)
    ]
    +
    o_p(1),
    \quad
    \tilde C_n
    =
    C+o_p(1),
\]
where
\[
    C
    :=
    \mathbb E[
        \nabla_{\mu\nu}^2
        \tilde{\ell}(X,\mu_0,\nu_0)
    ].
\]
Furthermore, dividing the asymptotic linear representation of the auxiliary parameter from Condition~\ref{supp:cond:gfm-auxiliary-linearity} by $\sqrt{n}$ yields 
$$
    \hat{\nu}_n - \nu_0 = \frac{1}{n} \sum_{i=1}^n \xi(X_i) + o_p(n^{-1/2}).
$$

We now substitute these probability limits back into \eqref{supp:eq:gfm-score-mvt}. By the Central Limit Theorem, $\frac{1}{n} \sum_{i=1}^n \xi(X_i) = O_p(n^{-1/2})$. Consequently, any cross-product between an $o_p(1)$ term and this sample average becomes $o_p(n^{-1/2})$. Applying these asymptotic orders allows us to simplify the expansion as follows:
\begin{align}
\label{supp:eq:gfm-score-first-order-expansion}
    & \frac{1}{n} \sum_{i=1}^n \nabla_\mu \tilde{\ell}(X_i, \hat{\mu}_n, \hat{\nu}_n) \nonumber \\
    &= \frac{1}{n} \sum_{i=1}^n \nabla_\mu \tilde{\ell}(X_i, \mu_0, \nu_0) + \tilde{J}_n (\hat{\mu}_n - \mu_0) + \tilde{C}_n (\hat{\nu}_n - \nu_0) \nonumber \\
    &= \frac{1}{n} \sum_{i=1}^n \nabla_\mu \tilde{\ell}(X_i, \mu_0, \nu_0) + \Big( \mathbb{E}[\nabla_{\mu\mu}^2 \tilde{\ell}(X, \mu_0, \nu_0)] + o_p(1) \Big) (\hat{\mu}_n - \mu_0) \nonumber\\
    & \quad +  \Big( C + o_p(1) \Big) \left( \frac{1}{n} \sum_{i=1}^n \xi(X_i) + o_p(n^{-1/2}) \right) \nonumber \\
    &= \frac{1}{n} \sum_{i=1}^n \nabla_\mu \tilde{\ell}(X_i, \mu_0, \nu_0) + \Big( \mathbb{E}[\nabla_{\mu\mu}^2 \tilde{\ell}(X, \mu_0, \nu_0)] + o_p(1) \Big) (\hat{\mu}_n - \mu_0) \nonumber \\
    &\quad + \frac{1}{n} \sum_{i=1}^n C \xi(X_i) + \underbrace{C \cdot o_p(n^{-1/2})}_{= o_p(n^{-1/2})} + \underbrace{o_p(1) \left( \frac{1}{n} \sum_{i=1}^n \xi(X_i) \right)}_{= o_p(1)O_p(n^{-1/2}) = o_p(n^{-1/2})} + \underbrace{o_p(1) o_p(n^{-1/2})}_{= o_p(n^{-1/2})} \nonumber \\
    &= \frac{1}{n} \sum_{i=1}^n \Big( \nabla_\mu \tilde{\ell}(X_i, \mu_0, \nu_0) + C \xi(X_i) \Big) + \Big( \mathbb{E}[\nabla_{\mu\mu}^2 \tilde{\ell}(X, \mu_0, \nu_0)] + o_p(1) \Big) (\hat{\mu}_n - \mu_0) + o_p(n^{-1/2}).
\end{align}

\textbf{Step 2: Consistency of \(\hat{\lambda}_n\) and expansion of
\(D h(\hat{\mu}_n)^\top\hat{\lambda}_n\).}

We first show that
\( \hat{\lambda}_n \xrightarrow{p} \lambda_0 . \) Let \( \hat{H}_n := D h(\hat{\mu}_n)^\top \in \mathbb{R}^{K\times P}. \) Since \(\hat{\mu}_n \xrightarrow{p} \mu_0\) and the true Jacobian
\(H=Dh(\mu_0)^\top\) has full column rank (Condition~\ref{supp:cond:gfm-geometric-constraint}), \(\hat{H}_n\) also has full
column rank with probability tending to one. Hence, on an event whose
probability tends to one, we may multiply both sides of
\eqref{supp:eq:gfm-empirical-foc} from the left by the left inverse
\((\hat{H}_n^\top\hat{H}_n)^{-1}\hat{H}_n^\top\). This gives
\begin{equation}
\label{supp:eq:gfm-lambda-hat-left-inverse}
    \hat{\lambda}_n
    =
    -
    (\hat{H}_n^\top \hat{H}_n)^{-1}\hat{H}_n^\top
    \left(
        \frac{1}{n}\sum_{i=1}^n
        \nabla_\mu \tilde{\ell}(X_i,\hat{\mu}_n,\hat{\nu}_n)
    \right).
\end{equation}

Similarly, applying the corresponding left inverse to the population
first-order condition \eqref{supp:eq:gfm-population-foc} yields
\begin{equation}
\label{supp:eq:gfm-lambda-zero-left-inverse}
    \lambda_0
    =
    -
    (H^\top H)^{-1}H^\top
    \mathbb{E}[\nabla_\mu \tilde{\ell}(X,\mu_0,\nu_0)].
\end{equation}

By the consistency of
\((\widehat\mu_n,\widehat\nu_n)\) and Lemma~\ref{supp:lem:gfm-local-ulln-differentiation}.(iii),
\(
    (\hat\mu_n,\hat\nu_n)
    \xrightarrow{p}
    (\mu_0,\nu_0).
\)
Since \(\mathcal K\) is a neighborhood of
\((\mu_0,\nu_0)\),
\[
    \mathcal P\bigl(
        (\hat\mu_n,\hat\nu_n)\in\mathcal K
    \bigr)
    \longrightarrow1.
\]
Therefore, Lemma~\ref{supp:lem:gfm-local-ulln-differentiation} yields
\begin{equation}
\label{supp:eq:gfm-empirical-score-convergence}
    \frac{1}{n}
    \sum_{i=1}^n
    \nabla_\mu
    \tilde{\ell}(X_i,\hat{\mu}_n,\hat{\nu}_n)
    \xrightarrow{p}
    \mathbb{E}[
        \nabla_\mu
        \tilde{\ell}(X,\mu_0,\nu_0)
    ].
\end{equation}
Moreover, by the continuous mapping theorem,
\(
    \hat{H}_n \xrightarrow{p} H,
\)
and therefore
\(
    (\hat{H}_n^\top\hat{H}_n)^{-1}\hat{H}_n^\top
    \xrightarrow{p}
    (H^\top H)^{-1}H^\top .
\)
Applying Slutsky's theorem to \eqref{supp:eq:gfm-lambda-hat-left-inverse} and
\eqref{supp:eq:gfm-empirical-score-convergence}, and comparing with \eqref{supp:eq:gfm-lambda-zero-left-inverse}, we obtain
\begin{equation}
\label{supp:eq:gfm-lambda-consistency}
    \hat{\lambda}_n
    =
    \lambda_0 + o_p(1).
\end{equation}

Using \eqref{supp:eq:gfm-lambda-consistency}, we expand the Lagrange multiplier term
\(Dh(\hat{\mu}_n)^\top \hat{\lambda}_n\) from the empirical first-order
condition. By definition, this term can be written as a linear combination of
the gradients of the individual constraints:
\begin{equation}
\label{supp:eq:gfm-multiplier-sum}
    Dh(\hat{\mu}_n)^\top \hat{\lambda}_n
    =
    \sum_{j=1}^{P}(\hat{\lambda}_n)_j \nabla h_j(\hat{\mu}_n).
\end{equation}
Since each \(\nabla h_j(\mu)\) is a \(K\)-dimensional vector-valued function, we
apply the mean value theorem componentwise. For each constraint
\(j\in\{1,\dots,P\}\) and each component \(l\in\{1,\dots,K\}\), there exists an
intermediate point \(\tilde{\mu}_{n,j,l}'\) on the line segment between
\(\mu_0\) and \(\hat{\mu}_n\) such that
\begin{equation}
\label{supp:eq:gfm-constraint-gradient-mvt}
    \nabla h_j(\hat{\mu}_n)
    =
    \nabla h_j(\mu_0)
    +
    A_{n,j}(\hat{\mu}_n-\mu_0),
\end{equation}
where \(A_{n,j}\in\mathbb{R}^{K\times K}\) is defined row-wise by
\[
    [A_{n,j}]_{l,\cdot}
    =
    [\nabla^2 h_j(\tilde{\mu}_{n,j,l}')]_{l,\cdot},
    \quad l=1,\dots,K .
\]

Substituting \eqref{supp:eq:gfm-constraint-gradient-mvt} into \eqref{supp:eq:gfm-multiplier-sum}, we obtain
\begin{align}
\label{supp:eq:gfm-multiplier-taylor-explicit}
    Dh(\hat{\mu}_n)^\top \hat{\lambda}_n
    &=
    \sum_{j=1}^{P}
    (\hat{\lambda}_n)_j
    \Big(
        \nabla h_j(\mu_0)
        +
        A_{n,j}(\hat{\mu}_n-\mu_0)
    \Big) \nonumber \\
    &=
    \underbrace{
        \sum_{j=1}^{P}(\hat{\lambda}_n)_j\nabla h_j(\mu_0)
    }_{=H\hat{\lambda}_n}
    +
    \left(
        \sum_{j=1}^{P}(\hat{\lambda}_n)_j A_{n,j}
    \right)
    (\hat{\mu}_n-\mu_0) \nonumber \\
    &=
    H\lambda_0
    +
    H(\hat{\lambda}_n-\lambda_0)
    +
    \left(
        \sum_{j=1}^{P}(\hat{\lambda}_n)_j A_{n,j}
    \right)
    (\hat{\mu}_n-\mu_0).
\end{align}
By the consistency of the primary estimator,
\(\hat{\mu}_n\xrightarrow{p}\mu_0\), each intermediate point satisfies
\(\tilde{\mu}_{n,j,l}'\xrightarrow{p}\mu_0\). Hence, by the continuity of the
second derivatives of \(h_j\),
\[
    A_{n,j}
    \xrightarrow{p}
    \nabla^2 h_j(\mu_0),
    \quad j=1,\dots,P.
\]
Combined with \(\hat{\lambda}_n\xrightarrow{p}\lambda_0\), this gives
\[
    \sum_{j=1}^{P}(\hat{\lambda}_n)_j A_{n,j}
    =
    \sum_{j=1}^{P}(\lambda_0)_j \nabla^2 h_j(\mu_0)
    +
    o_p(1).
\]
Thus, we get
\begin{equation}
\label{supp:eq:gfm-multiplier-first-order-expansion}
    Dh(\hat{\mu}_n)^\top \hat{\lambda}_n
    =
    H\lambda_0
    +
    H(\hat{\lambda}_n-\lambda_0)
    +
    \left(
        \sum_{j=1}^{P}(\lambda_0)_j\nabla^2 h_j(\mu_0)
        +
        o_p(1)
    \right)
    (\hat{\mu}_n-\mu_0).
\end{equation}

\textbf{Step 3: Taylor expansion of the feasibility condition \(h(\hat{\mu}_n)=0\).}

The empirical first-order condition \eqref{supp:eq:gfm-empirical-foc} gives only the
stationarity part of the KKT system. To obtain the full linearized KKT system,
we must also expand the feasibility condition \(h(\hat{\mu}_n)=0\), which gives
the tangent-space constraint for the first-order variation of \(\hat{\mu}_n\).

We now expand the geometric constraint \(h(\hat{\mu}_n)=0\) around \(\mu_0\). Because $h$ is a vector-valued function taking values in $\mathbb{R}^{P}$, we apply the multivariate mean value theorem componentwise. For each $j \in \{1, \dots, P\}$, there exists an intermediate point $\tilde{\mu}_{n,j}$ on the line segment between $\mu_0$ and $\hat{\mu}_n$ such that:
\begin{equation}
\label{supp:eq:gfm-constraint-mvt-component}
    0 = h_j(\hat{\mu}_n) = h_j(\mu_0) + \nabla h_j(\tilde{\mu}_{n,j})^\top (\hat{\mu}_n - \mu_0) = \nabla h_j(\tilde{\mu}_{n,j})^\top (\hat{\mu}_n - \mu_0).
\end{equation}
Stacking these $P$ equations yields $0 = \tilde{H}_n^\top (\hat{\mu}_n - \mu_0)$, where the matrix $$\tilde{H}_n := [\nabla h_1(\tilde{\mu}_{n,1}), \dots, \nabla h_{P}(\tilde{\mu}_{n,P})] \in \mathbb{R}^{K \times P}.$$ Since $h(\mu_0) = 0$ and the consistency $\hat{\mu}_n \xrightarrow{p} \mu_0$ implies $\tilde{\mu}_{n,j} \xrightarrow{p} \mu_0$ for all $j=1, \dots, P$, we have $\tilde{H}_n^\top = H^\top + o_p(1)$. Multiplying by $\sqrt{n}$ gives:
\begin{equation}
\label{supp:eq:gfm-feasibility-first-order-expansion}
    (H^\top + o_p(1)) \sqrt{n}(\hat{\mu}_n - \mu_0) = 0.
\end{equation}

\textbf{Step 4: Linearized stationarity equation and block KKT system.}
Substituting the expansions for the empirical score
\eqref{supp:eq:gfm-score-first-order-expansion} and the Lagrange multiplier term
\eqref{supp:eq:gfm-multiplier-first-order-expansion} into the empirical first-order condition
\eqref{supp:eq:gfm-empirical-foc}, we obtain the following linearized stationarity
equation:
\begin{align*}
0 &= \frac{1}{n} \sum_{i=1}^n \left( \nabla_\mu \tilde{\ell}(X_i, \mu_0, \nu_0) + C \xi(X_i) \right)+ \Big( \mathbb{E}[\nabla_{\mu\mu}^2 \tilde{\ell}(X, \mu_0, \nu_0)] + o_p(1) \Big) (\hat{\mu}_n - \mu_0)  \\
&+  H \lambda_0 + H (\hat{\lambda}_n - \lambda_0) + \Big( \sum_{j=1}^{P} (\lambda_0)_j \nabla^2 h_j(\mu_0) + o_p(1) \Big) (\hat{\mu}_n - \mu_0)  + o_p(n^{-1/2}).
\end{align*}

Using \eqref{supp:eq:gfm-population-foc}, i.e., 
$H \lambda_0 = -\mathbb{E}[\nabla_\mu \tilde{\ell}(X, \mu_0, \nu_0)]$ , we get 
\begin{align*}
&\Big( \mathbb{E}[\nabla_{\mu\mu}^2 \tilde{\ell}(X, \mu_0, \nu_0)] + \sum_{j=1}^{P} (\lambda_0)_j \nabla^2 h_j(\mu_0) + o_p(1) \Big) (\hat{\mu}_n - \mu_0) + H (\hat{\lambda}_n - \lambda_0) \\
&= -\left[ \frac{1}{n} \sum_{i=1}^n \nabla_\mu \tilde{\ell}(X_i, \mu_0, \nu_0) - \mathbb{E}[\nabla_\mu \tilde{\ell}(X, \mu_0, \nu_0)] + \frac{1}{n} \sum_{i=1}^n C \xi(X_i) \right] + o_p(n^{-1/2}).
\end{align*}

Consequently, the combined first-order conditions yield the following linearized system
\begin{equation}
\label{supp:eq:gfm-linearized-stationarity}
    (J+o_p(1))(\hat{\mu}_n-\mu_0)
    +
    H(\hat{\lambda}_n-\lambda_0)
    =
    -\frac1n\sum_{i=1}^n Z(X_i)
    +
    o_p(n^{-1/2}),
\end{equation}
Here \(J\) is defined in \eqref{supp:eq:gfm-hessian-J}, and
\begin{equation}
\label{supp:eq:gfm-influence-vector}
    Z(X_i)
    =
    \nabla_\mu\tilde{\ell}(X_i,\mu_0,\nu_0)
    -
    \mathbb E[
        \nabla_\mu\tilde{\ell}(X,\mu_0,\nu_0)
    ]
    +
    C\xi(X_i)
\end{equation}
is the effective score in \eqref{eq: effective_score},
evaluated at \(X_i\), where \(
C:=\mathbb E[
    \nabla_{\mu\nu}^2
    \tilde{\ell}(X,\mu_0,\nu_0)]\).
   
Note that by construction, $\mathbb{E}[Z(X_{i})] = 0$. Multiplying \eqref{supp:eq:gfm-linearized-stationarity} by \(\sqrt n\) and stacking it with the linearized feasibility equation \eqref{supp:eq:gfm-feasibility-first-order-expansion} gives the
linearized KKT system
\begin{equation}
\label{supp:eq:gfm-linearized-kkt-system}
    \underbrace{\begin{pmatrix} J + o_p(1) & H \\ H^\top + o_p(1) & 0 \end{pmatrix}}_{=: \hat{B}_n}
    \begin{pmatrix} \sqrt{n}(\hat{\mu}_n - \mu_0) \\ \sqrt{n}(\hat{\lambda}_n - \lambda_0) \end{pmatrix} 
    = \begin{pmatrix} -\frac{1}{\sqrt{n}} \sum_{i=1}^n Z(X_i) \\ 0 \end{pmatrix} + o_p(1).
\end{equation}

\textbf{Step 5: Derive the first-order stochastic expansion and the central limit theorem.}

By Condition~\ref{supp:cond:gfm-kkt-invertibility},
\(B\) is nonsingular. Since \(\widehat B_n\xrightarrow{p}B\), \(\widehat B_n\) is invertible with probability tending to one, and the continuous mapping theorem gives
\[
    \widehat B_n^{-1}=B^{-1}+o_p(1).
\]

Multiplying both sides of \eqref{supp:eq:gfm-linearized-kkt-system} by \(\hat B_n^{-1}\) gives
\[
    \begin{pmatrix}
        \sqrt n(\hat\mu_n-\mu_0)\\
        \sqrt n(\hat\lambda_n-\lambda_0)
    \end{pmatrix}
    =
    -\hat B_n^{-1}
    \begin{pmatrix}
        \frac1{\sqrt n}\sum_{i=1}^n Z(X_i)\\
        0
    \end{pmatrix}
    +o_p(1).
\]
Since \(Z(X_i)\) has a finite second moment,
\[
    \frac1{\sqrt n}\sum_{i=1}^nZ(X_i)=O_p(1).
\]
Therefore,
\[
    \begin{pmatrix}
        \sqrt n(\hat\mu_n-\mu_0)\\
        \sqrt n(\hat\lambda_n-\lambda_0)
    \end{pmatrix}
    =
    -B^{-1}
    \begin{pmatrix}
        \frac1{\sqrt n}\sum_{i=1}^n Z(X_i)\\
        0
    \end{pmatrix}
    +o_p(1).
\]
Equivalently,
\[
    \begin{pmatrix}
        \hat\mu_n-\mu_0\\
        \hat\lambda_n-\lambda_0
    \end{pmatrix}
    =
    -B^{-1}
    \begin{pmatrix}
        \frac1n\sum_{i=1}^n Z(X_i)\\
        0
    \end{pmatrix}
    +o_p(n^{-1/2}).
\]
Extracting the first \(K\) coordinates gives
\[
    \hat\mu_n-\mu_0
    =
    -B_{11}\frac1n\sum_{i=1}^{n} Z(X_i)
    +o_p(n^{-1/2}),
\]
which shows \eqref{eq: first_order_expansion}. In particular,
\[
    \hat\mu_n-\mu_0=O_p(n^{-1/2}),
    \quad
    \hat\lambda_n-\lambda_0=O_p(n^{-1/2}).
\]

It remains to derive the limiting distribution. Since \(Z(X_i)\) are i.i.d., centered, and have covariance matrix \(V=\operatorname{Var}(Z)\), the multivariate Central Limit Theorem gives
\[
    \frac1{\sqrt n}\sum_{i=1}^nZ(X_i)
    \xrightarrow{d}
    \mathcal N(0,V).
\]
Applying Slutsky's Theorem to the preceding joint stochastic expansion yields
\begin{equation}
\label{supp:eq:gfm-joint-clt}
    \begin{pmatrix}
        \sqrt n(\hat\mu_n-\mu_0)\\
        \sqrt n(\hat\lambda_n-\lambda_0)
    \end{pmatrix}
    \xrightarrow{d}
    \mathcal N\left(
        0,\,
        B^{-1}
        \begin{pmatrix}
            V&0\\
            0&0
        \end{pmatrix}
        (B^{-1})^\top
    \right).
\end{equation}

The asymptotic covariance matrix of the first \(K\) coordinates is \(B_{11}VB_{11}^\top\). Since \(B\), and hence \(B_{11}\), is symmetric, this equals \(B_{11}VB_{11}\). Therefore,
\(
    B_{11}VB_{11}^\top=B_{11}VB_{11}.
\)
Consequently,
\begin{equation}
\label{supp:eq:gfm-primary-clt}
    \sqrt n(\hat\mu_n-\mu_0)
    \xrightarrow{d}
    \mathcal N\left(0,B_{11}VB_{11}\right),
\end{equation}
which shows \eqref{eq: clt_result}.

\textbf{Step 6: Quadratic refinement.} Suppose now, in addition, that Condition~\ref{supp:cond:gfm-quadratic-refinement} holds. The first-order stochastic expansion obtained in Step 5 has already shown that
\[
    \hat{\mu}_n-\mu_0=O_p(n^{-1/2}),
    \quad
    \hat{\lambda}_n-\lambda_0=O_p(n^{-1/2}).
\]
Moreover, \eqref{supp:eq:gfm-auxiliary-quadratic-remainder} in Condition~\ref{supp:cond:gfm-quadratic-refinement} gives
\[
    \hat{\nu}_n-\nu_0=O_p(n^{-1/2}).
\]
Thus, all parameter perturbations appearing in the empirical KKT equations are of order \(O_p(n^{-1/2})\). We now revisit the expansions in Steps 1--3 under the strengthened smoothness and moment assumptions, retaining all terms up to first order and controlling the remaining terms at order \(O_p(n^{-1})\).

First, we revisit Step 1 and expand the empirical gradient \(\frac1n\sum_{i=1}^n
    \nabla_\mu\tilde{\ell}(X_i,\hat{\mu}_n,\hat{\nu}_n)\) around
\((\mu_0,\nu_0)\). For each \(i\), consider the vector-valued function
\[
    t
    \longmapsto
    \nabla_\mu\tilde{\ell}\!\left(
        X_i,\,
        \mu_0+t(\hat{\mu}_n-\mu_0),\,
        \nu_0+t(\hat{\nu}_n-\nu_0)
    \right),
    \quad 0\le t\le1.
\]
Its first derivative with respect to \(t\) is
\begin{align*}
&\nabla_{\mu\mu}^2\tilde{\ell}\!\left(
    X_i,\,
    \mu_0+t(\hat{\mu}_n-\mu_0),\,
    \nu_0+t(\hat{\nu}_n-\nu_0)
\right)
(\hat{\mu}_n-\mu_0)\\
&+
\nabla_{\mu\nu}^2\tilde{\ell}\!\left(
    X_i,\,
    \mu_0+t(\hat{\mu}_n-\mu_0),\,
    \nu_0+t(\hat{\nu}_n-\nu_0)
\right)
(\hat{\nu}_n-\nu_0).
\end{align*}
Applying the one-dimensional Taylor theorem with integral remainder along this path, and then averaging over \(i=1,\dots,n\), we get
\begin{align}
&\frac1n\sum_{i=1}^n
    \nabla_\mu\tilde{\ell}(X_i,\hat{\mu}_n,\hat{\nu}_n)
\nonumber\\
&=
\frac1n\sum_{i=1}^n
    \nabla_\mu\tilde{\ell}(X_i,\mu_0,\nu_0)\nonumber\\
&+
\left\{
    \frac1n\sum_{i=1}^n
    \nabla_{\mu\mu}^2\tilde{\ell}(X_i,\mu_0,\nu_0)
\right\}
(\hat{\mu}_n-\mu_0)\nonumber\\
&+
\left\{
    \frac1n\sum_{i=1}^n
    \nabla_{\mu\nu}^2\tilde{\ell}(X_i,\mu_0,\nu_0)
\right\}
(\hat{\nu}_n-\nu_0)
+
R_n,
\label{supp:eq:gfm-score-quadratic-taylor}
\end{align}
with the Taylor remainder
\begin{align}\label{supp:eq:gfm-score-remainder-integral}
R_n =
\frac1n\sum_{i=1}^n
\int_0^1(1-t)
\Bigg[
\nabla_{\mu\mu\mu}^3\tilde{\ell}\!\left(
    X_i,\,
    \mu_0+t(\hat{\mu}_n-\mu_0),\,
    \nu_0+t(\hat{\nu}_n-\nu_0)
\right)
[\hat{\mu}_n-\mu_0,\hat{\mu}_n-\mu_0]
\nonumber\\
\quad+
2\nabla_{\mu\mu\nu}^3\tilde{\ell}\!\left(
    X_i,\,
    \mu_0+t(\hat{\mu}_n-\mu_0),\,
    \nu_0+t(\hat{\nu}_n-\nu_0)
\right)
[\hat{\mu}_n-\mu_0,\hat{\nu}_n-\nu_0]
\nonumber\\
\quad +\nabla_{\mu\nu\nu}^3\tilde{\ell}\!\left(
    X_i,\,
    \mu_0+t(\hat{\mu}_n-\mu_0),\,
    \nu_0+t(\hat{\nu}_n-\nu_0)
\right)
[\hat{\nu}_n-\nu_0,\hat{\nu}_n-\nu_0]
\Bigg]dt .
\end{align}
Here, for example,
\[
    \nabla_{\mu\mu\nu}^3\tilde{\ell}(X,\mu,\nu)[u,v]
\]
denotes the contraction of the third derivative tensor with
\(u\in\mathbb R^K\) in its \(\mu\)-direction and
\(v\in\mathbb R^{D_Q}\) in its \(\nu\)-direction. The factor \(2\) in the mixed term arises from the two cross-derivative terms in the second derivative along the path.

Since \(\mathcal K\) is convex and
\[
    \mathcal P\bigl(
        (\hat{\mu}_n,\hat{\nu}_n)\in\mathcal K
    \bigr)
    \longrightarrow1,
\]
the entire line segment between
\((\mu_0,\nu_0)\) and
\((\hat{\mu}_n,\hat{\nu}_n)\) belongs to \(\mathcal K\) with probability tending to one. On this event, Condition~\ref{supp:cond:gfm-quadratic-refinement} and
\eqref{supp:eq:gfm-third-derivative-dominance} imply
\begin{align*}
&\Big\|
\nabla_{\mu\mu\mu}^3\tilde{\ell}(X_i,\mu,\nu)
[\hat{\mu}_n-\mu_0,\hat{\mu}_n-\mu_0]
\Big\|\le
D_3(X_i)\|\hat{\mu}_n-\mu_0\|^2,
\end{align*}
\begin{align*}
&\Big\|
\nabla_{\mu\mu\nu}^3\tilde{\ell}(X_i,\mu,\nu)
[\hat{\mu}_n-\mu_0,\hat{\nu}_n-\nu_0]
\Big\|\le
D_3(X_i)
\|\hat{\mu}_n-\mu_0\|
\|\hat{\nu}_n-\nu_0\|,
\end{align*}
and
\begin{align*}
&\Big\|
\nabla_{\mu\nu\nu}^3\tilde{\ell}(X_i,\mu,\nu)
[\hat{\nu}_n-\nu_0,\hat{\nu}_n-\nu_0]
\Big\|\le
D_3(X_i)\|\hat{\nu}_n-\nu_0\|^2
\end{align*}
uniformly over \((\mu,\nu)\in\mathcal K\). Therefore,
\begin{align*}
\|R_n\|
&\le
\frac1n\sum_{i=1}^n
\int_0^1(1-t)D_3(X_i)
\Bigl\{
    \|\hat{\mu}_n-\mu_0\|^2
    +2\|\hat{\mu}_n-\mu_0\|
       \|\hat{\nu}_n-\nu_0\|
    +\|\hat{\nu}_n-\nu_0\|^2
\Bigr\}dt\\
&=
\frac12
\left\{
    \frac1n\sum_{i=1}^nD_3(X_i)
\right\}
\left(
    \|\hat{\mu}_n-\mu_0\|
    +
    \|\hat{\nu}_n-\nu_0\|
\right)^2.
\end{align*}

Since \(D_3(X)\) is integrable by Condition~\ref{supp:cond:gfm-quadratic-refinement}, the law of large numbers gives
\[
    \frac1n\sum_{i=1}^nD_3(X_i)
    \xrightarrow{p}
    \mathbb E[D_3(X)],
\]
and hence
\(
    \frac1n\sum_{i=1}^nD_3(X_i)=O_p(1).
\)
Combining this with 
\(
    \hat{\mu}_n-\mu_0=O_p(n^{-1/2}) \text{ and }
    \hat{\nu}_n-\nu_0=O_p(n^{-1/2}),
\)
we get
\begin{equation}
    \|R_n\|=O_p(n^{-1}).
    \label{supp:eq:gfm-score-quadratic-remainder}
\end{equation}

Moreover, the Hessian and cross-Hessian are dominated by \(D(X)\), and
\(\mathbb E[D(X)^2]<\infty\). Applying Chebyshev's inequality to their finitely many coordinates, we get
\[
    \frac1n\sum_{i=1}^n
    \nabla_{\mu\mu}^2\tilde{\ell}(X_i,\mu_0,\nu_0)
    =
    \mathbb E[
        \nabla_{\mu\mu}^2\tilde{\ell}(X,\mu_0,\nu_0)
    ]
    +
    O_p(n^{-1/2})
\]
and
\[
    \frac1n\sum_{i=1}^n
    \nabla_{\mu\nu}^2\tilde{\ell}(X_i,\mu_0,\nu_0)
    =
    C+O_p(n^{-1/2}).
\]
Furthermore, Condition~\ref{supp:cond:gfm-quadratic-refinement} gives
\(
    \hat{\nu}_n-\nu_0
    =
    \frac1n\sum_{i=1}^n\xi(X_i)
    +
    O_p(n^{-1}).
\)
As in Step 1, the Central Limit Theorem gives
\(
    \frac1n\sum_{i=1}^n\xi(X_i)=O_p(n^{-1/2}).
\)
We now substitute these expansions into
\eqref{supp:eq:gfm-score-quadratic-taylor} and get a sharper version of \eqref{supp:eq:gfm-score-first-order-expansion} as follows.
\begin{align}
&\frac1n\sum_{i=1}^n
    \nabla_\mu\tilde{\ell}(X_i,\hat{\mu}_n,\hat{\nu}_n)
\nonumber\\
&=
\frac1n\sum_{i=1}^n
    \nabla_\mu\tilde{\ell}(X_i,\mu_0,\nu_0)
+
\left\{
    \mathbb E[
        \nabla_{\mu\mu}^2\tilde{\ell}(X,\mu_0,\nu_0)
    ]
    +
    O_p(n^{-1/2})
\right\}
(\hat{\mu}_n-\mu_0)
\nonumber\\
&\quad+
\left\{
    C+O_p(n^{-1/2})
\right\}
\left\{
    \frac1n\sum_{i=1}^n\xi(X_i)
    +
    O_p(n^{-1})
\right\}
+
O_p(n^{-1})
\nonumber\\
&=
\frac1n\sum_{i=1}^n
    \left\{
        \nabla_\mu\tilde{\ell}(X_i,\mu_0,\nu_0)
        +
        C\xi(X_i)
    \right\}
+
\mathbb E[
    \nabla_{\mu\mu}^2\tilde{\ell}(X,\mu_0,\nu_0)
]
(\hat{\mu}_n-\mu_0)
\nonumber\\
&\quad+
\underbrace{
    O_p(n^{-1/2})(\hat{\mu}_n-\mu_0)
}_{=O_p(n^{-1})}
+
\underbrace{
    O_p(n^{-1/2})
    \left(
        \frac1n\sum_{i=1}^n\xi(X_i)
    \right)
}_{=O_p(n^{-1})}
\nonumber\\
&\quad+
\underbrace{
    C O_p(n^{-1})
}_{=O_p(n^{-1})}
+
\underbrace{
    O_p(n^{-1/2})O_p(n^{-1})
}_{=O_p(n^{-3/2})}
+
O_p(n^{-1})
\nonumber\\
&=
\frac1n\sum_{i=1}^n
    \left\{
        \nabla_\mu\tilde{\ell}(X_i,\mu_0,\nu_0)
        +
        C\xi(X_i)
    \right\}
+
\mathbb E[
    \nabla_{\mu\mu}^2\tilde{\ell}(X,\mu_0,\nu_0)
]
(\hat{\mu}_n-\mu_0)
+
O_p(n^{-1}).
\label{supp:eq:gfm-score-quadratic-expansion}
\end{align}

Secondly, we revisit Step 2. Fix \(j\in\{1,\dots,P\}\). Since \(h_j: \mathcal{U}\to\mathbb R\), its gradient \(\nabla h_j\) is a vector-valued function from \(\mathcal{U}\) to \(\mathbb R^K\). Consider the one-dimensional path
\[
    t\longmapsto
    \nabla h_j\!\left(
        \mu_0+t(\hat{\mu}_n-\mu_0)
    \right),
    \quad 0\le t\le1.
\]
Its first and second derivatives with respect to \(t\) are
\[
    \nabla^2h_j\!\left(
        \mu_0+t(\hat{\mu}_n-\mu_0)
    \right)
    (\hat{\mu}_n-\mu_0)
\]
and
\[
    \nabla^3h_j\!\left(
        \mu_0+t(\hat{\mu}_n-\mu_0)
    \right)
    [\hat{\mu}_n-\mu_0,\hat{\mu}_n-\mu_0],
\]
respectively. Here,
\(\nabla^3h_j(\mu)[u,v]\in\mathbb R^K\) denotes the contraction of the third derivative tensor with the two vectors \(u\) and \(v\).

Applying the one-dimensional Taylor formula with integral remainder to this path gives
\begin{align}
    \nabla h_j(\hat{\mu}_n)
    &=
    \nabla h_j(\mu_0)
    +
    \nabla^2h_j(\mu_0)(\hat{\mu}_n-\mu_0)
    \nonumber\\
    &\quad+
    \int_0^1(1-t)
    \nabla^3h_j\!\left(
        \mu_0+t(\hat{\mu}_n-\mu_0)
    \right)
    [\hat{\mu}_n-\mu_0,\hat{\mu}_n-\mu_0]
    \,dt .
    \label{supp:eq:gfm-constraint-gradient-quadratic-taylor}
\end{align}

We now control the integral remainder. By Condition~\ref{supp:cond:gfm-quadratic-refinement}, the third derivative of \(h_j\) is locally bounded around \(\mu_0\). Hence, there exist \(r>0\) and a finite constant \(M_j>0\) such that
\[
    \sup_{\|\mu-\mu_0\|\le r}
    \|\nabla^3h_j(\mu)\|_{\mathrm{op}}
    \le M_j.
\]
Since \(\hat{\mu}_n\xrightarrow{p}\mu_0\) by Condition~\ref{supp:cond:gfm-primary-consistency},
\[
    \mathcal P\left(
        \|\hat{\mu}_n-\mu_0\|\le r
    \right)
    \longrightarrow1.
\]
On this event, the entire line segment
\[
    \left\{
        \mu_0+t(\hat{\mu}_n-\mu_0):
        0\le t\le1
    \right\}
\]
is contained in the ball
\(\{\mu:\|\mu-\mu_0\|\le r\}\). Therefore,
\begin{align*}
&\left\|
    \int_0^1(1-t)
    \nabla^3h_j\!\left(
        \mu_0+t(\hat{\mu}_n-\mu_0)
    \right)
    [\hat{\mu}_n-\mu_0,\hat{\mu}_n-\mu_0]
    \,dt
\right\|\\
&\quad\le
\int_0^1(1-t)
\left\|
    \nabla^3h_j\!\left(
        \mu_0+t(\hat{\mu}_n-\mu_0)
    \right)
    [\hat{\mu}_n-\mu_0,\hat{\mu}_n-\mu_0]
\right\|dt\\
&\quad\le
\int_0^1(1-t)
\left\|
    \nabla^3h_j\!\left(
        \mu_0+t(\hat{\mu}_n-\mu_0)
    \right)
\right\|_{\mathrm{op}}
\|\hat{\mu}_n-\mu_0\|^2dt\\
&\quad\le
M_j\|\hat{\mu}_n-\mu_0\|^2
\int_0^1(1-t)\,dt\\
&\quad=
\frac{M_j}{2}
\|\hat{\mu}_n-\mu_0\|^2.
\end{align*}

Since we have \(
    \|\hat{\mu}_n-\mu_0\|^2=O_p(n^{-1}),
\)
it follows that the integral remainder in
\eqref{supp:eq:gfm-constraint-gradient-quadratic-taylor} is \(O_p(n^{-1})\). Hence,
\begin{equation}
    \nabla h_j(\hat{\mu}_n)
    =
    \nabla h_j(\mu_0)
    +
    \nabla^2h_j(\mu_0)(\hat{\mu}_n-\mu_0)
    +
    O_p(n^{-1}).
    \label{supp:eq:gfm-constraint-gradient-quadratic-expansion}
\end{equation}

We now substitute \eqref{supp:eq:gfm-constraint-gradient-quadratic-expansion} into
\eqref{supp:eq:gfm-multiplier-sum}. Since \(P\) is fixed,
\begin{align*}
    Dh(\hat{\mu}_n)^\top\hat{\lambda}_n
    &=
    \sum_{j=1}^P
    (\hat{\lambda}_n)_j
    \nabla h_j(\hat{\mu}_n)\\
    &=
    \sum_{j=1}^P
    (\hat{\lambda}_n)_j
    \nabla h_j(\mu_0)+
    \sum_{j=1}^P
    (\hat{\lambda}_n)_j
    \nabla^2h_j(\mu_0)
    (\hat{\mu}_n-\mu_0)
    +
    O_p(n^{-1}).
\end{align*}
The last remainder remains \(O_p(n^{-1})\) because
\[
    \hat{\lambda}_n
    =
    \lambda_0+O_p(n^{-1/2})
    =
    O_p(1).
\]

The first sum satisfies
\[
    \sum_{j=1}^P
    (\hat{\lambda}_n)_j\nabla h_j(\mu_0)
    =
    H\hat{\lambda}_n
    =
    H\lambda_0
    +
    H(\hat{\lambda}_n-\lambda_0).
\]
For the second sum, write
\begin{align*}
&\sum_{j=1}^P
    (\hat{\lambda}_n)_j
    \nabla^2h_j(\mu_0)
    (\hat{\mu}_n-\mu_0)\\
&=
\left\{
    \sum_{j=1}^P
    (\lambda_0)_j
    \nabla^2h_j(\mu_0)
\right\}
(\hat{\mu}_n-\mu_0)+
\sum_{j=1}^P
\bigl\{
    (\hat{\lambda}_n)_j-(\lambda_0)_j
\bigr\}
\nabla^2h_j(\mu_0)
(\hat{\mu}_n-\mu_0).
\end{align*}
Since \(P\) is fixed and the matrices
\(\nabla^2h_j(\mu_0)\) are fixed,
\begin{align*}
&\left\|
\sum_{j=1}^P
\bigl\{
    (\hat{\lambda}_n)_j-(\lambda_0)_j
\bigr\}
\nabla^2h_j(\mu_0)
(\hat{\mu}_n-\mu_0)
\right\|\le
c
\|\hat{\lambda}_n-\lambda_0\|
\|\hat{\mu}_n-\mu_0\|
\end{align*}
for some finite constant \(c>0\). Since
\[
    \hat{\lambda}_n-\lambda_0=O_p(n^{-1/2}),
    \quad
    \hat{\mu}_n-\mu_0=O_p(n^{-1/2}),
\]
this cross-product is \(O_p(n^{-1})\). Combining the preceding expansions, we obtain
\begin{equation}
\label{supp:eq:gfm-multiplier-quadratic-expansion}    
Dh(\hat{\mu}_n)^\top\hat{\lambda}_n =
    H\lambda_0
    +
    H(\hat{\lambda}_n-\lambda_0)+
    \left\{
        \sum_{j=1}^P
        (\lambda_0)_j
        \nabla^2h_j(\mu_0)
    \right\}
    (\hat{\mu}_n-\mu_0)
    +
    O_p(n^{-1}), 
\end{equation}
which is a strengthened version of \eqref{supp:eq:gfm-multiplier-first-order-expansion}.

Thirdly, we revisit Step 3. A second-order Taylor expansion of the feasibility
condition gives
\[
    0
    =
    h(\hat{\mu}_n)-h(\mu_0)
    =
    H^\top(\hat{\mu}_n-\mu_0)
    +
    O_p\!\left(
        \|\hat{\mu}_n-\mu_0\|^2
    \right),
\]
and therefore we get
\begin{equation}
\label{supp:eq:gfm-feasibility-quadratic-expansion}
H^\top(\hat{\mu}_n-\mu_0)=O_p(n^{-1}),
\end{equation}
which is a strengthened version of \eqref{supp:eq:gfm-feasibility-first-order-expansion}.

Finally, substituting the refined empirical-gradient expansion
\eqref{supp:eq:gfm-score-quadratic-expansion}
and the expansion of the Lagrange multiplier term
\eqref{supp:eq:gfm-multiplier-quadratic-expansion}
into \eqref{supp:eq:gfm-empirical-foc}, and using
\eqref{supp:eq:gfm-population-foc}, yields
\[
    J(\hat{\mu}_n-\mu_0)
    +
    H(\hat{\lambda}_n-\lambda_0)
    =
    -\frac1n\sum_{i=1}^{n} Z(X_i)
    +
    O_p(n^{-1}).
\]
Combining this with the refined feasibility equation \eqref{supp:eq:gfm-feasibility-quadratic-expansion} gives
\[
    B
    \begin{pmatrix}
        \hat{\mu}_n-\mu_0\\
        \hat{\lambda}_n-\lambda_0
    \end{pmatrix}
    =
    \begin{pmatrix}
        -\frac1n\sum_{i=1}^{n} Z(X_i)\\
        0
    \end{pmatrix}
    +
    O_p(n^{-1}).
\]
Finally, multiplying by \(B^{-1}\) and extracting the first \(K\)
coordinates gives
\[
    \hat{\mu}_n-\mu_0
    =
    -B_{11}\frac1n\sum_{i=1}^{n}Z(X_i)
    +
    O_p(n^{-1}),
\]
which shows \eqref{eq: quadratic_stochastic_expansion}.
\end{proof}

\section{Proofs for the iterated \texttt{plug-in} estimators}
\label{supp:sec:plugin-estimator-proofs}

This subsection proves Theorems~\ref{thm:iterated-plugin-loading-expansion-clt} and~\ref{thm:iterated-plugin-noise-covariance-expansion}. We first derive a first-order loading expansion with an
\(\mathcal O_p(n^{-1})\) remainder using the generalized
Fr\'echet mean framework. We then apply
Lemma~\ref{supp:lem:real-residual-projection-loading-expansion}
to obtain the second-order covariance update. The covariance update preserves the first-order influence function, so we can iterate these arguments for any fixed number of updates.
Throughout this subsection, we write \(L_{\kappa,n}=L_{\kappa,n}^{\texttt{plug-in}}\),
with the latter defined in \eqref{eq:plugin-common-loading-first-order-term}.

\subsection{Strong consistency of iterated \texttt{plug-in} estimator}
\label{supp:sec:plugin-strong-consistency}

This subsection establishes the strong consistency of the \texttt{plug-in} estimator and of the iterated \texttt{plug-in}
estimator after any fixed number of updates. We first collect the identities needed to decompose and analyze the population profile loss. We then show that its unique minimizer at the true noise covariance \(\Sigma^{(0)}\) is \([\kappa^{(0)}]\), and verify the model-specific conditions required to apply the general consistency theorem in Appendix~\ref{sec: GFM}. Finally, starting from the strong consistency of the \texttt{cov}-estimator, we apply this result recursively, with the covariance estimate from the preceding iteration serving as the data-dependent auxiliary parameter. The consistency of each loading estimate is then propagated to the corresponding covariance update through the moment map \(\mathcal M\).

We recall the basic notation. For the \texttt{plug-in} estimator, the data, primary parameter, and auxiliary parameter spaces are
\[
    T=\mathbb C^p,
    \quad
    M=\mathbb{CP}^{p-1},
    \quad
    Q=\mathrm{SPD}(2),
\]
respectively. We equip \(M\) and \(Q\) with the metrics
\[
\begin{aligned}
    d_M\bigl([\kappa],[\widetilde\kappa]\bigr)
    &:=
    d\bigl([\kappa],[\widetilde\kappa]\bigr)
    =
    \sqrt{
        2-2\left|\kappa^*\widetilde\kappa\right|
    },\\
    d_Q(\Sigma_1,\Sigma_2)
    &:=
    \|\Sigma_1-\Sigma_2\|_{\operatorname{op}},
    \quad
    \Sigma_1,\Sigma_2\in\mathrm{SPD}(2).
\end{aligned}
\]
For \(y\in\mathbb C^p\), \([\kappa]\in M\), and
\(\Sigma\in Q\), denote the profiled loss
\[
    \rho(y,[\kappa],\Sigma)
    :=
    \|
        y-\widehat\phi(y,\kappa,\Sigma)\kappa
    \|_{\Sigma}^2,
\]
where \(\widehat\phi(y,\kappa,\Sigma)\) is the conditional minimizer $$\operatorname{argmin}_{\phi\in \mathbb{C}}\|y-\phi\kappa\|_{\Sigma}^{2}.$$ The true auxiliary parameter is
\(q_0=\Sigma^{(0)}\). Given an auxiliary estimator
\(\widehat q_n\), define
\[
    F_n(m,\widehat q_n)
    :=
    \frac1n\sum_{i=1}^n
    \rho(Y_i,m,\widehat q_n),
    \quad
    F(m)
    :=
    \mathbb E\bigl[\rho(Y_1,m,q_0)\bigr].
\]
The \texttt{plug-in} estimator corresponds to
\(\widehat q_n=\widehat\Sigma^{\texttt{cov}}\). Thus, the
\texttt{plug-in} estimator can be seen as a generalized \fre~mean with a random
cost function. 

\subsubsection{Common identities for the population profile loss}
\label{supp:sec:plugin-population-loss-identities}

This subsubsection collects basic identities for the profiled loss that are used throughout the consistency proofs. We first derive an algebraic decomposition of the profile loss, then separate the population objective into signal and noise components, and finally obtain a closed-form expression for the expected noise contribution under an arbitrary covariance matrix.

For $z,w\in\mathbb{C}^p$ and $\Sigma\in\operatorname{SPD}(2)$, we use the real inner product \[\langle z,w\rangle_\Sigma := \sum_{j=1}^{p}\operatorname{vec}(z_j)^\top\Sigma^{-1}\operatorname{vec}(w_j),\] so that $\langle z,z\rangle_\Sigma=\|z\|_\Sigma^2$.
\begin{lemma}
\label{supp:lem:plugin-profile-loss-decomposition}
For all \(t\in\mathbb C^p\),
\(\kappa\in\mathcal S^{2p-1}\), and
\(\Sigma\in\mathrm{SPD}(2)\),
\[
    \rho(t,[\kappa],\Sigma)
    =
    \langle t,t\rangle_\Sigma
    -
    \left\langle
        \widehat\phi(t, \kappa, \Sigma)\kappa,t
    \right\rangle_\Sigma.
\]
\end{lemma}

\begin{proof}[{Proof of Lemma~\ref{supp:lem:plugin-profile-loss-decomposition}}]
By definition of \(\rho\),
\begin{align*}
\rho(t, [\kappa], \Sigma) &= \sum_{j=1}^{p} \left\| \operatorname{vec}(t_{j}) - M(\kappa_{j})\operatorname{vec} \left( \widehat{\phi}(t, \kappa, \Sigma) \right) \right\|_\Sigma^2 \\
&= \langle t, t \rangle_\Sigma - 2 \left\langle \widehat{\phi}(t, \kappa, \Sigma)\kappa, t \right\rangle_\Sigma + \left\langle \widehat{\phi}(t, \kappa, \Sigma)\kappa, \widehat{\phi}(t,\kappa, \Sigma)\kappa \right\rangle_\Sigma.
\end{align*}

Using Lemma~\ref{supp:lem: phi-hat-formula}, we get
\begin{align*}
&\left\langle \widehat{\phi}(t, \kappa, \Sigma)\kappa, \widehat{\phi}(t, \kappa, \Sigma)\kappa \right\rangle_\Sigma \\
& = \sum_{j=1}^{p} \operatorname{vec} \left( \widehat{\phi}(t, \kappa, \Sigma) \right)^{\top} M(\kappa_{j})^{\top} \Sigma^{-1}~ M(\kappa_{j}) \operatorname{vec} \left( \widehat{\phi}(t, \kappa, \Sigma) \right) \\
& = \operatorname{vec} \left( \widehat{\phi}(t, \kappa, \Sigma) \right)^{\top} (\kappa \diamond_{\Sigma} \kappa) ~\operatorname{vec} \left( \widehat{\phi}(t, \kappa, \Sigma) \right) \\
& = \operatorname{vec} \left( \widehat{\phi}(t, \kappa, \Sigma) \right)^{\top} (\kappa \diamond_{\Sigma} \kappa) (\kappa \diamond_{\Sigma} \kappa)^{-1} (\kappa \bullet_\Sigma t) \\
& = \operatorname{vec} \left( \widehat{\phi}(t, \kappa, \Sigma) \right)^{\top} (\kappa \bullet_\Sigma t).
\end{align*}

It follows directly that
\[
    \operatorname{vec}\left(
        \widehat\phi(t, \kappa, \Sigma)
    \right)^\top
    (\kappa\bullet_\Sigma t)
    =
    \left\langle
        \widehat\phi(t, \kappa, \Sigma)\kappa,t
    \right\rangle_\Sigma
\]
and therefore
\[
    \rho(t,[\kappa],\Sigma)
    =
    \langle t,t\rangle_\Sigma
    -
    \left\langle
        \widehat\phi(t, \kappa, \Sigma)\kappa,t
    \right\rangle_\Sigma.
\]
\end{proof}

\begin{lemma}
\label{supp:lem:plugin-population-loss-decomposition}
Assume the model~\eqref{model:centered-complex-factor} with true
parameters
\(\kappa^{(0)}\in\mathcal S^{2p-1}\) and
\(\Sigma^{(0)}\in\mathrm{SPD}(2)\).
For
\(([\kappa],\Sigma)
\in\mathbb{CP}^{p-1}\times\mathrm{SPD}(2)\)
and any representative \(\kappa\in[\kappa]\),
\begin{equation}
\label{supp:eq:plugin-population-loss-decomposition}
\begin{aligned}
&\mathbb E\{\rho(Y_{1},[\kappa],\Sigma)\}
\\
&=\mathbb E\left\{
        \|
            \phi_1\kappa^{(0)}-
            \widehat\phi(\phi_1\kappa^{(0)}, \kappa,\Sigma)\kappa
        \|_{\Sigma}^2
    \right\}+
    \mathbb E\left\{
        \|
            \varepsilon_1-\widehat\phi(\varepsilon_1,\kappa,\Sigma)\kappa
        \|_{\Sigma}^2
    \right\}.
\end{aligned}
\end{equation}
\end{lemma}

\begin{proof}[Proof of Lemma~\ref{supp:lem:plugin-population-loss-decomposition}]
By Lemma~\ref{supp:lem: phi-hat-formula}, the mapping
\[
    t
    \longmapsto
    \widehat\phi(t, \kappa, \Sigma)
\]
is real-linear, and in particular additive. Since
\[
    Y_1
    =
    \phi_1\kappa^{(0)}+\varepsilon_1,
\]
we therefore have
\[
    \widehat\phi(Y_1,\kappa,\Sigma)
    =
    \widehat\phi(
        \phi_1\kappa^{(0)},\kappa,\Sigma
    )
    +
    \widehat\phi(
        \varepsilon_1,\kappa,\Sigma
    ).
\]
Therefore,
\begin{align*}
    Y_1-\widehat\phi(Y_1,\kappa,\Sigma)\kappa &=
    \left(
        \phi_1\kappa^{(0)}
        -
        \widehat\phi(\phi_1\kappa^{(0)}, \kappa,\Sigma)\kappa
    \right)+
    \left(
        \varepsilon_1
        -
        \widehat\phi(\varepsilon_1,\kappa,\Sigma)\kappa
    \right).
\end{align*}
Expanding the squared Mahalanobis norm gives the following
decomposition:
\begin{align*}
    \mathbb E\left[
        \|
            Y_1-\widehat\phi(Y_1,\kappa,\Sigma)\kappa
        \|_{\Sigma}^{2}
    \right] &=
    \mathbb E\left[
            \|\phi_1\kappa^{(0)}-\widehat\phi(\phi_1\kappa^{(0)}, \kappa,\Sigma)\kappa\|_{\Sigma}^{2}
    \right] \\
    &\quad
    +
    \mathbb E\left[
        \|
            \varepsilon_1-\widehat\phi(\varepsilon_1,\kappa,\Sigma)\kappa
        \|_{\Sigma}^{2}
    \right] \\
    &\quad
    +
    2\mathbb E\left[
        \left\langle
            \phi_1\kappa^{(0)}
            -
            \widehat\phi(\phi_1\kappa^{(0)}, \kappa,\Sigma)\kappa,
            \varepsilon_1
            -
            \widehat\phi(\varepsilon_1,\kappa,\Sigma)\kappa
        \right\rangle_\Sigma
    \right].
\end{align*}
Since \(\phi_1\) and \(\varepsilon_1\) are independent
and \(\mathbb E[\varepsilon_1]=0\), the third term vanishes. Thus we
get~\eqref{supp:eq:plugin-population-loss-decomposition}.
\end{proof}

\begin{lemma}
\label{supp:lem:plugin-expected-noise-loss}
Under the model~\eqref{model:centered-complex-factor} with true
parameters
\(\kappa^{(0)}\in\mathcal S^{2p-1}\) and
\(\Sigma^{(0)}\in\mathrm{SPD}(2)\), 
it holds for any \(([\kappa],\Sigma)
\in\mathbb{CP}^{p-1}\times\mathrm{SPD}(2)\)
and any representative \(\kappa\in[\kappa]\),
that
\[
\begin{aligned}
&\mathbb E\left[
    \rho(\varepsilon_{1}, [\kappa], \Sigma)
\right]=
p\operatorname{Tr}(\Sigma^{(0)}\Sigma^{-1})
-
\operatorname{Tr}\left(
    (\kappa\diamond_\Sigma\kappa)^{-1}
    \bigl(
        \kappa
        \diamond_{\Sigma(\Sigma^{(0)})^{-1}\Sigma}
        \kappa
    \bigr)
\right).
\end{aligned}
\]

In particular, we get
\[
    \mathbb E
        \|
            \varepsilon_1-\widehat\phi(
                \varepsilon_1,\kappa,\Sigma^{(0)}
            )\kappa
        \|_{\Sigma^{(0)}}^2
    =
    2p-2.
\]
\end{lemma}

\begin{proof}[Proof of Lemma~\ref{supp:lem:plugin-expected-noise-loss}]
Using Lemma~\ref{supp:lem:plugin-profile-loss-decomposition}, we obtain
\[
  \mathbb{E} \|\varepsilon_{1}- \widehat{\phi}(\varepsilon_{1}, \kappa, \Sigma)\kappa)\|_{\Sigma}^2 = \mathbb{E}\langle \varepsilon_{1}, \varepsilon_{1} \rangle_\Sigma - \mathbb{E}\langle \widehat{\phi}(\varepsilon_{1}, \kappa, \Sigma)\kappa, \varepsilon_{1} \rangle_\Sigma.
\]
First, we calculate the expectation of the squared norm of
\(\varepsilon_1\):
\begin{align*}
& \mathbb{E}[\langle \varepsilon_{1}, \varepsilon_{1} \rangle_\Sigma] \\
& =\mathbb{E}[ \sum_{j=1}^{p} \operatorname{vec}(\varepsilon_{1,j})^{\top} \Sigma^{-1} \operatorname{vec}(\varepsilon_{1,j}) ]  \\
& =\mathbb{E}[ \sum_{j=1}^{p} \operatorname{Tr} \left(\operatorname{vec}(\varepsilon_{1,j})^{\top} \Sigma^{-1} \operatorname{vec}(\varepsilon_{1,j})\right) ] \\
&=\mathbb{E}[\sum_{j=1}^{p} \operatorname{Tr} \left(\operatorname{vec}(\varepsilon_{1,j})\operatorname{vec}(\varepsilon_{1,j})^{\top} \Sigma^{-1} \right)] \\
& = \sum_{j=1}^{p} \operatorname{Tr} \left(\mathbb{E}[\operatorname{vec}(\varepsilon_{1,j})\operatorname{vec}(\varepsilon_{1,j})^{\top}] \Sigma^{-1} \right) = 
p \operatorname{Tr}(\Sigma^{(0)}\Sigma^{-1}).  
\end{align*}

We get the second term by using
Lemma~\ref{supp:lem: phi-hat-formula}, linearity, and the cyclic
property of the trace operator:
\begin{align*}
  & \mathbb{E}[\langle \widehat{\phi}(\varepsilon_1, \kappa, \Sigma)\kappa, \varepsilon_1 \rangle_\Sigma] \\
  &= \mathbb{E}\left[ \sum_{j=1}^{p} \operatorname{vec}(\varepsilon_{1,j})^\top \Sigma^{-1} M(\kappa_{j}) \operatorname{vec}(\widehat{\phi}(\varepsilon_1, \kappa, \Sigma)) \right] \\
  &= \mathbb{E}\left[ (\kappa \bullet_\Sigma \varepsilon_1)^\top (\kappa \diamond_{\Sigma} \kappa)^{-1} (\kappa \bullet_\Sigma \varepsilon_1) \right] \\
  & = \operatorname{Tr}\left( (\kappa \diamond_{\Sigma} \kappa)^{-1} \mathbb{E}\left[ (\kappa \bullet_\Sigma \varepsilon_1)(\kappa \bullet_\Sigma \varepsilon_1)^\top \right] \right).
\end{align*}
Since
\(\varepsilon_{1,1},\dots,\varepsilon_{1,p}
\sim\mathcal{CN}(0,\Sigma^{(0)})\)
are independent and mean \(0\),
\begin{align*}
  & \mathbb{E}\left[ (\kappa \bullet_\Sigma \varepsilon_1)(\kappa \bullet_\Sigma \varepsilon_1)^\top \right] \\
  &= \mathbb{E}\left[ \left( \sum_{j=1}^{p} M(\kappa_{j})^\top \Sigma^{-1} \operatorname{vec}(\varepsilon_{1,j}) \right) \left( \sum_{j=1}^{p} \operatorname{vec}(\varepsilon_{1,j})^\top \Sigma^{-1} M(\kappa_{j}) \right) \right] \\
  &= \sum_{j=1}^{p} M(\kappa_{j})^\top \Sigma^{-1} \left( \mathbb{E}[\operatorname{vec}(\varepsilon_{1,j})\operatorname{vec}(\varepsilon_{1,j})^\top] \right) \Sigma^{-1} M(\kappa_{j}) \\
  &= \sum_{j=1}^{p} M(\kappa_{j})^\top \Sigma^{-1} \Sigma^{(0)} \Sigma^{-1} M(\kappa_{j}) \\
  &= \kappa \diamond_{\Sigma(\Sigma^{(0)})^{-1}~\Sigma} \kappa.
\end{align*}
Consequently,
\[
\begin{aligned}
&\mathbb E\left[
    \left\langle
        \widehat\phi(\varepsilon_1,\kappa,\Sigma)\kappa,
        \varepsilon_1
    \right\rangle_\Sigma
\right]=
\operatorname{Tr}\left(
    (\kappa\diamond_\Sigma\kappa)^{-1}
    \bigl(
        \kappa
        \diamond_{\Sigma(\Sigma^{(0)})^{-1}\Sigma}
        \kappa
    \bigr)
\right),
\end{aligned}
\]
and therefore
\[
\begin{aligned}
&\mathbb E\left[
    \rho(\varepsilon_{1}, [\kappa], \Sigma)
\right]=
p\operatorname{Tr}(\Sigma^{(0)}\Sigma^{-1})
-
\operatorname{Tr}\left(
    (\kappa\diamond_\Sigma\kappa)^{-1}
    \bigl(
        \kappa
        \diamond_{\Sigma(\Sigma^{(0)})^{-1}\Sigma}
        \kappa
    \bigr)
\right).
\end{aligned}
\]
Substituting \(\Sigma=\Sigma^{(0)}\), we get
\[
    \mathbb E
        \|
            \varepsilon_1-\widehat\phi(
                \varepsilon_1,\kappa,\Sigma^{(0)}
            )\kappa
        \|_{\Sigma^{(0)}}^{2}
    =
    p\operatorname{Tr}(\mathrm{Id}_2)
    -
    \operatorname{Tr}(\mathrm{Id}_2)
    =
    2p-2.
\]
\end{proof}

\subsubsection{Identification of the population minimizer}
\label{supp:sec:plugin-population-identification}

This subsubsection identifies the unique minimizer of the population profile loss at the true noise covariance \(\Sigma^{(0)}\). We first obtain a quantitative lower bound for its signal component in terms of the projective separation from \([\kappa^{(0)}]\). Since the expected noise component is constant over \([\kappa]\) at \(\Sigma^{(0)}\), this bound shows that the population loss is uniquely minimized at \([\kappa^{(0)}]\).

The uniqueness of the population minimizer for the profile loss evaluated at the true noise covariance was established by \citet[Theorem~4.5]{wiechers2023drift}. We provide an alternative derivation of the underlying separation bound using an orthogonal decomposition and a Schur-complement argument.

\begin{lemma}
\label{supp:lem:plugin-population-signal-separation}
Assume model~\eqref{model:centered-complex-factor} with \(p\geq3\),
true parameters
\(\kappa^{(0)}\in\mathcal S^{2p-1}\) and
\(\Sigma^{(0)}\in\mathrm{SPD}(2)\). Then, for every
\([\kappa]\in\mathbb{CP}^{p-1}\), there exists a matrix
\[
    S=S([\kappa],\kappa^{(0)},\Sigma^{(0)})
    \in\mathrm{SPD}(2)
\]
such that
\[
\begin{aligned}
&\mathbb E\left[
    \rho(
        \phi_1\kappa^{(0)},
        [\kappa],
        \Sigma^{(0)}
    )
\right]
=
\left(
    \tilde\eta^2-\frac{\tilde\eta^4}{4}
\right)
\mathbb E\left[
    \operatorname{vec}(\phi_1)^\top
    S
    \operatorname{vec}(\phi_1)
\right],
\end{aligned}
\]
where
\[
    \tilde\eta
    =
    d\left(
        [\kappa],
        [\kappa^{(0)}]
    \right),
\]
and
\[
    \lambda_{\min}(S)
    \ge
    \frac{1}{
        \lambda_{\max}(\Sigma^{(0)})
    }.
\]
\end{lemma}

\begin{proof}[Proof of Lemma~\ref{supp:lem:plugin-population-signal-separation}]

If \(\widetilde\eta=0\), then
\([\kappa]=[\kappa^{(0)}]\), and hence
\[
    \rho(
        \phi_1\kappa^{(0)},
        [\kappa],
        \Sigma^{(0)}
    )
    =
    0
\]
for every \(\phi_1\). Therefore, the conclusion holds with
\[
    S
    =
    \lambda_{\max}(\Sigma^{(0)})^{-1}
    \mathrm{Id}_2.
\]

Henceforth, suppose that \(\widetilde\eta>0\). By applying
Lemma~\ref{supp:lem:plugin-profile-loss-decomposition}, we get
\[
\begin{aligned}
\rho(\phi_{1}\kappa^{(0)}, [\kappa], \Sigma^{(0)})
&=
\langle \phi_{1}\kappa^{(0)}, \phi_{1}\kappa^{(0)}
\rangle_{\Sigma^{(0)}}-
\left\langle
    \widehat{\phi}(\phi_{1}\kappa^{(0)}, \kappa, \Sigma^{(0)}
    )\kappa,
    \phi_{1}\kappa^{(0)}
\right\rangle_{\Sigma^{(0)}}.
\end{aligned}
\]

The first term simplifies to:
\[
    \langle
        \phi_{1}\kappa^{(0)},
        \phi_{1}\kappa^{(0)}
    \rangle_{\Sigma^{(0)}}
    =
    \operatorname{vec}(\phi_{1})^{\top}
    (\kappa^{(0)}
        \diamond_{\Sigma^{(0)}}
        \kappa^{(0)})
    \operatorname{vec}(\phi_{1}).
\]
For the second term, we get
\[
\begin{aligned}
\operatorname{vec}\big(
    \widehat{\phi}( \phi_{1}\kappa^{(0)},\kappa,\Sigma^{(0)}
    )
\big)
&=
(\kappa\diamond_{\Sigma^{(0)}}\kappa)^{-1}
\big(
    \kappa\bullet_{\Sigma^{(0)}}
    (\phi_{1}\kappa^{(0)})
\big)
\end{aligned}
\]
by Lemma~\ref{supp:lem: phi-hat-formula}. By the definition of the
bullet operator, we can see that
\[
    \kappa\bullet_{\Sigma^{(0)}}
    (\phi_{1}\kappa^{(0)})
    =
    (\kappa\diamond_{\Sigma^{(0)}}\kappa^{(0)})
    \operatorname{vec}(\phi_{1}),
\]
which yields
\[
\begin{aligned}
\operatorname{vec}\big(
    \widehat{\phi}(\phi_{1}\kappa^{(0)},\kappa,\Sigma^{(0)}
    )
\big)
&=
(\kappa\diamond_{\Sigma^{(0)}}\kappa)^{-1}
(\kappa\diamond_{\Sigma^{(0)}}\kappa^{(0)})
\operatorname{vec}(\phi_{1}).
\end{aligned}
\]
Substituting this explicit vector form back into the inner product of
the cross-term, we obtain:
\[
\begin{aligned}
&\left\langle
    \widehat{\phi}(\phi_{1}\kappa^{(0)},\kappa,\Sigma^{(0)}
    )\kappa,
    \phi_{1}\kappa^{(0)}
\right\rangle_{\Sigma^{(0)}}\\
&=
\operatorname{vec}(\phi_{1})^\top
(\kappa^{(0)}\diamond_{\Sigma^{(0)}}\kappa)
(\kappa\diamond_{\Sigma^{(0)}}\kappa)^{-1}
(\kappa\diamond_{\Sigma^{(0)}}\kappa^{(0)})
\operatorname{vec}(\phi_{1}).
\end{aligned}
\]

Finally, by combining these two components, we get
\[
\begin{aligned}
&\rho(\phi_{1}\kappa^{(0)},[\kappa],\Sigma^{(0)})\\
&=
\operatorname{vec}(\phi_{1})^\top
\Big[
(\kappa^{(0)}\diamond_{\Sigma^{(0)}}\kappa^{(0)})
-
(\kappa^{(0)}\diamond_{\Sigma^{(0)}}\kappa)
(\kappa\diamond_{\Sigma^{(0)}}\kappa)^{-1}
(\kappa\diamond_{\Sigma^{(0)}}\kappa^{(0)})
\Big]
\operatorname{vec}(\phi_{1}).
\end{aligned}
\]

Without loss of generality, assume that \(\kappa\) and
\(\kappa^{(0)}\) are aligned in their optimal position, so that
\((\kappa^{(0)})^*\kappa\) is real and non-negative. Denote
\((\kappa^{(0)})^*\kappa=\cos(\eta)\). Then,
\(\eta\) satisfies
\[
    \tilde{\eta}^2
    =
    \|\kappa-\kappa^{(0)}\|^2
    =
    2-2\cos(\eta)
    =
    4\sin^2(\eta/2),
\]
thus \(\sin(\eta/2)=\tilde{\eta}/2\). Moreover, we can orthogonally
decompose \(\kappa^{(0)}\) into a component parallel to \(\kappa\)
and a unit vector \(\kappa^\perp\) orthogonal to \(\kappa\) in
\(\mathbb C^p\):
\[
    \kappa^{(0)}
    =
    \cos(\eta)\kappa
    +
    \sin(\eta)\kappa^\perp.
\]
Since the operator
\[
    x\diamond_{\Sigma^{(0)}}y
    =
    \sum_{j=1}^{p}
    M(x_j)^\top
    (\Sigma^{(0)})^{-1}
    M(y_j)
\]
is \(\mathbb R\)-bilinear, we can substitute this decomposition
directly into the quadratic form. Using
\[
\begin{aligned}
\kappa^{(0)}\diamond_{\Sigma^{(0)}}\kappa
&=
\cos(\eta)
(\kappa\diamond_{\Sigma^{(0)}}\kappa)
+
\sin(\eta)
(\kappa^\perp\diamond_{\Sigma^{(0)}}\kappa),
\end{aligned}
\]
the middle term expands to:
\begin{align*}
&(\kappa^{(0)} \diamond_{\Sigma^{(0)}} \kappa) (\kappa \diamond_{\Sigma^{(0)}} \kappa)^{-1} (\kappa \diamond_{\Sigma^{(0)}} \kappa^{(0)} ) \\
&= \Big[ \cos(\eta)(\kappa \diamond_{\Sigma^{(0)}} \kappa) + \sin(\eta)(\kappa^\perp \diamond_{\Sigma^{(0)}} \kappa) \Big] (\kappa \diamond_{\Sigma^{(0)}} \kappa)^{-1} \Big[ \cos(\eta)(\kappa \diamond_{\Sigma^{(0)}} \kappa) + \sin(\eta)(\kappa \diamond_{\Sigma^{(0)}} \kappa^\perp) \Big] \\
&= \cos^2(\eta)(\kappa \diamond_{\Sigma^{(0)}} \kappa) \\
&\quad+ \cos(\eta)\sin(\eta)(\kappa^\perp \diamond_{\Sigma^{(0)}} \kappa + \kappa \diamond_{\Sigma^{(0)}} \kappa^\perp) \\
&\quad + \sin^2(\eta) (\kappa^\perp \diamond_{\Sigma^{(0)}} \kappa) (\kappa \diamond_{\Sigma^{(0)}} \kappa)^{-1} (\kappa \diamond_{\Sigma^{(0)}} \kappa^\perp).
\end{align*}
Subtracting this from
\[
\begin{aligned}
&\kappa^{(0)}\diamond_{\Sigma^{(0)}}\kappa^{(0)}\\
&=
\cos^2(\eta)
(\kappa\diamond_{\Sigma^{(0)}}\kappa)\\
&\quad+
\cos(\eta)\sin(\eta)
\left(
    \kappa^\perp\diamond_{\Sigma^{(0)}}\kappa
    +
    \kappa\diamond_{\Sigma^{(0)}}\kappa^\perp
\right)\\
&\quad+
\sin^2(\eta)
(\kappa^\perp\diamond_{\Sigma^{(0)}}\kappa^\perp),
\end{aligned}
\]
we obtain
\[
    \rho(\phi_1\kappa^{(0)},[\kappa],\Sigma^{(0)})
    =
    \sin^2(\eta)\,
    \operatorname{vec}(\phi_1)^\top
    S
    \operatorname{vec}(\phi_1),
\]
for
\[
\begin{aligned}
    S
    &:=
    (\kappa^\perp\diamond_{\Sigma^{(0)}}\kappa^\perp)-
    (\kappa^\perp\diamond_{\Sigma^{(0)}}\kappa)
    (\kappa\diamond_{\Sigma^{(0)}}\kappa)^{-1}
    (\kappa\diamond_{\Sigma^{(0)}}\kappa^\perp).
\end{aligned}
\]
Since
\[
    \sin^2(\eta)
    =
    4\sin^2(\eta/2)\cos^2(\eta/2)
    =
    \tilde{\eta}^2
    \left(1-\frac{\tilde{\eta}^2}{4}\right)
    =
    \tilde{\eta}^2-\frac{\tilde{\eta}^4}{4},
\]
this gives
\[
\begin{aligned}
\rho(\phi_1\kappa^{(0)},[\kappa],\Sigma^{(0)})
&=
\left(
    \tilde{\eta}^2-\frac{\tilde{\eta}^4}{4}
\right)
\operatorname{vec}(\phi_1)^\top
S
\operatorname{vec}(\phi_1).
\end{aligned}
\]

It remains to lower-bound the eigenvalues of \(S\). The matrix \(S\)
is the Schur complement of the upper-left block of
\[
    G
    =
    \begin{pmatrix}
        \kappa\diamond_{\Sigma^{(0)}}\kappa
        &
        \kappa\diamond_{\Sigma^{(0)}}\kappa^\perp
        \\
        \kappa^\perp\diamond_{\Sigma^{(0)}}\kappa
        &
        \kappa^\perp\diamond_{\Sigma^{(0)}}\kappa^\perp
    \end{pmatrix}.
\]
By the variational characterization of the Schur complement, for every
\(x\in\mathbb R^2\),
\[
\begin{aligned}
    x^\top Sx
    &=
    \min_{y\in\mathbb R^2}
    \begin{pmatrix}
        y\\x
    \end{pmatrix}^{\!\top}
    G
    \begin{pmatrix}
        y\\x
    \end{pmatrix} \ge
    \lambda_{\min}(G)\|x\|^2.
\end{aligned}
\]
Hence \(
    \lambda_{\min}(S)
    \ge
    \lambda_{\min}(G)
\).

Now write
\[
    G
    =
    B^\top
    \left\{
        \mathrm{Id}_p
        \otimes
        (\Sigma^{(0)})^{-1}
    \right\}
    B,
    \quad
    B
    :=
    \begin{pmatrix}
        M(\kappa_1) & M(\kappa^\perp_1)\\
        \vdots & \vdots\\
        M(\kappa_p) & M(\kappa^\perp_p)
    \end{pmatrix}.
\]
Since \(\kappa\) and \(\kappa^\perp\) are orthonormal in
\(\mathbb C^p\), we have
\[
    B^\top B=\mathrm{Id}_4.
\]
Therefore, for any \(w\in\mathbb R^4\),
\[
\begin{aligned}
    w^\top Gw
    &=
    (Bw)^\top
    \left\{
        \mathrm{Id}_p
        \otimes
        (\Sigma^{(0)})^{-1}
    \right\}
    (Bw)\\
    &\ge
    \lambda_{\min}\bigl(
        (\Sigma^{(0)})^{-1}
    \bigr)
    \|Bw\|^2\\
    &=
    \frac{1}{\lambda_{\max}(\Sigma^{(0)})}
    \|w\|^2.
\end{aligned}
\]
Thus,
\[
    \lambda_{\min}(S)
    \ge
    \lambda_{\min}(G)
    \ge
    \frac{1}{\lambda_{\max}(\Sigma^{(0)})}.
\]

Taking expectations in
\[
    \rho(\phi_1\kappa^{(0)},[\kappa],\Sigma^{(0)})
    =
    \sin^2(\eta)\,
    \operatorname{vec}(\phi_1)^\top
    S
    \operatorname{vec}(\phi_1)
\]
completes the proof.
\end{proof}

Using Lemma~\ref{supp:lem:plugin-population-signal-separation}, we verify that \([\kappa^{(0)}]\) is the unique minimizer of
\[
    F(m)
    =
    \mathbb E\{
        \rho(Y_{1},m,q_0)
    \}
\]
for \(q_0=\Sigma^{(0)}\).

\begin{theorem}
\label{supp:thm:plugin-population-identification}
Assume the centered drift model~\eqref{model:centered-complex-factor}
with \(p\geq3\), true parameters
\(\kappa^{(0)}\in\mathcal S^{2p-1}\) and
\(\Sigma^{(0)}\in\mathrm{SPD}(2)\), and
\(0<\mathbb E[|\phi_1|^2]<\infty\). Then
\([\kappa^{(0)}]\) is the unique minimizer of the population cost
function
\[
    \mathbb E\left\{
        \rho(Y_1,[\kappa],\Sigma^{(0)})
    \right\}
\]
over \(\mathbb{CP}^{p-1}\).
\end{theorem}

\begin{proof}[Proof of Theorem~\ref{supp:thm:plugin-population-identification}]
By Lemmas~\ref{supp:lem:plugin-population-loss-decomposition},
\ref{supp:lem:plugin-expected-noise-loss}, and
\ref{supp:lem:plugin-population-signal-separation}, we have
\[
\begin{aligned}
\mathbb E[
    \rho(Y_{1},[\kappa],\Sigma^{(0)})
]
&=
2p-2
+
\left(
    \tilde{\eta}^2
    -
    \frac{\tilde{\eta}^4}{4}
\right)
\mathbb E\left[
    \operatorname{vec}(\phi_1)^\top
    S
    \operatorname{vec}(\phi_1)
\right],
\end{aligned}
\]
where
\(
    \tilde{\eta}
    =
    d(
        [\kappa],
        [\kappa^{(0)}]
    )
\)
and \(S\in\mathrm{SPD}(2)\) satisfies
\[
    \lambda_{\min}(S)
    \ge
    \frac{1}{
        \lambda_{\max}(\Sigma^{(0)})
    }.
\]

Since
\[
\begin{aligned}
&\left(
    \tilde{\eta}^2
    -
    \frac{\tilde{\eta}^4}{4}
\right)
\mathbb E\left[
    \operatorname{vec}(\phi_1)^\top
    S
    \operatorname{vec}(\phi_1)
\right]\\
&\quad\ge
\left(
    \tilde{\eta}^2
    -
    \frac{\tilde{\eta}^4}{4}
\right)
\mathbb E\left[
    \frac{
        \|\operatorname{vec}(\phi_1)\|_2^2
    }{
        \lambda_{\max}(\Sigma^{(0)})
    }
\right]\\
&\quad=
\left(
    \tilde{\eta}^2
    -
    \frac{\tilde{\eta}^4}{4}
\right)
\frac{
    \mathbb E|\phi_1|^2
}{
    \lambda_{\max}(\Sigma^{(0)})
},
\end{aligned}
\]
we get
\[
\begin{aligned}
&\inf_{
    [\kappa]:
    d([\kappa],[\kappa^{(0)}])>\epsilon
}
\mathbb E[
    \rho(Y_{1},[\kappa],\Sigma^{(0)})
]>
\mathbb E[
    \rho(Y_{1},[\kappa^{(0)}],\Sigma^{(0)})
]
=
2p-2
\end{aligned}
\]
for any \(\epsilon>0\). This directly verifies that
\([\kappa^{(0)}]\) is the unique minimizer of the population cost
function
\[
    \mathbb E\{
        \rho(Y_{1},[\kappa],\Sigma^{(0)})
    \}
\]
over \(\mathbb{CP}^{p-1}\).
\end{proof}

\subsubsection{Verification of the GFM consistency conditions}
\label{supp:sec:plugin-consistency-conditions}

We next verify the model-specific conditions required to apply
Theorem~\ref{thm: BP-SC} to the \texttt{plug-in} estimator.

To regard each auxiliary covariance estimator as a
\(Q\)-valued random element for every \(n\), we adopt the following
convention throughout the consistency arguments below. Whenever a
covariance update is undefined or does not belong to
\(\mathrm{SPD}(2)\), we replace it by \(\mathrm{Id}_2\), retaining the
same notation for the resulting extended estimator. The argument below
shows recursively that such replacements occur only finitely often
almost surely. Hence they do not affect any of the asymptotic
conclusions.

\begin{lemma}
\label{supp:lem:plugin-gfm-consistency-conditions}
Assume the conditions of Theorem~\ref{thm:iterated-plugin-strong-consistency}. Let \(\widehat q_n\) be an \(\mathrm{SPD}(2)\)-valued random element satisfying
\[
    \|\widehat q_n-\Sigma^{(0)}\|_{\operatorname{op}}
    \longrightarrow0
    \quad\text{almost surely}.
\]
Then Conditions~\ref{supp:cond:gfm-auxiliary-consistency},
\ref{supp:cond:gfm-primary-polish},
\ref{supp:cond:gfm-heine-borel},
\ref{supp:cond:gfm-auxiliary-local-polish},
\ref{supp:cond:gfm-cost-regularity}, and
\ref{supp:cond:gfm-local-integrability}
hold for the profile loss with auxiliary input
\(\widehat q_n\).
Moreover, the empirical minimizer set is nonempty and compact
for every \(n\), and
Condition~\ref{supp:cond:gfm-eventual-boundedness} holds.
\end{lemma}

\begin{proof}[Proof of Lemma~\ref{supp:lem:plugin-gfm-consistency-conditions}]
The assumed convergence of \(\widehat q_n\) verifies
Condition~\ref{supp:cond:gfm-auxiliary-consistency}.
The remaining conditions concern the parameter spaces and
the profile loss, and do not depend on the construction of
\(\widehat q_n\). Since
\(\widehat\Sigma^{\texttt{cov}}\to\Sigma^{(0)}\) almost surely and
\(\Sigma^{(0)}\in\mathrm{SPD}(2)\), the unextended covariance update
belongs to \(\mathrm{SPD}(2)\) almost surely for all sufficiently large
\(n\). Consequently, the extended estimator agrees eventually almost
surely with the original update and converges to \(\Sigma^{(0)}\) in \(d_Q\). This verifies Condition~\ref{supp:cond:gfm-auxiliary-consistency}.

Conditions~\ref{supp:cond:gfm-primary-polish} and
\ref{supp:cond:gfm-heine-borel} are trivially satisfied due to the compactness of the complex projective space
\(\mathbb{CP}^{p-1}\).

The space \(\mathrm{SPD}(2)\) is an open subset of the three-dimensional normed vector space of real symmetric \(2\times2\) matrices. Since \(\Sigma^{(0)}\) is positive definite, choose
\[
    0<r_0<
    \lambda_{\min}(\Sigma^{(0)}).
\]
Then, for every symmetric matrix \(\Sigma\) satisfying
\[
    \|\Sigma-\Sigma^{(0)}\|_{\mathrm{op}}
    \le r_0,
\]
Weyl's inequality gives
\[
    \lambda_{\min}(\Sigma)
    \ge
    \lambda_{\min}(\Sigma^{(0)})-r_0
    >
    0.
\]
Hence
\[
    \overline{
        B_{d_Q}(\Sigma^{(0)},r_0)
    }
    \subset
    \mathrm{SPD}(2).
\]
This closed ball is compact since it is closed and bounded in a
finite-dimensional normed space. It is therefore separable and
complete, which verifies
Condition~\ref{supp:cond:gfm-auxiliary-local-polish}.

Next, we verify the measurability and continuity requirement of
Condition~\ref{supp:cond:gfm-cost-regularity}. For this purpose, we
introduce an extended function
\[
    \tilde{\rho}:
    \mathbb C^p
    \times
    \mathcal S^{2p-1}
    \times
    \mathrm{SPD}(2)
    \longrightarrow
    \mathbb R
\]
defined by dropping the projective equivalence:
\[
    \tilde{\rho}(t,\kappa,\Sigma)
    :=
    \|
        t-\mathfrak c\bigl(
            (\kappa\diamond_\Sigma\kappa)^{-1}
            (\kappa\bullet_\Sigma t)
        \bigr)\kappa
    \|_{\Sigma}^2
\]
for \(t\in\mathbb C^p\),
\(\kappa\in\mathcal S^{2p-1}\), and
\(\Sigma\in\mathrm{SPD}(2)\). This representation clearly
demonstrates that \(\tilde\rho\) is continuous, as it is constructed
entirely from basic algebraic operations, polynomials of the
components, and the matrix inversion of
\((\kappa\diamond_\Sigma\kappa)\), which is positive definite (and
thus non-singular) for any
\(\kappa\in\mathcal S^{2p-1}\) and
\(\Sigma\in\mathrm{SPD}(2)\); see
Lemma~\ref{supp:lem: diamond_invertible}.

The function \(\tilde\rho\) is invariant under global phase
rotations:
\[
    \tilde\rho(
        t,e^{i\lambda}\kappa,\Sigma
    )
    =
    \tilde\rho(t,\kappa,\Sigma)
\]
for any \(\lambda\in\mathbb R\) by
Lemma~\ref{supp:lem: phi-hat-equivariance}. Let
\[
    \pi:
    \mathcal S^{2p-1}
    \longrightarrow
    \mathbb{CP}^{p-1}
\]
be the standard quotient map. This map is open. Consequently,
\[
    \mathrm{Id}
    \times\pi
    \times\mathrm{Id}
\]
is a continuous open surjection and hence a quotient map. Since
\(\tilde\rho\) is continuous and constant on its fibers, the
universal property of quotient maps yields a unique continuous
function
\[
    \rho:
    \mathbb C^p
    \times
    \mathbb{CP}^{p-1}
    \times
    \mathrm{SPD}(2)
    \longrightarrow
    \mathbb R
\]
such that
\[
    \tilde\rho
    =
    \rho\circ
    (
        \mathrm{Id}
        \times\pi
        \times\mathrm{Id}
    ).
\]
In particular, \(\rho\) is measurable, verifying
Condition~\ref{supp:cond:gfm-cost-regularity}. The relationship
between the original cost function \(\rho\) and the extended loss
function \(\tilde\rho\) is summarized by the following commutative
diagram:
\[
\begin{tikzcd}[row sep=large, column sep=huge]
\mathbb C^p\times\mathcal S^{2p-1}\times\mathrm{SPD}(2)
    \arrow[dr,"\tilde\rho"]
    \arrow[d,"\mathrm{Id}\times\pi\times\mathrm{Id}"']
&\\
\mathbb C^p\times\mathbb{CP}^{p-1}\times\mathrm{SPD}(2)
    \arrow[r,"\rho"']
&
\mathbb R.
\end{tikzcd}
\]

Finally, we verify the integrability requirement for envelope
functions in
Condition~\ref{supp:cond:gfm-local-integrability}. For any
\(t\in\mathbb C^p\),
\(\kappa\in\mathcal S^{2p-1}\), and
\(\Sigma\in\mathrm{SPD}(2)\), the following inequalities hold:
\begin{equation}
\label{supp:eq:plugin-profile-loss-bound}
0
\le
\rho(t,[\kappa],\Sigma)
\le
\|t\|_\Sigma^{2}
\le
\lambda^{{\max}}(\Sigma^{-1})
\sum_{j=1}^{p}|t_j|^2.
\end{equation}

The first inequality is trivial since \(\rho\) is a squared
distance. The second inequality arises directly from the definition
of the profile loss:
\(\rho(t,[\kappa],\Sigma)\) minimizes the Mahalanobis distance over
all possible \(\phi\in\mathbb C\), which must be less than or equal
to the distance evaluated at \(\phi=0\). The third inequality
follows from the definition of the induced norm
\[
    \|t\|_\Sigma^{2}
    =
    \sum_{j=1}^{p}
    \operatorname{vec}(t_j)^\top
    \Sigma^{-1}
    \operatorname{vec}(t_j).
\]

Consequently, for any local neighborhood defined by
\(0<r<r_0\), using \(r_0\) from
Condition~\ref{supp:cond:gfm-auxiliary-local-polish}, and any primary
parameter \(m=[\kappa]\in M\), we obtain
\begin{equation}
\label{supp:eq:plugin-local-envelope-bound}
\begin{aligned}
0
&\le
\pi_{m,r}(Y_1)
\le
\Pi_{m,r}(Y_1)\\
&\le
\sup\left\{
    \lambda_{\max}(\Sigma^{-1})
    \,\middle|\,
    \Sigma\in
    \overline{
        B_{d_Q}(\Sigma^{(0)},r_0)
    }
\right\}
\sum_{j=1}^{p}|Y_{1,j}|^2.
\end{aligned}
\end{equation}
Since \(\overline{B_{d_Q}(\Sigma^{(0)},r_0)}\) is a compact subset of
\(\mathrm{SPD}(2)\), the continuous map
\[
    \Sigma
    \longmapsto
    \lambda_{\max}(\Sigma^{-1})
\]
attains a finite maximum. Thus, the local extrema are bounded by a
constant multiple of
\(\sum_{j=1}^{p}|Y_{1,j}|^2\). Under the
model~\eqref{model:centered-complex-factor} with assumption
\(0<\mathbb E[|\phi_1|^2]<\infty\), each \(Y_{1,j}\) has a finite
second moment. Thus, taking \(r_m=r_0/2\) for every \(m\in M\), both
\(\pi_{m,r}(Y_1)\) and \(\Pi_{m,r}(Y_1)\) are integrable for every
\(0<r\leq r_m\), verifying
Condition~\ref{supp:cond:gfm-local-integrability}.

Finally, for every \(n\), the empirical criterion
\[
    m\longmapsto F_n(m,\widehat q_n)
\]
is continuous on the compact space
\(M=\mathbb{CP}^{p-1}\).
It therefore attains its minimum, and its minimizer set
\(\widehat E_n\) is nonempty and compact.
Since \(M\) itself is bounded,
Condition~\ref{supp:cond:gfm-eventual-boundedness} holds.

The empirical criterion is jointly measurable in the
observations, the covariance input, and \(m\), and is
continuous in \(m\).
Its nonempty compact argmin set therefore admits a measurable
selection. At each iteration, we choose such a selection.
\end{proof}

\subsubsection{Proof of strong consistency for the \texttt{plug-in}
and iterated \texttt{plug-in} estimators}
\label{supp:sec:plugin-strong-consistency-proof}

This subsubsection combines the preceding population minimizer identification and regularity conditions to prove strong consistency. We first apply the
general BP consistency theorem to the one-step \texttt{plug-in}
estimator. We then show that consistency of a loading estimator is
preserved by the moment covariance update and apply the same argument
recursively to every fixed number of \texttt{plug-in} iterations.

\begin{proof}[Proof of
Theorem~\ref{thm:iterated-plugin-strong-consistency}]
By Theorem~\ref{supp:thm:plugin-population-identification}, the
population minimizer set is
\[
    E_0
    =
    \{[\kappa^{(0)}]\}.
\]

We first note that any strongly consistent loading estimator yields
a strongly consistent covariance update through the moment map
\(\mathcal M\). Let
\(\widetilde{[\kappa]}_n\) be any sequence of random elements satisfying
\[
    d\left(
        \widetilde{[\kappa]}_n,
        [\kappa^{(0)}]
    \right)
    \xrightarrow{\mathrm{a.s.}}
    0,
\]
and choose representatives
\(\widetilde\kappa_n\in\widetilde{[\kappa]}_n\) such that
\(\widetilde\kappa_n\to\kappa^{(0)}\) almost surely. 

Since
\[
    \mathcal U(\widetilde\kappa_n)
    \xrightarrow{\mathrm{a.s.}}
    U_0
\]
and \(U_0\) has full column rank,
\(\mathcal U(\widetilde\kappa_n)\) also has full column rank almost
surely for all sufficiently large \(n\). Consequently,
\[
\begin{aligned}
    \Pi_\perp(\widetilde\kappa_n)
    &=
    \mathrm{Id}_p
    -
    \mathcal U(\widetilde\kappa_n)
    \left\{
        \mathcal U(\widetilde\kappa_n)^\top
        \mathcal U(\widetilde\kappa_n)
    \right\}^{-1}
    \mathcal U(\widetilde\kappa_n)^\top\\
    &\xrightarrow{\mathrm{a.s.}}
    \mathrm{Id}_p
    -
    U_0(U_0^\top U_0)^{-1}U_0^\top
    =
    \Pi_{\perp,0}.
\end{aligned}
\]
Moreover, the strong law of large numbers gives
\[
    \widehat\Sigma^{\texttt{Her}}
    \xrightarrow{\mathrm{a.s.}}
    \Sigma^{\texttt{Her}},
    \quad
    \widehat\Sigma^{\texttt{Tra}}
    \xrightarrow{\mathrm{a.s.}}
    \Sigma^{\texttt{Tra}}.
\]
Consequently,
\[
\begin{aligned}
    \widehat\tau^2(\widetilde\kappa_n)
    &=
    \frac{1}{p-2}
    \operatorname{Tr}\left\{
        \Pi_\perp(\widetilde\kappa_n)
        \widehat\Sigma^{\texttt{Her}}
    \right\}
    \xrightarrow{\mathrm{a.s.}}
    \tau^2,\\
    \widehat\gamma(\widetilde\kappa_n)
    &=
    \frac{1}{p-2}
    \operatorname{Tr}\left\{
        \Pi_\perp(\widetilde\kappa_n)
        \widehat\Sigma^{\texttt{Tra}}
    \right\}
    \xrightarrow{\mathrm{a.s.}}
    \gamma.
\end{aligned}
\]
By the definition of the moment map in
\eqref{eq:moment-covariance-map}, it follows that
\begin{equation}
\label{supp:eq:plugin-moment-update-consistency}
    \mathcal M(\widetilde{[\kappa]}_n)
    \xrightarrow{\mathrm{a.s.}}
    \Sigma^{(0)}.
\end{equation}

We now consider the first \texttt{plug-in} update. By Theorem~\ref{thm:cov-estimator-strong-consistency},
\[
    \widehat\Sigma^{\texttt{cov}}
    \xrightarrow{\mathrm{a.s.}}
    \Sigma^{(0)}.
\]
Lemma~\ref{supp:lem:plugin-gfm-consistency-conditions}
therefore verifies the assumptions of
Theorem~\ref{thm: BP-SC} with
\(\widehat q_n=\widehat\Sigma^{\texttt{cov}}\).
Therefore, the empirical minimizer set defining
\(\widehat{[\kappa]}^{\texttt{plug-in}}\) is BP consistent with
\[
    E_0
    =
    \{[\kappa^{(0)}]\}.
\]
Since \(E_0\) is a singleton, Lemma~\ref{lem: singletone} yields
\[
    \widehat{[\kappa]}^{\texttt{plug-in}}
    =
    \widehat{[\kappa]}^{[1]}
    \xrightarrow{\mathrm{a.s.}}
    [\kappa^{(0)}].
\]
Applying \eqref{supp:eq:plugin-moment-update-consistency} then gives
\[
    \widehat\Sigma^{\texttt{plug-in}}
    =
    \widehat\Sigma^{[1]}
    =
    \mathcal M\left(
        \widehat{[\kappa]}^{[1]}
    \right)
    \xrightarrow{\mathrm{a.s.}}
    \Sigma^{(0)}.
\]

It remains to propagate this conclusion through the
subsequent iterations.
Let \(s\geq2\), and suppose that
\[
    \widehat\Sigma^{[s-1]}
    \xrightarrow{\mathrm{a.s.}}
    \Sigma^{(0)}.
\]
Applying
Lemma~\ref{supp:lem:plugin-gfm-consistency-conditions}
with
\(\widehat q_n=\widehat\Sigma^{[s-1]}\),
followed by Theorem~\ref{thm: BP-SC} and
Lemma~\ref{lem: singletone}, gives
\[
    \widehat{[\kappa]}^{[s]}
    \xrightarrow{\mathrm{a.s.}}
    [\kappa^{(0)}].
\]
Equation~\eqref{supp:eq:plugin-moment-update-consistency}
then yields
\[
    \widehat\Sigma^{[s]}
    \xrightarrow{\mathrm{a.s.}}
    \Sigma^{(0)}.
\]
Since the first update has already been treated, induction
proves these conclusions for every fixed \(s\geq1\).

For each such \(s\), loading consistency and the full column
rank of \(U_0\) imply that the unextended moment covariance
update is defined almost surely for all sufficiently large
\(n\).
Its convergence to \(\Sigma^{(0)}\in\mathrm{SPD}(2)\)
then implies eventual positive definiteness.
Consequently, the covariance replacements occur only
finitely often almost surely at each fixed iteration,
and the original updates satisfy the asserted conclusions.
\end{proof}

\subsection{Stochastic expansions and asymptotic normality of iterated \texttt{plug-in} estimator}
\label{supp:sec:plugin-asymptotic-expansions}
This subsection proves
Theorems~\ref{thm:iterated-plugin-loading-expansion-clt}
and~\ref{thm:iterated-plugin-noise-covariance-expansion}.
We first derive a first-order loading expansion with an
\(\mathcal O_p(n^{-1})\) remainder using the generalized
Fr\'echet mean framework. We then apply
Lemma~\ref{supp:lem:real-residual-projection-loading-expansion}
to obtain the second-order covariance update.
The covariance update preserves the first-order influence
function, allowing these arguments to be iterated for any
fixed number of updates.
Throughout this subsection, we write
\(L_{\kappa,n}=L_{\kappa,n}^{\texttt{plug-in}}\),
with the latter defined in
\eqref{eq:plugin-common-loading-first-order-term}.

\subsubsection{Differential identities for the profile loss}
\label{supp:sec:plugin-profile-loss-differential-identities}

In this subsubsection, we derive the differential identities for the
representative-level profile loss function. In particular, we calculate its gradient and Hessian with respect to the
real coordinates of \(\kappa\), as well as its derivative with respect to
\(\Sigma\) and the corresponding mixed derivative. These formulas will be
used below to verify the differentiability conditions for the generalized
Fréchet mean framework and to identify the matrices \(J\) and \(C_0\)
appearing in the stochastic expansion of a loading update.

\begin{lemma}\label{supp:lem: rho differentiation}
For \(y\in\mathbb C^p\), \(\kappa\in\mathbb C^p\setminus\{0\}\), and
\(\Sigma\in\mathrm{SPD}(2)\), define the extended cost function \[
    \tilde\rho(y,\kappa,\Sigma)
    :=
    \left\|
        y-\widehat\phi(y,\kappa,\Sigma)\kappa
    \right\|_\Sigma^2.
\]
Viewed as a function of
\(\widetilde{\operatorname{vec}}(\kappa)\) and
\(\operatorname{vech}(\Sigma)\),
\(\tilde\rho(y,\kappa,\Sigma)\) is infinitely continuously
differentiable on
\[
    \left(
        \mathbb R^{2p}\setminus\{0\}
    \right)
    \times
    \operatorname{vech}
    \left\{
        \mathrm{SPD}(2)
    \right\}.
\]
Moreover, the following statements hold.
\begin{itemize}

\item[(i)] The gradient with respect to
\(\widetilde{\operatorname{vec}}(\kappa)\) is
\begin{equation}
\label{supp:eq: explicit kappa gradient stacked}
    \nabla_{\widetilde{\operatorname{vec}}(\kappa)}
    \tilde\rho(y,\kappa,\Sigma)
    =
    -2
    \begin{pmatrix}
        \mathbf E\Sigma^{-1}
        \operatorname{vec}\bigl(\widehat\phi(y,\kappa,\Sigma)\bigr)
        \\
        \mathbf E\Sigma^{-1}J_2
        \operatorname{vec}\bigl(\widehat\phi(y,\kappa,\Sigma)\bigr)
    \end{pmatrix},
\end{equation}
where
\begin{equation}\label{supp:eq: E definition}
\mathbf E
    :=
    \begin{pmatrix}
        r_1 & \ldots & r_p
    \end{pmatrix}^\top \in\mathbb{R}^{p\times 2}, 
\end{equation}
$
    J_2
    :=
    \begin{pmatrix}
        0&-1\\
        1&0
    \end{pmatrix}
$
and
\(
    r_j
    :=
    \operatorname{vec}(y_j)
    -
    M(\kappa_j)
    \operatorname{vec}\bigl(\widehat\phi(y,\kappa,\Sigma)\bigr)
\)
for \(j=1,\ldots,p\).

\item[(ii)] Write
\[
    \Sigma
    =
    \begin{pmatrix}
        \sigma_{11} & \sigma_{12}\\
        \sigma_{12} & \sigma_{22}
    \end{pmatrix}.
\]
Then, for \(1\leq a\leq b\leq2\),
\begin{equation}
\label{supp:eq: explicit sigma gradient}
\frac{
    \partial\tilde\rho(y,\kappa,\Sigma)
}{
    \partial\sigma_{ab}
}
=
\operatorname{Tr}
\left[
    -\Sigma^{-1}
    \left(
        \sum_{j=1}^p r_jr_j^\top
    \right)
    \Sigma^{-1}
    \frac{\partial\Sigma}{\partial\sigma_{ab}}
\right].
\end{equation}

\item[(iii)] The Hessian with respect to
\(\widetilde{\operatorname{vec}}(\kappa)\) is
\begin{equation}
\label{supp:eq: explicit kappa Hessian}
\begin{aligned}
&\nabla^2_{\widetilde{\operatorname{vec}}(\kappa)}
\tilde\rho(y,\kappa,\Sigma)\\
=
&2\Big[
\left(M\bigl(\widehat\phi(y,\kappa,\Sigma)\bigr)^\top
\Sigma^{-1}
M\bigl(\widehat\phi(y,\kappa,\Sigma)\bigr)\right)
\otimes
\mathrm{Id}_p-
\mathbf Q^\top
(\kappa\diamond_\Sigma\kappa)^{-1}
\mathbf Q
\Big],
\end{aligned}
\end{equation}
where
\begin{equation}
\label{supp:eq: Q matrix definition}
    \mathbf Q
    :=
    \begin{pmatrix}
        q_{R,1}&\cdots&q_{R,p}&
        q_{I,1}&\cdots&q_{I,p}
    \end{pmatrix}
    \in\mathbb R^{2\times2p}
\end{equation} for
\[
    q_{\mathrm R,j}
    :=
    \Sigma^{-1}r_j
    -
    M(\kappa_j)^\top
    \Sigma^{-1}\operatorname{vec}\bigl(
        \widehat\phi(y,\kappa,\Sigma)
    \bigr),
\]
\[
    q_{\mathrm I,j}
    :=
    J_2^\top\Sigma^{-1}r_j
    -
    M(\kappa_j)^\top
    \Sigma^{-1}J_2\operatorname{vec}\bigl(
        \widehat\phi(y,\kappa,\Sigma)
    \bigr),
    \quad
    j=1,\ldots,p.
\]

\item[(iv)] Write
\(
    \operatorname{vech}(\Sigma)
    :=
    \begin{pmatrix}
        \sigma_{11}&\sigma_{12}&\sigma_{22}
    \end{pmatrix}^{\top}.
\)
Then
\begin{equation}
\label{supp:eq: explicit kappa Sigma mixed derivative symmetric}
\nabla^2_{
    \widetilde{\operatorname{vec}}(\kappa),
    \operatorname{vech}(\Sigma)
}
\tilde\rho(y,\kappa,\Sigma)
=
\begin{pmatrix}
    \mathbf C_\Sigma
    \left(
        \begin{pmatrix}
            1&0\\
            0&0
        \end{pmatrix}
    \right)
    &
    \mathbf C_\Sigma
    \left(
        \begin{pmatrix}
            0&1\\
            1&0
        \end{pmatrix}
    \right)
    &
    \mathbf C_\Sigma
    \left(
        \begin{pmatrix}
            0&0\\
            0&1
        \end{pmatrix}
    \right)
\end{pmatrix},
\end{equation}
where, for any symmetric \(\Delta\in\mathbb R^{2\times2}\),
\begin{equation}
\label{supp:eq: C Sigma definition}
\begin{aligned}
\mathbf C_\Sigma(\Delta)
:=
2\Bigg[
&
\begin{pmatrix}
\mathbf E\Sigma^{-1}\Delta\Sigma^{-1}
\operatorname{vec}
\bigl(
    \widehat\phi(y,\kappa,\Sigma)
\bigr)
\\
\mathbf E\Sigma^{-1}\Delta\Sigma^{-1}J_2
\operatorname{vec}
\bigl(
    \widehat\phi(y,\kappa,\Sigma)
\bigr)
\end{pmatrix}+
\mathbf Q^\top
(\kappa\diamond_\Sigma\kappa)^{-1}
\sum_{j=1}^p
M(\kappa_j)^\top
\Sigma^{-1}\Delta\Sigma^{-1}r_j
\Bigg]
\in\mathbb R^{2p}.
\end{aligned}
\end{equation}

\end{itemize}
\end{lemma}

\begin{proof}[Proof of Lemma~\ref{supp:lem: rho differentiation}]

First, let
\[
    A(\kappa,\Sigma)
    :=
    \kappa\diamond_\Sigma\kappa.
\]
For every \(\kappa\neq0\) and
\(\Sigma\in\mathrm{SPD}(2)\), the matrix
\(A(\kappa,\Sigma)\) is positive definite and hence invertible.
Since matrix inversion is \(C^\infty\) on the set of invertible
matrices, the representation
\[
    \operatorname{vec}
    \bigl(
        \widehat\phi(y,\kappa,\Sigma)
    \bigr)
    =
    A(\kappa,\Sigma)^{-1}
    (\kappa\bullet_\Sigma y)
\]
shows that
\(\widehat\phi(y,\kappa,\Sigma)\), the corresponding residuals, and
\(\widetilde\rho(y,\kappa,\Sigma)\) are \(C^\infty\) on the stated
domain.

\noindent\textit{Proof of (i).}
Write
\[
    \kappa_j=\kappa_j^{\mathrm R}+i\kappa_j^{\mathrm I},
    \quad
    \widetilde{\operatorname{vec}}(\kappa)
    =
    \begin{pmatrix}
        \kappa_1^{\mathrm R}&\cdots&\kappa_p^{\mathrm R}&
        \kappa_1^{\mathrm I}&\cdots&\kappa_p^{\mathrm I}
    \end{pmatrix}^{\top}.
\]
Recall that
\(
    \tilde\rho(y,\kappa,\Sigma)
    =
    \sum_{l=1}^p r_l^\top\Sigma^{-1}r_l.
\)
Since
\(
    \frac{\partial M(\kappa_l)}{\partial \kappa_j^{\mathrm R}}
    =
    \mathbf 1_{\{l=j\}}\mathrm{Id}_2,
\)
we have
\begin{align*}
    \frac{\partial r_l}{\partial \kappa_j^{\mathrm R}}
    &=
    -
    \mathbf 1_{\{l=j\}}
    \operatorname{vec}
    \left(
        \widehat\phi(y,\kappa,\Sigma)
    \right)
    -
    M(\kappa_l)
    \frac{\partial}{\partial \kappa_j^{\mathrm R}}
    \operatorname{vec}
    \left(
        \widehat\phi(y,\kappa,\Sigma)
    \right).
\end{align*}
Therefore,
\begin{align*}
    \frac{\partial\tilde\rho}{\partial \kappa_j^{\mathrm R}}
    &=
    2\sum_{l=1}^p
    r_l^\top\Sigma^{-1}
    \frac{\partial r_l}{\partial \kappa_j^{\mathrm R}}\\
    &=
    -2r_j^\top\Sigma^{-1}
    \operatorname{vec}
    \left(
        \widehat\phi(y,\kappa,\Sigma)
    \right)
    -
    2
    \left\{
        \sum_{l=1}^p
        M(\kappa_l)^\top
        \Sigma^{-1}r_l
    \right\}^{\top}
    \frac{\partial}{\partial \kappa_j^{\mathrm R}}
    \operatorname{vec}
    \left(
        \widehat\phi(y,\kappa,\Sigma)
    \right).
\end{align*}
Since
\(
    \operatorname{vec}\bigl(\widehat\phi(y,\kappa,\Sigma)\bigr)
    =
    (\kappa\diamond_\Sigma\kappa)^{-1}
    (\kappa\bullet_\Sigma y),
\)
we have
\begin{equation}
\label{supp:eq: profile phi first order condition}
\begin{aligned}
    \sum_{j=1}^p
    M(\kappa_j)^\top\Sigma^{-1}r_j
    &=
    (\kappa\bullet_\Sigma y)
    -
    (\kappa\diamond_\Sigma\kappa)
    \operatorname{vec}
    \bigl(
        \widehat\phi(y,\kappa,\Sigma)
    \bigr)\\
    &=
    (\kappa\bullet_\Sigma y)
    -
    (\kappa\diamond_\Sigma\kappa)
    (\kappa\diamond_\Sigma\kappa)^{-1}
    (\kappa\bullet_\Sigma y)
    =
    0.
\end{aligned}
\end{equation}
Hence,
\[
    \frac{\partial\tilde\rho}{\partial \kappa_j^{\mathrm R}}
    =
    -2r_j^\top\Sigma^{-1}
    \operatorname{vec}
    \left(
        \widehat\phi(y,\kappa,\Sigma)
    \right).
\]

Similarly, since
\(
    \frac{\partial M(\kappa_l)}{\partial \kappa_j^{\mathrm I}}
    =
    \mathbf 1_{\{l=j\}}J_2,
\)
we obtain
\begin{align*}
    \frac{\partial r_l}{\partial \kappa_j^{\mathrm I}}
    &=
    -
    \mathbf 1_{\{l=j\}}
    J_2
    \operatorname{vec}
    \left(
        \widehat\phi(y,\kappa,\Sigma)
    \right)
    -
    M(\kappa_l)
    \frac{\partial}{\partial \kappa_j^{\mathrm I}}
    \operatorname{vec}
    \left(
        \widehat\phi(y,\kappa,\Sigma)
    \right).
\end{align*}
Thus,
\begin{align*}
    \frac{\partial\tilde\rho}{\partial \kappa_j^{\mathrm I}}
    &=
    2\sum_{l=1}^p
    r_l^\top\Sigma^{-1}
    \frac{\partial r_l}{\partial \kappa_j^{\mathrm I}}\\
    &=
    -2r_j^\top\Sigma^{-1}J_2
    \operatorname{vec}
    \left(
        \widehat\phi(y,\kappa,\Sigma)
    \right)
    -
    2
    \left\{
        \sum_{l=1}^p
        M(\kappa_l)^\top
        \Sigma^{-1}r_l
    \right\}^{\top}
    \frac{\partial}{\partial \kappa_j^{\mathrm I}}
    \operatorname{vec}
    \left(
        \widehat\phi(y,\kappa,\Sigma)
    \right).
\end{align*}
Again, the last term vanishes by
\eqref{supp:eq: profile phi first order condition}, and therefore
\[
    \frac{\partial\tilde\rho}{\partial \kappa_j^{\mathrm I}}
    =
    -2r_j^\top\Sigma^{-1}J_2
    \operatorname{vec}
    \left(
        \widehat\phi(y,\kappa,\Sigma)
    \right).
\]

Stacking all these derivatives, we get 
\[
    \nabla_{\widetilde{\operatorname{vec}}(\kappa)}
    \tilde\rho(y,\kappa,\Sigma)
    =
    \begin{pmatrix}
        \nabla_{\kappa^{\mathrm R}}\tilde\rho\\
        \nabla_{\kappa^{\mathrm I}}\tilde\rho
    \end{pmatrix}
    =
    -2
    \begin{pmatrix}
        \mathbf E\Sigma^{-1}
        \operatorname{vec}
        \left(
            \widehat\phi(y,\kappa,\Sigma)
        \right)\\
        \mathbf E\Sigma^{-1}J_2
        \operatorname{vec}
        \left(
            \widehat\phi(y,\kappa,\Sigma)
        \right)
    \end{pmatrix},
\]
which shows \eqref{supp:eq: explicit kappa gradient stacked}.

\medskip
\noindent\textit{Proof of (ii).}
Write
\(
    \Sigma
    =
    \begin{pmatrix}
        \sigma_{11} & \sigma_{12}\\
        \sigma_{12} & \sigma_{22}
    \end{pmatrix}.
\) For each component \(\sigma_{ab}\), since
\(
    \tilde\rho(y,\kappa,\Sigma)
    =
    \sum_{j=1}^p r_j^\top\Sigma^{-1}r_j,
\)
we have
\begin{align*}
    \frac{\partial\tilde\rho}{\partial\sigma_{ab}}
    &=
    2\sum_{j=1}^p
    r_j^\top\Sigma^{-1}
    \frac{\partial r_j}{\partial\sigma_{ab}}
    +
    \sum_{j=1}^p
    r_j^\top
    \frac{\partial\Sigma^{-1}}{\partial\sigma_{ab}}
    r_j.
\end{align*}
Since \(y\) and \(\kappa\) do not depend on $\Sigma$,
\[
    \frac{\partial r_j}{\partial\sigma_{ab}}
    =
    -
    M(\kappa_j)
    \frac{\partial}{\partial\sigma_{ab}}
    \operatorname{vec}
    \left(
        \widehat\phi(y,\kappa,\Sigma)
    \right).
\]
Therefore,
\begin{align*}
    2\sum_{j=1}^p
    r_j^\top\Sigma^{-1}
    \frac{\partial r_j}{\partial\sigma_{ab}}
    &=
    -2\sum_{j=1}^p
    r_j^\top\Sigma^{-1}M(\kappa_j)
    \frac{\partial}{\partial\sigma_{ab}}
    \operatorname{vec}
    \left(
        \widehat\phi(y,\kappa,\Sigma)
    \right)\\
    &=
    -2
    \left\{
        \sum_{j=1}^p
        M(\kappa_j)^\top\Sigma^{-1}r_j
    \right\}^{\top}
    \frac{\partial}{\partial\sigma_{ab}}
    \operatorname{vec}
    \left(
        \widehat\phi(y,\kappa,\Sigma)
    \right)\\
    &=0
\end{align*}
by \eqref{supp:eq: profile phi first order condition}.

Using
\(
    \frac{\partial\Sigma^{-1}}{\partial\sigma_{ab}}
    =
    -
    \Sigma^{-1}
    \frac{\partial\Sigma}{\partial\sigma_{ab}}
    \Sigma^{-1},
\) and the cyclic property of the trace operator, we obtain
\begin{align*}
    \frac{\partial\tilde\rho}{\partial\sigma_{ab}}
    &=
    -
    \sum_{j=1}^p
    r_j^\top
    \Sigma^{-1}
    \frac{\partial\Sigma}{\partial\sigma_{ab}}
    \Sigma^{-1}
    r_j\\
    &=
    -
    \sum_{j=1}^p
    \operatorname{Tr}
    \left(
        \Sigma^{-1}
        r_jr_j^\top
        \Sigma^{-1}
        \frac{\partial\Sigma}{\partial\sigma_{ab}}
    \right)\\
    &=
    \operatorname{Tr}
    \left[
        -\Sigma^{-1}
        \left(
            \sum_{j=1}^p r_jr_j^\top
        \right)
        \Sigma^{-1}
        \frac{\partial\Sigma}{\partial\sigma_{ab}}
    \right].
\end{align*}

\medskip

\noindent\textit{Proof of (iii).}
We calculate the Hessian in two steps. First, we derive the coordinate
derivatives of
\(\operatorname{vec}(\widehat\phi)\) with respect to \(\kappa_l^{\mathrm R}\) and \(\kappa_l^{\mathrm I}\).
We then substitute these intermediate derivatives into \eqref{supp:eq: explicit kappa gradient stacked} and stack the resulting componentwise expressions.

Let \(q_{\mathrm R,j}\) and \(q_{\mathrm I,j}\) denote the \(j\)-th and
\((p+j)\)-th columns of \(\mathbf Q\), respectively, for
\(j=1,\ldots,p\). By \eqref{supp:eq: Q matrix definition},
\[
\begin{aligned}
    q_{\mathrm R,j}
    &=
    \Sigma^{-1}r_j
    -
    M(\kappa_j)^\top\Sigma^{-1}
    \operatorname{vec}(\widehat\phi),\\
    q_{\mathrm I,j}
    &=
    J_2^\top\Sigma^{-1}r_j
    -
    M(\kappa_j)^\top\Sigma^{-1}J_2
    \operatorname{vec}(\widehat\phi).
\end{aligned}
\]

We first calculate the derivatives of
\(
    \operatorname{vec}(\widehat\phi)
    =
    (\kappa\diamond_\Sigma\kappa)^{-1}
    (\kappa\bullet_\Sigma y).
\)
Since
\[
    \frac{\partial(\kappa\bullet_\Sigma y)}{\partial \kappa_l^{\mathrm R}}
    =
    \Sigma^{-1}\operatorname{vec}(y_l)\]
and 
\[
    \frac{\partial(\kappa\diamond_\Sigma\kappa)}{\partial \kappa_l^{\mathrm R}}
    =
    \Sigma^{-1}M(\kappa_l)
    +
    M(\kappa_l)^\top\Sigma^{-1},
\]
the matrix inverse derivative formula gives
\begin{align}\label{supp:eq: phi derivative r}
    \frac{\partial\operatorname{vec}(\widehat\phi)}
    {\partial \kappa_l^{\mathrm R}}
    & = (\kappa\diamond_\Sigma\kappa)^{-1}
    \Sigma^{-1}\operatorname{vec}(y_l)
    -
    (\kappa\diamond_\Sigma\kappa)^{-1}
    \left\{
        \Sigma^{-1}M(\kappa_l)
        +
        M(\kappa_l)^\top\Sigma^{-1}
    \right\}
    (\kappa\diamond_\Sigma\kappa)^{-1}
    (\kappa\bullet_\Sigma y)\nonumber \\ 
    &=
    (\kappa\diamond_\Sigma\kappa)^{-1}
    \Big[
        \Sigma^{-1}\operatorname{vec}(y_l)
        -
        \left\{
            \Sigma^{-1}M(\kappa_l)
            +
            M(\kappa_l)^\top\Sigma^{-1}
        \right\}
        \operatorname{vec}(\widehat\phi)
    \Big]\nonumber \\ 
    &=
    (\kappa\diamond_\Sigma\kappa)^{-1}
    \Big[
        \Sigma^{-1}
        \left\{
            \operatorname{vec}(y_l)
            -
            M(\kappa_l)\operatorname{vec}(\widehat\phi)
        \right\}
        -
        M(\kappa_l)^\top\Sigma^{-1}
        \operatorname{vec}(\widehat\phi)
    \Big]\nonumber \\ 
    &=
    (\kappa\diamond_\Sigma\kappa)^{-1}q_{\mathrm R,l}.
\end{align}

The corresponding derivative with respect to \(\kappa_l^{\mathrm I}\) is calculated in the
same way. In this case,
\[
    \frac{\partial(\kappa\bullet_\Sigma y)}{\partial \kappa_l^{\mathrm I}}
    =
    J_2^\top\Sigma^{-1}\operatorname{vec}(y_l)
\text{ and }
    \frac{\partial(\kappa\diamond_\Sigma\kappa)}{\partial \kappa_l^{\mathrm I}}
    =
    J_2^\top\Sigma^{-1}M(\kappa_l)
    +
    M(\kappa_l)^\top\Sigma^{-1}J_2.
\]
Hence,
\begin{align}\label{supp:eq: phi derivative c}
    \frac{\partial\operatorname{vec}(\widehat\phi)}
    {\partial \kappa_l^{\mathrm I}}
    &= (\kappa\diamond_\Sigma\kappa)^{-1}
    J_2^\top\Sigma^{-1}\operatorname{vec}(y_l)
    -
    (\kappa\diamond_\Sigma\kappa)^{-1}
    \left\{
        J_2^\top\Sigma^{-1}M(\kappa_l)
        +
        M(\kappa_l)^\top\Sigma^{-1}J_2
    \right\}
    (\kappa\diamond_\Sigma\kappa)^{-1}
    (\kappa\bullet_\Sigma y)\nonumber \\
    &=
    (\kappa\diamond_\Sigma\kappa)^{-1}
    \Big[
        J_2^\top\Sigma^{-1}\operatorname{vec}(y_l)
        -
        \left\{
            J_2^\top\Sigma^{-1}M(\kappa_l)
            +
            M(\kappa_l)^\top\Sigma^{-1}J_2
        \right\}
        \operatorname{vec}(\widehat\phi)
    \Big]\nonumber \\
    &=
    (\kappa\diamond_\Sigma\kappa)^{-1}
    \Big[
        J_2^\top\Sigma^{-1}
        \left\{
            \operatorname{vec}(y_l)
            -
            M(\kappa_l)\operatorname{vec}(\widehat\phi)
        \right\}
        -
        M(\kappa_l)^\top\Sigma^{-1}J_2
        \operatorname{vec}(\widehat\phi)
    \Big]\nonumber \\
    &=
    (\kappa\diamond_\Sigma\kappa)^{-1}q_{\mathrm I,l}.
\end{align}

We now use \eqref{supp:eq: phi derivative r} and
\eqref{supp:eq: phi derivative c} to differentiate the first-order formula.
For notational convenience, set
\(
    D_{\mathrm R}:=\mathrm{Id}_2,
    D_{\mathrm I}:=J_2
\) for \(\alpha\in\{\mathrm R,\mathrm I\}\). Then \eqref{supp:eq: explicit kappa gradient stacked} can be written as
\begin{equation}
\label{supp:eq: unified first derivative kappa}
    \frac{\partial\tilde\rho}{\partial\kappa_j^{\alpha}}
    =
    -2r_j^\top\Sigma^{-1}
    D_\alpha\operatorname{vec}(\widehat\phi).
\end{equation}
For \(\beta\in\{\mathrm R,\mathrm I\}\), we also get 
\[
    \frac{\partial r_j}{\partial\kappa_l^{\beta}}
    =
    -
    \mathbf 1_{\{j=l\}}
    D_\beta\operatorname{vec}(\widehat\phi)
    -
    M(\kappa_j)
    \frac{\partial\operatorname{vec}(\widehat\phi)}
    {\partial\kappa_l^{\beta}}.
\]
Differentiating \eqref{supp:eq: unified first derivative kappa} with respect to
\(\kappa_l^{\beta}\), we obtain
\begin{align*}
\frac{\partial^2\tilde\rho}
{\partial\kappa_j^{\alpha}\,\partial\kappa_l^{\beta}}
&=
-2
\left(
    \frac{\partial r_j}{\partial\kappa_l^{\beta}}
\right)^\top
\Sigma^{-1}
D_\alpha\operatorname{vec}(\widehat\phi)
-
2r_j^\top\Sigma^{-1}D_\alpha
\frac{\partial\operatorname{vec}(\widehat\phi)}
{\partial\kappa_l^{\beta}}\\
&=
2\times \mathbf 1_{\{j=l\}}
\left\{
    D_\beta\operatorname{vec}(\widehat\phi)
\right\}^\top
\Sigma^{-1}
D_\alpha\operatorname{vec}(\widehat\phi)\\
&\quad+
2
\left(
    \frac{\partial\operatorname{vec}(\widehat\phi)}
    {\partial\kappa_l^{\beta}}
\right)^\top
\left[
    M(\kappa_j)^\top\Sigma^{-1}
    D_\alpha\operatorname{vec}(\widehat\phi)
    -
    D_\alpha^\top\Sigma^{-1}r_j
\right]\\
&=
2\times \mathbf 1_{\{j=l\}}
\left\{
    D_\beta\operatorname{vec}(\widehat\phi)
\right\}^\top
\Sigma^{-1}
D_\alpha\operatorname{vec}(\widehat\phi)
-
2
\left(
    \frac{\partial\operatorname{vec}(\widehat\phi)}
    {\partial\kappa_l^{\beta}}
\right)^\top
q_{\alpha,j}.
\end{align*}
Here, the second equality follows by substituting the expression for
\(\partial r_j/\partial\kappa_l^{\beta}\) and collecting the two terms involving
\(\partial\operatorname{vec}(\widehat\phi)/\partial\kappa_l^{\beta}\). The last
equality follows from the definitions of \(q_{\mathrm R,j}\) and \(q_{\mathrm I,j}\), since
\[
    M(\kappa_j)^\top\Sigma^{-1}
    D_\alpha\operatorname{vec}(\widehat\phi)
    -
    D_\alpha^\top\Sigma^{-1}r_j
    =
    -q_{\alpha,j}.
\]

Substituting the intermediate derivatives
\eqref{supp:eq: phi derivative r} and
\eqref{supp:eq: phi derivative c}, it follows that
\begin{equation}
\label{supp:eq: kappa Hessian component}
\begin{aligned}
\frac{\partial^2\tilde\rho}
{\partial\kappa_j^{\alpha}\,\partial\kappa_l^{\beta}}
&=
2\times \mathbf 1_{\{j=l\}}
\left\{
    D_\beta\operatorname{vec}(\widehat\phi)
\right\}^\top
\Sigma^{-1}
D_\alpha\operatorname{vec}(\widehat\phi)-
2q_{\beta,l}^\top
(\kappa\diamond_\Sigma\kappa)^{-1}
q_{\alpha,j}.
\end{aligned}
\end{equation}

It remains to stack the componentwise Hessian entries in
\eqref{supp:eq: kappa Hessian component}. Since
\(
    M(\widehat\phi)
    =
    \begin{pmatrix}
        \operatorname{vec}(\widehat\phi)
        &
        J_2\operatorname{vec}(\widehat\phi)
    \end{pmatrix},
\)
the first term in \eqref{supp:eq: kappa Hessian component}, stacked over
\(\alpha,\beta\in\{\mathrm R,\mathrm I\}\) and \(j,l=1,\ldots,p\), is
\[
    2\left\{
        M(\widehat\phi)^\top
        \Sigma^{-1}
        M(\widehat\phi)
        \otimes
        \mathrm{Id}_p
    \right\}.
\]
Moreover, since the first \(p\) rows of \(\mathbf Q^\top\) are
\(q_{\mathrm R,1}^\top,\ldots,q_{\mathrm R,p}^\top\), and the remaining \(p\) rows are
\(q_{\mathrm I,1}^\top,\ldots,q_{\mathrm I,p}^\top\), the second term in
\eqref{supp:eq: kappa Hessian component} becomes
\[
    -2\mathbf Q^\top
    (\kappa\diamond_\Sigma\kappa)^{-1}
    \mathbf Q.
\]
Combining the two stacked terms gives
\[
\nabla^2_{\widetilde{\operatorname{vec}}(\kappa)}
\tilde\rho(y,\kappa,\Sigma)
=
2\left[
M\bigl(\widehat\phi(y,\kappa,\Sigma)\bigr)^\top
\Sigma^{-1}
M\bigl(\widehat\phi(y,\kappa,\Sigma)\bigr)
\otimes\mathrm{Id}_p
-
\mathbf Q^\top
(\kappa\diamond_\Sigma\kappa)^{-1}
\mathbf Q
\right],
\]
which shows \eqref{supp:eq: explicit kappa Hessian}.

\medskip
\noindent\textit{Proof of (iv).}
We derive the mixed derivative in two steps. First, we calculate the
directional derivatives of
\(\operatorname{vec}(\widehat\phi)\) and \(r_j\) with respect to
\(\Sigma\). We then use these intermediate calculations to differentiate
\eqref{supp:eq: explicit kappa gradient stacked}
with respect to \(\Sigma\).

Let \(\Delta\in\mathbb R^{2\times2}\) be symmetric. We first differentiate
\[
    \operatorname{vec}(\widehat\phi)
    =
    (\kappa\diamond_\Sigma\kappa)^{-1}
    (\kappa\bullet_\Sigma y)
\]
with respect to \(\Sigma\) in the direction \(\Delta\). Since
\[
\begin{aligned}
    D_\Sigma
    (\kappa\diamond_\Sigma\kappa)[\Delta]
    &=
    -
    \sum_{l=1}^p
    M(\kappa_l)^\top
    \Sigma^{-1}\Delta\Sigma^{-1}
    M(\kappa_l),\\
    D_\Sigma
    (\kappa\bullet_\Sigma y)[\Delta]
    &=
    -
    \sum_{l=1}^p
    M(\kappa_l)^\top
    \Sigma^{-1}\Delta\Sigma^{-1}
    \operatorname{vec}(y_l),
\end{aligned}
\]
the matrix inverse derivative formula gives
\begin{align}\label{supp:eq: phi Sigma directional derivative}
&D_\Sigma
\operatorname{vec}(\widehat\phi)[\Delta] \nonumber \\
&=
-
(\kappa\diamond_\Sigma\kappa)^{-1}
D_\Sigma(\kappa\diamond_\Sigma\kappa)[\Delta]
(\kappa\diamond_\Sigma\kappa)^{-1}
(\kappa\bullet_\Sigma y)+
(\kappa\diamond_\Sigma\kappa)^{-1}
D_\Sigma(\kappa\bullet_\Sigma y)[\Delta] \nonumber\\
&=
(\kappa\diamond_\Sigma\kappa)^{-1}
\sum_{l=1}^p
M(\kappa_l)^\top
\Sigma^{-1}\Delta\Sigma^{-1}
M(\kappa_l)
\operatorname{vec}(\widehat\phi)-
(\kappa\diamond_\Sigma\kappa)^{-1}
\sum_{l=1}^p
M(\kappa_l)^\top
\Sigma^{-1}\Delta\Sigma^{-1}
\operatorname{vec}(y_l)\nonumber \\
&=
-
(\kappa\diamond_\Sigma\kappa)^{-1}
\sum_{l=1}^p
M(\kappa_l)^\top
\Sigma^{-1}\Delta\Sigma^{-1}
\left\{
    \operatorname{vec}(y_l)
    -
    M(\kappa_l)\operatorname{vec}(\widehat\phi)
\right\} \nonumber \\
&=
-
(\kappa\diamond_\Sigma\kappa)^{-1}
\sum_{l=1}^p
M(\kappa_l)^\top
\Sigma^{-1}\Delta\Sigma^{-1}r_l.
\end{align}

We next use this result to calculate the directional derivative of the
residual. Since \(y_j\) and \(\kappa_j\) do not depend on \(\Sigma\),
\[
    D_\Sigma r_j[\Delta]
=
-
M(\kappa_j)
D_\Sigma
\operatorname{vec}(\widehat\phi)[\Delta].
\]
Using \eqref{supp:eq: phi Sigma directional derivative}, we obtain
\begin{equation}
\label{supp:eq: residual Sigma directional derivative}
\begin{aligned}
    D_\Sigma r_j[\Delta]
    =
    M(\kappa_j)
    (\kappa\diamond_\Sigma\kappa)^{-1}
    \sum_{l=1}^p
    M(\kappa_l)^\top
    \Sigma^{-1}\Delta\Sigma^{-1}r_l.
\end{aligned}
\end{equation}

We now substitute these intermediate derivatives into \eqref{supp:eq: explicit kappa gradient stacked}. For \(\alpha\in\{\mathrm R,\mathrm I\}\), recall that
\(
    D_{\mathrm R}=\mathrm{Id}_2,
    D_{\mathrm I}=J_2,
\)
and
\(
    \frac{\partial\tilde\rho}{\partial\kappa_j^{\alpha}}
    =
    -2r_j^\top
    \Sigma^{-1}
    D_\alpha
    \operatorname{vec}(\widehat\phi).
\)
Differentiating $\frac{\partial\tilde\rho}{\partial\kappa_j^{\alpha}}$ with respect to \(\Sigma\) in the direction \(\Delta\) gives
\begin{align*}
&D_\Sigma
\left(
    \frac{\partial\tilde\rho}{\partial\kappa_j^{\alpha}}
\right)[\Delta]\\
&=
-2
\left(
    D_\Sigma r_j[\Delta]
\right)^\top
\Sigma^{-1}
D_\alpha
\operatorname{vec}(\widehat\phi)-2r_j^\top
D_\Sigma \Sigma^{-1}[\Delta]
D_\alpha
\operatorname{vec}(\widehat\phi)-
2r_j^\top
\Sigma^{-1}D_\alpha
D_\Sigma
\operatorname{vec}(\widehat\phi)[\Delta]\\
&=
-2
\left(
    D_\Sigma r_j[\Delta]
\right)^\top
\Sigma^{-1}
D_\alpha
\operatorname{vec}(\widehat\phi)+2r_j^\top
\Sigma^{-1} \Delta \Sigma^{-1}
D_\alpha
\operatorname{vec}(\widehat\phi)-
2r_j^\top
\Sigma^{-1}D_\alpha
D_\Sigma
\operatorname{vec}(\widehat\phi)[\Delta].
\end{align*}
Here, the first equality follows from the product rule, while the second
equality uses
\[
    D_\Sigma\Sigma^{-1}[\Delta]
    =
    -\Sigma^{-1}\Delta\Sigma^{-1}.
\]

Substituting
\eqref{supp:eq: phi Sigma directional derivative} and
\eqref{supp:eq: residual Sigma directional derivative}, we obtain
\begin{equation}
\label{supp:eq: mixed derivative component}
\begin{aligned}
&D_\Sigma
\left(
    \frac{\partial\tilde\rho}{\partial\kappa_j^{\alpha}}
\right)[\Delta]\\
&=
2r_j^\top
\Sigma^{-1}\Delta\Sigma^{-1}
D_\alpha
\operatorname{vec}(\widehat\phi)\\
&\quad-
2
\left[
    \sum_{l=1}^p
    M(\kappa_l)^\top
    \Sigma^{-1}\Delta\Sigma^{-1}r_l
\right]^\top
(\kappa\diamond_\Sigma\kappa)^{-1}
M(\kappa_j)^\top
\Sigma^{-1}
D_\alpha
\operatorname{vec}(\widehat\phi)\\
&\quad+
2r_j^\top
\Sigma^{-1}D_\alpha
(\kappa\diamond_\Sigma\kappa)^{-1}
\sum_{l=1}^p
M(\kappa_l)^\top
\Sigma^{-1}\Delta\Sigma^{-1}r_l\\
&=
2r_j^\top
\Sigma^{-1}\Delta\Sigma^{-1}
D_\alpha
\operatorname{vec}(\widehat\phi)\\
&\quad+
2
\left[
    D_\alpha^\top\Sigma^{-1}r_j
    -
    M(\kappa_j)^\top
    \Sigma^{-1}
    D_\alpha
    \operatorname{vec}(\widehat\phi)
\right]^\top
(\kappa\diamond_\Sigma\kappa)^{-1}
\sum_{l=1}^p
M(\kappa_l)^\top
\Sigma^{-1}\Delta\Sigma^{-1}r_l\\
&=
2r_j^\top
\Sigma^{-1}\Delta\Sigma^{-1}
D_\alpha
\operatorname{vec}(\widehat\phi)
+
2q_{\alpha,j}^\top
(\kappa\diamond_\Sigma\kappa)^{-1}
\sum_{l=1}^p
M(\kappa_l)^\top
\Sigma^{-1}\Delta\Sigma^{-1}r_l.
\end{aligned}
\end{equation}
Here, the second equality follows from the symmetry of
\(\kappa\diamond_\Sigma\kappa\), which allows the last two terms in the
first expression to be collected. The final equality follows from the
definitions of \(q_{\mathrm R,j}\) and \(q_{\mathrm I,j}\), since
\[
    D_\alpha^\top\Sigma^{-1}r_j
    -
    M(\kappa_j)^\top
    \Sigma^{-1}
    D_\alpha
    \operatorname{vec}(\widehat\phi)
    =
    q_{\alpha,j}.
\]

Stacking
\eqref{supp:eq: mixed derivative component} first over
\(j=1,\ldots,p\) for \(\alpha=\mathrm R\), and then over \(j=1,\ldots,p\) for
\(\alpha=\mathrm I\), gives
\begin{align*}
&D_\Sigma
\left\{
    \nabla_{\widetilde{\operatorname{vec}}(\kappa)}
    \tilde\rho(y,\kappa,\Sigma)
\right\}[\Delta]\\
&=
2
\begin{pmatrix}
\mathbf E\Sigma^{-1}\Delta\Sigma^{-1}
\operatorname{vec}(\widehat\phi)
\\
\mathbf E\Sigma^{-1}\Delta\Sigma^{-1}J_2
\operatorname{vec}(\widehat\phi)
\end{pmatrix}+
2\mathbf Q^\top
(\kappa\diamond_\Sigma\kappa)^{-1}
\sum_{l=1}^p
M(\kappa_l)^\top
\Sigma^{-1}\Delta\Sigma^{-1}r_l\\
&=
\mathbf C_\Sigma(\Delta).
\end{align*}

Finally, we evaluate this directional derivative along the independent
coordinate directions of the symmetric matrix \(\Sigma\). The independent coordinates of the symmetric matrix \(\Sigma\) are
\(\sigma_{11},\sigma_{12},\sigma_{22}\), with corresponding directions
\[
    \begin{pmatrix}
        1&0\\
        0&0
    \end{pmatrix},
    \quad
    \begin{pmatrix}
        0&1\\
        1&0
    \end{pmatrix},
    \quad
    \begin{pmatrix}
        0&0\\
        0&1
    \end{pmatrix}.
\]
Evaluating \(\mathbf C_\Sigma(\Delta)\) at these three directions gives
\[
\nabla^2_{
    \widetilde{\operatorname{vec}}(\kappa),
    \operatorname{vech}(\Sigma)
}
\tilde\rho(y,\kappa,\Sigma)
=
\begin{pmatrix}
    \mathbf C_\Sigma
    \left(
        \begin{pmatrix}
            1&0\\
            0&0
        \end{pmatrix}
    \right)
    &
    \mathbf C_\Sigma
    \left(
        \begin{pmatrix}
            0&1\\
            1&0
        \end{pmatrix}
    \right)
    &
    \mathbf C_\Sigma
    \left(
        \begin{pmatrix}
            0&0\\
            0&1
        \end{pmatrix}
    \right)
\end{pmatrix},
\]
which shows
\eqref{supp:eq: explicit kappa Sigma mixed derivative symmetric}.
\end{proof}

\subsubsection{Verification of the GFM conditions}
\label{supp:sec:plugin-gfm-condition-verification}

The strong consistency result established above verifies Condition~\ref{supp:cond:gfm-primary-consistency}. Equation~\eqref{eq:covariance-influence-representation}, together with
Lemma~\ref{supp:lem:plugin-covariance-influence-moments} below, verifies Condition~\ref{supp:cond:gfm-auxiliary-linearity}; its \(\mathcal O_p(n^{-1})\) remainder also verifies the second requirement
of Condition~\ref{supp:cond:gfm-quadratic-refinement}. We now verify the remaining model-specific requirements in Conditions~\ref{supp:cond:gfm-geometric-constraint}, \ref{supp:cond:gfm-cost-differentiability}, and \ref{supp:cond:gfm-kkt-invertibility}, together with the first and third
requirements of Condition~\ref{supp:cond:gfm-quadratic-refinement}.

\begin{lemma}
\label{supp:lem:plugin-geometric-constraint}
For the \texttt{plug-in} estimator, set
\[
    \mathbb J_p
    :=
    J_2\otimes\mathrm{Id}_p
    =
    \begin{pmatrix}
        0&-\mathrm{Id}_p\\
        \mathrm{Id}_p&0
    \end{pmatrix},
\]
and define \(h:\mathbb R^{2p}\to\mathbb R^2\) by
\[
    h_1(\mu)
    =
    \frac12(\mu^\top\mu-1),
    \quad
    h_2(\mu)
    =
    (\mathbb J_p\mu_0)^\top\mu.
\]
Then Condition~\ref{supp:cond:gfm-geometric-constraint} holds with
\(K=2p\) and \(P=2\). In particular,
\begin{equation}
\label{supp:eq:plugin-H-definition}
    H
    :=
    Dh(\mu_0)^\top
    =
    \begin{pmatrix}
        \operatorname{Re}(\kappa^{(0)})
        &
        -\operatorname{Im}(\kappa^{(0)})
        \\
        \operatorname{Im}(\kappa^{(0)})
        &
        \operatorname{Re}(\kappa^{(0)})
    \end{pmatrix}
    \in\mathbb R^{2p\times2},
\end{equation}
and \(H\) has full column rank. Moreover, each \(h_j\) is three
times continuously differentiable on \(\mathbb R^{2p}\), and its
third derivative is identically zero. Hence, the first requirement in Condition~\ref{supp:cond:gfm-quadratic-refinement} also holds.
\end{lemma}

Let
\[
    \mathcal V_0
    :=
    \left\{
        [\kappa]\in\mathbb{CP}^{p-1}:
        \left|\kappa^*\kappa^{(0)}\right|>0
    \right\}.
\]
For \([\kappa]\in\mathcal V_0\), let \(\kappa\in[\kappa]\) be any
unit-norm representative and define
\[
    \kappa^{\mathrm{opt}}([\kappa])
    :=
    \frac{
        \kappa^*\kappa^{(0)}
    }{
        |\kappa^*\kappa^{(0)}|
    }\kappa,
    \quad
    \mathcal{F}([\kappa])
    :=
    \widetilde{\operatorname{vec}}
    \left(
        \kappa^{\mathrm{opt}}([\kappa])
    \right).
\]
Also define
\[
    \mathcal{G}(\Sigma):=\operatorname{vech}(\Sigma),
    \quad
    \mu_0
    :=
    \widetilde{\operatorname{vec}}(\kappa^{(0)}),
    \quad
    \nu_0
    :=
    \operatorname{vech}(\Sigma^{(0)}).
\]
Then
\[
    \mathcal{F}(\mathcal V_0)
    =
    \left\{
        \mu\in\mathbb R^{2p}:
        h(\mu)=0,\ 
        \mu_0^\top\mu>0
    \right\},
\]
which is an open subset of the constraint manifold
\(
S=\{\mu\in\mathbb R^{2p}:h(\mu)=0\}.
\)
Moreover, \(\mathcal{F}\) and \(\mathcal{G}\) are smooth diffeomorphisms onto their respective images.

\begin{proof}[Proof of Lemma~\ref{supp:lem:plugin-geometric-constraint}]
The map \(h\) is infinitely continuously differentiable. Moreover,
\(h_1\) is quadratic and \(h_2\) is linear, so
\[
    D^3h_1(\mu)
    =
    D^3h_2(\mu)
    =
    0
    \quad
    \text{for every }\mu\in\mathbb R^{2p}.
\]
Thus, their third derivatives are locally bounded.

Since
\[
    \mu_0
    =
    \begin{pmatrix}
        \operatorname{Re}(\kappa^{(0)})\\
        \operatorname{Im}(\kappa^{(0)})
    \end{pmatrix},
    \quad
    \mathbb J_p\mu_0
    =
    \begin{pmatrix}
        -\operatorname{Im}(\kappa^{(0)})\\
        \operatorname{Re}(\kappa^{(0)})
    \end{pmatrix},
\]
we obtain~\eqref{supp:eq:plugin-H-definition}.

Since \(\|\kappa^{(0)}\|=1\),
\[
    \operatorname{Re}(\kappa^{(0)})^\top \operatorname{Re}(\kappa^{(0)})
    +
    \operatorname{Im}(\kappa^{(0)})^\top \operatorname{Im}(\kappa^{(0)})
    =
    1.
\]
Moreover,
\[
    h_1(\mu_0)
    =
    \frac12
    \left(
        \operatorname{Re}(\kappa^{(0)})^\top \operatorname{Re}(\kappa^{(0)})
        +
        \operatorname{Im}(\kappa^{(0)})^\top \operatorname{Im}(\kappa^{(0)})
        -
        1
    \right)
    =
    0,
\]
and
\[
    h_2(\mu_0)
    =
    \begin{pmatrix}
        -\operatorname{Im}(\kappa^{(0)})\\
        \operatorname{Re}(\kappa^{(0)})
    \end{pmatrix}^{\!\top}
    \begin{pmatrix}
        \operatorname{Re}(\kappa^{(0)})\\
        \operatorname{Im}(\kappa^{(0)})
    \end{pmatrix}
    =
    -\operatorname{Im}(\kappa^{(0)})^\top \operatorname{Re}(\kappa^{(0)})
    +
    \operatorname{Re}(\kappa^{(0)})^\top \operatorname{Im}(\kappa^{(0)})
    =
    0.
\]
Thus \(h(\mu_0)=0\). Furthermore,
\[
\begin{aligned}
    H^\top H
    &=
    \begin{pmatrix}
        \operatorname{Re}(\kappa^{(0)})^\top \operatorname{Re}(\kappa^{(0)})
        +
        \operatorname{Im}(\kappa^{(0)})^\top \operatorname{Im}(\kappa^{(0)})
        &
        -\operatorname{Re}(\kappa^{(0)})^\top \operatorname{Im}(\kappa^{(0)})
        +
        \operatorname{Im}(\kappa^{(0)})^\top \operatorname{Re}(\kappa^{(0)})
        \\
        -\operatorname{Im}(\kappa^{(0)})^\top \operatorname{Re}(\kappa^{(0)})
        +
        \operatorname{Re}(\kappa^{(0)})^\top \operatorname{Im}(\kappa^{(0)})
        &
        \operatorname{Im}(\kappa^{(0)})^\top \operatorname{Im}(\kappa^{(0)})
        +
        \operatorname{Re}(\kappa^{(0)})^\top \operatorname{Re}(\kappa^{(0)})
    \end{pmatrix}\\
    &=
    \mathrm{Id}_2.
\end{aligned}
\]
Therefore, the two columns of \(H\) are orthonormal, and \(H\) has full column rank.
\end{proof}

\begin{lemma}[Moments of the covariance influence function]
\label{supp:lem:plugin-covariance-influence-moments}
Assume model~\eqref{model:centered-complex-factor} with \(p\geq3\),
true parameters
\(\kappa^{(0)}\in\mathcal S^{2p-1}\) and
\(\Sigma^{(0)}\in\mathrm{SPD}(2)\), and
\(0<\mathbb E[|\phi_1|^2]<\infty\). Suppose that
\(\operatorname{Re}(\kappa^{(0)})\) and
\(\operatorname{Im}(\kappa^{(0)})\) are linearly independent in
\(\mathbb R^p\). Let \(\xi_\Sigma\) be the covariance influence function
defined in \eqref{eq:covariance-influence-function}. Then
\begin{equation}
\label{supp:eq:plugin-covariance-influence-moments}
    \mathbb E
    \left[
        \xi_\Sigma(Y_1)
    \right]
    =
    0,
    \quad
    \mathbb E
    \left[
        \left\|
            \xi_\Sigma(Y_1)
        \right\|^2
    \right]
    <
    \infty.
\end{equation}
Moreover,
\begin{equation}
\label{supp:eq:plugin-covariance-influence-variance}
\operatorname{Var}
\left(
    \xi_\Sigma(Y_1)
\right)
=
\frac{1}{p-2}
\begin{pmatrix}
    2\sigma_x^4
    &
    2\sigma_x^2\sigma_{xy}
    &
    2\sigma_{xy}^2
    \\
    2\sigma_x^2\sigma_{xy}
    &
    \sigma_x^2\sigma_y^2+\sigma_{xy}^2
    &
    2\sigma_{xy}\sigma_y^2
    \\
    2\sigma_{xy}^2
    &
    2\sigma_{xy}\sigma_y^2
    &
    2\sigma_y^4
\end{pmatrix}
\end{equation}
where
$$
\Sigma^{(0)}=
\begin{pmatrix}\sigma_x^2&\sigma_{xy}\\\sigma_{xy}&\sigma_y^2\end{pmatrix}.
$$
\end{lemma}

\begin{proof}[Proof of
Lemma~\ref{supp:lem:plugin-covariance-influence-moments}]
Since
\[
    \Pi_{\perp,0}
    \operatorname{Re}(\kappa^{(0)})
    =
    \Pi_{\perp,0}
    \operatorname{Im}(\kappa^{(0)})
    =
    0,
\]
model~\eqref{model:centered-complex-factor} implies
\[
\begin{aligned}
    \Pi_{\perp,0}\operatorname{Re}(Y_1)
    &=
    \Pi_{\perp,0}\operatorname{Re}(\varepsilon_1),\\
    \Pi_{\perp,0}\operatorname{Im}(Y_1)
    &=
    \Pi_{\perp,0}\operatorname{Im}(\varepsilon_1).
\end{aligned}
\]
Consequently, the definition in
\eqref{eq:covariance-influence-function} can be written as
\begin{equation}
\label{supp:eq:plugin-covariance-influence-coordinate-form}
\xi_\Sigma(Y_1)
=
\begin{pmatrix}
    \dfrac{1}{p-2}
    \operatorname{Re}(Y_1)^\top
    \Pi_{\perp,0}
    \operatorname{Re}(Y_1)
    -
    \sigma_x^2
    \\[2mm]
    \dfrac{1}{p-2}
    \operatorname{Re}(Y_1)^\top
    \Pi_{\perp,0}
    \operatorname{Im}(Y_1)
    -
    \sigma_{xy}
    \\[2mm]
    \dfrac{1}{p-2}
    \operatorname{Im}(Y_1)^\top
    \Pi_{\perp,0}
    \operatorname{Im}(Y_1)
    -
    \sigma_y^2
\end{pmatrix}.
\end{equation}

The linear independence assumption gives
\[
    \operatorname{rank}(\Pi_{\perp,0})=p-2.
\]
Choose \(U_{\perp,0}\in\mathbb R^{p\times(p-2)}\) such that
\[
    U_{\perp,0}^\top U_{\perp,0}
    =
    \mathrm{Id}_{p-2},
    \quad
    U_{\perp,0}U_{\perp,0}^\top
    =
    \Pi_{\perp,0},
\]
and define
\[
    Z_R
    :=
    U_{\perp,0}^\top\operatorname{Re}(\varepsilon_1),
    \quad
    Z_I
    :=
    U_{\perp,0}^\top\operatorname{Im}(\varepsilon_1).
\]
Then
\[
    \operatorname{Cov}
    \begin{pmatrix}
        Z_R\\
        Z_I
    \end{pmatrix}
    =
    \Sigma^{(0)}
    \otimes
    \mathrm{Id}_{p-2}.
\]
Hence the vectors
\[
    \begin{pmatrix}
        Z_{R,l}\\
        Z_{I,l}
    \end{pmatrix},
    \quad
    l=1,\ldots,p-2,
\]
are independent and identically distributed as
\(\mathcal N_2(0,\Sigma^{(0)})\). It follows from
\eqref{supp:eq:plugin-covariance-influence-coordinate-form} that
\begin{equation}
\label{supp:eq:plugin-covariance-influence-iid-decomposition}
\xi_\Sigma(Y_1)
=
\frac{1}{p-2}
\sum_{l=1}^{p-2}
\begin{pmatrix}
    Z_{R,l}^2-\sigma_x^2\\
    Z_{R,l}Z_{I,l}-\sigma_{xy}\\
    Z_{I,l}^2-\sigma_y^2
\end{pmatrix}.
\end{equation}
This representation immediately gives
\[
    \mathbb E[\xi_\Sigma(Y_1)]=0.
\]

For a generic vector
\[
    \begin{pmatrix}Z_R^{(1)}\\Z_I^{(1)}\end{pmatrix}
    \sim
    \mathcal N_2(0,\Sigma^{(0)}),
\]
Isserlis' formula gives
\[
\operatorname{Var}
\begin{pmatrix}
    (Z_R^{(1)})^2-\sigma_x^2\\
    Z_R^{(1)}Z_I^{(1)}-\sigma_{xy}\\
    (Z_I^{(1)})^2-\sigma_y^2
\end{pmatrix}
=
\begin{pmatrix}
    2\sigma_x^4
    &
    2\sigma_x^2\sigma_{xy}
    &
    2\sigma_{xy}^2
    \\
    2\sigma_x^2\sigma_{xy}
    &
    \sigma_x^2\sigma_y^2+\sigma_{xy}^2
    &
    2\sigma_{xy}\sigma_y^2
    \\
    2\sigma_{xy}^2
    &
    2\sigma_{xy}\sigma_y^2
    &
    2\sigma_y^4
\end{pmatrix}.
\]
Using the independence of the summands in
\eqref{supp:eq:plugin-covariance-influence-iid-decomposition} proves
\eqref{supp:eq:plugin-covariance-influence-variance}. The entries of this
matrix are finite, so
\(\mathbb E[\|\xi_\Sigma(Y_1)\|^2]<\infty\), completing the proof.
\end{proof}

\begin{lemma}
\label{supp:lem:plugin-derivative-domination}
There exist a compact convex neighborhood \(\mathcal K\) of
\((\mu_0,\nu_0)\) and a constant \(C>0\) such that
\[
\begin{aligned}
&\sup_{(\mu,\nu)\in\mathcal K}
\left\|
    \nabla_\mu\tilde{\ell}(Y,\mu,\nu)
\right\|,~\sup_{(\mu,\nu)\in\mathcal K}
\left\|
    \nabla_{\mu\mu}^2\tilde{\ell}(Y,\mu,\nu)
\right\|,~\sup_{(\mu,\nu)\in\mathcal K}
\left\|
    \nabla_{\mu\nu}^2\tilde{\ell}(Y,\mu,\nu)
\right\|
\end{aligned}
\]
are bounded above by
\[
    D(Y):=C\texttt{max}~\{\|Y\|, \|Y\|^2\}
    \quad\text{a.s.}
\]
Consequently, Condition~\ref{supp:cond:gfm-cost-differentiability} holds for the \texttt{plug-in} estimator whenever \(\mathbb E[|\phi_1|^2]<\infty\). If, in addition, \(\mathbb E[|\phi_1|^4]<\infty\), then
\[
    \mathbb E[D(Y)^2]<\infty.
\]
\end{lemma}

\begin{proof}[Proof of Lemma~\ref{supp:lem:plugin-derivative-domination}]
The ambient extension is
\[
    \tilde{\ell}(y,\mu,\nu)
    =
    \tilde\rho(y,\kappa,\Sigma),
    \quad
    \mu
    =
    \widetilde{\operatorname{vec}}(\kappa),
    \quad
    \nu
    =
    \operatorname{vech}(\Sigma).
\]
It is twice continuously differentiable on
\[
    \left(
        \mathbb R^{2p}\setminus\{0\}
    \right)
    \times
    \operatorname{vech}
    \left\{
        \mathrm{SPD}(2)
    \right\},
\]
since \(\kappa\diamond_\Sigma\kappa\) is positive definite for every nonzero \(\kappa\) and every \(\Sigma\in\mathrm{SPD}(2)\), as shown in Lemma~\ref{supp:lem: diamond_invertible}.

Choose \(\mathcal K\) sufficiently small that there exist constants
\(0<c_1<c_2<\infty\) satisfying
\[
    c_1
    \le
    \|\kappa\|
    \le
    c_2,
    \quad
    c_1\mathrm{Id}_2
    \preceq
    \Sigma
    \preceq
    c_2\mathrm{Id}_2
\]
for every \((\mu,\nu)\in\mathcal K\). Then clearly
\[
    \sup_{(\mu,\nu)\in\mathcal K}
    \|\Sigma^{-1}\|_{\operatorname{op}}
    <
    \infty.
\]
Moreover, for every \(v\in\mathbb R^2\),
\begin{align*}
    v^\top
    (\kappa\diamond_\Sigma\kappa)
    v
    &=
    \sum_{j=1}^p
    \left\{
        M(\kappa_j)v
    \right\}^\top
    \Sigma^{-1}
    \left\{
        M(\kappa_j)v
    \right\}\ge
    \lambda_{\min}(\Sigma^{-1})
    \|\kappa\|^2
    \|v\|^2.
\end{align*}
Hence, we get
\[
    \sup_{(\mu,\nu)\in\mathcal K}
    \left\|
        (\kappa\diamond_\Sigma\kappa)^{-1}
    \right\|_{\operatorname{op}}
    <
    \infty.
\]

Since
\(
    \operatorname{vec}
    \bigl(
        \widehat\phi(y,\kappa,\Sigma)
    \bigr)
    =
    (\kappa\diamond_\Sigma\kappa)^{-1}
    (\kappa\bullet_\Sigma Y),
\)
the preceding bounds and the Cauchy--Schwarz inequality imply
\[
    \sup_{(\mu,\nu)\in\mathcal K}
    \left\|
        \operatorname{vec}
        \bigl(
            \widehat\phi(y,\kappa,\Sigma)
        \bigr)
    \right\|
    \le
    C\|Y\|.
\]
Since
\[
    r_j
    =
    \operatorname{vec}(Y_j)
    -
    M(\kappa_j)
    \operatorname{vec}
    \left(
        \widehat\phi(y,\kappa,\Sigma)
    \right),
\]
the matrix
\[
    \mathbf E
    =
    \begin{pmatrix}
        r_1&\cdots&r_p
    \end{pmatrix}^\top
    \in\mathbb R^{p\times2}
\]
satisfies
\[
    \sup_{(\mu,\nu)\in\mathcal K}
    \|\mathbf E\|_{F}
    \le
    C\|Y\|.
\]
By the definition of \(\mathbf Q\) in
\eqref{supp:eq: Q matrix definition}, we also have
\[
    \sup_{(\mu,\nu)\in\mathcal K}
    \|\mathbf Q\|_{F}
    \le
    C\|Y\|.
\]

Applying \eqref{supp:eq: explicit kappa gradient stacked} gives
\[
    \sup_{(\mu,\nu)\in\mathcal K}
    \left\|
        \nabla_\mu
        \tilde{\ell}(Y,\mu,\nu)
    \right\|
    \le
    C\|Y\|^2.
\]
Similarly, every term in \eqref{supp:eq: explicit kappa Hessian} is bounded by
a constant multiple of \(\|Y\|^2\), and hence
\[
    \sup_{(\mu,\nu)\in\mathcal K}
    \left\|
        \nabla_{\mu\mu}^2
        \tilde{\ell}(Y,\mu,\nu)
    \right\|
    \le
    C\|Y\|^2.
\]
Finally, every term in \eqref{supp:eq: C Sigma definition} is a product of two
quantities whose norms are bounded by a constant multiple of \(\|Y\|\).
Therefore,
\[
    \sup_{(\mu,\nu)\in\mathcal K}
    \left\|
        \nabla_{\mu\nu}^2
        \tilde{\ell}(Y,\mu,\nu)
    \right\|
    \le
    C\|Y\|^2.
\]

Under model~\eqref{model:centered-complex-factor},
\[
    \mathbb E\|Y\|^2
    =
    \mathbb E[|\phi_1|^2]
    +
    p\operatorname{Tr}
    \left(
        \Sigma^{(0)}
    \right)
    <
    \infty.
\]
Thus,
\[
    D(Y)
    =
    C\texttt{max}\{\|Y\|,\|Y\|^2\}
\]
is integrable, proving
Condition~\ref{supp:cond:gfm-cost-differentiability} for the
\texttt{plug-in} estimator.

If, in addition, \(\mathbb E[|\phi_1|^4]<\infty\), then, since
\[
    Y
    =
    \phi_1\kappa^{(0)}
    +
    \varepsilon_1
\]
and \(\|\kappa^{(0)}\|=1\),
\[
    \|Y\|^4
    \le
    8
    \left(
        |\phi_1|^4
        +
        \|\varepsilon_1\|^4
    \right).
\]
The Gaussian noise \(\varepsilon_1\) has a finite fourth moment, and hence
\[
    \mathbb E\|Y\|^4
    <
    \infty.
\]
Therefore,
\[
\begin{aligned}
    \mathbb E[D(Y)^2]
    &=
    C^2
    \mathbb E
    \left[
        \texttt{max}\{\|Y\|^2,\|Y\|^4\}
    \right]\\
    &\le
    C^2
    \mathbb E
    \left[
        \|Y\|^2+\|Y\|^4
    \right]
    <
    \infty.
\end{aligned}
\]
\end{proof}

\begin{lemma}
\label{supp:lem:plugin-population-kkt}
Assume model~\eqref{model:centered-complex-factor} with true parameters
\(\kappa^{(0)}\in\mathcal S^{2p-1}\) and
\(\Sigma^{(0)}\in\mathrm{SPD}(2)\), satisfying
\[
    0<\mathbb E[|\phi_1|^2]<\infty.
\]
For the \texttt{plug-in} estimator, the
population Lagrange multiplier is
\[
    \lambda_0=0.
\]

The expected Hessian of the constrained Lagrangian is
\begin{equation}
\label{supp:eq:plugin-J}
\begin{aligned}
    J
    =
    2\mathbb E\Bigg[
    &
    \left(
        M(\phi_1)
        \otimes
        \mathrm{Id}_p
    \right)^\top
    \left\{
        (\Sigma^{(0)})^{-1/2}
        \otimes
        \mathrm{Id}_p
    \right\}
    \left(
        \mathrm{Id}_{2p}
        -
        \mathcal P_0
    \right)
    \left\{
        (\Sigma^{(0)})^{-1/2}
        \otimes
        \mathrm{Id}_p
    \right\}
    \left(
        M(\phi_1)
        \otimes
        \mathrm{Id}_p
    \right)
    \Bigg],
\end{aligned}
\end{equation}
where 
\begin{equation}\label{supp:eq: P0 definition}
\mathcal P_0
    :=
    \left\{
        (\Sigma^{(0)})^{-1/2}
        \otimes
        \mathrm{Id}_p
    \right\}
    H
    \left(
        \kappa^{(0)}
        \diamond_{\Sigma^{(0)}}
        \kappa^{(0)}
    \right)^{-1}
    H^\top
    \left\{
        (\Sigma^{(0)})^{-1/2}
        \otimes
        \mathrm{Id}_p
    \right\}.    
\end{equation}

Moreover, 
\begin{equation}\label{supp:eq: null J = col H}
\operatorname{null}(J)=\operatorname{col}(H),
\end{equation} 
\(
    B
    =
    \begin{pmatrix}
        J&H\\
        H^\top&0
    \end{pmatrix}
\)
is nonsingular, with
\(
    B^{-1}
    =
    \begin{pmatrix}
        J^+&H\\
        H^\top&0
    \end{pmatrix},
\)
where \(J^+\) is the Moore--Penrose inverse of \(J\). In particular,
\(
    B_{11}=J^+.
\)
\end{lemma}

\begin{proof}[Proof of Lemma~\ref{supp:lem:plugin-population-kkt}]
The \texttt{plug-in} estimator has population auxiliary parameter
\[
    \nu_0
    =
    \operatorname{vech}
    \left(
        \Sigma^{(0)}
    \right).
\]
For every \(\kappa\neq0\), the ambient extension satisfies
\[
    \tilde\rho(y,\kappa,\Sigma)
    =
    \rho\left(
        y,
        [\kappa/\|\kappa\|],
        \Sigma
    \right).
\]
Indeed, rescaling \(\kappa\) is absorbed by a bijective rescaling
of the profiled scalar coefficient. Hence the population loss
decomposition and the expected noise-loss identity below hold
throughout an ambient neighborhood of \(\mu_0\).
By Lemma~\ref{supp:lem:plugin-population-loss-decomposition},
\[
\begin{aligned}
    \mathbb E
    \left[
        \tilde{\ell}(Y_1,\mu,\nu_0)
    \right]
    &=
    \mathbb E
    \left[
        \tilde{\ell}(
            \phi_1\kappa^{(0)},
            \mu,\nu_0
        )
    \right]
    +
    \mathbb E
    \left[
        \tilde{\ell}(
            \varepsilon_1,
            \mu,\nu_0
        )
    \right].
\end{aligned}
\]
Moreover, Lemma~\ref{supp:lem:plugin-expected-noise-loss} gives
\[
    \mathbb E
    \left[
        \tilde{\ell}(
            \varepsilon_1,\mu,\nu_0
        )
    \right]
    =
    2p-2.
\]
Thus, the expected noise contribution is constant in \(\mu\). By
Lemma~\ref{supp:lem:plugin-derivative-domination}, differentiation and
expectation may be interchanged, so
\[
    \mathbb E
    \left[
        \nabla_\mu
        \tilde{\ell}(
            \varepsilon_1,\mu_0,\nu_0
        )
    \right]
    =
    0
\]
and
\[
    \mathbb E
    \left[
        \nabla_{\mu\mu}^2
        \tilde{\ell}(
            \varepsilon_1,\mu_0,\nu_0
        )
    \right]
    =
    0.
\]

We first calculate the population Lagrange multiplier. At
\(
    y=\phi_1\kappa^{(0)},
    \kappa=\kappa^{(0)},
    \Sigma=\Sigma^{(0)},
\)
we have
    \[
    \widehat\phi
    \left(
        \phi_1\kappa^{(0)},
        \kappa^{(0)},
        \Sigma^{(0)}
    \right)
    =
    \phi_1.
\]
Consequently,
\[
    r_j
    =
    \operatorname{vec}
    \left(
        \phi_1\kappa_j^{(0)}
    \right)
    -
    M(\kappa_j^{(0)})
    \operatorname{vec}(\phi_1)
    =
    0,
    \quad
    j=1,\ldots,p.
\]
Hence, \(\mathbf E=0\), and
\eqref{supp:eq: explicit kappa gradient stacked} gives
\[
    \nabla_\mu
    \tilde{\ell}
    \left(
        \phi_1\kappa^{(0)},
        \mu_0,
        \nu_0
    \right)
    =
    0.
\]
Combining this with the vanishing expected noise gradient yields
\[
    \mathbb E
    \left[
        \nabla_\mu
        \tilde{\ell}(Y_1,\mu_0,\nu_0)
    \right]
    =
    0.
\]
Therefore, the population KKT equation reduces to
\(
    H\lambda_0=0.
\)
Since \(H\) has full column rank,
\(
    \lambda_0=0.
\)

We now calculate \(J\). Since \(\lambda_0=0\), the constraint-Hessian
term in the constrained Lagrangian vanishes. Together with the vanishing
expected noise Hessian, this gives
\[
    J
    =
    \mathbb E
    \left[
        \nabla_{\mu\mu}^2
        \tilde{\ell}(
            \phi_1\kappa^{(0)},
            \mu_0,\nu_0
        )
    \right].
\]
To evaluate this Hessian using
\eqref{supp:eq: explicit kappa Hessian}, we first calculate \(\mathbf Q\) at
the true parameter.

Since \(\mathbf E=0\) at the true parameter, \eqref{supp:eq: Q matrix definition} gives
\begin{equation}
\label{supp:eq: Q signal true parameter expanded}
\mathbf Q^\top
=
-
\begin{pmatrix}
\operatorname{vec}(\phi_1)^\top
(\Sigma^{(0)})^{-1}
M(\kappa_1^{(0)})
\\
\vdots
\\
\operatorname{vec}(\phi_1)^\top
(\Sigma^{(0)})^{-1}
M(\kappa_p^{(0)})
\\
\operatorname{vec}(\phi_1)^\top
J_2^\top
(\Sigma^{(0)})^{-1}
M(\kappa_1^{(0)})
\\
\vdots
\\
\operatorname{vec}(\phi_1)^\top
J_2^\top
(\Sigma^{(0)})^{-1}
M(\kappa_p^{(0)})
\end{pmatrix}.
\end{equation}

Note that
\[
    M(\phi_1)^\top
    =
    \begin{pmatrix}
        \operatorname{vec}(\phi_1)^\top
        \\
        \left\{
            J_2\operatorname{vec}(\phi_1)
        \right\}^\top
    \end{pmatrix}
    =
    \begin{pmatrix}
        \operatorname{vec}(\phi_1)^\top
        \\
        \operatorname{vec}(\phi_1)^\top J_2^\top
    \end{pmatrix}.
\]
Therefore,
\[
\begin{aligned}
    \left(
        M(\phi_1)\otimes\mathrm{Id}_p
    \right)^\top
    =
    M(\phi_1)^\top\otimes\mathrm{Id}_p=
    \begin{pmatrix}
        \operatorname{Re}(\phi_1)\mathrm{Id}_p
        &
        \operatorname{Im}(\phi_1)\mathrm{Id}_p
        \\
        -\operatorname{Im}(\phi_1)\mathrm{Id}_p
        &
        \operatorname{Re}(\phi_1)\mathrm{Id}_p
    \end{pmatrix}.
\end{aligned}
\]
Moreover,
\[
    H
    =
    \begin{pmatrix}
        \operatorname{Re}(\kappa^{(0)})
        &
        -\operatorname{Im}(\kappa^{(0)})
        \\
        \operatorname{Im}(\kappa^{(0)})
        &
        \operatorname{Re}(\kappa^{(0)})
    \end{pmatrix}.
\]
In particular, for each \(j=1,\ldots,p\), the \(j\)-th and
\((p+j)\)-th rows of \(H\) satisfy
\[
    \begin{pmatrix}
        H_{j,\cdot}
        \\
        H_{p+j,\cdot}
    \end{pmatrix}
    =
    \begin{pmatrix}
        \operatorname{Re}(\kappa_j^{(0)})
        &
        -\operatorname{Im}(\kappa_j^{(0)})
        \\
        \operatorname{Im}(\kappa_j^{(0)})
        &
        \operatorname{Re}(\kappa_j^{(0)})
    \end{pmatrix}
    =
    M(\kappa_j^{(0)}).
\]
Therefore, the corresponding two rows of
\[
    \left\{
        (\Sigma^{(0)})^{-1}
        \otimes
        \mathrm{Id}_p
    \right\}
    H
\]
are
\begin{equation}
\label{supp:eq: weighted H row pair}
    \begin{pmatrix}
        \left[
            \left\{
                (\Sigma^{(0)})^{-1}
                \otimes
                \mathrm{Id}_p
            \right\}
            H
        \right]_{j,\cdot}
        \\
        \left[
            \left\{
                (\Sigma^{(0)})^{-1}
                \otimes
                \mathrm{Id}_p
            \right\}
            H
        \right]_{p+j,\cdot}
    \end{pmatrix}
    =
    (\Sigma^{(0)})^{-1}
    M(\kappa_j^{(0)}).
\end{equation}

The \(j\)-th row of
\[
    \left(
        M(\phi_1)\otimes\mathrm{Id}_p
    \right)^\top
    \left\{
        (\Sigma^{(0)})^{-1}
        \otimes
        \mathrm{Id}_p
    \right\}
    H
\]
is consequently
\begin{align*}
&
\operatorname{Re}(\phi_1)
\left[
    \left\{
        (\Sigma^{(0)})^{-1}
        \otimes
        \mathrm{Id}_p
    \right\}
    H
\right]_{j,\cdot}
+
\operatorname{Im}(\phi_1)
\left[
    \left\{
        (\Sigma^{(0)})^{-1}
        \otimes
        \mathrm{Id}_p
    \right\}
    H
\right]_{p+j,\cdot}\\
&=
\begin{pmatrix}
    \operatorname{Re}(\phi_1)&\operatorname{Im}(\phi_1)
\end{pmatrix}
(\Sigma^{(0)})^{-1}
M(\kappa_j^{(0)})\\
&=
\operatorname{vec}(\phi_1)^\top
(\Sigma^{(0)})^{-1}
M(\kappa_j^{(0)}),
\end{align*}
where the second equality follows from
\eqref{supp:eq: weighted H row pair}. Similarly, its \((p+j)\)-th row is
\begin{align*}
&
-\operatorname{Im}(\phi_1)
\left[
    \left\{
        (\Sigma^{(0)})^{-1}
        \otimes
        \mathrm{Id}_p
    \right\}
    H
\right]_{j,\cdot}
+
\operatorname{Re}(\phi_1)
\left[
    \left\{
        (\Sigma^{(0)})^{-1}
        \otimes
        \mathrm{Id}_p
    \right\}
    H
\right]_{p+j,\cdot}\\
&=
\begin{pmatrix}
    -\operatorname{Im}(\phi_1)&\operatorname{Re}(\phi_1)
\end{pmatrix}
(\Sigma^{(0)})^{-1}
M(\kappa_j^{(0)})\\
&=
\left\{
    J_2\operatorname{vec}(\phi_1)
\right\}^\top
(\Sigma^{(0)})^{-1}
M(\kappa_j^{(0)})\\
&=
\operatorname{vec}(\phi_1)^\top
J_2^\top
(\Sigma^{(0)})^{-1}
M(\kappa_j^{(0)}).
\end{align*}
Stacking these rows over \(j=1,\ldots,p\) gives
\begin{align*}
&
\left(
    M(\phi_1)\otimes\mathrm{Id}_p
\right)^\top
\left\{
    (\Sigma^{(0)})^{-1}
    \otimes
    \mathrm{Id}_p
\right\}
H=
\begin{pmatrix}
\operatorname{vec}(\phi_1)^\top
(\Sigma^{(0)})^{-1}
M(\kappa_1^{(0)})
\\
\vdots
\\
\operatorname{vec}(\phi_1)^\top
(\Sigma^{(0)})^{-1}
M(\kappa_p^{(0)})
\\
\operatorname{vec}(\phi_1)^\top
J_2^\top
(\Sigma^{(0)})^{-1}
M(\kappa_1^{(0)})
\\
\vdots
\\
\operatorname{vec}(\phi_1)^\top
J_2^\top
(\Sigma^{(0)})^{-1}
M(\kappa_p^{(0)})
\end{pmatrix}.
\end{align*}
Comparing this expression with
\eqref{supp:eq: Q signal true parameter expanded}, we obtain
\begin{equation}
\label{supp:eq: Q signal true parameter}
\begin{aligned}
    \mathbf Q^\top
    =
    -
    \left(
        M(\phi_1)\otimes\mathrm{Id}_p
    \right)^\top
    \left\{
        (\Sigma^{(0)})^{-1}
        \otimes
        \mathrm{Id}_p
    \right\}
    H.
\end{aligned}
\end{equation}

The same row-pair representation of \(H\) gives
\begin{align}
    &H^\top
    \left\{
        (\Sigma^{(0)})^{-1}
        \otimes
        \mathrm{Id}_p
    \right\}
    H\\
    &=
    \sum_{j=1}^p
    M(\kappa_j^{(0)})^\top
    (\Sigma^{(0)})^{-1}
    M(\kappa_j^{(0)})\notag\\
    &=
    \kappa^{(0)}
    \diamond_{\Sigma^{(0)}}
    \kappa^{(0)}.
\label{supp:eq: diamond H identity}
\end{align}

We next verify the asserted property of \(\mathcal P_0\). By
\eqref{supp:eq: diamond H identity}, \(\mathcal P_0\) is symmetric. Moreover,
\begin{align*}
    \mathcal P_0^2
    &=
    \left\{
        (\Sigma^{(0)})^{-1/2}
        \otimes
        \mathrm{Id}_p
    \right\}
    H
    \left(
        \kappa^{(0)}
        \diamond_{\Sigma^{(0)}}
        \kappa^{(0)}
    \right)^{-1}
    H^\top
    \left\{
        (\Sigma^{(0)})^{-1}
        \otimes
        \mathrm{Id}_p
    \right\}
    H\\
    &\quad\times
    \left(
        \kappa^{(0)}
        \diamond_{\Sigma^{(0)}}
        \kappa^{(0)}
    \right)^{-1}
    H^\top
    \left\{
        (\Sigma^{(0)})^{-1/2}
        \otimes
        \mathrm{Id}_p
    \right\}\\
    &=
    \left\{
        (\Sigma^{(0)})^{-1/2}
        \otimes
        \mathrm{Id}_p
    \right\}
    H
    \left(
        \kappa^{(0)}
        \diamond_{\Sigma^{(0)}}
        \kappa^{(0)}
    \right)^{-1}
    H^\top
    \left\{
        (\Sigma^{(0)})^{-1/2}
        \otimes
        \mathrm{Id}_p
    \right\}\\
    &=
    \mathcal P_0.
\end{align*}
Thus, \(\mathcal P_0\) is the orthogonal projection onto
\[
    \operatorname{col}
    \left(
        \left\{
            (\Sigma^{(0)})^{-1/2}
            \otimes
            \mathrm{Id}_p
        \right\}
        H
    \right).
\]

We can now evaluate the Hessian in
\eqref{supp:eq: explicit kappa Hessian}. Its first term satisfies
\begin{align*}
&
M(\phi_1)^\top
(\Sigma^{(0)})^{-1}
M(\phi_1)
\otimes
\mathrm{Id}_p\\
&=
\left\{
    (\Sigma^{(0)})^{-1/2}
    M(\phi_1)
    \otimes
    \mathrm{Id}_p
\right\}^{\top}
\left\{
    (\Sigma^{(0)})^{-1/2}
    M(\phi_1)
    \otimes
    \mathrm{Id}_p
\right\}.
\end{align*}
Using \eqref{supp:eq: Q signal true parameter}, its second term satisfies
\begin{align*}
&
\mathbf Q^\top
\left(
    \kappa^{(0)}
    \diamond_{\Sigma^{(0)}}
    \kappa^{(0)}
\right)^{-1}
\mathbf Q\\
&=
\left\{
    (\Sigma^{(0)})^{-1/2}
    M(\phi_1)
    \otimes
    \mathrm{Id}_p
\right\}^{\top}
\mathcal P_0
\left\{
    (\Sigma^{(0)})^{-1/2}
    M(\phi_1)
    \otimes
    \mathrm{Id}_p
\right\}.
\end{align*}
Using the following Kronecker-product identity,
\[
    \left\{
        (\Sigma^{(0)})^{-1/2}
        M(\phi_1)
        \otimes
        \mathrm{Id}_p
    \right\}
    =
    \left\{
        (\Sigma^{(0)})^{-1/2}
        \otimes
        \mathrm{Id}_p
    \right\}
    \left(
        M(\phi_1)
        \otimes
        \mathrm{Id}_p
    \right),
\]
we obtain, upon substitution into
\eqref{supp:eq: explicit kappa Hessian},
\begin{align*}
&
\nabla_{\mu\mu}^2
\tilde{\ell}
\left(
    \phi_1\kappa^{(0)},
    \mu_0,
    \nu_0
\right)\\
&=
2
\left(
    M(\phi_1)
    \otimes
    \mathrm{Id}_p
\right)^\top
\left\{
    (\Sigma^{(0)})^{-1/2}
    \otimes
    \mathrm{Id}_p
\right\}\left(
    \mathrm{Id}_{2p}
    -
    \mathcal P_0
\right)\\
&\quad\left\{
    (\Sigma^{(0)})^{-1/2}
    \otimes
    \mathrm{Id}_p
\right\}
\left(
    M(\phi_1)
    \otimes
    \mathrm{Id}_p
\right).
\end{align*}
Taking expectations yields~\eqref {supp:eq:plugin-J}.

Because \(\mathrm{Id}_{2p}-\mathcal P_0\) is an orthogonal
projection, \(J\) is symmetric and positive semidefinite. Moreover, for
every \(v\in\mathbb R^{2p}\),
\begin{equation}
\label{supp:eq: J quadratic form true plugin}
\begin{aligned}
    v^\top Jv
    =
    2\mathbb E
    \left[
        \left\|
            \left(
                \mathrm{Id}_{2p}
                -
                \mathcal P_0
            \right)
            \left\{
                (\Sigma^{(0)})^{-1/2}
                \otimes
                \mathrm{Id}_p
            \right\}
            \left(
                M(\phi_1)
                \otimes
                \mathrm{Id}_p
            \right)
            v
        \right\|^2
    \right].
\end{aligned}
\end{equation}

We now determine the null space of \(J\). For every
\(\phi\in\mathbb C\),
\begin{equation}
\label{supp:eq: complex multiplication preserves H}
    \left(
        M(\phi)\otimes\mathrm{Id}_p
    \right)
    H
    =
    HM(\phi).
\end{equation}
Hence, whenever \(\phi\neq0\),
\[
    \left(
        M(\phi)\otimes\mathrm{Id}_p
    \right)
    \operatorname{col}(H)
    =
    \operatorname{col}(H),
\]
because \(M(\phi)\) is invertible.

Since
\[
    \operatorname{null}
    \left(
        \mathrm{Id}_{2p}-\mathcal P_0
    \right)
    =
    \operatorname{col}
    \left(
        \left\{
            (\Sigma^{(0)})^{-1/2}
            \otimes
            \mathrm{Id}_p
        \right\}
        H
    \right),
\]
for every \(\phi\neq0\), we have
\begin{align*}
&
\left(
    \mathrm{Id}_{2p}-\mathcal P_0
\right)
\left\{
    (\Sigma^{(0)})^{-1/2}
    M(\phi)
    \otimes
    \mathrm{Id}_p
\right\}
v
=
0\\
&\quad\Longleftrightarrow\quad
\left\{
    (\Sigma^{(0)})^{-1/2}
    M(\phi)
    \otimes
    \mathrm{Id}_p
\right\}
v
\in
\operatorname{col}
\left(
    \left\{
        (\Sigma^{(0)})^{-1/2}
        \otimes
        \mathrm{Id}_p
    \right\}
    H
\right)\\
&\quad\Longleftrightarrow\quad
\left(
    M(\phi)\otimes\mathrm{Id}_p
\right)v
\in
\operatorname{col}(H)\\
&\quad\Longleftrightarrow\quad
v\in\operatorname{col}(H).
\end{align*}

It follows immediately that
\(\operatorname{col}(H)\subset\operatorname{null}(J)\). Conversely, if
\(v^\top Jv=0\), then the nonnegative integrand in
\eqref{supp:eq: J quadratic form true plugin} is zero almost surely. Since
\[
    \mathcal{P}(|\phi_1|>0)>0
\]
follows from \(\mathbb E[|\phi_1|^2]>0\), the preceding equivalences imply
that \(v\in\operatorname{col}(H)\). Therefore,
\[
    \operatorname{null}(J)
    =
    \operatorname{col}(H)
    =
    \operatorname{span}
    \left\{
        \mu_0,\mathbb J_p\mu_0
    \right\}.
\] This proves~\eqref{supp:eq: null J = col H}. In particular, \(J\) is positive definite on the tangent space
\[
    T_{\mu_0}S
    =
    \left\{
        v\in\mathbb R^{2p}:H^\top v=0
    \right\}.
\]

Finally, since \(J\) is symmetric and
\[
    H^\top H
    =
    \mathrm{Id}_2,
\]
we have
\[
    JJ^+
    =
    J^+J
    =
    \mathrm{Id}_{2p}
    -
    HH^\top,
    \quad
    JH=0,
    \quad
    J^+H=0,
    \quad
    H^\top J^+=0.
\]
It follows by direct block multiplication that
\[
\begin{pmatrix}
    J&H\\
    H^\top&0
\end{pmatrix}
\begin{pmatrix}
    J^+&H\\
    H^\top&0
\end{pmatrix}
=
\mathrm{Id}_{2p+2}.
\]
Hence, \(B\) is nonsingular,
\(
    B^{-1}
    =
    \begin{pmatrix}
        J^+&H\\
        H^\top&0
    \end{pmatrix},
\)
and \(B_{11}=J^+\).
\end{proof}

\begin{lemma}
\label{supp:lem:plugin-third-derivative-domination}
Let \(\mathcal K\) be the compact convex neighborhood chosen in
Lemma~\ref{supp:lem:plugin-derivative-domination}. For almost every
\(Y\), the map
\[
    (\mu,\nu)
    \longmapsto
    \tilde{\ell}(Y,\mu,\nu)
\]
is three times continuously differentiable on \(\mathcal K\).
Moreover, there exists a constant \(C_3>0\) such that
\begin{align*}
&\texttt{max}\left\{
    \sup_{(\mu,\nu)\in\mathcal K}
    \left\|
        \nabla_{\mu\mu\mu}^3\tilde{\ell}(Y,\mu,\nu)
    \right\|,
    \sup_{(\mu,\nu)\in\mathcal K}
    \left\|
        \nabla_{\mu\mu\nu}^3\tilde{\ell}(Y,\mu,\nu)
    \right\|,
    \sup_{(\mu,\nu)\in\mathcal K}
    \left\|
        \nabla_{\mu\nu\nu}^3\tilde{\ell}(Y,\mu,\nu)
    \right\|
\right\}\\
&\le
D_3(Y)
:=
C_3\|Y\|^2
\quad\text{a.s.}    
\end{align*}

Consequently,
\(
    \mathbb E[D_3(Y)]<\infty
\)
whenever \(\mathbb E[|\phi_1|^2]<\infty\).  In particular, if \(\mathbb E[|\phi_1|^4]<\infty\), then this result, together with Lemma~\ref{supp:lem:plugin-derivative-domination}, verifies the third requirement of
Condition~\ref{supp:cond:gfm-quadratic-refinement} for the
\texttt{plug-in} estimator.
\end{lemma}

\begin{proof}[Proof of Lemma~\ref{supp:lem:plugin-third-derivative-domination}]
Recall that
\[
    \tilde{\ell}(y,\mu,\nu)
    =
    \tilde\rho(y,\kappa,\Sigma),
    \quad
    \mu
    =
    \widetilde{\operatorname{vec}}(\kappa),
    \quad
    \nu
    =
    \operatorname{vech}(\Sigma).
\]
Since \(\mathcal K\) is contained in
\[
    \left(
        \mathbb R^{2p}\setminus\{0\}
    \right)
    \times
    \operatorname{vech}
    \left\{
        \mathrm{SPD}(2)
    \right\},
\]
Lemma~\ref{supp:lem: rho differentiation} implies that, for every \(Y\), the
map
\[
    (\mu,\nu)
    \longmapsto
    \tilde{\ell}(y,\mu,\nu)
\]
is three times continuously differentiable on \(\mathcal K\).

Moreover,
\[
    \operatorname{vec}
    \bigl(
        \widehat\phi(y,\kappa,\Sigma)
    \bigr)
    =
    (\kappa\diamond_\Sigma\kappa)^{-1}
    (\kappa\bullet_\Sigma y)
\]
is real-linear in \(Y\), with coefficients that are infinitely
continuously differentiable in \((\mu,\nu)\). It follows that there
exists an infinitely continuously differentiable matrix-valued map
\(A(\mu,\nu)\) such that
\[
    \tilde{\ell}(Y,\mu,\nu)
    =
    \widetilde{\operatorname{vec}}(Y)^\top
    A(\mu,\nu)
    \widetilde{\operatorname{vec}}(Y).
\]
For every multi-index \(\alpha\) of order three,
\[
    \partial^\alpha
    \tilde{\ell}(Y,\mu,\nu)
    =
    \widetilde{\operatorname{vec}}(Y)^\top
    \partial^\alpha A(\mu,\nu)
    \widetilde{\operatorname{vec}}(Y).
\]
Since \(A\) is infinitely continuously differentiable and
\(\mathcal K\) is compact,
\[
    \sup_{(\mu,\nu)\in\mathcal K}
    \left\|
        \partial^\alpha A(\mu,\nu)
    \right\|_{\operatorname{op}}
    <
    \infty
\]
for every such \(\alpha\). There are only finitely many third-order
partial derivatives, and
\(
    \|\widetilde{\operatorname{vec}}(Y)\|=\|Y\|.
\)
Therefore, there exists a constant \(C_3>0\) such that
\[
\texttt{max}\left\{
    \sup_{(\mu,\nu)\in\mathcal K}
    \left\|
        \nabla_{\mu\mu\mu}^3\tilde{\ell}(Y,\mu,\nu)
    \right\|,
    \sup_{(\mu,\nu)\in\mathcal K}
    \left\|
        \nabla_{\mu\mu\nu}^3\tilde{\ell}(Y,\mu,\nu)
    \right\|,
    \sup_{(\mu,\nu)\in\mathcal K}
    \left\|
        \nabla_{\mu\nu\nu}^3\tilde{\ell}(Y,\mu,\nu)
    \right\|
\right\}
\le
C_3\|Y\|^2
\]
almost surely.

Finally, under model~\eqref{model:centered-complex-factor},
\[
    \mathbb E\|Y\|^2
    =
    \mathbb E[|\phi_1|^2]
    +
    p\operatorname{Tr}
    \left(
        \Sigma^{(0)}
    \right)
    <
    \infty.
\]
Hence, with \(D_3(Y)=C_3\|Y\|^2\),
\[
    \mathbb E[D_3(Y)]
    =
    C_3\mathbb E\|Y\|^2
    <
    \infty,
\]
which shows the result.
\end{proof}

\subsubsection{A loading update with an asymptotically linear
covariance input}
\label{supp:sec:plugin-loading-update}

In this subsubsection, we apply
Theorem~\ref{thm: clt_gfm} to a covariance estimator
\(\widehat\Sigma_n\) whose first-order influence function is the map
\[
    y\longmapsto\xi_\Sigma(y),
    \quad y\in\mathbb C^p,
\]
defined in~\eqref{eq:covariance-influence-function}. We then derive the stochastic expansion of the optimally
aligned loading estimator \(\widehat\kappa^{\texttt{opt}}_{n}\) obtained
by plugging \(\widehat\Sigma_n\) into the profile loss.

\begin{theorem}[Loading expansion for an asymptotically linear
covariance input]
\label{supp:thm:plugin-loading-update-transfer}
Assume model~\eqref{model:centered-complex-factor} with \(p\geq3\),
\[
    \kappa^{(0)}\in\mathcal S^{2p-1},
    \quad
    \Sigma^{(0)}\in\mathrm{SPD}(2),
    \quad
    0<\mathbb E[|\phi_1|^4]<\infty,
\]
and suppose that
\(\operatorname{Re}(\kappa^{(0)})\) and
\(\operatorname{Im}(\kappa^{(0)})\) are linearly independent in
\(\mathbb R^p\).

Let \(\widehat\Sigma_n\) be a random symmetric \(2\times2\) matrix such that
\[
    \mathcal{P}
    \left(
        \widehat\Sigma_n\in\mathrm{SPD}(2)
    \right)
    \longrightarrow 1
\]
and
\begin{equation}
\label{supp:eq:plugin-generic-covariance-input-expansion}
\operatorname{vech}
\left(
    \widehat\Sigma_n-\Sigma^{(0)}
\right)
=
\frac1n
\sum_{i=1}^n
\xi_\Sigma(Y_i)
+
\mathcal O_p(n^{-1}),
\end{equation}
where \(\xi_\Sigma\) is defined in
\eqref{eq:covariance-influence-function}. On the event
\(\{\widehat\Sigma_n\in\mathrm{SPD}(2)\}\), let
\begin{equation}
\label{supp:eq:plugin-generic-loading-update}
\widehat{[\kappa]}_{n}
\in
\operatorname*{argmin}_{[\kappa]\in\mathbb{CP}^{p-1}}
\frac1n
\sum_{i=1}^n
\rho
\left(
    Y_i,
    [\kappa],
    \widehat\Sigma_n
\right)
\end{equation}
be a measurable minimizer, defined arbitrarily on the complementary
event. Suppose that
\begin{equation}
\label{supp:eq:plugin-generic-loading-consistency}
d
\left(
    \widehat{[\kappa]}_{n},
    [\kappa^{(0)}]
\right)
\xrightarrow{p}
0.
\end{equation}
On the event
\[
    d\left(
        \widehat{[\kappa]}_n,
        [\kappa^{(0)}]
    \right)
    <
    \sqrt{2},
\]
let
\(\widehat\kappa_n^{\texttt{opt}}\in
\widehat{[\kappa]}_n\)
be the unique unit-norm representative satisfying
\[
    \left(
        \widehat\kappa_n^{\texttt{opt}}
    \right)^*
    \kappa^{(0)}
    >
    0.
\]
Define \(\widehat\kappa_n^{\texttt{opt}}\) arbitrarily on the complementary event. Then
\begin{equation}
\label{supp:eq:plugin-generic-loading-complex-expansion}
\widehat\kappa^{\texttt{opt}}_{n}
=
\kappa^{(0)}
+
L_{\kappa,n}
+
\mathcal O_p(n^{-1}),
\end{equation}
where we write
\(L_{\kappa,n}:=L_{\kappa,n}^{\texttt{plug-in}}\)
for the first-order term defined in
\eqref{eq:plugin-common-loading-first-order-term}. Moreover,
\begin{equation}
\label{supp:eq:plugin-common-loading-perturbation-order}
L_{\kappa,n}
=
\mathcal O_p(n^{-1/2}).
\end{equation}
\end{theorem}

\begin{proof}[Proof of
Theorem~\ref{supp:thm:plugin-loading-update-transfer}]
On the event
\(\{\widehat\Sigma_n\notin\mathrm{SPD}(2)\}\), replace
\(\widehat\Sigma_n\) by \(\mathrm{Id}_2\) and choose a measurable minimizer of the corresponding empirical objective. Since the modified and original estimators differ only on an event
whose probability tends to zero, the modified covariance estimator
continues to satisfy
\eqref{supp:eq:plugin-generic-covariance-input-expansion}, and all
subsequent stochastic expansions transfer back to the original
estimators. We retain the same notation for the resulting modified covariance and
loading estimators.

Let
\[
    \widehat\mu_n
    :=
    \widetilde{\operatorname{vec}}
    \left(
        \widehat\kappa^{\texttt{opt}}_{n}
    \right),
    \quad
    \widehat\nu_n
    :=
    \operatorname{vech}
    \left(
        \widehat\Sigma_n
    \right).
\]

Theorem~\ref{supp:thm:plugin-population-identification} shows that
\([\kappa^{(0)}]\) is the unique population minimizer. Under the local
coordinate map \(\mathcal{F}\), the corresponding mapped population minimizer is
\[
    \mu_0
    =
    \widetilde{\operatorname{vec}}
    \left(
        \kappa^{(0)}
    \right).
\]

We apply the quadratic stochastic expansion in
Theorem~\ref{thm: clt_gfm}, taking
\[
    X_i=Y_i,
    \quad
    \widehat q_n=\widehat\Sigma_n,
    \quad
    \widehat m_n=\widehat{[\kappa]}_{n},
\]
and using the local primary and auxiliary coordinates
\[
    \mu
    =
    \widetilde{\operatorname{vec}}(\kappa),
    \quad
    \nu
    =
    \operatorname{vech}(\Sigma).
\]

On the event
\[
    d\left(
        \widehat{[\kappa]}_n,
        [\kappa^{(0)}]
    \right)<\sqrt{2},
\]
the definition of the projective distance and the choice of the
optimal-position representative give
\[
    \|\widehat\mu_n-\mu_0\|
    =
    \|\widehat\kappa_n^{\texttt{opt}}-\kappa^{(0)}\|
    =
    d\left(
        \widehat{[\kappa]}_n,
        [\kappa^{(0)}]
    \right).
\]
By \eqref{supp:eq:plugin-generic-loading-consistency}, this event
has probability tending to one and the projective distance
converges to zero in probability. Hence
\(\widehat\mu_n\xrightarrow{p}\mu_0\), verifying
Condition~\ref{supp:cond:gfm-primary-consistency}.

Next,
\eqref{supp:eq:plugin-generic-covariance-input-expansion} gives
\[
    \widehat\nu_n-\nu_0
    =
    \frac1n
    \sum_{i=1}^n
    \xi_\Sigma(Y_i)
    +
    \mathcal O_p(n^{-1}).
\]
By
Lemma~\ref{supp:lem:plugin-covariance-influence-moments},
\[
    \mathbb E
    \left[
        \xi_\Sigma(Y_1)
    \right]
    =
    0,
    \quad
    \mathbb E
    \left[
        \left\|
            \xi_\Sigma(Y_1)
        \right\|^2
    \right]
    <
    \infty.
\]
Therefore,
\[
    \sqrt n
    \left(
        \widehat\nu_n-\nu_0
    \right)
    =
    \frac1{\sqrt n}
    \sum_{i=1}^n
    \xi_\Sigma(Y_i)
    +
    \mathcal O_p(n^{-1/2}).
\]
Thus,
Condition~\ref{supp:cond:gfm-auxiliary-linearity} and the
auxiliary-estimator requirement in
Condition~\ref{supp:cond:gfm-quadratic-refinement} both hold with
influence function \(\xi_\Sigma\). No independence between
\(\widehat\Sigma_n\) and \(Y_1,\ldots,Y_n\) is required.

Lemma~\ref{supp:lem:plugin-geometric-constraint} verifies
Condition~\ref{supp:cond:gfm-geometric-constraint}, together with the
corresponding third-order requirement in
Condition~\ref{supp:cond:gfm-quadratic-refinement}.
Lemma~\ref{supp:lem:plugin-derivative-domination} verifies
Condition~\ref{supp:cond:gfm-cost-differentiability}; under
\(\mathbb E[|\phi_1|^4]<\infty\), its dominating variable has a finite
second moment. The remaining third-order requirements follow from
Lemma~\ref{supp:lem:plugin-third-derivative-domination}.

Finally,
Lemma~\ref{supp:lem:plugin-population-kkt} verifies
Condition~\ref{supp:cond:gfm-kkt-invertibility} and gives
\[
    \lambda_0=0,
    \quad
    B_{11}=J^+.
\]
Since \(\lambda_0=0\), the expected Hessian of the constrained
Lagrangian appearing in Theorem~\ref{thm: clt_gfm} reduces to
\[
    J=\mathbb E
    \left[
        \nabla_{\mu\mu}^2
        \tilde{\ell}(Y_1,\mu_0,\nu_0)
    \right].
\]
Moreover, the population KKT equation gives
\[
    \mathbb E
    \left[
        \nabla_\mu
        \tilde{\ell}(Y_1,\mu_0,\nu_0)
    \right]
    =
    0.
\]

Under the local coordinates
\[
    \mu
    =
    \widetilde{\operatorname{vec}}(\kappa),
    \quad
    \nu
    =
    \operatorname{vech}(\Sigma),
\]
for every \(y\in\mathbb C^p\),
\[
    \nabla_\mu
    \tilde{\ell}(y,\mu_0,\nu_0)
    =
    \left.
    \nabla_\kappa
    \rho(y,[\kappa],\Sigma)
    \right|_{
        \kappa=\kappa^{(0)},\,
        \Sigma=\Sigma^{(0)}
    }
\]
and
\[
    C_0
    =
    \mathbb E
    \left[
        \nabla_{\mu\nu}^2
        \tilde{\ell}(Y_1,\mu_0,\nu_0)
    \right].
\]
Define the effective-score function
\(Z:\mathbb C^p\to\mathbb R^{2p}\) by
\[
    Z(y)
    :=
    \nabla_\mu\tilde{\ell}(y,\mu_0,\nu_0)
    +
    C_0\xi_\Sigma(y)
    =
    \psi_\kappa^{\texttt{plug-in}}(y),
\]
where the last equality follows from the preceding coordinate
identities and \eqref{eq:plugin-common-loading-first-order-term}.
The population KKT equation and
Lemma~\ref{supp:lem:plugin-covariance-influence-moments} give
\[
    \mathbb E[Z(Y_1)]
    =
    0.
\]
Moreover,
Lemmas~\ref{supp:lem:plugin-derivative-domination} and
\ref{supp:lem:plugin-covariance-influence-moments} imply
\[
    \mathbb E
    \left[
        \|Z(Y_1)\|^2
    \right]
    <
    \infty.
\]

All the conditions for the quadratic refinement in
Theorem~\ref{thm: clt_gfm} therefore hold. Applying
\eqref{eq: quadratic_stochastic_expansion} and using
\(B_{11}=J^+\), we obtain
\begin{align*}
    \widehat\mu_n-\mu_0
    &=
    -
    J^+
    \frac1n
    \sum_{i=1}^n
    Z(Y_i)
    +
    \mathcal O_p(n^{-1})
    \\
    &=
    -
    J^+
    \frac1n
    \sum_{i=1}^n
    \left\{
        \nabla_\mu
        \tilde{\ell}(Y_i,\mu_0,\nu_0)
        +
        C_0\xi_\Sigma(Y_i)
    \right\}
    +
    \mathcal O_p(n^{-1}).
\end{align*}
By the coordinate identity above and the definition in
\eqref{eq:plugin-common-loading-first-order-term},
\[
    -
    J^+
    \frac1n
    \sum_{i=1}^n
    Z(Y_i)
    =
    \widetilde{\operatorname{vec}}
    \left(
        L_{\kappa,n}
    \right).
\]
Since \(\widetilde{\operatorname{vec}}\) is a linear isometry, it
follows that
\[
    \widehat\kappa^{\texttt{opt}}_{n}
    =
    \kappa^{(0)}
    +
    L_{\kappa,n}
    +
    \mathcal O_p(n^{-1}),
\]
which proves
\eqref{supp:eq:plugin-generic-loading-complex-expansion}.

It remains to establish the order of \(L_{\kappa,n}\). Since
\(Z(Y_1),\ldots,Z(Y_n)\) are independent, centered, and square
integrable, their cross terms have zero expectation. Therefore,
\begin{align*}
&\mathbb E
\left[
    \left\|
        \widetilde{\operatorname{vec}}
        \left(
            L_{\kappa,n}
        \right)
    \right\|^2
\right]
=
\mathbb E
\left[
    \left\|
        J^+
        \frac1n
        \sum_{i=1}^n
        Z(Y_i)
    \right\|^2
\right]
\\
&\leq
\|J^+\|_{\operatorname{op}}^2
\mathbb E
\left[
    \left\|
        \frac1n
        \sum_{i=1}^n
        Z(Y_i)
    \right\|^2
\right]
=
\frac{
    \|J^+\|_{\operatorname{op}}^2
}{
    n
}
\mathbb E
\left[
    \|Z(Y_1)\|^2
\right].
\end{align*}
Since
\(
\|\widetilde{\operatorname{vec}}(L_{\kappa,n})\|
=
\|L_{\kappa,n}\|
\),
Markov's inequality now yields
\[
    L_{\kappa,n}
    =
    \mathcal O_p(n^{-1/2}),
\]
which proves
\eqref{supp:eq:plugin-common-loading-perturbation-order}.
\end{proof}

\subsubsection{Second-order expansion of the covariance moment map}
\label{supp:sec:plugin-covariance-moment-map-expansion}

We apply Lemma~\ref{supp:lem:real-residual-projection-loading-expansion} to the first-order loading perturbation
\(L_{\kappa,n}=L_{\kappa,n}^{\texttt{plug-in}}\)
defined in \eqref{eq:plugin-common-loading-first-order-term}.
This gives the projection expansion needed to obtain the
second-order covariance update.

Suppose that \(U_0\) has full column rank and that
\(\widehat\kappa_n^{\texttt{opt}}\in\mathcal S^{2p-1}\) satisfies
\begin{equation}
\label{supp:eq:plugin-common-loading-input-expansion}
    \widehat\kappa_n^{\texttt{opt}}
    =
    \kappa^{(0)}
    +
    L_{\kappa,n}
    +
    R_{\kappa,n},
    \quad
    L_{\kappa,n}=\mathcal O_p(n^{-1/2}),
    \quad
    R_{\kappa,n}=\mathcal O_p(n^{-1}).
\end{equation}
Set
\(\widehat U_n=\mathcal U(\widehat\kappa_n^{\texttt{opt}})\),
and define \(\widehat\Pi_{\perp,n}\) as in
Lemma~\ref{supp:lem:real-residual-projection-loading-expansion}.
Write \(A_{\Pi,n}^{\texttt{plug-in}}\) for the matrix
\(A_{\Pi,n}\) in that lemma with
\(L_{\kappa,n}=L_{\kappa,n}^{\texttt{plug-in}}\).
Consequently,
\begin{align}
    \widehat\Pi_{\perp,n}-\Pi_{\perp,0}
    &=
    -A_{\Pi,n}^{\texttt{plug-in}}+R_{\Pi,n},
    &
    \|R_{\Pi,n}\|_{\operatorname{op}}
    &=
    \mathcal O_p(n^{-1}),
    \label{supp:eq:plugin-residual-projection-expansion}
    \\
    \widehat\Pi_{\perp,n}\kappa^{(0)}
    &=
    -\Pi_{\perp,0}L_{\kappa,n}+r_{\Pi,n},
    &
    \|r_{\Pi,n}\|
    &=
    \mathcal O_p(n^{-1}),
    \label{supp:eq:plugin-projected-true-loading-expansion}
\end{align}
where \(r_{\Pi,n}=R_{\Pi,n}\kappa^{(0)}\).
Define the second-order noise terms by
\begin{equation}
\label{eq:plugin-noise-second-order-terms}
\begin{aligned}
Q_{\tau,n}^{\texttt{plug-in}}
&:=
\frac{1}{p-2}\Bigl\{
-\operatorname{Tr}\bigl(A_{\Pi,n}^{\texttt{plug-in}}E_H\bigr)
\\
&\quad\quad
+\mathbb E[|\phi_1|^2]
\bigl\|\Pi_{\perp,0}L_{\kappa,n}^{\texttt{plug-in}}\bigr\|^2
\Bigr\},
\\[6pt]
Q_{\gamma,n}^{\texttt{plug-in}}
&:=
\frac{1}{p-2}\Bigl\{
-\operatorname{Tr}\bigl(A_{\Pi,n}^{\texttt{plug-in}}E_T\bigr)
\\
&\quad\quad
+\mathbb E[\phi_1^2]
\bigl(\Pi_{\perp,0}L_{\kappa,n}^{\texttt{plug-in}}\bigr)^\top
\bigl(\Pi_{\perp,0}L_{\kappa,n}^{\texttt{plug-in}}\bigr)
\Bigr\},
\end{aligned}
\end{equation}
where, explicitly,
\[
\begin{aligned}
A_{\Pi,n}^{\texttt{plug-in}}
&:=
\mathcal U\bigl(
\Pi_{\perp,0}L_{\kappa,n}^{\texttt{plug-in}}
\bigr)
(U_0^\top U_0)^{-1}U_0^\top
\\
&\quad+
U_0(U_0^\top U_0)^{-1}
\mathcal U\bigl(
\Pi_{\perp,0}L_{\kappa,n}^{\texttt{plug-in}}
\bigr)^\top.
\end{aligned}
\]

We now propagate these expansions through the covariance moment map.

\begin{lemma}[Covariance moment update induced by a loading expansion]
\label{supp:lem:plugin-covariance-moment-map-expansion}
Assume model~\eqref{model:centered-complex-factor} with \(p\geq3\),
\[
    \kappa^{(0)}\in\mathcal S^{2p-1},
    \quad
    \Sigma^{(0)}\in\mathrm{SPD}(2),
    \quad
    0<\mathbb E[|\phi_1|^4]<\infty,
\]
and suppose that
\(\operatorname{Re}(\kappa^{(0)})\) and
\(\operatorname{Im}(\kappa^{(0)})\) are linearly independent in
\(\mathbb R^p\). Let \(\widehat\kappa_n^{\texttt{opt}}\) satisfy
\eqref{supp:eq:plugin-common-loading-input-expansion}
with \(L_{\kappa,n}=L_{\kappa,n}^{\texttt{plug-in}}\), and let
\(\widehat\Pi_{\perp,n}\) be as above.

Define
\begin{equation}
\label{supp:eq:plugin-covariance-moment-update}
\begin{aligned}
    \widehat\tau_n^2
    :=
    \widehat\tau^2
    \left(
        \widehat\kappa^{\texttt{opt}}_{n}
    \right),\quad
    \widehat\gamma_n:=
    \widehat\gamma
    \left(
        \widehat\kappa^{\texttt{opt}}_{n}
    \right),\quad
    \widehat\Sigma_n
    :=
    \mathcal M
    \left(
        \widehat\kappa^{\texttt{opt}}_{n}
    \right).
\end{aligned}
\end{equation}
These quantities may be defined arbitrarily on the event that
\(\widehat U_n\) does not have full column rank.

Then
\begin{align}
    \widehat\tau_n^2-\tau^2
    &=
    L_{\tau,n}^{\texttt{cov}}
    +
    Q_{\tau,n}^{\texttt{plug-in}}
    +
    \mathcal O_p(n^{-3/2}),
    \label{supp:eq:plugin-tau-second-order-expansion}
    \\
    \widehat\gamma_n-\gamma
    &=
    L_{\gamma,n}^{\texttt{cov}}
    +
    Q_{\gamma,n}^{\texttt{plug-in}}
    +
    \mathcal O_p(n^{-3/2}),
    \label{supp:eq:plugin-gamma-second-order-expansion}
\end{align}
where \(L_{\tau,n}^{\texttt{cov}}\) and
\(L_{\gamma,n}^{\texttt{cov}}\) are defined in
\eqref{eq:cov-noise-first-order-terms}, while
\(Q_{\tau,n}^{\texttt{plug-in}}\) and
\(Q_{\gamma,n}^{\texttt{plug-in}}\) are defined in
\eqref{eq:plugin-noise-second-order-terms}.

Moreover,
\[
    Q_{\tau,n}^{\texttt{plug-in}},
    Q_{\gamma,n}^{\texttt{plug-in}}
    =
    \mathcal O_p(n^{-1}).
\]

To express the corresponding expansion in covariance coordinates,
define
\(Q_{\Sigma,n}^{\texttt{plug-in}}\in\mathbb R^3\) by
\begin{equation}
\label{supp:eq:plugin-covariance-second-order-term}
    Q_{\Sigma,n}^{\texttt{plug-in}}
    :=
    \frac{1}{2}
    \operatorname{vech}
    \begin{pmatrix}
        Q_{\tau,n}^{\texttt{plug-in}}
        +\operatorname{Re}
        \left\{
            Q_{\gamma,n}^{\texttt{plug-in}}
        \right\}
        &
        \operatorname{Im}
        \left\{
            Q_{\gamma,n}^{\texttt{plug-in}}
        \right\}
        \\
        \operatorname{Im}
        \left\{
            Q_{\gamma,n}^{\texttt{plug-in}}
        \right\}
        &
        Q_{\tau,n}^{\texttt{plug-in}}
        -\operatorname{Re}
        \left\{
            Q_{\gamma,n}^{\texttt{plug-in}}
        \right\}
    \end{pmatrix}.
\end{equation}
Then
\begin{equation}
\label{supp:eq:plugin-covariance-second-order-expansion}
    \operatorname{vech}
    \left(
        \widehat\Sigma_n-\Sigma^{(0)}
    \right)
    =
    \frac{1}{n}
    \sum_{i=1}^n
    \xi_\Sigma(Y_i)
    +
    Q_{\Sigma,n}^{\texttt{plug-in}}
    +
    \mathcal O_p(n^{-3/2}).
\end{equation}
In particular,
\begin{equation}
\label{supp:eq:plugin-covariance-first-order-stability}
    \operatorname{vech}
    \left(
        \widehat\Sigma_n-\Sigma^{(0)}
    \right)
    =
    \frac{1}{n}
    \sum_{i=1}^n
    \xi_\Sigma(Y_i)
    +
    \mathcal O_p(n^{-1}).
\end{equation}
Consequently,
\(\widehat\Sigma_n\in\mathrm{SPD}(2)\) with probability tending to
one.
\end{lemma}

\begin{proof}[Proof of
Lemma~\ref{supp:lem:plugin-covariance-moment-map-expansion}]
By Lemma~\ref{supp:lem:cov-moment-bounds}, applied with \(k=2\),
\[
    \|E_H\|_{F}=\mathcal O_p(n^{-1/2}),
    \quad
    \|E_T\|_{F}=\mathcal O_p(n^{-1/2}).
\]
By
Lemma~\ref{supp:lem:real-residual-projection-loading-expansion},
\(\widehat U_n\) has full column rank with probability tending
to one. It suffices to establish the expansions on this event,
since the definitions on its complement do not affect the
asserted stochastic orders.

On this event, \(\widehat\Pi_{\perp,n}\) is a rank-\((p-2)\)
real orthogonal projection. Hence the population moment
identities give the exact decompositions
\begin{align*}
    \widehat\tau_n^2-\tau^2
    &=\frac{1}{p-2}
    \operatorname{Tr}
    \left(
        \widehat\Pi_{\perp,n}E_H
    \right)
    +
    \frac{\mathbb E[|\phi_1|^2]}{p-2}
    (\kappa^{(0)})^*
    \widehat\Pi_{\perp,n}
    \kappa^{(0)},
    \\
    \widehat\gamma_n-\gamma
    &=\frac{1}{p-2}
    \operatorname{Tr}
    \left(
        \widehat\Pi_{\perp,n}E_T
    \right)
    +
    \frac{\mathbb E[\phi_1^2]}{p-2}
    (\kappa^{(0)})^\top
    \widehat\Pi_{\perp,n}
    \kappa^{(0)}.
\end{align*}

Since \(p\) is fixed,
\[
    \|E_H\|_*
    \leq
    \sqrt p\,\|E_H\|_F
    =
    \mathcal O_p(n^{-1/2}),
    \quad
    \|E_T\|_*
    \leq
    \sqrt p\,\|E_T\|_F
    =
    \mathcal O_p(n^{-1/2}).
\]
Therefore, by
\eqref{supp:eq:plugin-residual-projection-expansion},
\begin{align*}
    \operatorname{Tr}
    \left(
        \widehat\Pi_{\perp,n}E_H
    \right)
    &=
    \operatorname{Tr}
    \left(
        \Pi_{\perp,0}E_H
    \right)
    -
    \operatorname{Tr}
    \left(
        A_{\Pi,n}^{\texttt{plug-in}}E_H
    \right)
    +
    \operatorname{Tr}
    \left(
        R_{\Pi,n}E_H
    \right)
    \\
    &=
    \operatorname{Tr}
    \left(
        \Pi_{\perp,0}E_H
    \right)
    -
    \operatorname{Tr}
    \left(
        A_{\Pi,n}^{\texttt{plug-in}}E_H
    \right)
    +
    \mathcal O_p(n^{-3/2}),
\end{align*}
where the final equality follows from
\[
\begin{aligned}
    \left|
        \operatorname{Tr}
        \left(
            R_{\Pi,n}E_H
        \right)
    \right|
    &\leq
    \|R_{\Pi,n}\|_{\operatorname{op}}
    \|E_H\|_*
    \\
    &=
    \mathcal O_p(n^{-1})
    \mathcal O_p(n^{-1/2})
    =
    \mathcal O_p(n^{-3/2}).
\end{aligned}
\]

The same argument gives
\begin{align*}
    \operatorname{Tr}
    \left(
        \widehat\Pi_{\perp,n}E_T
    \right)
    &={}
    \operatorname{Tr}
    \left(
        \Pi_{\perp,0}E_T
    \right)
    -
    \operatorname{Tr}
    \left(
        A_{\Pi,n}^{\texttt{plug-in}}E_T
    \right)
    +
    \mathcal O_p(n^{-3/2}).
\end{align*}

Next, by \eqref{supp:eq:plugin-projected-true-loading-expansion},
\[
    \widehat\Pi_{\perp,n}\kappa^{(0)}
    =
    -\Pi_{\perp,0}L_{\kappa,n}+r_{\Pi,n},
    \quad
    \|r_{\Pi,n}\|=\mathcal O_p(n^{-1}).
\]
Since \(\widehat\Pi_{\perp,n}\) is real, symmetric, and idempotent,
\begin{align*}
    (\kappa^{(0)})^*
    \widehat\Pi_{\perp,n}
    \kappa^{(0)}
    &=
    \left\|
        \widehat\Pi_{\perp,n}\kappa^{(0)}
    \right\|^2
    \\
    &=
    \left\|
        \Pi_{\perp,0}L_{\kappa,n}
    \right\|^2
    +
    \mathcal O_p(n^{-3/2}),
\end{align*}
and, similarly,
\begin{align*}
    (\kappa^{(0)})^\top
    \widehat\Pi_{\perp,n}
    \kappa^{(0)}
    &={}
    \left(
        \Pi_{\perp,0}L_{\kappa,n}
    \right)^\top
    \left(
        \Pi_{\perp,0}L_{\kappa,n}
    \right)
    +
    \mathcal O_p(n^{-3/2}).
\end{align*}
Indeed, by
\eqref{supp:eq:plugin-common-loading-input-expansion}
and
\eqref{supp:eq:plugin-projected-true-loading-expansion},
\[
    \|\Pi_{\perp,0}L_{\kappa,n}\|\,\|r_{\Pi,n}\|
    =
    \mathcal O_p(n^{-3/2}),
    \quad
    \|r_{\Pi,n}\|^2
    =
    \mathcal O_p(n^{-2}).
\]
These bounds control both the cross terms and the quadratic
remainder terms in the two expansions.

Substituting these expansions into the exact decompositions gives
\eqref{supp:eq:plugin-tau-second-order-expansion} and
\eqref{supp:eq:plugin-gamma-second-order-expansion}, with the
second-order terms in
\eqref{eq:plugin-noise-second-order-terms}. Moreover,
\[
    \|A_{\Pi,n}^{\texttt{plug-in}}\|_{\operatorname{op}}
    =
    \mathcal O_p(n^{-1/2}),
\]
so the two trace products in
\eqref{eq:plugin-noise-second-order-terms} are
\(\mathcal O_p(n^{-1})\); the quadratic loading terms have the same
order. Therefore,
\(Q_{\tau,n}^{\texttt{plug-in}},
Q_{\gamma,n}^{\texttt{plug-in}}=\mathcal O_p(n^{-1})\).

Finally, the transformation from \((\tau^2,\gamma)\) to \(\Sigma\) is
linear. By the definitions of \(\xi_\Sigma\),
\(L_{\tau,n}^{\texttt{cov}}\), and
\(L_{\gamma,n}^{\texttt{cov}}\),
\[
    \frac{1}{n}\sum_{i=1}^n\xi_\Sigma(Y_i)
    =
    \frac{1}{2}
    \begin{pmatrix}
        L_{\tau,n}^{\texttt{cov}}
        +\operatorname{Re}
        \left\{
            L_{\gamma,n}^{\texttt{cov}}
        \right\}
        \\
        \operatorname{Im}
        \left\{
            L_{\gamma,n}^{\texttt{cov}}
        \right\}
        \\
        L_{\tau,n}^{\texttt{cov}}
        -\operatorname{Re}
        \left\{
            L_{\gamma,n}^{\texttt{cov}}
        \right\}
    \end{pmatrix}.
\]
Thus
\eqref{supp:eq:plugin-covariance-second-order-expansion} follows from
\eqref{supp:eq:plugin-tau-second-order-expansion} and
\eqref{supp:eq:plugin-gamma-second-order-expansion}. Since
\(Q_{\Sigma,n}^{\texttt{plug-in}}=\mathcal O_p(n^{-1})\), this also
proves \eqref{supp:eq:plugin-covariance-first-order-stability}. In
particular, \(\widehat\Sigma_n\xrightarrow{p}\Sigma^{(0)}\). Since
\(\Sigma^{(0)}\in\mathrm{SPD}(2)\), it follows that
\(\widehat\Sigma_n\in\mathrm{SPD}(2)\) with probability tending to
one.
\end{proof}

\subsubsection{Recursive proof of the common plug-in expansions}
\label{supp:sec:plugin-recursive-expansions}

We combine
Theorem~\ref{supp:thm:plugin-loading-update-transfer} and
Lemma~\ref{supp:lem:plugin-covariance-moment-map-expansion} to prove
Theorems~\ref{thm:iterated-plugin-loading-expansion-clt} and
\ref{thm:iterated-plugin-noise-covariance-expansion} jointly by
induction over the plug-in iterations.

\begin{proof}[Proofs of
Theorems~\ref{thm:iterated-plugin-loading-expansion-clt} and
\ref{thm:iterated-plugin-noise-covariance-expansion}]
Fix an integer \(K\geq1\), independent of \(n\). For this proof, write
\[
    \widehat\Sigma^{[0]}
    :=
    \widehat\Sigma^{\texttt{cov}}.
\]
By \eqref{eq:covariance-influence-representation},
\[
    \operatorname{vech}
    \left(
        \widehat\Sigma^{[0]}-\Sigma^{(0)}
    \right)
    =
    \frac1n
    \sum_{i=1}^n
    \xi_\Sigma(Y_i)
    +
    \mathcal O_p(n^{-1}).
\]
Moreover,
Theorem~\ref{thm:cov-estimator-strong-consistency} implies that
\[
    \mathcal{P}
    \left(
        \widehat\Sigma^{[0]}\in\mathrm{SPD}(2)
    \right)
    \longrightarrow1.
\]
Thus, the covariance-input hypotheses of
Theorem~\ref{supp:thm:plugin-loading-update-transfer}
hold at iteration \(0\).

More generally, suppose that, for some fixed \(s\geq1\), the covariance
input from iteration \(s-1\) satisfies
\[
    \operatorname{vech}
    \left(
        \widehat\Sigma^{[s-1]}-\Sigma^{(0)}
    \right)
    =
    \frac1n
    \sum_{i=1}^n
    \xi_\Sigma(Y_i)
    +
    \mathcal O_p(n^{-1}),
    \quad
    \mathcal{P}
    \left(
        \widehat\Sigma^{[s-1]}\in\mathrm{SPD}(2)
    \right)
    \longrightarrow1.
\]
Theorem~\ref{thm:iterated-plugin-strong-consistency} gives the required
consistency of \(\widehat{[\kappa]}^{[s]}\). Therefore, applying
Theorem~\ref{supp:thm:plugin-loading-update-transfer} with
\[
    \widehat\Sigma_n
    =
    \widehat\Sigma^{[s-1]},
    \quad
    \widehat{[\kappa]}_{n}
    =
    \widehat{[\kappa]}^{[s]},
\]
yields
\[
    \widehat\kappa^{[s],\texttt{opt}}
    =
    \kappa^{(0)}
    +
    L_{\kappa,n}
    +
    \mathcal O_p(n^{-1}),
    \quad
    L_{\kappa,n}
    =
    \mathcal O_p(n^{-1/2}).
\]
Thus, the common first-order covariance expansion at iteration \(s-1\)
is propagated into the common first-order loading expansion at
iteration \(s\).

Applying
Lemma~\ref{supp:lem:plugin-covariance-moment-map-expansion} to this
loading expansion gives
\eqref{eq:iterated-plugin-tau-second-order-expansion} and
\eqref{eq:iterated-plugin-gamma-second-order-expansion} at iteration
\(s\). The same lemma also yields
\[
\begin{aligned}
    \operatorname{vech}
    \left(
        \widehat\Sigma^{[s]}-\Sigma^{(0)}
    \right)
    &=
    \frac1n
    \sum_{i=1}^n
    \xi_\Sigma(Y_i)
    +
    Q_{\Sigma,n}^{\texttt{plug-in}}
    +
    \mathcal O_p(n^{-3/2})
    \\
    &=
    \frac1n
    \sum_{i=1}^n
    \xi_\Sigma(Y_i)
    +
    \mathcal O_p(n^{-1}),
\end{aligned}
\]
Moreover,
\(\widehat\Sigma^{[s]}\in\mathrm{SPD}(2)\) with probability tending to
one. Thus, \(\widehat\Sigma^{[s]}\) satisfies the covariance-input
conditions of
Theorem~\ref{supp:thm:plugin-loading-update-transfer}, supplying the
induction hypothesis for iteration \(s+1\).

Starting with the covariance input at iteration \(0\) and applying the preceding implication successively for \(s=1,\ldots,K\), we obtain the
asserted loading and covariance expansions. Specifically,
simultaneously for \(s=1,\ldots,K\),
\begin{align*}
    \widehat\kappa^{[s],\texttt{opt}}
    &=
    \kappa^{(0)}
    +
    L_{\kappa,n}
    +
    \mathcal O_p(n^{-1}),
    \\
    \widehat\tau^{2,[s]}-\tau^2
    &=
    L_{\tau,n}^{\texttt{cov}}
    +
    Q_{\tau,n}^{\texttt{plug-in}}
    +
    \mathcal O_p(n^{-3/2}),
    \\
    \widehat\gamma^{[s]}-\gamma
    &=
    L_{\gamma,n}^{\texttt{cov}}
    +
    Q_{\gamma,n}^{\texttt{plug-in}}
    +
    \mathcal O_p(n^{-3/2}).
\end{align*}
Applying the linear isometry
\(\widetilde{\operatorname{vec}}\) to the loading expansion and using
the definition of \(L_{\kappa,n}\) in
\eqref{eq:plugin-common-loading-first-order-term} gives
\eqref{eq:iterated-plugin-common-expansion}.

The first-order loading term is the same \(L_{\kappa,n}\) at every
iteration. The second-order noise terms
\(Q_{\tau,n}^{\texttt{plug-in}}\) and
\(Q_{\gamma,n}^{\texttt{plug-in}}\), defined in \eqref{eq:plugin-noise-second-order-terms}, are also common to all
iterations. This proves the stochastic expansions in
Theorems~\ref{thm:iterated-plugin-loading-expansion-clt} and
\ref{thm:iterated-plugin-noise-covariance-expansion}.

Finally, the summands defining \(L_{\kappa,n}\) are i.i.d., centered,
and square integrable, as shown in the proof of
Theorem~\ref{supp:thm:plugin-loading-update-transfer}. After applying
\(\widetilde{\operatorname{vec}}\) and multiplying the loading
expansion by \(\sqrt n\), its remainder becomes
\(\mathcal O_p(n^{-1/2})=o_p(1)\). The multivariate central limit
theorem and Slutsky's theorem therefore yield
\eqref{eq:iterated-plugin-common-clt}, with
\(V^{\texttt{plug-in}}\) defined in
\eqref{eq:plugin-asymptotic-covariance}.
\end{proof}

\section{Proofs and calculations for the asymptotic efficiency comparison}
\label{supp:sec:asymptotic-efficiency-comparison}

This section provides the detailed calculations underlying
Theorem~\ref{thm:trace-comparison}. Throughout, we use the phase
convention and the quantities
\(d_\kappa\), \(d_\Sigma\), \(\rho_\phi\), and \(\lambda_\phi\)
introduced in
Subsection~\ref{subsec:asymptotic-efficiency-comparison} of the main
text. We first derive the trace of
\(V^{\texttt{cov}}\). We then calculate the trace of
\(V^{\texttt{plug-in}}\) by reducing its effective-score covariance to
canonical block form. Finally, we compare the resulting explicit trace
formulas to prove the isotropic equality, the strict efficiency gain
under second-order circularity, and the phase-uniform cutoff result in
Theorem~\ref{thm:trace-comparison}. Since all fixed \texttt{plug-in} iterations
have the same first-order asymptotic covariance matrix, the \texttt{plug-in}
calculation applies to every fixed iteration.

\subsection{Trace of the covariance-estimator asymptotic covariance}
\label{supp:sec:trace-cov-asymptotic-covariance}

We first derive the explicit trace of the asymptotic covariance matrix
\(V^{\texttt{cov}}\).

\begin{proposition}[Trace of \(V^{\texttt{cov}}\)]
\label{supp:prop:trace-cov-asymptotic-covariance}
Under the conditions of
Theorem~\ref{thm:cov-loading-expansion-clt},
\begin{equation}
\label{supp:eq:trace-cov-asymptotic-covariance}
\begin{aligned}
\operatorname{Tr}
\left(
    V^{\texttt{cov}}
\right)
&=
\frac{
    \operatorname{Tr}(\Sigma^{(0)})
}{
    \mathbb E[|\phi_1|^2]
}(p-1)
+
\bigg\{\frac{
    \operatorname{Tr}(\Sigma^{(0)})
}{
    \mathbb E[|\phi_1|^2]
}\bigg\}^{2}
\left\{
    (p-1)
    +
    d_\Sigma^2
    \left(
        1-d_\kappa^2
    \right)
\right\}.
\end{aligned}
\end{equation}
\end{proposition}

\begin{proof}[Proof of
Proposition~\ref{supp:prop:trace-cov-asymptotic-covariance}]
By
Proposition~\ref{supp:lem:explicit-cov-loading-asymptotic-covariance},
\[
V^{\texttt{cov}}
=
\frac{
    1
}{
    2\{\mathbb E[|\phi_1|^2]\}^2
}
\begin{pmatrix}
    \operatorname{Re}(\Gamma_H+\Gamma_T)
    &
    \operatorname{Im}(-\Gamma_H+\Gamma_T)
    \\
    \operatorname{Im}(\Gamma_H+\Gamma_T)
    &
    \operatorname{Re}(\Gamma_H-\Gamma_T)
\end{pmatrix}.
\]
Taking the trace and using the fact that \(\Gamma_H\) is Hermitian gives
\[
\begin{aligned}
\operatorname{Tr}
\left(
    V^{\texttt{cov}}
\right)
&=
\frac{
    \operatorname{Tr}
    \left\{
        \operatorname{Re}(\Gamma_H+\Gamma_T)
    \right\}
    +
    \operatorname{Tr}
    \left\{
        \operatorname{Re}(\Gamma_H-\Gamma_T)
    \right\}
}{
    2\{\mathbb E[|\phi_1|^2]\}^2
}
\\
&=
\frac{
    \operatorname{Tr}(\Gamma_H)
}{
    \{\mathbb E[|\phi_1|^2]\}^2
}.
\end{aligned}
\]
From \eqref{supp:eq:cov-loading-Gamma-H},
\[
\Gamma_H
=
\left\{
    \mathbb E[|\phi_1|^2]\tau^2
    +
    \tau^4
\right\}
P_{\perp,0}
+
|\gamma|^2
P_{\perp,0}
\overline{\kappa^{(0)}}
(\kappa^{(0)})^\top
P_{\perp,0}.
\]
Since \(P_{\perp,0}\) is an orthogonal projection of rank \(p-1\),
\[
    \operatorname{Tr}(P_{\perp,0})
    =
    p-1.
\]
Moreover,
\begin{align*}
&\operatorname{Tr}
\left\{
    P_{\perp,0}
    \overline{\kappa^{(0)}}
    (\kappa^{(0)})^\top
    P_{\perp,0}
\right\}
\\
&\quad=
(\kappa^{(0)})^\top
P_{\perp,0}^2
\overline{\kappa^{(0)}}
\\
&\quad=
(\kappa^{(0)})^\top
P_{\perp,0}
\overline{\kappa^{(0)}}
\\
&\quad=
1-
\left|
    (\kappa^{(0)})^\top\kappa^{(0)}
\right|^2
=
1-d_\kappa^2.
\end{align*}
where the last equality follows from
\eqref{eq:def-d-kappa}.

By \eqref{eq:tau-gamma} and \eqref{eq:def-d-sigma},
\[
    \tau^2
    =
    \operatorname{Tr}(\Sigma^{(0)}),
    \quad
    |\gamma|^2
    =
    d_\Sigma^2
    \{\operatorname{Tr}(\Sigma^{(0)})\}^2.
\]
Consequently,
\[
\begin{aligned}
\operatorname{Tr}(\Gamma_H)
&=
(p-1)
\left[
    \mathbb E[|\phi_1|^2]
    \operatorname{Tr}(\Sigma^{(0)})
    +
    \{\operatorname{Tr}(\Sigma^{(0)})\}^2
\right]
+
d_\Sigma^2
\{\operatorname{Tr}(\Sigma^{(0)})\}^2
\left(
    1-d_\kappa^2
\right).
\end{aligned}
\]
Dividing by
\(\{\mathbb E[|\phi_1|^2]\}^2\)
proves
\eqref{supp:eq:trace-cov-asymptotic-covariance}.
\end{proof}

\subsection{Canonical coordinates for the plug-in calculation}
\label{supp:sec:plugin-canonical-coordinates}

To evaluate \(V^{\texttt{plug-in}}\), we first rotate the common
real--imaginary coordinate system so that the noise covariance is
diagonal. We show that this simultaneous rotation leaves the full
plug-in effective score, and hence \(V^{\texttt{plug-in}}\), unchanged.
We then use the resulting diagonal-noise coordinates to construct the
canonical block representation used in the subsequent calculations.

\subsubsection{Rotation invariance and diagonal noise coordinates}
\label{supp:sec:plugin-rotation-invariance}

Let
\(\lambda_1^{(0)}\geq\lambda_2^{(0)}>0\)
denote the eigenvalues of \(\Sigma^{(0)}\). Choose
\(\theta_0\in\mathbb R\) such that
\[
\widetilde\Sigma^{(0)}
:=
M(e^{-i\theta_0})
\Sigma^{(0)}
M(e^{i\theta_0})
=
\begin{pmatrix}
    \lambda_1^{(0)} & 0\\
    0 & \lambda_2^{(0)}
\end{pmatrix}.
\]
When \(\lambda_1^{(0)}=\lambda_2^{(0)}\), we take
\(\theta_0=0\). Define
\[
    \widetilde Y_i
    :=
    e^{-i\theta_0}Y_i,
    \quad
    \widetilde\phi_i
    :=
    e^{-i\theta_0}\phi_i,
    \quad
    \widetilde\varepsilon_i
    :=
    e^{-i\theta_0}\varepsilon_i.
\]
Then
\[
    \widetilde Y_i
    =
    \widetilde\phi_i\kappa^{(0)}
    +
    \widetilde\varepsilon_i,
    \quad
    \widetilde\varepsilon_{i,j}
    \sim
    \mathcal{CN}
    \left(
        0,\widetilde\Sigma^{(0)}
    \right).
\]

\begin{lemma}[Rotation invariance of the plug-in asymptotic covariance]
\label{supp:lem:plugin-rotation-invariance}
Under the simultaneous rotation above,
\[
    \widetilde V^{\texttt{plug-in}}
    =
    V^{\texttt{plug-in}}.
\]
Moreover, the quantities
\(d_\kappa,d_\Sigma,\rho_\phi\), and \(\lambda_\phi\)
are unchanged.
\end{lemma}

\begin{proof}[Proof of
Lemma~\ref{supp:lem:plugin-rotation-invariance}]
For \(t,\alpha\in\mathbb C\), applied coordinatewise when
\(t\in\mathbb C^p\), let
\[
    \widetilde t
    :=
    e^{-i\theta_0}t,
    \quad
    \widetilde\alpha
    :=
    e^{-i\theta_0}\alpha.
\]
The simultaneous rotation preserves the weighted norm:
\[
    \left\|
        \widetilde t-\widetilde\alpha\kappa
    \right\|_{\widetilde\Sigma^{(0)}}^2
    =
    \left\|
        t-\alpha\kappa
    \right\|_{\Sigma^{(0)}}^2.
\]
Consequently,
\[
\widehat\phi
\left(
    \widetilde t,\kappa,\widetilde\Sigma^{(0)}
\right)
=
e^{-i\theta_0}
\widehat\phi
\left(
    t,\kappa,\Sigma^{(0)}
\right),
\]
and hence
\[
    \rho
    \left(
        \widetilde t,[\kappa],\widetilde\Sigma^{(0)}
    \right)
    =
    \rho
    \left(
        t,[\kappa],\Sigma^{(0)}
    \right).
\]

Write
\[
    \mu_0
    :=
    \widetilde{\operatorname{vec}}
    \left(
        \kappa^{(0)}
    \right),
    \quad
    \nu_0
    :=
    \operatorname{vech}
    \left(
        \Sigma^{(0)}
    \right),
    \quad
    \widetilde\nu_0
    :=
    \operatorname{vech}
    \left(
        \widetilde\Sigma^{(0)}
    \right).
\]
Let \(\mathsf T_{\theta_0}\) denote the invertible linear
transformation satisfying
\[
\operatorname{vech}
\left\{
    M(e^{-i\theta_0})
    \Delta
    M(e^{i\theta_0})
\right\}
=
\mathsf T_{\theta_0}
\operatorname{vech}(\Delta)
\]
for every symmetric \(2\times2\) matrix \(\Delta\).
Thus,
\[
    \widetilde\nu_0
    =
    \mathsf T_{\theta_0}\nu_0.
\]

Under the local coordinates
\(
\mu=\widetilde{\operatorname{vec}}(\kappa)
\)
and
\(
\nu=\operatorname{vech}(\Sigma)
\),
the invariance of the profile loss gives
\[
    \tilde{\ell}
    \left(
        e^{-i\theta_0}t,\mu,\mathsf T_{\theta_0}\nu
    \right)
    =
    \tilde{\ell}(t,\mu,\nu).
\]
Differentiating with respect to \(\mu\) and \(\nu\) yields
\[
\begin{aligned}
&\nabla_\mu
\widetilde\ell
\left(
    \widetilde Y_1,\mu_0,\widetilde\nu_0
\right)
=
\nabla_\mu
\tilde{\ell}
\left(
    Y_1,\mu_0,\nu_0
\right),
\\
&\widetilde J
=
J,
\quad
(\tilde{J})^+
=
J^+,
\quad
\widetilde C_0
=
C_0\mathsf T_{\theta_0}^{-1}.
\end{aligned}
\]
The covariance influence function transforms equivariantly:
\[
    \widetilde\xi_\Sigma
    \left(
        \widetilde Y_1
    \right)
    =
    \mathsf T_{\theta_0}\xi_\Sigma(Y_1).
\]
Therefore, the full effective score is unchanged:
\[
\begin{aligned}
&
\nabla_\mu
\widetilde\ell
\left(
    \widetilde Y_1,\mu_0,\widetilde\nu_0
\right)
+
\widetilde C_0
\widetilde\xi_\Sigma
\left(
    \widetilde Y_1
\right)
=
\nabla_\mu
\tilde{\ell}
\left(
    Y_1,\mu_0,\nu_0
\right)
+
C_0\xi_\Sigma(Y_1).
\end{aligned}
\]
Using the definition of \(V^{\texttt{plug-in}}\) in
\eqref{eq:plugin-asymptotic-covariance}, we conclude that
\[
\begin{aligned}
\widetilde V^{\texttt{plug-in}}
&=
\widetilde J^+
\operatorname{Var}
\left[
    \nabla_\mu\widetilde\ell
    \left(
        \widetilde Y_1,\mu_0,\widetilde\nu_0
    \right)
    +
    \widetilde C_0
    \widetilde\xi_\Sigma
    \left(
        \widetilde Y_1
    \right)
\right]
\widetilde J^+
\\
&=
J^+
\operatorname{Var}
\left[
    \nabla_\mu\tilde{\ell}
    \left(
        Y_1,\mu_0,\nu_0
    \right)
    +
    C_0\xi_\Sigma(Y_1)
\right]
J^+
=
V^{\texttt{plug-in}}.
\end{aligned}
\]

It remains to consider the scalar parameters used in the trace
comparison. The simultaneous rotation leaves the second absolute
moment of the factor unchanged, while both nonconjugate second moments
acquire the same phase:
\[
\begin{aligned}
\mathbb E
\left[
    |\widetilde\phi_1|^2
\right]
&=
\mathbb E
\left[
    |\phi_1|^2
\right],
\\
\mathbb E
\left[
    \widetilde\phi_1^2
\right]
&=
e^{-2i\theta_0}
\mathbb E
\left[
    \phi_1^2
\right],
\\
\mathbb E
\left[
    \widetilde\varepsilon_{1,1}^2
\right]
&=
e^{-2i\theta_0}
\mathbb E
\left[
    \varepsilon_{1,1}^2
\right].
\end{aligned}
\]
Thus, \(\rho_\phi\) and \(\lambda_\phi\) are unchanged.
The loading \(\kappa^{(0)}\) itself is not altered, so \(d_\kappa\)
is unchanged, while \(d_\Sigma\) is unchanged because the eigenvalues
of the noise covariance are preserved.
\end{proof}

In the rotated coordinates,
\[
    \mathbb E
    \left[
        \widetilde\varepsilon_{1,1}^2
    \right]
    =
    \lambda_1^{(0)}-\lambda_2^{(0)}
    \geq0.
\]
Hence, when
\(
\mathbb E[\widetilde\phi_1^2]\neq0
\)
and
\(
\lambda_1^{(0)}>\lambda_2^{(0)}
\),
the relative phase alignment reduces to
\[
    \lambda_\phi
    =
    \frac{
        \operatorname{Re}
        \left\{
            \mathbb E[\widetilde\phi_1^2]
        \right\}
    }{
        \left|
            \mathbb E[\widetilde\phi_1^2]
        \right|
    }.
\]
If either nonconjugate second moment is zero, we retain the convention
\(\lambda_\phi=0\).

By Lemma~\ref{supp:lem:plugin-rotation-invariance}, we may suppress all
tildes and assume throughout the remaining calculation that
\begin{equation}
\label{supp:eq:plugin-diagonal-noise-covariance}
\Sigma^{(0)}
=
\begin{pmatrix}
    \lambda_1^{(0)} & 0\\
    0 & \lambda_2^{(0)}
\end{pmatrix}.
\end{equation}
In the following subsubsection, we use these coordinates to obtain a
canonical block representation of \(J^+\) and the plug-in effective-score
covariance, and hence of \(V^{\texttt{plug-in}}\).

\subsubsection{Canonical loading coordinates and tangent-space decomposition}
\label{supp:sec:plugin-canonical-loading-coordinates}

In the diagonal noise coordinates introduced in
Subsubsection~\ref{supp:sec:plugin-rotation-invariance}, we next choose
an orthogonal basis adapted to the normal and horizontal spaces at
\([\kappa^{(0)}]\). This basis separates one distinguished two-dimensional real horizontal block, consisting of complex perturbations whose real and imaginary parts lie in
\[
    \operatorname{span}_{\mathbb R}
    \{\operatorname{Re}(\kappa^{(0)}),
      \operatorname{Im}(\kappa^{(0)})\},
\]
from the remaining \(p-2\) two-dimensional real horizontal blocks.

We assume that
\(\operatorname{Re}(\kappa^{(0)})\) and
\(\operatorname{Im}(\kappa^{(0)})\) are linearly independent.
Consequently,
\[
    d_\kappa\in[0,1),
    \quad
    r
    :=
    \left\|
        \operatorname{Re}(\kappa^{(0)})
    \right\|^2
    =
    \frac{1+d_\kappa}{2}
    \in
    \left[
        \frac12,1
    \right).
\]
Under the phase convention in
\eqref{eq:def-d-kappa},
\[
    \left\|
        \operatorname{Im}(\kappa^{(0)})
    \right\|^2
    =
    1-r,
    \quad
    \operatorname{Re}(\kappa^{(0)})^\top
    \operatorname{Im}(\kappa^{(0)})
    =
    0.
\]
Therefore, the vectors
\[
    u
    :=
    \frac{
        \operatorname{Re}(\kappa^{(0)})
    }{
        \sqrt r
    },
    \quad
    v
    :=
    \frac{
        \operatorname{Im}(\kappa^{(0)})
    }{
        \sqrt{1-r}
    }
\]
are orthonormal in \(\mathbb R^p\). Complete them to an orthonormal
basis
\[
    u,v,w_1,\ldots,w_{p-2}
\]
of \(\mathbb R^p\).

Let \(\mathscr S_p\in\mathbb R^{2p\times2p}\) be the permutation
matrix characterized by
\[
    \mathscr S_p
    \widetilde{\operatorname{vec}}(z)
    =
    \operatorname{vec}(z),
    \quad
    z\in\mathbb C^p.
\]
Thus, \(\mathscr S_p\) changes the grouped real--imaginary ordering
into the coordinatewise ordering. In particular, for every
\(B\in\mathbb R^{2\times2}\),
\begin{equation}
\label{supp:eq:permutation-kronecker-identity}
    \mathscr S_p
    \left(
        B\otimes\mathrm{Id}_p
    \right)
    \mathscr S_p^\top
    =
    \mathrm{Id}_p\otimes B.
\end{equation}

Write
\[
    J_2
    :=
    M(\mathrm i)
    =
    \begin{pmatrix}
        0&-1\\
        1&0
    \end{pmatrix}.
\]
For \(z=(z_1,\ldots,z_p)^\top\in\mathbb C^p\), extend the scalar
notation \(M\) by defining
\[
    M(z)
    :=
    \begin{pmatrix}
        M(z_1)\\
        \vdots\\
        M(z_p)
    \end{pmatrix}
    \in\mathbb R^{2p\times2}.
\] Recalling that \(H\in\mathbb R^{2p\times2}\) is the constraint
Jacobian at the true loading in the grouped coordinates, its
coordinate-wise representation is
\begin{equation}
\label{supp:eq:plugin-canonical-normal-block}
\begin{aligned}
    \mathscr S_pH
    &=
    M(\kappa^{(0)})=
    \sqrt r
    \left(
        u\otimes\mathrm{Id}_2
    \right)
    +
    \sqrt{1-r}
    \left(
        v\otimes J_2
    \right).
\end{aligned}
\end{equation}
The two columns of \(M(\kappa^{(0)})\) represent respectively the
radial and phase directions generated by
\(\kappa^{(0)}\) and \(\mathrm i\kappa^{(0)}\).

Define
\begin{align*}
    N_0
    &:=
    \sqrt{1-r}
    \left(
        u\otimes\mathrm{Id}_2
    \right)
    -
    \sqrt r
    \left(
        v\otimes J_2
    \right),\\
    N_l
    &:=
    w_l\otimes\mathrm{Id}_2,
    \quad
    l=1,\ldots,p-2,
\end{align*}
and set
\begin{equation}
\label{supp:eq:plugin-canonical-orthogonal-matrix}
    \mathsf Q_0
    :=
    \left(
        M(\kappa^{(0)}),
        N_0,
        N_1,
        \ldots,
        N_{p-2}
    \right)
    \in
    \mathbb R^{2p\times2p}.
\end{equation}
Since
\(u,v,w_1,\ldots,w_{p-2}\) are orthonormal and
\(J_2^\top J_2=\mathrm{Id}_2\), we have
\[
    M(\kappa^{(0)})^\top
    M(\kappa^{(0)})
    =
    N_0^\top N_0
    =
    \mathrm{Id}_2,
    \quad
    M(\kappa^{(0)})^\top N_0
    =
    0_2,
\]
and
\[
    M(\kappa^{(0)})^\top N_l
    =
    N_0^\top N_l
    =
    0_2,
    \quad
    N_l^\top N_m
    =
    \delta_{lm}\mathrm{Id}_2.
\]
Consequently,
\[
    \mathsf Q_0^\top\mathsf Q_0
    =
    \mathrm{Id}_{2p},
\]
so \(\mathsf Q_0\) is orthogonal.

For the tangent-space interpretation, define
\[
    N
    :=
    \left(
        N_0,N_1,\ldots,N_{p-2}
    \right)
    \in
    \mathbb R^{2p\times(2p-2)}.
\]
The preceding orthogonality relations give
\[
    N^\top N
    =
    \mathrm{Id}_{2p-2},
    \quad
    M(\kappa^{(0)})^\top N
    =
    0,
\]
and hence
\begin{equation}
\label{supp:eq:plugin-canonical-normal-tangent-projections}
    NN^\top
    =
    \mathrm{Id}_{2p}
    -
    M(\kappa^{(0)})
    M(\kappa^{(0)})^\top.
\end{equation}
Moreover, for every \(z\in\mathbb C^p\),
\[
    M(\kappa^{(0)})^\top
    \operatorname{vec}(z)
    =
    \operatorname{vec}
    \left(
        (\kappa^{(0)})^*z
    \right).
\]
It follows that the horizontal representative of
\(T_{[\kappa^{(0)}]}\mathbb{CP}^{p-1}\) satisfies
\begin{equation}
\label{supp:eq:plugin-canonical-horizontal-space}
\begin{aligned}
&
\left\{
    \widetilde{\operatorname{vec}}(z):
    z\in\mathbb C^p,\ 
    (\kappa^{(0)})^*z=0
\right\}
=
\operatorname{col}
\left(
    \mathscr S_p^\top N
\right).
\end{aligned}
\end{equation}
Equivalently, every horizontal perturbation \(z\) has a unique
representation
\begin{equation}
\label{supp:eq:plugin-canonical-tangent-decomposition}
    \mathscr S_p
    \widetilde{\operatorname{vec}}(z)
    =
    N_0a_0
    +
    \sum_{l=1}^{p-2}N_la_l,
    \quad
    a_0,\ldots,a_{p-2}\in\mathbb R^2.
\end{equation}

Thus, under the coordinate transformation
\[
    A
    \longmapsto
    \mathsf Q_0^\top
    \mathscr S_p
    A
    \mathscr S_p^\top
    \mathsf Q_0
\]
the first \(2\times2\) block corresponds to the constrained radial
and phase directions, the \(N_0\)-block corresponds to the horizontal complex line in \(\operatorname{span}_{\mathbb C}\{u,v\}\), and the \(N_l\)-blocks correspond to the \(p-2\) transverse horizontal complex lines. This decomposition will be used below to calculate \(J^+\) and the covariance of the effective
score block by block.

\subsection{Trace of the plug-in asymptotic covariance}
\label{supp:sec:trace-plugin-asymptotic-covariance}

In this subsection, we derive an explicit expression for
\(\operatorname{Tr}(V^{\texttt{plug-in}})\) to compare the asymptotic
efficiencies of the \texttt{plug-in} and \texttt{cov} estimators.
To this end, we obtain canonical block representations of the
population curvature and the effective-score covariance, together
with auxiliary matrix identities that are also used in the
subsequent analysis of the maximum-aligned spectrum estimators.

\subsubsection{Canonical form of the population curvature}
\label{supp:sec:plugin-population-curvature}

We now use the canonical coordinates constructed in
Subsubsection~\ref{supp:sec:plugin-canonical-loading-coordinates} to
block-diagonalize the population curvature \(J\) and its
Moore--Penrose inverse \(J^+\). For a positive definite matrix
\(W\in\mathbb R^{2\times2}\), define
\begin{equation}
\label{supp:eq:plugin-A-W-definition}
    A_W
    :=
    \mathbb E
    \left[
        M(\phi_1)^\top
        W
        M(\phi_1)
    \right].
\end{equation}
We denote 
\begin{align}\label{supp:eq:D-def} D := \kappa^{(0)} \diamond_{\Sigma^{(0)}} \kappa^{(0)} \end{align} and \begin{align} S &:= (1-r)(\Sigma^{(0)})^{-1} + rJ_2^\top(\Sigma^{(0)})^{-1}J_2 \nonumber\\ &\quad- r(1-r) \left\{ (\Sigma^{(0)})^{-1} - J_2^\top(\Sigma^{(0)})^{-1}J_2 \right\} D^{-1} \left\{ (\Sigma^{(0)})^{-1} - J_2^\top(\Sigma^{(0)})^{-1}J_2 \right\}. \label{supp: eq:S-def} \end{align} for $ J_2 := M(\mathrm{i}) = \begin{pmatrix} 0&-1\\ 1&0 \end{pmatrix}$.

\begin{lemma}[Block decomposition of \(J\) and \(J^+\)]
\label{supp:lem:plugin-curvature-block-decomposition}
Under the conditions of
Theorem~\ref{thm:iterated-plugin-loading-expansion-clt}, the matrices
\(J\) and \(J^+\) satisfy
\begin{equation}
\label{supp:eq:plugin-J-block-decomposition}
\begin{aligned}
    \mathsf Q_0^\top
    \mathscr S_p
    J
    \mathscr S_p^\top
    \mathsf Q_0
    =
    \operatorname{diag}
    \left(
        0_2,\,
        2A_S,\,
        \mathrm{Id}_{p-2}
        \otimes
        \left\{
            2A_{(\Sigma^{(0)})^{-1}}
        \right\}
    \right).
\end{aligned}
\end{equation}
Consequently,
\begin{equation}
\label{supp:eq:plugin-J-plus-block-decomposition}
\begin{aligned}
    \mathsf Q_0^\top
    \mathscr S_p
    J^+
    \mathscr S_p^\top
    \mathsf Q_0
    =
    \operatorname{diag}
    \left(
        0_2,\,
        \frac12A_S^{-1},\,
        \mathrm{Id}_{p-2}
        \otimes
        \left\{
            \frac12
            A_{(\Sigma^{(0)})^{-1}}^{-1}
        \right\}
    \right).
\end{aligned}
\end{equation}
\end{lemma}

\begin{proof}
By Lemma~\ref{supp:lem:plugin-population-kkt}, the population
curvature can be written as
\begin{equation}
\label{supp:eq:plugin-J-weighted-projection-representation}
    J
    =
    2\mathbb E
    \left[
        \left\{
            M(\phi_1)^\top
            \otimes
            \mathrm{Id}_p
        \right\}
        \mathcal W_0
        \left\{
            M(\phi_1)
            \otimes
            \mathrm{Id}_p
        \right\}
    \right],
\end{equation}
where
\begin{equation}
\label{supp:eq:plugin-W0-representation}
\begin{aligned}
    \mathcal W_0
    &:=
    \left\{
        (\Sigma^{(0)})^{-1}
        \otimes
        \mathrm{Id}_p
    \right\}\\
    &\quad-
    \left\{
        (\Sigma^{(0)})^{-1}
        \otimes
        \mathrm{Id}_p
    \right\}
    H D^{-1}H^\top
    \left\{
        (\Sigma^{(0)})^{-1}
        \otimes
        \mathrm{Id}_p
    \right\}.
\end{aligned}
\end{equation}

We first record how complex scalar multiplication acts in the
canonical basis. Since \(M(z)\) commutes with \(J_2=M(\mathrm i)\),
the definitions of \(M(\kappa^{(0)})\), \(N_0\), and \(N_l\) imply
that, for every \(z\in\mathbb C\),
\[
    \left\{
        \mathrm{Id}_p\otimes M(z)
    \right\}
    \mathsf Q_0
    =
    \mathsf Q_0
    \left\{
        \mathrm{Id}_p\otimes M(z)
    \right\}.
\]
Since \(\mathsf Q_0\) is orthogonal,
\begin{equation}
\label{supp:eq:plugin-scalar-multiplication-block-action}
    \mathsf Q_0^\top
    \left\{
        \mathrm{Id}_p\otimes M(z)
    \right\}
    \mathsf Q_0
    =
    \mathrm{Id}_p\otimes M(z).
\end{equation}
Using \eqref{supp:eq:permutation-kronecker-identity} together with
\(\mathscr S_p^\top\mathscr S_p=\mathrm{Id}_{2p}\) and
\(\mathsf Q_0\mathsf Q_0^\top=\mathrm{Id}_{2p}\), we obtain

\begin{equation} \label{supp:eq: J transformed representation} 
\begin{aligned}
&\mathsf Q_0^\top
    \mathscr S_pJ\mathscr S_p^\top
    \mathsf Q_0\\
    &=
    2\mathbb E
    \left[
        \mathsf Q_0^\top
        \mathscr S_p
        \left\{
            M(\phi_1)^\top\otimes\mathrm{Id}_p
        \right\}
        \mathcal W_0
        \left\{
            M(\phi_1)\otimes\mathrm{Id}_p
        \right\}
        \mathscr S_p^\top
        \mathsf Q_0
    \right]\\
    &=
    2\mathbb E
    \left[
        \mathsf Q_0^\top
        \mathscr S_p
        \left\{
            M(\phi_1)^\top\otimes\mathrm{Id}_p
        \right\}
        \mathscr S_p^\top
        (\mathscr S_p\mathcal W_0\mathscr S_p^\top)
        \mathscr S_p
        \left\{
            M(\phi_1)\otimes\mathrm{Id}_p
        \right\}
        \mathscr S_p^\top
        \mathsf Q_0
    \right]\\
    &=
    2\mathbb E
    \left[
        \mathsf Q_0^\top
        \left\{
            \mathrm{Id}_p\otimes M(\phi_1)
        \right\}^\top
        (\mathscr S_p\mathcal W_0\mathscr S_p^\top)
        \left\{
            \mathrm{Id}_p\otimes M(\phi_1)
        \right\}
        \mathsf Q_0
    \right]\\
    &=
    2\mathbb E
    \left[
        \left\{
            \mathsf Q_0^\top
            \left(
                \mathrm{Id}_p\otimes M(\phi_1)
            \right)
            \mathsf Q_0
        \right\}^\top
        \left(
            \mathsf Q_0^\top
            \mathscr S_p\mathcal W_0\mathscr S_p^\top
            \mathsf Q_0
        \right)
        \left\{
            \mathsf Q_0^\top
            \left(
                \mathrm{Id}_p\otimes M(\phi_1)
            \right)
            \mathsf Q_0
        \right\}
    \right] \\
    & =
    2\mathbb E \left[
        \left\{
            \mathrm{Id}_p\otimes M(\phi_1)
        \right\}^\top
        \left(
            \mathsf Q_0^\top
            \mathscr S_p\mathcal W_0\mathscr S_p^\top
            \mathsf Q_0
        \right)
        \left\{
            \mathrm{Id}_p\otimes M(\phi_1)
        \right\}
    \right]
\end{aligned}
\end{equation}
We next calculate the canonical block representation of
\(\mathcal W_0\). By
\eqref{supp:eq:permutation-kronecker-identity} and
\(\mathscr S_pH=M(\kappa^{(0)})\),
\begin{align*}
    \mathscr S_p
    \mathcal W_0
    \mathscr S_p^\top
    &=
    \mathrm{Id}_p
    \otimes
    (\Sigma^{(0)})^{-1}-
    \left\{
        \mathrm{Id}_p
        \otimes
        (\Sigma^{(0)})^{-1}
    \right\}
    M(\kappa^{(0)})
    D^{-1}
    M(\kappa^{(0)})^\top
    \left\{
        \mathrm{Id}_p
        \otimes
        (\Sigma^{(0)})^{-1}
    \right\}.
\end{align*}
In particular,
\[
    \left(
        \mathscr S_p
        \mathcal W_0
        \mathscr S_p^\top
    \right)
    M(\kappa^{(0)})
    =
    0_{2p\times2}.
\]

To evaluate
\(
    \mathsf Q_0^\top
    \mathscr S_p
    \mathcal W_0
    \mathscr S_p^\top
    \mathsf Q_0
\),
we first compute the blockwise inner products induced by
\(\mathrm{Id}_p\otimes(\Sigma^{(0)})^{-1}\). For
\(a,b\in\mathbb R^p\) and
\(B,C,E\in\mathbb R^{2\times2}\), the mixed-product property of the
Kronecker product gives
\begin{align}
    &(a\otimes B)^\top
    \left(
        \mathrm{Id}_p\otimes C
    \right)
    (b\otimes E)
    =
    \left(
        a^\top\otimes B^\top
    \right)
    \left(
        \mathrm{Id}_p\otimes C
    \right)
    (b\otimes E)
    =
    (a^\top b)B^\top CE.
    \label{supp:eq:kronecker-block-product}
\end{align}
We apply
\eqref{supp:eq:kronecker-block-product} throughout with
\(C=(\Sigma^{(0)})^{-1}\). Since
\(u,v,w_1,\ldots,w_{p-2}\) form an orthonormal basis of
\(\mathbb R^p\),
\[
    u^\top u=v^\top v=1,
    \quad
    u^\top v=0,
\]
and, for \(l,m=1,\ldots,p-2\),
\[
    u^\top w_l=v^\top w_l=0,
    \quad
    w_l^\top w_m=\delta_{lm}.
\]

For the \(N_0\)-block, expanding \(N_0\) on both sides and applying
\eqref{supp:eq:kronecker-block-product} term by term yields
\begin{align*}
    &N_0^\top
    \left\{
        \mathrm{Id}_p\otimes(\Sigma^{(0)})^{-1}
    \right\}
    N_0\\
    &\quad=
    (1-r)(u^\top u)(\Sigma^{(0)})^{-1}
    +
    r(v^\top v)
    J_2^\top(\Sigma^{(0)})^{-1}J_2\\
    &\quad-
    \sqrt{r(1-r)}(u^\top v)
    (\Sigma^{(0)})^{-1}J_2
    -
    \sqrt{r(1-r)}(v^\top u)
    J_2^\top(\Sigma^{(0)})^{-1}\\
    &\quad=
    (1-r)(\Sigma^{(0)})^{-1}
    +
    rJ_2^\top(\Sigma^{(0)})^{-1}J_2.
\end{align*}
Similarly, using
\[
    M(\kappa^{(0)})
    =
    \sqrt r
    \left(
        u\otimes\mathrm{Id}_2
    \right)
    +
    \sqrt{1-r}
    \left(
        v\otimes J_2
    \right),
\]
we obtain
\begin{align*}
    &M(\kappa^{(0)})^\top
    \left\{
        \mathrm{Id}_p\otimes(\Sigma^{(0)})^{-1}
    \right\}
    N_0\\
    &\quad=
    \sqrt{r(1-r)}(u^\top u)(\Sigma^{(0)})^{-1}
    -
    r(u^\top v)(\Sigma^{(0)})^{-1}J_2\\
    &\quad+
    (1-r)(v^\top u)
    J_2^\top(\Sigma^{(0)})^{-1}
    -
    \sqrt{r(1-r)}(v^\top v)
    J_2^\top(\Sigma^{(0)})^{-1}J_2\\
    &\quad=
    \sqrt{r(1-r)}
    \left\{
        (\Sigma^{(0)})^{-1}
        -
        J_2^\top(\Sigma^{(0)})^{-1}J_2
    \right\}.
\end{align*}

Next, since
\(N_l=w_l\otimes\mathrm{Id}_2\), the orthogonality of \(w_l\) to both
\(u\) and \(v\) gives
\begin{align*}
    &M(\kappa^{(0)})^\top
    \left\{
        \mathrm{Id}_p\otimes(\Sigma^{(0)})^{-1}
    \right\}
    N_l\\
    &\quad=
    \sqrt r\,(u^\top w_l)(\Sigma^{(0)})^{-1}
    +
    \sqrt{1-r}\,(v^\top w_l)
    J_2^\top(\Sigma^{(0)})^{-1}
    =
    0_2,
\end{align*}
and
\begin{align*}
    &N_0^\top
    \left\{
        \mathrm{Id}_p\otimes(\Sigma^{(0)})^{-1}
    \right\}
    N_l\\
    &\quad=
    \sqrt{1-r}\,(u^\top w_l)(\Sigma^{(0)})^{-1}
    -
    \sqrt r\,(v^\top w_l)
    J_2^\top(\Sigma^{(0)})^{-1}
    =
    0_2.
\end{align*}
Finally, for \(l,m=1,\ldots,p-2\),
\begin{align*}
    &N_l^\top
    \left\{
        \mathrm{Id}_p\otimes(\Sigma^{(0)})^{-1}
    \right\}
    N_m\\
    &\quad=
    (w_l^\top w_m)(\Sigma^{(0)})^{-1}
    =
    \delta_{lm}(\Sigma^{(0)})^{-1}.
\end{align*}
Substitution into
\eqref{supp:eq:plugin-W0-representation} shows that
\begin{equation}
\label{supp:eq:plugin-W0-block-decomposition}
    \mathsf Q_0^\top
    \mathscr S_p
    \mathcal W_0
    \mathscr S_p^\top
    \mathsf Q_0
    =
    \operatorname{diag}
    \left(
        0_2,\,
        S,\,
        \mathrm{Id}_{p-2}
        \otimes
        (\Sigma^{(0)})^{-1}
    \right),
\end{equation}
where the \(N_0\)-block is exactly the matrix \(S\) defined in
\eqref{supp: eq:S-def}.

Using \eqref{supp:eq: J transformed representation} and \eqref{supp:eq:plugin-W0-block-decomposition}, we get
\begin{align*}
&\mathsf Q_0^\top\mathscr S_pJ\mathscr S_p^\top \mathsf Q_0\\
&=2\mathbb E
\left[
    \left\{
        \mathrm{Id}_p\otimes M(\phi_1)
    \right\}^\top
    \operatorname{diag}
    \left(
        0_2,
        S,
        \mathrm{Id}_{p-2}\otimes(\Sigma^{(0)})^{-1}
    \right)
    \left\{
        \mathrm{Id}_p\otimes M(\phi_1)
    \right\}
\right]\\
&=\operatorname{diag}
\left(
    0_2,
    2\mathbb E
    \left[
        M(\phi_1)^\top S M(\phi_1)
    \right],
    \mathrm{Id}_{p-2}\otimes
    2\mathbb E
    \left[
        M(\phi_1)^\top
        (\Sigma^{(0)})^{-1}
        M(\phi_1)
    \right]
\right)\\
&=\operatorname{diag}
\left(
    0_2,
    2A_S,
    \mathrm{Id}_{p-2}\otimes
    \left(2A_{(\Sigma^{(0)})^{-1}}\right)
\right).
\end{align*}

Finally, \(\mathcal W_0\) is positive semidefinite with
\[
    \operatorname{null}(\mathcal W_0)
    =
    \operatorname{col}(H).
\]
Hence its restrictions to the \(N_0\)-block and the transverse
\(N_l\)-blocks are positive definite. Consequently,
\(S\), \(A_S\), and
\(A_{(\Sigma^{(0)})^{-1}}\) are positive definite. Since
\(\mathscr S_p\) and \(\mathsf Q_0\) are orthogonal, the equivariance of the
Moore--Penrose inverse under orthogonal transformations and
\eqref{supp:eq:plugin-J-block-decomposition} yield
\eqref{supp:eq:plugin-J-plus-block-decomposition}.
\end{proof}

\subsubsection{Canonical form of the Covariance of the effective score}
\label{supp:sec:plugin-effective-score-covariance}

We next calculate the covariance of the effective score
\[
    \nabla_\mu
    \tilde{\ell}(Y_1,\mu_0,\nu_0)
    +
    C_0\xi_\Sigma(Y_1)
\]
appearing in
\eqref{eq:plugin-asymptotic-covariance}.

For \(W\in\{S,(\Sigma^{(0)})^{-1}\}\), define
\begin{equation}
\label{supp:eq:plugin-G-W-definition}
    G_W
    :=
    \mathbb E
    \left[
        M
        \left(
            \widehat\phi
            \left(
                \varepsilon_1,
                \kappa^{(0)},
                \Sigma^{(0)}
            \right)
        \right)^\top
        W
        M
        \left(
            \widehat\phi
            \left(
                \varepsilon_1,
                \kappa^{(0)},
                \Sigma^{(0)}
            \right)
        \right)
    \right].
\end{equation}
Also, let
\begin{equation}
\label{supp:eq:plugin-beta-kappa-Sigma-definition}
    \beta_{\kappa,\Sigma}
    :=
    \frac{
        4d_\kappa
        \sqrt{1-d_\kappa^2}\,
        d_\Sigma
        (1-d_\Sigma^2)
    }{
        (1-d_\kappa^2d_\Sigma^2)^2
    },
    \quad
    e_1
    :=
    \begin{pmatrix}
        1\\
        0
    \end{pmatrix}.
\end{equation}

\begin{lemma}[Covariance of the plug-in effective score]
\label{supp:lem:plugin-effective-score-covariance}
Under the conditions of
Theorem~\ref{thm:iterated-plugin-loading-expansion-clt}, the two components of the effective score are uncorrelated:
\begin{equation}
\label{supp:eq:plugin-effective-score-cross-covariance}
    \operatorname{Cov}
    \left\{
        \nabla_\mu
        \tilde{\ell}(Y_1,\mu_0,\nu_0),
        \xi_\Sigma(Y_1)
    \right\}
    =
    0_{2p\times3}.
\end{equation}
Moreover, the profile-score covariance satisfies
\begin{equation}
\label{supp:eq:plugin-profile-score-block-covariance}
\begin{aligned}
&
\mathsf Q_0^\top
\mathscr S_p
\operatorname{Var}
\left\{
    \nabla_\mu
    \tilde{\ell}(Y_1,\mu_0,\nu_0)
\right\}
\mathscr S_p^\top
\mathsf Q_0\\
&=
\operatorname{diag}
\left(
    0_2,\,
    4
    \left(
        A_S+G_S
    \right),\,
    \mathrm{Id}_{p-2}
    \otimes
    4
    \left\{
        A_{(\Sigma^{(0)})^{-1}}
        +
        G_{(\Sigma^{(0)})^{-1}}
    \right\}
\right).
\end{aligned}
\end{equation}

Let
\[
    \Omega_\xi
    :=
    \operatorname{Var}
    \left\{
        \xi_\Sigma(Y_1)
    \right\}.
\]
Then the contribution arising from the covariance input satisfies
\begin{equation}
\label{supp:eq:plugin-C-Omega-C-block-form}
\begin{aligned}
&
\mathsf Q_0^\top
\mathscr S_p
C_0\Omega_\xi C_0^\top
\mathscr S_p^\top
\mathsf Q_0=
\operatorname{diag}
\left(
    0_2,\,
    \frac{
        4\beta_{\kappa,\Sigma}^2
    }{
        p-2
    }
    e_1e_1^\top,\,
    \mathrm{Id}_{p-2}
    \otimes
    0_2
\right).
\end{aligned}
\end{equation}
Consequently,
\begin{equation}
\label{supp:eq:plugin-effective-score-block-covariance}
\begin{aligned}
&
\mathsf Q_0^\top
\mathscr S_p
\operatorname{Var}
\left\{
    \nabla_\mu
    \tilde{\ell}(Y_1,\mu_0,\nu_0)
    +
    C_0\xi_\Sigma(Y_1)
\right\}
\mathscr S_p^\top
\mathsf Q_0\\
&=
\operatorname{diag}
\Bigg(
    0_2,\,
    4
    \left\{
        A_S
        +
        G_S
        +
        \frac{
            \beta_{\kappa,\Sigma}^2
        }{
            p-2
        }
        e_1e_1^\top
    \right\},
    \mathrm{Id}_{p-2}
    \otimes
    4
    \left\{
        A_{(\Sigma^{(0)})^{-1}}
        +
        G_{(\Sigma^{(0)})^{-1}}
    \right\}
\Bigg).
\end{aligned}
\end{equation}
\end{lemma}

\begin{proof}
We first calculate the covariance of the profile-score component.
By the real-linearity, and in particular the additivity, of
\(\widehat\phi(y,\kappa,\Sigma)\) in \(y\),
\begin{equation}
\label{supp:eq:plugin-phi-signal-noise-decomposition}
\begin{aligned}
    \widehat\phi
    \left(
        Y_1,
        \kappa^{(0)},
        \Sigma^{(0)}
    \right)
    &=
    \phi_1
    +
    \widehat\phi
    \left(
        \varepsilon_1,
        \kappa^{(0)},
        \Sigma^{(0)}
    \right).
\end{aligned}
\end{equation}
Consequently, the profiled residual at the true parameter satisfies
\[
\begin{aligned}
&
Y_1
-
\widehat\phi
\left(
    Y_1,
    \kappa^{(0)},
    \Sigma^{(0)}
\right)
\kappa^{(0)}
=
\varepsilon_1
-
\widehat\phi
\left(
    \varepsilon_1,
    \kappa^{(0)},
    \Sigma^{(0)}
\right)
\kappa^{(0)}.
\end{aligned}
\]
Substituting this identity into
\eqref{supp:eq: explicit kappa gradient stacked} and using the
definition of \(M(\cdot)\), we obtain
\begin{equation}
\label{supp:eq:plugin-kappa-gradient-whitened-residual-form}
\begin{aligned}
&
\nabla_{\widetilde{\operatorname{vec}}(\kappa)}
\widetilde\rho
\left(
    Y_1,
    \kappa^{(0)},
    \Sigma^{(0)}
\right)\\
&=
-2
\left[
    \left\{
        (\Sigma^{(0)})^{-1/2}
        M
        \left(
            \widehat\phi
            \left(
                Y_1,
                \kappa^{(0)},
                \Sigma^{(0)}
            \right)
        \right)
    \right\}
    \otimes
    \mathrm{Id}_p
\right]^\top\\
&\quad\times
\left\{
    (\Sigma^{(0)})^{-1/2}
    \otimes
    \mathrm{Id}_p
\right\}
\widetilde{\operatorname{vec}}
\left[
    \varepsilon_1
    -
    \widehat\phi
    \left(
        \varepsilon_1,
        \kappa^{(0)},
        \Sigma^{(0)}
    \right)
    \kappa^{(0)}
\right].
\end{aligned}
\end{equation}

We also have
\begin{align}
\label{supp:eq:plugin-phihat-epsilon-true-parameter}
&
\operatorname{vec}
\left(
    \widehat\phi
    \left(
        \varepsilon_1,
        \kappa^{(0)},
        \Sigma^{(0)}
    \right)
\right)\notag\\
&=
\left(
    \kappa^{(0)}
    \diamond_{\Sigma^{(0)}}
    \kappa^{(0)}
\right)^{-1}
\left(
    \kappa^{(0)}
    \bullet_{\Sigma^{(0)}}
    \varepsilon_1
\right)\notag\\
&=
\left(
    \kappa^{(0)}
    \diamond_{\Sigma^{(0)}}
    \kappa^{(0)}
\right)^{-1}
H^\top
\left\{
    (\Sigma^{(0)})^{-1}
    \otimes
    \mathrm{Id}_p
\right\}
\widetilde{\operatorname{vec}}(\varepsilon_1).
\end{align}
Consequently,
\begin{align}
&
\left\{
    (\Sigma^{(0)})^{-1/2}
    \otimes
    \mathrm{Id}_p
\right\}
\widetilde{\operatorname{vec}}
\left[
    \varepsilon_1
    -
    \widehat\phi
    \left(
        \varepsilon_1,
        \kappa^{(0)},
        \Sigma^{(0)}
    \right)
    \kappa^{(0)}
\right]\notag\\
&=
\left(
    \mathrm{Id}_{2p}
    -
    \mathcal P_0
\right)
\left\{
    (\Sigma^{(0)})^{-1/2}
    \otimes
    \mathrm{Id}_p
\right\}
\widetilde{\operatorname{vec}}(\varepsilon_1).
\label{supp:eq:plugin-whitened-profiled-epsilon-residual}
\end{align}
Combining
\eqref{supp:eq:plugin-phi-signal-noise-decomposition},
\eqref{supp:eq:plugin-kappa-gradient-whitened-residual-form}, and
\eqref{supp:eq:plugin-whitened-profiled-epsilon-residual}, we obtain
\begin{equation}
\label{supp:eq:plugin-profile-score-direct}
\begin{aligned}
&
\nabla_\mu
\tilde{\ell}(Y_1,\mu_0,\nu_0)\\
&=
-2
\left[
    \left\{
        (\Sigma^{(0)})^{-1/2}
        M
        \left(
            \phi_1
            +
            \widehat\phi
            \left(
                \varepsilon_1,
                \kappa^{(0)},
                \Sigma^{(0)}
            \right)
        \right)
    \right\}
    \otimes
    \mathrm{Id}_p
\right]^\top\\
&\quad\times
\left(
    \mathrm{Id}_{2p}
    -
    \mathcal P_0
\right)
\left\{
    (\Sigma^{(0)})^{-1/2}
    \otimes
    \mathrm{Id}_p
\right\}
\widetilde{\operatorname{vec}}(\varepsilon_1).
\end{aligned}
\end{equation}

Both
\[
    \operatorname{vec}
    \left(
        \widehat\phi
        \left(
            \varepsilon_1,
            \kappa^{(0)},
            \Sigma^{(0)}
        \right)
    \right)
\quad\text{and}\quad
    \left(
        \mathrm{Id}_{2p}
        -
        \mathcal P_0
    \right)
    \left\{
        (\Sigma^{(0)})^{-1/2}
        \otimes
        \mathrm{Id}_p
    \right\}
    \widetilde{\operatorname{vec}}(\varepsilon_1)
\]
are linear transformations of the Gaussian vector
\(\widetilde{\operatorname{vec}}(\varepsilon_1)\), and hence are
jointly Gaussian. Their cross-covariance is
\begin{align*}
&
\operatorname{Cov}
\Bigg[
    \operatorname{vec}
    \left(
        \widehat\phi
        \left(
            \varepsilon_1,
            \kappa^{(0)},
            \Sigma^{(0)}
        \right)
    \right),
    \left(
        \mathrm{Id}_{2p}
        -
        \mathcal P_0
    \right)
    \left\{
        (\Sigma^{(0)})^{-1/2}
        \otimes
        \mathrm{Id}_p
    \right\}
    \widetilde{\operatorname{vec}}(\varepsilon_1)
\Bigg]
\\
&=
\left(
    \kappa^{(0)}
    \diamond_{\Sigma^{(0)}}
    \kappa^{(0)}
\right)^{-1}
H^\top
\left\{
    (\Sigma^{(0)})^{-1}
    \otimes
    \mathrm{Id}_p
\right\}
\operatorname{Var}
\left\{
    \widetilde{\operatorname{vec}}(\varepsilon_1)
\right\}
\\
&\quad\times
\left\{
    (\Sigma^{(0)})^{-1/2}
    \otimes
    \mathrm{Id}_p
\right\}
\left(
    \mathrm{Id}_{2p}
    -
    \mathcal P_0
\right)
\\
&=
\left(
    \kappa^{(0)}
    \diamond_{\Sigma^{(0)}}
    \kappa^{(0)}
\right)^{-1}
H^\top
\left\{
    (\Sigma^{(0)})^{-1}
    \otimes
    \mathrm{Id}_p
\right\}
\left\{
    \Sigma^{(0)}
    \otimes
    \mathrm{Id}_p
\right\}
\\
&\quad\times
\left\{
    (\Sigma^{(0)})^{-1/2}
    \otimes
    \mathrm{Id}_p
\right\}
\left(
    \mathrm{Id}_{2p}
    -
    \mathcal P_0
\right)
\\
&=
D^{-1}
H^\top
\left\{
    (\Sigma^{(0)})^{-1/2}
    \otimes
    \mathrm{Id}_p
\right\}
\left(
    \mathrm{Id}_{2p}
    -
    \mathcal P_0
\right),
\end{align*}
where we used
\[
    D
    =
    \kappa^{(0)}
    \diamond_{\Sigma^{(0)}}
    \kappa^{(0)}
    =
    H^\top
    \left\{
        (\Sigma^{(0)})^{-1}
        \otimes
        \mathrm{Id}_p
    \right\}
    H
\]
and
\[
    \operatorname{Var}
    \left\{
        \widetilde{\operatorname{vec}}(\varepsilon_1)
    \right\}
    =
    \Sigma^{(0)}
    \otimes
    \mathrm{Id}_p.
\]

To evaluate the remaining product, recall that
\[
    \mathcal P_0
    =
    \left\{
        (\Sigma^{(0)})^{-1/2}
        \otimes
        \mathrm{Id}_p
    \right\}
    H
    D^{-1}
    H^\top
    \left\{
        (\Sigma^{(0)})^{-1/2}
        \otimes
        \mathrm{Id}_p
    \right\}.
\]
Therefore,
\begin{align*}
&
H^\top
\left\{
    (\Sigma^{(0)})^{-1/2}
    \otimes
    \mathrm{Id}_p
\right\}
\mathcal P_0
\\
&=
H^\top
\left\{
    (\Sigma^{(0)})^{-1}
    \otimes
    \mathrm{Id}_p
\right\}
H
D^{-1}
H^\top
\left\{
    (\Sigma^{(0)})^{-1/2}
    \otimes
    \mathrm{Id}_p
\right\}
\\
&=
D
D^{-1}
H^\top
\left\{
    (\Sigma^{(0)})^{-1/2}
    \otimes
    \mathrm{Id}_p
\right\}
\\
&=
H^\top
\left\{
    (\Sigma^{(0)})^{-1/2}
    \otimes
    \mathrm{Id}_p
\right\}.
\end{align*}
It follows that
\begin{align*}
&
H^\top
\left\{
    (\Sigma^{(0)})^{-1/2}
    \otimes
    \mathrm{Id}_p
\right\}
\left(
    \mathrm{Id}_{2p}
    -
    \mathcal P_0
\right)
\\
&=
H^\top
\left\{
    (\Sigma^{(0)})^{-1/2}
    \otimes
    \mathrm{Id}_p
\right\}
-
H^\top
\left\{
    (\Sigma^{(0)})^{-1/2}
    \otimes
    \mathrm{Id}_p
\right\}
\mathcal P_0
\\
&=
0_{2\times2p}.
\end{align*}
Consequently,
\[
\operatorname{Cov}
\Bigg[
    \operatorname{vec}
    \left(
        \widehat\phi
        \left(
            \varepsilon_1,
            \kappa^{(0)},
            \Sigma^{(0)}
        \right)
    \right),
    \left(
        \mathrm{Id}_{2p}
        -
        \mathcal P_0
    \right)
    \left\{
        (\Sigma^{(0)})^{-1/2}
        \otimes
        \mathrm{Id}_p
    \right\}
    \widetilde{\operatorname{vec}}(\varepsilon_1)
\Bigg]
=
0_{2\times2p}.
\]
Since the two random vectors are jointly Gaussian, they are therefore
independent.

Finally, \(\phi_1\) is independent of \(\varepsilon_1\). Together with
the independence established above, this implies that
\[
    \left(
        \phi_1,
        \widehat\phi
        \left(
            \varepsilon_1,
            \kappa^{(0)},
            \Sigma^{(0)}
        \right)
    \right)
\]
is independent of the whitened profiled residual.

Conditional on this pair, the matrix multiplying the whitened
profiled residual in
\eqref{supp:eq:plugin-profile-score-direct} is fixed. The conditional
mean of the score is therefore zero, while its conditional covariance
is obtained by replacing the covariance of the whitened profiled
residual by
\(\mathrm{Id}_{2p}-\mathcal P_0\). Hence, the law of total covariance
gives
\begin{align}
&
\operatorname{Var}
\left\{
    \nabla_\mu
    \tilde{\ell}(Y_1,\mu_0,\nu_0)
\right\}\notag\\
&=
4\mathbb E
\Bigg[
    \left[
        \left\{
            (\Sigma^{(0)})^{-1/2}
            M
            \left(
                \phi_1
                +
                \widehat\phi
                \left(
                    \varepsilon_1,
                    \kappa^{(0)},
                    \Sigma^{(0)}
                \right)
            \right)
        \right\}
        \otimes
        \mathrm{Id}_p
    \right]^\top\notag\\
&\hspace{18mm}\times
    \left(
        \mathrm{Id}_{2p}
        -
        \mathcal P_0
    \right)
    \left[
        \left\{
            (\Sigma^{(0)})^{-1/2}
            M
            \left(
                \phi_1
                +
                \widehat\phi
                \left(
                    \varepsilon_1,
                    \kappa^{(0)},
                    \Sigma^{(0)}
                \right)
            \right)
        \right\}
        \otimes
        \mathrm{Id}_p
    \right]
\Bigg].
\label{supp:eq:plugin-profile-score-covariance}
\end{align}

Since \(M(\cdot)\) is real-linear,
\[
\begin{aligned}
&M
\left(
    \phi_1
    +
    \widehat\phi
    \left(
        \varepsilon_1,
        \kappa^{(0)},
        \Sigma^{(0)}
    \right)
\right)=
M(\phi_1)
+
M
\left(
    \widehat\phi
    \left(
        \varepsilon_1,
        \kappa^{(0)},
        \Sigma^{(0)}
    \right)
\right).
\end{aligned}
\]
Furthermore, \(\phi_1\) is independent of \(\varepsilon_1\), while
\eqref{supp:eq:plugin-phihat-epsilon-true-parameter} implies
\[
    \mathbb E
    \left[
        M
        \left(
            \widehat\phi
            \left(
                \varepsilon_1,
                \kappa^{(0)},
                \Sigma^{(0)}
            \right)
        \right)
    \right]
    =
    0.
\]
Therefore, the two cross terms in
\eqref{supp:eq:plugin-profile-score-covariance} vanish. The quadratic
term involving \(M(\phi_1)\) equals \(2J\) by definition. Consequently,
\begin{align}
&
\operatorname{Var}
\left\{
    \nabla_\mu
    \tilde{\ell}(Y_1,\mu_0,\nu_0)
\right\}\notag\\
&=
2J\notag\\
&\quad+
4\mathbb E
\Bigg[
    \left\{
        (\Sigma^{(0)})^{-1/2}
        M
        \left(
            \widehat\phi
            \left(
                \varepsilon_1,
                \kappa^{(0)},
                \Sigma^{(0)}
            \right)
        \right)
        \otimes
        \mathrm{Id}_p
    \right\}^{\top}\notag\\
&\hspace{18mm}\times
    \left(
        \mathrm{Id}_{2p}
        -
        \mathcal P_0
    \right)
    \left\{
        (\Sigma^{(0)})^{-1/2}
        M
        \left(
            \widehat\phi
            \left(
                \varepsilon_1,
                \kappa^{(0)},
                \Sigma^{(0)}
            \right)
        \right)
        \otimes
        \mathrm{Id}_p
    \right\}
\Bigg].
\label{supp:eq:plugin-profile-score-covariance-decomposed}
\end{align}

Using the definition of \(\mathcal W_0\), the second term in
\eqref{supp:eq:plugin-profile-score-covariance-decomposed} can be
written as
\[
4\mathbb E
\left[
    \left\{
        M
        \left(
            \widehat\phi
            \left(
                \varepsilon_1,
                \kappa^{(0)},
                \Sigma^{(0)}
            \right)
        \right)^\top
        \otimes
        \mathrm{Id}_p
    \right\}
    \mathcal W_0
    \left\{
        M
        \left(
            \widehat\phi
            \left(
                \varepsilon_1,
                \kappa^{(0)},
                \Sigma^{(0)}
            \right)
        \right)
        \otimes
        \mathrm{Id}_p
    \right\}
\right].
\]

Since
\[
    \mathscr S_p^\top\mathscr S_p
    =
    \mathsf Q_0\mathsf Q_0^\top
    =
    \mathrm{Id}_{2p},
\]
we have
\begin{align}
&
\mathsf Q_0^\top
\mathscr S_p
\mathbb E
\left[
    \left\{
        M
        \left(
            \widehat\phi
            \left(
                \varepsilon_1,
                \kappa^{(0)},
                \Sigma^{(0)}
            \right)
        \right)^\top
        \otimes
        \mathrm{Id}_p
    \right\}
    \mathcal W_0
    \left\{
        M
        \left(
            \widehat\phi
            \left(
                \varepsilon_1,
                \kappa^{(0)},
                \Sigma^{(0)}
            \right)
        \right)
        \otimes
        \mathrm{Id}_p
    \right\}
\right]
\mathscr S_p^\top
\mathsf Q_0
\notag\\
&=
\mathbb E
\Bigg[
    \mathsf Q_0^\top
    \mathscr S_p
    \left\{
        M
        \left(
            \widehat\phi
            \left(
                \varepsilon_1,
                \kappa^{(0)},
                \Sigma^{(0)}
            \right)
        \right)^\top
        \otimes
        \mathrm{Id}_p
    \right\}
    \mathcal W_0
\notag\\
&\hspace{35mm}\times
    \left\{
        M
        \left(
            \widehat\phi
            \left(
                \varepsilon_1,
                \kappa^{(0)},
                \Sigma^{(0)}
            \right)
        \right)
        \otimes
        \mathrm{Id}_p
    \right\}
    \mathscr S_p^\top
    \mathsf Q_0
\Bigg]
\notag\\
&=
\mathbb E
\Bigg[
    \mathsf Q_0^\top
    \mathscr S_p
    \left\{
        M
        \left(
            \widehat\phi
            \left(
                \varepsilon_1,
                \kappa^{(0)},
                \Sigma^{(0)}
            \right)
        \right)^\top
        \otimes
        \mathrm{Id}_p
    \right\}
    \mathscr S_p^\top
\notag\\
&\hspace{18mm}\times
    \left(
        \mathscr S_p
        \mathcal W_0
        \mathscr S_p^\top
    \right)
    \mathscr S_p
    \left\{
        M
        \left(
            \widehat\phi
            \left(
                \varepsilon_1,
                \kappa^{(0)},
                \Sigma^{(0)}
            \right)
        \right)
        \otimes
        \mathrm{Id}_p
    \right\}
    \mathscr S_p^\top
    \mathsf Q_0
\Bigg]
\notag\\
&=
\mathbb E
\Bigg[
    \mathsf Q_0^\top
    \left\{
        \mathrm{Id}_p
        \otimes
        M
        \left(
            \widehat\phi
            \left(
                \varepsilon_1,
                \kappa^{(0)},
                \Sigma^{(0)}
            \right)
        \right)
    \right\}^{\top}
    \left(
        \mathscr S_p
        \mathcal W_0
        \mathscr S_p^\top
    \right)
\notag\\
&\hspace{35mm}\times
    \left\{
        \mathrm{Id}_p
        \otimes
        M
        \left(
            \widehat\phi
            \left(
                \varepsilon_1,
                \kappa^{(0)},
                \Sigma^{(0)}
            \right)
        \right)
    \right\}
    \mathsf Q_0
\Bigg]
\notag\\
&=
\mathbb E
\Bigg[
    \left[
        \mathsf Q_0^\top
        \left\{
            \mathrm{Id}_p
            \otimes
            M
            \left(
                \widehat\phi
                \left(
                    \varepsilon_1,
                    \kappa^{(0)},
                    \Sigma^{(0)}
                \right)
            \right)
        \right\}
        \mathsf Q_0
    \right]^\top
\notag\\
&\hspace{18mm}\times
    \left(
        \mathsf Q_0^\top
        \mathscr S_p
        \mathcal W_0
        \mathscr S_p^\top
        \mathsf Q_0
    \right)
    \left[
        \mathsf Q_0^\top
        \left\{
            \mathrm{Id}_p
            \otimes
            M
            \left(
                \widehat\phi
                \left(
                    \varepsilon_1,
                    \kappa^{(0)},
                    \Sigma^{(0)}
                \right)
            \right)
        \right\}
        \mathsf Q_0
    \right]
\Bigg]
\notag\\
&=
\mathbb E
\Bigg[
    \left\{
        \mathrm{Id}_p
        \otimes
        M
        \left(
            \widehat\phi
            \left(
                \varepsilon_1,
                \kappa^{(0)},
                \Sigma^{(0)}
            \right)
        \right)
    \right\}^{\top}
    \operatorname{diag}
    \left(
        0_2,\,
        S,\,
        \mathrm{Id}_{p-2}
        \otimes
        (\Sigma^{(0)})^{-1}
    \right)
\notag\\
&\hspace{35mm}\times
    \left\{
        \mathrm{Id}_p
        \otimes
        M
        \left(
            \widehat\phi
            \left(
                \varepsilon_1,
                \kappa^{(0)},
                \Sigma^{(0)}
            \right)
        \right)
    \right\}
\Bigg]
\notag\\
&=
\operatorname{diag}
\Bigg(
    0_2,\,
    \mathbb E
    \left[
        M
        \left(
            \widehat\phi
            \left(
                \varepsilon_1,
                \kappa^{(0)},
                \Sigma^{(0)}
            \right)
        \right)^\top
        S
        M
        \left(
            \widehat\phi
            \left(
                \varepsilon_1,
                \kappa^{(0)},
                \Sigma^{(0)}
            \right)
        \right)
    \right],
\notag\\
&\hspace{27mm}
    \mathrm{Id}_{p-2}
    \otimes
    \mathbb E
    \left[
        M
        \left(
            \widehat\phi
            \left(
                \varepsilon_1,
                \kappa^{(0)},
                \Sigma^{(0)}
            \right)
        \right)^\top
        (\Sigma^{(0)})^{-1}
        M
        \left(
            \widehat\phi
            \left(
                \varepsilon_1,
                \kappa^{(0)},
                \Sigma^{(0)}
            \right)
        \right)
    \right]
\Bigg)
\notag\\
&=
\operatorname{diag}
\left(
    0_2,\,
    G_S,\,
    \mathrm{Id}_{p-2}
    \otimes
    G_{(\Sigma^{(0)})^{-1}}
\right).
\label{supp:eq:plugin-middle-variance-block-decomposition}
\end{align}

Combining this identity with
\eqref{supp:eq:plugin-J-block-decomposition} and
\eqref{supp:eq:plugin-profile-score-covariance-decomposed} gives
\eqref{supp:eq:plugin-profile-score-block-covariance}.

We now establish
\eqref{supp:eq:plugin-effective-score-cross-covariance}. By
Lemma~\ref{supp:lem:plugin-covariance-influence-moments},
\(\xi_\Sigma(Y_1)\) is a function of the Gaussian noise coordinates
in the orthogonal complement of
\[
    \operatorname{span}
    \left\{
        \operatorname{Re}(\kappa^{(0)}),
        \operatorname{Im}(\kappa^{(0)})
    \right\}.
\]
On the other hand,
\[
    \widehat\phi
    \left(
        \varepsilon_1,
        \kappa^{(0)},
        \Sigma^{(0)}
    \right)
\]
depends only on the complementary Gaussian coordinates associated
with
\[
    \operatorname{span}
    \left\{
        \operatorname{Re}(\kappa^{(0)}),
        \operatorname{Im}(\kappa^{(0)})
    \right\},
\]
and is therefore independent of the noise coordinates entering
\(\xi_\Sigma(Y_1)\).

Conditional on
\[
    \phi_1
    \quad\text{and}\quad
    \widehat\phi
    \left(
        \varepsilon_1,
        \kappa^{(0)},
        \Sigma^{(0)}
    \right),
\]
the score in
\eqref{supp:eq:plugin-profile-score-direct} is linear in the remaining
centered Gaussian noise coordinates, whereas
\(\xi_\Sigma(Y_1)\) is centered quadratic in those coordinates.
Hence, their covariance vanishes by the zero third moments of
centered Gaussian random variables:
\[
    \operatorname{Cov}
    \left\{
        \nabla_\mu
        \tilde{\ell}(Y_1,\mu_0,\nu_0),
        \xi_\Sigma(Y_1)
    \right\}
    =
    0_{2p\times3}.
\]

It remains to calculate $\mathsf Q_0^\top
\mathscr S_p
C_0\Omega_\xi C_0^\top
\mathscr S_p^\top
\mathsf Q_0$. Fix a symmetric perturbation
\[
    \Delta
    =
    \begin{pmatrix}
        \Delta_{11}&\Delta_{12}\\
        \Delta_{12}&\Delta_{22}
    \end{pmatrix}.
\]
By the definition of \(C_0\) and the linearity of
\(\mathbf C_{\Sigma^{(0)}}(\Delta)\) in \(\Delta\),
\[
    C_0\operatorname{vech}(\Delta)
    =
    \mathbb E
    \left[
        \left.
        \mathbf C_{\Sigma^{(0)}}(\Delta)
        \right|_{
            \kappa=\kappa^{(0)},\,y=Y_1
        }
    \right].
\]
We first calculate \(\mathscr S_p C_0\operatorname{vech}(\Delta)\). By
\eqref{supp:eq: C Sigma definition}, the \(j\)-th consecutive
two-dimensional block is
\begin{align}
&
\left[
    \mathscr S_p
    \mathbf C_{\Sigma^{(0)}}(\Delta)
\right]_j
\nonumber\\
&\quad=
2
\begin{pmatrix}
r_j^\top(\Sigma^{(0)})^{-1}
\Delta(\Sigma^{(0)})^{-1}
\operatorname{vec}
\left\{
    \widehat\phi
    \left(
        Y_1,\kappa^{(0)},\Sigma^{(0)}
    \right)
\right\}
\\[1mm]
r_j^\top(\Sigma^{(0)})^{-1}
\Delta(\Sigma^{(0)})^{-1}J_2
\operatorname{vec}
\left\{
    \widehat\phi
    \left(
        Y_1,\kappa^{(0)},\Sigma^{(0)}
    \right)
\right\}
\end{pmatrix}
\nonumber\\
&\quad+
2
\begin{pmatrix}
q_{\mathrm R,j}^\top\\
q_{\mathrm I,j}^\top
\end{pmatrix}
\left(
    \kappa^{(0)}
    \diamond_{\Sigma^{(0)}}
    \kappa^{(0)}
\right)^{-1}
\sum_{l=1}^p
M(\kappa_l^{(0)})^\top
(\Sigma^{(0)})^{-1}
\Delta(\Sigma^{(0)})^{-1}r_l.
\label{supp:eq:plugin-Sp-C-Sigma-direct}
\end{align}
Here, all \(r_j,q_{\mathrm R,j}\), and \(q_{\mathrm I,j}\) are evaluated at
\(
    \kappa=\kappa^{(0)},
    \Sigma=\Sigma^{(0)},
    y=Y_1.
\) Now we write
\[
\mathbf r
=
\begin{pmatrix}
r_1\\
\vdots\\
r_p
\end{pmatrix}
\] 

By the definition of the profiled coefficient,
\begin{align*}
&
\operatorname{vec}
\left\{
\widehat\phi
\left(
Y_1,\kappa^{(0)},\Sigma^{(0)}
\right)
\right\}=
\left(
\kappa^{(0)}
\diamond_{\Sigma^{(0)}}
\kappa^{(0)}
\right)^{-1}
M(\kappa^{(0)})^\top
\left\{
\mathrm{Id}_p\otimes(\Sigma^{(0)})^{-1}
\right\}
\operatorname{vec}(Y_1),
\end{align*}
where
\[
M(\kappa^{(0)})^\top
\left\{
\mathrm{Id}_p\otimes(\Sigma^{(0)})^{-1}
\right\}
M(\kappa^{(0)})
=
\kappa^{(0)}
\diamond_{\Sigma^{(0)}}
\kappa^{(0)}.
\]
Since
\[
\operatorname{vec}(Y_1)
=
M(\kappa^{(0)})\operatorname{vec}(\phi_1)
+
\operatorname{vec}(\varepsilon_1),
\]
it follows that
\[
\operatorname{vec}
\left\{
\widehat\phi
\left(
Y_1,\kappa^{(0)},\Sigma^{(0)}
\right)
\right\}
=
\operatorname{vec}(\phi_1)
+
\operatorname{vec}(\widehat\phi_{\varepsilon,1}),
\]
where
\begin{equation}
\operatorname{vec}(\widehat\phi_{\varepsilon,1})
=
\left(
\kappa^{(0)}
\diamond_{\Sigma^{(0)}}
\kappa^{(0)}
\right)^{-1}
M(\kappa^{(0)})^\top
\left\{
\mathrm{Id}_p\otimes(\Sigma^{(0)})^{-1}
\right\}
\operatorname{vec}(\varepsilon_1).
\label{supp:eq:plugin-profile-noise-coefficient}
\end{equation}
Consequently,
\begin{align}
\mathbf r
&=
\operatorname{vec}(\varepsilon_1)
-
M(\kappa^{(0)})
\operatorname{vec}(\widehat\phi_{\varepsilon,1})
\nonumber\\
&=
\left[
\mathrm{Id}_{2p}
-
M(\kappa^{(0)})
\left(
\kappa^{(0)}
\diamond_{\Sigma^{(0)}}
\kappa^{(0)}
\right)^{-1}
M(\kappa^{(0)})^\top
\left\{
\mathrm{Id}_p\otimes(\Sigma^{(0)})^{-1}
\right\}
\right]
\operatorname{vec}(\varepsilon_1).
\label{supp:eq:plugin-profile-residual-linear-form}
\end{align}
Thus, both
\(\operatorname{vec}(\widehat\phi_{\varepsilon,1})\) and \(\mathbf{r}\) are linear transformations of
\(\operatorname{vec}(\varepsilon_1)\).

Since
\(
\operatorname{Cov}
\left\{
\operatorname{vec}(\varepsilon_1)
\right\}
=
\mathrm{Id}_p\otimes\Sigma^{(0)},
\)
equation \eqref{supp:eq:plugin-profile-noise-coefficient} gives
\begin{align*}
&
\operatorname{Cov}
\left\{
\operatorname{vec}(\widehat\phi_{\varepsilon,1})
\right\}\\
&=
\left(
\kappa^{(0)}
\diamond_{\Sigma^{(0)}}
\kappa^{(0)}
\right)^{-1}
M(\kappa^{(0)})^\top
\left\{
\mathrm{Id}_p\otimes(\Sigma^{(0)})^{-1}
\right\}
\left\{
\mathrm{Id}_p\otimes\Sigma^{(0)}
\right\}
\left\{
\mathrm{Id}_p\otimes(\Sigma^{(0)})^{-1}
\right\}
M(\kappa^{(0)})
\left(
\kappa^{(0)}
\diamond_{\Sigma^{(0)}}
\kappa^{(0)}
\right)^{-1}\\
&=
\left(
\kappa^{(0)}
\diamond_{\Sigma^{(0)}}
\kappa^{(0)}
\right)^{-1}.
\end{align*}
Similarly,
\begin{align*}
&
\operatorname{Cov}
\left\{
\operatorname{vec}(\widehat\phi_{\varepsilon,1}),
\operatorname{vec}(\varepsilon_1)
\right\}\\
&=
\left(
\kappa^{(0)}
\diamond_{\Sigma^{(0)}}
\kappa^{(0)}
\right)^{-1}
M(\kappa^{(0)})^\top
\left\{
\mathrm{Id}_p\otimes(\Sigma^{(0)})^{-1}
\right\}
\left\{
\mathrm{Id}_p\otimes\Sigma^{(0)}
\right\}\\
&=
\left(
\kappa^{(0)}
\diamond_{\Sigma^{(0)}}
\kappa^{(0)}
\right)^{-1}
M(\kappa^{(0)})^\top.
\end{align*}
Therefore,
\begin{align*}
&
\operatorname{Cov}
\left\{
\operatorname{vec}(\widehat\phi_{\varepsilon,1}),\mathbf{r}
\right\}\\
&=
\operatorname{Cov}
\left\{
\operatorname{vec}(\widehat\phi_{\varepsilon,1}),
\operatorname{vec}(\varepsilon_1)
\right\}
-
\operatorname{Cov}
\left\{
\operatorname{vec}(\widehat\phi_{\varepsilon,1})
\right\}
M(\kappa^{(0)})^\top\\
&=
\left(
\kappa^{(0)}
\diamond_{\Sigma^{(0)}}
\kappa^{(0)}
\right)^{-1}
M(\kappa^{(0)})^\top
-
\left(
\kappa^{(0)}
\diamond_{\Sigma^{(0)}}
\kappa^{(0)}
\right)^{-1}
M(\kappa^{(0)})^\top\\
&=0.
\end{align*}

Since
\(\operatorname{vec}(\widehat\phi_{\varepsilon,1})\) and $\mathbf{r}$ are jointly Gaussian, their zero cross-covariance
also implies that they are independent.

By \eqref{supp:eq:plugin-profile-residual-linear-form} and
\[
\operatorname{Cov}
\{
\operatorname{vec}(\varepsilon_1)
\}
=
\mathrm{Id}_p\otimes\Sigma^{(0)},
\]
we have
\begin{align*}
\operatorname{Cov}(\mathbf{r})&
=
\left[
\mathrm{Id}_{2p}
-
M(\kappa^{(0)})
\left(
\kappa^{(0)}
\diamond_{\Sigma^{(0)}}
\kappa^{(0)}
\right)^{-1}
M(\kappa^{(0)})^\top
\{
\mathrm{Id}_p\otimes(\Sigma^{(0)})^{-1}
\}
\right]\\
&
\{
\mathrm{Id}_p\otimes\Sigma^{(0)}
\}
\left[
\mathrm{Id}_{2p}
-
\{
\mathrm{Id}_p\otimes(\Sigma^{(0)})^{-1}
\}
M(\kappa^{(0)})
\left(
\kappa^{(0)}
\diamond_{\Sigma^{(0)}}
\kappa^{(0)}
\right)^{-1}
M(\kappa^{(0)})^\top
\right].
\end{align*}
Using
\[
M(\kappa^{(0)})^\top
\{
\mathrm{Id}_p\otimes(\Sigma^{(0)})^{-1}
\}
M(\kappa^{(0)})
=
\kappa^{(0)}
\diamond_{\Sigma^{(0)}}
\kappa^{(0)},
\]
the preceding expression reduces to
\begin{align*}
&
\mathrm{Id}_p\otimes\Sigma^{(0)}
-
2M(\kappa^{(0)})
\left(
\kappa^{(0)}
\diamond_{\Sigma^{(0)}}
\kappa^{(0)}
\right)^{-1}
M(\kappa^{(0)})^\top\\
&
+
M(\kappa^{(0)})
\left(
\kappa^{(0)}
\diamond_{\Sigma^{(0)}}
\kappa^{(0)}
\right)^{-1}
M(\kappa^{(0)})^\top
\{
\mathrm{Id}_p\otimes(\Sigma^{(0)})^{-1}
\}
M(\kappa^{(0)})
\left(
\kappa^{(0)}
\diamond_{\Sigma^{(0)}}
\kappa^{(0)}
\right)^{-1}
M(\kappa^{(0)})^\top\\
&=
\mathrm{Id}_p\otimes\Sigma^{(0)}
-
M(\kappa^{(0)})
\left(
\kappa^{(0)}
\diamond_{\Sigma^{(0)}}
\kappa^{(0)}
\right)^{-1}
M(\kappa^{(0)})^\top.
\end{align*}
Therefore,
\begin{align}
&
\operatorname{Cov}(\mathbf{r})
=
\mathrm{Id}_p\otimes\Sigma^{(0)}
-
M(\kappa^{(0)})
\left(
\kappa^{(0)}
\diamond_{\Sigma^{(0)}}
\kappa^{(0)}
\right)^{-1}
M(\kappa^{(0)})^\top.
\label{supp:eq:plugin-profile-residual-covariance}
\end{align}

At the true parameter, the definitions of \(q_{\mathrm R,j}\) and \(q_{\mathrm I,j}\)
give
\begin{align*}
q_{\mathrm R,j}^\top
&=
r_j^\top(\Sigma^{(0)})^{-1}
-
\operatorname{vec}
\left\{
    \widehat\phi
    \left(
        Y_1,\kappa^{(0)},\Sigma^{(0)}
    \right)
\right\}^{\!\top}
(\Sigma^{(0)})^{-1}M(\kappa_j^{(0)}),\\
q_{\mathrm I,j}^\top
&=
r_j^\top(\Sigma^{(0)})^{-1}J_2
-
\operatorname{vec}
\left\{
    \widehat\phi
    \left(
        Y_1,\kappa^{(0)},\Sigma^{(0)}
    \right)
\right\}^{\!\top}
J_2^\top(\Sigma^{(0)})^{-1}M(\kappa_j^{(0)}).
\end{align*}

For every \(l=1,\ldots,p\), independence of \(\phi_1\) and
\(\varepsilon_1\), the centeredness of the residuals, and the zero
cross-covariance established above give
\begin{align*}
&
\mathbb E
\left[
\operatorname{vec}
\{
\widehat\phi
(
Y_1,\kappa^{(0)},\Sigma^{(0)}
)
\}
r_l^\top
\right]=
\mathbb E
[
\operatorname{vec}(\phi_1)
]
\mathbb E[r_l]^\top
+
\mathbb E
[
\operatorname{vec}(\widehat\phi_{\varepsilon,1})r_l^\top
]
=0.
\end{align*}
Consequently, after substituting the definitions of \(q_{\mathrm R,j}\) and
\(q_{\mathrm I,j}\) into \eqref{supp:eq:plugin-Sp-C-Sigma-direct}, every term containing
the fitted factor has zero expectation. 

\begin{align}
&
\mathscr S_p
C_0\operatorname{vech}(\Delta)
=
2\mathbb E
\left[
\begin{pmatrix}
r_1^\top(\Sigma^{(0)})^{-1}\\
r_1^\top(\Sigma^{(0)})^{-1}J_2\\
\vdots\\
r_p^\top(\Sigma^{(0)})^{-1}\\
r_p^\top(\Sigma^{(0)})^{-1}J_2
\end{pmatrix}
b_\Delta
\right]
\label{supp:eq:plugin-Sp-C0-after-expectation}
\end{align}
for 
\[
b_\Delta
:=
D^{-1}
M(\kappa^{(0)})^\top
\{
\mathrm{Id}_p\otimes
(\Sigma^{(0)})^{-1}
\Delta
(\Sigma^{(0)})^{-1}
\}
\mathbf r.
\]

Now we calculate 
\begin{align}
&
\mathsf Q_0^\top\mathscr S_p
C_0\operatorname{vech}(\Delta)=
\begin{pmatrix}
M(\kappa^{(0)})^\top
\mathscr S_pC_0\operatorname{vech}(\Delta)
\\[1mm]
N_0^\top
\mathscr S_pC_0\operatorname{vech}(\Delta)
\\[1mm]
N_1^\top
\mathscr S_pC_0\operatorname{vech}(\Delta)
\\
\vdots
\\
N_{p-2}^\top
\mathscr S_pC_0\operatorname{vech}(\Delta)
\end{pmatrix},
\label{supp:eq:plugin-U-C0-block-decomposition}
\end{align}
by calculating these two-dimensional blocks separately.

Fix a \(2p\times2\) block column of \(\mathsf Q_0\), denoted by \(X\), and
write
\[
    X
    =
    \begin{pmatrix}
        B_1\\
        \vdots\\
        B_p
    \end{pmatrix},
    \quad
    B_j\in\mathbb R^{2\times2}.
\]
By definition of \(\mathsf Q_0\), every \(B_j\) is a linear combination
of \(\mathrm{Id}_2\) and \(J_2\). Thus, we may write
\[
    B_j
    =
    \alpha_j\mathrm{Id}_2+\beta_jJ_2.
\]
We first establish a \(2\times2\) identity that will be applied to
these row blocks. Let
\[
    B=\alpha\mathrm{Id}_2+\beta J_2,
    \quad
    a,b\in\mathbb R^2.
\]
Since \(J_2^\top=-J_2\) and \(J_2^2=-\mathrm{Id}_2\),
\begin{align*}
B^\top
\begin{pmatrix}
    b^\top a\\
    (J_2b)^\top a
\end{pmatrix}
&=
\begin{pmatrix}
    \alpha b^\top a+\beta(J_2b)^\top a\\
    -\beta b^\top a+\alpha(J_2b)^\top a
\end{pmatrix}.
\end{align*}
On the other hand, using
\(\operatorname{Tr}(Cab^\top)=b^\top Ca\), we obtain
\begin{align*}
\operatorname{Tr}(B^\top ab^\top)
&=
b^\top B^\top a
=
(Bb)^\top a
=
\alpha b^\top a+\beta(J_2b)^\top a,
\\
\operatorname{Tr}(J_2^\top B^\top ab^\top)
&=
b^\top J_2^\top B^\top a
=
(BJ_2b)^\top a
=
-\beta b^\top a+\alpha(J_2b)^\top a.
\end{align*}
Consequently,
\begin{equation}
B^\top
\begin{pmatrix}
    b^\top a\\
    (J_2b)^\top a
\end{pmatrix}
=
\begin{pmatrix}
    \operatorname{Tr}(B^\top ab^\top)\\[1mm]
    \operatorname{Tr}(J_2^\top B^\top ab^\top)
\end{pmatrix}.
\label{supp:eq:plugin-two-dimensional-block-trace-identity}
\end{equation}

Now set
\[
    a_j=(\Sigma^{(0)})^{-1}r_j,
    \quad j=1,\ldots,p.
\]
Since \((\Sigma^{(0)})^{-1}\) is symmetric, for every
\(b\in\mathbb R^2\),
\[
    b^\top a_j
    =
    r_j^\top(\Sigma^{(0)})^{-1}b,
    \quad
    (J_2b)^\top a_j
    =
    r_j^\top(\Sigma^{(0)})^{-1}J_2b.
\]
Applying \eqref{supp:eq:plugin-two-dimensional-block-trace-identity} to
\(B_j\) and summing over \(j=1,\ldots,p\) therefore gives
\begin{align}
&
X^\top
\begin{pmatrix}
    r_1^\top(\Sigma^{(0)})^{-1}b\\
    r_1^\top(\Sigma^{(0)})^{-1}J_2b\\
    \vdots\\
    r_p^\top(\Sigma^{(0)})^{-1}b\\
    r_p^\top(\Sigma^{(0)})^{-1}J_2b
\end{pmatrix}
=
\begin{pmatrix}
\operatorname{Tr}
\left[
    X^\top
    \left\{
        \mathrm{Id}_p\otimes(\Sigma^{(0)})^{-1}
    \right\}
    \mathbf r b^\top
\right]
\\[3mm]
\operatorname{Tr}
\left[
    J_2^\top X^\top
    \left\{
        \mathrm{Id}_p\otimes(\Sigma^{(0)})^{-1}
    \right\}
    \mathbf r b^\top
\right]
\end{pmatrix},
\quad b\in\mathbb R^2.
\label{supp:eq:plugin-stacked-block-trace-identity}
\end{align}

Applying \eqref{supp:eq:plugin-stacked-block-trace-identity} with
\(b=b_\Delta\) to \eqref{supp:eq:plugin-Sp-C0-after-expectation} yields
\begin{align}
&
X^\top\mathscr S_p
C_0\operatorname{vech}(\Delta)
=
2
\begin{pmatrix}
\operatorname{Tr}
\left[
    X^\top
    \mathbb E
    \left\{
        \left(
            \mathrm{Id}_p\otimes(\Sigma^{(0)})^{-1}
        \right)
        \mathbf r b_\Delta^\top
    \right\}
\right]
\\[3mm]
\operatorname{Tr}
\left[
    J_2^\top X^\top
    \mathbb E
    \left\{
        \left(
            \mathrm{Id}_p\otimes(\Sigma^{(0)})^{-1}
        \right)
        \mathbf r b_\Delta^\top
    \right\}
\right]
\end{pmatrix}.
\label{supp:eq:plugin-X-block-C0-before-cross-moment}
\end{align}

Since \(\mathbf r\) is centered, we have
\begin{align*}
&
\mathbb E
\left[
\{
\mathrm{Id}_p\otimes(\Sigma^{(0)})^{-1}
\}
\mathbf r b_\Delta^\top
\right]
\\
&=
\{
\mathrm{Id}_p\otimes(\Sigma^{(0)})^{-1}
\}
\operatorname{Cov}(\mathbf r)
\{
\mathrm{Id}_p\otimes
(\Sigma^{(0)})^{-1}
\Delta
(\Sigma^{(0)})^{-1}
\}
M(\kappa^{(0)})D^{-1}
\\
&=
\{
\mathrm{Id}_p\otimes(\Sigma^{(0)})^{-1}
\}
\operatorname{Cov}(\mathbf r)
\{
\mathrm{Id}_p\otimes(\Sigma^{(0)})^{-1}
\}
\{
\mathrm{Id}_p\otimes
\Delta(\Sigma^{(0)})^{-1}
\}
M(\kappa^{(0)})D^{-1}.
\end{align*}
By \eqref{supp:eq:plugin-profile-residual-covariance},
\begin{align*}
&
\{
\mathrm{Id}_p\otimes(\Sigma^{(0)})^{-1}
\}
\operatorname{Cov}(\mathbf r)
\{
\mathrm{Id}_p\otimes(\Sigma^{(0)})^{-1}
\}
\\
&=
\mathrm{Id}_p\otimes(\Sigma^{(0)})^{-1}
-
\{
\mathrm{Id}_p\otimes(\Sigma^{(0)})^{-1}
\}
M(\kappa^{(0)})D^{-1}M(\kappa^{(0)})^\top
\{
\mathrm{Id}_p\otimes(\Sigma^{(0)})^{-1}
\}
\\
&=
\mathscr S_p\mathcal W_0\mathscr S_p^\top.
\end{align*}
Consequently,
\begin{align}
&
\mathbb E
\left[
\{
\mathrm{Id}_p\otimes(\Sigma^{(0)})^{-1}
\}
\mathbf r b_\Delta^\top
\right]
=
\left(
\mathscr S_p\mathcal W_0\mathscr S_p^\top
\right)
\{
\mathrm{Id}_p\otimes
\Delta(\Sigma^{(0)})^{-1}
\}
M(\kappa^{(0)})D^{-1}.
\label{supp:eq:plugin-stacked-residual-cross-moment-C0}
\end{align}

Finally, substituting
\eqref{supp:eq:plugin-stacked-residual-cross-moment-C0} into
\eqref{supp:eq:plugin-X-block-C0-before-cross-moment} gives
\begin{align}
&
X^\top\mathscr S_p
C_0\operatorname{vech}(\Delta)
\nonumber\\
&=
2
\begin{pmatrix}
\operatorname{Tr}
\left[
    X^\top
    \left(
        \mathscr S_p\mathcal W_0\mathscr S_p^\top
    \right)
    \left\{
        \mathrm{Id}_p\otimes
        \Delta(\Sigma^{(0)})^{-1}
    \right\}
    M(\kappa^{(0)})D^{-1}
\right]
\\[4mm]
\operatorname{Tr}
\left[
    J_2^\top X^\top
    \left(
        \mathscr S_p\mathcal W_0\mathscr S_p^\top
    \right)
    \left\{
        \mathrm{Id}_p\otimes
        \Delta(\Sigma^{(0)})^{-1}
    \right\}
    M(\kappa^{(0)})D^{-1}
\right]
\end{pmatrix}.
\label{supp:eq:plugin-U-block-C0-direct}
\end{align}

Now we calculate $X^\top\mathscr S_p
C_0\operatorname{vech}(\Delta)$ for $X=M(\kappa^{(0)}), N_{0}, \ldots, N_{p-2}$. First, the null-space identity proved above and the symmetry of
\(\mathcal W_0\) give
\[
M(\kappa^{(0)})^\top
\left(
\mathscr S_p\mathcal W_0\mathscr S_p^\top
\right)
=
0_{2\times2p}.
\]
Consequently,
\begin{align}
&
M(\kappa^{(0)})^\top
\mathscr S_pC_0\operatorname{vech}(\Delta)
\nonumber\\
&=
2
\begin{pmatrix}
\operatorname{Tr}
\left[
M(\kappa^{(0)})^\top
\left(
\mathscr S_p\mathcal W_0\mathscr S_p^\top
\right)
\{
\mathrm{Id}_p\otimes
\Delta(\Sigma^{(0)})^{-1}
\}
M(\kappa^{(0)})D^{-1}
\right]
\\[3mm]
\operatorname{Tr}
\left[
J_2^\top M(\kappa^{(0)})^\top
\left(
\mathscr S_p\mathcal W_0\mathscr S_p^\top
\right)
\{
\mathrm{Id}_p\otimes
\Delta(\Sigma^{(0)})^{-1}
\}
M(\kappa^{(0)})D^{-1}
\right]
\end{pmatrix}
\nonumber\\
&=
0_{2\times1}.
\label{supp:eq:plugin-M-block-C0-zero}
\end{align}

For \(l=1,\ldots,p-2\),
\eqref{supp:eq:plugin-W0-block-decomposition} gives
\[
N_l^\top
\left(
\mathscr S_p\mathcal W_0\mathscr S_p^\top
\right)
=
(\Sigma^{(0)})^{-1}N_l^\top.
\]
Moreover, \eqref{supp:eq:kronecker-block-product} and the orthogonality of
\(w_l\) to \(u\) and \(v\) give
\begin{align*}
&
N_l^\top
\{
\mathrm{Id}_p\otimes
\Delta(\Sigma^{(0)})^{-1}
\}
M(\kappa^{(0)})
\\
&=
\sqrt r\,(w_l^\top u)
\Delta(\Sigma^{(0)})^{-1}
+
\sqrt{1-r}\,(w_l^\top v)
\Delta(\Sigma^{(0)})^{-1}J_2
=
0_2.
\end{align*}
Therefore,
\begin{align}
&
N_l^\top
\mathscr S_pC_0\operatorname{vech}(\Delta)
\nonumber\\
&=
2
\begin{pmatrix}
\operatorname{Tr}
\left[
(\Sigma^{(0)})^{-1}N_l^\top
\{
\mathrm{Id}_p\otimes
\Delta(\Sigma^{(0)})^{-1}
\}
M(\kappa^{(0)})D^{-1}
\right]
\\[3mm]
\operatorname{Tr}
\left[
J_2^\top(\Sigma^{(0)})^{-1}N_l^\top
\{
\mathrm{Id}_p\otimes
\Delta(\Sigma^{(0)})^{-1}
\}
M(\kappa^{(0)})D^{-1}
\right]
\end{pmatrix}
\nonumber\\
&=
0_{2\times1},
\quad
l=1,\ldots,p-2.
\label{supp:eq:plugin-Nl-block-C0-zero}
\end{align}

It remains to calculate the \(N_0\)-block. By
\eqref{supp:eq:plugin-W0-block-decomposition},
\[
N_0^\top
\left(
\mathscr S_p\mathcal W_0\mathscr S_p^\top
\right)
=
SN_0^\top.
\]
Thus, \eqref{supp:eq:plugin-stacked-residual-cross-moment-C0} and the blockwise
trace identity give
\begin{align}
&
N_0^\top
\mathscr S_pC_0\operatorname{vech}(\Delta)=
2
\begin{pmatrix}
\operatorname{Tr}
\left[
SN_0^\top
\{
\mathrm{Id}_p\otimes
\Delta(\Sigma^{(0)})^{-1}
\}
M(\kappa^{(0)})D^{-1}
\right]
\\[3mm]
\operatorname{Tr}
\left[
J_2^\top SN_0^\top
\{
\mathrm{Id}_p\otimes
\Delta(\Sigma^{(0)})^{-1}
\}
M(\kappa^{(0)})D^{-1}
\right]
\end{pmatrix}.
\label{supp:eq:plugin-N0-block-C0-traces}
\end{align}

Direct multiplication using the definitions of
\(M(\kappa^{(0)})\) and \(N_0\) gives
\begin{align}
&
N_0^\top
\{
\mathrm{Id}_p\otimes
\Delta(\Sigma^{(0)})^{-1}
\}
M(\kappa^{(0)})
\nonumber\\
&=
\sqrt{r(1-r)}
\left[
\Delta(\Sigma^{(0)})^{-1}
-
J_2^\top\Delta(\Sigma^{(0)})^{-1}J_2
\right]
\nonumber\\
&=
\sqrt{r(1-r)}
\begin{pmatrix}
\displaystyle
\frac{\Delta_{11}}{\lambda_1^{(0)}}
-
\frac{\Delta_{22}}{\lambda_2^{(0)}}
&
\displaystyle
\Delta_{12}
\left(
\frac1{\lambda_1^{(0)}}
+
\frac1{\lambda_2^{(0)}}
\right)
\\[4mm]
\displaystyle
\Delta_{12}
\left(
\frac1{\lambda_1^{(0)}}
+
\frac1{\lambda_2^{(0)}}
\right)
&
\displaystyle
-\frac{\Delta_{11}}{\lambda_1^{(0)}}
+
\frac{\Delta_{22}}{\lambda_2^{(0)}}
\end{pmatrix}.
\label{supp:eq:plugin-N0-Delta-product}
\end{align}

By Lemma~\ref{supp:lem:plugin-canonical-forms-S-D}, we have
\[
S
=
\frac{
2
}{
\operatorname{Tr}(\Sigma^{(0)})
(1-d_\kappa^2d_\Sigma^2)
}
\begin{pmatrix}
1+d_\kappa d_\Sigma&0\\
0&1-d_\kappa d_\Sigma
\end{pmatrix}
\]
and
\[
D^{-1}
=
\left(
\kappa^{(0)}
\diamond_{\Sigma^{(0)}}
\kappa^{(0)}
\right)^{-1}
=
\frac{
\operatorname{Tr}(\Sigma^{(0)})
(1-d_\Sigma^2)
}{
2(1-d_\kappa^2d_\Sigma^2)
}
\begin{pmatrix}
1+d_\kappa d_\Sigma&0\\
0&1-d_\kappa d_\Sigma
\end{pmatrix}.
\]
It follows from \eqref{supp:eq:plugin-N0-Delta-product} that $SN_0^\top
\{
\mathrm{Id}_p\otimes
\Delta(\Sigma^{(0)})^{-1}
\}
M(\kappa^{(0)})D^{-1}$is
\begin{align}
&
\begin{pmatrix}
1+d_\kappa d_\Sigma&0\\
0&1-d_\kappa d_\Sigma
\end{pmatrix}
\begin{pmatrix}
\displaystyle
\frac{\Delta_{11}}{\lambda_1^{(0)}}
-
\frac{\Delta_{22}}{\lambda_2^{(0)}}
&
\displaystyle
\Delta_{12}
\left(
\frac1{\lambda_1^{(0)}}
+
\frac1{\lambda_2^{(0)}}
\right)
\\[4mm]
\displaystyle
\Delta_{12}
\left(
\frac1{\lambda_1^{(0)}}
+
\frac1{\lambda_2^{(0)}}
\right)
&
\displaystyle
-\frac{\Delta_{11}}{\lambda_1^{(0)}}
+
\frac{\Delta_{22}}{\lambda_2^{(0)}}
\end{pmatrix}
\begin{pmatrix}
1+d_\kappa d_\Sigma&0\\
0&1-d_\kappa d_\Sigma
\end{pmatrix}
\label{supp:eq:plugin-N0-common-trace-matrix}
\end{align}
with constant factor $$\frac{
\sqrt{r(1-r)}(1-d_\Sigma^2)
}{
(1-d_\kappa^2d_\Sigma^2)^2
}.$$

The matrix in \eqref{supp:eq:plugin-N0-common-trace-matrix} is symmetric.
Since \(J_2^\top\) is skew-symmetric, its contribution to the second
trace in \eqref{supp:eq:plugin-N0-block-C0-traces} is zero. On the other hand,
its trace is
\begin{align*}
&
\frac{
\sqrt{r(1-r)}(1-d_\Sigma^2)
}{
(1-d_\kappa^2d_\Sigma^2)^2
}
\left[
(1+d_\kappa d_\Sigma)^2
-
(1-d_\kappa d_\Sigma)^2
\right]
\left(
\frac{\Delta_{11}}{\lambda_1^{(0)}}
-
\frac{\Delta_{22}}{\lambda_2^{(0)}}
\right)
\\
&=
\frac{
2d_\kappa\sqrt{1-d_\kappa^2}\,
d_\Sigma(1-d_\Sigma^2)
}{
(1-d_\kappa^2d_\Sigma^2)^2
}
\left(
\frac{\Delta_{11}}{\lambda_1^{(0)}}
-
\frac{\Delta_{22}}{\lambda_2^{(0)}}
\right)
\\
&=
\frac{\beta_{\kappa,\Sigma}}2
\left(
\frac{\Delta_{11}}{\lambda_1^{(0)}}
-
\frac{\Delta_{22}}{\lambda_2^{(0)}}
\right),
\end{align*}
where we used
\[
    2\sqrt{r(1-r)}=\sqrt{1-d_\kappa^2}
\]
and the definition of \(\beta_{\kappa,\Sigma}\) in
\eqref{supp:eq:plugin-beta-kappa-Sigma-definition}.
Accounting for the factor \(2\) in
\eqref{supp:eq:plugin-N0-block-C0-traces}, we therefore obtain
\begin{equation}
N_0^\top
\mathscr S_pC_0\operatorname{vech}(\Delta)
=
\beta_{\kappa,\Sigma}e_1
\left(
\frac{\Delta_{11}}{\lambda_1^{(0)}}
-
\frac{\Delta_{22}}{\lambda_2^{(0)}}
\right),
\quad
e_1=
\begin{pmatrix}
1\\
0
\end{pmatrix}.
\label{supp:eq:plugin-cross-derivative}
\end{equation}

Combining \eqref{supp:eq:plugin-M-block-C0-zero},
\eqref{supp:eq:plugin-Nl-block-C0-zero}, and
\eqref{supp:eq:plugin-cross-derivative}, we obtain
\begin{align*}
\mathsf Q_0^\top\mathscr S_p
C_0\operatorname{vech}(\Delta)
&=
\begin{pmatrix}
    0_{2\times1}\\[1mm]
    \beta_{\kappa,\Sigma}e_1\\[1mm]
    0_{2(p-2)\times1}
\end{pmatrix}
\left(
    \frac{\Delta_{11}}{\lambda_1^{(0)}}
    -
    \frac{\Delta_{22}}{\lambda_2^{(0)}}
\right)
=
\begin{pmatrix}
    0_{2\times1}\\[1mm]
    \beta_{\kappa,\Sigma}e_1\\[1mm]
    0_{2(p-2)\times1}
\end{pmatrix}
\begin{pmatrix}
        (\lambda_1^{(0)})^{-1} & 
        0 &
        -(\lambda_2^{(0)})^{-1}
    \end{pmatrix}\operatorname{vech}(\Delta).
\end{align*}
Since \(\Delta\) is arbitrary, we conclude that
\begin{equation}
    \mathsf Q_0^\top\mathscr S_pC_0
    =
    \begin{pmatrix}
        0_{2\times3}\\[1mm]
        \beta_{\kappa,\Sigma}e_1 \begin{pmatrix}
        (\lambda_1^{(0)})^{-1} & 
        0 &
        -(\lambda_2^{(0)})^{-1}
    \end{pmatrix}\\
        0_{2(p-2)\times3}
    \end{pmatrix}.
    \label{supp:eq:plugin-C0-block-form}
\end{equation}

By Lemma~\ref{supp:lem:plugin-covariance-influence-moments},
\[
    \Omega_\xi
    :=
    \operatorname{Var}
    \left\{
        \xi_{\Sigma}(Y_1)
    \right\}
    =
    \frac1{p-2}
    \begin{pmatrix}
        2(\lambda_1^{(0)})^2&0&0\\
        0&\lambda_1^{(0)}\lambda_2^{(0)}&0\\
        0&0&2(\lambda_2^{(0)})^2
    \end{pmatrix}.
\]
Therefore,
\[
    \begin{pmatrix}
        (\lambda_1^{(0)})^{-1} & 
        0 &
        -(\lambda_2^{(0)})^{-1}
    \end{pmatrix}\Omega_\xi \begin{pmatrix}
        (\lambda_1^{(0)})^{-1} \\ 
        0 \\
        -(\lambda_2^{(0)})^{-1}
    \end{pmatrix}
    =
    \frac1{p-2}
    \left(
        \frac{2(\lambda_1^{(0)})^2}
        {(\lambda_1^{(0)})^2}
        +
        \frac{2(\lambda_2^{(0)})^2}
        {(\lambda_2^{(0)})^2}
    \right)
    =
    \frac4{p-2}.
\]
Using \eqref{supp:eq:plugin-C0-block-form}, we conclude that
\begin{align*}
&
\mathsf Q_0^\top\mathscr S_p
C_0\Omega_\xi C_0^\top
\mathscr S_p^\top \mathsf Q_0=
\operatorname{diag}
    \left(
        0_2,
\left\{
    \frac{4\beta_{\kappa,\Sigma}^2}{p-2}
    e_1e_1^\top
\right\},
        \mathrm{Id}_{p-2}
        \otimes
        0_2
    \right).
\end{align*}
\end{proof}

\subsubsection{Trace formula for the plug-in estimator}
\label{supp:sec:plugin-trace-formula}

We now combine the block decompositions of the population curvature
and the effective-score covariance. This yields the full asymptotic
covariance matrix of the \texttt{plug-in} estimator in the canonical
coordinates introduced above. Its trace then follows from the
orthogonal invariance of the trace.

\begin{lemma}[Block decomposition and trace of
\(V^{\texttt{plug-in}}\)]
\label{supp:lem:plugin-asymptotic-covariance-block-decomposition}
Under the assumptions of
Theorem~\ref{thm:iterated-plugin-loading-expansion-clt}, and with the
notation introduced in
Subsubsections~\ref{supp:sec:plugin-canonical-loading-coordinates},
\ref{supp:sec:plugin-population-curvature}, and
\ref{supp:sec:plugin-effective-score-covariance}, the asymptotic
covariance matrix \(V^{\texttt{plug-in}}\) satisfies
\begin{align}
&
\mathsf Q_0^\top\mathscr S_p
V^{\texttt{plug-in}}
\mathscr S_p^\top \mathsf Q_0
\nonumber\\
&=
\operatorname{diag}
\Bigg(
    0_2,
    \,
    A_S^{-1}
    +
    A_S^{-1}G_SA_S^{-1}
    +
    \frac{1}{p-2}
    \frac{
        16d_\kappa^2(1-d_\kappa^2)
        d_\Sigma^2(1-d_\Sigma^2)^2
    }{
        (1-d_\kappa^2d_\Sigma^2)^4
    }
    A_S^{-1}e_1e_1^\top A_S^{-1},
    \nonumber\\
&\hspace{38mm}
    \mathrm{Id}_{p-2}
    \otimes
    \left\{
        A_{(\Sigma^{(0)})^{-1}}^{-1}
        +
        A_{(\Sigma^{(0)})^{-1}}^{-1}
        G_{(\Sigma^{(0)})^{-1}}
        A_{(\Sigma^{(0)})^{-1}}^{-1}
    \right\}
\Bigg).
\label{supp:eq:plugin-asymptotic-covariance-block-decomposition}
\end{align}
Consequently,
\begin{align}
&
\operatorname{Tr}
\left(
    V^{\texttt{plug-in}}
\right)
\nonumber\\
&=
\operatorname{Tr}
\left(
    A_S^{-1}
    +
    A_S^{-1}G_SA_S^{-1}
\right)
\nonumber\\
&\quad+
(p-2)
\operatorname{Tr}
\left\{
    A_{(\Sigma^{(0)})^{-1}}^{-1}
    +
    A_{(\Sigma^{(0)})^{-1}}^{-1}
    G_{(\Sigma^{(0)})^{-1}}
    A_{(\Sigma^{(0)})^{-1}}^{-1}
\right\}
\nonumber\\
&\quad+
\frac{1}{p-2}
\frac{
    16d_\kappa^2(1-d_\kappa^2)
    d_\Sigma^2(1-d_\Sigma^2)^2
}{
    (1-d_\kappa^2d_\Sigma^2)^4
}
e_1^\top A_S^{-2}e_1.
\label{supp:eq:plugin-trace-formula}
\end{align}
\end{lemma}

\begin{proof}
Define the effective score
\[
    Z(Y_1)
    :=
    \nabla_\mu
    \tilde{\ell}(Y_1,\mu_0,\nu_0)
    +
    C_0\xi_\Sigma(Y_1),
\]
and write
\[
    \Omega_Z
    :=
    \operatorname{Var}
    \left\{
        Z(Y_1)
    \right\}.
\]
By the definition of \(V^{\texttt{plug-in}}\) in
\eqref{eq:plugin-asymptotic-covariance},
\[
    V^{\texttt{plug-in}}
    =
    J^+\Omega_ZJ^+.
\]

Since \(\mathsf Q_0\) and \(\mathscr S_p\) are orthogonal,
\[
    \mathscr S_p^\top
    \mathsf Q_0\mathsf Q_0^\top
    \mathscr S_p
    =
    \mathrm{Id}_{2p}.
\]
Inserting this identity on both sides of \(\Omega_Z\) gives
\begin{align*}
&
\mathsf Q_0^\top\mathscr S_p
V^{\texttt{plug-in}}
\mathscr S_p^\top \mathsf Q_0
\\
&=
\mathsf Q_0^\top\mathscr S_p
J^+\Omega_ZJ^+
\mathscr S_p^\top \mathsf Q_0
\\
&=
\mathsf Q_0^\top\mathscr S_p
J^+
\left(
    \mathscr S_p^\top
    \mathsf Q_0\mathsf Q_0^\top
    \mathscr S_p
\right)
\Omega_Z
\left(
    \mathscr S_p^\top
    \mathsf Q_0\mathsf Q_0^\top
    \mathscr S_p
\right)
J^+
\mathscr S_p^\top \mathsf Q_0
\\
&=
\left(
    \mathsf Q_0^\top\mathscr S_p
    J^+
    \mathscr S_p^\top \mathsf Q_0
\right)
\left(
    \mathsf Q_0^\top\mathscr S_p
    \Omega_Z
    \mathscr S_p^\top \mathsf Q_0
\right)
\left(
    \mathsf Q_0^\top\mathscr S_p
    J^+
    \mathscr S_p^\top \mathsf Q_0
\right).
\end{align*}

By
\eqref{supp:eq:plugin-J-plus-block-decomposition},
\begin{align*}
&
\mathsf Q_0^\top\mathscr S_p
J^+
\mathscr S_p^\top \mathsf Q_0
=
\operatorname{diag}
\left(
    0_2,
    \frac12A_S^{-1},
    \mathrm{Id}_{p-2}
    \otimes
    \left\{
        \frac12
        A_{(\Sigma^{(0)})^{-1}}^{-1}
    \right\}
\right).
\end{align*}
Moreover, the effective-score covariance decomposition established in
Lemma~\ref{supp:lem:plugin-effective-score-covariance} gives
\begin{align*}
&
\mathsf Q_0^\top\mathscr S_p
\Omega_Z
\mathscr S_p^\top \mathsf Q_0
\\
&=
\operatorname{diag}
\Bigg(
    0_2,
    \,
    4
    \left\{
        A_S
        +
        G_S
        +
        \frac{\beta_{\kappa,\Sigma}^2}{p-2}
        e_1e_1^\top
    \right\},
    \nonumber\\
&\hspace{35mm}
    \mathrm{Id}_{p-2}
    \otimes
    4
    \left\{
        A_{(\Sigma^{(0)})^{-1}}
        +
        G_{(\Sigma^{(0)})^{-1}}
    \right\}
\Bigg).
\end{align*}

The product of these three block-diagonal matrices is again
block diagonal. Its \(N_0\)-block is
\begin{align*}
&
\left(
    \frac12A_S^{-1}
\right)
4
\left\{
    A_S
    +
    G_S
    +
    \frac{\beta_{\kappa,\Sigma}^2}{p-2}
    e_1e_1^\top
\right\}
\left(
    \frac12A_S^{-1}
\right)
\\
&=
A_S^{-1}
+
A_S^{-1}G_SA_S^{-1}
+
\frac{\beta_{\kappa,\Sigma}^2}{p-2}
A_S^{-1}e_1e_1^\top A_S^{-1}.
\end{align*}
For each \(l=1,\ldots,p-2\), the corresponding \(N_l\)-block is
\begin{align*}
&
\left(
    \frac12
    A_{(\Sigma^{(0)})^{-1}}^{-1}
\right)
4
\left\{
    A_{(\Sigma^{(0)})^{-1}}
    +
    G_{(\Sigma^{(0)})^{-1}}
\right\}
\left(
    \frac12
    A_{(\Sigma^{(0)})^{-1}}^{-1}
\right)
\\
&=
A_{(\Sigma^{(0)})^{-1}}^{-1}
+
A_{(\Sigma^{(0)})^{-1}}^{-1}
G_{(\Sigma^{(0)})^{-1}}
A_{(\Sigma^{(0)})^{-1}}^{-1}.
\end{align*}
Finally,
\[
    \beta_{\kappa,\Sigma}^2
    =
    \frac{
        16d_\kappa^2(1-d_\kappa^2)
        d_\Sigma^2(1-d_\Sigma^2)^2
    }{
        (1-d_\kappa^2d_\Sigma^2)^4
    }.
\]
Substitution gives
\eqref{supp:eq:plugin-asymptotic-covariance-block-decomposition}.

Since \(\mathsf Q_0\) and \(\mathscr S_p\) are orthogonal,
\begin{align*}
    \operatorname{Tr}
    \left(
        V^{\texttt{plug-in}}
    \right)
    &=
    \operatorname{Tr}
    \left(
        \mathsf Q_0^\top\mathscr S_p
        V^{\texttt{plug-in}}
        \mathscr S_p^\top \mathsf Q_0
    \right).
\end{align*}
Taking the trace of the block-diagonal matrix in
\eqref{supp:eq:plugin-asymptotic-covariance-block-decomposition} and
using
\[
    \operatorname{Tr}
    \left(
        A_S^{-1}e_1e_1^\top A_S^{-1}
    \right)
    =
    e_1^\top A_S^{-2}e_1
\]
proves \eqref{supp:eq:plugin-trace-formula}.
\end{proof}

To evaluate the two-dimensional blocks in
Lemma~\ref{supp:lem:plugin-asymptotic-covariance-block-decomposition}
explicitly, we first express
\(\Sigma^{(0)}\), \(D\), and \(S\) in the oriented eigenbasis of
\(\Sigma^{(0)}\).

\begin{lemma}[Canonical forms of
\(\Sigma^{(0)}\), \(D\), and \(S\)]
\label{supp:lem:plugin-canonical-forms-S-D}
Work in an oriented orthonormal eigenbasis of
\(\Sigma^{(0)}\), with ordered eigenvalues
\[
\lambda_1^{(0)}
\geq
\lambda_2^{(0)}
>
0.
\]
Let
\(
Z
:=
\operatorname{diag}(1,-1)\). Then we have
\begin{align}
\Sigma^{(0)}
&=
\frac{\operatorname{Tr}(\Sigma^{(0)})}{2}
\left(
    \mathrm{Id}_2+d_\Sigma Z
\right),
\notag\\
(\Sigma^{(0)})^{-1}
&=
\frac{2}{
    \operatorname{Tr}(\Sigma^{(0)})
    (1-d_\Sigma^2)
}
\left(
    \mathrm{Id}_2-d_\Sigma Z
\right).
\label{supp:eq:plugin-Sigma-canonical-forms}
\end{align}

\begin{align}
D
&=
\frac{2}{
    \operatorname{Tr}(\Sigma^{(0)})
    (1-d_\Sigma^2)
}
\left(
    \mathrm{Id}_2-d_\kappa d_\Sigma Z
\right),
\notag\\
D^{-1}
&=
\frac{
    \operatorname{Tr}(\Sigma^{(0)})
    (1-d_\Sigma^2)
}{
    2(1-d_\kappa^2d_\Sigma^2)
}
\left(
    \mathrm{Id}_2+d_\kappa d_\Sigma Z
\right).
\label{supp:eq:plugin-D-and-D-inverse-explicit}
\end{align}

\begin{align}
S
&=
\frac{2}{
    \operatorname{Tr}(\Sigma^{(0)})
    (1-d_\kappa^2d_\Sigma^2)
}
\left(
    \mathrm{Id}_2+d_\kappa d_\Sigma Z
\right),
\notag\\
S^{-1}
&=
\frac{\operatorname{Tr}(\Sigma^{(0)})}{2}
\left(
    \mathrm{Id}_2-d_\kappa d_\Sigma Z
\right).
\label{supp:eq:plugin-S-and-S-inverse-explicit}
\end{align}
\end{lemma}

\begin{proof}
Since
\[
\operatorname{Tr}(\Sigma^{(0)})
=
\lambda_1^{(0)}+\lambda_2^{(0)}
\]
and
\[
d_\Sigma
=
\frac{
    \lambda_1^{(0)}-\lambda_2^{(0)}
}{
    \lambda_1^{(0)}+\lambda_2^{(0)}
},
\]
we have
\[
\lambda_1^{(0)}
=
\frac{\operatorname{Tr}(\Sigma^{(0)})}{2}
(1+d_\Sigma),
\quad
\lambda_2^{(0)}
=
\frac{\operatorname{Tr}(\Sigma^{(0)})}{2}
(1-d_\Sigma).
\]
Therefore, in the chosen eigenbasis,
\[
\Sigma^{(0)}
=
\frac{\operatorname{Tr}(\Sigma^{(0)})}{2}
\left(
    \mathrm{Id}_2+d_\Sigma Z
\right).
\]

Since \(Z^2=\mathrm{Id}_2\),
\[
\left(
    \mathrm{Id}_2+d_\Sigma Z
\right)
\left(
    \mathrm{Id}_2-d_\Sigma Z
\right)
=
(1-d_\Sigma^2)\mathrm{Id}_2.
\]
It follows that
\[
(\Sigma^{(0)})^{-1}
=
\frac{2}{
    \operatorname{Tr}(\Sigma^{(0)})
    (1-d_\Sigma^2)
}
\left(
    \mathrm{Id}_2-d_\Sigma Z
\right).
\]
This shows \eqref{supp:eq:plugin-Sigma-canonical-forms}.

Next, using
\[
J_2^\top J_2=\mathrm{Id}_2,
\quad
J_2^\top ZJ_2=-Z,
\]
we obtain
\begin{align*}
J_2^\top(\Sigma^{(0)})^{-1}J_2
&=
\frac{2}{
    \operatorname{Tr}(\Sigma^{(0)})
    (1-d_\Sigma^2)
}
J_2^\top
\left(
    \mathrm{Id}_2-d_\Sigma Z
\right)
J_2\\
&=
\frac{2}{
    \operatorname{Tr}(\Sigma^{(0)})
    (1-d_\Sigma^2)
}
\left(
    J_2^\top J_2
    -
    d_\Sigma J_2^\top ZJ_2
\right)\\
&=
\frac{2}{
    \operatorname{Tr}(\Sigma^{(0)})
    (1-d_\Sigma^2)
}
\left(
    \mathrm{Id}_2+d_\Sigma Z
\right).
\end{align*}

Substituting the preceding expressions into the definition of \(D\)
gives
\begin{align*}
D
&=
\frac{2}{
    \operatorname{Tr}(\Sigma^{(0)})
    (1-d_\Sigma^2)
}
\left[
    r\left(
        \mathrm{Id}_2-d_\Sigma Z
    \right)
    +
    (1-r)\left(
        \mathrm{Id}_2+d_\Sigma Z
    \right)
\right]\\
&=
\frac{2}{
    \operatorname{Tr}(\Sigma^{(0)})
    (1-d_\Sigma^2)
}
\left[
    \{r+(1-r)\}\mathrm{Id}_2
    +
    \{-r+(1-r)\}d_\Sigma Z
\right]\\
&=
\frac{2}{
    \operatorname{Tr}(\Sigma^{(0)})
    (1-d_\Sigma^2)
}
\left[
    \mathrm{Id}_2+(1-2r)d_\Sigma Z
\right].
\end{align*}
Recalling that \(d_\kappa=2r-1\), we have
\(1-2r=-d_\kappa\), and hence
\[
D
=
\frac{2}{
    \operatorname{Tr}(\Sigma^{(0)})
    (1-d_\Sigma^2)
}
\left(
    \mathrm{Id}_2-d_\kappa d_\Sigma Z
\right).
\]

For any \(x\) satisfying \(|x|<1\),
\[
\left(
    \mathrm{Id}_2-xZ
\right)^{-1}
=
\frac{1}{1-x^2}
\left(
    \mathrm{Id}_2+xZ
\right).
\]
Since \(0\leq d_\Sigma<1\) and
\(|d_\kappa|\leq1\),
\(
1-d_\kappa^2d_\Sigma^2>0.
\) Applying the preceding identity with
\(x=d_\kappa d_\Sigma\) yields
\[
D^{-1}
=
\frac{
    \operatorname{Tr}(\Sigma^{(0)})
    (1-d_\Sigma^2)
}{
    2(1-d_\kappa^2d_\Sigma^2)
}
\left(
    \mathrm{Id}_2+d_\kappa d_\Sigma Z
\right).
\]
Together, the preceding two identities show
\eqref{supp:eq:plugin-D-and-D-inverse-explicit}.

To calculate \(S\), first observe that
\begin{align*}
&(1-r)(\Sigma^{(0)})^{-1}
+
rJ_2^\top(\Sigma^{(0)})^{-1}J_2\\
&\quad=
\frac{2}{
    \operatorname{Tr}(\Sigma^{(0)})
    (1-d_\Sigma^2)
}
\left[
    (1-r)\left(
        \mathrm{Id}_2-d_\Sigma Z
    \right)
    +
    r\left(
        \mathrm{Id}_2+d_\Sigma Z
    \right)
\right]\\
&\quad=
\frac{2}{
    \operatorname{Tr}(\Sigma^{(0)})
    (1-d_\Sigma^2)
}
\left[
    \mathrm{Id}_2+(2r-1)d_\Sigma Z
\right]\\
&\quad=
\frac{2}{
    \operatorname{Tr}(\Sigma^{(0)})
    (1-d_\Sigma^2)
}
\left(
    \mathrm{Id}_2+d_\kappa d_\Sigma Z
\right).
\end{align*}
Moreover,
\begin{align*}
&(\Sigma^{(0)})^{-1}
-
J_2^\top(\Sigma^{(0)})^{-1}J_2\\
&\quad=
\frac{2}{
    \operatorname{Tr}(\Sigma^{(0)})
    (1-d_\Sigma^2)
}
\left[
    \left(
        \mathrm{Id}_2-d_\Sigma Z
    \right)
    -
    \left(
        \mathrm{Id}_2+d_\Sigma Z
    \right)
\right]\\
&\quad=
-\frac{4d_\Sigma}{
    \operatorname{Tr}(\Sigma^{(0)})
    (1-d_\Sigma^2)
}
Z.
\end{align*}

Since
\[
r(1-r)
=
\frac{1+d_\kappa}{2}
\frac{1-d_\kappa}{2}
=
\frac{1-d_\kappa^2}{4},
\]
substituting the preceding identities and the expression for
\(D^{-1}\) into the definition of \(S\) gives
\begin{align*}
S
&=
\frac{2}{
    \operatorname{Tr}(\Sigma^{(0)})
    (1-d_\Sigma^2)
}
\left(
    \mathrm{Id}_2+d_\kappa d_\Sigma Z
\right)\\
&\quad-
\frac{1-d_\kappa^2}{4}
\left\{
-\frac{4d_\Sigma}{
    \operatorname{Tr}(\Sigma^{(0)})
    (1-d_\Sigma^2)
}
Z
\right\}\\
&\quad\quad\times
\left\{
\frac{
    \operatorname{Tr}(\Sigma^{(0)})
    (1-d_\Sigma^2)
}{
    2(1-d_\kappa^2d_\Sigma^2)
}
\left(
    \mathrm{Id}_2+d_\kappa d_\Sigma Z
\right)
\right\}\\
&\quad\quad\times
\left\{
-\frac{4d_\Sigma}{
    \operatorname{Tr}(\Sigma^{(0)})
    (1-d_\Sigma^2)
}
Z
\right\}.
\end{align*}
The scalar coefficient in the second term is
\[
\frac{
    2(1-d_\kappa^2)d_\Sigma^2
}{
    \operatorname{Tr}(\Sigma^{(0)})
    (1-d_\Sigma^2)
    (1-d_\kappa^2d_\Sigma^2)
},
\]
and, since \(Z^2=\mathrm{Id}_2\),
\[
Z
\left(
    \mathrm{Id}_2+d_\kappa d_\Sigma Z
\right)
Z
=
\mathrm{Id}_2+d_\kappa d_\Sigma Z.
\]
Consequently,
\begin{align*}
S
&=
\frac{2}{
    \operatorname{Tr}(\Sigma^{(0)})
    (1-d_\Sigma^2)
}
\left(
    \mathrm{Id}_2+d_\kappa d_\Sigma Z
\right)-
\frac{
    2(1-d_\kappa^2)d_\Sigma^2
}{
    \operatorname{Tr}(\Sigma^{(0)})
    (1-d_\Sigma^2)
    (1-d_\kappa^2d_\Sigma^2)
}
\left(
    \mathrm{Id}_2+d_\kappa d_\Sigma Z
\right)\\
&=
\frac{2}{
    \operatorname{Tr}(\Sigma^{(0)})
    (1-d_\Sigma^2)
}
\left(
    1-
    \frac{
        (1-d_\kappa^2)d_\Sigma^2
    }{
        1-d_\kappa^2d_\Sigma^2
    }
\right)
\left(
    \mathrm{Id}_2+d_\kappa d_\Sigma Z
\right).
\end{align*}
Since
\[
1-
\frac{
    (1-d_\kappa^2)d_\Sigma^2
}{
    1-d_\kappa^2d_\Sigma^2
}
=
\frac{
    1-d_\Sigma^2
}{
    1-d_\kappa^2d_\Sigma^2
},
\]
we obtain
\[
S
=
\frac{2}{
    \operatorname{Tr}(\Sigma^{(0)})
    (1-d_\kappa^2d_\Sigma^2)
}
\left(
    \mathrm{Id}_2+d_\kappa d_\Sigma Z
\right).
\]

Finally, applying the same inverse identity \[
\left(
    \mathrm{Id}_2-xZ
\right)^{-1}
=
\frac{1}{1-x^2}
\left(
    \mathrm{Id}_2+xZ
\right).
\]
with
\(x=-d_\kappa d_\Sigma\) gives
\begin{align*}
S^{-1}
&=
\frac{
    \operatorname{Tr}(\Sigma^{(0)})
    (1-d_\kappa^2d_\Sigma^2)
}{2}
\frac{1}{
    1-d_\kappa^2d_\Sigma^2
}
\left(
    \mathrm{Id}_2-d_\kappa d_\Sigma Z
\right)\\
&=
\frac{\operatorname{Tr}(\Sigma^{(0)})}{2}
\left(
    \mathrm{Id}_2-d_\kappa d_\Sigma Z
\right).
\end{align*}
Thus, we get \eqref{supp:eq:plugin-S-and-S-inverse-explicit}.
\end{proof}

We next propagate the preceding canonical matrix forms through the
second-moment operators defining
\(A_S\), \(A_{(\Sigma^{(0)})^{-1}}\),
\(G_S\), and \(G_{(\Sigma^{(0)})^{-1}}\).
The resulting formulas provide all the moment blocks needed for the
trace calculation.
\begin{lemma}[Canonical forms of the second-moment blocks]
\label{supp:lem:plugin-canonical-second-moment-blocks}
Recall that
\[
    A_W
    :=
    \mathbb E
    \left[
        M(\phi_1)^\top W M(\phi_1)
    \right]
\]
and
\[
    G_W
    :=
    \mathbb E
    \left[
        M(\widehat{\phi}_{\varepsilon,1})^\top
        W
        M(\widehat{\phi}_{\varepsilon,1})
    \right],
    \quad
    \widehat{\phi}_{\varepsilon,1}
    :=
    \widehat\phi
    \left(
        \varepsilon_1,
        \kappa^{(0)},
        \Sigma^{(0)}
    \right).
\]
When \(\mathbb E(\phi_1^2)\neq0\), define
\begin{equation}
\label{supp:eq:plugin-Q-phi-definition}
Q_\phi
:=
\frac{
    \mathbb E
    \left[
        M(\phi_1)^\top ZM(\phi_1)
    \right]
}{
    |\mathbb E(\phi_1^2)|
}
=
\frac{1}{|\mathbb E(\phi_1^2)|}
\begin{pmatrix}
    \operatorname{Re}\mathbb E(\phi_1^2)
    &
    -\operatorname{Im}\mathbb E(\phi_1^2)
    \\
    -\operatorname{Im}\mathbb E(\phi_1^2)
    &
    -\operatorname{Re}\mathbb E(\phi_1^2)
\end{pmatrix}.
\end{equation}
Then
\begin{equation}
\label{supp:eq:plugin-Q-phi-properties}
    Q_\phi^2=\mathrm{Id}_2,
    \quad
    \operatorname{Tr}(Q_\phi)=0.
\end{equation}
If \(d_\Sigma>0\), then additionally
\[
    \frac12\operatorname{Tr}(ZQ_\phi)
    =
    \lambda_\phi.
\]
Moreover,
\begin{align}
A_S
&=
\frac{
    2\mathbb E[|\phi_1|^2]
}{
    \operatorname{Tr}(\Sigma^{(0)})
    (1-d_\kappa^2d_\Sigma^2)
}
\left(
    \mathrm{Id}_2
    +
    d_\kappa d_\Sigma\rho_\phi Q_\phi
\right),
\label{supp:eq:plugin-A-S}
\\
A_{(\Sigma^{(0)})^{-1}}
&=
\frac{
    2\mathbb E[|\phi_1|^2]
}{
    \operatorname{Tr}(\Sigma^{(0)})
    (1-d_\Sigma^2)
}
\left(
    \mathrm{Id}_2
    -
    d_\Sigma\rho_\phi Q_\phi
\right),
\label{supp:eq:plugin-A-Sigma-inverse}
\\
G_S
&=
\frac{
    2(1-d_\Sigma^2)
}{
    (1-d_\kappa^2d_\Sigma^2)^2
}
\left(
    \mathrm{Id}_2
    +
    d_\kappa^2d_\Sigma^2Z
\right),
\label{supp:eq:plugin-G-S}
\\
G_{(\Sigma^{(0)})^{-1}}
&=
\frac{2}{
    1-d_\kappa^2d_\Sigma^2
}
\left(
    \mathrm{Id}_2
    -
    d_\kappa d_\Sigma^2Z
\right).
\label{supp:eq:plugin-G-Sigma-inverse}
\end{align}
When \(\mathbb E(\phi_1^2)=0\), the first two identities are
understood with the convention
\(\rho_\phi Q_\phi=0_2\).
\end{lemma}

\begin{proof}
We first identify the matrix associated with the second complex moment
of \(\phi_1\). For \(z=x+\mathrm{i}y\),
\[
    M(z)^\top M(z)
    =
    |z|^2\mathrm{Id}_2,
\]
and
\[
    M(z)^\top ZM(z)
    =
    \begin{pmatrix}
        x^2-y^2&-2xy\\
        -2xy&y^2-x^2
    \end{pmatrix}.
\]
Consequently,
\[
\mathbb E
\left[
    M(\phi_1)^\top ZM(\phi_1)
\right]
=
\begin{pmatrix}
    \operatorname{Re}\mathbb E(\phi_1^2)
    &
    -\operatorname{Im}\mathbb E(\phi_1^2)
    \\
    -\operatorname{Im}\mathbb E(\phi_1^2)
    &
    -\operatorname{Re}\mathbb E(\phi_1^2)
\end{pmatrix}.
\]
When \(\mathbb E(\phi_1^2)\neq0\), this gives
\eqref{supp:eq:plugin-Q-phi-definition}. If \(d_\Sigma>0\), then, in the present oriented eigenbasis,
definition~\eqref{eq:def-lambda-phi} reduces to
\[
    \lambda_\phi
    =
    \frac{
        \Re\mathbb E(\phi_1^2)
    }{
        |\mathbb E(\phi_1^2)|
    }.
\]
It follows directly from
\eqref{supp:eq:plugin-Q-phi-definition} that
\[
    Q_\phi^2=\mathrm{Id}_2,
    \quad
    \operatorname{Tr}(Q_\phi)=0,
    \quad
    \frac12\operatorname{Tr}(ZQ_\phi)=\lambda_\phi,
\]
which proves
\eqref{supp:eq:plugin-Q-phi-properties}.

More generally, if
\(
W=w(\mathrm{Id}_2+vZ)
\),
then
\begin{align}
\mathbb E
\left[
    M(\xi)^\top WM(\xi)
\right]
=
w\mathbb E[|\xi|^2]\mathrm{Id}_2
+
wv
\begin{pmatrix}
    \operatorname{Re}\mathbb E(\xi^2)
    &
    -\operatorname{Im}\mathbb E(\xi^2)
    \\
    -\operatorname{Im}\mathbb E(\xi^2)
    &
    -\operatorname{Re}\mathbb E(\xi^2)
\end{pmatrix}.
\label{supp:eq:plugin-generic-second-moment-block}
\end{align}
In particular,
\begin{equation}
\label{supp:eq:plugin-generic-A-block}
A_W
=
w\mathbb E[|\phi_1|^2]
\left(
    \mathrm{Id}_2+v\rho_\phi Q_\phi
\right).
\end{equation}

We first apply this identity to \(W=S\). By
Lemma~\ref{supp:lem:plugin-canonical-forms-S-D},
\[
S
=
\frac{2}{
    \operatorname{Tr}(\Sigma^{(0)})
    (1-d_\kappa^2d_\Sigma^2)
}
\left(
    \mathrm{Id}_2+d_\kappa d_\Sigma Z
\right).
\]
Therefore,
\[
A_S
=
\frac{
    2\mathbb E[|\phi_1|^2]
}{
    \operatorname{Tr}(\Sigma^{(0)})
    (1-d_\kappa^2d_\Sigma^2)
}
\left(
    \mathrm{Id}_2
    +
    d_\kappa d_\Sigma\rho_\phi Q_\phi
\right),
\]
which proves
\eqref{supp:eq:plugin-A-S}.

Similarly, Lemma~\ref{supp:lem:plugin-canonical-forms-S-D} gives
\[
(\Sigma^{(0)})^{-1}
=
\frac{2}{
    \operatorname{Tr}(\Sigma^{(0)})
    (1-d_\Sigma^2)
}
\left(
    \mathrm{Id}_2-d_\Sigma Z
\right).
\]
It follows that
\begin{align*}
A_{(\Sigma^{(0)})^{-1}}
=
\frac{
    2\mathbb E[|\phi_1|^2]
}{
    \operatorname{Tr}(\Sigma^{(0)})
    (1-d_\Sigma^2)
}
\left(
    \mathrm{Id}_2-d_\Sigma\rho_\phi Q_\phi
\right),
\end{align*}
which proves
\eqref{supp:eq:plugin-A-Sigma-inverse}.

It remains to calculate the two \(G\)-matrices. We have
\[
\operatorname{vec}(\widehat{\phi}_{\varepsilon,1})
=
D^{-1}
M(\kappa^{(0)})^\top
\left\{
    \mathrm{Id}_p
    \otimes
    (\Sigma^{(0)})^{-1}
\right\}
\operatorname{vec}(\varepsilon_1).
\]
Since \(\varepsilon_1\) is centered,
\(\widehat{\phi}_{\varepsilon,1}\) is centered. Moreover,
\[
\operatorname{Cov}
\left\{
    \operatorname{vec}(\varepsilon_1)
\right\}
=
\mathrm{Id}_p\otimes\Sigma^{(0)}.
\]
Since \(D\) and \(D^{-1}\) are symmetric, it follows that
\begin{align*}
\operatorname{Cov}
\left\{
    \operatorname{vec}(\widehat{\phi}_{\varepsilon,1})
\right\}
&=
D^{-1}
M(\kappa^{(0)})^\top
\left\{
    \mathrm{Id}_p
    \otimes
    (\Sigma^{(0)})^{-1}
\right\}
\left\{
    \mathrm{Id}_p
    \otimes
    \Sigma^{(0)}
\right\}
\\
&\quad\times
\left\{
    \mathrm{Id}_p
    \otimes
    (\Sigma^{(0)})^{-1}
\right\}
M(\kappa^{(0)})
D^{-1}
\\
&=
D^{-1}
M(\kappa^{(0)})^\top
\left\{
    \mathrm{Id}_p
    \otimes
    (\Sigma^{(0)})^{-1}
\right\}
M(\kappa^{(0)})
D^{-1}
\\
&=
D^{-1}DD^{-1}
\\
&=
D^{-1}.
\end{align*}
By Lemma~\ref{supp:lem:plugin-canonical-forms-S-D},
\begin{align*}
\operatorname{Cov}
\left\{
    \operatorname{vec}(\widehat{\phi}_{\varepsilon,1})
\right\}
&=
\frac{
    \operatorname{Tr}(\Sigma^{(0)})
    (1-d_\Sigma^2)
}{
    2(1-d_\kappa^2d_\Sigma^2)
}
\left(
    \mathrm{Id}_2+d_\kappa d_\Sigma Z
\right)
\\
&=
\frac{
    \operatorname{Tr}(\Sigma^{(0)})
    (1-d_\Sigma^2)
}{
    2(1-d_\kappa^2d_\Sigma^2)
}
\begin{pmatrix}
    1+d_\kappa d_\Sigma&0\\
    0&1-d_\kappa d_\Sigma
\end{pmatrix}.
\end{align*}
It follows that
\begin{align*}
\mathbb E
\left[
    |\widehat{\phi}_{\varepsilon,1}|^2
\right]
&=
\frac{
    \operatorname{Tr}(\Sigma^{(0)})
    (1-d_\Sigma^2)
}{
    1-d_\kappa^2d_\Sigma^2
}.
\end{align*}
Similarly,
\begin{align*}
\mathbb E
\left(
    \widehat{\phi}_{\varepsilon,1}^2
\right)
&=
\frac{
    d_\kappa d_\Sigma
    \operatorname{Tr}(\Sigma^{(0)})
    (1-d_\Sigma^2)
}{
    1-d_\kappa^2d_\Sigma^2
}
=
d_\kappa d_\Sigma\,
\mathbb E
\left[
    |\widehat{\phi}_{\varepsilon,1}|^2
\right].
\end{align*}
In particular,
\(\mathbb E(\widehat{\phi}_{\varepsilon,1}^2)\) is real, and hence
\begin{align*}
&\mathbb E
\left[
    M(\widehat{\phi}_{\varepsilon,1})^\top
    Z
    M(\widehat{\phi}_{\varepsilon,1})
\right]
=
\begin{pmatrix}
    \operatorname{Re}\mathbb E(\widehat{\phi}_{\varepsilon,1}^2)
    &
    -\operatorname{Im}\mathbb E(\widehat{\phi}_{\varepsilon,1}^2)
    \\
    -\operatorname{Im}\mathbb E(\widehat{\phi}_{\varepsilon,1}^2)
    &
    -\operatorname{Re}\mathbb E(\widehat{\phi}_{\varepsilon,1}^2)
\end{pmatrix}
=
d_\kappa d_\Sigma
\mathbb E
\left[
    |\widehat{\phi}_{\varepsilon,1}|^2
\right]
Z.
\end{align*}

Consequently, if
\(
W=w(\mathrm{Id}_2+vZ)
\),
then
\eqref{supp:eq:plugin-generic-second-moment-block} and
\[
    \mathbb E
    \left[
        \widehat{\phi}_{\varepsilon,1}^2
    \right]
    =
    d_\kappa d_\Sigma
    \mathbb E
    \left[
        |\widehat{\phi}_{\varepsilon,1}|^2
    \right]
\]
give
\begin{align}
G_W
&=
\mathbb E
\left[
    M(\widehat{\phi}_{\varepsilon,1})^\top
    W
    M(\widehat{\phi}_{\varepsilon,1})
\right]
\nonumber\\
&=
w\mathbb E
\left[
    |\widehat{\phi}_{\varepsilon,1}|^2
\right]
\left(
    \mathrm{Id}_2
    +
    vd_\kappa d_\Sigma Z
\right)
\nonumber\\
&=
\frac{
    w\operatorname{Tr}(\Sigma^{(0)})
    (1-d_\Sigma^2)
}{
    1-d_\kappa^2d_\Sigma^2
}
\left(
    \mathrm{Id}_2
    +
    vd_\kappa d_\Sigma Z
\right).
\label{supp:eq:plugin-generic-G-block}
\end{align}

We first apply
\eqref{supp:eq:plugin-generic-G-block}
to \(W=S\). By
Lemma~\ref{supp:lem:plugin-canonical-forms-S-D},
\[
    S
    =
    \frac{2}{
        \operatorname{Tr}(\Sigma^{(0)})
        (1-d_\kappa^2d_\Sigma^2)
    }
    \left(
        \mathrm{Id}_2+d_\kappa d_\Sigma Z
    \right).
\]
Thus,
\begin{align*}
G_S
&=
\frac{
    2(1-d_\Sigma^2)
}{
    (1-d_\kappa^2d_\Sigma^2)^2
}
\left(
    \mathrm{Id}_2
    +
    d_\kappa^2d_\Sigma^2 Z
\right).
\end{align*}

Similarly,
\[
    (\Sigma^{(0)})^{-1}
    =
    \frac{2}{
        \operatorname{Tr}(\Sigma^{(0)})
        (1-d_\Sigma^2)
    }
    \left(
        \mathrm{Id}_2-d_\Sigma Z
    \right)
\]
by
Lemma~\ref{supp:lem:plugin-canonical-forms-S-D}. Therefore,
\eqref{supp:eq:plugin-generic-G-block} gives
\begin{align*}
G_{(\Sigma^{(0)})^{-1}}
&=
\frac{2}{
    1-d_\kappa^2d_\Sigma^2
}
\left(
    \mathrm{Id}_2
    -
    d_\kappa d_\Sigma^2 Z
\right).
\end{align*}

If \(\mathbb E(\phi_1^2)=0\), then \(\rho_\phi=0\), and the same
calculation gives
\[
    \mathbb E
    \left[
        M(\phi_1)^\top ZM(\phi_1)
    \right]
    =
    0_2.
\]
Thus, the first two identities remain valid under the convention
\(\rho_\phi Q_\phi=0_2\).
\end{proof}

With the four second-moment blocks now in canonical form, the remaining
calculation reduces to an elementary trace identity for a pair of
\(2\times2\) matrices. We record this identity before proving the final
trace formula.

\begin{lemma}[Entries and traces of a two-dimensional covariance block]
\label{supp:lem:selected-entries-two-dimensional-block}
Let
\[
Z
:=
\begin{pmatrix}
1&0\\
0&-1
\end{pmatrix}, \quad
Q
:=
\begin{pmatrix}
\lambda&q\\
q&-\lambda
\end{pmatrix}
\quad
e_1
:=
\begin{pmatrix}
1\\
0
\end{pmatrix},
\quad
e_2
:=
\begin{pmatrix}
0\\
1
\end{pmatrix}
\]
where \(\lambda\in[-1,1]\), and \(q\in\mathbb R\) satisfy \(
q^2=1-\lambda^2
\).

For \(a>0\), \(|t|<1\), \(g\geq0\), and \(u\in[-1,1]\), define
\[
A
:=
a\left(\mathrm{Id}_2+tQ\right),
\quad
G
:=
g\left(\mathrm{Id}_2+uZ\right).
\]
Then
\begin{equation}
\label{supp:eq:two-dimensional-block-inverse}
A^{-1}
=
\frac{1}{a(1-t^2)}
\begin{pmatrix}
1-t\lambda&-tq\\
-tq&1+t\lambda
\end{pmatrix}.
\end{equation}
Consequently,
\begin{equation}
\label{supp:eq:two-dimensional-block-inverse-trace}
\operatorname{Tr}(A^{-1})
=
\frac{2}{a(1-t^2)}.
\end{equation}
The diagonal entries of
\(A^{-1}+A^{-1}GA^{-1}\) are
\begin{align}
e_1^\top
\left(
A^{-1}+A^{-1}GA^{-1}
\right)e_1
&=
\frac{1-t\lambda}{a(1-t^2)}
\nonumber\\
&\quad+
\frac{
g
\left\{
(1+u)(1-t\lambda)^2
+
(1-u)t^2(1-\lambda^2)
\right\}
}{
a^2(1-t^2)^2
},
\label{supp:eq:selected-entry-first-coordinate}
\\
e_2^\top
\left(
A^{-1}+A^{-1}GA^{-1}
\right)e_2
&=
\frac{1+t\lambda}{a(1-t^2)}
\nonumber\\
&\quad+
\frac{
g
\left\{
(1+u)t^2(1-\lambda^2)
+
(1-u)(1+t\lambda)^2
\right\}
}{
a^2(1-t^2)^2
}.
\label{supp:eq:selected-entry-second-coordinate}
\end{align}
In particular,
\begin{equation}
\label{supp:eq:plugin-trace-two-dimensional-block}
\operatorname{Tr}
\left(
A^{-1}+A^{-1}GA^{-1}
\right)
=
\frac{2}{a(1-t^2)}
+
\frac{
2g\left(1+t^2-2tu\lambda\right)
}{
a^2(1-t^2)^2
}.
\end{equation}
\end{lemma}

\begin{proof}
Since \(q^2=1-\lambda^2\), we have
\[
Q^2
=
\begin{pmatrix}
\lambda&q\\
q&-\lambda
\end{pmatrix}^2
=
\left(
\lambda^2+q^2
\right)
\mathrm{Id}_2
=
\mathrm{Id}_2.
\]
Moreover,
\[
\operatorname{Tr}(Q)=0,
\quad
\operatorname{Tr}(ZQ)=2\lambda.
\]
Hence,
\[
\left(
\mathrm{Id}_2+tQ
\right)
\left(
\mathrm{Id}_2-tQ
\right)
=
\left(
1-t^2
\right)
\mathrm{Id}_2.
\]
Since \(|t|<1\), it follows that
\[
A^{-1}
=
\frac{
\mathrm{Id}_2-tQ
}{
a(1-t^2)
}
=
\frac{1}{a(1-t^2)}
\begin{pmatrix}
1-t\lambda&-tq\\
-tq&1+t\lambda
\end{pmatrix},
\]
which proves
\eqref{supp:eq:two-dimensional-block-inverse}. Taking the trace gives
\[
\operatorname{Tr}(A^{-1})
=
\frac{
(1-t\lambda)+(1+t\lambda)
}{
a(1-t^2)
}
=
\frac{2}{a(1-t^2)},
\]
proving
\eqref{supp:eq:two-dimensional-block-inverse-trace}.

Next,
\[
G
=
g
\begin{pmatrix}
1+u&0\\
0&1-u
\end{pmatrix}.
\]
Direct multiplication gives
\begin{align*}
e_1^\top A^{-1}GA^{-1}e_1
&=
\frac{
g
\left\{
(1+u)(1-t\lambda)^2
+
(1-u)t^2q^2
\right\}
}{
a^2(1-t^2)^2
}
\\
&=
\frac{
g
\left\{
(1+u)(1-t\lambda)^2
+
(1-u)t^2(1-\lambda^2)
\right\}
}{
a^2(1-t^2)^2
},
\\
e_2^\top A^{-1}GA^{-1}e_2
&=
\frac{
g
\left\{
(1+u)t^2q^2
+
(1-u)(1+t\lambda)^2
\right\}
}{
a^2(1-t^2)^2
}
\\
&=
\frac{
g
\left\{
(1+u)t^2(1-\lambda^2)
+
(1-u)(1+t\lambda)^2
\right\}
}{
a^2(1-t^2)^2
}.
\end{align*}
Combining these identities with
\[
e_1^\top A^{-1}e_1
=
\frac{1-t\lambda}{a(1-t^2)},
\quad
e_2^\top A^{-1}e_2
=
\frac{1+t\lambda}{a(1-t^2)}
\]
proves
\eqref{supp:eq:selected-entry-first-coordinate} and
\eqref{supp:eq:selected-entry-second-coordinate}.

\eqref{supp:eq:plugin-trace-two-dimensional-block} is directly obtained from \eqref{supp:eq:selected-entry-first-coordinate} and
\eqref{supp:eq:selected-entry-second-coordinate}.
\end{proof}

Combining the block trace representation in
Lemma~\ref{supp:lem:plugin-asymptotic-covariance-block-decomposition},
the canonical moment formulas in
Lemma~\ref{supp:lem:plugin-canonical-second-moment-blocks}, and the
two-dimensional identity in
Lemma~\ref{supp:lem:selected-entries-two-dimensional-block}
gives the following explicit expression for the trace of the
\texttt{plug-in} asymptotic covariance matrix.

\begin{proposition}[Trace of \(V^{\texttt{plug-in}}\)]
\label{supp:prop:trace-vplugin}
Under the assumptions of Theorem~\ref{thm:iterated-plugin-loading-expansion-clt}, the asymptotic covariance matrix of the optimally aligned \texttt{plug-in} estimator satisfies
\begin{align}
&
\operatorname{Tr}
\left(
    V^{\texttt{plug-in}}
\right)
\nonumber\\
&=
\frac{
    \operatorname{Tr}(\Sigma^{(0)})
}{
    \mathbb E[|\phi_1|^2]
}
\Bigg\{
\frac{
    1-d_\kappa^2d_\Sigma^2
}{
    1-d_\kappa^2d_\Sigma^2\rho_\phi^2
}
+
(p-2)
\frac{
    1-d_\Sigma^2
}{
    1-\rho_\phi^2d_\Sigma^2
}
\Bigg\}
\nonumber\\
&+
\left\{
\frac{
    \operatorname{Tr}(\Sigma^{(0)})
}{
    \mathbb E[|\phi_1|^2]
}
\right\}^{\!2}
\Bigg[
\frac{
    (1-d_\Sigma^2)
    \left(
        1+d_\kappa^2d_\Sigma^2\rho_\phi^2
        -2d_\kappa^3d_\Sigma^3
        \rho_\phi\lambda_\phi
    \right)
}{
    \left(
        1-d_\kappa^2d_\Sigma^2\rho_\phi^2
    \right)^2
}
\nonumber\\
&\quad\quad+
(p-2)
\frac{
    (1-d_\Sigma^2)^2
    \left(
        1+\rho_\phi^2d_\Sigma^2
        -2d_\kappa\rho_\phi
        d_\Sigma^3\lambda_\phi
    \right)
}{
    (1-d_\kappa^2d_\Sigma^2)
    (1-\rho_\phi^2d_\Sigma^2)^2
}
\nonumber\\
&\quad\quad+
\frac{4}{p-2}
\frac{
    d_\kappa^2(1-d_\kappa^2)
    d_\Sigma^2(1-d_\Sigma^2)^2
}{
    (1-d_\kappa^2d_\Sigma^2)^2
}
\frac{
    1+d_\kappa^2d_\Sigma^2\rho_\phi^2
    -2d_\kappa d_\Sigma
    \rho_\phi\lambda_\phi
}{
    (1-d_\kappa^2d_\Sigma^2\rho_\phi^2)^2
}
\Bigg].
\label{supp:eq:plugin-explicit-trace-formula}
\end{align}
\end{proposition}
Since \(\operatorname{Tr}(\Sigma^{(0)})/
\mathbb E[|\phi_1|^2]=\eta^{-1}\), the coefficients of
\(\eta^{-1}\) and \(\eta^{-2}\) in
\eqref{supp:eq:plugin-explicit-trace-formula} are
\(A^{\texttt{plug-in}}\) and \(R^{\texttt{plug-in}}\),
respectively, as used in Theorem~\ref{thm:trace-comparison}.

\begin{proof}
We first assume that
\(d_\Sigma>0\) and
\(\mathbb E(\phi_1^2)\neq0\).
The remaining boundary cases are treated at the end.

By \eqref{supp:eq:plugin-trace-formula}, it suffices to evaluate the
four trace terms determined by the two moment-block pairs
\[
    (A_S,G_S)
    \quad\text{and}\quad
    \left(
        A_{(\Sigma^{(0)})^{-1}},
        G_{(\Sigma^{(0)})^{-1}}
    \right),
\]
and the remaining rank-one term
\[
    \frac{\beta_{\kappa,\Sigma}^2}{p-2}
    e_1^\top A_S^{-2}e_1.
\]
The canonical forms of the four moment blocks are given in
\eqref{supp:eq:plugin-A-S}--\eqref{supp:eq:plugin-G-Sigma-inverse}.
We first evaluate the four trace terms by applying
Lemma~\ref{supp:lem:selected-entries-two-dimensional-block}
to the two moment-block pairs.

For \((A_S,G_S)\), the quantities in
Lemma~\ref{supp:lem:selected-entries-two-dimensional-block} are
\[
a
=
\frac{2\mathbb E[|\phi_1|^2]}
{\operatorname{Tr}(\Sigma^{(0)})
(1-d_\kappa^2d_\Sigma^2)},
\quad
t=d_\kappa d_\Sigma\rho_\phi,
\]
and
\[
g
=
\frac{2(1-d_\Sigma^2)}
{(1-d_\kappa^2d_\Sigma^2)^2},
\quad
u=d_\kappa^2d_\Sigma^2,
\quad
\lambda=\lambda_\phi
\]
by Lemma~\ref{supp:lem:plugin-canonical-second-moment-blocks}. Therefore, Lemma~\ref{supp:lem:selected-entries-two-dimensional-block} gives
\begin{align*}
\operatorname{Tr}(A_S^{-1})
&=
\frac{2}{
\dfrac{2\mathbb E[|\phi_1|^2]}
{\operatorname{Tr}(\Sigma^{(0)})
(1-d_\kappa^2d_\Sigma^2)}
\left(
1-d_\kappa^2d_\Sigma^2\rho_\phi^2
\right)
}\\
&=
\frac{
\operatorname{Tr}(\Sigma^{(0)})
(1-d_\kappa^2d_\Sigma^2)
}{
\mathbb E[|\phi_1|^2]
(1-d_\kappa^2d_\Sigma^2\rho_\phi^2)
}
\end{align*}
and
\begin{align*}
&\operatorname{Tr}
\left(
A_S^{-1}G_SA_S^{-1}
\right)\\
&\quad=
\frac{
2
\dfrac{2(1-d_\Sigma^2)}
{(1-d_\kappa^2d_\Sigma^2)^2}
\left\{
1+
(d_\kappa d_\Sigma\rho_\phi)^2
-
2(d_\kappa d_\Sigma\rho_\phi)
(d_\kappa^2d_\Sigma^2)\lambda_\phi
\right\}
}{
\left\{
\dfrac{2\mathbb E[|\phi_1|^2]}
{\operatorname{Tr}(\Sigma^{(0)})
(1-d_\kappa^2d_\Sigma^2)}
\right\}^2
\left(
1-d_\kappa^2d_\Sigma^2\rho_\phi^2
\right)^2
}\\
&\quad=
\frac{
\{\operatorname{Tr}(\Sigma^{(0)})\}^2
(1-d_\Sigma^2)
}{
\{\mathbb E[|\phi_1|^2]\}^2
(1-d_\kappa^2d_\Sigma^2\rho_\phi^2)^2
}\times
\left(
1+d_\kappa^2d_\Sigma^2\rho_\phi^2
-
2d_\kappa^3d_\Sigma^3
\rho_\phi\lambda_\phi
\right).
\end{align*}

For
\((A_{(\Sigma^{(0)})^{-1}},G_{(\Sigma^{(0)})^{-1}})\), the corresponding quantities
are
\[
a
=
\frac{2\mathbb E[|\phi_1|^2]}
{\operatorname{Tr}(\Sigma^{(0)})(1-d_\Sigma^2)},
\quad
t=-d_\Sigma\rho_\phi,
\]
and
\[
g
=
\frac{2}{1-d_\kappa^2d_\Sigma^2},
\quad
u=-d_\kappa d_\Sigma^2,
\quad
\lambda=\lambda_\phi
\] by Lemma~\ref{supp:lem:plugin-canonical-second-moment-blocks}. Thus, Lemma \ref{supp:lem:selected-entries-two-dimensional-block} yields
\begin{align*}
\operatorname{Tr}(A_{(\Sigma^{(0)})^{-1}}^{-1})
&=
\frac{2}{
\dfrac{2\mathbb E[|\phi_1|^2]}
{\operatorname{Tr}(\Sigma^{(0)})(1-d_\Sigma^2)}
\left(
1-\rho_\phi^2d_\Sigma^2
\right)
}=
\frac{
\operatorname{Tr}(\Sigma^{(0)})(1-d_\Sigma^2)
}{
\mathbb E[|\phi_1|^2]
(1-\rho_\phi^2d_\Sigma^2)
}
\end{align*}
and
\begin{align*}
&\operatorname{Tr}
\left(
A_{(\Sigma^{(0)})^{-1}}^{-1}
G_{(\Sigma^{(0)})^{-1}}
A_{(\Sigma^{(0)})^{-1}}^{-1}
\right)\\
&\quad=
\frac{
2
\dfrac{2}{1-d_\kappa^2d_\Sigma^2}
\left\{
1+
(-d_\Sigma\rho_\phi)^2
-
2(-d_\Sigma\rho_\phi)
(-d_\kappa d_\Sigma^2)\lambda_\phi
\right\}
}{
\left\{
\dfrac{2\mathbb E[|\phi_1|^2]}
{\operatorname{Tr}(\Sigma^{(0)})(1-d_\Sigma^2)}
\right\}^2
\left(
1-\rho_\phi^2d_\Sigma^2
\right)^2
}\\
&\quad=
\frac{
\{\operatorname{Tr}(\Sigma^{(0)})\}^2
(1-d_\Sigma^2)^2
}{
\{\mathbb E[|\phi_1|^2]\}^2
(1-d_\kappa^2d_\Sigma^2)
(1-\rho_\phi^2d_\Sigma^2)^2
}\times
\left(
1+\rho_\phi^2d_\Sigma^2
-
2d_\kappa\rho_\phi d_\Sigma^3\lambda_\phi
\right).
\end{align*}

Finally, recall from \eqref{supp:eq:plugin-trace-formula} that
\begin{align*}
&
\operatorname{Tr}
\left(
V^{\texttt{plug-in}}
\right)
\\
&=
\operatorname{Tr}(A_S^{-1})
+
\operatorname{Tr}
\left(
A_S^{-1}G_SA_S^{-1}
\right)\\
&+
(p-2)\operatorname{Tr}(A_{(\Sigma^{(0)})^{-1}}^{-1})
+
(p-2)
\operatorname{Tr}
\left(
A_{(\Sigma^{(0)})^{-1}}^{-1}
G_{(\Sigma^{(0)})^{-1}}
A_{(\Sigma^{(0)})^{-1}}^{-1}
\right)\\
&+
\frac{\beta_{\kappa,\Sigma}^2}{p-2}
e_1^\top A_S^{-2}e_1.
\end{align*}
Substituting the preceding four identities into the first four terms
gives
\begin{align*}
&
\operatorname{Tr}
\left(
V^{\texttt{plug-in}}
\right)
-
\frac{\beta_{\kappa,\Sigma}^2}{p-2}
e_1^\top A_S^{-2}e_1
\\
&=
\frac{
\operatorname{Tr}(\Sigma^{(0)})
(1-d_\kappa^2d_\Sigma^2)
}{
\mathbb E[|\phi_1|^2]
(1-d_\kappa^2d_\Sigma^2\rho_\phi^2)
}\\
&+
\frac{
\{\operatorname{Tr}(\Sigma^{(0)})\}^2
(1-d_\Sigma^2)
}{
\{\mathbb E[|\phi_1|^2]\}^2
(1-d_\kappa^2d_\Sigma^2\rho_\phi^2)^2
}
\left(
1+d_\kappa^2d_\Sigma^2\rho_\phi^2
-
2d_\kappa^3d_\Sigma^3
\rho_\phi\lambda_\phi
\right)\\
&+
(p-2)
\frac{
\operatorname{Tr}(\Sigma^{(0)})(1-d_\Sigma^2)
}{
\mathbb E[|\phi_1|^2]
(1-\rho_\phi^2d_\Sigma^2)
}\\
&+
(p-2)
\frac{
\{\operatorname{Tr}(\Sigma^{(0)})\}^2
(1-d_\Sigma^2)^2
}{
\{\mathbb E[|\phi_1|^2]\}^2
(1-d_\kappa^2d_\Sigma^2)
(1-\rho_\phi^2d_\Sigma^2)^2
}
\left(
1+\rho_\phi^2d_\Sigma^2
-
2d_\kappa\rho_\phi d_\Sigma^3\lambda_\phi
\right).
\end{align*}
It remains to calculate the last term.
From Lemma~\ref{supp:lem:plugin-canonical-second-moment-blocks}, we obtain
\begin{align*}
A_S^{-1}
&=
\frac{
\operatorname{Tr}(\Sigma^{(0)})
(1-d_\kappa^2d_\Sigma^2)
}{
2\mathbb E[|\phi_1|^2]
}
\frac{
\mathrm{Id}_2-d_\kappa d_\Sigma\rho_\phi Q_\phi
}{
1-d_\kappa^2d_\Sigma^2\rho_\phi^2
}.
\end{align*}
Moreover, the definitions of \(Q_\phi\) and \(\lambda_\phi\) give
\(
e_1^\top Q_\phi e_1
=
\lambda_\phi.
\)
Therefore,
\begin{align*}
e_1^\top A_S^{-2}e_1
&=
\frac14
\left\{
\frac{\operatorname{Tr}(\Sigma^{(0)})}
{\mathbb E[|\phi_1|^2]}
\right\}^{\!2}
(1-d_\kappa^2d_\Sigma^2)^2
\times
\frac{
1+d_\kappa^2d_\Sigma^2\rho_\phi^2
-2d_\kappa d_\Sigma\rho_\phi\lambda_\phi
}{
(1-d_\kappa^2d_\Sigma^2\rho_\phi^2)^2
}.
\end{align*}
On the other hand,
\[
\beta_{\kappa,\Sigma}^2
=
\frac{
16d_\kappa^2(1-d_\kappa^2)
d_\Sigma^2(1-d_\Sigma^2)^2
}{
(1-d_\kappa^2d_\Sigma^2)^4
}.
\]
Substituting the preceding two identities gives
\begin{align*}
&
\frac{\beta_{\kappa,\Sigma}^2}{p-2}
e_1^\top A_S^{-2}e_1
\\
&=
\frac{4}{p-2}
\left\{
\frac{\operatorname{Tr}(\Sigma^{(0)})}
{\mathbb E[|\phi_1|^2]}
\right\}^{\!2}
\frac{
d_\kappa^2(1-d_\kappa^2)
d_\Sigma^2(1-d_\Sigma^2)^2
}{
(1-d_\kappa^2d_\Sigma^2)^2
}\times
\frac{
1+d_\kappa^2d_\Sigma^2\rho_\phi^2
-2d_\kappa d_\Sigma\rho_\phi\lambda_\phi
}{
(1-d_\kappa^2d_\Sigma^2\rho_\phi^2)^2
}.
\end{align*}
Combining these identities and rearranging the resulting terms according
to their order in
\(\operatorname{Tr}(\Sigma^{(0)})/\mathbb E[|\phi_1|^2]\)
gives \eqref{supp:eq:plugin-explicit-trace-formula}.

If \(\mathbb E(\phi_1^2)=0\), then \(\rho_\phi=0\), and we interpret
\[
\rho_\phi Q_\phi=0.
\]
Thus, every term involving \(Q_\phi\) or \(\lambda_\phi\) disappears.
If \(d_\Sigma=0\), all such terms likewise disappear because they are
multiplied by a positive power of \(d_\Sigma\). Therefore, in either
case, the preceding calculation remains valid under the stated
convention for \(\lambda_\phi\).
\end{proof}

Proposition~\ref{supp:prop:trace-vplugin} provides the closed-form
quantity used in the asymptotic efficiency comparison below.

\subsection{Trace Comparison of optimally aligned loading estimators}
\label{supp:sec:proof-trace-comparison}

We now prove Theorem~\ref{thm:trace-comparison}. The proof combines the
explicit covariance and trace formulas obtained above. In particular, the
isotropic case follows from the corresponding covariance block forms, while
the anisotropic cases follow by comparing
\eqref{supp:eq:trace-cov-asymptotic-covariance} and
\eqref{supp:eq:plugin-explicit-trace-formula}.

\begin{proof}[Proof of Theorem~\ref{thm:trace-comparison}]

\medskip
\noindent\emph{Proof of (i).}
Suppose that \(d_\Sigma=0\). Then \(\Sigma^{(0)}\) is proportional to
\(\mathrm{Id}_2\), and hence \(\gamma=0\). Therefore,
\eqref{supp:eq:cov-loading-Gamma-H} and
\eqref{supp:eq:cov-loading-Gamma-T} reduce to
\[
    \Gamma_H
    =
    \left\{
        \mathbb E[|\phi_1|^2]
        \operatorname{Tr}(\Sigma^{(0)})
        +
        \{\operatorname{Tr}(\Sigma^{(0)})\}^2
    \right\}
    P_{\perp,0},
    \quad
    \Gamma_T=0.
\]
Substituting these identities into
\eqref{supp:eq:explicit-cov-loading-asymptotic-covariance} gives
\begin{align*}
V^{\texttt{cov}}
&=
\frac12
\left\{
    \frac{\operatorname{Tr}(\Sigma^{(0)})}
    {\mathbb E[|\phi_1|^2]}
    +
    \frac{\{\operatorname{Tr}(\Sigma^{(0)})\}^2}
    {\{\mathbb E[|\phi_1|^2]\}^2}
\right\}
\begin{pmatrix}
    \operatorname{Re}(P_{\perp,0})
    &
    -\operatorname{Im}(P_{\perp,0})
    \\
    \operatorname{Im}(P_{\perp,0})
    &
    \operatorname{Re}(P_{\perp,0})
\end{pmatrix}.
\end{align*}
Since \(P_{\perp,0}
=\mathrm{Id}_p-\kappa^{(0)}(\kappa^{(0)})^*\), the definition of
\(\mathscr S_p\) yields
\begin{align*}
&\mathscr S_p
\begin{pmatrix}
    \operatorname{Re}(P_{\perp,0})
    &
    -\operatorname{Im}(P_{\perp,0})
    \\
    \operatorname{Im}(P_{\perp,0})
    &
    \operatorname{Re}(P_{\perp,0})
\end{pmatrix}
\mathscr S_p^\top
=
\mathrm{Id}_{2p}
-
M(\kappa^{(0)})M(\kappa^{(0)})^\top.
\end{align*}
The first two columns of \(\mathsf Q_0\) are the columns of
\(M(\kappa^{(0)})\), while the remaining columns form an orthonormal
basis of their orthogonal complement. Consequently,
\begin{align*}
&
\mathsf{Q}_{0}^\top\mathscr S_p
V^{\texttt{cov}}
\mathscr S_p^\top \mathsf Q_0
\\
&=
\frac12
\left\{
    \frac{\operatorname{Tr}(\Sigma^{(0)})}
    {\mathbb E[|\phi_1|^2]}
    +
    \frac{\{\operatorname{Tr}(\Sigma^{(0)})\}^2}
    {\{\mathbb E[|\phi_1|^2]\}^2}
\right\}
\operatorname{diag}
\left(
    0_2,\,
    \mathrm{Id}_2,\,
    \mathrm{Id}_{p-2}\otimes\mathrm{Id}_2
\right).
\end{align*}

We next calculate the corresponding blocks of
\(V^{\texttt{plug-in}}\). Setting \(d_\Sigma=0\) in
\eqref{supp:eq:plugin-A-S}--
\eqref{supp:eq:plugin-G-Sigma-inverse} gives
\[
    A_S
    =
    A_{(\Sigma^{(0)})^{-1}}
    =
    \frac{2\mathbb E[|\phi_1|^2]}
    {\operatorname{Tr}(\Sigma^{(0)})}
    \mathrm{Id}_2,
    \quad
    G_S
    =
    G_{(\Sigma^{(0)})^{-1}}
    =2\mathrm{Id}_2.
\]
It follows that
\begin{align*}
&A_S^{-1}
+A_S^{-1}G_SA_S^{-1}
\\
&\quad=
A_{(\Sigma^{(0)})^{-1}}^{-1}
+
A_{(\Sigma^{(0)})^{-1}}^{-1}
G_{(\Sigma^{(0)})^{-1}}
A_{(\Sigma^{(0)})^{-1}}^{-1}
\\
&\quad=
\frac12
\left\{
    \frac{\operatorname{Tr}(\Sigma^{(0)})}
    {\mathbb E[|\phi_1|^2]}
    +
    \frac{\{\operatorname{Tr}(\Sigma^{(0)})\}^2}
    {\{\mathbb E[|\phi_1|^2]\}^2}
\right\}
\mathrm{Id}_2.
\end{align*}
Moreover, the rank-one term in
\eqref{supp:eq:plugin-asymptotic-covariance-block-decomposition}
vanishes because its coefficient contains \(d_\Sigma^2\). Hence,
Lemma~\ref{supp:lem:plugin-asymptotic-covariance-block-decomposition}
gives
\begin{align*}
&
\mathsf{Q}_{0}^\top\mathscr S_p
V^{\texttt{plug-in}}
\mathscr S_p^\top \mathsf Q_0
\\
&=
\frac12
\left\{
    \frac{\operatorname{Tr}(\Sigma^{(0)})}
    {\mathbb E[|\phi_1|^2]}
    +
    \frac{\{\operatorname{Tr}(\Sigma^{(0)})\}^2}
    {\{\mathbb E[|\phi_1|^2]\}^2}
\right\}
\operatorname{diag}
\left(
    0_2,\,
    \mathrm{Id}_2,\,
    \mathrm{Id}_{p-2}\otimes\mathrm{Id}_2
\right).
\end{align*}
Thus,
\[
\mathsf{Q}_{0}^\top\mathscr S_p
V^{\texttt{cov}}
\mathscr S_p^\top \mathsf Q_0
=
\mathsf{Q}_{0}^\top\mathscr S_p
V^{\texttt{plug-in}}
\mathscr S_p^\top \mathsf Q_0.
\]
Since \(\mathsf Q_0\) and \(\mathscr S_p\) are orthogonal, this proves
\[
V^{\texttt{cov}}
=
V^{\texttt{plug-in}}.
\]

\medskip
\noindent\emph{Proof of (ii).}
Recall that
\[
\eta
=
\frac{
\mathbb E[|\phi_1|^2]
}{
\operatorname{Tr}(\Sigma^{(0)})
}.
\]
By
\eqref{supp:eq:trace-cov-asymptotic-covariance},
\begin{align*}
\operatorname{Tr}
\left(
V^{\texttt{cov}}
\right)
&=
(p-1)\eta^{-1}
+
\left\{
p-1+d_\Sigma^2(1-d_\kappa^2)
\right\}
\eta^{-2}.
\end{align*}
Moreover,
\eqref{supp:eq:plugin-explicit-trace-formula} gives
\begin{align*}
\operatorname{Tr}
\left(
V^{\texttt{plug-in}}
\right)
&=
\Bigg\{
\frac{
1-d_\kappa^2d_\Sigma^2
}{
1-d_\kappa^2d_\Sigma^2\rho_\phi^2
}
+
(p-2)
\frac{
1-d_\Sigma^2
}{
1-d_\Sigma^2\rho_\phi^2
}
\Bigg\}
\eta^{-1}
+
O(\eta^{-2}),
\end{align*}
where the remainder is uniform over
\(\lambda_\phi\in[-1,1]\). Indeed, the coefficient of
\(\eta^{-2}\) in
\eqref{supp:eq:plugin-explicit-trace-formula}
is continuous in \(\lambda_\phi\) on the compact interval
\([-1,1]\), while all of its denominators are strictly positive.

Dividing the preceding two expansions gives
\begin{align*}
&
\frac{
\operatorname{Tr}
\left(
V^{\texttt{plug-in}}
\right)
}{
\operatorname{Tr}
\left(
V^{\texttt{cov}}
\right)
}
\\
&=
\frac{
\displaystyle
\frac{
1-d_\kappa^2d_\Sigma^2
}{
1-d_\kappa^2d_\Sigma^2\rho_\phi^2
}
+
\displaystyle
(p-2)
\frac{
1-d_\Sigma^2
}{
1-d_\Sigma^2\rho_\phi^2
}
+
O(\eta^{-1})
}{
p-1+O(\eta^{-1})
}
\\
&=
\frac{1}{p-1}
\left\{
\frac{
1-d_\kappa^2d_\Sigma^2
}{
1-d_\kappa^2d_\Sigma^2\rho_\phi^2
}
+
(p-2)
\frac{
1-d_\Sigma^2
}{
1-d_\Sigma^2\rho_\phi^2
}
\right\}
+
O(\eta^{-1}),
\end{align*}
uniformly over \(\lambda_\phi\in[-1,1]\). This proves
\eqref{eq:high-snr-trace-ratio}.

We next establish the bounds for the leading-order ratio. Direct
calculation gives
\begin{align}
&
\frac{
1-d_\kappa^2d_\Sigma^2
}{
1-d_\kappa^2d_\Sigma^2\rho_{\phi}^{2}
}
-
\frac{
1-d_\Sigma^2
}{
1-d_\Sigma^2\rho_{\phi}^{2}
}
\nonumber\\
&=
\frac{
d_\Sigma^2(1-d_\kappa^2)(1-\rho_{\phi}^{2})
}{
(1-d_\kappa^2d_\Sigma^2\rho_{\phi}^{2})
(1-d_\Sigma^2\rho_{\phi}^{2})
}
\geq0,
\label{supp:eq:high-snr-shrinkage-factor-ordering}
\end{align}
and
\begin{align}
&
1-
\frac{
1-d_\kappa^2d_\Sigma^2
}{
1-d_\kappa^2d_\Sigma^2\rho_{\phi}^{2}
}
\nonumber\\
&=
\frac{
d_\kappa^2d_\Sigma^2(1-\rho_{\phi}^{2})
}{
1-d_\kappa^2d_\Sigma^2\rho_{\phi}^{2}
}
\geq0.
\label{supp:eq:high-snr-upper-factor-below-one}
\end{align}
The leading-order ratio can be written as the convex combination
\begin{align*}
&
\frac{p-2}{p-1}
\frac{
1-d_\Sigma^2
}{
1-d_\Sigma^2\rho_{\phi}^{2}
}
+
\frac{1}{p-1}
\frac{
1-d_\kappa^2d_\Sigma^2
}{
1-d_\kappa^2d_\Sigma^2\rho_{\phi}^{2}
}.
\end{align*}
Therefore,
\eqref{supp:eq:high-snr-shrinkage-factor-ordering} and
\eqref{supp:eq:high-snr-upper-factor-below-one}
give
\eqref{eq:high-snr-leading-ratio-bounds}.

If \(\rho_{\phi}<1\), then
\eqref{supp:eq:high-snr-shrinkage-factor-ordering}
is strict because \(d_\Sigma>0\) and \(d_\kappa<1\).
Since \(p\geq3\), both weights in the preceding convex combination are
strictly positive. Hence, the leading-order ratio lies strictly between
the two bounds in
\eqref{eq:high-snr-leading-ratio-bounds}. The final bound by one is
strict whenever \(d_\kappa>0\), by
\eqref{supp:eq:high-snr-upper-factor-below-one}.

Now fix \(\rho_\phi<1\). The leading-order ratio lies strictly inside
the interval
\[
\left(
\frac{
1-d_\Sigma^2
}{
1-d_\Sigma^2\rho_\phi^2
},
\,
\frac{
1-d_\kappa^2d_\Sigma^2
}{
1-d_\kappa^2d_\Sigma^2\rho_\phi^2
}
\right).
\]
Because the \(O(\eta^{-1})\) remainder in
\eqref{eq:high-snr-trace-ratio} is uniform over
\(\lambda_\phi\in[-1,1]\), the full trace ratio remains inside this
interval for all sufficiently large \(\eta\), uniformly over
\(\lambda_\phi\). This proves the asserted high-signal-to-noise
inequalities.

If \(\rho_\phi=1\), both factors in the leading-order ratio equal one.
Thus, the coefficients of \(\eta^{-1}\) in the two trace formulas are
both equal to \(p-1\), and their comparison is determined by the terms
of order \(\eta^{-2}\).

Finally, as \(d_\kappa\uparrow1\),
\[
\frac{
1-d_\kappa^2d_\Sigma^2
}{
1-d_\kappa^2d_\Sigma^2\rho_\phi^2
}
\longrightarrow
\frac{
1-d_\Sigma^2
}{
1-d_\Sigma^2\rho_\phi^2
}.
\]
Hence, the upper and lower bounds in \eqref{eq:high-snr-leading-ratio-bounds} coincide in the limit. The leading-order variance ratio therefore also converges to this common value as \(d_\kappa\uparrow 1 \).

\medskip
\noindent\emph{Proof of (iii).}
Suppose that \(d_\Sigma>0\) and \(\rho_\phi=0\). As shown in
\eqref{supp:eq:trace-cov-asymptotic-covariance},
the trace of \(V^{\texttt{cov}}\) does not depend on \(\rho_\phi\) and
is given by
\begin{align*}
\operatorname{Tr}(V^{\texttt{cov}})
&=
(p-1)
\frac{
\operatorname{Tr}(\Sigma^{(0)})
}{
\mathbb E[|\phi_1|^2]
}
+
\left\{
p-1
+
d_\Sigma^2
\left(
1-d_\kappa^2
\right)
\right\}
\left\{
\frac{
\operatorname{Tr}(\Sigma^{(0)})
}{
\mathbb E[|\phi_1|^2]
}
\right\}^{\!2}.
\end{align*}

By contrast, setting \(\rho_\phi=0\) in
\eqref{supp:eq:plugin-explicit-trace-formula} gives
\begin{align*}
\operatorname{Tr}
\left(
V^{\texttt{plug-in}}
\right)
&=
\left\{
1-d_\kappa^2d_\Sigma^2
+
(p-2)
\left(
1-d_\Sigma^2
\right)
\right\}
\frac{
\operatorname{Tr}(\Sigma^{(0)})
}{
\mathbb E[|\phi_1|^2]
}
\\
&\quad+
\Bigg\{
1-d_\Sigma^2
+
\frac{
(p-2)
\left(
1-d_\Sigma^2
\right)^2
}{
1-d_\kappa^2d_\Sigma^2
}
\\
&\quad\quad+
\frac{
4d_\kappa^2
\left(
1-d_\kappa^2
\right)
d_\Sigma^2
\left(
1-d_\Sigma^2
\right)^2
}{
(p-2)
\left(
1-d_\kappa^2d_\Sigma^2
\right)^2
}
\Bigg\}
\left\{
\frac{
\operatorname{Tr}(\Sigma^{(0)})
}{
\mathbb E[|\phi_1|^2]
}
\right\}^{\!2}.
\end{align*}
Subtracting the second identity from the first gives
\begin{align}
&\operatorname{Tr}(V^{\texttt{cov}})
-
\operatorname{Tr}
\left(
V^{\texttt{plug-in}}
\right)
\nonumber\\
&=
d_\Sigma^2
\left(
p-2+d_\kappa^2
\right)
\frac{
\operatorname{Tr}(\Sigma^{(0)})
}{
\mathbb E[|\phi_1|^2]
}
\nonumber\\
&\quad+
\Bigg[
\frac{
d_\Sigma^2
\left[
(p-2)
\left(
2-d_\kappa^2-d_\Sigma^2
\right)
+
\left(
2-d_\kappa^2
\right)
\left(
1-d_\kappa^2d_\Sigma^2
\right)
\right]
}{
1-d_\kappa^2d_\Sigma^2
}
\nonumber\\
&\quad-
\frac{
4d_\kappa^2
\left(
1-d_\kappa^2
\right)
d_\Sigma^2
\left(
1-d_\Sigma^2
\right)^2
}{
(p-2)
\left(
1-d_\kappa^2d_\Sigma^2
\right)^2
}
\Bigg]
\left\{
\frac{
\operatorname{Tr}(\Sigma^{(0)})
}{
\mathbb E[|\phi_1|^2]
}
\right\}^{\!2}.
\label{supp:eq:cov-plugin-circular-trace-difference}
\end{align}
The first term on the right-hand side is strictly positive. We next show that the expression inside the square brackets is also strictly positive. Equivalently, it is enough to show
\begin{align}
&\frac{
d_\Sigma^2
\left[
(p-2)
\left(
2-d_\kappa^2-d_\Sigma^2
\right)
+
\left(
2-d_\kappa^2
\right)
\left(
1-d_\kappa^2d_\Sigma^2
\right)
\right]
}{
1-d_\kappa^2d_\Sigma^2
}
>
\frac{
4d_\kappa^2
\left(
1-d_\kappa^2
\right)
d_\Sigma^2
\left(
1-d_\Sigma^2
\right)^2
}{
(p-2)
\left(
1-d_\kappa^2d_\Sigma^2
\right)^2
}.
\label{supp:eq:cov-gain-exceeds-plugin-penalty}
\end{align}
Multiplying both sides by
\[
\frac{
(p-2)
\left(
1-d_\kappa^2d_\Sigma^2
\right)^2
}{
d_\Sigma^2
}
\]
shows that
\eqref{supp:eq:cov-gain-exceeds-plugin-penalty} is equivalent to
\begin{align}
&(p-2)
\left[
(p-2)
\left(
2-d_\kappa^2-d_\Sigma^2
\right)
+
\left(
2-d_\kappa^2
\right)
\left(
1-d_\kappa^2d_\Sigma^2
\right)
\right]\left(
1-d_\kappa^2d_\Sigma^2
\right)
>
4d_\kappa^2
\left(
1-d_\kappa^2
\right)
\left(
1-d_\Sigma^2
\right)^2.
\label{supp:eq:cleared-cov-plugin-comparison}
\end{align}
We now bound the left-hand side of
\eqref{supp:eq:cleared-cov-plugin-comparison} from below. Expanding only
the outer factor \(p-2\), it can be written as
\begin{align*}
    &(p-2)^2
    \left(
        2-d_\kappa^2-d_\Sigma^2
    \right)
    \left(
        1-d_\kappa^2d_\Sigma^2
    \right)
    +
    (p-2)
    \left(
        2-d_\kappa^2
    \right)
    \left(
        1-d_\kappa^2d_\Sigma^2
    \right)^2.
\end{align*}
Since
\[
    2-d_\kappa^2-d_\Sigma^2>0
    \quad\text{and}\quad
    1-d_\kappa^2d_\Sigma^2>0,
\]
the first term is strictly positive. Hence the left-hand side of
\eqref{supp:eq:cleared-cov-plugin-comparison} is strictly larger than
\[
    (p-2)
    \left(
        2-d_\kappa^2
    \right)
    \left(
        1-d_\kappa^2d_\Sigma^2
    \right)^2.
\]
Since \(p-2\geq1\), it is therefore strictly larger than
\begin{equation}
    \left(
        2-d_\kappa^2
    \right)
    \left(
        1-d_\kappa^2d_\Sigma^2
    \right)^2.
\label{supp:eq:lower-bound-cleared-comparison}
\end{equation}

It remains to compare
\eqref{supp:eq:lower-bound-cleared-comparison} with the right-hand side
of \eqref{supp:eq:cleared-cov-plugin-comparison}. First,
\begin{align*}
    1-d_\kappa^2d_\Sigma^2
    -
    \left(
        1-d_\Sigma^2
    \right)
    &=
    d_\Sigma^2
    \left(
        1-d_\kappa^2
    \right)
    >0,
\end{align*}
so that
\[
    \left(
        1-d_\kappa^2d_\Sigma^2
    \right)^2
    >
    \left(
        1-d_\Sigma^2
    \right)^2.
\]
Second,
\begin{align*}
    2-d_\kappa^2
    -
    4d_\kappa^2
    \left(
        1-d_\kappa^2
    \right)
    &=
    4d_\kappa^4-5d_\kappa^2+2
    \\
    &=
    4
    \left(
        d_\kappa^2-\frac58
    \right)^2
    +
    \frac7{16}
    >0.
\end{align*}
Therefore,
\[
    2-d_\kappa^2
    >
    4d_\kappa^2
    \left(
        1-d_\kappa^2
    \right).
\]
Combining these two inequalities yields
\begin{align*}
    &
    \left(
        2-d_\kappa^2
    \right)
    \left(
        1-d_\kappa^2d_\Sigma^2
    \right)^2
    \\
    &\quad>
    4d_\kappa^2
    \left(
        1-d_\kappa^2
    \right)
    \left(
        1-d_\kappa^2d_\Sigma^2
    \right)^2
    \\
    &\quad \geq
    4d_\kappa^2
    \left(
        1-d_\kappa^2
    \right)
    \left(
        1-d_\Sigma^2
    \right)^2.
\end{align*}
Together with
\eqref{supp:eq:lower-bound-cleared-comparison}, this shows
\eqref{supp:eq:cleared-cov-plugin-comparison}, and hence
\eqref{supp:eq:cov-gain-exceeds-plugin-penalty}. It now follows from
\eqref{supp:eq:cov-plugin-circular-trace-difference} that
\[
\operatorname{Tr}
\left(
V^{\texttt{plug-in}}
\right)
<
\operatorname{Tr}
\left(
V^{\texttt{cov}}
\right).
\]
\end{proof}

\section{Inconsistency of the joint profile likelihood estimator}
\label{supp:sec:joint-profile-likelihood-inconsistency}

This section proves Theorem~\ref{thm: PL strong inconsistency}. We
first show that the true parameter pair is not a stationary point of
the population profile-likelihood objective. We then establish that
both the empirical and population minimizer sets are contained in a
common compact subset of the joint parameter space. Finally, we apply
the general BP-consistency theorem on this compact restriction and
combine the resulting set convergence with population
nonstationarity to show that the empirical minimizers are eventually
bounded away from the true parameter almost surely.

We use the following notation throughout this section. For
\(y\in\mathbb C^p\), \([\kappa]\in\mathbb{CP}^{p-1}\), and
\(\Sigma\in\mathrm{SPD}(2)\), define
\begin{equation}
\label{supp:eq:pl-single-observation-loss}
    \rho^{\texttt{PL}}(y,[\kappa],\Sigma)
    :=
    \rho(y,[\kappa],\Sigma)
    +
    p\log\det(\Sigma).
\end{equation}
The corresponding empirical and population objectives are
\begin{equation}
\label{supp:eq:pl-empirical-population-objectives}
\begin{aligned}
    F_n^{\texttt{PL}}([\kappa],\Sigma)
    &:=
    \frac1n\sum_{i=1}^n
    \rho^{\texttt{PL}}(Y_i,[\kappa],\Sigma),\\
    F^{\texttt{PL}}([\kappa],\Sigma)
    &:=
    \mathbb E\!\left[
        \rho^{\texttt{PL}}(Y_1,[\kappa],\Sigma)
    \right].
\end{aligned}
\end{equation}
Accordingly, let
\begin{equation}
\label{supp:eq:pl-empirical-population-minimizer-sets}
\begin{aligned}
    \widehat E_n^{\texttt{PL}}
    &:=
    \operatorname*{argmin}_{
        ([\kappa],\Sigma)
        \in
        \mathbb{CP}^{p-1}\times\mathrm{SPD}(2)
    }
    F_n^{\texttt{PL}}([\kappa],\Sigma),\\
    E^{\texttt{PL}}
    &:=
    \operatorname*{argmin}_{
        ([\kappa],\Sigma)
        \in
        \mathbb{CP}^{p-1}\times\mathrm{SPD}(2)
    }
    F^{\texttt{PL}}([\kappa],\Sigma).
\end{aligned}
\end{equation}

\subsection{Nonstationarity of the population profile likelihood}
\label{supp:sec:pl-population-nonstationarity}

We show that the true parameter pair is not a minimizer of the population profile-likelihood objective.

\begin{theorem}
\label{supp:thm:pl-true-parameter-not-minimizer}
Assume the centered drift model~\eqref{model:centered-complex-factor},
with true parameters
\(\kappa^{(0)}\in\mathcal S^{2p-1}\) and
\(\Sigma^{(0)}\in\mathrm{SPD}(2)\), and suppose that
\(0<\mathbb E[|\phi_1|^2]<\infty\).
Then \(([\kappa^{(0)}],\Sigma^{(0)})\) is not a stationary
point, and hence is not a local or global minimizer, of
\(F^{\texttt{PL}}\).
In particular, if \(p\ge2\), then
\[
    F^{\texttt{PL}}\left(
        [\kappa^{(0)}],
        \frac{p-1}{p}\Sigma^{(0)}
    \right)
    <
    F^{\texttt{PL}}([\kappa^{(0)}],\Sigma^{(0)}).
\]
\end{theorem}

\begin{proof}
For every \(t>0\),
\[
\begin{aligned}
    \rho(y,[\kappa],t\Sigma)
    =
    \min_{\phi\in\mathbb C}
    \|y-\phi\kappa\|_{t\Sigma}^2=
    \frac1t
    \min_{\phi\in\mathbb C}
    \|y-\phi\kappa\|_\Sigma^2
    =
    \frac1t\rho(y,[\kappa],\Sigma).
\end{aligned}
\]
At the true loading, the signal contribution to the profiled
loss vanishes. By
Lemmas~\ref{supp:lem:plugin-population-loss-decomposition}
and~\ref{supp:lem:plugin-expected-noise-loss},
\[
    \mathbb E\!\left[
        \rho(Y_1,[\kappa^{(0)}],\Sigma^{(0)})
    \right]
    =
    p\operatorname{Tr}(\mathrm{Id}_2)
    -
    \operatorname{Tr}(\mathrm{Id}_2)
    =
    2p-2.
\]
Since
\[
    \log\det(t\Sigma^{(0)})
    =
    2\log t+\log\det(\Sigma^{(0)}),
\]
we obtain
\[
    F^{\texttt{PL}}([\kappa^{(0)}],t\Sigma^{(0)})
    =
    \frac{2p-2}{t}
    +
    2p\log t
    +
    p\log\det(\Sigma^{(0)}).
\]
Consequently,
\[
    \left.
    \frac{d}{dt}
    F^{\texttt{PL}}([\kappa^{(0)}],t\Sigma^{(0)})
    \right|_{t=1}
    =
    -(2p-2)+2p
    =
    2>0.
\]
Thus, for \(t<1\) sufficiently close to \(1\),
\[
    F^{\texttt{PL}}([\kappa^{(0)}],t\Sigma^{(0)})
    <
    F^{\texttt{PL}}([\kappa^{(0)}],\Sigma^{(0)}).
\]
The true parameter pair is therefore not stationary and is
not a local or global minimizer.

For \(p\ge2\), choosing \(t=(p-1)/p>0\) gives explicitly
\[
\begin{aligned}
    F^{\texttt{PL}}\left(
        [\kappa^{(0)}],
        \frac{p-1}{p}\Sigma^{(0)}
    \right)
    &=
    F^{\texttt{PL}}([\kappa^{(0)}],\Sigma^{(0)})
    +
    2+2p\log\left(1-\frac1p\right)\\
    &<
    F^{\texttt{PL}}([\kappa^{(0)}],\Sigma^{(0)}),
\end{aligned}
\]
where the strict inequality follows from
\(\log(1-x)<-x\) for \(0<x<1\).
\end{proof}

\subsection{Compact containment of profile-likelihood minimizers}
\label{supp:sec:pl-compact-containment}

This subsection establishes compact containment of the
profile-likelihood minimizers. Since the joint profile-likelihood
parameter space
\(\mathbb{CP}^{p-1}\times\mathrm{SPD}(2)\) is not compact, we first
show that both the empirical and population minimizers are contained
in \(\mathbb{CP}^{p-1}\times K\) for some compact set
\(K\subset\mathrm{SPD}(2)\). This allows us to apply the generalized
Fr\'echet mean consistency theorem on the restricted parameter
space. For this purpose, we express the profiled residual as a
quadratic form, derive a lower bound in terms of the eigenvalues of
\(\Sigma\), and show that the objective diverges as these eigenvalues
approach \(0\) or \(\infty\).

\begin{lemma}
\label{supp:lem:pl-residual-quadratic-form}
For any \(y\in\mathbb C^p\),
\(\kappa\in\mathcal S^{2p-1}\), and
\(\Sigma\in\mathrm{SPD}(2)\),
\begin{equation}
\label{supp:eq:pl-residual-quadratic-form}
    \rho(y,[\kappa],\Sigma) 
    =
    \operatorname{vec}(y)^\top
    \Pi_\Sigma(\kappa)
    \operatorname{vec}(y),
\end{equation}
where
\begin{equation}
\label{supp:eq:pl-weighted-residual-matrix}
\begin{aligned}
    \Pi_\Sigma(\kappa)
    &:=
    \mathrm{Id}_p\otimes\Sigma^{-1}\\
    &\quad-
    (\mathrm{Id}_p\otimes\Sigma^{-1})
    M(\kappa)
    \left\{
        M(\kappa)^\top
        (\mathrm{Id}_p\otimes\Sigma^{-1})
        M(\kappa)
    \right\}^{-1}
    M(\kappa)^\top
    (\mathrm{Id}_p\otimes\Sigma^{-1}),
\end{aligned}
\end{equation}
and
\[
    M(\kappa)
    =
    \left(
        M(\kappa_1)^\top,\dots,
        M(\kappa_p)^\top
    \right)^\top
    \in\mathbb R^{2p\times2}.
\]
Moreover, \(\Pi_\Sigma(\kappa)\succeq0\).
\end{lemma}

\begin{proof}[Proof of Lemma~\ref{supp:lem:pl-residual-quadratic-form}]

We express the profile loss \(\rho(y,[\kappa],\Sigma)\) as a quadratic form. Let \(W=\mathrm{Id}_p\otimes\Sigma^{-1}\). Then we can write $$\kappa \diamond_\Sigma \kappa = M(\kappa)^\top W M(\kappa)$$ and $$\kappa \bullet_\Sigma y = M(\kappa)^\top W \operatorname{vec}(y).$$ Consequently,
\begin{equation*}
  \operatorname{vec}(\widehat{\phi}(y, \kappa, \Sigma)) = (M(\kappa)^\top W M(\kappa))^{-1} M(\kappa)^\top W \operatorname{vec}(y).
\end{equation*}

Using Lemma~\ref{supp:lem:plugin-profile-loss-decomposition} and the above formula, we get
\begin{align}\label{supp:eq: residual expansion}
  & \rho(y,[\kappa],\Sigma) \nonumber\\
  & = \langle y, y \rangle_\Sigma - \left\langle \widehat{\phi}(y, \kappa, \Sigma)\kappa, y \right\rangle_\Sigma \nonumber
  \\
  &= \operatorname{vec}(y)^\top W \operatorname{vec}(y) - \operatorname{vec}(\widehat{\phi}(y, \kappa, \Sigma))^\top M(\kappa)^\top W \operatorname{vec}(y). 
\end{align}
Substituting $\operatorname{vec}(\widehat{\phi}(y, \kappa, \Sigma)) = (M(\kappa)^\top W M(\kappa))^{-1} M(\kappa)^\top W \operatorname{vec}(y)$ 
into \eqref{supp:eq: residual expansion}, we get
\begin{align*}
  \rho(y,[\kappa],\Sigma) 
  &= \operatorname{vec}(y)^\top W \operatorname{vec}(y) - \operatorname{vec}(y)^\top W M(\kappa) (M(\kappa)^\top W M(\kappa))^{-1} M(\kappa)^\top W \operatorname{vec}(y) \\
  &= \operatorname{vec}(y)^\top \left[ W - W M(\kappa) (M(\kappa)^\top W M(\kappa))^{-1} M(\kappa)^\top W \right] \operatorname{vec}(y).
\end{align*}

Let \(
    U:=W^{1/2}M(\kappa).
\)
Since
\(
    M(\kappa)^\top M(\kappa)
    =
    \sum_{j=1}^p
    M(\kappa_j)^\top M(\kappa_j)
    =
    \sum_{j=1}^p|\kappa_j|^2\mathrm{Id}_2
    =
    \mathrm{Id}_2,
\)
the matrix \(M(\kappa)\), and hence \(U\), has full column rank.
Therefore,
\[
    P_U
    :=
    U(U^\top U)^{-1}U^\top
\]
is the orthogonal projection onto the column space of \(U\), and
\[
    \Pi_\Sigma(\kappa)
    =
    W^{1/2}
    (\mathrm{Id}_{2p}-P_U)
    W^{1/2}.
\]
Consequently, \( \Pi_\Sigma(\kappa) \succeq0. \)

\end{proof}

\begin{lemma}
\label{supp:lem:pl-trace-projection-lower-bound}
For any \(\kappa\in\mathcal S^{2p-1}\) and
\(\Sigma\in\mathrm{SPD}(2)\) with eigenvalues
\(\lambda_1\ge\lambda_2>0\), let
\(\Pi_\Sigma(\kappa)\in\mathbb R^{2p\times2p}\) be defined as
in~\eqref{supp:eq:pl-weighted-residual-matrix}. Then
\begin{equation}
\label{supp:eq:pl-trace-projection-lower-bound}
    \operatorname{Tr}\bigl(\Pi_\Sigma(\kappa)\bigr)
    \ge
    \frac{p-2}{\lambda_2}
    +
    \frac{p}{\lambda_1} 
    ~\text{for}~ p \geq 2.
\end{equation}

\end{lemma}

\begin{proof}[Proof of Lemma~\ref{supp:lem:pl-trace-projection-lower-bound}]
To establish the lower bound for $\operatorname{Tr}(\Pi_\Sigma(\kappa))$, we decompose $\Pi_\Sigma(\kappa)$ into two parts, $\Pi_\Sigma(\kappa) = W - H$, where:
\begin{align*}
  W &= \mathrm{Id}_{p} \otimes \Sigma^{-1}, \\
  H &= (\mathrm{Id}_{p} \otimes \Sigma^{-1}) M(\kappa) \big(M(\kappa)^\top (\mathrm{Id}_{p} \otimes \Sigma^{-1}) M(\kappa)\big)^{-1} M(\kappa)^\top (\mathrm{Id}_{p} \otimes \Sigma^{-1}).
\end{align*}

For the matrix $W$, because it is a block diagonal matrix consisting of $p$ copies of $\Sigma^{-1}$, and the eigenvalues of $\Sigma^{-1}$ are $1/\lambda_2 \ge 1/\lambda_1$, its trace is simply computed as:
\begin{equation}\label{supp:eq: trace of W}
  \operatorname{Tr}(W) = \operatorname{Tr}(\mathrm{Id}_{p} \otimes \Sigma^{-1}) = p \cdot \operatorname{Tr}(\Sigma^{-1}) = \frac{p}{\lambda_1} + \frac{p}{\lambda_2}.
\end{equation}

Next, we derive an upper bound for $\operatorname{Tr}(H)$. Let
$W^{1/2}:=\mathrm{Id}_{p}\otimes\Sigma^{-1/2}$ so that
$W=W^{1/2}W^{1/2}$. Then $H$ can be rewritten as
\[
  H
  =
  W^{1/2}
  \left[
      W^{1/2}M(\kappa)
      \left\{
          (W^{1/2}M(\kappa))^\top
          (W^{1/2}M(\kappa))
      \right\}^{-1}
      (W^{1/2}M(\kappa))^\top
  \right]
  W^{1/2}.
\]
Set $U:=W^{1/2}M(\kappa)\in\mathbb R^{2p\times 2}$. The matrix inside the brackets is the standard orthogonal projection matrix onto the column space of $U$, namely
$P_U:=U(U^\top U)^{-1}U^\top$. Hence $H=W^{1/2}P_UW^{1/2}$.

By cyclicity of the trace, $\operatorname{Tr}(H)=\operatorname{Tr}(W^{1/2}P_UW^{1/2})=\operatorname{Tr}(P_UW)$. Since $P_U$ is a positive semidefinite orthogonal projection with $\operatorname{Tr}(P_U)=2$, we have
$$\operatorname{Tr}(P_UW)\le \operatorname{Tr}(P_U)\|W\|_{\mathrm{op}}=2\|W\|_{\mathrm{op}}.$$
Since $W=\mathrm{Id}_{p}\otimes\Sigma^{-1}$, it follows that
$\|W\|_{\mathrm{op}}=\|\Sigma^{-1}\|_{\mathrm{op}}=1/\lambda_2$. Therefore,
$\operatorname{Tr}(H)\le 2/\lambda_2$.

Finally, since
$\operatorname{Tr}(W)
=p/\lambda_1+p/\lambda_2$, we obtain
\[
    \operatorname{Tr}(\Pi_\Sigma(\kappa))
    =
    \operatorname{Tr}(W)-\operatorname{Tr}(H)
    \ge
    \left(
        \frac{p}{\lambda_1}
        +
        \frac{p}{\lambda_2}
    \right)
    -
    \frac{2}{\lambda_2}
    =
    \frac{p}{\lambda_1}
    +
    \frac{p-2}{\lambda_2}.
\]
This shows the desired bound.
\end{proof}

\begin{lemma}
\label{supp:lem:pl-second-moment-spectral-lower-bound}
Assume the model~\eqref{model:centered-complex-factor} with
\(\kappa^{(0)}\in\mathcal S^{2p-1}\),
\(\Sigma^{(0)}\in\mathrm{SPD}(2)\), and
\(\mathbb E[|\phi_1|^2]<\infty\). Let
\[
   S_Y^{(0)}
    :=
    \mathbb E\left[
        \operatorname{vec}(Y_1)
        \operatorname{vec}(Y_1)^\top
    \right],
    \quad
    \widehat S_Y
    :=
    \frac1n\sum_{i=1}^n
    \operatorname{vec}(Y_i)
    \operatorname{vec}(Y_i)^\top.
\]
Let \(\lambda_2^{(0)}>0\) denote the minimum eigenvalue of
\(\Sigma^{(0)}\). Then
\[
    \lambda_{\min}(S_{Y}^{(0)})
    \ge
    \lambda_2^{(0)},
\]
and
\[
    \liminf_{n\to\infty}
    \lambda_{\min}(\widehat S_Y)
    \ge
    \lambda_2^{(0)}
    \quad\text{almost surely}.
\]
\end{lemma}

\begin{proof}
Under the model~\eqref{model:centered-complex-factor},
\[
    \operatorname{vec}(Y_1)
    =
    M(\kappa^{(0)})
    \operatorname{vec}(\phi_1)
    +
    \operatorname{vec}(\varepsilon_1).
\]
Since
\(
    M(\kappa^{(0)})^\top M(\kappa^{(0)})
    =
    \mathrm{Id}_2,
\)
we have
\(
    \left\|
        M(\kappa^{(0)})\operatorname{vec}(\phi_1)
    \right\|^2
    =
    |\phi_1|^2.
\)
Furthermore,
\(
    \mathbb E[\|\operatorname{vec}(\varepsilon_1)\|^2]
    =
    p\operatorname{Tr}(\Sigma^{(0)})
    <
    \infty.
\)
It follows that
\[
\begin{aligned}
    \mathbb E[\|Y_1\|^2]
    &=
    \mathbb E[
        \|\operatorname{vec}(Y_1)\|^2
    ]\le
    2\mathbb E[|\phi_1|^2]
    +
    2p\operatorname{Tr}(\Sigma^{(0)})
    <
    \infty.
\end{aligned}
\]
In particular, \(S_{Y}^{(0)}\) is well-defined.

Since \(\phi_1\) and \(\varepsilon_1\) are independent and
\(
    \mathbb E[
        \operatorname{vec}(\varepsilon_1)
    ]
    =
    0,
\)
the cross terms vanish, and hence
\begin{align}
   S_Y^{(0)}
    &=
    M(\kappa^{(0)})
    \mathbb E\left[
        \operatorname{vec}(\phi_1)
        \operatorname{vec}(\phi_1)^\top
    \right]
    M(\kappa^{(0)})^\top +
    \mathbb E\left[
        \operatorname{vec}(\varepsilon_1)
        \operatorname{vec}(\varepsilon_1)^\top
    \right].
\label{supp:eq: pop covariance decomposition}
\end{align}
Since
\(
    \varepsilon_{1,j}
    \stackrel{\mathrm{i.i.d.}}{\sim}
    \mathcal{CN}(0,\Sigma^{(0)}),
\)
we have
\(
    \mathbb E\left[
        \operatorname{vec}(\varepsilon_1)
        \operatorname{vec}(\varepsilon_1)^\top
    \right]
    =
    \mathrm{Id}_p\otimes\Sigma^{(0)}.
\)
The first term in~\eqref{supp:eq: pop covariance decomposition} is
positive semidefinite. Therefore,
\[
   S_Y^{(0)}
    \succeq
    \mathrm{Id}_p\otimes\Sigma^{(0)},
\]
which gives
\begin{equation}
\label{supp:eq: pop min eigenvalue}
    \lambda_{\min}(S_{Y}^{(0)})
    \ge
    \lambda_{\min}\left(
        \mathrm{Id}_p\otimes\Sigma^{(0)}
    \right)
    =
    \lambda_2^{(0)}.
\end{equation}

Since \(Y_1\) has a finite second moment, every entry of
\(
    \operatorname{vec}(Y_1)
    \operatorname{vec}(Y_1)^\top
\) is integrable by the Cauchy--Schwarz inequality. Hence, the
entrywise strong law of large numbers yields
\[
    \widehat S_Y
    \xrightarrow{\mathrm{a.s.}}
   S_Y^{(0)}
\]
entrywise. Since the matrix dimension is fixed, this convergence also
holds in the operator norm.

Finally, Weyl's inequality gives
\[
    \left|
        \lambda_{\min}(\widehat S_Y)
        -
        \lambda_{\min}(S_{Y}^{(0)})
    \right|
    \le
    \left\|
        \widehat S_Y-S_{Y}^{(0)}
    \right\|_{\mathrm{op}}
    \xrightarrow{\mathrm{a.s.}}
    0.
\]
Hence
\[
    \lambda_{\min}(\widehat S_Y)
    \xrightarrow{\mathrm{a.s.}}
    \lambda_{\min}(S_{Y}^{(0)}),
\]
and, by~\eqref{supp:eq: pop min eigenvalue},
\begin{equation}
\label{supp:eq: emp min eigenvalue a.s.}
    \liminf_{n\to\infty}
    \lambda_{\min}(\widehat S_Y)
    \ge
    \lambda_2^{(0)}
    \quad\text{almost surely}.
\end{equation}
\end{proof}

\begin{lemma}
\label{supp:lem:pl-objective-lower-bound}
Assume the model~\eqref{model:centered-complex-factor}, with true
parameter
\(\kappa^{(0)}\in\mathcal S^{2p-1}\) and
\(\Sigma^{(0)}\in\mathrm{SPD}(2)\), with
\(0<\mathbb E[|\phi_1|^2]<\infty\) and \(p\ge3\).
Let
\(\lambda_1(\Sigma)\ge\lambda_2(\Sigma)>0\) be the eigenvalues of
\(\Sigma\), and let \(\lambda_2^{(0)}>0\) be the minimum eigenvalue
of \(\Sigma^{(0)}\). Then, for every nonempty
\(\mathcal A\subset\mathrm{SPD}(2)\), the empirical profile
likelihood cost function almost surely satisfies
\begin{equation}
\label{supp:eq:pl-empirical-objective-lower-bound}
\begin{aligned}
&\liminf_{n\to\infty}
\inf_{\substack{
    [\kappa]\in\mathbb{CP}^{p-1}\\
    \Sigma\in\mathcal A
}}
F_n^{\texttt{PL}}([\kappa],\Sigma)\\
&\quad\ge
\inf_{\Sigma\in\mathcal A}
\left[
    \lambda_2^{(0)}
    \left\{
        \frac{p-2}{\lambda_2(\Sigma)}
        +
        \frac{p}{\lambda_1(\Sigma)}
    \right\}
    +
    p\log\left\{
        \lambda_1(\Sigma)\lambda_2(\Sigma)
    \right\}
\right].
\end{aligned}
\end{equation}
Furthermore, the identical lower bound holds for the population
cost function:
\begin{equation}
\label{supp:eq:pl-population-objective-lower-bound}
\begin{aligned}
&\inf_{\substack{
    [\kappa]\in\mathbb{CP}^{p-1}\\
    \Sigma\in\mathcal A
}}
F^{\texttt{PL}}([\kappa],\Sigma)\\
&\quad\ge
\inf_{\Sigma\in\mathcal A}
\left[
    \lambda_2^{(0)}
    \left\{
        \frac{p-2}{\lambda_2(\Sigma)}
        +
        \frac{p}{\lambda_1(\Sigma)}
    \right\}
    +
    p\log\left\{
        \lambda_1(\Sigma)\lambda_2(\Sigma)
    \right\}
\right].
\end{aligned}
\end{equation}
\end{lemma}

\begin{proof}[Proof of Lemma~\ref{supp:lem:pl-objective-lower-bound}]
We begin by evaluating the empirical profile loss at a single
observation \(y\):
\[
    \rho^{\texttt{PL}}(y,[\kappa],\Sigma)
    =
    \rho(y,[\kappa],\Sigma)
    +
    p\log\det(\Sigma).
\]
Since
\(
    \det(\Sigma)
    =
    \lambda_1(\Sigma)\lambda_2(\Sigma),
\)
the log-determinant term is
\(
    \log\det(\Sigma)
    =
    \log\left(
        \lambda_1(\Sigma)\lambda_2(\Sigma)
    \right).
\)
By Lemma~\ref{supp:lem:pl-residual-quadratic-form},
\[
    \rho(y,[\kappa],\Sigma)
    =
    \operatorname{vec}(y)^\top
    \Pi_\Sigma(\kappa)
    \operatorname{vec}(y),
\]
where \(\Pi_\Sigma(\kappa)\) is defined in
\eqref{supp:eq:pl-weighted-residual-matrix}. Therefore,
\begin{align}
&\frac1n
\sum_{i=1}^n
 \rho(Y_{i},[\kappa],\Sigma) =
\frac1n
\sum_{i=1}^n
\operatorname{vec}(Y_i)^\top
\Pi_\Sigma(\kappa)
\operatorname{vec}(Y_i)
\nonumber\\
&\quad=
\frac1n
\sum_{i=1}^n
\operatorname{Tr}\left(
    \Pi_\Sigma(\kappa)
    \operatorname{vec}(Y_i)
    \operatorname{vec}(Y_i)^\top
\right) =
\operatorname{Tr}\left(
    \Pi_\Sigma(\kappa)
    \widehat S_Y
\right),
\label{supp:eq: trace representation}
\end{align}
where \(\widehat S_Y\) is the empirical second-moment matrix
defined in Lemma~\ref{supp:lem:pl-second-moment-spectral-lower-bound}. Hence
\[
    F_n^{\texttt{PL}}([\kappa],\Sigma)
    =
    \operatorname{Tr}\left(
        \Pi_\Sigma(\kappa)\widehat S_Y
    \right)
    +
    p\log\left(
        \lambda_1(\Sigma)\lambda_2(\Sigma)
    \right).
\]

Since both
\(\Pi_\Sigma(\kappa)\) and \(\widehat S_Y\) are symmetric positive
semidefinite,
\(
    \operatorname{Tr}(AB)
    \ge
    \lambda_{\min}(B)\operatorname{Tr}(A)
\)
for \(A,B\succeq0\). Thus, by
Lemma~\ref{supp:lem:pl-trace-projection-lower-bound},
\begin{align}
\operatorname{Tr}\left(
    \Pi_\Sigma(\kappa)\widehat S_Y
\right)
&\ge
\lambda_{\min}(\widehat S_Y)
\operatorname{Tr}\left(
    \Pi_\Sigma(\kappa)
\right)
\ge
\lambda_{\min}(\widehat S_Y)
\left\{
    \frac{p-2}{\lambda_2(\Sigma)}
    +
    \frac{p}{\lambda_1(\Sigma)}
\right\}.
\label{supp:eq: trace inequality empirical}
\end{align}
Consequently,
\begin{align}\label{supp:eq: empirical cost pre-limit bound}
F_n^{\texttt{PL}}([\kappa],\Sigma)
\ge{}&
\lambda_{\min}(\widehat S_Y)
\left\{
    \frac{p-2}{\lambda_2(\Sigma)}
    +
    \frac{p}{\lambda_1(\Sigma)}
\right\}+
p\log\left(
    \lambda_1(\Sigma)\lambda_2(\Sigma)
\right).
\end{align}

Now fix a nonempty set
\(\mathcal A\subset\mathrm{SPD}(2)\).
Taking the infimum in
\eqref{supp:eq: empirical cost pre-limit bound} over
\([\kappa]\in\mathbb{CP}^{p-1}\) and
\(\Sigma\in\mathcal A\), we obtain
\begin{align}\label{supp:eq:empirical-infimum-pre-limit-bound}
\inf_{\substack{
    [\kappa]\in\mathbb{CP}^{p-1}\\
    \Sigma\in\mathcal A
}}
F_n^{\texttt{PL}}([\kappa],\Sigma)\ge
\inf_{\Sigma\in\mathcal A}
\Bigg[
    \lambda_{\min}(\widehat S_Y)
    \left\{
        \frac{p-2}{\lambda_2(\Sigma)}
        +
        \frac{p}{\lambda_1(\Sigma)}
    \right\}
    +
    p\log\left(
        \lambda_1(\Sigma)\lambda_2(\Sigma)
    \right)
\Bigg].
\end{align}

For \(a>0\), define
\[
    \Psi_{\mathcal A}(a)
    :=
    \inf_{\Sigma\in\mathcal A}
    \Bigg[
        a
        \left\{
            \frac{p-2}{\lambda_2(\Sigma)}
            +
            \frac{p}{\lambda_1(\Sigma)}
        \right\}
        +
        p\log\left(
            \lambda_1(\Sigma)\lambda_2(\Sigma)
        \right)
    \Bigg].
\]
As the pointwise infimum of affine functions of \(a\),
\(\Psi_{\mathcal A}\) is concave and nondecreasing. Since \(p\ge3\),
the expression inside the infimum is bounded below uniformly over
\(\Sigma\in\mathrm{SPD}(2)\), while
\(\mathcal A\neq\varnothing\) gives a finite upper bound. Thus
\(\Psi_{\mathcal A}\) is finite-valued and continuous on
\((0,\infty)\).

By Lemma~\ref{supp:lem:pl-second-moment-spectral-lower-bound}, almost surely,
\[
    \liminf_{n\to\infty}
    \lambda_{\min}(\widehat S_Y)
    \ge
    \lambda_2^{(0)}.
\]
Hence, for every
\(\delta\in(0,\lambda_2^{(0)})\), almost surely for sufficiently
large \(n\),
\[
    \lambda_{\min}(\widehat S_Y)
    \ge
    \lambda_2^{(0)}-\delta.
\]
Since \(\Psi_{\mathcal A}\) is nondecreasing,
\eqref{supp:eq:empirical-infimum-pre-limit-bound} gives
\begin{align*}
&\liminf_{n\to\infty}
\inf_{\substack{
    [\kappa]\in\mathbb{CP}^{p-1}\\
    \Sigma\in\mathcal A
}}
F_n^{\texttt{PL}}([\kappa],\Sigma)\ge
\Psi_{\mathcal A}\left(
    \lambda_2^{(0)}-\delta
\right)
\end{align*}
almost surely. Letting \(\delta\downarrow0\) and using the continuity
of \(\Psi_{\mathcal A}\), we obtain
\begin{align*}
&\liminf_{n\to\infty}
\inf_{\substack{
    [\kappa]\in\mathbb{CP}^{p-1}\\
    \Sigma\in\mathcal A
}}
F_n^{\texttt{PL}}([\kappa],\Sigma)\ge
\inf_{\Sigma\in\mathcal A}
\left(
    \lambda_2^{(0)}
    \left\{
        \frac{p-2}{\lambda_2(\Sigma)}
        +
        \frac{p}{\lambda_1(\Sigma)}
    \right\}
    +
    p\log\left(
        \lambda_1(\Sigma)\lambda_2(\Sigma)
    \right)
\right)
\end{align*}
almost surely. This shows~\eqref{supp:eq:pl-empirical-objective-lower-bound}.

For the population objective, taking expectations in the quadratic
representation gives
\begin{align}
\mathbb E\left[
    \operatorname{vec}(Y_1)^\top
    \Pi_\Sigma(\kappa)
    \operatorname{vec}(Y_1)
\right]
&=
\operatorname{Tr}\left(
    \Pi_\Sigma(\kappa)S_{Y}^{(0)}
\right)
\nonumber\\
&\ge
\lambda_{\min}(S_{Y}^{(0)})
\operatorname{Tr}\left(
    \Pi_\Sigma(\kappa)
\right)
\nonumber\\
&\ge
\lambda_{\min}(S_{Y}^{(0)})
\left\{
    \frac{p-2}{\lambda_2(\Sigma)}
    +
    \frac{p}{\lambda_1(\Sigma)}
\right\}.
\label{supp:eq: trace inequality population}
\end{align}
By Lemma~\ref{supp:lem:pl-second-moment-spectral-lower-bound},
\[
    \lambda_{\min}(S_{Y}^{(0)})
    \ge
    \lambda_2^{(0)}.
\]
Therefore, for every
\([\kappa]\in\mathbb{CP}^{p-1}\) and
\(\Sigma\in\mathrm{SPD}(2)\),
\begin{align*}
F^{\texttt{PL}}([\kappa],\Sigma)
\ge{}&
\lambda_2^{(0)}
\left\{
    \frac{p-2}{\lambda_2(\Sigma)}
    +
    \frac{p}{\lambda_1(\Sigma)}
\right\}+
p\log\left(
    \lambda_1(\Sigma)\lambda_2(\Sigma)
\right).
\end{align*}
Taking the infimum over
\([\kappa]\in\mathbb{CP}^{p-1}\) and
\(\Sigma\in\mathcal A\) shows
\eqref{supp:eq:pl-population-objective-lower-bound}.

\end{proof}

\begin{lemma}
\label{supp:lem:pl-compact-containment}
Assume the model~\eqref{model:centered-complex-factor}, with true
parameter
\(\kappa^{(0)}\in\mathcal S^{2p-1}\) and
\(\Sigma^{(0)}\in\mathrm{SPD}(2)\), with
\(0<\mathbb E[|\phi_1|^2]<\infty\) and \(p\ge3\). Then there exists
a compact subset \(K\subset\mathrm{SPD}(2)\) such that, almost surely
for sufficiently large \(n\),
\[
    \operatorname*{argmin}_{
        ([\kappa],\Sigma)
        \in
        \mathbb{CP}^{p-1}\times\mathrm{SPD}(2)
    }
    F_n^{\texttt{PL}}([\kappa],\Sigma)
    \subset
    \mathbb{CP}^{p-1}\times K.
\]
Furthermore,
\[
    \operatorname*{argmin}_{
        ([\kappa],\Sigma)
        \in
        \mathbb{CP}^{p-1}\times\mathrm{SPD}(2)
    }
    F^{\texttt{PL}}([\kappa],\Sigma)
    \subset
    \mathbb{CP}^{p-1}\times K.
\]
The population minimizer set is nonempty and compact.
Almost surely, the empirical minimizer set is also nonempty
and compact for all sufficiently large \(n\).
\end{lemma}

\begin{proof}[Proof of Lemma~\ref{supp:lem:pl-compact-containment}]
First, we establish an almost-sure upper bound for the empirical
objective at the true parameter. Since
\[
    0
    \le
    \rho(Y_1,[\kappa^{(0)}],\Sigma^{(0)})
    \le
    \|Y_{1}\|^{2}_{\Sigma^{(0)}}
    \le
     \lambda_{\max}((\Sigma^{(0)})^{-1})\|Y_1\|^2,
\]
and \(\mathbb E[\|Y_1\|^2]<\infty\), the profiled residual term is
integrable.
The log-determinant term is
a deterministic constant at \(\Sigma^{(0)}\), so the corresponding
profile cost is integrable. Hence, by the strong law of large numbers,
\[
    F_n^{\texttt{PL}}
    ([\kappa^{(0)}],\Sigma^{(0)})
    \xrightarrow{\mathrm{a.s.}}
    F^{\texttt{PL}}
    ([\kappa^{(0)}],\Sigma^{(0)}).
\]
Let
\[
    C:=
    F^{\texttt{PL}}
    ([\kappa^{(0)}],\Sigma^{(0)})+1.
\]
Then, almost surely for sufficiently large \(n\),
\begin{equation}\label{supp:eq: cost upper bound}
    F_n^{\texttt{PL}}
    ([\kappa^{(0)}],\Sigma^{(0)})
    \le
    C.
\end{equation}

For \(\lambda_1\ge\lambda_2>0\), define
\[
    L(\lambda_1,\lambda_2)
    =
    g_1(\lambda_1)+g_2(\lambda_2),
\]
where
\[
    g_1(x)
    :=
    p\log x+\frac{p\lambda_2^{(0)}}{x},
    \quad
    g_2(x)
    :=
    p\log x+\frac{(p-2)\lambda_2^{(0)}}{x}.
\]
Since \(p\ge3\), both reciprocal coefficients are positive, and
\[
    g_j(x)\longrightarrow+\infty
    \quad
    \text{as }x\downarrow0
    \text{ or }x\uparrow\infty,
    \quad j=1,2.
\]
Moreover, each \(g_j\) is bounded below on \((0,\infty)\).
Consequently, the sublevel set
\begin{equation}\label{supp:eq: L sublevel set}
    \Lambda_{C+1}
    :=
    \left\{
        (\lambda_1,\lambda_2):
        \lambda_1\ge\lambda_2>0,\ 
        L(\lambda_1,\lambda_2)\le C+1
    \right\}
\end{equation}
is bounded above and bounded away from zero. Together with the
continuity of \(L\), this shows that
\(\Lambda_{C+1}\) is closed and bounded in \(\mathbb R^2\), and
hence compact. In particular, there exist
constants \(0<a<b<\infty\) such that
\[
    a
    \le
    \lambda_2
    \le
    \lambda_1
    \le
    b
\]
for every
\((\lambda_1,\lambda_2)\in\Lambda_{C+1}\).

Define
\[
    K
    :=
    \left\{
        \Sigma\in\mathrm{SPD}(2):
        \bigl(
            \lambda_1(\Sigma),
            \lambda_2(\Sigma)
        \bigr)
        \in\Lambda_{C+1}
    \right\}.
\]
The ordered eigenvalue map on the finite-dimensional vector space of
real symmetric \(2\times2\) matrices is continuous. Since
\(\Lambda_{C+1}\) is compact and its second coordinate is bounded
below by \(a>0\), the set \(K\) is closed in that ambient vector
space. Moreover, every \(\Sigma\in K\) satisfies
\[
    a
    \le
    \lambda_{\min}(\Sigma)
    \le
     \lambda_{\max}(\Sigma)
    \le
    b.
\]
Thus \(K\) is bounded, and therefore compact.

Applying~\eqref{supp:eq:pl-population-objective-lower-bound} with
\(\mathcal A=\{\Sigma^{(0)}\}\), we obtain
\[
\begin{aligned}
    L\bigl(
        \lambda_1(\Sigma^{(0)}),
        \lambda_2(\Sigma^{(0)})
    \bigr)
    &\le
    \inf_{[\kappa]\in\mathbb{CP}^{p-1}}
    F^{\texttt{PL}}([\kappa],\Sigma^{(0)})\le
    F^{\texttt{PL}}
    ([\kappa^{(0)}],\Sigma^{(0)})
    =
    C-1
    <
    C+1.
\end{aligned}
\]
Hence \(\Sigma^{(0)}\in K\).

Applying~\eqref{supp:eq:pl-empirical-objective-lower-bound} with
\(
    \mathcal A = \mathrm{SPD}(2)\setminus K,
\)
we obtain, almost surely,
\[
\begin{aligned}
&\liminf_{n\to\infty}
\inf_{\substack{
    [\kappa]\in\mathbb{CP}^{p-1}\\
    \Sigma\in\mathrm{SPD}(2)\setminus K
}}
F_n^{\texttt{PL}}([\kappa],\Sigma)\ge
\inf_{\Sigma\in\mathrm{SPD}(2)\setminus K}
L\bigl(
    \lambda_1(\Sigma),
    \lambda_2(\Sigma)
\bigr)
\ge
C+1.
\end{aligned}
\]
The first inequality follows from
\eqref{supp:eq:pl-empirical-objective-lower-bound},
whereas the second follows from the definition of \(K\). Hence,
almost surely for sufficiently large \(n\),
\[
\begin{aligned}
&\inf_{\substack{
    [\kappa]\in\mathbb{CP}^{p-1}\\
    \Sigma\in\mathrm{SPD}(2)\setminus K
}}
F_n^{\texttt{PL}}([\kappa],\Sigma)
>
C\ge
F_n^{\texttt{PL}}
([\kappa^{(0)}],\Sigma^{(0)}),
\end{aligned}
\]
where the second inequality follows from
\eqref{supp:eq: cost upper bound}.

Since the continuous function \(F_n^{\texttt{PL}}\) attains its
minimum on the compact set
\(\mathbb{CP}^{p-1}\times K\), and every point outside this set has
strictly larger cost than
\(([\kappa^{(0)}],\Sigma^{(0)})\in\mathbb{CP}^{p-1}\times K\), this
minimum is also global. Consequently, almost surely for sufficiently
large \(n\),
\[
    \argmin_{([\kappa],\Sigma)
        \in
        \mathbb{CP}^{p-1}\times\mathrm{SPD}(2)}
    F_n^{\texttt{PL}}([\kappa],\Sigma)
    \subset
    \mathbb{CP}^{p-1}\times K.
\]

For the population objective, applying
\eqref{supp:eq:pl-population-objective-lower-bound} with
\(\mathcal A=\mathrm{SPD}(2)\setminus K\) gives
\[
\begin{aligned}
&\inf_{\substack{
    [\kappa]\in\mathbb{CP}^{p-1}\\
    \Sigma\in\mathrm{SPD}(2)\setminus K
}}
F^{\texttt{PL}}([\kappa],\Sigma)\ge
\inf_{\Sigma\in\mathrm{SPD}(2)\setminus K}
L\bigl(
    \lambda_1(\Sigma),
    \lambda_2(\Sigma)
\bigr)
\ge
C+1
>
C-1
=
F^{\texttt{PL}}
([\kappa^{(0)}],\Sigma^{(0)}).
\end{aligned}
\]
Thus no population minimizer can have covariance outside \(K\).

Moreover, \(F^{\texttt{PL}}\) is continuous on
\(\mathbb{CP}^{p-1}\times K\). Indeed, compactness of \(K\) implies
that
\[
    \sup_{\Sigma\in K}
     \lambda_{\max}(\Sigma^{-1})
    <
    \infty
    \quad\text{and}\quad
    \sup_{\Sigma\in K}
    |\log\det(\Sigma)|
    <
    \infty.
\]
Together with
\[
    0
    \le
    \rho(Y_{1}, [\kappa], \Sigma) \le \|Y_1\|_\Sigma^2,
\]
this gives an integrable envelope of the form
\[
    C_K(1+\|Y_1\|^2).
\]
Hence continuity follows from the dominated convergence theorem.
Hence \(F^{\texttt{PL}}\) attains its minimum on
the compact set \(\mathbb{CP}^{p-1}\times K\), and therefore
\[
    \argmin_{([\kappa],\Sigma)
        \in
        \mathbb{CP}^{p-1}\times\mathrm{SPD}(2)}
    F^{\texttt{PL}}([\kappa],\Sigma)
    \subset
    \mathbb{CP}^{p-1}\times K.
\]
\end{proof}

\subsection{Proof of Theorem~\ref{thm: PL strong inconsistency}}
\label{supp:sec:pl-strong-inconsistency-proof}

This subsection completes the proof of
Theorem~\ref{thm: PL strong inconsistency} using the two ingredients
established above. First,
Lemma~\ref{supp:lem:pl-compact-containment} confines both the
empirical and population profile-likelihood minimizers to
\(\mathbb{CP}^{p-1}\times K\) for some compact
\(K\subset\mathrm{SPD}(2)\). This allows us to apply the general
BP-consistency theorem on the restricted parameter space. Second,
combining this convergence with
Theorem~\ref{supp:thm:pl-true-parameter-not-minimizer}, which
excludes
\(([\kappa^{(0)}],\Sigma^{(0)})\) from the population minimizer set,
shows that the empirical profile-likelihood minimizers are eventually
separated from the true parameter almost surely.

\begin{theorem}
\label{supp:thm:pl-bp-consistency}
Assume the centered drift model~\eqref{model:centered-complex-factor},
with true parameter
\(\kappa^{(0)}\in\mathcal S^{2p-1}\) and
\(\Sigma^{(0)}\in\mathrm{SPD}(2)\), with
\[
    0<\mathbb E[|\phi_1|^2]<\infty
    \quad\text{and}\quad
    p\ge3.
\]
Then \(\widehat E_n^{\texttt{PL}}\) is BP strongly consistent with
\(E^{\texttt{PL}}\).
\end{theorem}

\begin{proof}[Proof of Theorem~\ref{supp:thm:pl-bp-consistency}]
The proof proceeds by restricting the optimization problem to a compact subset, thereby enabling application of Theorem~\ref{thm: BP-SC}.

By Lemma~\ref{supp:lem:pl-compact-containment}, there
exists a compact set \(K\subset\mathrm{SPD}(2)\) such that
\[
    E^{\texttt{PL}}
    \subset
    \mathbb{CP}^{p-1}\times K
\]
and, almost surely for sufficiently large \(n\),
\[
    \widehat E_n^{\texttt{PL}}
    \subset
    \mathbb{CP}^{p-1}\times K.
\]
The proof of that lemma also shows that the population minimum is
attained. Hence
\[
    E^{\texttt{PL}}
    =
    \argmin_{\substack{
        [\kappa]\in\mathbb{CP}^{p-1}\\
        \Sigma\in K
    }}
    F^{\texttt{PL}}([\kappa],\Sigma).
\]
For every \(n\), define the restricted empirical minimizer set by
\[
    \widetilde E_n^{\texttt{PL}}
    :=
    \argmin_{\substack{
        [\kappa]\in\mathbb{CP}^{p-1}\\
        \Sigma\in K
    }}
    F_n^{\texttt{PL}}([\kappa],\Sigma).
\]
Since \(F_n^{\texttt{PL}}\) is continuous on the compact set
\(\mathbb{CP}^{p-1}\times K\), the set
\(\widetilde E_n^{\texttt{PL}}\) is nonempty for every \(n\).
Moreover, by Lemma~\ref{supp:lem:pl-compact-containment}, almost surely
for sufficiently large \(n\),
\[
    \widehat E_n^{\texttt{PL}}
    =
    \widetilde E_n^{\texttt{PL}}.
\]
We now regard the restricted profile-likelihood problem as a
generalized \fre~mean problem with
\[
    T=\mathbb C^p,
    \quad
    M=\mathbb{CP}^{p-1}\times K,
    \quad
    Q=\{q_\ast\}.
\]
We equip \(M\) with the product metric
\[
\begin{aligned}
    &d_M\left(
        ([\kappa],\Sigma),
        ([\widetilde\kappa],\widetilde\Sigma)
    \right):=
    \left\{
        d\left([\kappa],[\widetilde\kappa]\right)^2
        +
        \|\Sigma-\widetilde\Sigma\|_F^2
    \right\}^{1/2}.
\end{aligned}
\]
There is no nontrivial auxiliary parameter, so we set
\(\widehat q_n=q_0=q_\ast\) for every \(n\). We use the profile loss
\(\rho^{\texttt{PL}}\) as the cost function in
Theorem~\ref{thm: BP-SC}, viewing it as a function on
\(T\times M\times Q\) that is constant in the singleton coordinate
\(q_\ast\). To verify that \(\widetilde E_n^{\texttt{PL}}\) is BP
consistent with \(E^{\texttt{PL}}\), we must verify
Conditions~\ref{supp:cond:gfm-auxiliary-consistency},
\ref{supp:cond:gfm-primary-polish},
\ref{supp:cond:gfm-heine-borel},
\ref{supp:cond:gfm-auxiliary-local-polish},
\ref{supp:cond:gfm-cost-regularity}, and
\ref{supp:cond:gfm-local-integrability}.

Condition~\ref{supp:cond:gfm-auxiliary-consistency} is trivially
satisfied because \(\widehat q_n=q_0=q_\ast\) for every \(n\).
Condition~\ref{supp:cond:gfm-primary-polish} and
Condition~\ref{supp:cond:gfm-heine-borel} are trivially satisfied
because $\mathbb{CP}^{p-1}$ and $K$ are both compact metric spaces;
hence, their product space $M$ is compact, making it separable,
complete, and fulfilling the Heine--Borel property.
Condition~\ref{supp:cond:gfm-auxiliary-local-polish} is trivially
satisfied because the auxiliary space is a singleton.

For Condition~\ref{supp:cond:gfm-cost-regularity}, the profile loss
function is given by
\[
    \rho^{\texttt{PL}}(y,[\kappa],\Sigma)
    =
    \rho(y, [\kappa], \Sigma)
    +
    p\log\det(\Sigma).
\]
The proof of Lemma~\ref{supp:lem:plugin-gfm-consistency-conditions}
establishes that \(\rho\) is continuous on
\(\mathbb C^p\times\mathbb{CP}^{p-1}\times\mathrm{SPD}(2)\).
Since \(\Sigma\mapsto p\log\det(\Sigma)\) is continuous,
\(\rho^{\texttt{PL}}\) is continuous, and hence measurable,
on \(T\times M\times Q\).

For Condition~\ref{supp:cond:gfm-local-integrability}, we bound the
local infima and extrema of the cost function. Since $K$ is a compact
subset of $\mathrm{SPD}(2)$, the continuous functions
$\log\det(\Sigma)$ and $ \lambda_{\max}(\Sigma^{-1})$ attain finite global extrema on $K$. Using the profile-loss bound established in
Lemma~\ref{supp:lem:plugin-gfm-consistency-conditions},
\[
    0
    \le
    \rho(Y_{1}, [\kappa], \Sigma)
    \le
    \left(
        \sup_{\Sigma\in K}
         \lambda_{\max}(\Sigma^{-1})
    \right)\|Y_{1}\|^{2}.
\]
Thus, we get
\[
\begin{aligned}
    p\inf_{\Sigma\in K}\log\det(\Sigma)
    &\le
    \rho^{\texttt{PL}}(Y_1,[\kappa],\Sigma)\\
    &\le
    \left(
        \sup_{\Sigma\in K}
         \lambda_{\max}(\Sigma^{-1})
    \right)
    \|Y_{1}\|^{2}
    +
    p\sup_{\Sigma\in K}\log\det(\Sigma)
\end{aligned}
\]
for all $[\kappa]\in\mathbb{CP}^{p-1}$ and $\Sigma\in K$.

Under model~\eqref{model:centered-complex-factor} with
$0<\mathbb{E}[|\phi_1|^2]<\infty$, the upper bound has a finite
expectation. Consequently, the local infimum $\pi_{m,r}(Y_1)$ and
supremum $\Pi_{m,r}(Y_1)$ are bounded by integrable functions,
satisfying Condition~\ref{supp:cond:gfm-local-integrability}.

Since \(M=\mathbb{CP}^{p-1}\times K\) is compact, the restricted
empirical minimizer sets \(\widetilde E_n^{\texttt{PL}}\) are bounded
for every \(n\), so the eventual
boundedness assumption of Theorem~\ref{thm: BP-SC} is satisfied.
We have therefore verified all assumptions of
Theorem~\ref{thm: BP-SC}. It follows that the restricted empirical
minimizer sets \(\widetilde E_n^{\texttt{PL}}\) are BP strongly
consistent with
\(E^{\texttt{PL}}\).

Finally, the restricted and unrestricted empirical minimizer sets
coincide almost surely for all sufficiently large \(n\) by
Lemma~\ref{supp:lem:pl-compact-containment}. Altering
finitely many terms does not affect BP consistency. Therefore,
\(\widehat E_n^{\texttt{PL}}\) is BP strongly consistent with
\(E^{\texttt{PL}}\).
\end{proof}

We now prove the main inconsistency result,
Theorem~\ref{thm: PL strong inconsistency}.

\begin{proof}[Proof of Theorem~\ref{thm: PL strong inconsistency}]
Let \(E^{\texttt{PL}}\) denote the minimizer set of the population
cost function \(F^{\texttt{PL}}\). By
Lemma~\ref{supp:lem:pl-compact-containment},
\(E^{\texttt{PL}}\) is a nonempty subset of the compact set
\(\mathbb{CP}^{p-1}\times K\). Since \(F^{\texttt{PL}}\) is
continuous on this compact set, \(E^{\texttt{PL}}\), as its minimizer
set, is closed. Hence \(E^{\texttt{PL}}\) is compact.

By
Theorem~\ref{supp:thm:pl-true-parameter-not-minimizer},
\[
    ([\kappa^{(0)}],\Sigma^{(0)})
    \notin
    E^{\texttt{PL}}.
\]
Therefore, since \(E^{\texttt{PL}}\) is compact and the true
parameter is a single point outside \(E^{\texttt{PL}}\),
\[
    \delta_0
    :=
    \inf_{y\in E^{\texttt{PL}}}
    d_M\left(
        ([\kappa^{(0)}],\Sigma^{(0)}),
        y
    \right)
    >0.
\]

By Theorem~\ref{supp:thm:pl-bp-consistency},
\(\widehat E_n^{\texttt{PL}}\) is BP strongly consistent with
\(E^{\texttt{PL}}\), so
\[
    \sup_{x\in\widehat E_n^{\texttt{PL}}}
    \inf_{y\in E^{\texttt{PL}}}
    d_M(x,y)
    \longrightarrow0
    \quad\text{almost surely}.
\]
Hence, almost surely for all sufficiently large \(n\),
\[
    \sup_{x\in\widehat E_n^{\texttt{PL}}}
    \inf_{y\in E^{\texttt{PL}}}
    d_M(x,y)
    <
    \frac{\delta_0}{2}.
\]
For every \(x\in\widehat E_n^{\texttt{PL}}\), the triangle
inequality therefore gives
\[
\begin{aligned}
    d_M\left(
        x,
        ([\kappa^{(0)}],\Sigma^{(0)})
    \right)
    &\ge
    \inf_{y\in E^{\texttt{PL}}}
    d_M\left(
        y,
        ([\kappa^{(0)}],\Sigma^{(0)})
    \right)
    -
    \inf_{y\in E^{\texttt{PL}}}d_M(x,y)\\
    &>
    \frac{\delta_0}{2}.
\end{aligned}
\]
Consequently, almost surely for all sufficiently large \(n\),
\[
    \inf_{x\in\widehat E_n^{\texttt{PL}}}
    d_M\left(
        x,
        ([\kappa^{(0)}],\Sigma^{(0)})
    \right)
    \ge
    \frac{\delta_0}{2}
    >0.
\]
Thus the empirical profile-likelihood minimizer set is eventually
bounded away from the true parameter almost surely.
\end{proof}

\section{Proofs for spectrum--wave recovery by the maximum method}
\label{supp:sec:maximum-method}
This section establishes the properties of the maximum method
and proves strong consistency and asymptotic normality of the
maximum-aligned estimators.
We also derive explicit asymptotic covariance matrices for the recovered spectrum and compare their traces for the \texttt{cov} and \texttt{plug-in} estimators.

\subsection{Characterization and equivariance of the maximum method}
\label{supp:sec:maximum-method-characterization}
We characterize the maximizing phases and establish the resulting
orthogonality of the spectrum and wave.
When \(\kappa^\top\kappa\neq0\), we also show that the recovered
components are invariant under changes of the loading representative,
up to a common sign.

\begin{proof}[Proof of Lemma~\ref{lem: max method projection}]
Using $\text{Re}(e^{i\lambda}\kappa) = \frac{1}{2}(e^{i\lambda}\kappa+e^{-i\lambda}\bar{\kappa})$, we can notice that maximizing $\|\operatorname{Re}(e^{i\lambda}\kappa)\|^2$ over $\lambda \in [0, 2\pi)$ is equivalent to maximizing:
\begin{equation}
\label{supp:eq:maximum-method-norm-expansion}
\left\| \frac{e^{i\lambda}\kappa + e^{-i\lambda}\bar{\kappa}}{2} \right\|^2 = \frac{1}{4} \left( e^{i\lambda}\kappa + e^{-i\lambda}\bar{\kappa} \right)^* \left( e^{i\lambda}\kappa + e^{-i\lambda}\bar{\kappa} \right).
\end{equation}
Expanding the inner product yields:
\begin{align*}
\left\| \operatorname{Re}(e^{i\lambda}\kappa) \right\|^2 &= \frac{1}{4} \left( e^{-i\lambda}\kappa^* + e^{i\lambda}\kappa^{\top} \right) \left( e^{i\lambda}\kappa + e^{-i\lambda}\bar{\kappa} \right) \\
&= \frac{1}{4} \left( e^{-i\lambda}e^{i\lambda}\kappa^*\kappa + e^{-i\lambda}e^{-i\lambda}\kappa^*\bar{\kappa} + e^{i\lambda}e^{i\lambda}\kappa^{\top}\kappa + e^{i\lambda}e^{-i\lambda}\kappa^{\top}\bar{\kappa} \right).
\end{align*}
Since $\kappa\in \mathcal{S}^{2p-1}$, we have $\kappa^*\kappa = 1$. Consequently, $\kappa^{\top}\bar{\kappa} = (\kappa^*\kappa)^{\top} = 1$. Furthermore, observing that $\kappa^*\bar{\kappa} = \overline{\kappa^{\top}\kappa}$, the expression simplifies to:
\begin{align*}
\left\| \operatorname{Re}(e^{i\lambda}\kappa) \right\|^2 &= \frac{1}{4} \left( 1 + e^{-2i\lambda}\overline{\kappa^{\top}\kappa} + e^{2i\lambda}\kappa^{\top}\kappa + 1 \right) \\
&= \frac{1}{2} + \frac{1}{4} \left( e^{2i\lambda}\kappa^{\top}\kappa + \overline{e^{2i\lambda}\kappa^{\top}\kappa} \right).
\end{align*}

If $\kappa^{\top}\kappa = 0$, the expression evaluates to the constant $1/2$, meaning all $\lambda \in [0, 2\pi)$ maximize the expression. 
If $\kappa^{\top}\kappa \neq 0$, choose the unique \(\beta\in[0,2\pi)\) satisfying
\[
    e^{i\beta}
    =
    \frac{\kappa^\top\kappa}{|\kappa^\top\kappa|}.
\]
We can write $\kappa^{\top}\kappa = a e^{i\beta}$, where $a = |\kappa^{\top}\kappa| > 0$. Substituting this polar representation into our objective function gives:
\begin{align*}
e^{2i\lambda}\kappa^{\top}\kappa + \overline{e^{2i\lambda}\kappa^{\top}\kappa} &= e^{2i\lambda}(a e^{i\beta}) + \overline{e^{2i\lambda}(a e^{i\beta})} \\
&= a e^{i(2\lambda + \beta)} + a e^{-i(2\lambda + \beta)} \\
& = 2a \cos(2\lambda + \beta).
\end{align*}

Thus, our original maximization problem is equivalent to
\[
\operatorname{arg\,max}_{\lambda \in [0, 2\pi)} \left[ \frac{1}{2} + \frac{a}{2} \cos(2\lambda + \beta) \right] = \operatorname{arg\,max}_{\lambda \in [0, 2\pi)} \cos(2\lambda + \beta).
\]

The expression $\cos(2\lambda + \beta)$ is maximized exactly when $2\lambda + \beta \equiv 0 \pmod{2\pi}$. Therefore, any maximizing phase $\lambda^{\texttt{max}}(\kappa)$ must satisfy:
\[
2\lambda^{\texttt{max}}(\kappa) \equiv -\beta \pmod{2\pi} \implies \lambda^{\texttt{max}}(\kappa) \equiv -\frac{\operatorname{Arg}(\kappa^{\top}\kappa)}{2} \pmod{\pi}.
\]
This confirms that the optimal phase $\lambda^{\texttt{max}}(\kappa)$ is unique up to a shift of $\pi$.

To establish the properties of the resulting components $I^{\texttt{max}}$ and $W^{\texttt{max}}$, we utilize the complex algebraic identity for $e^{i\lambda^{\texttt{max}}}\kappa = I^{\texttt{max}} + iW^{\texttt{max}}$:
\[
(e^{i\lambda^{\texttt{max}}}\kappa)^{\top} (e^{i\lambda^{\texttt{max}}}\kappa) = (I^{\texttt{max}} + iW^{\texttt{max}})^{\top} (I^{\texttt{max}} + iW^{\texttt{max}}) = \|I^{\texttt{max}}\|^2 - \|W^{\texttt{max}}\|^2 + 2i\langle I^{\texttt{max}}, W^{\texttt{max}} \rangle.
\]
Substituting
\[
    2\lambda^{\texttt{max}}
    \equiv
    -\beta
    =
    -\operatorname{Arg}(\kappa^\top\kappa)
    \pmod{2\pi}
\]directly into the left-hand side yields:
\[
e^{2i\lambda^{\texttt{max}}} (\kappa^{\top} \kappa) = e^{-i\beta} (a e^{i\beta}) = a.
\]
By equating the real and imaginary parts of both expressions, we obtain:
\[
\|I^{\texttt{max}}\|^2 - \|W^{\texttt{max}}\|^2 = a > 0 \quad \text{and} \quad 2\langle I^{\texttt{max}}, W^{\texttt{max}} \rangle = 0.
\]
This immediately establishes that $\langle I^{\texttt{max}}, I^{\texttt{max}} \rangle > \langle W^{\texttt{max}}, W^{\texttt{max}} \rangle$ and $\langle I^{\texttt{max}}, W^{\texttt{max}} \rangle = 0$.

Finally, suppose $\kappa'$ is equivalent to $\kappa$ ($\kappa' \sim \kappa$), meaning $\kappa' = e^{i\alpha}\kappa$ for some $\alpha \in \mathbb{R}$. We evaluate the argmax condition for $\kappa'$ by first computing $(\kappa')^{\top} \kappa'$:
\[
(\kappa')^{\top} \kappa' = (e^{i\alpha}\kappa)^{\top} (e^{i\alpha}\kappa) = e^{2i\alpha} (\kappa^{\top}\kappa).
\]
Taking the argument of both sides, we obtain:
\[
\operatorname{Arg}((\kappa')^{\top} \kappa') \equiv \operatorname{Arg}(\kappa^{\top}\kappa) + 2\alpha \pmod{2\pi}.
\]
Recall that any maximizing phase $\lambda^{\texttt{max}}(\kappa')$ must satisfy $2\lambda^{\texttt{max}}(\kappa') + \operatorname{Arg}((\kappa')^{\top} \kappa') \equiv 0 \pmod{2\pi}$. Substituting the derived argument into this condition yields:
\begin{align*}
2\lambda^{\texttt{max}}(\kappa') + \operatorname{Arg}(\kappa^{\top}\kappa) + 2\alpha &\equiv 0 \pmod{2\pi} \\
2\lambda^{\texttt{max}}(\kappa') &\equiv -\operatorname{Arg}(\kappa^{\top}\kappa) - 2\alpha \pmod{2\pi} \\
\lambda^{\texttt{max}}(\kappa') &\equiv -\frac{\operatorname{Arg}(\kappa^{\top}\kappa)}{2} - \alpha \pmod{\pi}.
\end{align*}
Since we previously established that $\lambda^{\texttt{max}}(\kappa) \equiv -\frac{\operatorname{Arg}(\kappa^{\top}\kappa)}{2} \pmod{\pi}$, this directly yields the equivariance property:
\[
\lambda^{\texttt{max}}(\kappa') \equiv \lambda^{\texttt{max}}(\kappa) - \alpha \pmod{\pi}.
\]
Because the optimal phase is congruent modulo $\pi$, we can write $\lambda^{\texttt{max}}(\kappa') = \lambda^{\texttt{max}}(\kappa) - \alpha + k\pi$ for some integer $k$. Consequently, the phase-aligned vector satisfies:
\begin{align*}
e^{i\lambda^{\texttt{max}}(\kappa')}\kappa' &= e^{i(\lambda^{\texttt{max}}(\kappa) - \alpha + k\pi)} (e^{i\alpha}\kappa) \\
&= e^{ik\pi} e^{i\lambda^{\texttt{max}}(\kappa)} e^{-i\alpha} e^{i\alpha} \kappa \\
&= \pm e^{i\lambda^{\texttt{max}}(\kappa)}\kappa.
\end{align*}
This confirms that the resulting components $I^{\texttt{max}}$ and $W^{\texttt{max}}$ are invariant under the equivalence relation up to a global sign, completing the proof.
\end{proof}

\subsection{Strong consistency of the maximum-aligned estimators}
\label{supp:sec:maximum-method-consistency}
We prove strong consistency of the maximum-aligned loading, spectrum, and wave estimators. The argument uses local continuity of the maximum-alignment map and eventual stability of the sign convention.

Throughout this section, fix an arbitrary integer \(K\geq1\).
We use \(s=0\) to denote the \texttt{cov} estimator, so that
\[
    \widehat{[\kappa]}^{[0]}
    :=
    \widehat{[\kappa]}^{\texttt{cov}}.
\]
The same convention applies to the optimally positioned and
maximum-aligned representatives and their spectrum and wave
components.

\begin{proof}[
Proof of
Theorem~\ref{thm:maximum-aligned-strong-consistency}
]

By
Theorems~\ref{thm:cov-estimator-strong-consistency}
and~\ref{thm:iterated-plugin-strong-consistency}, with probability one,
simultaneously for \(s=0,\ldots,K\),
\[
\widehat{[\kappa]}^{[s]}
\longrightarrow
[\kappa^{(0)}].
\]
Since \(K\) is fixed, it suffices to prove that the maximum-alignment
map is continuous at \([\kappa^{(0)}]\).

Let \([\kappa_n]\to[\kappa^{(0)}]\) in
\(\mathbb{CP}^{p-1}\), and choose representatives such that
\[
\kappa_n\longrightarrow\kappa^{(0)}
\quad\text{in }\mathcal S^{2p-1}.
\]
Writing
\(\kappa^{(0)}=I^{(0)}+iW^{(0)}\), the maximum-alignment conditions give
\begin{align*}
(\kappa^{(0)})^\top\kappa^{(0)}
&=
\left\|I^{(0)}\right\|^2
-
\left\|W^{(0)}\right\|^2
+
2i
\left\langle
I^{(0)},W^{(0)}
\right\rangle
\\
&=
\left\|I^{(0)}\right\|^2
-
\left\|W^{(0)}\right\|^2
>
0.
\end{align*}
Consequently,
\[
\kappa_n^\top\kappa_n
\longrightarrow
(\kappa^{(0)})^\top\kappa^{(0)}>0,
\]
so \(\kappa_n^\top\kappa_n\neq0\) for all sufficiently large \(n\).
Moreover,
\[
\frac{
\kappa_n^\top\kappa_n
}{
|\kappa_n^\top\kappa_n|
}
\longrightarrow1,
\]
and hence
\[
\operatorname{Arg}
\left(
\kappa_n^\top\kappa_n
\right)
\longrightarrow0
\pmod{2\pi}.
\]

By Lemma~\ref{lem: max method projection}, the maximizing phase
satisfies
\[
\lambda^{\texttt{max}}(\kappa_n)
\equiv
-\frac{1}{2}
\operatorname{Arg}
\left(
\kappa_n^\top\kappa_n
\right)
\pmod{\pi}.
\]
Choosing the local branch converging to zero gives
\[
\lambda^{\texttt{max}}(\kappa_n)\longrightarrow0.
\]
Therefore,
\[
e^{i\lambda^{\texttt{max}}(\kappa_n)}\kappa_n
\longrightarrow
\kappa^{(0)}.
\]
Since
\([H^\top I^{(0)}]_{j_c}>0\), we have
\[
\left[H^\top\operatorname{Re}
\{e^{i\lambda^{\texttt{max}}(\kappa_n)}\kappa_n\}\right]_{j_c}
\longrightarrow [H^\top I^{(0)}]_{j_c}>0.
\]
Thus, for all sufficiently large \(n\), the sign-selection multiplier
in \eqref{eq:maximum-alignment-map} equals \(1\). The phase is unique
modulo \(\pi\), and the sign convention resolves the corresponding
common-sign ambiguity. Hence \(\kappa_n^{\texttt{max}}\) is uniquely
defined for all sufficiently large \(n\) and
\[
\kappa_n^{\texttt{max}}
\longrightarrow
\kappa^{(0)}.
\]
This proves the continuity of the maximum-alignment map at
\([\kappa^{(0)}]\).

Applying this result to
\(\widehat{[\kappa]}^{[s]}\), \(s=0,\ldots,K\), yields, almost surely,
\[
\widehat\kappa^{[s],\texttt{max}}
\longrightarrow
\kappa^{(0)}
\]
simultaneously over \(s=0,\ldots,K\). Taking real and imaginary parts
gives
\[
\widehat I^{[s],\texttt{max}}
\longrightarrow
I^{(0)},
\quad
\widehat W^{[s],\texttt{max}}
\longrightarrow
W^{(0)},
\]
which proves the result.
\end{proof}

\subsection{Stochastic expansions and asymptotic normality of maximum-aligned loading estimators}
\label{supp:sec:maximum-method-asymptotics}
We compute the Jacobian of the maximum-alignment map and apply
its local Taylor expansion to the optimally positioned loading
estimators. This yields stochastic expansions with \(\mathcal O_p(n^{-1})\) remainders and the corresponding central limit theorems for the
maximum-aligned representatives.

\begin{theorem}[Joint asymptotics of the maximum-aligned spectrum and wave]
\label{supp:thm:maximum-aligned-joint-expansion-clt}
Assume the conditions of
Theorem~\ref{thm:maximum-aligned-strong-consistency}
and \(\mathbb E[|\phi_1|^4]<\infty\).
Fix an integer \(K\geq1\), and write
\[
\widehat\kappa^{[s],\texttt{max}}
=
\widehat I^{[s],\texttt{max}}
+i\widehat W^{[s],\texttt{max}},
\qquad s=0,\ldots,K,
\]
where
\(\widehat\kappa^{[0],\texttt{max}}
=\widehat\kappa^{\texttt{cov,max}}\).

The Jacobian of \(\mathcal A^{\texttt{max}}\) at
\(\kappa^{(0)}\), in grouped real coordinates, is
\begin{equation}
\label{supp:eq:maximum-alignment-joint-jacobian}
A^{\texttt{max}}
:=
\mathrm{Id}_{2p}
-
\frac{1}{d_\kappa}
\begin{pmatrix}
-W^{(0)}\\
I^{(0)}
\end{pmatrix}
\begin{pmatrix}
W^{(0)}\\
I^{(0)}
\end{pmatrix}^{\top}
\in\mathbb R^{2p\times2p}.
\end{equation}
The following stochastic expansions hold simultaneously for
\(s=0,\ldots,K\):
\begin{equation}
\label{supp:eq:maximum-aligned-joint-common-expansion}
\begin{aligned}
\begin{pmatrix}
\widehat I^{[0],\texttt{max}}-I^{(0)}\\
\widehat W^{[0],\texttt{max}}-W^{(0)}
\end{pmatrix}
&=
A^{\texttt{max}}
\widetilde{\operatorname{vec}}
\left(L_{\kappa,n}^{\texttt{cov}}\right)
+\mathcal O_p(n^{-1}),
\\
\begin{pmatrix}
\widehat I^{[s],\texttt{max}}-I^{(0)}\\
\widehat W^{[s],\texttt{max}}-W^{(0)}
\end{pmatrix}
&=
A^{\texttt{max}}
\widetilde{\operatorname{vec}}
\left(L_{\kappa,n}^{\texttt{plug-in}}\right)
+\mathcal O_p(n^{-1}),
\qquad s=1,\ldots,K.
\end{aligned}
\end{equation}
Consequently,
\begin{align}
\sqrt n
\begin{pmatrix}
\widehat I^{[0],\texttt{max}}-I^{(0)}\\
\widehat W^{[0],\texttt{max}}-W^{(0)}
\end{pmatrix}
&\xrightarrow{d}
N_{2p}\left(0,V^{\texttt{cov,max}}\right),
\label{supp:eq:maximum-aligned-joint-cov-clt}
\\
\sqrt n
\begin{pmatrix}
\widehat I^{[s],\texttt{max}}-I^{(0)}\\
\widehat W^{[s],\texttt{max}}-W^{(0)}
\end{pmatrix}
&\xrightarrow{d}
N_{2p}\left(0,V^{\texttt{plug-in,max}}\right),
\qquad s=1,\ldots,K,
\label{supp:eq:maximum-aligned-joint-common-clt}
\end{align}
where
\begin{equation}
\label{supp:eq:maximum-aligned-joint-asymptotic-covariances}
V^{m,\texttt{max}}
:=
A^{\texttt{max}}V^m(A^{\texttt{max}})^\top,
\qquad
m\in\{\texttt{cov},\texttt{plug-in}\}.
\end{equation}
\end{theorem}

\begin{proof}[
Proof of
Theorem~\ref{supp:thm:maximum-aligned-joint-expansion-clt}
]
By Theorem~\ref{thm:maximum-aligned-strong-consistency}, with
probability one,
\[
\widehat\kappa^{[s],\texttt{max}}
\longrightarrow
\kappa^{(0)}
\]
simultaneously for \(s=0,\ldots,K\). Moreover, the consistency of the
optimally aligned representatives gives
\[
\widehat\kappa^{[s],\texttt{opt}}
\longrightarrow
\kappa^{(0)}
\]
simultaneously for \(s=0,\ldots,K\). We work on this probability-one
event.

Since
\[
(\kappa^{(0)})^\top\kappa^{(0)}
=
d_\kappa>0,
\quad
[H^\top I^{(0)}]_{j_c}>0,
\]
there exists \(\delta>0\) such that, whenever
\(\|\kappa-\kappa^{(0)}\|<\delta\),

\(
\kappa^\top\kappa\neq0
\)
and the local branch
\[
\lambda^{\texttt{max}}(\kappa)
=
-\frac{1}{2}
\operatorname{Arg}(\kappa^\top\kappa),
\quad
\lambda^{\texttt{max}}(\kappa^{(0)})=0,
\]
is well-defined and twice continuously differentiable. By decreasing
\(\delta\) if necessary, the central component in
\eqref{eq:maximum-alignment-map} remains positive throughout this
neighborhood, so that \(s^{\texttt{max}}(\kappa)=1\). Hence, locally,
\[
\mathcal A^{\texttt{max}}(\kappa)
=
e^{i\lambda^{\texttt{max}}(\kappa)}\kappa,
\]
and \(\mathcal A^{\texttt{max}}\) is twice continuously differentiable.

We now compute its Jacobian. Write
\[
\kappa=x+iy,
\quad
u
:=
\widetilde{\operatorname{vec}}(\kappa)
=
\begin{pmatrix}
x\\
y
\end{pmatrix}.
\]
In real coordinates,
\[
\widetilde{\operatorname{vec}}
\left(
\mathcal A^{\texttt{max}}(\kappa)
\right)
=
R\left(\lambda^{\texttt{max}}(\kappa)\right)u,
\quad
R(\lambda)
:=
\begin{pmatrix}
\cos(\lambda)\mathrm{Id}_p
&
-\sin(\lambda)\mathrm{Id}_p
\\
\sin(\lambda)\mathrm{Id}_p
&
\cos(\lambda)\mathrm{Id}_p
\end{pmatrix}.
\]
The product and chain rules give
\begin{align*}
\frac{\partial}{
\partial u^\top
}
\left\{
R\left(\lambda^{\texttt{max}}(\kappa)\right)u
\right\}
&=
R\left(\lambda^{\texttt{max}}(\kappa)\right)
+
\left[
R'\left(\lambda^{\texttt{max}}(\kappa)\right)u
\right]
\left(\frac{
\partial\lambda^{\texttt{max}}(\kappa)
}{
\partial u
}\right)^{\top}
.
\end{align*}

It remains to calculate the derivative of the maximizing phase. Since
\[
\kappa^\top\kappa
=
x^\top x-y^\top y+2ix^\top y,
\]
we have locally
\[
\lambda^{\texttt{max}}(\kappa)
=
-\frac{1}{2}
\operatorname{atan2}
\left(
2x^\top y,\,
x^\top x-y^\top y
\right).
\]
Recall that
\[
\frac{\partial}{\partial b}
\operatorname{atan2}(b,a)
=
\frac{a}{a^2+b^2},
\quad
\frac{\partial}{\partial a}
\operatorname{atan2}(b,a)
=
-\frac{b}{a^2+b^2}.
\]
Moreover,
\[
\frac{\partial(2x^\top y)}{\partial x}
=
2y,
\quad
\frac{\partial(x^\top x-y^\top y)}{\partial x}
=
2x,
\]
and
\[
\frac{\partial(2x^\top y)}{\partial y}
=
2x,
\quad
\frac{\partial(x^\top x-y^\top y)}{\partial y}
=
-2y.
\]
Therefore, the chain rule gives
\begin{align*}
\frac{
\partial\lambda^{\texttt{max}}(\kappa)
}{
\partial x
}
&=
-\frac{1}{2}
\Bigg[
\frac{
x^\top x-y^\top y
}{
\left(x^\top x-y^\top y\right)^2
+
4(x^\top y)^2
}
(2y)
-
\frac{
2x^\top y
}{
\left(x^\top x-y^\top y\right)^2
+
4(x^\top y)^2
}
(2x)
\Bigg]
\\
&=
-
\frac{
\left(x^\top x-y^\top y\right)y
-
2(x^\top y)x
}{
\left(x^\top x-y^\top y\right)^2
+
4(x^\top y)^2
},
\end{align*}
and
\begin{align*}
\frac{
\partial\lambda^{\texttt{max}}(\kappa)
}{
\partial y
}
&=
-\frac{1}{2}
\Bigg[
\frac{
x^\top x-y^\top y
}{
\left(x^\top x-y^\top y\right)^2
+
4(x^\top y)^2
}
(2x)
+
\frac{
2x^\top y
}{
\left(x^\top x-y^\top y\right)^2
+
4(x^\top y)^2
}
(2y)
\Bigg]
\\
&=
-
\frac{
\left(x^\top x-y^\top y\right)x
+
2(x^\top y)y
}{
\left(x^\top x-y^\top y\right)^2
+
4(x^\top y)^2
}.
\end{align*}

At \(
x=I^{(0)},
y=W^{(0)}\),
the maximum-alignment conditions give
\[
x^\top y=0,
\quad
x^\top x-y^\top y=d_\kappa.
\]
Hence,
\[
\left.
\frac{
\partial\lambda^{\texttt{max}}(\kappa)
}{
\partial x
}
\right|_{\kappa=\kappa^{(0)}}
=
-\frac{W^{(0)}}{d_\kappa},
\quad
\left.
\frac{
\partial\lambda^{\texttt{max}}(\kappa)
}{
\partial y
}
\right|_{\kappa=\kappa^{(0)}}
=
-\frac{I^{(0)}}{d_\kappa}.
\]
Consequently,
\[
\left.
\frac{
\partial\lambda^{\texttt{max}}(\kappa)
}{
\partial u
}
\right|_{\kappa=\kappa^{(0)}}
=
-\frac{1}{d_\kappa}
\begin{pmatrix}
W^{(0)}
\\
I^{(0)}
\end{pmatrix}.
\]

At the true parameter,
\[
\lambda^{\texttt{max}}(\kappa^{(0)})=0,
\quad
R(0)=\mathrm{Id}_{2p},
\quad \text{and} \quad
R'(0)
\begin{pmatrix}
I^{(0)}
\\
W^{(0)}
\end{pmatrix}
=
\begin{pmatrix}
-W^{(0)}
\\
I^{(0)}
\end{pmatrix}.
\]
Substituting these identities into the product-rule expansion gives
\begin{align*}
\left.
\frac{\partial}{
\partial u^\top
}
\widetilde{\operatorname{vec}}
\left(
\mathcal A^{\texttt{max}}(\kappa)
\right)
\right|_{\kappa=\kappa^{(0)}}
&=
\mathrm{Id}_{2p}
-
\frac{1}{d_\kappa}
\begin{pmatrix}
-W^{(0)}
\\
I^{(0)}
\end{pmatrix}
\begin{pmatrix}
W^{(0)}
\\
I^{(0)}
\end{pmatrix}^{\top}
\\
&=
A^{\texttt{max}}.
\end{align*}

Since \(\mathcal A^{\texttt{max}}\) is twice continuously differentiable near
\(\kappa^{(0)}\) and
\(\mathcal A^{\texttt{max}}(\kappa^{(0)})=\kappa^{(0)}\), its local Taylor
expansion is
\begin{align*}
&
\widetilde{\operatorname{vec}}
\left(
\mathcal A^{\texttt{max}}(\kappa)
\right)
-
\widetilde{\operatorname{vec}}
\left(
\kappa^{(0)}
\right)
=
A^{\texttt{max}}
\left\{
\widetilde{\operatorname{vec}}(\kappa)
-
\widetilde{\operatorname{vec}}
\left(
\kappa^{(0)}
\right)
\right\}
+
\mathcal O
\left(
\left\|
\kappa-\kappa^{(0)}
\right\|^2
\right).
\end{align*}

For \(s=0\),
Theorem~\ref{thm:cov-loading-expansion-clt} gives
\[
\widehat\kappa^{[0],\texttt{opt}}
=
\kappa^{(0)}
+
L_{\kappa,n}^{\texttt{cov}}
+
\mathcal O_p(n^{-1}),
\]
whereas, simultaneously for \(s=1,\ldots,K\),
Theorem~\ref{thm:iterated-plugin-loading-expansion-clt} gives
\[
\widehat\kappa^{[s],\texttt{opt}}
=
\kappa^{(0)}
+
L_{\kappa,n}^{\texttt{plug-in}}
+
\mathcal O_p(n^{-1}).
\]
Since \(L_{\kappa,n}^{\texttt{cov}}\) and \(L_{\kappa,n}^{\texttt{plug-in}}\) are \(\mathcal O_p(n^{-1/2})\), substituting these expansions directly into the local Taylor expansion of \(\mathcal A^{\texttt{max}}\) gives
\begin{align*}
&
\widetilde{\operatorname{vec}}
\left(
\widehat\kappa^{[0],\texttt{max}}
\right)
-
\widetilde{\operatorname{vec}}
\left(
\kappa^{(0)}
\right)
=
A^{\texttt{max}}
\widetilde{\operatorname{vec}}
\left(
L_{\kappa,n}^{\texttt{cov}}
\right)
+
\mathcal O_p(n^{-1}),
\end{align*}
and, simultaneously for \(s=1,\ldots,K\),
\begin{align*}
&
\widetilde{\operatorname{vec}}
\left(
\widehat\kappa^{[s],\texttt{max}}
\right)
-
\widetilde{\operatorname{vec}}
\left(
\kappa^{(0)}
\right)
=
A^{\texttt{max}}
\widetilde{\operatorname{vec}}
\left(
L_{\kappa,n}^{\texttt{plug-in}}
\right)
+
\mathcal O_p(n^{-1}).
\end{align*}
This proves
\eqref{supp:eq:maximum-aligned-joint-common-expansion}.

Finally, the central limit theorems in
Theorems~\ref{thm:cov-loading-expansion-clt}
and~\ref{thm:iterated-plugin-loading-expansion-clt}, together with the
preceding expansions and Slutsky's theorem, yield
\eqref{supp:eq:maximum-aligned-joint-cov-clt} and \eqref{supp:eq:maximum-aligned-joint-common-clt}, with
\[
V^{\texttt{cov,max}}
=
A^{\texttt{max}}
V^{\texttt{cov}}
(A^{\texttt{max}})^{\top},
\quad
V^{\texttt{plug-in,max}}
=
A^{\texttt{max}}
V^{\texttt{plug-in}}
(A^{\texttt{max}})^{\top}.
\]
This completes the proof.
\end{proof}

\begin{proof}[
Proof of Theorem~\ref{thm:maximum-aligned-loading-expansion-clt}
]
By Theorem~\ref{supp:thm:maximum-aligned-joint-expansion-clt},
the Jacobian of the spectrum map is given by the first \(p\) rows
of \(A^{\texttt{max}}\). These rows are
\[
A_I^{\texttt{max}}
=
\begin{pmatrix}
\mathrm{Id}_p+
\dfrac{W^{(0)}(W^{(0)})^\top}{d_\kappa}
&
\dfrac{W^{(0)}(I^{(0)})^\top}{d_\kappa}
\end{pmatrix},
\]
which proves \eqref{eq:maximum-alignment-jacobian}.

Taking the first \(p\) components in
\eqref{supp:eq:maximum-aligned-joint-common-expansion}
gives \eqref{eq:maximum-aligned-common-expansion}.
Applying the same coordinate projection to
\eqref{supp:eq:maximum-aligned-joint-cov-clt}
and
\eqref{supp:eq:maximum-aligned-joint-common-clt}
gives \eqref{eq:maximum-aligned-cov-clt}
and \eqref{eq:maximum-aligned-common-clt}.
The corresponding covariance matrices are the upper-left
\(p\times p\) blocks of \(V^{m,\texttt{max}}\), namely
\[
V_I^{m,\texttt{max}}
=
A_I^{\texttt{max}}V^m(A_I^{\texttt{max}})^\top,
\qquad
m\in\{\texttt{cov},\texttt{plug-in}\}.
\]
This proves \eqref{eq:maximum-aligned-asymptotic-covariances}.
\end{proof}

\subsection{Explicit form of the maximum-aligned \texttt{cov}-spectrum covariance}
\label{supp:subsec:cov-maximum-spectrum-covariance}
We derive explicit expressions for
\(V_I^{\texttt{cov,max}}\) and its trace.

\begin{proposition}[Maximum-aligned \texttt{cov} spectrum covariance]
\label{supp:prop:cov-maximum-spectrum-covariance}
Under the conditions of
Theorem~\ref{thm:maximum-aligned-loading-expansion-clt}, let
\[
r
:=
\|I^{(0)}\|^2
=
\frac{1+d_\kappa}{2},
\quad
1-r
=
\|W^{(0)}\|^2.
\]
Then
\[
\sqrt n
\left(
\widehat I^{[0],\texttt{max}}-I^{(0)}
\right)
\xrightarrow{d}
\mathcal N_p
\left(
0,V_I^{\texttt{cov,max}}
\right),
\]
where
\begin{align}
V_I^{\texttt{cov,max}}
&=
\frac{\operatorname{Tr}(\Sigma^{(0)})}
{2\mathbb E[|\phi_1|^2]}
\Bigg[
(1+d_\Sigma\rho_\phi\lambda_\phi)\Pi_{\perp,0}
+
\frac{1-r}{r}
(1-d_\kappa d_\Sigma\rho_\phi\lambda_\phi)
I^{(0)}(I^{(0)})^\top
\nonumber\\
&\hspace{31mm}
+
\frac{r}{(1-r)d_\kappa^2}
(1+d_\kappa d_\Sigma\rho_\phi\lambda_\phi)
W^{(0)}(W^{(0)})^\top
\Bigg]
\nonumber\\
&\quad-
\frac{
\operatorname{Im}
\left\{
\gamma\mathbb E[\overline{\phi_1}^{\,2}]
\right\}
}{
2\{\mathbb E[|\phi_1|^2]\}^2
}
\left\{
I^{(0)}(W^{(0)})^\top
+
W^{(0)}(I^{(0)})^\top
\right\}
\nonumber\\
&\quad+\frac{1}{2}
\left\{
\frac{\operatorname{Tr}(\Sigma^{(0)})}
{\mathbb E[|\phi_1|^2]}
\right\}^{2}
\Bigg[
(1+d_\kappa d_\Sigma^2)\Pi_{\perp,0}
+
\frac{1-r}{r}
\{1+d_\Sigma^2(1-2d_\kappa^2)\}
I^{(0)}(I^{(0)})^\top
\nonumber\\
&\hspace{31mm}
+
\frac{r}{(1-r)d_\kappa^2}
(1+d_\Sigma^2)
W^{(0)}(W^{(0)})^\top
\Bigg].
\label{supp:eq:cov-maximum-spectrum-covariance}
\end{align}
Consequently,
\begin{align}
&\operatorname{Tr}
\left(
    V_I^{\texttt{cov,max}}
\right)\nonumber\\
&=
\frac{
    \operatorname{Tr}(\Sigma^{(0)})
}{
    \mathbb E[|\phi_1|^2]
}
\Bigg[
\frac{p-2}{2}
(1+d_\Sigma\rho_\phi\lambda_\phi)
+
\frac{1-d_\kappa}{4}
(1-d_\kappa d_\Sigma\rho_\phi\lambda_\phi)
+
\frac{1+d_\kappa}{4d_\kappa^2}
(1+d_\kappa d_\Sigma\rho_\phi\lambda_\phi)
\Bigg]
\nonumber\\
&+
\left\{
\frac{
    \operatorname{Tr}(\Sigma^{(0)})
}{
    \mathbb E[|\phi_1|^2]
}
\right\}^{\!2}
\Bigg[
\frac{p-2}{2}
(1+d_\kappa d_\Sigma^2)
+
\frac{1-d_\kappa}{4}
\left\{
    1+d_\Sigma^2(1-2d_\kappa^2)
\right\}
+
\frac{1+d_\kappa}{4d_\kappa^2}
(1+d_\Sigma^2)
\Bigg].
\label{supp:eq:trace-cov-maximum-spectrum-covariance}
\end{align}
\end{proposition}

\begin{proof}[Proof of Proposition~\ref{supp:prop:cov-maximum-spectrum-covariance}]
The asserted central limit theorem follows by Theorem~\ref{thm:maximum-aligned-loading-expansion-clt}. Its asymptotic
covariance is
\begin{equation}
\label{supp:eq:cov-maximum-spectrum-sandwich}
V_I^{\texttt{cov,max}}
=
A_{I}^{\texttt{max}}
V^{\texttt{cov}}
(A_{I}^{\texttt{max}})^\top.
\end{equation}
We have
\[
A_{I}^{\texttt{max}}
=
\begin{pmatrix}
F&G
\end{pmatrix},
\quad
F
:=
\mathrm{Id}_p
+
\frac{W^{(0)}(W^{(0)})^\top}{d_\kappa},
\quad
G
:=
\frac{W^{(0)}(I^{(0)})^\top}{d_\kappa}.
\]
Substituting
\eqref{supp:eq:explicit-cov-loading-asymptotic-covariance}
into
\eqref{supp:eq:cov-maximum-spectrum-sandwich}
gives
\begin{align}
2\{\mathbb E[|\phi_1|^2]\}^2
V_I^{\texttt{cov,max}}
&=
F\operatorname{Re}(\Gamma_H+\Gamma_T)F^\top
+
F\operatorname{Im}(-\Gamma_H+\Gamma_T)G^\top
\nonumber\\
&\quad+
G\operatorname{Im}(\Gamma_H+\Gamma_T)F^\top
+
G\operatorname{Re}(\Gamma_H-\Gamma_T)G^\top.
\label{supp:eq:cov-maximum-spectrum-block-product}
\end{align}

Under the maximum-alignment convention,
\[
\kappa^{(0)}
=
I^{(0)}+iW^{(0)},
\quad
(I^{(0)})^\top W^{(0)}
=
0,
\quad
(\kappa^{(0)})^\top\kappa^{(0)}
=
d_\kappa.
\]
Consequently,
\[
P_{\perp,0}
=
\mathrm{Id}_p
-
I^{(0)}(I^{(0)})^\top
-
W^{(0)}(W^{(0)})^\top
+
i
\left\{
I^{(0)}(W^{(0)})^\top
-
W^{(0)}(I^{(0)})^\top
\right\},
\]
and
\[
P_{\perp,0}\overline{\kappa^{(0)}}
=
\overline{\kappa^{(0)}}
-
d_\kappa\kappa^{(0)}
=
(1-d_\kappa)I^{(0)}
-
i(1+d_\kappa)W^{(0)}.
\]
Substituting these identities and
\eqref{supp:eq:cov-loading-Gamma-H}--\eqref{supp:eq:cov-loading-Gamma-T}
into
\eqref{supp:eq:cov-maximum-spectrum-block-product},
and collecting the four matrix components, yields
\begin{align*}
V_I^{\texttt{cov,max}}
&=
\frac{1}{2\{\mathbb E[|\phi_1|^2]\}^2}
\Bigg[
\Bigg\{
\mathbb E[|\phi_1|^2]\tau^2
+
\tau^4
+
\operatorname{Re}
\left\{
\gamma\mathbb E[\overline{\phi_1}^{\,2}]
\right\}
+
|\gamma|^2d_\kappa
\Bigg\}
\Pi_{\perp,0}
\\
&\quad+
\frac{1-r}{r}
\Bigg\{
\mathbb E[|\phi_1|^2]\tau^2
+
\tau^4
+
|\gamma|^2(1-2d_\kappa^2)
-
d_\kappa
\operatorname{Re}
\left\{
\gamma\mathbb E[\overline{\phi_1}^{\,2}]
\right\}
\Bigg\}
I^{(0)}(I^{(0)})^\top
\\
&\quad-
\operatorname{Im}
\left\{
\gamma\mathbb E[\overline{\phi_1}^{\,2}]
\right\}
\left\{
I^{(0)}(W^{(0)})^\top
+
W^{(0)}(I^{(0)})^\top
\right\}
\\
&\quad+
\frac{r}{(1-r)d_\kappa^2}
\Bigg\{
\mathbb E[|\phi_1|^2]\tau^2
+
\tau^4
+
|\gamma|^2
+
d_\kappa
\operatorname{Re}
\left\{
\gamma\mathbb E[\overline{\phi_1}^{\,2}]
\right\}
\Bigg\}
W^{(0)}(W^{(0)})^\top
\Bigg].
\end{align*}
Here we used
\begin{align*}
&
|\gamma|^2(1-d_\kappa^2)
+
d_\kappa
\operatorname{Re}
\left\{
\gamma\mathbb E[\overline{\phi_1}^{\,2}]
+
|\gamma|^2d_\kappa
\right\}
=
|\gamma|^2
+
d_\kappa
\operatorname{Re}
\left\{
\gamma\mathbb E[\overline{\phi_1}^{\,2}]
\right\},
\\
&
|\gamma|^2(1-d_\kappa^2)
-
d_\kappa
\operatorname{Re}
\left\{
\gamma\mathbb E[\overline{\phi_1}^{\,2}]
+
|\gamma|^2d_\kappa
\right\}
=
|\gamma|^2(1-2d_\kappa^2)
-
d_\kappa
\operatorname{Re}
\left\{
\gamma\mathbb E[\overline{\phi_1}^{\,2}]
\right\},
\end{align*}
as well as
\[
\operatorname{Im}
\left\{
\gamma\mathbb E[\overline{\phi_1}^{\,2}]
+
|\gamma|^2d_\kappa
\right\}
=
\operatorname{Im}
\left\{
\gamma\mathbb E[\overline{\phi_1}^{\,2}]
\right\}.
\]

We now rewrite this covariance matrix in terms of the dimensionless model
parameters. By \eqref{eq:def-d-sigma},
\[
\tau^2
=
\operatorname{Tr}(\Sigma^{(0)}),
\quad
|\gamma|^2
=
d_\Sigma^2
\{\operatorname{Tr}(\Sigma^{(0)})\}^2.
\]
Moreover, by
\eqref{eq:def-rho-phi} and
\eqref{eq:def-lambda-phi},
\[
\operatorname{Re}
\left\{
\gamma\mathbb E[\overline{\phi_1}^{\,2}]
\right\}
=
d_\Sigma\rho_\phi\lambda_\phi
\operatorname{Tr}(\Sigma^{(0)})
\mathbb E[|\phi_1|^2].
\]
Substitution of these three identities into the preceding covariance
matrix, followed by grouping the terms proportional to
\[
\frac{\operatorname{Tr}(\Sigma^{(0)})}
{\mathbb E[|\phi_1|^2]}
\quad\text{and}\quad
\left\{
\frac{\operatorname{Tr}(\Sigma^{(0)})}
{\mathbb E[|\phi_1|^2]}
\right\}^{\!2},
\]
gives
\eqref{supp:eq:cov-maximum-spectrum-covariance}.

It remains to calculate its trace. Since
\[
\operatorname{Tr}(\Pi_{\perp,0})=p-2,
\quad
\operatorname{Tr}
\left\{
I^{(0)}(I^{(0)})^\top
\right\}
=r,
\quad
\operatorname{Tr}
\left\{
W^{(0)}(W^{(0)})^\top
\right\}
=1-r,
\]
and
\[
\operatorname{Tr}
\left[
I^{(0)}(W^{(0)})^\top
+
W^{(0)}(I^{(0)})^\top
\right]
=
2(I^{(0)})^\top W^{(0)}
=
0,
\]
taking the trace of
\eqref{supp:eq:cov-maximum-spectrum-covariance}
gives
\begin{align*}
&
\operatorname{Tr}
\left(
    V_I^{\texttt{cov,max}}
\right)
\\
&=
\frac{
    \operatorname{Tr}(\Sigma^{(0)})
}{
    \mathbb E[|\phi_1|^2]
}
\Bigg\{
\frac{p-2}{2}
(1+d_\Sigma\rho_\phi\lambda_\phi)
+
\frac{1-r}{2}
(1-d_\kappa d_\Sigma\rho_\phi\lambda_\phi)
+
\frac{r}{2d_\kappa^2}
(1+d_\kappa d_\Sigma\rho_\phi\lambda_\phi)
\Bigg\}
\\
&\quad+
\left\{
\frac{
    \operatorname{Tr}(\Sigma^{(0)})
}{
    \mathbb E[|\phi_1|^2]
}
\right\}^{\!2}
\Bigg\{
\frac{p-2}{2}
(1+d_\kappa d_\Sigma^2)
+
\frac{1-r}{2}
\left\{
    1+d_\Sigma^2(1-2d_\kappa^2)
\right\}
+
\frac{r}{2d_\kappa^2}
(1+d_\Sigma^2)
\Bigg\}.
\end{align*}
Finally, substituting
\(
r=(1+d_\kappa)/2
\)
gives
\eqref{supp:eq:trace-cov-maximum-spectrum-covariance}.
\end{proof}

\subsection{Explicit form of the maximum-aligned \texttt{plug-in} spectrum covariance}
\label{supp:subsec:plugin-maximum-spectrum-covariance}
We derive the asymptotic covariance and its trace for the maximum-aligned \texttt{plug-in} spectrum estimator. We first determine how  $A_{I}^{\texttt{max}}$ acts on the canonical coordinates used in Lemma~\ref{supp:lem:plugin-asymptotic-covariance-block-decomposition}.

\begin{lemma}[Maximum-alignment action in canonical coordinates]
\label{supp:lem:plugin-maximum-spectrum-coordinate-action}
Let \(u,v,w_1,\ldots,w_{p-2}\), \(\mathscr S_p\), and \(\mathsf Q_0\) be the
canonical coordinates introduced in
Subsubsection~\ref{supp:sec:plugin-canonical-loading-coordinates}. Then
\begin{align}
A_{I}^{\texttt{max}}\mathscr S_p^\top \mathsf Q_0
=
\Bigg(
I^{(0)},\,0_p
\ \Big|\
\sqrt{1-r}\,u,\,
\frac{\sqrt r}{d_\kappa}v
\Big|\
w_1,\,0_p
\ \Big|\
\cdots
\ \Big|\
w_{p-2},\,0_p
\Bigg).
\label{supp:eq:plugin-maximum-spectrum-coordinate-action}
\end{align}
\end{lemma}

\begin{proof} In the proof of Proposition~\ref{supp:prop:cov-maximum-spectrum-covariance}, we have shown that
\[
A_{I}^{\texttt{max}}
=
\begin{pmatrix}
F&G
\end{pmatrix},
\quad
F
=
\mathrm{Id}_p+
\frac{W^{(0)}(W^{(0)})^\top}{d_\kappa},
\quad
G
=
\frac{W^{(0)}(I^{(0)})^\top}{d_\kappa}.
\]
Moreover,
\[
\mathscr S_p^\top M(\kappa^{(0)})
=
\begin{pmatrix}
I^{(0)}&-W^{(0)}\\
W^{(0)}&I^{(0)}
\end{pmatrix},
\quad
\mathscr S_p^\top N_0
=
\begin{pmatrix}
\sqrt{1-r}\,u&\sqrt r\,v\\
-\sqrt r\,v&\sqrt{1-r}\,u
\end{pmatrix},
\]
and, for \(l=1,\ldots,p-2\),
\[
\mathscr S_p^\top N_l
=
\begin{pmatrix}
w_l&0_p\\
0_p&w_l
\end{pmatrix}.
\]
Using
\[
I^{(0)}=\sqrt r\,u,
\quad
W^{(0)}=\sqrt{1-r}\,v,
\quad
u^\top v=u^\top w_l=v^\top w_l=0,
\]
direct multiplication gives
\[
A_{I}^{\texttt{max}}\mathscr S_p^\top M(\kappa^{(0)})
=
\begin{pmatrix}
I^{(0)}&0_p
\end{pmatrix},
\]
and
\[
A_{I}^{\texttt{max}}\mathscr S_p^\top N_0
=
\begin{pmatrix}
\sqrt{1-r}\,u&
\dfrac{\sqrt r}{d_\kappa}v
\end{pmatrix},
\quad
A_{I}^{\texttt{max}}\mathscr S_p^\top N_l
=
\begin{pmatrix}
w_l&0_p
\end{pmatrix}.
\]
Combining these block products proves
\eqref{supp:eq:plugin-maximum-spectrum-coordinate-action}.
\end{proof}

For \(e_1=(1,0)^\top\) and \(e_2=(0,1)^\top\), denote the two nonzero
covariance blocks in
\eqref{supp:eq:plugin-asymptotic-covariance-block-decomposition} by
\begin{equation}
\label{supp:eq:plugin-maximum-spectrum-blocks}
\begin{aligned}
B_S^{\texttt{plug-in}}
&:=
A_S^{-1}
+
A_S^{-1}G_SA_S^{-1}
+
\frac{1}{p-2}
\frac{
16d_\kappa^2(1-d_\kappa^2)
d_\Sigma^2(1-d_\Sigma^2)^2
}{
(1-d_\kappa^2d_\Sigma^2)^4
}
A_S^{-1}e_1e_1^\top A_S^{-1},
\\
B_\perp^{\texttt{plug-in}}
&:=
A_{(\Sigma^{(0)})^{-1}}^{-1}
+
A_{(\Sigma^{(0)})^{-1}}^{-1}
G_{(\Sigma^{(0)})^{-1}}
A_{(\Sigma^{(0)})^{-1}}^{-1}.
\end{aligned}
\end{equation}

\begin{proposition}[Maximum-aligned \texttt{plug-in} spectrum covariance]
\label{supp:prop:plugin-maximum-spectrum-covariance}
Under the conditions of
Theorem~\ref{thm:maximum-aligned-loading-expansion-clt}, for every
\(s=1,\ldots,K\),
\[
\sqrt n
\left(
\widehat I^{[s],\texttt{max}}-I^{(0)}
\right)
\xrightarrow{d}
\mathcal N_p
\left(
0,V_I^{\texttt{plug-in,max}}
\right),
\]
where
\begin{align}
V_I^{\texttt{plug-in,max}}
&=
\left(
e_1^\top B_\perp^{\texttt{plug-in}}e_1
\right)
\Pi_{\perp,0}
+
\frac{1-r}{r}
\left(
e_1^\top B_S^{\texttt{plug-in}}e_1
\right)
I^{(0)}(I^{(0)})^\top
\nonumber\\
&\quad+
\frac{1}{d_\kappa}
\left(
e_1^\top B_S^{\texttt{plug-in}}e_2
\right)
\left\{
I^{(0)}(W^{(0)})^\top
+
W^{(0)}(I^{(0)})^\top
\right\}
\nonumber\\
&\quad+
\frac{r}{(1-r)d_\kappa^2}
\left(
e_2^\top B_S^{\texttt{plug-in}}e_2
\right)
W^{(0)}(W^{(0)})^\top.
\label{supp:eq:plugin-maximum-spectrum-covariance}
\end{align}
Consequently,
\begin{align}
&
\operatorname{Tr}
\left(
    V_I^{\texttt{plug-in,max}}
\right)
\nonumber\\
&=
\frac{
    \operatorname{Tr}(\Sigma^{(0)})
}{
    \mathbb E[|\phi_1|^2]
}
\Bigg[
\frac{
    (p-2)(1-d_\Sigma^2)
    (1+d_\Sigma\rho_\phi\lambda_\phi)
}{
    2(1-d_\Sigma^2\rho_\phi^2)
}
\nonumber\\
&\quad+
\frac{
    (1-d_\kappa)
    (1-d_\kappa^2d_\Sigma^2)
    (1-d_\kappa d_\Sigma\rho_\phi\lambda_\phi)
}{
    4(1-d_\kappa^2d_\Sigma^2\rho_\phi^2)
}
\nonumber\\
&\quad+
\frac{
    (1+d_\kappa)
    (1-d_\kappa^2d_\Sigma^2)
    (1+d_\kappa d_\Sigma\rho_\phi\lambda_\phi)
}{
    4d_\kappa^2
    (1-d_\kappa^2d_\Sigma^2\rho_\phi^2)
}
\Bigg]
\nonumber\\
&\quad+
\left\{
\frac{
    \operatorname{Tr}(\Sigma^{(0)})
}{
    \mathbb E[|\phi_1|^2]
}
\right\}^{\!2}
\Bigg[
\frac{
    (p-2)(1-d_\Sigma^2)^2
}{
    2(1-d_\kappa^2d_\Sigma^2)
    (1-d_\Sigma^2\rho_\phi^2)^2
}
\nonumber\\
&\quad\quad\times
\Bigl\{
    (1-d_\kappa d_\Sigma^2)
    (1+d_\Sigma\rho_\phi\lambda_\phi)^2
    +
    (1+d_\kappa d_\Sigma^2)
    d_\Sigma^2\rho_\phi^2
    (1-\lambda_\phi^2)
\Bigr\}
\nonumber\\
&\quad+
\frac{
    (1-d_\kappa)(1-d_\Sigma^2)
}{
    4(1-d_\kappa^2d_\Sigma^2\rho_\phi^2)^2
}
\nonumber\\
&\quad\quad\times
\Bigl\{
    (1+d_\kappa^2d_\Sigma^2)
    (1-d_\kappa d_\Sigma\rho_\phi\lambda_\phi)^2
    +
    (1-d_\kappa^2d_\Sigma^2)
    d_\kappa^2d_\Sigma^2\rho_\phi^2
    (1-\lambda_\phi^2)
\Bigr\}
\nonumber\\
&\quad+
\frac{
    (1+d_\kappa)(1-d_\Sigma^2)
}{
    4d_\kappa^2
    (1-d_\kappa^2d_\Sigma^2\rho_\phi^2)^2
}
\nonumber\\
&\quad\quad\times
\Bigl\{
    (1-d_\kappa^2d_\Sigma^2)
    (1+d_\kappa d_\Sigma\rho_\phi\lambda_\phi)^2
    +
    (1+d_\kappa^2d_\Sigma^2)
    d_\kappa^2d_\Sigma^2\rho_\phi^2
    (1-\lambda_\phi^2)
\Bigr\}
\nonumber\\
&\quad+
\frac{2}{p-2}
\frac{
    d_\kappa^2(1-d_\kappa^2)
    d_\Sigma^2(1-d_\Sigma^2)^2
}{
    (1-d_\kappa^2d_\Sigma^2)^2
    (1-d_\kappa^2d_\Sigma^2\rho_\phi^2)^2
}
\nonumber\\
&\quad\quad\times
\Bigl\{
    (1-d_\kappa)
    (1-d_\kappa d_\Sigma\rho_\phi\lambda_\phi)^2
    +
    (1+d_\kappa)
    d_\Sigma^2\rho_\phi^2
    (1-\lambda_\phi^2)
\Bigr\}
\Bigg].
\label{supp:eq:trace-plugin-maximum-spectrum-covariance}
\end{align}
\end{proposition}

\begin{proof}
Applying Theorem~\ref{thm:maximum-aligned-loading-expansion-clt} gives
\[
\sqrt n
\left(
\widehat I^{[s],\texttt{max}}-I^{(0)}
\right)
\xrightarrow{d}
\mathcal N_p
\left(
0,V_I^{\texttt{plug-in,max}}
\right),
\]
where
\[
V_I^{\texttt{plug-in,max}}
=
A_{I}^{\texttt{max}}
V^{\texttt{plug-in}}
(A_{I}^{\texttt{max}})^\top .
\]
By
Lemma~\ref{supp:lem:plugin-asymptotic-covariance-block-decomposition},
\[
V^{\texttt{plug-in}}
=
\mathscr S_p^\top \mathsf Q_0
\operatorname{diag}
\left(
0_2,\,
B_S^{\texttt{plug-in}},\,
\mathrm{Id}_{p-2}\otimes B_\perp^{\texttt{plug-in}}
\right)
\mathsf Q_{0}^\top\mathscr S_p.
\]
It follows that
\begin{align*}
V_I^{\texttt{plug-in,max}}
&=
\left(
A_{I}^{\texttt{max}}\mathscr S_p^\top \mathsf Q_0
\right)
\operatorname{diag}
\left(
0_2,\,
B_S^{\texttt{plug-in}},\,
\mathrm{Id}_{p-2}\otimes B_\perp^{\texttt{plug-in}}
\right)
\left(
A_{I}^{\texttt{max}}\mathscr S_p^\top \mathsf Q_0
\right)^\top.
\end{align*}

The first \(2\times2\) block makes no contribution. By
Lemma~\ref{supp:lem:plugin-maximum-spectrum-coordinate-action}, the
\(N_0\)-block contributes
\[
\begin{pmatrix}
\sqrt{1-r}\,u&
\dfrac{\sqrt r}{d_\kappa}v
\end{pmatrix}
B_S^{\texttt{plug-in}}
\begin{pmatrix}
\sqrt{1-r}\,u&
\dfrac{\sqrt r}{d_\kappa}v
\end{pmatrix}^{\!\top},
\]
whereas each \(N_l\)-block contributes
\[
\left(
e_1^\top B_\perp^{\texttt{plug-in}}e_1
\right)
w_lw_l^\top.
\]
Since
\[
u=\frac{I^{(0)}}{\sqrt r},
\quad
v=\frac{W^{(0)}}{\sqrt{1-r}},
\quad
\sum_{l=1}^{p-2}w_lw_l^\top
=
\Pi_{\perp,0},
\]
expanding these block contributions gives
\eqref{supp:eq:plugin-maximum-spectrum-covariance}.

We next calculate its trace. Because
\[
\operatorname{Tr}(\Pi_{\perp,0})=p-2,
\quad
\left\|I^{(0)}\right\|^2=r,
\quad
\left\|W^{(0)}\right\|^2=1-r,
\]
and
\[
\operatorname{Tr}
\left\{
I^{(0)}(W^{(0)})^\top
+
W^{(0)}(I^{(0)})^\top
\right\}
=
2(I^{(0)})^\top W^{(0)}
=
0,
\]
we obtain
\begin{align}
\operatorname{Tr}
\left(
V_I^{\texttt{plug-in,max}}
\right)
&=
(p-2)e_1^\top B_\perp^{\texttt{plug-in}}e_1
+
(1-r)e_1^\top B_S^{\texttt{plug-in}}e_1
+
\frac{r}{d_\kappa^2}
e_2^\top B_S^{\texttt{plug-in}}e_2
\nonumber\\
&=
(p-2)e_1^\top B_\perp^{\texttt{plug-in}}e_1
+
\frac{1-d_\kappa}{2}
e_1^\top B_S^{\texttt{plug-in}}e_1
+
\frac{1+d_\kappa}{2d_\kappa^2}
e_2^\top B_S^{\texttt{plug-in}}e_2.
\label{supp:eq:plugin-maximum-spectrum-trace-block-form}
\end{align}

It remains to eliminate the \(B\)-block notation from
\eqref{supp:eq:plugin-maximum-spectrum-trace-block-form}. Recall that
\begin{align*}
A_S
&=
\frac{
2\mathbb E[|\phi_1|^2]
}{
\operatorname{Tr}(\Sigma^{(0)})
(1-d_\kappa^2d_\Sigma^2)
}
\left(
\mathrm{Id}_2+
d_\kappa d_\Sigma\rho_\phi Q_\phi
\right),
\\
G_S
&=
\frac{
2(1-d_\Sigma^2)
}{
(1-d_\kappa^2d_\Sigma^2)^2
}
\left(
\mathrm{Id}_2+
d_\kappa^2d_\Sigma^2 Z
\right),
\end{align*}
and
\begin{align*}
A_{(\Sigma^{(0)})^{-1}}
&=
\frac{
2\mathbb E[|\phi_1|^2]
}{
\operatorname{Tr}(\Sigma^{(0)})
(1-d_\Sigma^2)
}
\left(
\mathrm{Id}_2-
d_\Sigma\rho_\phi Q_\phi
\right),
\\
G_{(\Sigma^{(0)})^{-1}}
&=
\frac{2}{
1-d_\kappa^2d_\Sigma^2
}
\left(
\mathrm{Id}_2-
d_\kappa d_\Sigma^2Z
\right).
\end{align*}
Therefore, Lemma~\ref{supp:lem:selected-entries-two-dimensional-block}
applies to the \(S\)-block with
\[
a
=
\frac{
2\mathbb E[|\phi_1|^2]
}{
\operatorname{Tr}(\Sigma^{(0)})
(1-d_\kappa^2d_\Sigma^2)
},
\quad
t=d_\kappa d_\Sigma\rho_\phi,
\quad
g=
\frac{
2(1-d_\Sigma^2)
}{
(1-d_\kappa^2d_\Sigma^2)^2
},
\quad
u=d_\kappa^2d_\Sigma^2,
\]
and to the perpendicular block with
\[
a
=
\frac{
2\mathbb E[|\phi_1|^2]
}{
\operatorname{Tr}(\Sigma^{(0)})
(1-d_\Sigma^2)
},
\quad
t=-d_\Sigma\rho_\phi,
\quad
g=
\frac{2}{
1-d_\kappa^2d_\Sigma^2
},
\quad
u=-d_\kappa d_\Sigma^2.
\]
In both applications,
Lemma~\ref{supp:lem:selected-entries-two-dimensional-block}
is applied with \(Q_\phi\) and \(\lambda_\phi\) in place of its
generic matrix \(Q\) and scalar \(\lambda\), respectively.

The \(A^{-1}\) terms in
Lemma~\ref{supp:lem:selected-entries-two-dimensional-block}
give the terms proportional to
\[
\frac{
\operatorname{Tr}(\Sigma^{(0)})
}{
2\mathbb E[|\phi_1|^2]
}
\]
in
\eqref{supp:eq:trace-plugin-maximum-spectrum-covariance}.
Similarly, the \(A^{-1}GA^{-1}\) terms give the first three terms
proportional to
\[
\frac{
\{\operatorname{Tr}(\Sigma^{(0)})\}^2
}{
2\{\mathbb E[|\phi_1|^2]\}^2
}.
\]

It remains only to calculate the contribution of the rank-one correction
in \(B_S^{\texttt{plug-in}}\). By
\eqref{supp:eq:two-dimensional-block-inverse},
\begin{align*}
&
\frac{1}{p-2}
\frac{
16d_\kappa^2(1-d_\kappa^2)
d_\Sigma^2(1-d_\Sigma^2)^2
}{
(1-d_\kappa^2d_\Sigma^2)^4
}
\\
&\quad\times
\Bigg\{
\frac{1-d_\kappa}{2}
e_1^\top
A_S^{-1}e_1e_1^\top A_S^{-1}e_1
+
\frac{1+d_\kappa}{2d_\kappa^2}
e_2^\top
A_S^{-1}e_1e_1^\top A_S^{-1}e_2
\Bigg\}
\\
&=
\frac{2}{p-2}
\left\{
\frac{
\operatorname{Tr}(\Sigma^{(0)})
}{
\mathbb E[|\phi_1|^2]
}
\right\}^{\!2}
\frac{
d_\kappa^2(1-d_\kappa^2)
d_\Sigma^2(1-d_\Sigma^2)^2
}{
(1-d_\kappa^2d_\Sigma^2)^2
(1-d_\kappa^2d_\Sigma^2\rho_\phi^2)^2
}
\\
&\quad\times
\Bigl\{
(1-d_\kappa)
(1-d_\kappa d_\Sigma\rho_\phi\lambda_\phi)^2
+
(1+d_\kappa)
d_\Sigma^2\rho_\phi^2
(1-\lambda_\phi^2)
\Bigr\}.
\end{align*}
Combining the preceding three contributions in
\eqref{supp:eq:plugin-maximum-spectrum-trace-block-form}
gives the explicit expression
\eqref{supp:eq:trace-plugin-maximum-spectrum-covariance}.

If \(\rho_\phi=0\), then the terms involving \(Q_\phi\) vanish and the
same result follows under the convention \(\lambda_\phi=0\). The case
\(d_\Sigma=0\) follows similarly. This completes the proof.
\end{proof}

\subsection{Trace comparison for the maximum-aligned spectrum estimators}
\label{supp:sec:maximum-spectrum-trace-comparison}
We use the preceding trace formulas to compare the asymptotic
efficiencies of the two spectrum estimators. We establish equality under isotropic noise, high-SNR bounds uniform in \(\lambda_\phi\), and strict improvement for the \texttt{plug-in} estimator under anisotropic noise when \(\mathbb E[\phi_1^2]=0\).

For the proof, it is convenient to work with the normalized trace
difference
\begin{equation}
\label{supp:eq:maximum-spectrum-normalized-trace-difference}
\Delta_I(\rho,\lambda)
:=
\frac{
2\mathbb E[|\phi_1|^2]
}{
\operatorname{Tr}(\Sigma^{(0)})
}
\left\{
\operatorname{Tr}
\left(
V_I^{\texttt{plug-in,max}}(\rho,\lambda)
\right)
-
\operatorname{Tr}
\left(
V_I^{\texttt{cov,max}}(\rho,\lambda)
\right)
\right\},
\end{equation}
Here the right-hand side is evaluated using the scalar trace
expressions in
\eqref{supp:eq:trace-cov-maximum-spectrum-covariance}
and
\eqref{supp:eq:trace-plugin-maximum-spectrum-covariance},
with \(\rho_\phi=\rho\), \(\lambda_\phi=\lambda\), and
\(p,d_\kappa,d_\Sigma,\eta\) held fixed.

\begin{proof}[Proof of Theorem~\ref{thm:maximum-spectrum-trace-comparison}]

\medskip
\noindent\emph{Proof of (i).}
Recall that
\[
V_I^{\texttt{cov,max}}
=
A_I^{\texttt{max}}
V^{\texttt{cov}}
(A_I^{\texttt{max}})^\top
\]
and
\[
V_I^{\texttt{plug-in,max}}
=
A_I^{\texttt{max}}
V^{\texttt{plug-in}}
(A_I^{\texttt{max}})^\top.
\]
If \(d_\Sigma=0\), then
Theorem~\ref{thm:trace-comparison} gives
\[
V^{\texttt{cov}}
=
V^{\texttt{plug-in}}.
\]
It follows immediately that
\[
V_I^{\texttt{cov,max}}
=
V_I^{\texttt{plug-in,max}}.
\]

\medskip
\noindent\emph{Proof of (ii).}
Suppose that \(d_\Sigma>0\), and recall that
\[
\eta
=
\frac{\mathbb E[|\phi_1|^2]}
{\operatorname{Tr}(\Sigma^{(0)})}.
\]
For \(m\in\{\texttt{cov},\texttt{plug-in}\}\), write
\(A_I^m(\rho,\lambda)\) and \(R_I^m(\rho,\lambda)\)
for the coefficients in the theorem evaluated at
\(\rho_\phi=\rho\) and \(\lambda_\phi=\lambda\),
with \(p,d_\kappa,d_\Sigma\) held fixed.

By
\eqref{supp:eq:trace-cov-maximum-spectrum-covariance}
and
\eqref{supp:eq:trace-plugin-maximum-spectrum-covariance},
the traces satisfy
\begin{align*}
\operatorname{Tr}
\left(V_I^{\texttt{cov,max}}(\rho,\lambda)\right)
&=
A_I^{\texttt{cov}}(\rho,\lambda)\eta^{-1}
+
R_I^{\texttt{cov}}(\rho,\lambda)\eta^{-2},
\\
\operatorname{Tr}
\left(V_I^{\texttt{plug-in,max}}(\rho,\lambda)\right)
&=
A_I^{\texttt{plug-in}}(\rho,\lambda)\eta^{-1}
+
R_I^{\texttt{plug-in}}(\rho,\lambda)\eta^{-2}.
\end{align*}
For fixed \(\rho\in[0,1]\), the remainder coefficients
are continuous in \(\lambda\in[-1,1]\), and all their
denominators are strictly positive. They are therefore
uniformly bounded over \(\lambda\in[-1,1]\).

Moreover, the definition of \(A_I^{\texttt{cov}}\) gives
\[
A_I^{\texttt{cov}}(\rho,\lambda)
\geq
\frac{p-2}{2}(1-d_\Sigma)
>0,
\qquad
(\rho,\lambda)\in[0,1]\times[-1,1].
\]
Thus, \(A_I^{\texttt{cov}}(\rho,\lambda)\) is bounded
away from zero uniformly over \(\lambda\in[-1,1]\).
It follows that
\begin{align*}
&
\frac{
\operatorname{Tr}
\left(V_I^{\texttt{plug-in,max}}(\rho,\lambda)\right)
}{
\operatorname{Tr}
\left(V_I^{\texttt{cov,max}}(\rho,\lambda)\right)
}
\\
&=
\frac{
A_I^{\texttt{plug-in}}(\rho,\lambda)
+
R_I^{\texttt{plug-in}}(\rho,\lambda)\eta^{-1}
}{
A_I^{\texttt{cov}}(\rho,\lambda)
+
R_I^{\texttt{cov}}(\rho,\lambda)\eta^{-1}
}
\\
&=
\frac{
A_I^{\texttt{plug-in}}(\rho,\lambda)
}{
A_I^{\texttt{cov}}(\rho,\lambda)
}
+
O(\eta^{-1}),
\end{align*}
uniformly over \(\lambda\in[-1,1]\).
Taking \(\rho=\rho_\phi\) and \(\lambda=\lambda_\phi\)
proves \eqref{eq:high-snr-maximum-spectrum-trace-ratio}.

We next establish the bounds for the leading ratio.
By the definition of \(A_I^{\texttt{plug-in}}\), the ratio
\[
\frac{
A_I^{\texttt{plug-in}}(\rho,\lambda)
}{
A_I^{\texttt{cov}}(\rho,\lambda)
}
\]
is a weighted average of
\[
\frac{1-d_\Sigma^2}{1-d_\Sigma^2\rho^2}
\quad\text{and}\quad
\frac{1-d_\kappa^2d_\Sigma^2}
{1-d_\kappa^2d_\Sigma^2\rho^2},
\]
with respective weights
\[
\frac{
(p-2)(1+d_\Sigma\rho\lambda)
}{
2A_I^{\texttt{cov}}(\rho,\lambda)
}
\quad\text{and}\quad
1-
\frac{
(p-2)(1+d_\Sigma\rho\lambda)
}{
2A_I^{\texttt{cov}}(\rho,\lambda)
}.
\]
Both weights are strictly positive.

By \eqref{supp:eq:high-snr-shrinkage-factor-ordering},
\[
\frac{1-d_\Sigma^2}{1-d_\Sigma^2\rho^2}
\leq
\frac{1-d_\kappa^2d_\Sigma^2}
{1-d_\kappa^2d_\Sigma^2\rho^2}.
\]
Therefore,
\[
\frac{1-d_\Sigma^2}{1-d_\Sigma^2\rho^2}
\leq
\frac{
A_I^{\texttt{plug-in}}(\rho,\lambda)
}{
A_I^{\texttt{cov}}(\rho,\lambda)
}
\leq
\frac{1-d_\kappa^2d_\Sigma^2}
{1-d_\kappa^2d_\Sigma^2\rho^2},
\]
which proves
\eqref{eq:high-snr-maximum-spectrum-leading-ratio-bounds}.

If \(\rho<1\), the ordering of the two shrinkage factors
is strict. Since both weights are strictly positive,
the leading ratio lies strictly between these factors.
In addition, since \(d_\kappa>0\) and \(d_\Sigma>0\),
\[
\frac{1-d_\kappa^2d_\Sigma^2}
{1-d_\kappa^2d_\Sigma^2\rho^2}
<1.
\]

Now fix \(\rho=\rho_\phi<1\). Define
\[
\delta
:=
\frac12
\min_{\lambda\in[-1,1]}
\min
\left\{
\frac{
A_I^{\texttt{plug-in}}(\rho,\lambda)
}{
A_I^{\texttt{cov}}(\rho,\lambda)
}
-
\frac{1-d_\Sigma^2}{1-d_\Sigma^2\rho^2},
\;
\frac{1-d_\kappa^2d_\Sigma^2}
{1-d_\kappa^2d_\Sigma^2\rho^2}
-
\frac{
A_I^{\texttt{plug-in}}(\rho,\lambda)
}{
A_I^{\texttt{cov}}(\rho,\lambda)
}
\right\}.
\]
Both differences are strictly positive and continuous
in \(\lambda\). Compactness of \([-1,1]\) therefore gives
\(\delta>0\).

The remainder in \eqref{eq:high-snr-maximum-spectrum-trace-ratio} is uniform over \(\lambda\in[-1,1]\).
Hence, for all sufficiently large \(\eta\), its absolute
value is smaller than \(\delta\) uniformly over this interval. The full trace ratio therefore remains strictly between the same lower and upper bounds, uniformly over \(\lambda_\phi\in[-1,1]\).

Finally, if \(\rho_\phi=1\), both shrinkage factors equal
one. The leading ratio is therefore one, and the comparison is determined by the terms of order \(\eta^{-2}\) in the two trace formulas.

\medskip
\noindent\emph{Proof of (iii).}
Suppose that \(d_\Sigma>0\) and \(\rho_\phi=0\). In this case, \(
\mathbb E[\phi_1^2]=0,
\) so \(\lambda_\phi\) is set equal to zero by convention. More generally, inspection of
\eqref{supp:eq:trace-cov-maximum-spectrum-covariance}
and
\eqref{supp:eq:trace-plugin-maximum-spectrum-covariance}
shows that \(\lambda\) appears only through \(\rho\lambda\) and
\(\rho^2(1-\lambda^2)\). Therefore,
\[
\Delta_I(0,\lambda)
=
\Delta_I(0,0)
\]
for every \(\lambda\in[-1,1]\).

Under the assumptions of the theorem,
\[
0<d_\kappa<1,
\quad
0<d_\Sigma<1,
\quad
p\geq3,
\quad
\frac{
\operatorname{Tr}(\Sigma^{(0)})
}{
\mathbb E[|\phi_1|^2]
}
>0.
\]
Substituting \(\rho=0\) into
\eqref{supp:eq:trace-cov-maximum-spectrum-covariance}
and
\eqref{supp:eq:trace-plugin-maximum-spectrum-covariance},
and simplifying, gives
\begin{equation}
\label{supp:eq:maximum-spectrum-circular-trace-difference}
\begin{aligned}
\Delta_I(0,\lambda)
&=
-d_\Sigma^2
\Bigg[
p-2
+
\frac{
1+d_\kappa+d_\kappa^2-d_\kappa^3
}{2}
\\
&\quad+
\frac{
\operatorname{Tr}(\Sigma^{(0)})
}{
\mathbb E[|\phi_1|^2]
}
\Bigg\{
(p-2)
\frac{
2+2d_\kappa-d_\kappa^2
-
(1+2d_\kappa+d_\kappa^3)d_\Sigma^2
+
d_\kappa d_\Sigma^4
}{
1-d_\kappa^2d_\Sigma^2
}
\\
&\quad\quad+
\frac{1}{2}
\Bigg(
(1-d_\kappa)
\left(
2-3d_\kappa^2+d_\kappa^2d_\Sigma^2
\right)
+
\frac{1+d_\kappa}{d_\kappa^2}
\left(
2+d_\kappa^2-d_\kappa^2d_\Sigma^2
\right)
\Bigg)
\\
&\quad\quad-
\frac{4}{p-2}
\frac{
d_\kappa^2(1-d_\kappa)^2(1+d_\kappa)
(1-d_\Sigma^2)^2
}{
(1-d_\kappa^2d_\Sigma^2)^2
}
\Bigg\}
\Bigg],
\quad
\lambda\in[-1,1].
\end{aligned}
\end{equation}

Clearly,
\[
p-2
+
\frac{
1+d_\kappa+d_\kappa^2-d_\kappa^3
}{2}
>0.
\]
Moreover,
\begin{align*}
&
2+2d_\kappa-d_\kappa^2
-
(1+2d_\kappa+d_\kappa^3)d_\Sigma^2
+
d_\kappa d_\Sigma^4
\\
&=
(1-d_\kappa)(1+d_\kappa)^2
+
(1-d_\Sigma^2)
\left(
1+d_\kappa+d_\kappa^3-d_\kappa d_\Sigma^2
\right)
>0.
\end{align*}
Thus, the first term inside the inner braces in
\eqref{supp:eq:maximum-spectrum-circular-trace-difference}
is strictly positive.

For the second term, we have
\[
(1-d_\kappa)
\left(
2-3d_\kappa^2+d_\kappa^2d_\Sigma^2
\right)
\geq -1
\]
and
\[
\frac{1+d_\kappa}{d_\kappa^2}
\left(
2+d_\kappa^2-d_\kappa^2d_\Sigma^2
\right)
\geq
\frac{2(1+d_\kappa)}{d_\kappa^2}
\geq4.
\]
Therefore,
\[
\frac{1}{2}
\Bigg(
(1-d_\kappa)
\left(
2-3d_\kappa^2+d_\kappa^2d_\Sigma^2
\right)
+
\frac{1+d_\kappa}{d_\kappa^2}
\left(
2+d_\kappa^2-d_\kappa^2d_\Sigma^2
\right)
\Bigg)
\geq
\frac{3}{2}.
\]
Moreover,
\[
\frac{
1-d_\Sigma^2
}{
1-d_\kappa^2d_\Sigma^2
}
\leq1,
\]
and hence
\begin{align*}
&
\frac{4}{p-2}
\frac{
d_\kappa^2(1-d_\kappa)^2(1+d_\kappa)
(1-d_\Sigma^2)^2
}{
(1-d_\kappa^2d_\Sigma^2)^2
}
\\
&\leq
4d_\kappa^2(1-d_\kappa)^2(1+d_\kappa)
\\
&\leq
8\{d_\kappa(1-d_\kappa)\}^2
\\
&\leq
\frac{1}{2}.
\end{align*}
It follows that the expression inside the inner braces in
\eqref{supp:eq:maximum-spectrum-circular-trace-difference}
is strictly positive. Since \(d_\Sigma>0\) and
\[
\frac{
\operatorname{Tr}(\Sigma^{(0)})
}{
\mathbb E[|\phi_1|^2]
}
>0,
\]
\eqref{supp:eq:maximum-spectrum-circular-trace-difference} yields
\[
\Delta_I(0,\lambda)<0
\]
for every \(\lambda\in[-1,1]\). In particular,
\[
\Delta_I(0,0)<0.
\]
By
\eqref{supp:eq:maximum-spectrum-normalized-trace-difference},
this is equivalent to
\[
\operatorname{Tr}
\left(
V_I^{\texttt{plug-in,max}}
\right)
<
\operatorname{Tr}
\left(
V_I^{\texttt{cov,max}}
\right).
\]
\end{proof}

\subsection{Estimation of the spectrum covariance and confidence intervals}
\label{supp:subsec:spectrum-covariance-estimation}

We estimate \(V_I^{\texttt{cov,max}}\) and
\(V_I^{\texttt{plug-in,max}}\) by substituting estimated parameters into
\eqref{supp:eq:cov-maximum-spectrum-covariance} and
\eqref{supp:eq:plugin-maximum-spectrum-covariance}, respectively.
We work in Helmert coordinates and suppress the method index.
For \texttt{cov}, we use
\((\widehat\kappa,\widehat\Sigma)
=(\widehat\kappa^{\texttt{cov,max}},\widehat\Sigma^{\texttt{cov}})\).
For \texttt{plug-in}, we use
\((\widehat\kappa,\widehat\Sigma)
=(\widehat\kappa^{[s],\texttt{max}},\widehat\Sigma^{[s]})\),
for fixed \(s\geq1\).
In both cases, \(\widehat\kappa\) is a maximum-aligned, unit-norm loading,
and \(\widehat\Sigma=\mathcal M(\widehat\kappa)\) is its corresponding
moment covariance estimate. The following parameter estimates are
computed separately for each method.
Set \(\widehat I=\operatorname{Re}(\widehat\kappa)\),
\(\widehat W=\operatorname{Im}(\widehat\kappa)\),
\(\widehat d_\kappa=\widehat\kappa^\top\widehat\kappa\), and
\[
\widehat\tau^2=\operatorname{Tr}(\widehat\Sigma),
\quad
\widehat\gamma=\widehat\Sigma_{11}-\widehat\Sigma_{22}
+2i\widehat\Sigma_{12}.
\]
We estimate \(m_2:=\mathbb E[|\phi_1|^2]\) and \(c_2:=\mathbb E[\phi_1^2]\) from the empirical moments after subtracting their estimated noise contributions:
\begin{equation}
\label{supp:eq:spectrum-noise-corrected-moments}
\widehat m_2=\widehat\kappa^*(\widehat\Sigma^{\texttt{Her}}-\widehat\tau^2\mathrm{Id}_p)\widehat\kappa,\qquad
\widehat c_2=\widehat\kappa^*(\widehat\Sigma^{\texttt{Tra}}-\widehat\gamma\mathrm{Id}_p)\overline{\widehat\kappa}.
\end{equation}
For interval construction, we first replace \(\widehat m_2\) by \(\max\{\widehat m_2,\widehat\tau^2/n\}\). Set \(\epsilon_n=1-\min\{0.005,n^{-1}\}\). If \(|\widehat c_2|>\epsilon_n\widehat m_2\), we replace \(\widehat c_2\) by \(\epsilon_n\widehat m_2\,\widehat c_2/|\widehat c_2|\), preserving its phase. We use the same notation for these adjusted estimates in all subsequent calculations. Under the assumptions of Theorem~\ref{thm:maximum-aligned-loading-expansion-clt}, these adjustments preserve consistency because \(\epsilon_n\to1\) and \(\widehat\tau^2/n\to0\) in probability.
The remaining scalar estimates are
\[
\widehat d_\Sigma=\frac{|\widehat\gamma|}{\widehat\tau^2},
\quad
\widehat\rho_\phi=\frac{|\widehat c_2|}{\widehat m_2},
\quad
\widehat\lambda_\phi=
\frac{\operatorname{Re}(\widehat\gamma\,\overline{\widehat c_2})}
     {|\widehat\gamma|\,|\widehat c_2|},
\quad
\widehat\eta=\frac{\widehat m_2}{\widehat\tau^2},
\]
with \(\widehat\lambda_\phi=0\) if
\(\widehat\gamma\widehat c_2=0\).
We also use \(\widehat r=(1+\widehat d_\kappa)/2\) and
\(\widehat\Pi_\perp=\Pi_\perp(\widehat\kappa)\).

For \texttt{cov}, we substitute these estimates into
\eqref{supp:eq:cov-maximum-spectrum-covariance}.
In particular, the term
\(\operatorname{Im}\{\gamma\mathbb E[\overline{\phi_1}^{\,2}]\}\)
is estimated by
\(\operatorname{Im}(\widehat\gamma\,\overline{\widehat c_2})\).
For \texttt{plug-in}, we evaluate
\eqref{supp:eq:plugin-maximum-spectrum-covariance} using the estimated canonical blocks. In these blocks, we replace
\(\mathbb E[\phi_1^2]\) in
\eqref{supp:eq:plugin-Q-phi-definition} by
\(\widehat c_2\overline{\widehat\gamma}/|\widehat\gamma|\),
the estimated second moment in diagonal noise coordinates.
If \(\widehat\gamma=0\), we use \(\widehat c_2\) directly;
if \(\widehat c_2=0\), we set
\(\widehat\rho_\phi\widehat Q_\phi=0_2\).

Under the assumptions of
Theorem~\ref{thm:maximum-aligned-loading-expansion-clt},
the empirical moment estimates are consistent. The covariance formulas depend continuously on these moments, including when $\gamma=0$ or $c_2=0$, because all terms that involve an unidentified phase then vanish.
Thus $\widehat V_I\to V_I$ in probability. Moreover, $V_I$ is positive definite for $0<d_\kappa<1$, and $(H^\top V_IH)_{jj}>0$ for every original-grid coordinate. The pointwise intervals below therefore have asymptotic coverage $0.95$ by Slutsky's theorem.

Let \(\widehat V_I\) denote the resulting covariance estimate for the selected method. The pointwise \(95\%\) confidence interval at coordinate
\(j\) of the original frequency grid is
\[
(H^\top\widehat I)_j
\ \pm\ z_{0.975}
\sqrt{\frac{(H^\top\widehat V_I H)_{jj}}{n}}.
\]

\section{Simulation and numerical details}
\label{supp:sec:simulation-numerical-details}

We present additional simulation results and describe how we selected the smoothing bandwidth for each ENDOR dataset.

\subsection{Numerical comparison of residual projections}
\label{sec:triangle-projector-comparison}
We compare the covariance estimators based on \(P_\perp\) and \(\Pi_\perp\) described in Section~\ref{supp:sec:projected-covariance-definiteness},
focusing on positive definiteness and the accuracy of the first plug-in loading update.

\begin{figure}[tbp]
\centering
\includegraphics[width=\linewidth]{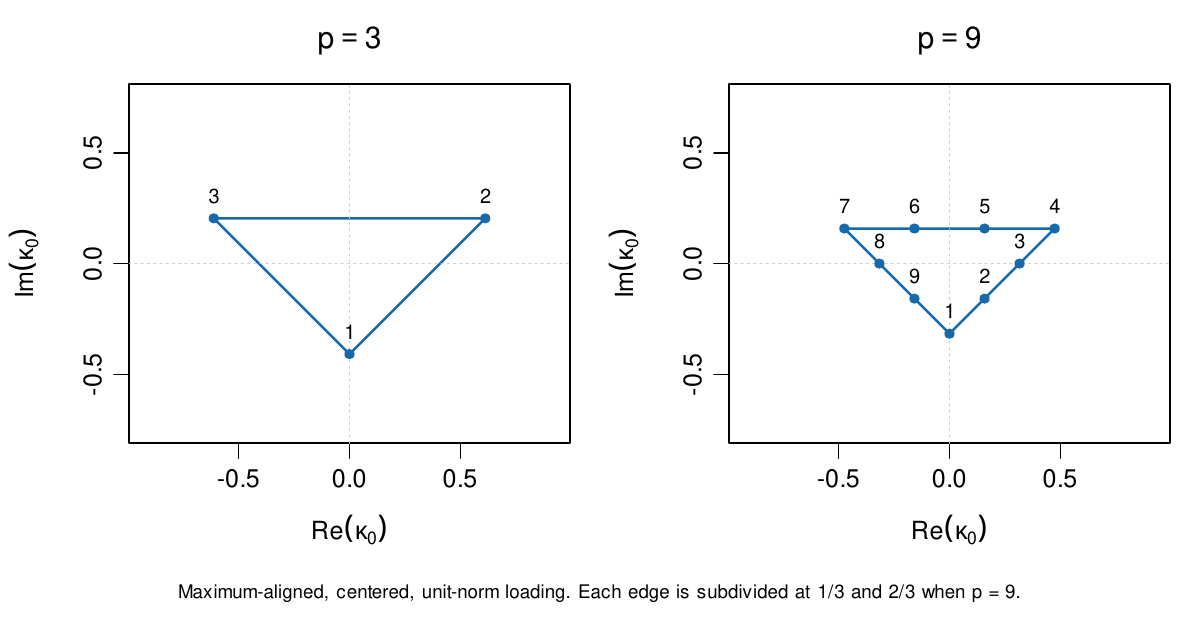}
\caption{True loading coordinates for \(p=3\) and \(p=9\).
Labels indicate the coordinate order.}
\label{fig:triangle-projector-kappa}
\end{figure}

\medskip
\noindent\textit{Simulation Setting.}
We generate observations directly from
model~\eqref{model:centered-complex-factor} in \(\mathbb C^p\), with
\[
\Sigma^{(0)}
=\frac{1}{\mathrm{CN}+1}
\begin{pmatrix}\mathrm{CN}&0\\0&1\end{pmatrix},
\quad
\operatorname{Tr}(\Sigma^{(0)})=1.
\]
Thus, the condition number of $\Sigma^{(0)}$ is $\mathrm{CN}$. For \(p=3\), take the triangle vertices \(0,1+i,-1+i\). For \(p=9\), add the points at one-third and two-thirds of each edge. List these points in cyclic order as \(z_1,\ldots,z_p\), and set
\[
\kappa_j^{(0)}
=\frac{z_j-\bar z}
{\left\{\sum_{\ell=1}^p|z_\ell-\bar z|^2\right\}^{1/2}},
\quad
\bar z=\frac1p\sum_{\ell=1}^p z_\ell.
\]
In both dimensions, \(\|\kappa^{(0)}\|=1\) and
\(d_\kappa=1/2\). Figure~\ref{fig:triangle-projector-kappa} shows the loading coordinates.

\begin{figure}[tbp]
\centering
\includegraphics[width=\linewidth]{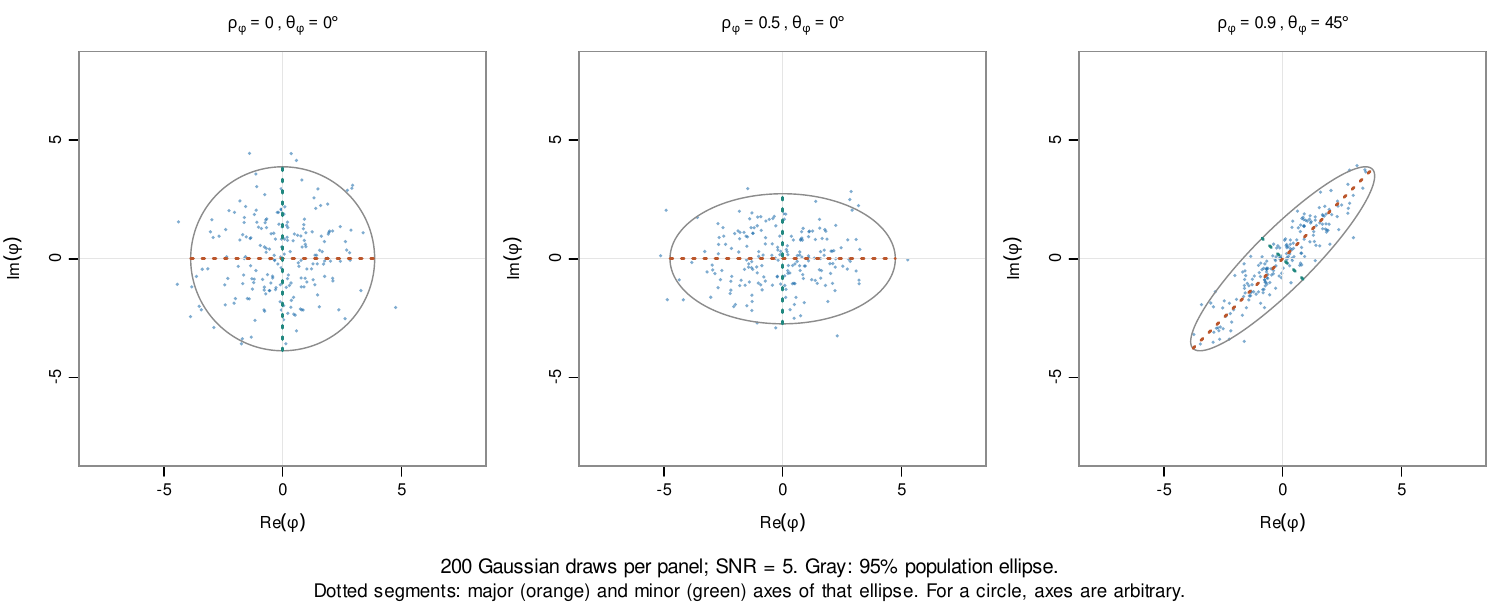}
\caption{Gaussian score samples at \(\eta=5\), with 200 points per panel.
Dotted segments show the semiaxes of the population 95\% Gaussian ellipse.}
\label{fig:triangle-projector-scores}
\end{figure}

The factor scores $\phi_{i}$ are independent \(\mathcal{CN}(0,\Omega_\phi)\)
variables, independent of the errors, with
\[
\Omega_\phi
=\frac{\eta}{2}
\begin{pmatrix}
1+\rho_\phi\cos(2\mu)&\rho_\phi\sin(2\mu)\\
\rho_\phi\sin(2\mu)&1-\rho_\phi\cos(2\mu)
\end{pmatrix}.
\]
Here \(\mu\) is the signed direction of the score major axis. Thus \(\mathbb E[|\phi_1|^2]=\eta\) and
\(\mathbb E[\phi_1^2]=\eta\rho_\phi e^{2i\mu}\). Under \eqref{eq:def-lambda-phi}, \(\lambda_\phi=\cos(2\mu)\) when
\(\mathrm{CN}>1\) and \(\rho_\phi>0\), and \(\lambda_\phi=0\) otherwise. Representative score samples are shown in Figure~\ref{fig:triangle-projector-scores}.

\begin{center}
\small
\begin{tabular}{ll}
\hline
Parameter & Values \\
\hline
\(p\) & \(3,9\) \\
\(n\) & \(100,200,400,800\) \\
\(\mathrm{CN}\) & \(1,3,5,10,20,50\) \\
\(\eta\) & \(1,3,5,10,20\) \\
\(\rho_\phi\) & \(0,0.25,0.5,0.75,0.9\) \\
\(\mu\) & \(-45^\circ,-30^\circ,-15^\circ,0^\circ,15^\circ,30^\circ,45^\circ\) \\
\hline
\end{tabular}
\end{center}
When \(\rho_\phi=0\), the score distribution does not depend on \(\mu\), so we use only \(\mu=0\). Together with the seven directions for each of the four positive values of \(\rho_\phi\), this gives \(1+4\times7=29\) distinct \((\rho_\phi,\mu)\) settings. Crossing these with two values of \(p\), four sample sizes, six noise condition numbers, and five values of \(\eta\)
gives \(2\times4\times6\times5\times29=6{,}960\) configurations.

\medskip
\noindent\textit{Estimation and evaluation.}
Both methods use the same samples and the common initial estimate \(\widehat\kappa^{\texttt{cov}}\). We compute the first plug-in loading update in \eqref{eq:iterated-plugin-initialization}, with covariance input
\(\mathcal M(\widehat\kappa^{\texttt{cov}})\) from
\eqref{eq:moment-covariance-map} for \(\Pi_\perp\), or
\(\widehat\Sigma_P(\widehat\kappa^{\texttt{cov}})\) from
\eqref{supp:eq:P-one-sided-covariance} for \(P_\perp\).

We record the frequency of
\(\lambda_{\min}(\widehat\Sigma_P)\leq0\) before correction. For the loading update, a non-SPD input with eigenvalues
\(\widehat\lambda_j\) is corrected by replacing them with
\[
\max\left\{
\widehat\lambda_j,
\sqrt{\epsilon_{\mathrm{mach}}}\max_k|\widehat\lambda_k|
\right\},
\]
where \(\epsilon_{\mathrm{mach}}\) is machine precision. SPD inputs are unchanged. The \(\Pi_\perp\) estimate is almost surely SPD in all settings by Theorem~\ref{thm:moment-covariance-finite-sample-definiteness}, since
\(n\geq2p\).

Loading accuracy is measured by the squared projective distance
\(d^2([\kappa^{(0)}] ,[\widehat\kappa])\) in
\eqref{eq:projective-distance}.
Let \(\widehat\kappa_{P,b}\) and \(\widehat\kappa_{\Pi,b}\)
denote the loading estimates from the \(b\)th paired replication. All \(R=200\) paired fits succeeded in each configuration. We report the empirical MSE ratio
\[
\widehat r
=
\frac{
\sum_{b=1}^R
d^2([\kappa^{(0)}],[\widehat\kappa_{P,b}])
}{
\sum_{b=1}^R
d^2([\kappa^{(0)}],[\widehat\kappa_{\Pi,b}])
},
\]
where values above one favor \(\Pi_\perp\).
Its Monte Carlo standard error accounts for pairing:
\[
\widehat{\operatorname{MCSE}}(\widehat r)
=
\frac{
\operatorname{sd}_{b=1,\ldots,R}
\left\{
d^2([\kappa^{(0)}],[\widehat\kappa_{P,b}])
-
\widehat r\,
d^2([\kappa^{(0)}],[\widehat\kappa_{\Pi,b}])
\right\}
}{
\sqrt R\,
\left\{
R^{-1}\sum_{b=1}^R
d^2([\kappa^{(0)}],[\widehat\kappa_{\Pi,b}])
\right\}
},
\]
where the sample standard deviation uses divisor \(R-1\).

\medskip
\noindent\textit{Results and conclusion.}
Tables~\ref{tab:rep-s1-profile01-p3}--\ref{tab:rep-s1-profile29-p9}
report results for
\((\rho_\phi,\mu)=(0,0)\), \((0.5,0)\), and \((0.9,45^\circ)\).
For each fixed \((p,\rho_\phi,\mu)\), the following summary
averages the non-SPD frequency and empirical MSE ratio
equally over the 120 combinations of
\(\mathrm{CN}\), \(\eta\), and \(n\).

\begin{center}
\small
\begin{tabular}{rrrrr}
\hline
$p$ & $\rho_\phi$ & $\mu$
& Mean non-SPD (\%) & Mean loading MSE ratio \\
\hline
3 & 0.00 & $0^\circ$  & 3.721 & 1.0528 \\
9 & 0.00 & $0^\circ$  & 0.217 & 1.0008 \\
3 & 0.50 & $0^\circ$  & 3.650 & 1.0490 \\
9 & 0.50 & $0^\circ$  & 0.196 & 1.0021 \\
3 & 0.90 & $45^\circ$ & 1.896 & 1.0340 \\
9 & 0.90 & $45^\circ$ & 0.004 & 1.0026 \\
\hline
\end{tabular}
\end{center}

As \(p\) increases, retaining \(p-1\) rather than \(p-2\) complex residual directions makes a smaller relative difference to the residual dimension. Consistent with this, the loading MSE ratios are close to one at \(p=9\). Even at \(p=3\), where the additional direction represents a larger fraction of the residual space, \(P_\perp\) provides no clear overall MSE advantage: the mean ratios exceed one in all three settings summarized above. Although \(P_\perp\) yields smaller MSE in some configurations, these results show little overall benefit from replacing \(\Pi_\perp\).

The covariance estimates also differ in finite-sample definiteness. At \(\mathrm{CN}=50\), averaging equally over all 29 \((\rho_\phi,\mu)\) settings, five values of \(\eta\), and four sample sizes gives mean non-SPD frequencies of 15.003\% for \(p=3\) and 0.809\% for \(p=9\). The corresponding mean loading MSE ratios are 1.1501 and 1.0072. Thus, strong noise anisotropy can cause the \(P_\perp\)-based estimate to require correction before the loading update.
The reported MSE comparisons already allow this correction, yet show no consistent advantage for \(P_\perp\).
Under the conditions of Theorem~\ref{thm:moment-covariance-finite-sample-definiteness}, the \(\Pi_\perp\)-based estimate is almost surely positive definite and requires no such correction.

The choice of \(\Pi_\perp\) also simplifies the asymptotic analysis.
By annihilating both \(\kappa\) and \(\overline\kappa\), it eliminates the first-order contribution of loading estimation error from the expansions of both \(\widehat\tau^2\) and
\(\widehat\gamma\). For \(P_\perp\), this contribution generally remains in the expansion of \(\widehat\gamma_P\). Together, the definiteness guarantee, the simpler expansions, and the observed loading MSEs support retaining \(\Pi_\perp\)
for covariance estimation and the subsequent plug-in loading update.
The empirical comparison concerns the configurations studied and does not establish a uniform MSE ordering.

\begingroup
\begin{table}[p]
\centering
\caption{Non-SPD frequencies and loading MSE ratios: $p=3$, $\rho_\phi=0$, $\mu=0^\circ$, $s=1$.}
\label{tab:rep-s1-profile01-p3}
\begingroup
\fontsize{8.5}{10}\selectfont
\setlength{\tabcolsep}{4pt}
\renewcommand{\arraystretch}{1.03}
\begin{tabular*}{\linewidth}{@{\extracolsep{\fill}}rrccccc@{}}
\toprule
$\mathrm{CN}$ & $n$ & $\eta=1$ & $\eta=3$ & $\eta=5$ & $\eta=10$ & $\eta=20$ \\
\midrule
\multicolumn{7}{@{}l}{\rule{0pt}{13pt}\textbf{(a) Non-SPD frequency (\%)}} \\[2pt]
1 & 100 & 0.0\,(0.00) & 0.0\,(0.00) & 0.0\,(0.00) & 0.0\,(0.00) & 0.0\,(0.00) \\
 & 200 & 0.0\,(0.00) & 0.0\,(0.00) & 0.0\,(0.00) & 0.0\,(0.00) & 0.0\,(0.00) \\
 & 400 & 0.0\,(0.00) & 0.0\,(0.00) & 0.0\,(0.00) & 0.0\,(0.00) & 0.0\,(0.00) \\
 & 800 & 0.0\,(0.00) & 0.0\,(0.00) & 0.0\,(0.00) & 0.0\,(0.00) & 0.0\,(0.00) \\
\addlinespace[2.2pt]
3 & 100 & 0.0\,(0.00) & 0.0\,(0.00) & 0.0\,(0.00) & 0.0\,(0.00) & 0.5\,(0.50) \\
 & 200 & 0.0\,(0.00) & 0.0\,(0.00) & 0.0\,(0.00) & 0.0\,(0.00) & 0.0\,(0.00) \\
 & 400 & 0.0\,(0.00) & 0.0\,(0.00) & 0.0\,(0.00) & 0.0\,(0.00) & 0.0\,(0.00) \\
 & 800 & 0.0\,(0.00) & 0.0\,(0.00) & 0.0\,(0.00) & 0.0\,(0.00) & 0.0\,(0.00) \\
\addlinespace[2.2pt]
5 & 100 & 0.0\,(0.00) & 0.0\,(0.00) & 0.0\,(0.00) & 0.0\,(0.00) & 2.0\,(0.99) \\
 & 200 & 0.0\,(0.00) & 0.0\,(0.00) & 0.0\,(0.00) & 0.0\,(0.00) & 0.0\,(0.00) \\
 & 400 & 0.0\,(0.00) & 0.0\,(0.00) & 0.0\,(0.00) & 0.0\,(0.00) & 0.0\,(0.00) \\
 & 800 & 0.0\,(0.00) & 0.0\,(0.00) & 0.0\,(0.00) & 0.0\,(0.00) & 0.0\,(0.00) \\
\addlinespace[2.2pt]
10 & 100 & 0.0\,(0.00) & 0.0\,(0.00) & 0.5\,(0.50) & 2.0\,(0.99) & 10.5\,(2.17) \\
 & 200 & 0.0\,(0.00) & 0.0\,(0.00) & 0.0\,(0.00) & 0.0\,(0.00) & 3.5\,(1.30) \\
 & 400 & 0.0\,(0.00) & 0.0\,(0.00) & 0.0\,(0.00) & 0.0\,(0.00) & 0.0\,(0.00) \\
 & 800 & 0.0\,(0.00) & 0.0\,(0.00) & 0.0\,(0.00) & 0.0\,(0.00) & 0.0\,(0.00) \\
\addlinespace[2.2pt]
20 & 100 & 0.0\,(0.00) & 6.5\,(1.75) & 7.5\,(1.87) & 16.5\,(2.63) & 30.5\,(3.26) \\
 & 200 & 0.0\,(0.00) & 0.0\,(0.00) & 1.5\,(0.86) & 13.0\,(2.38) & 20.5\,(2.86) \\
 & 400 & 0.0\,(0.00) & 0.0\,(0.00) & 0.0\,(0.00) & 1.0\,(0.71) & 7.5\,(1.87) \\
 & 800 & 0.0\,(0.00) & 0.0\,(0.00) & 0.0\,(0.00) & 0.0\,(0.00) & 3.5\,(1.30) \\
\addlinespace[2.2pt]
50 & 100 & 8.5\,(1.98) & 20.5\,(2.86) & 28.0\,(3.18) & 34.0\,(3.36) & 34.5\,(3.37) \\
 & 200 & 1.0\,(0.71) & 9.5\,(2.08) & 17.5\,(2.69) & 22.0\,(2.94) & 35.0\,(3.38) \\
 & 400 & 0.0\,(0.00) & 8.5\,(1.98) & 12.5\,(2.34) & 22.0\,(2.94) & 29.0\,(3.22) \\
 & 800 & 0.0\,(0.00) & 2.0\,(0.99) & 3.0\,(1.21) & 15.0\,(2.53) & 17.0\,(2.66) \\
\midrule
\multicolumn{7}{@{}l}{\rule{0pt}{13pt}\textbf{(b) Loading MSE ratio ($P_\perp/\Pi_\perp$)}} \\[2pt]
1 & 100 & 1.024\,(0.010) & 1.001\,(0.009) & 1.009\,(0.010) & 0.994\,(0.010) & 1.030\,(0.013) \\
 & 200 & 1.021\,(0.007) & 0.996\,(0.007) & 1.000\,(0.006) & 1.020\,(0.006) & 1.004\,(0.008) \\
 & 400 & 0.996\,(0.005) & 1.000\,(0.004) & 0.994\,(0.005) & 0.997\,(0.006) & 1.001\,(0.007) \\
 & 800 & 1.009\,(0.004) & 1.003\,(0.003) & 0.999\,(0.003) & 1.001\,(0.005) & 1.015\,(0.005) \\
\addlinespace[2.2pt]
3 & 100 & 0.959\,(0.016) & 0.981\,(0.012) & 1.006\,(0.011) & 1.003\,(0.014) & 0.992\,(0.014) \\
 & 200 & 0.994\,(0.012) & 0.999\,(0.010) & 1.013\,(0.011) & 1.008\,(0.012) & 1.019\,(0.011) \\
 & 400 & 0.987\,(0.009) & 0.997\,(0.010) & 0.989\,(0.009) & 0.990\,(0.009) & 1.017\,(0.010) \\
 & 800 & 0.954\,(0.012) & 0.987\,(0.012) & 0.998\,(0.012) & 1.015\,(0.009) & 1.010\,(0.010) \\
\addlinespace[2.2pt]
5 & 100 & 0.984\,(0.017) & 1.017\,(0.018) & 0.997\,(0.017) & 1.020\,(0.016) & 1.050\,(0.016) \\
 & 200 & 0.996\,(0.017) & 0.986\,(0.013) & 1.005\,(0.015) & 1.030\,(0.014) & 1.047\,(0.012) \\
 & 400 & 0.994\,(0.018) & 1.002\,(0.015) & 0.999\,(0.012) & 1.030\,(0.014) & 1.022\,(0.015) \\
 & 800 & 0.958\,(0.016) & 0.998\,(0.014) & 1.035\,(0.015) & 1.009\,(0.015) & 1.015\,(0.012) \\
\addlinespace[2.2pt]
10 & 100 & 0.983\,(0.022) & 1.058\,(0.021) & 1.041\,(0.018) & 1.083\,(0.022) & 1.119\,(0.030) \\
 & 200 & 1.017\,(0.023) & 1.043\,(0.020) & 1.070\,(0.025) & 1.057\,(0.026) & 1.129\,(0.033) \\
 & 400 & 0.966\,(0.022) & 1.034\,(0.019) & 1.070\,(0.026) & 1.046\,(0.020) & 1.047\,(0.025) \\
 & 800 & 0.999\,(0.023) & 0.984\,(0.019) & 1.045\,(0.022) & 1.074\,(0.025) & 1.074\,(0.023) \\
\addlinespace[2.2pt]
20 & 100 & 1.027\,(0.033) & 1.061\,(0.032) & 1.139\,(0.036) & 1.086\,(0.032) & 1.261\,(0.048) \\
 & 200 & 1.052\,(0.028) & 1.070\,(0.026) & 1.093\,(0.026) & 1.120\,(0.035) & 1.136\,(0.035) \\
 & 400 & 1.025\,(0.029) & 1.057\,(0.034) & 1.046\,(0.027) & 1.077\,(0.027) & 1.148\,(0.034) \\
 & 800 & 1.017\,(0.026) & 1.081\,(0.029) & 1.089\,(0.028) & 1.061\,(0.031) & 1.142\,(0.037) \\
\addlinespace[2.2pt]
50 & 100 & 1.032\,(0.047) & 1.318\,(0.074) & 1.337\,(0.061) & 1.253\,(0.059) & 1.171\,(0.052) \\
 & 200 & 1.066\,(0.035) & 1.235\,(0.057) & 1.114\,(0.043) & 1.100\,(0.043) & 1.096\,(0.042) \\
 & 400 & 1.102\,(0.044) & 1.166\,(0.036) & 1.244\,(0.050) & 1.138\,(0.043) & 1.163\,(0.043) \\
 & 800 & 1.114\,(0.035) & 1.086\,(0.032) & 1.157\,(0.036) & 1.256\,(0.051) & 1.235\,(0.052) \\
\bottomrule
\end{tabular*}
\par\vspace{7pt}
\begin{minipage}{\linewidth}
\fontsize{8}{9.5}\selectfont
\textit{For all six tables.}
Each configuration uses 200 paired datasets, with
\(d_\kappa=0.5\) and \(\operatorname{Tr}(\Sigma^{(0)})=1\).
Parentheses contain Monte Carlo standard errors (percentage points
in panel~(a)). Non-SPD frequencies are measured before correction;
loading MSEs use corrected covariance inputs and the squared
projective distance \(d^2\). Ratios above one favor \(\Pi_\perp\).
\end{minipage}
\endgroup
\end{table}
\clearpage

\begin{table}[p]
\centering
\caption{Non-SPD frequencies and loading MSE ratios: $p=9$, $\rho_\phi=0$, $\mu=0^\circ$, $s=1$.}
\label{tab:rep-s1-profile01-p9}
\begingroup
\fontsize{8.5}{10}\selectfont
\setlength{\tabcolsep}{4pt}
\renewcommand{\arraystretch}{1.03}
\begin{tabular*}{\linewidth}{@{\extracolsep{\fill}}rrccccc@{}}
\toprule
$\mathrm{CN}$ & $n$ & $\eta=1$ & $\eta=3$ & $\eta=5$ & $\eta=10$ & $\eta=20$ \\
\midrule
\multicolumn{7}{@{}l}{\rule{0pt}{13pt}\textbf{(a) Non-SPD frequency (\%)}} \\[2pt]
1 & 100 & 0.0\,(0.00) & 0.0\,(0.00) & 0.0\,(0.00) & 0.0\,(0.00) & 0.0\,(0.00) \\
 & 200 & 0.0\,(0.00) & 0.0\,(0.00) & 0.0\,(0.00) & 0.0\,(0.00) & 0.0\,(0.00) \\
 & 400 & 0.0\,(0.00) & 0.0\,(0.00) & 0.0\,(0.00) & 0.0\,(0.00) & 0.0\,(0.00) \\
 & 800 & 0.0\,(0.00) & 0.0\,(0.00) & 0.0\,(0.00) & 0.0\,(0.00) & 0.0\,(0.00) \\
\addlinespace[2.2pt]
3 & 100 & 0.0\,(0.00) & 0.0\,(0.00) & 0.0\,(0.00) & 0.0\,(0.00) & 0.0\,(0.00) \\
 & 200 & 0.0\,(0.00) & 0.0\,(0.00) & 0.0\,(0.00) & 0.0\,(0.00) & 0.0\,(0.00) \\
 & 400 & 0.0\,(0.00) & 0.0\,(0.00) & 0.0\,(0.00) & 0.0\,(0.00) & 0.0\,(0.00) \\
 & 800 & 0.0\,(0.00) & 0.0\,(0.00) & 0.0\,(0.00) & 0.0\,(0.00) & 0.0\,(0.00) \\
\addlinespace[2.2pt]
5 & 100 & 0.0\,(0.00) & 0.0\,(0.00) & 0.0\,(0.00) & 0.0\,(0.00) & 0.0\,(0.00) \\
 & 200 & 0.0\,(0.00) & 0.0\,(0.00) & 0.0\,(0.00) & 0.0\,(0.00) & 0.0\,(0.00) \\
 & 400 & 0.0\,(0.00) & 0.0\,(0.00) & 0.0\,(0.00) & 0.0\,(0.00) & 0.0\,(0.00) \\
 & 800 & 0.0\,(0.00) & 0.0\,(0.00) & 0.0\,(0.00) & 0.0\,(0.00) & 0.0\,(0.00) \\
\addlinespace[2.2pt]
10 & 100 & 0.0\,(0.00) & 0.0\,(0.00) & 0.0\,(0.00) & 0.0\,(0.00) & 0.0\,(0.00) \\
 & 200 & 0.0\,(0.00) & 0.0\,(0.00) & 0.0\,(0.00) & 0.0\,(0.00) & 0.0\,(0.00) \\
 & 400 & 0.0\,(0.00) & 0.0\,(0.00) & 0.0\,(0.00) & 0.0\,(0.00) & 0.0\,(0.00) \\
 & 800 & 0.0\,(0.00) & 0.0\,(0.00) & 0.0\,(0.00) & 0.0\,(0.00) & 0.0\,(0.00) \\
\addlinespace[2.2pt]
20 & 100 & 0.0\,(0.00) & 0.0\,(0.00) & 0.0\,(0.00) & 0.0\,(0.00) & 0.5\,(0.50) \\
 & 200 & 0.0\,(0.00) & 0.0\,(0.00) & 0.0\,(0.00) & 0.0\,(0.00) & 0.0\,(0.00) \\
 & 400 & 0.0\,(0.00) & 0.0\,(0.00) & 0.0\,(0.00) & 0.0\,(0.00) & 0.0\,(0.00) \\
 & 800 & 0.0\,(0.00) & 0.0\,(0.00) & 0.0\,(0.00) & 0.0\,(0.00) & 0.0\,(0.00) \\
\addlinespace[2.2pt]
50 & 100 & 0.0\,(0.00) & 0.5\,(0.50) & 0.5\,(0.50) & 6.0\,(1.68) & 12.5\,(2.34) \\
 & 200 & 0.0\,(0.00) & 0.0\,(0.00) & 0.0\,(0.00) & 0.5\,(0.50) & 4.5\,(1.47) \\
 & 400 & 0.0\,(0.00) & 0.0\,(0.00) & 0.0\,(0.00) & 0.5\,(0.50) & 0.5\,(0.50) \\
 & 800 & 0.0\,(0.00) & 0.0\,(0.00) & 0.0\,(0.00) & 0.0\,(0.00) & 0.0\,(0.00) \\
\midrule
\multicolumn{7}{@{}l}{\rule{0pt}{13pt}\textbf{(b) Loading MSE ratio ($P_\perp/\Pi_\perp$)}} \\[2pt]
1 & 100 & 1.000\,(0.001) & 0.999\,(0.001) & 1.001\,(0.001) & 1.001\,(0.001) & 1.000\,(0.002) \\
 & 200 & 0.999\,(0.001) & 1.000\,(0.001) & 0.999\,(0.001) & 1.000\,(0.001) & 1.002\,(0.001) \\
 & 400 & 1.000\,(0.000) & 1.000\,(0.000) & 1.000\,(0.001) & 1.001\,(0.001) & 1.001\,(0.001) \\
 & 800 & 0.999\,(0.000) & 1.000\,(0.000) & 1.000\,(0.000) & 1.000\,(0.000) & 1.000\,(0.001) \\
\addlinespace[2.2pt]
3 & 100 & 1.001\,(0.004) & 1.000\,(0.001) & 1.000\,(0.001) & 1.000\,(0.001) & 1.001\,(0.002) \\
 & 200 & 0.998\,(0.001) & 1.000\,(0.001) & 1.000\,(0.001) & 0.999\,(0.001) & 1.002\,(0.002) \\
 & 400 & 1.000\,(0.001) & 1.000\,(0.001) & 1.001\,(0.001) & 1.000\,(0.001) & 0.999\,(0.001) \\
 & 800 & 0.999\,(0.001) & 0.999\,(0.001) & 0.999\,(0.001) & 1.000\,(0.001) & 1.000\,(0.001) \\
\addlinespace[2.2pt]
5 & 100 & 0.998\,(0.003) & 1.001\,(0.002) & 1.002\,(0.002) & 1.002\,(0.002) & 1.001\,(0.002) \\
 & 200 & 0.998\,(0.002) & 0.998\,(0.002) & 1.000\,(0.001) & 0.999\,(0.001) & 1.003\,(0.002) \\
 & 400 & 0.998\,(0.001) & 1.000\,(0.001) & 0.999\,(0.001) & 0.999\,(0.001) & 1.000\,(0.001) \\
 & 800 & 0.998\,(0.001) & 0.998\,(0.001) & 0.999\,(0.001) & 1.001\,(0.001) & 1.001\,(0.001) \\
\addlinespace[2.2pt]
10 & 100 & 0.984\,(0.003) & 1.000\,(0.003) & 1.000\,(0.002) & 1.007\,(0.003) & 1.011\,(0.004) \\
 & 200 & 0.991\,(0.003) & 1.001\,(0.002) & 1.000\,(0.002) & 1.000\,(0.002) & 1.002\,(0.002) \\
 & 400 & 0.994\,(0.003) & 0.999\,(0.002) & 0.999\,(0.002) & 1.002\,(0.002) & 1.003\,(0.002) \\
 & 800 & 0.997\,(0.002) & 0.999\,(0.002) & 1.001\,(0.002) & 1.002\,(0.002) & 1.001\,(0.002) \\
\addlinespace[2.2pt]
20 & 100 & 0.984\,(0.005) & 0.996\,(0.004) & 1.002\,(0.004) & 1.010\,(0.004) & 1.011\,(0.006) \\
 & 200 & 0.987\,(0.005) & 0.993\,(0.005) & 1.002\,(0.003) & 1.003\,(0.004) & 1.005\,(0.004) \\
 & 400 & 1.001\,(0.003) & 1.003\,(0.003) & 0.997\,(0.003) & 1.002\,(0.003) & 0.996\,(0.003) \\
 & 800 & 0.990\,(0.004) & 1.004\,(0.003) & 1.007\,(0.003) & 1.006\,(0.003) & 1.002\,(0.004) \\
\addlinespace[2.2pt]
50 & 100 & 0.968\,(0.009) & 0.993\,(0.006) & 0.996\,(0.009) & 1.051\,(0.018) & 1.059\,(0.021) \\
 & 200 & 0.992\,(0.007) & 1.006\,(0.006) & 1.009\,(0.006) & 1.012\,(0.006) & 1.030\,(0.013) \\
 & 400 & 0.982\,(0.007) & 0.999\,(0.006) & 1.001\,(0.005) & 1.008\,(0.005) & 1.005\,(0.006) \\
 & 800 & 0.989\,(0.007) & 0.995\,(0.005) & 1.002\,(0.006) & 1.001\,(0.006) & 1.008\,(0.005) \\
\bottomrule
\end{tabular*}
\endgroup
\end{table}
\clearpage

\begin{table}[p]
\centering
\caption{Non-SPD frequencies and loading MSE ratios: $p=3$, $\rho_\phi=0.5$, $\mu=0^\circ$, $s=1$.}
\label{tab:rep-s1-profile12-p3}
\begingroup
\fontsize{8.5}{10}\selectfont
\setlength{\tabcolsep}{4pt}
\renewcommand{\arraystretch}{1.03}
\begin{tabular*}{\linewidth}{@{\extracolsep{\fill}}rrccccc@{}}
\toprule
$\mathrm{CN}$ & $n$ & $\eta=1$ & $\eta=3$ & $\eta=5$ & $\eta=10$ & $\eta=20$ \\
\midrule
\multicolumn{7}{@{}l}{\rule{0pt}{13pt}\textbf{(a) Non-SPD frequency (\%)}} \\[2pt]
1 & 100 & 0.0\,(0.00) & 0.0\,(0.00) & 0.0\,(0.00) & 0.0\,(0.00) & 0.0\,(0.00) \\
 & 200 & 0.0\,(0.00) & 0.0\,(0.00) & 0.0\,(0.00) & 0.0\,(0.00) & 0.0\,(0.00) \\
 & 400 & 0.0\,(0.00) & 0.0\,(0.00) & 0.0\,(0.00) & 0.0\,(0.00) & 0.0\,(0.00) \\
 & 800 & 0.0\,(0.00) & 0.0\,(0.00) & 0.0\,(0.00) & 0.0\,(0.00) & 0.0\,(0.00) \\
\addlinespace[2.2pt]
3 & 100 & 0.0\,(0.00) & 0.0\,(0.00) & 0.0\,(0.00) & 0.0\,(0.00) & 0.0\,(0.00) \\
 & 200 & 0.0\,(0.00) & 0.0\,(0.00) & 0.0\,(0.00) & 0.0\,(0.00) & 0.0\,(0.00) \\
 & 400 & 0.0\,(0.00) & 0.0\,(0.00) & 0.0\,(0.00) & 0.0\,(0.00) & 0.0\,(0.00) \\
 & 800 & 0.0\,(0.00) & 0.0\,(0.00) & 0.0\,(0.00) & 0.0\,(0.00) & 0.0\,(0.00) \\
\addlinespace[2.2pt]
5 & 100 & 0.0\,(0.00) & 0.0\,(0.00) & 0.0\,(0.00) & 0.0\,(0.00) & 1.0\,(0.71) \\
 & 200 & 0.0\,(0.00) & 0.0\,(0.00) & 0.0\,(0.00) & 0.0\,(0.00) & 0.0\,(0.00) \\
 & 400 & 0.0\,(0.00) & 0.0\,(0.00) & 0.0\,(0.00) & 0.0\,(0.00) & 0.0\,(0.00) \\
 & 800 & 0.0\,(0.00) & 0.0\,(0.00) & 0.0\,(0.00) & 0.0\,(0.00) & 0.0\,(0.00) \\
\addlinespace[2.2pt]
10 & 100 & 0.0\,(0.00) & 0.0\,(0.00) & 0.0\,(0.00) & 3.0\,(1.21) & 7.0\,(1.81) \\
 & 200 & 0.0\,(0.00) & 0.0\,(0.00) & 0.0\,(0.00) & 2.0\,(0.99) & 1.5\,(0.86) \\
 & 400 & 0.0\,(0.00) & 0.0\,(0.00) & 0.0\,(0.00) & 0.0\,(0.00) & 0.0\,(0.00) \\
 & 800 & 0.0\,(0.00) & 0.0\,(0.00) & 0.0\,(0.00) & 0.0\,(0.00) & 0.0\,(0.00) \\
\addlinespace[2.2pt]
20 & 100 & 1.0\,(0.71) & 3.5\,(1.30) & 7.0\,(1.81) & 12.0\,(2.30) & 22.0\,(2.94) \\
 & 200 & 0.0\,(0.00) & 0.5\,(0.50) & 4.0\,(1.39) & 5.0\,(1.54) & 14.0\,(2.46) \\
 & 400 & 0.0\,(0.00) & 0.0\,(0.00) & 0.5\,(0.50) & 2.0\,(0.99) & 7.0\,(1.81) \\
 & 800 & 0.0\,(0.00) & 0.0\,(0.00) & 0.0\,(0.00) & 0.5\,(0.50) & 2.5\,(1.11) \\
\addlinespace[2.2pt]
50 & 100 & 6.0\,(1.68) & 27.5\,(3.17) & 21.5\,(2.91) & 29.5\,(3.23) & 41.5\,(3.49) \\
 & 200 & 5.5\,(1.62) & 10.5\,(2.17) & 22.5\,(2.96) & 31.0\,(3.28) & 33.0\,(3.33) \\
 & 400 & 1.0\,(0.71) & 6.5\,(1.75) & 14.5\,(2.50) & 20.0\,(2.84) & 33.0\,(3.33) \\
 & 800 & 0.0\,(0.00) & 3.5\,(1.30) & 5.0\,(1.54) & 11.0\,(2.22) & 19.0\,(2.78) \\
\midrule
\multicolumn{7}{@{}l}{\rule{0pt}{13pt}\textbf{(b) Loading MSE ratio ($P_\perp/\Pi_\perp$)}} \\[2pt]
1 & 100 & 1.011\,(0.009) & 1.004\,(0.009) & 0.988\,(0.009) & 0.991\,(0.009) & 1.014\,(0.009) \\
 & 200 & 1.008\,(0.006) & 1.006\,(0.006) & 0.999\,(0.006) & 1.001\,(0.008) & 1.000\,(0.008) \\
 & 400 & 1.002\,(0.005) & 0.997\,(0.005) & 0.993\,(0.004) & 1.005\,(0.004) & 1.016\,(0.005) \\
 & 800 & 0.996\,(0.003) & 1.001\,(0.003) & 1.002\,(0.003) & 1.004\,(0.003) & 0.999\,(0.003) \\
\addlinespace[2.2pt]
3 & 100 & 1.003\,(0.013) & 1.009\,(0.014) & 1.016\,(0.011) & 1.040\,(0.013) & 1.035\,(0.015) \\
 & 200 & 0.987\,(0.009) & 1.000\,(0.010) & 0.999\,(0.008) & 1.004\,(0.010) & 1.030\,(0.012) \\
 & 400 & 1.005\,(0.011) & 1.011\,(0.010) & 1.011\,(0.010) & 1.000\,(0.008) & 1.006\,(0.008) \\
 & 800 & 1.007\,(0.011) & 0.999\,(0.009) & 1.016\,(0.010) & 1.018\,(0.007) & 1.008\,(0.007) \\
\addlinespace[2.2pt]
5 & 100 & 0.993\,(0.018) & 1.004\,(0.016) & 1.004\,(0.014) & 1.023\,(0.013) & 1.043\,(0.018) \\
 & 200 & 0.987\,(0.015) & 1.023\,(0.014) & 1.003\,(0.014) & 1.020\,(0.012) & 1.037\,(0.015) \\
 & 400 & 0.992\,(0.014) & 1.002\,(0.012) & 1.014\,(0.013) & 1.021\,(0.010) & 1.014\,(0.010) \\
 & 800 & 1.014\,(0.012) & 0.976\,(0.012) & 1.010\,(0.010) & 1.001\,(0.009) & 1.018\,(0.011) \\
\addlinespace[2.2pt]
10 & 100 & 1.039\,(0.023) & 1.037\,(0.020) & 1.033\,(0.019) & 1.051\,(0.020) & 1.032\,(0.027) \\
 & 200 & 1.028\,(0.023) & 1.018\,(0.015) & 1.033\,(0.019) & 1.054\,(0.018) & 1.064\,(0.018) \\
 & 400 & 1.068\,(0.021) & 1.007\,(0.019) & 1.066\,(0.017) & 1.022\,(0.013) & 1.055\,(0.018) \\
 & 800 & 1.016\,(0.021) & 1.021\,(0.018) & 1.072\,(0.018) & 1.046\,(0.012) & 1.035\,(0.017) \\
\addlinespace[2.2pt]
20 & 100 & 1.032\,(0.025) & 1.065\,(0.026) & 1.094\,(0.029) & 1.094\,(0.030) & 1.129\,(0.030) \\
 & 200 & 1.027\,(0.021) & 1.042\,(0.022) & 1.097\,(0.039) & 1.083\,(0.023) & 1.123\,(0.035) \\
 & 400 & 1.065\,(0.021) & 1.081\,(0.021) & 1.071\,(0.022) & 1.040\,(0.019) & 1.059\,(0.025) \\
 & 800 & 1.055\,(0.022) & 1.061\,(0.027) & 1.049\,(0.023) & 1.074\,(0.023) & 1.083\,(0.026) \\
\addlinespace[2.2pt]
50 & 100 & 1.199\,(0.069) & 1.242\,(0.056) & 1.169\,(0.036) & 1.225\,(0.046) & 1.143\,(0.054) \\
 & 200 & 1.355\,(0.088) & 1.133\,(0.047) & 1.229\,(0.050) & 1.161\,(0.048) & 1.122\,(0.038) \\
 & 400 & 1.150\,(0.045) & 1.119\,(0.045) & 1.119\,(0.034) & 1.115\,(0.038) & 1.171\,(0.037) \\
 & 800 & 1.073\,(0.026) & 1.105\,(0.034) & 1.148\,(0.031) & 1.100\,(0.035) & 1.143\,(0.032) \\
\bottomrule
\end{tabular*}
\endgroup
\end{table}
\clearpage

\begin{table}[p]
\centering
\caption{Non-SPD frequencies and loading MSE ratios: $p=9$, $\rho_\phi=0.5$, $\mu=0^\circ$, $s=1$.}
\label{tab:rep-s1-profile12-p9}
\begingroup
\fontsize{8.5}{10}\selectfont
\setlength{\tabcolsep}{4pt}
\renewcommand{\arraystretch}{1.03}
\begin{tabular*}{\linewidth}{@{\extracolsep{\fill}}rrccccc@{}}
\toprule
$\mathrm{CN}$ & $n$ & $\eta=1$ & $\eta=3$ & $\eta=5$ & $\eta=10$ & $\eta=20$ \\
\midrule
\multicolumn{7}{@{}l}{\rule{0pt}{13pt}\textbf{(a) Non-SPD frequency (\%)}} \\[2pt]
1 & 100 & 0.0\,(0.00) & 0.0\,(0.00) & 0.0\,(0.00) & 0.0\,(0.00) & 0.0\,(0.00) \\
 & 200 & 0.0\,(0.00) & 0.0\,(0.00) & 0.0\,(0.00) & 0.0\,(0.00) & 0.0\,(0.00) \\
 & 400 & 0.0\,(0.00) & 0.0\,(0.00) & 0.0\,(0.00) & 0.0\,(0.00) & 0.0\,(0.00) \\
 & 800 & 0.0\,(0.00) & 0.0\,(0.00) & 0.0\,(0.00) & 0.0\,(0.00) & 0.0\,(0.00) \\
\addlinespace[2.2pt]
3 & 100 & 0.0\,(0.00) & 0.0\,(0.00) & 0.0\,(0.00) & 0.0\,(0.00) & 0.0\,(0.00) \\
 & 200 & 0.0\,(0.00) & 0.0\,(0.00) & 0.0\,(0.00) & 0.0\,(0.00) & 0.0\,(0.00) \\
 & 400 & 0.0\,(0.00) & 0.0\,(0.00) & 0.0\,(0.00) & 0.0\,(0.00) & 0.0\,(0.00) \\
 & 800 & 0.0\,(0.00) & 0.0\,(0.00) & 0.0\,(0.00) & 0.0\,(0.00) & 0.0\,(0.00) \\
\addlinespace[2.2pt]
5 & 100 & 0.0\,(0.00) & 0.0\,(0.00) & 0.0\,(0.00) & 0.0\,(0.00) & 0.0\,(0.00) \\
 & 200 & 0.0\,(0.00) & 0.0\,(0.00) & 0.0\,(0.00) & 0.0\,(0.00) & 0.0\,(0.00) \\
 & 400 & 0.0\,(0.00) & 0.0\,(0.00) & 0.0\,(0.00) & 0.0\,(0.00) & 0.0\,(0.00) \\
 & 800 & 0.0\,(0.00) & 0.0\,(0.00) & 0.0\,(0.00) & 0.0\,(0.00) & 0.0\,(0.00) \\
\addlinespace[2.2pt]
10 & 100 & 0.0\,(0.00) & 0.0\,(0.00) & 0.0\,(0.00) & 0.0\,(0.00) & 0.0\,(0.00) \\
 & 200 & 0.0\,(0.00) & 0.0\,(0.00) & 0.0\,(0.00) & 0.0\,(0.00) & 0.0\,(0.00) \\
 & 400 & 0.0\,(0.00) & 0.0\,(0.00) & 0.0\,(0.00) & 0.0\,(0.00) & 0.0\,(0.00) \\
 & 800 & 0.0\,(0.00) & 0.0\,(0.00) & 0.0\,(0.00) & 0.0\,(0.00) & 0.0\,(0.00) \\
\addlinespace[2.2pt]
20 & 100 & 0.0\,(0.00) & 0.0\,(0.00) & 0.0\,(0.00) & 0.0\,(0.00) & 0.0\,(0.00) \\
 & 200 & 0.0\,(0.00) & 0.0\,(0.00) & 0.0\,(0.00) & 0.0\,(0.00) & 0.0\,(0.00) \\
 & 400 & 0.0\,(0.00) & 0.0\,(0.00) & 0.0\,(0.00) & 0.0\,(0.00) & 0.0\,(0.00) \\
 & 800 & 0.0\,(0.00) & 0.0\,(0.00) & 0.0\,(0.00) & 0.0\,(0.00) & 0.0\,(0.00) \\
\addlinespace[2.2pt]
50 & 100 & 0.0\,(0.00) & 0.0\,(0.00) & 1.0\,(0.71) & 3.0\,(1.21) & 12.5\,(2.34) \\
 & 200 & 0.0\,(0.00) & 0.0\,(0.00) & 0.0\,(0.00) & 1.0\,(0.71) & 5.5\,(1.62) \\
 & 400 & 0.0\,(0.00) & 0.0\,(0.00) & 0.0\,(0.00) & 0.0\,(0.00) & 0.5\,(0.50) \\
 & 800 & 0.0\,(0.00) & 0.0\,(0.00) & 0.0\,(0.00) & 0.0\,(0.00) & 0.0\,(0.00) \\
\midrule
\multicolumn{7}{@{}l}{\rule{0pt}{13pt}\textbf{(b) Loading MSE ratio ($P_\perp/\Pi_\perp$)}} \\[2pt]
1 & 100 & 0.999\,(0.001) & 1.000\,(0.001) & 0.999\,(0.001) & 1.002\,(0.001) & 1.000\,(0.001) \\
 & 200 & 0.999\,(0.001) & 1.000\,(0.001) & 0.998\,(0.001) & 1.001\,(0.001) & 1.000\,(0.001) \\
 & 400 & 1.000\,(0.000) & 1.000\,(0.000) & 0.999\,(0.001) & 1.001\,(0.001) & 1.000\,(0.001) \\
 & 800 & 1.000\,(0.000) & 1.001\,(0.000) & 1.000\,(0.000) & 1.000\,(0.000) & 1.000\,(0.000) \\
\addlinespace[2.2pt]
3 & 100 & 1.000\,(0.002) & 0.999\,(0.001) & 1.002\,(0.001) & 0.999\,(0.001) & 1.002\,(0.002) \\
 & 200 & 1.002\,(0.001) & 1.001\,(0.001) & 1.000\,(0.001) & 1.001\,(0.001) & 1.002\,(0.001) \\
 & 400 & 1.002\,(0.001) & 1.001\,(0.001) & 0.999\,(0.001) & 0.998\,(0.001) & 1.001\,(0.001) \\
 & 800 & 1.001\,(0.001) & 0.999\,(0.001) & 1.000\,(0.001) & 1.000\,(0.001) & 1.000\,(0.001) \\
\addlinespace[2.2pt]
5 & 100 & 0.993\,(0.002) & 1.003\,(0.002) & 1.003\,(0.002) & 1.004\,(0.002) & 1.000\,(0.002) \\
 & 200 & 1.000\,(0.001) & 1.000\,(0.001) & 1.002\,(0.001) & 1.002\,(0.001) & 1.001\,(0.002) \\
 & 400 & 1.001\,(0.001) & 1.001\,(0.001) & 1.000\,(0.001) & 1.003\,(0.001) & 1.001\,(0.002) \\
 & 800 & 1.001\,(0.001) & 1.001\,(0.001) & 1.001\,(0.001) & 1.001\,(0.001) & 1.001\,(0.001) \\
\addlinespace[2.2pt]
10 & 100 & 0.993\,(0.003) & 1.001\,(0.002) & 1.004\,(0.003) & 1.002\,(0.003) & 0.996\,(0.003) \\
 & 200 & 0.996\,(0.002) & 1.002\,(0.002) & 1.002\,(0.002) & 1.003\,(0.002) & 1.001\,(0.002) \\
 & 400 & 0.998\,(0.002) & 1.000\,(0.001) & 1.002\,(0.002) & 1.002\,(0.002) & 1.005\,(0.003) \\
 & 800 & 0.999\,(0.002) & 1.002\,(0.001) & 0.998\,(0.001) & 1.000\,(0.001) & 0.999\,(0.001) \\
\addlinespace[2.2pt]
20 & 100 & 0.987\,(0.004) & 1.002\,(0.004) & 1.001\,(0.003) & 1.009\,(0.004) & 1.015\,(0.004) \\
 & 200 & 1.001\,(0.003) & 1.002\,(0.002) & 0.999\,(0.003) & 1.008\,(0.003) & 1.010\,(0.004) \\
 & 400 & 1.002\,(0.003) & 1.003\,(0.002) & 1.005\,(0.003) & 1.001\,(0.003) & 1.004\,(0.003) \\
 & 800 & 0.993\,(0.002) & 1.003\,(0.002) & 0.998\,(0.002) & 1.001\,(0.002) & 1.004\,(0.002) \\
\addlinespace[2.2pt]
50 & 100 & 0.980\,(0.007) & 0.999\,(0.005) & 1.020\,(0.011) & 1.020\,(0.008) & 1.050\,(0.017) \\
 & 200 & 0.991\,(0.005) & 1.006\,(0.005) & 1.004\,(0.004) & 1.014\,(0.006) & 1.037\,(0.013) \\
 & 400 & 0.998\,(0.005) & 0.998\,(0.004) & 1.003\,(0.004) & 1.007\,(0.003) & 1.019\,(0.006) \\
 & 800 & 1.006\,(0.004) & 1.004\,(0.003) & 1.000\,(0.004) & 1.006\,(0.004) & 1.005\,(0.003) \\
\bottomrule
\end{tabular*}
\endgroup
\end{table}
\clearpage

\begin{table}[p]
\centering
\caption{Non-SPD frequencies and loading MSE ratios: $p=3$, $\rho_\phi=0.9$, $\mu=45^\circ$, $s=1$.}
\label{tab:rep-s1-profile29-p3}
\begingroup
\fontsize{8.5}{10}\selectfont
\setlength{\tabcolsep}{4pt}
\renewcommand{\arraystretch}{1.03}
\begin{tabular*}{\linewidth}{@{\extracolsep{\fill}}rrccccc@{}}
\toprule
$\mathrm{CN}$ & $n$ & $\eta=1$ & $\eta=3$ & $\eta=5$ & $\eta=10$ & $\eta=20$ \\
\midrule
\multicolumn{7}{@{}l}{\rule{0pt}{13pt}\textbf{(a) Non-SPD frequency (\%)}} \\[2pt]
1 & 100 & 0.0\,(0.00) & 0.0\,(0.00) & 0.0\,(0.00) & 0.0\,(0.00) & 0.0\,(0.00) \\
 & 200 & 0.0\,(0.00) & 0.0\,(0.00) & 0.0\,(0.00) & 0.0\,(0.00) & 0.0\,(0.00) \\
 & 400 & 0.0\,(0.00) & 0.0\,(0.00) & 0.0\,(0.00) & 0.0\,(0.00) & 0.0\,(0.00) \\
 & 800 & 0.0\,(0.00) & 0.0\,(0.00) & 0.0\,(0.00) & 0.0\,(0.00) & 0.0\,(0.00) \\
\addlinespace[2.2pt]
3 & 100 & 0.0\,(0.00) & 0.0\,(0.00) & 0.0\,(0.00) & 0.0\,(0.00) & 0.0\,(0.00) \\
 & 200 & 0.0\,(0.00) & 0.0\,(0.00) & 0.0\,(0.00) & 0.0\,(0.00) & 0.0\,(0.00) \\
 & 400 & 0.0\,(0.00) & 0.0\,(0.00) & 0.0\,(0.00) & 0.0\,(0.00) & 0.0\,(0.00) \\
 & 800 & 0.0\,(0.00) & 0.0\,(0.00) & 0.0\,(0.00) & 0.0\,(0.00) & 0.0\,(0.00) \\
\addlinespace[2.2pt]
5 & 100 & 0.0\,(0.00) & 0.0\,(0.00) & 0.0\,(0.00) & 0.0\,(0.00) & 0.0\,(0.00) \\
 & 200 & 0.0\,(0.00) & 0.0\,(0.00) & 0.0\,(0.00) & 0.0\,(0.00) & 0.0\,(0.00) \\
 & 400 & 0.0\,(0.00) & 0.0\,(0.00) & 0.0\,(0.00) & 0.0\,(0.00) & 0.0\,(0.00) \\
 & 800 & 0.0\,(0.00) & 0.0\,(0.00) & 0.0\,(0.00) & 0.0\,(0.00) & 0.0\,(0.00) \\
\addlinespace[2.2pt]
10 & 100 & 0.0\,(0.00) & 0.0\,(0.00) & 0.0\,(0.00) & 0.0\,(0.00) & 1.0\,(0.71) \\
 & 200 & 0.0\,(0.00) & 0.0\,(0.00) & 0.0\,(0.00) & 0.0\,(0.00) & 0.0\,(0.00) \\
 & 400 & 0.0\,(0.00) & 0.0\,(0.00) & 0.0\,(0.00) & 0.0\,(0.00) & 0.0\,(0.00) \\
 & 800 & 0.0\,(0.00) & 0.0\,(0.00) & 0.0\,(0.00) & 0.0\,(0.00) & 0.0\,(0.00) \\
\addlinespace[2.2pt]
20 & 100 & 0.0\,(0.00) & 1.5\,(0.86) & 4.0\,(1.39) & 8.0\,(1.92) & 10.0\,(2.13) \\
 & 200 & 0.0\,(0.00) & 0.5\,(0.50) & 0.5\,(0.50) & 1.5\,(0.86) & 3.0\,(1.21) \\
 & 400 & 0.0\,(0.00) & 0.0\,(0.00) & 0.0\,(0.00) & 0.0\,(0.00) & 1.0\,(0.71) \\
 & 800 & 0.0\,(0.00) & 0.0\,(0.00) & 0.0\,(0.00) & 0.0\,(0.00) & 0.0\,(0.00) \\
\addlinespace[2.2pt]
50 & 100 & 7.0\,(1.81) & 17.0\,(2.66) & 20.5\,(2.86) & 25.5\,(3.09) & 23.0\,(2.98) \\
 & 200 & 4.5\,(1.47) & 9.5\,(2.08) & 11.0\,(2.22) & 13.5\,(2.42) & 20.0\,(2.84) \\
 & 400 & 1.0\,(0.71) & 3.5\,(1.30) & 5.5\,(1.62) & 10.0\,(2.13) & 15.0\,(2.53) \\
 & 800 & 0.0\,(0.00) & 1.5\,(0.86) & 1.0\,(0.71) & 2.5\,(1.11) & 5.0\,(1.54) \\
\midrule
\multicolumn{7}{@{}l}{\rule{0pt}{13pt}\textbf{(b) Loading MSE ratio ($P_\perp/\Pi_\perp$)}} \\[2pt]
1 & 100 & 1.003\,(0.011) & 0.997\,(0.008) & 1.001\,(0.006) & 0.999\,(0.006) & 1.001\,(0.005) \\
 & 200 & 1.017\,(0.008) & 0.983\,(0.006) & 1.002\,(0.005) & 0.999\,(0.005) & 0.999\,(0.005) \\
 & 400 & 1.007\,(0.005) & 1.009\,(0.004) & 1.003\,(0.004) & 1.001\,(0.003) & 0.997\,(0.003) \\
 & 800 & 0.997\,(0.004) & 1.000\,(0.003) & 1.001\,(0.002) & 0.995\,(0.002) & 1.002\,(0.002) \\
\addlinespace[2.2pt]
3 & 100 & 1.021\,(0.013) & 0.991\,(0.012) & 0.970\,(0.010) & 1.015\,(0.010) & 0.995\,(0.010) \\
 & 200 & 0.995\,(0.013) & 0.981\,(0.011) & 0.981\,(0.008) & 0.994\,(0.008) & 1.007\,(0.007) \\
 & 400 & 1.001\,(0.013) & 0.983\,(0.009) & 1.013\,(0.009) & 0.988\,(0.006) & 0.992\,(0.005) \\
 & 800 & 0.955\,(0.012) & 0.977\,(0.008) & 0.984\,(0.008) & 0.995\,(0.006) & 1.006\,(0.005) \\
\addlinespace[2.2pt]
5 & 100 & 1.000\,(0.025) & 0.997\,(0.016) & 1.003\,(0.013) & 0.989\,(0.012) & 1.017\,(0.013) \\
 & 200 & 0.956\,(0.019) & 0.986\,(0.014) & 0.988\,(0.012) & 0.999\,(0.011) & 0.995\,(0.010) \\
 & 400 & 0.973\,(0.015) & 1.001\,(0.014) & 0.997\,(0.011) & 1.015\,(0.011) & 1.006\,(0.009) \\
 & 800 & 0.928\,(0.018) & 0.980\,(0.015) & 0.968\,(0.010) & 0.990\,(0.007) & 1.004\,(0.008) \\
\addlinespace[2.2pt]
10 & 100 & 0.975\,(0.049) & 1.020\,(0.022) & 1.004\,(0.022) & 1.017\,(0.017) & 1.036\,(0.023) \\
 & 200 & 0.975\,(0.043) & 1.039\,(0.018) & 1.028\,(0.021) & 0.983\,(0.016) & 1.054\,(0.026) \\
 & 400 & 0.914\,(0.039) & 0.986\,(0.018) & 1.031\,(0.022) & 1.013\,(0.013) & 1.004\,(0.011) \\
 & 800 & 0.995\,(0.021) & 0.992\,(0.019) & 0.988\,(0.013) & 1.009\,(0.013) & 1.002\,(0.012) \\
\addlinespace[2.2pt]
20 & 100 & 0.999\,(0.096) & 1.079\,(0.041) & 1.089\,(0.036) & 1.118\,(0.033) & 1.111\,(0.046) \\
 & 200 & 1.202\,(0.124) & 1.044\,(0.032) & 1.044\,(0.032) & 1.048\,(0.022) & 1.092\,(0.035) \\
 & 400 & 1.117\,(0.048) & 1.019\,(0.019) & 0.994\,(0.020) & 0.995\,(0.026) & 1.061\,(0.024) \\
 & 800 & 1.128\,(0.039) & 1.078\,(0.025) & 1.060\,(0.026) & 0.999\,(0.016) & 0.996\,(0.014) \\
\addlinespace[2.2pt]
50 & 100 & 1.422\,(0.198) & 1.208\,(0.069) & 1.165\,(0.044) & 1.084\,(0.032) & 1.150\,(0.038) \\
 & 200 & 1.565\,(0.187) & 1.200\,(0.045) & 1.067\,(0.031) & 1.071\,(0.027) & 1.076\,(0.023) \\
 & 400 & 1.286\,(0.088) & 1.107\,(0.042) & 1.048\,(0.025) & 1.104\,(0.029) & 1.076\,(0.024) \\
 & 800 & 1.145\,(0.063) & 1.143\,(0.043) & 1.092\,(0.029) & 1.056\,(0.028) & 1.035\,(0.022) \\
\bottomrule
\end{tabular*}
\endgroup
\end{table}
\clearpage

\begin{table}[p]
\centering
\caption{Non-SPD frequencies and loading MSE ratios: $p=9$, $\rho_\phi=0.9$, $\mu=45^\circ$, $s=1$.}
\label{tab:rep-s1-profile29-p9}
\begingroup
\fontsize{8.5}{10}\selectfont
\setlength{\tabcolsep}{4pt}
\renewcommand{\arraystretch}{1.03}
\begin{tabular*}{\linewidth}{@{\extracolsep{\fill}}rrccccc@{}}
\toprule
$\mathrm{CN}$ & $n$ & $\eta=1$ & $\eta=3$ & $\eta=5$ & $\eta=10$ & $\eta=20$ \\
\midrule
\multicolumn{7}{@{}l}{\rule{0pt}{13pt}\textbf{(a) Non-SPD frequency (\%)}} \\[2pt]
1 & 100 & 0.0\,(0.00) & 0.0\,(0.00) & 0.0\,(0.00) & 0.0\,(0.00) & 0.0\,(0.00) \\
 & 200 & 0.0\,(0.00) & 0.0\,(0.00) & 0.0\,(0.00) & 0.0\,(0.00) & 0.0\,(0.00) \\
 & 400 & 0.0\,(0.00) & 0.0\,(0.00) & 0.0\,(0.00) & 0.0\,(0.00) & 0.0\,(0.00) \\
 & 800 & 0.0\,(0.00) & 0.0\,(0.00) & 0.0\,(0.00) & 0.0\,(0.00) & 0.0\,(0.00) \\
\addlinespace[2.2pt]
3 & 100 & 0.0\,(0.00) & 0.0\,(0.00) & 0.0\,(0.00) & 0.0\,(0.00) & 0.0\,(0.00) \\
 & 200 & 0.0\,(0.00) & 0.0\,(0.00) & 0.0\,(0.00) & 0.0\,(0.00) & 0.0\,(0.00) \\
 & 400 & 0.0\,(0.00) & 0.0\,(0.00) & 0.0\,(0.00) & 0.0\,(0.00) & 0.0\,(0.00) \\
 & 800 & 0.0\,(0.00) & 0.0\,(0.00) & 0.0\,(0.00) & 0.0\,(0.00) & 0.0\,(0.00) \\
\addlinespace[2.2pt]
5 & 100 & 0.0\,(0.00) & 0.0\,(0.00) & 0.0\,(0.00) & 0.0\,(0.00) & 0.0\,(0.00) \\
 & 200 & 0.0\,(0.00) & 0.0\,(0.00) & 0.0\,(0.00) & 0.0\,(0.00) & 0.0\,(0.00) \\
 & 400 & 0.0\,(0.00) & 0.0\,(0.00) & 0.0\,(0.00) & 0.0\,(0.00) & 0.0\,(0.00) \\
 & 800 & 0.0\,(0.00) & 0.0\,(0.00) & 0.0\,(0.00) & 0.0\,(0.00) & 0.0\,(0.00) \\
\addlinespace[2.2pt]
10 & 100 & 0.0\,(0.00) & 0.0\,(0.00) & 0.0\,(0.00) & 0.0\,(0.00) & 0.0\,(0.00) \\
 & 200 & 0.0\,(0.00) & 0.0\,(0.00) & 0.0\,(0.00) & 0.0\,(0.00) & 0.0\,(0.00) \\
 & 400 & 0.0\,(0.00) & 0.0\,(0.00) & 0.0\,(0.00) & 0.0\,(0.00) & 0.0\,(0.00) \\
 & 800 & 0.0\,(0.00) & 0.0\,(0.00) & 0.0\,(0.00) & 0.0\,(0.00) & 0.0\,(0.00) \\
\addlinespace[2.2pt]
20 & 100 & 0.0\,(0.00) & 0.0\,(0.00) & 0.0\,(0.00) & 0.0\,(0.00) & 0.0\,(0.00) \\
 & 200 & 0.0\,(0.00) & 0.0\,(0.00) & 0.0\,(0.00) & 0.0\,(0.00) & 0.0\,(0.00) \\
 & 400 & 0.0\,(0.00) & 0.0\,(0.00) & 0.0\,(0.00) & 0.0\,(0.00) & 0.0\,(0.00) \\
 & 800 & 0.0\,(0.00) & 0.0\,(0.00) & 0.0\,(0.00) & 0.0\,(0.00) & 0.0\,(0.00) \\
\addlinespace[2.2pt]
50 & 100 & 0.0\,(0.00) & 0.0\,(0.00) & 0.0\,(0.00) & 0.0\,(0.00) & 0.5\,(0.50) \\
 & 200 & 0.0\,(0.00) & 0.0\,(0.00) & 0.0\,(0.00) & 0.0\,(0.00) & 0.0\,(0.00) \\
 & 400 & 0.0\,(0.00) & 0.0\,(0.00) & 0.0\,(0.00) & 0.0\,(0.00) & 0.0\,(0.00) \\
 & 800 & 0.0\,(0.00) & 0.0\,(0.00) & 0.0\,(0.00) & 0.0\,(0.00) & 0.0\,(0.00) \\
\midrule
\multicolumn{7}{@{}l}{\rule{0pt}{13pt}\textbf{(b) Loading MSE ratio ($P_\perp/\Pi_\perp$)}} \\[2pt]
1 & 100 & 0.997\,(0.001) & 0.999\,(0.001) & 1.000\,(0.001) & 1.000\,(0.001) & 1.000\,(0.001) \\
 & 200 & 1.001\,(0.001) & 1.000\,(0.000) & 1.000\,(0.000) & 1.000\,(0.000) & 1.000\,(0.000) \\
 & 400 & 0.999\,(0.000) & 1.000\,(0.000) & 1.000\,(0.000) & 1.000\,(0.000) & 1.000\,(0.000) \\
 & 800 & 1.000\,(0.000) & 1.000\,(0.000) & 1.000\,(0.000) & 1.000\,(0.000) & 1.000\,(0.000) \\
\addlinespace[2.2pt]
3 & 100 & 1.016\,(0.003) & 1.002\,(0.001) & 1.001\,(0.001) & 1.001\,(0.001) & 1.000\,(0.001) \\
 & 200 & 1.004\,(0.001) & 1.001\,(0.001) & 1.002\,(0.001) & 1.000\,(0.001) & 1.000\,(0.001) \\
 & 400 & 1.002\,(0.001) & 1.000\,(0.001) & 0.999\,(0.001) & 1.000\,(0.001) & 1.000\,(0.001) \\
 & 800 & 1.000\,(0.001) & 1.000\,(0.001) & 0.999\,(0.001) & 1.000\,(0.001) & 0.999\,(0.000) \\
\addlinespace[2.2pt]
5 & 100 & 1.013\,(0.003) & 1.008\,(0.002) & 1.002\,(0.002) & 1.000\,(0.001) & 0.999\,(0.001) \\
 & 200 & 1.014\,(0.002) & 1.001\,(0.002) & 1.001\,(0.001) & 1.000\,(0.001) & 0.998\,(0.001) \\
 & 400 & 1.003\,(0.002) & 1.000\,(0.001) & 1.000\,(0.001) & 0.999\,(0.001) & 1.001\,(0.001) \\
 & 800 & 1.000\,(0.001) & 1.000\,(0.001) & 1.000\,(0.001) & 1.000\,(0.001) & 1.000\,(0.001) \\
\addlinespace[2.2pt]
10 & 100 & 1.009\,(0.005) & 1.005\,(0.003) & 1.000\,(0.003) & 1.002\,(0.002) & 1.002\,(0.003) \\
 & 200 & 1.013\,(0.003) & 1.006\,(0.002) & 1.001\,(0.002) & 0.997\,(0.002) & 1.003\,(0.002) \\
 & 400 & 0.997\,(0.003) & 1.002\,(0.002) & 0.999\,(0.001) & 0.999\,(0.001) & 1.003\,(0.002) \\
 & 800 & 0.998\,(0.002) & 1.001\,(0.001) & 0.999\,(0.001) & 1.000\,(0.001) & 1.001\,(0.001) \\
\addlinespace[2.2pt]
20 & 100 & 0.963\,(0.015) & 1.009\,(0.004) & 1.006\,(0.003) & 1.007\,(0.003) & 1.008\,(0.005) \\
 & 200 & 0.998\,(0.008) & 1.002\,(0.003) & 1.004\,(0.003) & 1.006\,(0.003) & 1.001\,(0.003) \\
 & 400 & 1.000\,(0.003) & 0.999\,(0.002) & 1.002\,(0.002) & 0.996\,(0.002) & 1.000\,(0.002) \\
 & 800 & 1.001\,(0.003) & 1.003\,(0.002) & 0.999\,(0.002) & 1.002\,(0.001) & 1.002\,(0.002) \\
\addlinespace[2.2pt]
50 & 100 & 1.019\,(0.051) & 0.998\,(0.006) & 1.013\,(0.006) & 1.033\,(0.010) & 1.029\,(0.008) \\
 & 200 & 0.998\,(0.009) & 1.009\,(0.005) & 1.010\,(0.005) & 1.014\,(0.005) & 1.016\,(0.007) \\
 & 400 & 1.008\,(0.005) & 1.002\,(0.004) & 1.009\,(0.004) & 1.005\,(0.004) & 1.019\,(0.005) \\
 & 800 & 1.002\,(0.005) & 1.007\,(0.004) & 1.005\,(0.004) & 1.006\,(0.003) & 1.009\,(0.004) \\
\bottomrule
\end{tabular*}
\endgroup
\end{table}
\clearpage

\endgroup

\include{p_vs_pi_tables_all}

\subsection{Additional diagnostics for the simulation study in Section~\ref{subsec:endor-simulation-analysis}}
\label{subsec:endor-simulation-additional-results}
\medskip
\noindent\textit{Parameter settings.}
We use the simulation model described in Section~\ref{subsec:endor-simulation-analysis}, with \(p=70\), \(\operatorname{tr}(\Sigma^{(0)})=1\), and baseline parameters \((\mathrm{CN},d_\kappa,a,\lambda_\phi,\eta)=(6,0.9,45^\circ,0,50)\). We vary one parameter at a time over the following grids, keeping the remaining parameters at their baseline values:
\[
\begin{aligned}
\mathrm{CN} &\in \{1,2,5,6,10,20,32,50\},\\
d_\kappa &\in \{0.01,0.05,0.25,0.45,0.65,0.85,0.9,0.95,0.97,0.995\},\\
a &\in \{5^\circ,10^\circ,15^\circ,30^\circ,45^\circ,60^\circ,90^\circ,125^\circ,180^\circ,240^\circ,270^\circ,300^\circ,330^\circ,360^\circ\},\\
\lambda_\phi &\in \{-1,-0.8,-0.6,-0.4,-0.2,0,0.2,0.4,0.6,0.8,1\},\\
\eta &\in \{5,10,20,50,100,150,500\}.
\end{aligned}
\]
Counting the shared baseline only once gives $46$ distinct parameter configurations. Each configuration is evaluated at \(n\in\{100,200,400,800\}\), giving $184$ conditions, with \(R=200\) independent datasets per condition. The confidence-interval comparison uses \(\eta\in\{5,10,20,50\}\) at all four sample sizes, with the remaining parameters fixed at baseline. The bootstrap procedures use \(B=1000\) replications per stage.

\medskip
\noindent\textit{Covariance bias.}
We examine covariance estimation using the same Monte Carlo datasets and parameter settings as in Section~\ref{subsec:endor-simulation-analysis}. For the iterated \texttt{plug-in} estimator, we use \(\widehat\Sigma^{[s]}=\mathcal M(\widehat{[\kappa]}^{[s]})\), \(s=1,\ldots,4\).

\begin{figure}[p]
\centering
\includegraphics[width=\linewidth,height=0.7\textheight]{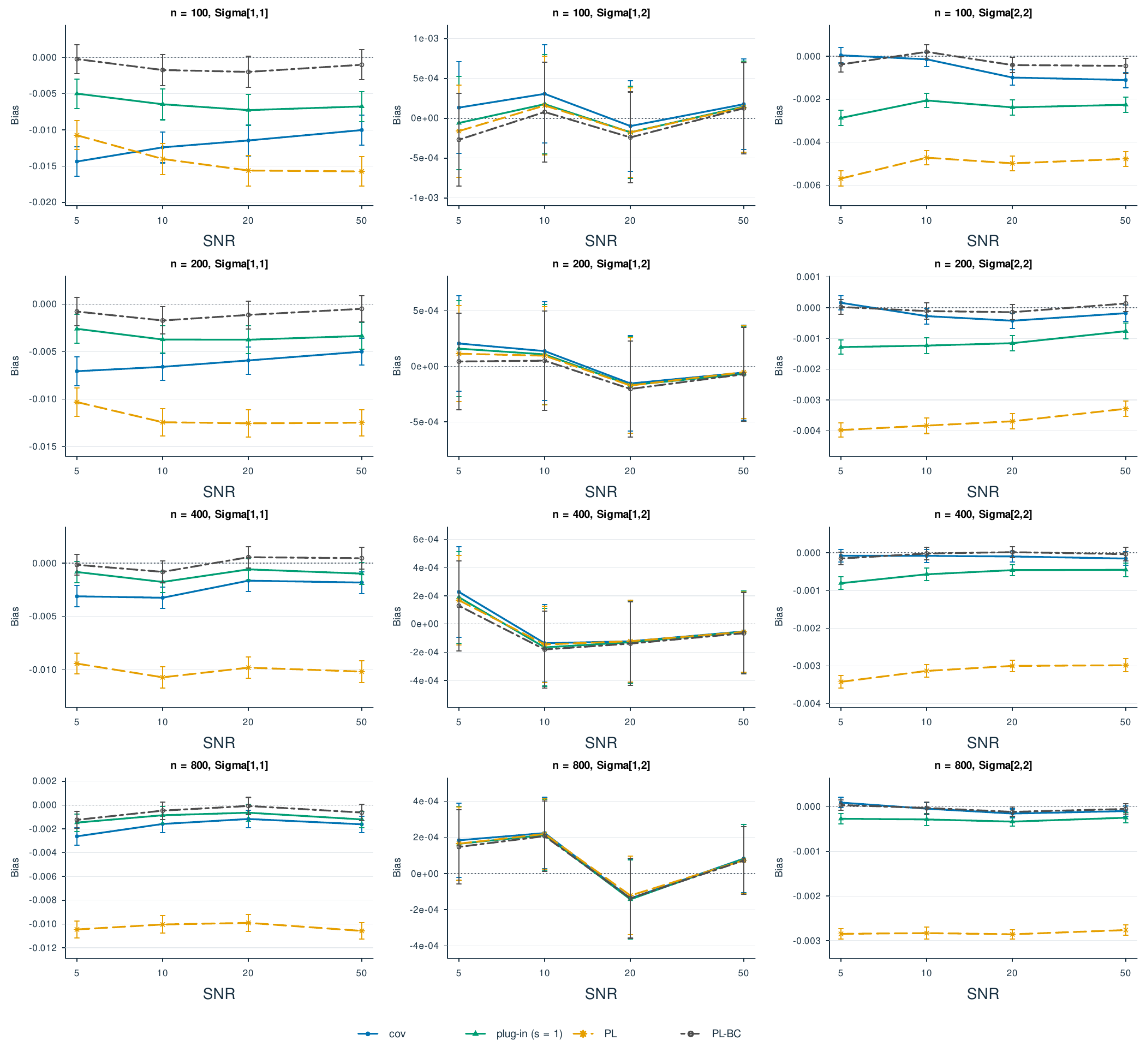}
\caption{Empirical covariance bias under the main-text simulation settings. Rows correspond to \(n=100,200,400,800\), and columns to entries \((1,1)\), \((1,2)\), and \((2,2)\). Each panel varies \(\eta\in\{5, 10, 20, 50\}\), with the remaining parameters at their baseline values. Points show empirical biases across \(R=200\) independent datasets, and error bars show \(\pm1.96\) MCSE. Horizontal dotted lines indicate zero bias.}
\label{fig:endor-supp-covariance-bias}
\end{figure}

Figure~\ref{fig:endor-supp-covariance-bias} shows the empirical biases at \(\eta\in\{5,10, 20, 50\}\), with the remaining parameters at their baseline values. The diagonal entries of \(\widehat\Sigma^{\texttt{PL}}\) exhibit downward bias that remains visible at \(n=800\). This pattern is consistent with the joint inconsistency result in Theorem~\ref{thm: PL strong inconsistency}. Bootstrap correction substantially reduces the diagonal bias, although small residual empirical biases remain. 

The \texttt{cov} and \texttt{plug-in} estimators have different finite-sample bias patterns, as described in Theorems~\ref{thm:cov-noise-covariance-expansion} and \ref{thm:iterated-plugin-noise-covariance-expansion}. In the displayed settings, the first loading update reduces the magnitude of the bias in the \((1,1)\) entry but generally increases it in the \((2,2)\) entry. Over the sample sizes shown, the empirical biases approximately scale as $n^{-1}$. This pattern agrees with the scale of the second-order stochastic terms in Theorems~\ref{thm:cov-noise-covariance-expansion} and \ref{thm:iterated-plugin-noise-covariance-expansion}. The effect of SNR is also component-specific, so lower SNR does not uniformly increase covariance bias.

\medskip
\noindent\textit{Numerical comparison of cov and one-step plug-in.}
We extend the numerical comparison to all 184 main-text conditions, consisting of 46 parameter settings at each sample size, with \(R=200\) datasets per condition. Define the empirical componentwise and trace biases by
\[
\widehat b_{jk}^{(m)}=\frac{1}{R}\sum_{r=1}^{R}\bigl(\widehat\Sigma_{jk}^{(m,r)}-\Sigma_{jk}^{(0)}\bigr), \qquad \widehat b_{\mathrm{tr}}^{(m)}=\frac{1}{R}\sum_{r=1}^{R}\operatorname{tr}\bigl(\widehat\Sigma^{(m,r)}-\Sigma^{(0)}\bigr).
\]

Table~\ref{tab:endor-covariance-bias-improvement} compares \texttt{cov} with \texttt{plug-in} at $s=1$. The latter has smaller absolute trace bias in 182 of the 184 conditions (98.9\%), including all 180 conditions with $\mathrm{CN}>1$. The two exceptions have $\mathrm{CN}=1$, at $n=100$ and $400$; their differences in absolute trace bias are only $2.49\times10^{-7}$ and $1.57\times10^{-8}$, both smaller than one paired MCSE. The componentwise results differ: absolute bias is smaller in 183 conditions for the $(1,1)$ entry but in only 16 conditions for the $(2,2)$ entry. The trace comparison therefore concerns the bias in estimating the total noise variance, rather than improvement in every covariance entry.

Table~\ref{tab:endor-covariance-trace-bias} examines the sample-size dependence of the paired trace-bias difference. At \(\eta=50\), \(n(\widehat b_{\mathrm{tr}}^{[1]}-\widehat b_{\mathrm{tr}}^{\texttt{cov}})\) equals \(0.211,0.216,0.216,0.209\) for \(n=100,200,400,800\), respectively. Averaging over all 46 parameter settings gives $0.293,0.290,0.291,0.291$ for $n=100,200,400,800$, respectively. These results support an approximately \(n^{-1}\)-order difference. The empirical covariance biases are numerically very similar for \(s=1,2,3,4\), with only small changes after the first plug-in step, consistent with Theorem~\ref{thm:iterated-plugin-noise-covariance-expansion}.

\begin{table}[tbp]
\centering
\small
\setlength{\tabcolsep}{5pt}
\caption{Main-text conditions in which \texttt{plug-in} with $s=1$ has smaller absolute empirical covariance bias than \texttt{cov}. Entries give the number and percentage of the 184 conditions, with $R=200$ independent datasets per condition. The comparison is $|\widehat b_{jk}^{[1]}|<|\widehat b_{jk}^{\mathrm{cov}}|$ for each covariance entry and $|\widehat b_{\mathrm{tr}}^{[1]}|<|\widehat b_{\mathrm{tr}}^{\mathrm{cov}}|$ for the trace.}
\label{tab:endor-covariance-bias-improvement}
\begin{tabular}{@{}lrrrr@{}}
\toprule
Method & \((1,1)\) & \((1,2)\) & \((2,2)\) & Trace \\
\midrule
\texttt{plug-in}, $s=1$ & 183 (99.5\%) & 92 (50.0\%) & 16 (8.7\%) & 182 (98.9\%) \\
\bottomrule
\end{tabular}
\end{table}

\begin{table}[tbp]
\centering
\small
\setlength{\tabcolsep}{5pt}
\caption{Trace bias of \texttt{cov} and \texttt{plug-in} as a function of sample size and SNR, with the other parameters fixed at their baseline values. The two method columns report \(10^3\widehat b_{\mathrm{tr}}^{(m)}\pm10^3\widehat{\operatorname{MCSE}}(\widehat b_{\mathrm{tr}}^{(m)})\). The last column reports \(n\widehat D_{\mathrm{tr}}\) followed by its paired MCSE, where \(\widehat D_{\mathrm{tr}}=\widehat b_{\mathrm{tr}}^{[1]}-\widehat b_{\mathrm{tr}}^{\texttt{cov}}\); this column is not multiplied by \(10^3\). All MCSEs are computed across \(R=200\) datasets. The true noise trace is one.}
\label{tab:endor-covariance-trace-bias}
\begin{tabular}{@{}rrccc@{}}
\toprule
\(\eta\) & \(n\) & \texttt{cov} & \texttt{plug-in}, \(s=1\) & \(n\widehat D_{\mathrm{tr}}\) \\
\midrule
50 & 100 & $-11.129 \pm 1.070$ & $-9.020 \pm 1.073$ & $+0.211 \pm 0.004$ \\
 & 200 & $-5.175 \pm 0.690$ & $-4.098 \pm 0.690$ & $+0.216 \pm 0.003$ \\
 & 400 & $-1.977 \pm 0.546$ & $-1.438 \pm 0.547$ & $+0.216 \pm 0.004$ \\
 & 800 & $-1.711 \pm 0.364$ & $-1.449 \pm 0.364$ & $+0.209 \pm 0.004$ \\
\midrule
20 & 100 & $-12.450 \pm 1.114$ & $-9.647 \pm 1.119$ & $+0.280 \pm 0.005$ \\
 & 200 & $-6.343 \pm 0.740$ & $-4.890 \pm 0.744$ & $+0.291 \pm 0.006$ \\
 & 400 & $-1.744 \pm 0.521$ & $-1.047 \pm 0.521$ & $+0.279 \pm 0.005$ \\
 & 800 & $-1.320 \pm 0.374$ & $-0.973 \pm 0.375$ & $+0.277 \pm 0.006$ \\
\midrule
10 & 100 & $-12.552 \pm 1.107$ & $-8.517 \pm 1.101$ & $+0.404 \pm 0.008$ \\
 & 200 & $-6.868 \pm 0.716$ & $-4.945 \pm 0.716$ & $+0.385 \pm 0.007$ \\
 & 400 & $-3.330 \pm 0.522$ & $-2.339 \pm 0.523$ & $+0.396 \pm 0.008$ \\
 & 800 & $-1.628 \pm 0.372$ & $-1.137 \pm 0.374$ & $+0.393 \pm 0.007$ \\
\midrule
5 & 100 & $-14.327 \pm 1.048$ & $-7.861 \pm 1.051$ & $+0.647 \pm 0.012$ \\
 & 200 & $-6.896 \pm 0.778$ & $-3.870 \pm 0.782$ & $+0.605 \pm 0.013$ \\
 & 400 & $-3.190 \pm 0.521$ & $-1.641 \pm 0.521$ & $+0.620 \pm 0.013$ \\
 & 800 & $-2.531 \pm 0.376$ & $-1.742 \pm 0.377$ & $+0.632 \pm 0.014$ \\
\bottomrule
\end{tabular}
\end{table}

\medskip
\noindent\textit{Persistence of PL trace bias.}
Table~\ref{tab:endor-pl-covariance-bias} quantifies the persistent bias of the PL noise-trace estimate. At \(\eta=50\), the empirical trace bias is approximately \(-0.01316\) at \(n=400\) and \(-0.01334\) at \(n=800\), with MCSEs \(0.00054\) and \(0.00036\), respectively. 

Across all 46 main-text parameter settings at $n=800$, the empirical PL trace bias is negative in every setting, ranging from $-0.01557$ to $-0.00410$, with a median of $-0.01274$. Thus, negative bias remains visible over the sample sizes considered; these finite-sample results alone do not identify its limiting value.

\begin{table}[tbp]
\centering
\small
\setlength{\tabcolsep}{5pt}
\caption{Persistent negative trace bias of \texttt{PL} as sample size increases. Entries report \(10^3\widehat b_{\mathrm{tr}}^{\texttt{PL}}\pm10^3\widehat{\operatorname{MCSE}}(\widehat b_{\mathrm{tr}}^{\texttt{PL}})\), computed across \(R=200\) datasets. Columns vary the SNR, with all remaining parameters at their baseline values. The true noise trace is one.}
\label{tab:endor-pl-covariance-bias}
\begin{tabular}{@{}rcccc@{}}
\toprule
\(n\) & \(\eta=50\) & \(\eta=20\) & \(\eta=10\) & \(\eta=5\) \\
\midrule
100 & $-20.505 \pm 1.050$ & $-20.591 \pm 1.091$ & $-18.729 \pm 1.106$ & $-16.429 \pm 1.030$ \\
200 & $-15.768 \pm 0.683$ & $-16.238 \pm 0.745$ & $-16.265 \pm 0.710$ & $-14.286 \pm 0.776$ \\
400 & $-13.161 \pm 0.537$ & $-12.805 \pm 0.511$ & $-13.849 \pm 0.518$ & $-12.851 \pm 0.510$ \\
800 & $-13.341 \pm 0.357$ & $-12.761 \pm 0.365$ & $-12.869 \pm 0.371$ & $-13.304 \pm 0.372$ \\
\bottomrule
\end{tabular}
\end{table}

\begin{figure}[tbp]
\centering
\includegraphics[width=\linewidth,height=0.7\textheight]{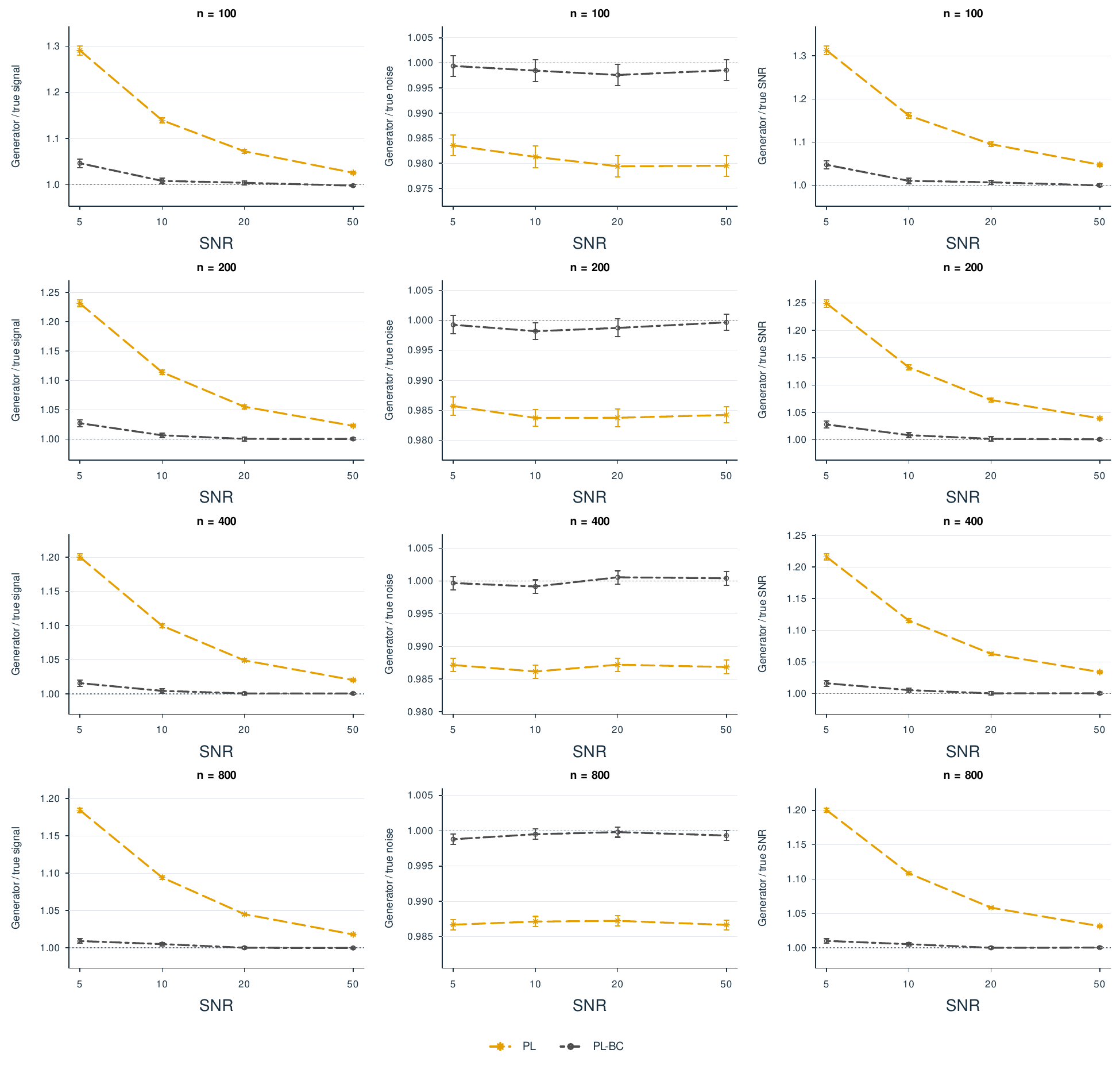}
\caption{Bootstrap generating parameters under the same simulation settings as Figure~\ref{fig:endor-supp-covariance-bias}. Rows correspond to \(n=100,200,400,800\). Columns show the signal energy ratio \(Q_\phi\), noise trace ratio \(Q_\Sigma\), and SNR ratio \(Q_\eta\). Ratios are calculated within each dataset and then averaged across \(R=200\) datasets. Error bars show \(\pm1.96\) MCSE, and horizontal dotted lines indicate one.}
\label{fig:endor-supp-signal-strength}
\end{figure}

\medskip
\noindent\textit{Signal strength overestimation in \texttt{PL} and \texttt{PL-BC}.}
Figure~\ref{fig:endor-supp-signal-strength} examines the signal, noise, and SNR used to generate bootstrap samples. For each method and dataset, we calculate
\[
Q_\phi=\frac{\sum_{i=1}^{n}|\widehat\phi_i|^2}{\sum_{i=1}^{n}|\phi_i|^2},
\quad
Q_\Sigma=\frac{\operatorname{tr}(\widehat\Sigma)}{\operatorname{tr}(\Sigma^{(0)})},
\quad
Q_\eta=\frac{Q_\phi}{Q_\Sigma}.
\]
For \texttt{PL-BC}, \(\widehat\phi_i\) and \(\widehat\Sigma\) are the corrected parameters used in the final bootstrap stage. The signal denominator is the realized score energy in each simulated dataset.

The \texttt{PL} procedure overestimates signal energy and underestimates the noise trace, producing an inflated SNR in the bootstrap generating model. The relative signal overestimation becomes substantially larger as \(\eta\) decreases and remains visible at the largest sample size. Bootstrap correction largely restores the noise trace and greatly reduces signal inflation. However, at low SNR, the corrected score energy still exceeds the true energy on average. Consequently, an inflated bootstrap SNR can remain even when covariance bias has been substantially corrected. This pattern is consistent with the residual undercoverage of \texttt{PL-BC} reported in the main text.

\medskip
\noindent\textit{Algorithm stability.} For \texttt{PL}, we implement the alternating conditional updates of \citet[SI Appendix, Section~4.1]{pokern2021statistical}, including a covariance update in each score--loading cycle. In our simulations, both the fixed-covariance and joint \texttt{PL} solvers use the stopping criteria and tolerances specified in Algorithm~\ref{supp:alg:fixed-covariance-loading}. These numerical stopping criteria differ from those in the authors' released implementation.

Across the 184 main-text simulation conditions, all fits for each method satisfied the stopping criteria within the iteration limit, with no recorded numerical fit failures. As shown in Table~\ref{tab:endor-algorithm-stability}, the mean iteration count is 9.4--9.5 for the individual \texttt{plug-in} steps and 9.6 for PL. The largest observed count is 103, below the limit of 1000.

\medskip
\noindent\textit{Computational cost.}
Tables~\ref{tab:endor-supp-fit-time-snr} and~\ref{tab:endor-supp-ci-time-snr} extend the baseline timing comparison in Table~\ref{tab:endor-computation-time} to other SNR values, with the remaining parameters fixed at their baseline values. Point estimation and confidence-interval construction are reported separately. Across the displayed settings, asymptotic interval construction for \texttt{cov} and one-step \texttt{plug-in} takes milliseconds, whereas the bootstrap procedures for \texttt{PL} and \texttt{PL-BC} take tens to hundreds of seconds with \(B=1000\) replications per stage.

\begin{table}[tbp]
\centering
\caption{Inner iteration counts across the 184 main-text simulation conditions, with $R=200$ datasets per condition (36,800 fits per method). For plug-in step $s$, counts refer only to the fixed-covariance loading fit at that step and exclude preceding steps. The final score--loading--score pass and bootstrap refits are excluded.}
\label{tab:endor-algorithm-stability}
\begin{tabular}{lrrr}
\toprule
Method & Mean & 95th percentile & Maximum \\
\midrule
\texttt{plug-in}, $s=1$ & 9.4 & 19 & 90 \\
\texttt{plug-in}, $s=2$ & 9.5 & 20 & 97 \\
\texttt{plug-in}, $s=3$ & 9.5 & 20 & 97 \\
\texttt{plug-in}, $s=4$ & 9.5 & 20 & 97 \\
PL & 9.6 & 20 & 103 \\
\bottomrule
\end{tabular}
\end{table}

\begin{table}[p]
\centering
\caption{Point-estimation times in milliseconds (mean \(\pm\) one Monte Carlo standard error over \(R=200\) datasets) across the full SNR sweep, with all remaining parameters at their baseline values. \texttt{Plug-in} timings include \texttt{cov} initialization and all updates through step \(s\). Bias correction and interval construction are excluded. }
\label{tab:endor-supp-fit-time-snr}
\begingroup
\scriptsize
\setlength{\tabcolsep}{1.5pt}
\renewcommand{\arraystretch}{1.0}
\begin{tabular}{@{}crrrrrrr@{}}
\toprule
\multirow{2}{*}{$\eta$} & \multirow{2}{*}{$n$} & \multirow{2}{*}{\texttt{cov}} & \multicolumn{4}{c}{\texttt{plug-in}} & \multirow{2}{*}{\texttt{PL}} \\
\cmidrule(lr){4-7}
& & & $s=1$ & $s=2$ & $s=3$ & $s=4$ & \\
\midrule
500 & 100 & $3.70\pm0.06$ & $8.21\pm0.11$ & $12.69\pm0.15$ & $17.13\pm0.20$ & $21.50\pm0.24$ & $97.10\pm0.82$ \\
 & 200 & $3.88\pm0.04$ & $20.79\pm1.32$ & $49.23\pm3.42$ & $91.53\pm5.38$ & $102.52\pm5.57$ & $98.07\pm0.61$ \\
 & 400 & $4.49\pm0.05$ & $69.52\pm4.85$ & $95.51\pm5.09$ & $110.08\pm5.07$ & $122.30\pm5.12$ & $108.83\pm0.98$ \\
 & 800 & $25.19\pm2.32$ & $196.05\pm6.22$ & $231.14\pm6.33$ & $268.37\pm6.30$ & $304.19\pm6.45$ & $178.98\pm2.80$ \\
\midrule
150 & 100 & $3.65\pm0.05$ & $8.34\pm0.09$ & $12.93\pm0.14$ & $17.45\pm0.18$ & $21.99\pm0.22$ & $96.98\pm0.80$ \\
 & 200 & $3.83\pm0.04$ & $24.74\pm1.61$ & $59.61\pm4.21$ & $97.14\pm5.68$ & $104.31\pm5.70$ & $98.03\pm0.67$ \\
 & 400 & $4.56\pm0.05$ & $56.85\pm3.62$ & $115.50\pm5.64$ & $131.24\pm5.66$ & $142.99\pm5.71$ & $108.57\pm0.91$ \\
 & 800 & $19.67\pm2.33$ & $191.34\pm6.59$ & $227.06\pm6.72$ & $264.63\pm6.60$ & $302.39\pm6.88$ & $178.85\pm2.57$ \\
\midrule
100 & 100 & $3.68\pm0.06$ & $8.51\pm0.10$ & $13.23\pm0.15$ & $17.86\pm0.19$ & $22.51\pm0.24$ & $98.44\pm0.91$ \\
 & 200 & $4.14\pm0.17$ & $34.56\pm2.58$ & $74.40\pm5.15$ & $94.25\pm5.22$ & $101.70\pm5.25$ & $98.82\pm0.60$ \\
 & 400 & $4.59\pm0.05$ & $71.51\pm5.07$ & $96.47\pm5.12$ & $111.90\pm5.11$ & $125.71\pm5.14$ & $110.15\pm0.96$ \\
 & 800 & $21.69\pm2.13$ & $186.59\pm6.44$ & $222.75\pm6.58$ & $261.77\pm6.71$ & $300.63\pm6.98$ & $178.86\pm2.44$ \\
\midrule
50 & 100 & $3.65\pm0.06$ & $8.83\pm0.11$ & $13.93\pm0.16$ & $18.96\pm0.21$ & $28.30\pm0.50$ & $99.19\pm0.81$ \\
 & 200 & $5.17\pm0.34$ & $35.30\pm2.33$ & $83.09\pm5.58$ & $98.91\pm5.46$ & $107.23\pm5.51$ & $100.16\pm0.61$ \\
 & 400 & $4.63\pm0.06$ & $74.27\pm5.26$ & $94.68\pm5.21$ & $110.20\pm5.14$ & $126.29\pm5.20$ & $110.28\pm1.05$ \\
 & 800 & $28.67\pm1.74$ & $226.28\pm6.86$ & $265.73\pm6.88$ & $306.28\pm7.23$ & $346.29\pm7.36$ & $183.44\pm2.87$ \\
\midrule
20 & 100 & $3.66\pm0.05$ & $9.90\pm0.11$ & $15.96\pm0.17$ & $27.75\pm0.58$ & $36.92\pm0.53$ & $99.24\pm0.82$ \\
 & 200 & $4.17\pm0.15$ & $35.76\pm2.40$ & $94.03\pm5.58$ & $103.66\pm5.68$ & $115.63\pm5.72$ & $101.29\pm0.59$ \\
 & 400 & $5.43\pm0.30$ & $111.83\pm5.51$ & $129.67\pm5.60$ & $147.56\pm5.71$ & $164.32\pm5.73$ & $113.23\pm0.91$ \\
 & 800 & $24.12\pm2.67$ & $210.32\pm7.81$ & $258.02\pm8.16$ & $305.69\pm8.39$ & $356.90\pm8.77$ & $193.44\pm3.09$ \\
\midrule
10 & 100 & $3.68\pm0.06$ & $11.99\pm0.13$ & $23.45\pm0.43$ & $38.07\pm0.80$ & $47.47\pm0.89$ & $103.90\pm1.00$ \\
 & 200 & $3.98\pm0.04$ & $25.77\pm2.19$ & $55.20\pm3.83$ & $70.50\pm3.85$ & $84.17\pm3.85$ & $109.01\pm0.68$ \\
 & 400 & $4.62\pm0.06$ & $39.54\pm2.66$ & $62.57\pm2.85$ & $85.34\pm2.92$ & $107.04\pm2.97$ & $121.71\pm0.95$ \\
 & 800 & $57.11\pm4.38$ & $207.78\pm7.03$ & $267.71\pm7.13$ & $328.05\pm7.27$ & $387.90\pm7.42$ & $210.00\pm2.92$ \\
\midrule
5 & 100 & $3.72\pm0.06$ & $19.07\pm0.45$ & $40.57\pm0.71$ & $58.08\pm1.04$ & $74.95\pm1.46$ & $118.56\pm1.19$ \\
 & 200 & $4.10\pm0.06$ & $93.42\pm5.50$ & $116.36\pm5.45$ & $137.94\pm5.43$ & $160.21\pm5.44$ & $123.17\pm1.06$ \\
 & 400 & $4.77\pm0.07$ & $115.17\pm5.21$ & $150.02\pm5.14$ & $182.27\pm5.10$ & $214.64\pm5.06$ & $139.45\pm1.40$ \\
 & 800 & $64.86\pm3.92$ & $256.61\pm6.57$ & $342.14\pm6.85$ & $426.16\pm7.30$ & $512.17\pm8.35$ & $246.80\pm3.37$ \\
\bottomrule
\end{tabular}
\endgroup
\end{table}

\begin{table}[p]
\centering
\caption{Confidence-interval construction times in seconds (mean \(\pm\) one Monte Carlo standard error over \(R=200\) datasets), with all parameters other than SNR at their baseline values. Original fitting is excluded. Bootstrap timings sum data generation, refitting, and alignment over \(B=1000\) draws for \texttt{PL} and \(3B\) draws for \texttt{PL-BC}, including its two correction stages. }
\label{tab:endor-supp-ci-time-snr}
\begingroup
\scriptsize
\setlength{\tabcolsep}{3pt}
\renewcommand{\arraystretch}{1.0}
\begin{tabular}{@{}crrrrr@{}}
\toprule
$\eta$ & $n$ & \texttt{cov} & \texttt{plug-in} ($s=1$) & \texttt{PL} & \texttt{PL-BC} \\
\midrule
50 & 100 & $0.00278\pm0.00004$ & $0.00273\pm0.00004$ & $83.67\pm1.13$ & $246.00\pm3.03$ \\
 & 200 & $0.00280\pm0.00004$ & $0.00265\pm0.00004$ & $101.50\pm2.01$ & $297.49\pm5.80$ \\
 & 400 & $0.00279\pm0.00004$ & $0.00275\pm0.00004$ & $106.92\pm1.28$ & $315.37\pm3.39$ \\
 & 800 & $0.00698\pm0.00239$ & $0.00453\pm0.00007$ & $161.09\pm2.32$ & $472.90\pm6.23$ \\
\midrule
20 & 100 & $0.00280\pm0.00004$ & $0.00269\pm0.00004$ & $68.66\pm0.47$ & $197.56\pm1.65$ \\
 & 200 & $0.00539\pm0.00006$ & $0.00272\pm0.00006$ & $77.32\pm0.59$ & $227.41\pm1.27$ \\
 & 400 & $0.00274\pm0.00004$ & $0.00280\pm0.00004$ & $110.72\pm2.44$ & $317.46\pm5.97$ \\
 & 800 & $0.00743\pm0.00269$ & $0.01069\pm0.00118$ & $166.29\pm4.12$ & $493.52\pm11.83$ \\
\midrule
10 & 100 & $0.00315\pm0.00008$ & $0.00407\pm0.00012$ & $88.31\pm1.15$ & $262.76\pm3.44$ \\
 & 200 & $0.00271\pm0.00004$ & $0.00262\pm0.00004$ & $111.13\pm2.07$ & $330.73\pm6.18$ \\
 & 400 & $0.00286\pm0.00004$ & $0.00279\pm0.00003$ & $130.22\pm1.65$ & $389.41\pm4.60$ \\
 & 800 & $0.00750\pm0.00290$ & $0.00909\pm0.00045$ & $186.57\pm2.75$ & $556.67\pm7.66$ \\
\midrule
5 & 100 & $0.00292\pm0.00007$ & $0.00289\pm0.00007$ & $93.86\pm1.04$ & $283.31\pm3.20$ \\
 & 200 & $0.00298\pm0.00006$ & $0.00309\pm0.00008$ & $117.95\pm2.09$ & $357.66\pm6.31$ \\
 & 400 & $0.00304\pm0.00007$ & $0.00294\pm0.00005$ & $138.12\pm1.65$ & $406.20\pm4.36$ \\
 & 800 & $0.00698\pm0.00237$ & $0.00443\pm0.00007$ & $209.47\pm3.13$ & $633.13\pm7.83$ \\
\bottomrule
\end{tabular}
\endgroup
\end{table}

\subsection{Bandwidth selection for ENDOR spectrum smoothing in Section~\ref{subsec:pnas-endor}}
\label{supp:subsec:endor-bandwidth-selection}
For each of the two D2 ENDOR datasets, we selected a common bandwidth by leave-one-out cross-validation over frequencies. We used the four distinct maximum-spectrum estimates from \texttt{averaging}, \texttt{cov}, \texttt{plug-in}, and \texttt{PL}. We did not include the \texttt{PL-BC} spectrum separately because it coincides with the \texttt{PL} spectrum estimate. For bandwidth selection only, each real-valued spectrum was centered and scaled to unit Euclidean norm.

At each candidate bandwidth, a Gaussian-weighted local quadratic fit to the remaining frequencies predicted the value at each omitted frequency. We selected the bandwidth that minimized the summed squared prediction error, averaged equally over the four spectra. The numerical search used 81 logarithmically spaced bandwidths between half the median frequency spacing and the full frequency span. These bounds were $0.05$ and $35.9$ MHz for both datasets. We refined each local minimum detected on this grid by bounded optimization in the log bandwidth between the neighboring grid points. We retained the bandwidth with the smallest criterion value among the evaluated candidates.

The selected bandwidths were $0.12507184$ MHz for $g_y$ and $0.15288284$ MHz for $g_x$, both in the interior of the search range. Each bandwidth was then held fixed across methods within its dataset, and smoothing was applied on the original unit complex-loading scale. The additional normalization used for bandwidth selection therefore does not change the scale of the displayed spectra.

\end{document}